\pdfoutput=1
\documentclass[letterpaper]{cernyrep}
\usepackage[T1]{fontenc}
\usepackage{lmodern}
\usepackage{graphicx}
\usepackage{rotating}
\usepackage{amsmath}
\usepackage{amssymb}
\usepackage[dvipsnames]{xcolor}
\usepackage{booktabs}
\usepackage{arydshln}
\usepackage{lscape}
\usepackage{floatpag}
\floatpagestyle{plain}
\usepackage{axodraw}
\usepackage{lineno}
\usepackage{graphpap}
\usepackage{cite}
\usepackage{import}
\usepackage{enumitem}

\usepackage{caption}
\usepackage{subcaption}
\usepackage{siunitx}
\usepackage{commath}
\usepackage{xspace}

\newcommand{\eV}{\ensuremath{\text{e\kern-0.15ex{}V}}\xspace}
\newcommand{\MeV}{\ensuremath{\text{M\eV}}\xspace}
\newcommand{\GeV}{\ensuremath{\text{G\eV}}\xspace}
\newcommand{\TeV}{\ensuremath{\text{T\eV}}\xspace}

\makeatletter
\renewcommand{\thechapter}{\@Roman\c@chapter}
\makeatother

\newcommand{\NLO}[1]{N${}^{#1}$LO}

\newcommand{\NLOH}[1]{N${}^{#1}$LO${}_{\rm HEFT}$}
\newcommand{\NLOQ}[1]{N${}^{#1}$LO${}_{\rm QCD}$}
\newcommand{\NLOE}[1]{N${}^{#1}$LO${}_{\rm EW}$}
\newcommand{\NLOHone}{NLO${}_{\rm HEFT}$}
\newcommand{\NLOQone}{NLO${}_{\rm QCD}$}
\newcommand{\NLOEone}{NLO${}_{\rm EW}$}
\newcommand{\NLOQE}[2]{N${}^{(#1,#2)}$LO${}_{{\rm QCD}\times{\rm EW}}$}
\newcommand{\LOQ}{LO${}_{\rm QCD}$}
\newcommand{\xs}{$\sigma$}
\newcommand{\tb}{\bar{t}}
\newcommand{\bb}{\bar{b}}
\newcommand{\qb}{\bar{q}}

\newcommand{\WishListMGaMC}{\textsc{Madgraph5}\_a\textsc{MC@NLO}\xspace}

\newcommand{\be}{\begin{equation}}
\newcommand{\ee}{\end{equation}}
\newcommand{\bea}{\begin{eqnarray}}
\newcommand{\eea}{\end{eqnarray}}
\newcommand{\bi}{\begin{itemize}}
\newcommand{\ei}{\end{itemize}}
\newcommand{\ben}{\begin{enumerate}}
\newcommand{\een}{\end{enumerate}}
\newcommand{\la}{\left\langle}
\newcommand{\ra}{\right\rangle}

\newcommand{\lp}{\left(}
\newcommand{\rp}{\right)}

\newcommand{\gsim}{\mathrel{\rlap{\lower4pt\hbox{\hskip1pt$\sim$}}
    \raise1pt\hbox{$>$}}}         
\newcommand{\lsim}{\mathrel{\rlap{\lower4pt\hbox{\hskip1pt$\sim$}}
    \raise1pt\hbox{$<$}}}         

\newcommand{\draft}[1]{}

\newcommand{\beq}{\begin{equation}}  
\newcommand{\eeq}{\end{equation}}  

\newcommand{\lapprox}{\lower .7ex\hbox{$\;\stackrel{\textstyle <}{\sim}\;$}}
\newcommand{\gapprox}{\lower .7ex\hbox{$\;\stackrel{\textstyle >}{\sim}\;$}}

\newcommand{\eps}{\epsilon}
\newcommand{\Obs}{O}
\newcommand{\dbox}{\hbox{d}}

\newcommand{\hjetscompGoSam}{G\protect\scalebox{0.8}{O}S\protect\scalebox{0.8}{AM}\xspace}
\newcommand{\hjetscompHej}{H\protect\scalebox{0.8}{EJ}\xspace}
\newcommand{\hjetscompHerwig}{\protect\textsf{Herwig~7.1}\xspace}
\newcommand{\hjetscompHNNLO}{H\protect\scalebox{0.8}{NNLO}\xspace}
\newcommand{\hjetscompHqT}{H\protect\scalebox{0.8}{Q}T\xspace}
\newcommand{\hjetscompResbos}{R\protect\scalebox{0.8}{ES}B\protect\scalebox{0.8}{OS}2\xspace}
\newcommand{\hjetscompMGaMC}{\protect\texttt{Mad\-graph5\_\-aMC@NLO}\xspace}
\newcommand{\hjetscompPowheg}{\protect\texttt{POWHEG}\xspace}
\newcommand{\hjetscompPowhegBox}{\protect\texttt{POWHEG-BOX}\xspace}
\newcommand{\hjetscompPythia}{P\protect\scalebox{0.8}{YTHIA}\xspace}
\newcommand{\hjetscompSherpa}{S\protect\scalebox{0.8}{HERPA}\xspace}

\newcommand{\hjetscompMinlo}{M\protect\scalebox{0.8}{I}N\protect\scalebox{0.8}{LO}\xspace}
\newcommand{\hjetscompLoopsim}{L\protect\scalebox{0.8}{OOP}S\protect\scalebox{0.8}{IM}\xspace}
\newcommand{\hjetscompLOPS}{L\protect\scalebox{0.8}{O}P\protect\scalebox{0.8}{S}\xspace}
\newcommand{\hjetscompNLOPS}{N\protect\scalebox{0.8}{LO}P\protect\scalebox{0.8}{S}\xspace}
\newcommand{\hjetscompNNLOPS}{N\protect\scalebox{0.8}{NLO}P\protect\scalebox{0.8}{S}\xspace}
\newcommand{\hjetscompMCatNLO}{M\protect\scalebox{0.8}{C}@N\protect\scalebox{0.8}{LO}\xspace}
\newcommand{\hjetscompSMCatNLO}{S-M\protect\scalebox{0.8}{C}@N\protect\scalebox{0.8}{LO}\xspace}
\newcommand{\hjetscompMEPSatNLO}{M\protect\scalebox{0.8}{E}P\protect\scalebox{0.8}{S}@N\protect\scalebox{0.8}{LO}\xspace}

\newcommand{\hjetscompgev}{\ensuremath{\text{\:GeV}}\xspace}
\newcommand{\hjetscomptev}{\ensuremath{\text{\:TeV}}\xspace}
\newcommand{\hjetscompantikt}{\ensuremath{\text{anti-}k_\text{T}}\xspace}

\newcommand{\Sherpa}{S\protect\scalebox{0.8}{HERPA}\xspace}
\newcommand{\Powheg}{P\protect\scalebox{0.8}{OWHEG}\xspace}
\newcommand{\CSS}{C\protect\scalebox{0.8}{SS}\xspace}
\newcommand{\Comix}{C\protect\scalebox{0.8}{OMIX}\xspace}
\newcommand{\Amegic}{A\protect\scalebox{0.8}{MEGIC++}\xspace}
\newcommand{\MCatNLO}{M\protect\scalebox{0.8}{C}@N\protect\scalebox{0.8}{LO}\xspace}
\newcommand{\MEPS}{M\scalebox{0.8}{E}P\scalebox{0.8}{S}\xspace}
\newcommand{\MEPSatNLO}{M\scalebox{0.8}{E}P\scalebox{0.8}{S}@N\protect\scalebox{0.8}{LO}\xspace}

\newcommand{\OpenLoops}{O\protect\scalebox{0.8}{PEN}L\protect\scalebox{0.8}{OOPS}\xspace}
\newcommand{\Herwig}{H\protect\scalebox{0.8}{ERWIG}7\xspace}
\newcommand{\Matchbox}{M\protect\scalebox{0.8}{ATCHBOX}\xspace}
\newcommand{\MGaMC}{M\protect\scalebox{0.8}{AD}G\protect\scalebox{0.8}{RAPH}5\_aMC@NLO\xspace}

\newcommand{\MadGraphfour}{M\protect\scalebox{0.8}{AD}G\protect\scalebox{0.8}{RAPH}4\xspace}
\newcommand{\CVolver}{CV\protect\scalebox{0.8}{OLVER}\xspace}
\newcommand{\ColorFull}{C\protect\scalebox{0.8}{OLOR}F\protect\scalebox{0.8}{ULL}\xspace}
\newcommand{\pt}{\ensuremath{p_{T}}\xspace}
\newcommand\sss{\mathchoice%
{\displaystyle}%
{\scriptstyle}%
{\scriptscriptstyle}%
{\scriptscriptstyle}%
}
\newcommand\MSB{\ifmmode {\overline{\rm MS}} \else $\overline{\rm MS}$\fi}
\newcommand\MINLO{{\tt MiNLO}}
\newcommand\muf{\mu_{\sss\rm F}}
\newcommand\mur{\mu_{\sss\rm R}}
\newcommand\KRA{K_{\scriptscriptstyle \rm R}}
\newcommand\KFA{K_{\scriptscriptstyle \rm F}}

\newcommand{\GOSAM}{G\protect\scalebox{0.8}{O}S\protect\scalebox{0.8}{AM}\xspace}
\newcommand{\POWHEGBOX}{P\protect\scalebox{0.8}{OWHEG} B\protect\scalebox{0.8}{OX}\xspace}
\newcommand{\QGRAF}{Q\protect\scalebox{0.8}{GRAF}\xspace}
\newcommand{\FORM}{F\protect\scalebox{0.8}{ORM}\xspace}

\newcommand{\GOLEM}{G\protect\scalebox{0.8}{OLEM}\xspace}
\newcommand{\NINJA}{N\protect\scalebox{0.8}{INJA}\xspace}
\newcommand{\SPINNEY}{S\protect\scalebox{0.8}{PINNEY}\xspace}
\newcommand{\ONELOOP}{O\protect\scalebox{0.8}{NE}LO\protect\scalebox{0.8}{OP}\xspace}
\newcommand{\MCFM}{M\protect\scalebox{0.8}{CFM}\xspace}

\begin{document}

{\centering{\LARGE{\bf{Les Houches 2015: Physics at TeV Colliders\\ Standard Model Working Group Report \par }}}}

\pagenumbering{roman}

\vspace{0.7cm}
\leftline{\bf Conveners}


\noindent \emph{Higgs physics: SM issues} \\
         J. Bendavid (CMS), \,
         M.~Grazzini (Theory), \,
         K. Tackmann (ATLAS), \,
         C.~Williams (Theory) \\
\vspace{0.1cm}

\noindent \emph{SM: Loops and Multilegs}  \\	
         S. Badger (Theory), \,
         A. Denner (Theory), \,
         J. Huston (ATLAS), \,
         J. Thaler (Jets contact) \\
\vspace{0.1cm}

\noindent \emph{Tools and Monte Carlos} \\	
         V. Ciulli (CMS), \,
         R. Frederix (Theory), \,
         M. Sch{\"o}nherr (Theory) \\
\vspace{1.0cm}

{\leftline{\bf{Abstract}}}

\vspace{0.5cm}
This Report summarizes the proceedings of the 2015 Les Houches workshop 
on Physics at TeV Colliders. Session 1 dealt with 
(I) new developments relevant for high precision Standard Model calculations, 
(II) the new PDF4LHC parton distributions, 
(III) issues in the theoretical description of the production of Standard 
Model Higgs bosons and how to relate experimental measurements, 
(IV) a host of phenomenological studies essential for comparing LHC data 
from Run I with theoretical predictions and projections for future 
measurements in Run II, 
and (V) new developments in Monte Carlo event generators.

\vspace{0.5cm}

{\leftline{\bf{Acknowledgements}}}

\vspace{0.5cm}
We would like to thank the organizers (G.~Belanger, F.~Boudjema, P.~Gras,
D.~Guadagnoli, S.~Gascon, J.~P.~Guillet, B. ~Herrmann, S. ~Kraml, G.~H.~de
Monchenault, G.~Moreau, E.~Pilon,  P.~Slavich and D.~Zerwas) and the Les
Houches staff for the stimulating environment always present at Les Houches. We
thank the Labex ENIGMASS for support.



\newpage
{\centering{\bf{Authors\par}}}
\begin{flushleft}
J.~R.~Andersen$^{1}$,
S.~Badger$^{2}$,
K.~Becker$^{3}$,
M.~Bell$^{4}$,
J.~Bellm$^{1}$,
J.~Bendavid$^{5}$,
E.~Bothmann$^{6}$,
R.~Boughezal$^{7}$,
J.~Butterworth$^{4}$,
S.~Carrazza$^{8}$,
M.~Chiesa$^{9}$,
L.~Cieri$^{10}$,
V.~Ciulli$^{11, 12}$,
A.~Denner$^{13}$,
M.~Duehrssen-Debling$^{14}$,
G.~Falmagne$^{15}$,
S.~Forte$^{16}$,
P.~Francavilla$^{17}$,
R.~Frederix$^{18}$,
M.~Freytsis$^{19}$,
J.~Gao$^{7}$,
P.~Gras$^{20}$,
M.~Grazzini$^{10}$,
N.~Greiner$^{10}$,
D.~Grellscheid$^{1}$,
G.~Heinrich$^{21}$,
G.~Hesketh$^{4}$,
S.~H{\"o}che$^{22}$,
L.~Hofer$^{23}$,
T.-J.~Hou$^{24}$,
A.~Huss$^{25}$,
J.~Huston$^{26}$,
J.~Isaacson$^{26}$,
A.~Jueid$^{27}$,
S.~Kallweit$^{28}$,
D.~Kar$^{29}$,
Z.~Kassabov$^{16, 30}$,
V.~Konstantinides$^{4}$,
F.~Krauss$^{1}$,
S.~Kuttimalai$^{1}$,
A.~Lazapoulos$^{25}$,
P.~Lenzi$^{12}$,
Y.~Li$^{31}$,
J.M.~Lindert$^{10}$,
X.~Liu$^{32}$,
G.~Luisoni$^{8}$,
L.~L{\"o}nnblad$^{33}$,
P.~Maierh{\"o}fer$^{34}$,
D.~Ma{\^i}tre$^{1}$,
A.~C.~Marini$^{35}$,
G.~Montagna$^{9, 36}$,
M.~Moretti$^{37}$,
P.~M.~Nadolsky$^{24}$,
G.~Nail$^{38, 39}$,
D.~Napoletano$^{1}$,
O.~Nicrosini$^{9}$,
C.~Oleari$^{40, 41}$,
D.~Pagani$^{42}$,
C.~Pandini$^{43}$,
L.~Perrozzi$^{44}$,
F.~Petriello$^{7, 45}$,
F.~Piccinini$^{9}$,
S.~Pl{\"a}tzer$^{1, 38}$,
I.~Pogrebnyak$^{26}$,
S.~Pozzorini$^{10}$,
S.~Prestel$^{22}$,
C.~Reuschle$^{46}$,
J.~Rojo$^{47}$,
L.~Russo$^{12, 48}$,
P.~Schichtel$^{1}$,
M.~Sch{\"o}nherr$^{10}$,
S.~Schumann$^{6}$,
A.~Si{\'o}dmok$^{8}$,
P.~Skands$^{49}$,
D.~Soper$^{19}$,
G.~Soyez$^{50}$,
P.~Sun$^{26}$,
F.~J.~Tackmann$^{51}$,
K.~Tackmann$^{51}$,
E.~Takasugi$^{21, 52}$,
J.~Thaler$^{35}$,
S.~Uccirati$^{30}$,
U.~Utku$^{4}$,
L.~Viliani$^{11, 12, 14}$,
E.~Vryonidou$^{42}$,
B.~T.~Wang$^{24}$,
B.~Waugh$^{4}$,
M.~A.~Weber$^{8, 52}$,
C.~Williams$^{53}$,
J.~Winter$^{26}$,
K.~P.~Xie$^{26}$,
C.-P.~Yuan$^{26}$,
F.~Yuan$^{54}$,
K.~Zapp$^{8}$,
M.~Zaro$^{55, 56}$,
\end{flushleft}
\begin{itemize}
\item[$^{1}$] Institute for Particle Physics Phenomenology, University of Durham, Durham DH1 3LE, UK
\item[$^{2}$] Higgs Centre for Theoretical Physics, School of Physics and Astronomy, The University of Edinburgh, Edinburgh EH9 3JZ, Scotland, UK
\item[$^{3}$] Department of Physics, Oxford University, Denys Wilkinson Building, Keble Road, Oxford OX1 3RH, UK
\item[$^{4}$] Department of Physics and Astronomy, University College London, Gower Street, London WC1E 6BT, UK
\item[$^{5}$] California Institute of Technology, Pasadena, USA
\item[$^{6}$] II. Physikalisches Institut, Universit\"at G\"ottingen, 37077 G\"ottingen, Germany
\item[$^{7}$] High Energy Physics Division, Argonne National Laboratory, Argonne, IL 60439, USA
\item[$^{8}$] CERN, TH Department, CH--1211 Geneva, Switzerland
\item[$^{9}$] INFN, Sezione di Pavia, Via A. Bassi 6, 27100 Pavia, Italy
\item[$^{10}$] Physik--Institut, Universit\"at Z\"urich, Winterthurerstrasse 190, CH--8057 Z\"urich, Switzerland
\item[$^{11}$] Dipartimento di Fisica e Astronomia, Universit\`a di Firenze, I-50019 Sesto Fiorentino, Florence, Italy
\item[$^{12}$] INFN, Sezione di Firenze, Firenze, Italy
\item[$^{13}$] Universit\"{a}t W\"{u}rzburg, Institut f\"{u}r Theoretische Physik und Astrophysik, D-97074 W\"{u}rz\-burg, Germany
\item[$^{14}$] CERN, EP Department, CH--1211 Geneva, Switzerland
\item[$^{15}$] ENS Cachan, Universit\'e Paris--Saclay, 61 avenue du Pr\'esident Wilson, Cachan, F--94230, France
\item[$^{16}$] TIF Lab, Dipartimento di Fisica, Universit\`a di Milano and INFN, Sezione di Milano, Via Celoria 16, I-20133 Milano, Italy
\item[$^{17}$] Laboratoire de Physique Nucl\'{e}aire et de Hautes Energies, Institut Lagrange de Paris, CNRS, Paris, France
\item[$^{18}$] Physik Department T31, Technische Universit\"at M\"unchen, 85748 Garching, Germany
\item[$^{19}$] Institute of Theoretical Science, University of Oregon, Eugene, OR 97403-5203, USA
\item[$^{20}$] CEA/IRFU Saclay, France
\item[$^{21}$] Max Planck Institute for Physics, F\"ohringer Ring 6, 80805 M\"unchen, Germany
\item[$^{22}$] SLAC National Accelerator Laboratory, Menlo Park, CA 94025, USA
\item[$^{23}$] Department de F\'isica Qu\`antica i Astrof\'isica (FQA), Institut de Ci\`encies del Cosmos (ICCUB), Universitat de Barcelona (UB), Mart\'i Franqu\`es 1, E-08028 Barcelona, Spain
\item[$^{24}$] Department of Physics, Southern Methodist University, Dallas, TX 75275-0181, USA
\item[$^{25}$] Institute for Theoretical Physics, ETH, CH--8093 Z\"urich, Switzerland
\item[$^{26}$] Department of Physics and Astronomy, Michigan State University, East Lansing, MI 48824, USA
\item[$^{27}$] D\'epartement des Math\'ematiques, Universit\'e Abdelmalek Essaadi, Tanger, Morocco
\item[$^{28}$] PRISMA Cluster of Excellence, Institute of Physics, Johannes Gutenberg University, D-55099 Mainz, Germany
\item[$^{29}$] School of Physics, University of the Witwatersrand, Johannesburg, Wits 2050, South Africa
\item[$^{30}$] Dipartimento di Fisica, Universit\`a di Torino and INFN, Sezione di Torino, Via Pietro Giuria 1, I-10125 Torino, Italy
\item[$^{31}$] Fermilab, PO Box 500, Batavia, IL 60510, USA
\item[$^{32}$] Maryland Center for Fundamental Physics, University of Maryland, College Park, Maryland 20742, USA
\item[$^{33}$] Dept. of Astronomy and Theoretical Physics, Lund University, Sweden
\item[$^{34}$] Physikalisches Institut, Albert-Ludwigs-Universit\"{a}t Freiburg, 79104 Freiburg, Germany
\item[$^{35}$] Center for Theoretical Physics, Massachusetts Institute of Technology, Cambridge, MA 02139, USA
\item[$^{36}$] Dipartimento di Fisica, Universit\`a di Pavia, Pavia, Italy
\item[$^{37}$] Dipartimento di Fisica, Universit\`a di Ferrara and INFN, Sezione di Ferrara, Italy
\item[$^{38}$] Particle Physics Group, School of Physics and Astronomy, University of Manchester, Manchester M13 9PL, UK
\item[$^{39}$] Institut f\"{u}r Theoretische Physik, Karlsruhe Institute of Technology, 76131 Karlsruhe, Germany
\item[$^{40}$] Universit\`a di Milano-Bicocca, Milano, Italy
\item[$^{41}$] INFN, Sezione di Milano-Bicocca, Piazza della Scienza 3, 20126 Milano, Italy
\item[$^{42}$] Centre for Cosmology, Particle Physics and Phenomenology (CP3), Universit\'e catholique de Louvain, B-1348 Louvain-la-Neuve, Belgium
\item[$^{43}$] Laboratoire de Physique Nucl\'{e}aire et de Hautes Energies, UPMC and Universit\'{e} Paris-Diderot and CNRS/IN2P3, Paris, France
\item[$^{44}$] Institute for Particle Physics, ETH, CH--8093 Z\"urich, Switzerland
\item[$^{45}$] Department of Physics \& Astronomy, Northwestern University, Evanston, IL 60208, USA
\item[$^{46}$] HEP Theory Group, Department of Physics, Florida State University, Tallahassee, USA
\item[$^{47}$] Rudolf Peierls Centre for Theoretical Physics, 1 Keble Road, University of Oxford, OX1 3NP Oxford, United Kingdom
\item[$^{48}$] Universit\`{a} di Siena, Siena, Italy
\item[$^{49}$] School of Physics and Astronomy, Monash University, VIC-3800, Australia
\item[$^{50}$] IPhT, CEA Saclay, CNRS UMR 3681, F-91191 Gif-sur-Yvette, France
\item[$^{51}$] Deutsches Elektronen-Synchrotron (DESY), D-22607 Hamburg, Germany
\item[$^{52}$] University of California, Los Angeles, USA
\item[$^{53}$] Department of Physics, University at Buffalo, The State University of New York, Buffalo 14260, USA
\item[$^{54}$] Nuclear Science Division, Lawrence Berkeley National Laboratory, Berkeley, CA 94720, USA
\item[$^{55}$] Sorbonne Universit\'{e}s, UPMC Univ. Paris 06, UMR 7589, LPTHE, F-75005, Paris, France
\item[$^{56}$] CNRS, UMR 7589, LPTHE, F-75005, Paris, France
\end{itemize}


\newpage
\tableofcontents


\newpage
\pagenumbering{arabic}
\setcounter{footnote}{0}

The 2015 Les Houches workshop saw great progress, both in terms of the
development of precision calculations as well as the analysis of high
statistics 8 TeV LHC data. The term of the workshop also saw the first data at
13 TeV, and the calculation of inclusive Higgs boson production at N$^3$LO in
perturbative QCD. There has been considerable progress for differential predictions
at N$^2$LO in QCD and automated electroweak corrections at NLO.

In these proceedings, we report on the progress with the Les Houches high
precision wish list, as well as the development of new tools, such as the
PDF4LHC15 PDFs, and simplified cross sections for Higgs boson measurements.
Detailed phenomenological studies, such as the comparison of predictions for
Higgs boson + jets in gluon-gluon fusion were carried out. The 2015 workshop
continued the emphasis on understanding electroweak corrections that started in
the 2013 workshop. As the increased energy and luminosity allow more
measurements to reach the TeV scale, greater attention must be paid to the
impact of EW corrections. Finally, the systematics of quark/gluon tagging have
been investigated, and, following a long tradition, a new Les Houches variable,
the Les Houches Angularity, has been defined.

As the LHC turns back on, we eagerly await the high statistics data at 13 TeV
that will dominate the work at Les Houches 2017.


\newpage

\chapter{NLO automation and (N)NLO techniques}
\label{cha:nnlo}

\renewcommand{\arraystretch}{1.2}
\section{Update on the precision Standard Model wish list 
  \texorpdfstring{\protect\footnote{
     S.~Badger, A.~Denner, J.~Huston
  }}{}}

Identifying key observables and processes that require improved theoretical input has been
a key part of the Les Houches programme. In this contribution we summarise progress since the previous
report in 2013 and explore the possibilities for further advancements during Run II.

\subsection{Introduction}

The 2013 Les Houches report introduced a new high precision wish list which was considered
an ``extremely ambitious'' list of processes that could be used to fully exploit
the Run II data \cite{Andersen:2014efa}. In fact there has been remarkable theoretical advances in NNLO computations
for QCD corrections and automated NLO electroweak corrections and many processes are now available
in some form. Of course there still remains a number of outstanding items but the wish list certainly deserves
to be updated to include recent progress.

An undoubted highlight in precision QCD predictions since the 2013 report has
been the completion of the total inclusive Higgs production cross-section at
\NLO3 \cite{Anastasiou:2015ema}. There has also been remarkable progress for
$2\rightarrow 2$ predictions at NNLO in QCD where a large number of items from the
2013 wish list have now been completed. The focus here has been on producing fully
differential predictions, and progress has largely been possible thanks to the
perfection of infrared subtraction methods.

Automated tools for NLO QCD are now familiar in experimental analyses where parton-shower
matching and matrix element improved multi-jet merging techniques are becoming a standard
level of theoretical precision. The automation of full SM corrections including mixed
electroweak predictions has also seen major improvements.

It is clear that computations at this level of complexity require careful
verification.  The importance of reproducing known results with alternative
techniques and implementations cannot be stressed enough. Though there are
still challenges ahead to make state-of-the-art $2\to2$ predictions publicly
available to experimental analyses, the theoretical framework is in place to
ensure this can be achieved during the course of Run II and keep theoretical
uncertainties in line with the experiments.

Root Ntuples have been a useful tool for complicated final states at NLO and
allow for extremely flexible re-weighting and analysis. The cost for this is
the large disk space required to store the event information.  A feasibility
study using \textsc{Root NTuples} to store the much larger NNLO events in $e^+ e^- \to
3$ jets is described in Sec.\ \ref{cha:mc}.\ref{sec:NNLOntuples}.  ApplGrid \cite{Carli:2010rw} and FastNLO
\cite{Kluge:2006xs} offer a simpler, but slightly less flexible, method to
distribute NNLO predictions. The latter option is likely to be used heavily in
precision PDF fits.

\subsection{Developments in theoretical methods}

Precision predictions require a long chain of different tools and methods all
of which demand detailed and highly technical computations. In this section we
attempt to summarise the current state of the art and point out some
bottlenecks which will need to be overcome in order to complete the still
unknown processes on the wish list.

Computational methods for the amplitude level ingredients have seen substantial
progress in the last few years. In order to be as pedagogical as possible we
begin by defining the various components that are required to construct
infrared finite observables at fixed order. Scattering amplitudes are
generally decomposed into a basis of integrals together with rational coefficients,
\begin{equation}
  A^{(L)}_{2\to n} = \sum (\text{coefficients})_i (\text{integrals})_i,
\end{equation}
one must then remove infrared singularities to obtain a finite cross-section,
\begin{equation}
  d\sigma_{2\to n} \text{\NLO{k}} = {\rm IR}_k(A^{k}_{2\to n}, A^{k-1}_{2\to n+1},\cdots, A^{0}_{2\to n+k}).
\end{equation}
where the function ${\rm IR}_k$ represents an infrared subtraction technique. Ultra-violet renormalisation
must also be performed but since it presents no technical difficulties we ignore it in our review.
We briefly review the three main components -- the computation of the loop integrals, the computation of
their rational coefficients and methods for infrared subtraction.

\subsubsection{Loop integrals}

A major result since the last report has been the completion of the two-loop
master integrals for vector boson pair production
\cite{Henn:2014lfa,Caola:2014lpa,Gehrmann:2014bfa,Papadopoulos:2014hla}
employing Henn's canonical differential equation technique \cite{Henn:2013pwa}.
The technique has also been used to obtain complete NLO QCD corrections the
$H\to Z\gamma$ decay retaining the full top and Z mass dependence of the
two-loop integrals \cite{Bonciani:2015eua}. Other notable examples include the
off-shell integrals for ladder topologies at three loops \cite{DiVita:2014pza}
and two-loop master integrals for mixed QCD/electroweak corrections in
Drell-Yan production \cite{Bonciani:2016ypc}.

Improvements in the understanding of the basis of multiple poly-logarithms through Symbol and Hopf
algebra's (see e.g. \cite{Duhr:2014woa} for a review) has led to a high degree of
automation for these integral computations. This is clearly a necessary step
in order to apply such techniques to phenomenologically relevant cases, most
notably in the case of $pp\to H$ at \NLO3.

There has also been developments for the direct evaluation of Feynman integrals.
Panzer's HyperInt \cite{Panzer:2014caa} and Bogner's MPL \cite{Bogner:2015nda} packages
have focused mainly on zero and one scale integrals with a high number of loops (see for example \cite{vonManteuffel:2015gxa})
but the algorithms employed have potential applications to a wider class of
two-loop integrals with more scales.

Direct numerical evaluation remains a powerful technique and the sector decomposition
algorithm has seen a number of optimisations implemented into the publicly available codes~\cite{Smirnov:2015mct,Smirnov:2013eza,Borowka:2015mxa,Borowka:2012yc}.
Though performing a full integration over the phase space using numerical methods is
extremely computationally intensive, very recently a complete computation of $pp\to HH$ via
gluon fusion at NLO including the full top mass dependence has been completed~\cite{Borowka:2016ehy}.

Prospects for going beyond $2\to2$ scattering at two loops are also improving.
Very recently an analytic basis of five-point master integrals for the planar
sector massless QCD\cite{Gehrmann:2015bfy} and off-shell integrals needed for
$pp\to H+2j$ \cite{Papadopoulos:2015jft} have been completed. Extending these
techniques to complete the non-planar sector definitely requires additional work
but is now a realistic short term goal.

The most difficult processes on the wish list from the point of view
of the loop integrals are those with many internal mass scales like
$t\tb+j$ and full $m_t$ corrections to $H+j$ due to the potential
appearance of elliptic functions.  The mathematics of these special
functions still remains to be fully understood and is an active area
of research\footnote{For some recent studies of elliptic functions and
  elliptic multiple zeta values see
  \cite{Broedel:2014vla,Adams:2015gva,Bloch:2013tra}. A more complete
  set of references can be found in the recent review
  \cite{Bogner:2016qbf}.}.  Nevertheless, developments give hope that
such processes will eventually be possible but for the time being
remain out of reach for phenomenology.  Numerical methods offer an
alternative solution to this problem, and presently have been the most
successful for example in $pp\to t\tb$~\cite{Czakon:2015owf} and
$pp\to HH$~\cite{Borowka:2016ehy}.  Nevertheless, an analytic solution
to the problem would allow for more flexible phenomenology.

\subsubsection{Loop amplitudes and integrands}

Following the completion of the two-loop integrals mentioned above, a complete set of helicity amplitudes has been obtained by two independent groups for $pp\to VV'$ \cite{%
Caola:2014iua,Caola:2015ila,%
vonManteuffel:2015msa,Gehrmann:2015ora}. A good deal of effort was spent to
ensure stable numerical evaluation of these amplitudes requiring knowledge of
the underlying basis of multiple poly-logarithms. This has resulted in a public
code for numerical evaluation available from
\url{http://vvamp.hepforge.org/}. Both approaches relied heavily on
efficient implementations of integration-by-parts reduction identities
\cite{Smirnov:2008iw,Smirnov:2013dia,Smirnov:2014hma,Lee:2012cn,Studerus:2009ye,vonManteuffel:2012np}.

The expressions, generated using Feynman diagrams, are quite lengthy and in many places
contain spurious poles, but this has not hindered full NNLO predictions where the major
computational effort is in the evaluations of double and single unresolved infrared divergent
components.

\subsubsection{Generalised unitarity and integrand reduction}

Extending the current multi-loop methods to higher multiplicity still represents
a serious challenge. The increased complexity in the kinematics, and large amount
of gauge redundancy in the traditional Feynman diagram approach, has been solved numerically through on-shell
and recursive off-shell methods at one-loop. This breakthrough has led to the development
of the now commonly used automated one-loop codes~\cite{Berger:2008sj,Cascioli:2011va,Bevilacqua:2011xh,Badger:2012pg,Actis:2016mpe,Cullen:2014yla,Alwall:2014hca}.

The $D$-dimensional generalised unitarity cuts algorithm
\cite{Bern:1994zx,Bern:1994cg,Britto:2004nc,Giele:2008ve,Forde:2007mi} has now
been extended to multi-loop integrands using integrand reduction
\cite{Ossola:2006us}\footnote{We do not attempt a complete review of integrand reduction here. Further information can be found in the review article \cite{Ellis:2011cr} and references therein.} and elements of computational algebraic geometry~%
\cite{Mastrolia:2011pr,Badger:2012dp,Zhang:2012ce,Kleiss:2012yv,Feng:2012bm,Mastrolia:2012an,Mastrolia:2013kca,Badger:2013gxa,Mastrolia:2016dhn}.
In contrast to the one-loop case, the basis of integrals obtained through this
method is not currently known analytically and is much larger than the set of
basis functions defined by standard integration-by-parts identities.
Nevertheless, new results in non-supersymmetric theories have been obtained for
five gluon scattering amplitudes with all positive helicities
\cite{Badger:2013gxa,Badger:2015lda}.  The maximal unitarity method
\cite{Kosower:2011ty}, which incorporates integration-by-parts (IBP) identities, has been applied to a
variety of two-loop examples in four dimensions
\cite{Larsen:2012sx,CaronHuot:2012ab,Johansson:2012zv,Johansson:2013sda,Johansson:2015ava}.
This approach can be seen as an extension of the generalised unitarity methods
of Britto, Cachazo and Feng~\cite{Britto:2004nc} and Forde~\cite{Forde:2007mi}.
Efficient algorithms to generate unitarity compatible IBP identities are a key ingredient
in both approaches and have been the focus of on-going investigations
\cite{Gluza:2010ws,Schabinger:2011dz,Ita:2015tya,Larsen:2015ped}.

Steady progress towards $2\to 3$ processes on the wish list like $pp\to H+2j$,
$pp\to3j$ appears to be on course though there is clearly a long way to fully
differential NNLO predictions.

\subsubsection{Infrared subtraction methods for differential cross-sections}

The construction of fully differential NNLO cross-sections for $2\to2$
processes has been a major theoretical challenge over the last years.
This programme has been a major success with many different approaches now
applied to LHC processes.

\begin{itemize}
\item Antenna subtraction \cite{GehrmannDeRidder:2005cm,Currie:2013vh}:\\
  Analytically integrated counter-terms for hadronic initial and final states. Almost
  completely local, requires averaging over azimuthal angles. Applied to $pp\to 2j$,
  $pp\to Z+j$ and $pp\to W+j$.

\item Sector Decomposition \cite{Czakon:2010td,Czakon:2011ve,Boughezal:2011jf}:\\
  Fully local counter-terms using extension of the FKS approach at NLO \cite{Frixione:1995ms}.
  Numerically integrated counter-terms for hadronic initial and final states.
  Recently formulated in a four-dimensional setting \cite{Czakon:2014oma}. Applied to $pp\to H+j$~\cite{Boughezal:2015dra,Caola:2015wna}
  and $pp\to t\tb$~\cite{Czakon:2016ckf} processes.

\item $q_T$ \cite{Catani:2007vq}:\\
  Phase-space slicing approach for colour-less final states applied to many
  $pp\to VV'$ processes. An extension for $t\tb$ final states has been recently
  proposed \cite{Bonciani:2015sha}.

\item $N$-jettiness \cite{Boughezal:2015aha,Boughezal:2015eha,Gaunt:2015pea}:\\
  Extension of the $q_T$ method to general hadronic final states
  matching to soft-collinear effective theory (SCET)
  below the $N$-jettiness cut-off parameter. Applied to $pp\to H+j$, $pp\to V+j$ and Drell-Yan including
  resummation and parton shower effects \cite{Alioli:2015toa}.

\item ColorFull \cite{DelDuca:2015zqa}:\\
  Fully local counter-terms extending the Catani-Seymour dipole method \cite{Catani:1996vz}.
  Analytically integrated for infrared poles, numerical integration for finite parts.
  Currently developed for hadronic final states such as $H\to b\bb$ \cite{DelDuca:2015zqa} and $e^+e^-\to$ jets \cite{DelDuca:2016csb}.
\end{itemize}

With the success of these methods for $2\to2$ applications it is an obvious question as to whether
these techniques also apply to higher multiplicity. The extension to $2\to3$ for most methods is
clear though the reality of dealing with an increasingly large number of counter-terms and
a more complicated phase space is not to be underestimated. It is clear that in the near future
such applications will be attempted based on the current technology.

The reverse unitarity method
\cite{Anastasiou:2002yz,Anastasiou:2002qz,Anastasiou:2003yy,Anastasiou:2003ds}
is a powerful method for fully inclusive observables (and rapidity
distributions).  It has been utilised in the complete computation of the
inclusive $pp\to H$ cross-section at \NLOQ3~\cite{Anastasiou:2015vya} in the infinite top-mass limit. This computation
has been carried out in many stages starting with expansions around the soft limit
\cite{Anastasiou:2014vaa,Anastasiou:2014lda,Li:2014afw}.  There are 5 main
components: triple-virtual, squared real-virtual, double-virtual-real,
double-real-virtual and triple-real, most of which have been verified by
independent groups.  The reverse unitarity method has then been used to obtain
each component of the triple-virtual
\cite{Baikov:2009bg,Gehrmann:2010ue,Gehrmann:2010tu}, squared real-virtual
\cite{Anastasiou:2013mca,Kilgore:2013gba}, double-virtual-real
\cite{Duhr:2013msa,Li:2013lsa,Duhr:2014nda,Dulat:2014mda}, double-real-virtual
\cite{Li:2014bfa,Anastasiou:2015yha} and triple-real radiation
\cite{Anastasiou:2013srw} as an expansion in the dimensional regularisation
parameter. The poles of these separate contributions cancel analytically when
summed together and combined with the counter-terms for UV poles
\cite{Tarasov:1980au,Larin:1993tp,vanRitbergen:1997va,Czakon:2004bu} and
initial state infrared singularities
\cite{Vogt:2004mw,Moch:2004pa,Anastasiou:2012kq,Hoschele:2012xc,Buehler:2013fha}.
A complete result for the $qq'\to H$ channel has also been obtained
\cite{Anzai:2015wma}.

The analytic result for Higgs production at threshold \cite{Anastasiou:2014vaa}
has given access to a number of additional predictions in the threshold limit
for Drell-Yan \cite{Ahmed:2014cla,Ahmed:2014uya}, $H\to b\bb$
\cite{Ahmed:2014cha}, $V'\to VH$ \cite{Kumar:2014uwa}.

\subsection{The precision wish list}

Following the 2013 report we break the list of precision observables into
three sections: Higgs, vector bosons and top quarks.

Corrections are defined with respect to the leading order, and we organise
the perturbative expansion into QCD corrections, electroweak (EW) corrections and
mixed QCD$\times$EW,
\begin{equation}
  d\sigma_X = d\sigma_X^{\rm LO} \left(1 +
      \sum_{k=1} \alpha_s^k d\sigma_X^{\delta \text{\NLOQ{k}}}
    + \sum_{k=1} \alpha^k d\sigma_X^{\delta \text{\NLOE{k}}}
    + \sum_{k,l=1} \alpha_s^k \alpha^l d\sigma_X^{\delta \text{\NLOQE{k}{l}}}
    \right).
  \label{eq:dsigmapertexp}
\end{equation}
We explicitly separate the mixed QCD and EW corrections to distinguish between additive predictions
QCD+EW and mixed predictions QCD$\times$EW. The definition above only applies in the case where the leading order
process contains only one unique power in each coupling constant. For example, in the case of $q\qb\to q\qb Z$ two
leading order processes exist: via gluon exchange of $\mathcal{O}(\alpha_s^2\alpha)$, via electroweak boson 
exchange of $\mathcal{O}(\alpha^3)$ and the interference $\mathcal{O}(\alpha_s\alpha^2)$. 
In these cases it is customary to classify the Born process with highest power in 
$\alpha_s$, and typically the largest cross section, as the leading order and 
label the others as subleading Born processes. The above classification is then, 
unless otherwise stated, understood with respect to the leading Born process.

In the following we attempt to give a current snap-shot of the state of
perturbative computations at \NLOQ2~or higher and electroweak corrections at
\NLOEone. The main aim is to summarise computations that appeared in the 2013
wish list and that have now been completed. There are obvious difficulties in compiling lists
in this way which make it difficult to address every possible relevant computation. Specific
approximations and extensions to fixed order are often necessary when comparing theory to data.

Following the 2013 wish list we clarify that it is desirable to have a prediction that combines all the known corrections.
For example d\xs~\NLOQ2+\NLOEone~refers to a single code that produced differential predictions including
$\mathcal{O}(\alpha_s^2)$ and $\mathcal{O}(\alpha)$. In most cases this is a non-trivial task and when considered
in combination with decays can lead to a large number of different sub-processes.

\hspace{0.5cm}\textit{Electroweak corrections}

Complete higher order corrections in the SM can quickly become technically complicated in comparison
to the better known corrections in QCD. An introductory guide to the nomenclature
and applications in which such corrections are important was presented in the last report so we do not
repeat it here \cite{Andersen:2014efa}. As a basic rule of thumb $\alpha_s^2 \sim \alpha$, so corrections at \NLOQ2~and \NLOEone~
are desirable together. Moreover, for energy scales that are large compared to
the W-boson mass EW corrections are enhanced by large logarithms (often called
Sudakov logarithms). There has been progress towards a complete automation of
\NLOEone~corrections within one-loop programs such as
\textsc{OpenLoops}, \textsc{GoSam}, \textsc{Recola} and \WishListMGaMC which has led to the completion
of many items from the 2013 list. A detailed comparison of these tools as well as comparisons with the Sudakov
approximation for a variety of processes is presented in Sec.\ \ref{cha:pheno}.\ref{sec:SMewcomparison}.

\hspace{0.5cm}\textit{Heavy top effective Higgs interactions and finite mass effects}

Many calculations of SM processes involving Higgs bosons use the effective
gluon interaction in the $m_t\to\infty$ limit. At high energy hadron colliders
this is by far the most dominant production process, yet at high transverse momentum the
approximation will break down. The data collected during Run II will certainly
probe the region where the approximation is invalid and finite mass effects are
an important addition. Complete top mass corrections are extremely difficult
since the occurring multi-loop integrals are thought to contain elliptic functions. As
mentioned above, though there is active research in this area, phenomenological
applications are only a realistic goal in a longer time frame. In the short
term there are many approaches to approximate finite mass effects, either by
re-weighting with a full leading order or including partial results where
available.  Expansions in $m_t$ have also been successfully computed. It is
clear that in all cases where the effective theory is used some approximate
treatment of finite mass effects is desirable.

We list processes in the wish list as \NLOH{k} + \NLOQ{l} when re-weighting including the full top mass dependence up to
order $l$ has been performed. For example, consider the differential cross section at NLO in the effective theory:
\begin{equation}
  d\sigma_n^{\text{NLO}_\text{HEFT}}
  = d\sigma_n^{\text{LO}_\text{HEFT}} + d\sigma_n^{\text{V}_\text{HEFT}} + d\sigma_{n+1}^{\text{R}_\text{HEFT}},
  \label{eq:xsHEFT}
\end{equation}
where we have grouped together the different contributions to the NLO cross section: leading order (LO), virtual (V) and real radiation (R).
The details of any infrared subtraction method will not affect the fact that there is an $n$ particle and $n+1$ particle phase space.
Including finite mass effects at LO can be simply achieved by
replacing parts of the calculation with the those from the full theory
where they are known:
\begin{equation}
  d\sigma_n^{\text{NLO}_\text{HEFT}+\text{LO}_\text{QCD}}
  = d\sigma_n^{\text{LO}_\text{QCD}} + d\sigma_n^{\text{V}_\text{HEFT}} + d\sigma_{n+1}^{\text{R}_\text{HEFT}}
  \label{eq:Heffreweight1}
\end{equation}
or by re-weighting the NLO corrections,
\begin{equation}
  d\sigma_n^{\text{NLO}_\text{HEFT}+\text{LO}_\text{QCD}}
  = \frac{d\sigma_n^{\text{LO}_\text{QCD}}}{d\sigma_n^{\text{LO}_\text{HEFT}}} \left(
  d\sigma_n^{\text{LO}_\text{HEFT}} + d\sigma_n^{\text{V}_\text{HEFT}} \right)
      + \frac{d\sigma_{n+1}^{\text{LO}_\text{QCD}}}{d\sigma_{n+1}^{\text{LO}_\text{HEFT}}} d\sigma_{n+1}^{\text{R}_\text{HEFT}}.
  \label{eq:Heffreweight2}
\end{equation}
Clearly more care is required in the latter case to ensure infrared finiteness is preserved. In either case the QCD correction
will proceed through a heavy quark loop and can be taken in $N_f=5$ ($m_t$ only), $N_f=4$ ($m_t$ and $m_b$ only) and so on.
For totally inclusive quantities the distinction between the two approaches above is not important,
the approach taken in the case of $pp\to H$ recently presents a good overview of this procedure \cite{Anastasiou:2016cez}. While the above constructions 
to incorporate quark mass effects into NLO calculations of Higgs production 
in gluon fusion is expected to work well for top quarks, it fails for bottom 
quarks. The reason therefore lies in the smallness of the bottom quark mass 
and the invalidity of the HEFT expressions in this case.
A more extensive discussion on this topic is presented in Sec.\ \ref{cha:higgs}.\ref{sec:Hmt}.

\hspace{0.5cm}\textit{Resummation}

We do not attempt a complete classification of all possible resummation
procedures that have been considered or applied to the processes in the list.
In many cases precision measurements will require additional treatment beyond
fixed order and since resummed predictions always match onto fixed order
outside of the divergent region it would be desirable if all predictions were available
this way. Since this is not feasible, some specific cases are highlighted in
addition to the fixed order.

There are several important kinematic regions where perturbative predictions
are expected to break down. Totally inclusive cross-sections often have large
contributions from soft-gluon emission in which higher order logarithms can be
computed analytically. The $q_T$ and $N$-jettiness subtraction methods naturally
match on to resummations of soft gluons, in the latter case through
soft-collinear effective theory. A recent study using the $q_T$ method has been
applied in the case of $pp\to ZZ$ and $pp\to W^+ W^-$ \cite{Grazzini:2015wpa} where further details and
references can be found. $0$-jettiness resummations within SCET have been
considered recently in Drell-Yan production \cite{Alioli:2015toa} using the
\textsc{Geneva} Monte Carlo framework which also matches the results to a
parton shower \cite{Alioli:2012fc}.

Observables with additional restrictions on jet transverse momenta can also
introduce large logarithms and jet veto resummations have been studied
extensively in the case of $pp\to H$ and $pp\to H+j$ \cite{Banfi:2015pju}.

With increasing precision of both experimental data and fixed order
calculations other regions may also begin to play a role. A method for the
resummation of logarithms from small jet radii has been developed and applied
in the case of $pp\to H$ \cite{Dasgupta:2014yra,Banfi:2015pju,Dasgupta:2016bnd}.

These represent only a tiny fraction of the currently available tools and predictions
with resummed logarithms. For further information the interested reader may refer to~\cite{Luisoni:2015xha}
and references therein.

\hspace{0.5cm}\textit{Parton showering}

As in the case of resummation - we refrain from listing all processes in the wish list to be desired with
matching to a parton shower (PS). \NLOQone + PS predictions can be considered fully automated within \WishListMGaMC,
\textsc{Sherpa}, \textsc{POWHEG} and \textsc{Herwig7}. There have been many recent efforts in matching \NLOQ2 corrections to parton showers
for singlet production processes. There are good prospects for extending these techniques to $2\to2$ processes in the
short term - nevertheless we have refrained from adding these processes to the 2013 wish list.

\hspace{0.5cm}\textit{Decay sub-processes}

The description of decay sub-processes is incomplete though we do list a few
notable cases. Ideally all on-shell (factorised) decays would be available up
to the order of the core process. In some cases this is potentially an
insufficient approximation and full off-shell decays including background
interference would be desirable but are often prohibitive. The $t\tb$ final state
is an obvious example where the off-shell decay to $WWb\bb$ at NNLO is beyond the scope
of current theoretical methods.

Decays in the context of electroweak corrections are usually much more
complicated. Full off-shell effects at NLO are expected to be small, but higher order
corrections within factorisable contributions to the decay can be important.
The case of vector boson pair production is particularly important given the
completion of the \NLOQ2 computation, and corrections are known at NLO  within the double pole
approximation \cite{Billoni:2013aba} and beyond \cite{Biedermann:2016yvs,Biedermann:2016guo}.

\subsection{Higgs boson associated processes}

\begin{table}
  \begin{center}
  \scalebox{0.92}{
  \begin{tabular}{lll}
    \hline
    \multicolumn{1}{c}{process} & \multicolumn{1}{c}{known} & \multicolumn{1}{c}{desired} \\
    \hline
    $pp\to H$ &
      \begin{tabular}{cl}
        \xs & \NLOH3+\NLOQ2$\left(\tfrac{1}{m_t^6}\right)$\\
        d\xs & \NLOEone \\
        d\xs & \NLOH2+\NLOQone+PS
      \end{tabular}
      &
      \begin{tabular}{cl}
        & \\
        d\xs & \NLOH3+\NLOQ2\\&+\NLOEone+\NLOQE11
      \end{tabular} \\
    \hline
    $pp\to H+j$ &
      \begin{tabular}{cl}
        d\xs & \NLOH2 \\
        d\xs & \NLOEone
      \end{tabular}
      &
      \begin{tabular}{cl}
         d\xs & \NLOH2+\NLOQone+\NLOEone\\
      \end{tabular} \\
    \hline
    $pp\to H+2j$ &
      \begin{tabular}{cl}
        d\xs & \NLOHone+\LOQ \\
        d\xs & \NLOQ2(VBF$^{*}$) \\
        d\xs & \NLOEone(VBF)
      \end{tabular}
      &
      \begin{tabular}{cl}
        d\xs & \NLOH2+\LOQ+\NLOEone\\
        d\xs & \NLOQ2(VBF)+\NLOEone(VBF) \\
      \end{tabular} \\
    \hline
    $pp\to H+3j$ &
      \begin{tabular}{cl}
        d\xs & \NLOHone \\
        d\xs & \NLOEone
      \end{tabular}
      &
      \begin{tabular}{cl}
        d\xs & \NLOQone+\LOQ+\NLOEone \\
      \end{tabular} \\
    \hline
    $pp\to H+V$ &
      \begin{tabular}{cl}
        d\xs & \NLOQ2 \\
      \end{tabular}
      &
      \begin{tabular}{cl}
        d\xs & \NLOQ2 + NLO${}_{gg\to HV}$ + \NLOEone \\
      \end{tabular} \\
    \hline
    $pp\to HH$ &
      \begin{tabular}{cl}
        d\xs & \NLOH2 \\
        d\xs & \NLOQone
      \end{tabular}
      &
      \begin{tabular}{cl}
         d\xs & \NLOH2+\NLOQone+\NLOEone\\
      \end{tabular} \\
    \hline
    $pp\to H+t\tb$ &
      \begin{tabular}{cl}
        d\xs & \NLOQone \\
        d\xs & \NLOEone
      \end{tabular}
      &
      \begin{tabular}{cl}
        d\xs & \NLOQone+\NLOEone
      \end{tabular} \\
    \hline
      \begin{tabular}{l}
        $pp\to H+t$ \\
        $pp\to H+\tb$ \\
      \end{tabular}
    &
      \begin{tabular}{cl}
        d\xs & \NLOQone
      \end{tabular}
      &
      \begin{tabular}{cl}
        d\xs & \NLOQone+\NLOEone
      \end{tabular} \\
    \hline
  \end{tabular}
  }
  \caption{Precision wish list: Higgs boson final states. VBF$^{*}$ refers to the computation
  using the factorised 'projection-to-Born' approximation \cite{Cacciari:2015jma}.}
  \label{tab:wlH}
  \end{center}
\end{table}

\begin{itemize}[leftmargin=2cm]
  \item[$H$:] Given the high importance of this
    process it has been investigated extensively in the
    literature. The calculation of the total
    inclusive cross section at \NLOH3\cite{Anastasiou:2015ema} reduces theoretical
    uncertainties to the 5\% level, the remaining uncertainties being dominated by PDF and finite mass
    effects. Several resummation schemes have been considered and a
    comprehensive phenomenological study has been presented recently
    \cite{Anastasiou:2016cez}. The most precise evaluation of the cross section
    combines the effective theory with:
    \begin{itemize}
      \item complete mass dependence at NLO including top, bottom and charm loops.
      \item ${m_H}/{m_t}$ corrections at NNLO.
      \item electro-weak corrections at NLO.
      \item re-scaling of the \NLOH3 with the complete \LOQ~top loop.
    \end{itemize}
    The NNLO+PS computation \cite{Hamilton:2013fea} has been extended to include finite top
    and bottom mass corrections at NLO \cite{Hamilton:2015nsa}.
    To match the data precisely it would be desirable to have fiducial predictions,
    requiring a fully differential computation.
  \item[$H+j$:] Known through to \NLOQ2 in the infinite top mass limit
    \cite{Chen:2014gva,Boughezal:2015dra,Boughezal:2015aha,Caola:2015wna}.
    Finite top mass effects are expected to play an important role for
    $p_T>150-200$ GeV.
  \item[$H+\geq 2j$:] QCD corrections are an essential background to Higgs
    production in vector boson fusion (VBF). VBF production of a Higgs boson has recently been computed
    through to \NLOQ2 accuracy \cite{Cacciari:2015jma}.
  \item[$VH$:] Associated production of a Higgs boson with a vector boson is important for pinning down the EW couplings of the Higgs.
    First predictions at \NLOQ2 have been available for some time \cite{Brein:2003wg,Brein:2011vx}. Differential predictions at
    \NLOQ2 to $WH$ \cite{Ferrera:2011bk} and $ZH$
    \cite{Ferrera:2014lca} have now been completed including effects
    from gluon initiated processes. These results have been studied more recently using the $N$-jettiness subtraction scheme within MCFM including a full set of decays \cite{Campbell:2016jau}. A parton shower matched prediction using the MiNLO procedure in \textsc{POWHEG} has very recently been completed \cite{Astill:2016hpa}.
    The total inclusive cross-section has been considered in the threshold limit at \NLOQ3, extracted from the inclusive Higgs cross-section \cite{Kumar:2014uwa}.
  \item[$HH$:] The \NLOQ2
    corrections were originally computed in the infinite top mass limit
    \cite{deFlorian:2013jea} and have since been improved with threshold resummation to NNLL \cite{deFlorian:2015moa}. Finite top mass corrections are very important
    here and power corrections to $\mathcal{O}({1}/{m_t^8})$ have been computed \cite{Grigo:2013rya,Grigo:2015dia}. Though the convergence is seen to be relatively slow, improvements can be made by normalising to LO result in the full theory. A complete computation at NLO including all finite quark mass effects has very recently been
    achieved using numerical integration methods~\cite{Borowka:2016ehy}.
  \item[$t\bar{t}H$:] \NLOEone~corrections have been considered within the
    automated\\\WishListMGaMC framework
    \cite{Frixione:2014qaa,Frixione:2015zaa}.  Moreover, NLO QCD corrections
    have been calculated for the process including the top quark
    decays~\cite{Denner:2015yca}.
\end{itemize}

\subsection{Vector boson associated processes}

The numerous decay channels for vector bosons and the possible inclusion
of full off-shell corrections versus factorised decays in the narrow width approximation
make vector boson processes complicated to classify.
A full range of decays in the narrow width approximation would be a desirable minimum precision.

\begin{table}
  \centering
  \begin{tabular}{lll}
    \hline
    \multicolumn{1}{c}{process} & \multicolumn{1}{c}{known} & \multicolumn{1}{c}{desired} \\
    \hline
    $pp\to V$ &
      \begin{tabular}{cl}
        \xs & \NLOQ3($z\to0$) \\
        d\xs & \NLOQ2 \\
        d\xs & \NLOEone \\
        d\xs & \NLOQE11
      \end{tabular}
      &
      \begin{tabular}{cl}
        d\xs & \NLOQ3 + \NLOE2 \\&+ \NLOQE11 + decays
      \end{tabular} \\
    \hline
    $pp\to VV'$ &
      \begin{tabular}{cl}
        d\xs & \NLOQ2 + decays\\
        d\xs & \NLOEone \\
      \end{tabular}
      &
      \begin{tabular}{cl}
        d\xs & \NLOQ2 + \NLOEone + decays
      \end{tabular} \\
    \hline
    $pp\to V+j$ &
      \begin{tabular}{cl}
        d\xs & \NLOQ2 \\
      \end{tabular}
      &
      \begin{tabular}{cl}
        d\xs & \NLOQ2 + \NLOEone + decays \\
      \end{tabular} \\
    \hline
    $pp\to V+2j$ &
      \begin{tabular}{cl}
        d\xs & \NLOQone + decays \\
        d\xs & \NLOEone + decays
      \end{tabular}
      &
      \begin{tabular}{cl}
        d\xs & \NLOQ2 + \NLOEone + decays \\
      \end{tabular} \\
    \hline
    $pp\to VV'+1,2j$ &
      \begin{tabular}{cl}
        d\xs & \NLOQone + decays \\
        d\xs & \NLOEone
      \end{tabular}
      &
      \begin{tabular}{cl}
        d\xs & \NLOQone + \NLOEone + decays \\
      \end{tabular} \\
    \hline
    $pp\to VV'V''$ &
      \begin{tabular}{cl}
        d\xs & \NLOQone \\
        d\xs & \NLOEone
      \end{tabular}
      &
      \begin{tabular}{cl}
        d\xs & \NLOQone + \NLOEone + decays \\
      \end{tabular} \\
    \hline
    $pp\to \gamma\gamma$ &
      \begin{tabular}{cl}
        d\xs & \NLOQ2
      \end{tabular}
      &
      \begin{tabular}{cl}
        d\xs & \NLOQ2 + \NLOEone
      \end{tabular} \\
    \hline
    $pp\to \gamma\gamma+j$ &
      \begin{tabular}{cl}
        d\xs & \NLOQone
      \end{tabular}
      &
      \begin{tabular}{cl}
        d\xs & \NLOQ2 + \NLOEone
      \end{tabular} \\
    \hline
  \end{tabular}
  \caption{Precision wish list: vector boson final
    states. $V=W,Z$ and $V',V''=W,Z,\gamma$.}
  \label{tab:wlV}
\end{table}

\begin{itemize}[leftmargin=2cm]
  \item[$V$:] Inclusive cross-sections and rapidity distributions in the
    threshold limit have been extracted from the $pp\to H$ results
    \cite{Ahmed:2014cla,Ahmed:2014uya}. Parton shower matched \NLOQ2
    computations using both the MiNLO method \cite{Karlberg:2014qua}, SCET
    resummation \cite{Alioli:2015toa} and via the UN${}^2$LOPS technique \cite{Hoeche:2014aia}. Completing the inclusive \NLOQ3 computation
    beyond the threshold limit is an important step for phenomenological studies.
    The dominant factorisable corrections at $\mathcal{O}(\alpha_s \alpha)$ (\NLOQE11) are also now available \cite{Dittmaier:2015rxo}.
  \item[$V+j$:] Both $Z+j$~\cite{Ridder:2015dxa,Boughezal:2015ded,Boughezal:2016isb,Boughezal:2016yfp} and
    $W+j$~\cite{Boughezal:2015dva,Boughezal:2016dtm,Boughezal:2016yfp} have recently been completed through
    to \NLOQ2 including leptonic decays.
  \item[$V+\geq2j$:] While fixed order \NLOQone~computations of $V+\geq2$ jet final states have been known for many years
    recent progress has been made for \NLOEone~corrections~\cite{Denner:2014ina} including merging and showering \cite{Kallweit:2014xda,Kallweit:2015dum}.
  \item[$VV',V'\gamma$:] Complete \NLOQ2 are now available for $WW$ \cite{Gehrmann:2014fva,Grazzini:2016ctr},
  $ZZ$ \cite{Cascioli:2014yka,Grazzini:2015hta},
  $Z\gamma$  \cite{Grazzini:2015nwa},
  $W\gamma$   \cite{Grazzini:2015nwa} using the $q_T$ subtraction method. These results should become publicly
  available in the near future including all on-shell leptonic
  decays. There has been good progress in calculating 
  NLO EW corrections for complete processes including decays~\cite{Denner:2014bna,Denner:2015fca,Biedermann:2016yvs,Biedermann:2016guo}. Thereby also the new automated approaches have been employed and a range of decays has
  been considered. Very recently the $WZ$ final state has been computed at \NLOQ2~\cite{Grazzini:2016swo}.
  The remaining $VV'$ processes, together with other colourless final states, will be available in the forthcoming \textsc{Matrix} Monte Carlo
  program based on $q_T$ subtraction.
\item[$\gamma\gamma, \gamma\gamma+j$:] This process remains an important
  ingredient in Higgs measurements at Run II. Originally computed at \NLOQ2
  with $q_T$ subtraction \cite{Cieri:2015rqa}, it has recently been re-computed
  \cite{Campbell:2016yrh} using the $N$-jettiness subtraction implemented
  within MCFM. The $q_T$ resummation at NNLL requested on the 2013 wish list
  are also now available \cite{Cieri:2015rqa}. Given the recent excitement in
  di-photon production a detailed understanding of these processes at high $q_T$ will
  be important in the coming years. Prospects for \NLOQ3 corrections remain
  closely connected with differential Higgs and Drell-Yan production at \NLOQ3.
  At high transverse momentum it may also be interesting to have \NLOQ2
  predictions for $\gamma\gamma+j$. Given that this is of equivalent complexity
  to $3j$ production we add this process to the wish list.

\end{itemize}

\subsection{Top quark and jet associated processes}

\begin{table}
  \centering
  \begin{tabular}{lll}
    \hline
    \multicolumn{1}{c}{process} & \multicolumn{1}{c}{known} & \multicolumn{1}{c}{desired} \\
    \hline
    $pp\to t\tb$ &
      \begin{tabular}{cl}
        d\xs & \NLOQ2 \\
        d\xs & \NLOQone + decays \\
        d\xs & \NLOEone \\
      \end{tabular}
      &
      \begin{tabular}{cl}
        d\xs & \NLOQ2 + \NLOEone + decays
      \end{tabular} \\
    \hline
    $pp\to t\tb+j$ &
      \begin{tabular}{cl}
        d\xs & \NLOQone + decays \\
        d\xs & \NLOEone
      \end{tabular}
      &
      \begin{tabular}{cl}
        d\xs & \NLOQ2 + \NLOEone + decays
      \end{tabular} \\
    \hline
    $pp\to t\tb+2j$ &
      \begin{tabular}{cl}
        d\xs & \NLOQone + on-shell decays
      \end{tabular}
      &
      \begin{tabular}{cl}
        d\xs & \NLOQone + \NLOEone + decays
      \end{tabular} \\
    \hline
    $pp\to t\tb+V$ &
      \begin{tabular}{cl}
        d\xs & \NLOQone \\
        d\xs & \NLOEone
      \end{tabular}
      &
      \begin{tabular}{cl}
        d\xs & \NLOQone + \NLOEone + decays
      \end{tabular} \\
    \hline
    $pp\to t/\tb$ &
      \begin{tabular}{cl}
        d\xs & \NLOQ2 ($t$-channel)
      \end{tabular}
      &
      \begin{tabular}{cl}
        d\xs & \NLOQ2 + \NLOEone + decays
      \end{tabular} \\
    \hline
    \hline
    $pp\to 2j$ &
      \begin{tabular}{cl}
        d\xs & \NLOQ2(gg,qq) \\
        d\xs & \NLOEone
      \end{tabular}
      &
      \begin{tabular}{cl}
        d\xs & \NLOQ2 + \NLOEone
      \end{tabular} \\
    \hline
    $pp\to j+\gamma$ &
      \begin{tabular}{cl}
        d\xs & \NLOQone \\
        d\xs & \NLOEone
      \end{tabular}
      &
      \begin{tabular}{cl}
        d\xs & \NLOQ2 + \NLOEone
      \end{tabular} \\
    \hline
    $pp\to 3j$ &
      \begin{tabular}{cl}
        d\xs & \NLOQone
      \end{tabular}
      &
      \begin{tabular}{cl}
        d\xs & \NLOQ2 + \NLOEone
      \end{tabular} \\
    \hline
  \end{tabular}
  \caption{Precision wish list: top quark and jet final states}
  \label{tab:wlTJ}
\end{table}

\begin{itemize}[leftmargin=2cm]
  \item[$t\tb$:] Another major achievement since the 2013 report has been the completion of fully differential
    predictions for $t\tb$ production at \NLOQ2~\cite{Czakon:2015owf,Czakon:2016ckf}. An on-shell description
    of the top quarks has been taken and it would be beneficial to extend these to studies including a full set
    of decays in the narrow width approximation in order to make fully fiducial predictions.
  \item[$t\tb V$:] \NLOQone~corrections to $t\tb Z$ including a full range of decays have been considered recently \cite{Rontsch:2014cca,Rontsch:2015una}.
    This will help to improve the constraints on anomalous EW couplings in the top quark sector during Run II. \NLOEone\ corrections have also been
    computed within the automated \WishListMGaMC framework \cite{Frixione:2015zaa}.
  \item[$2j$:] This process is now almost fully complete at \NLOQ2. The leading colour pure gluonic corrections known previously \cite{Ridder:2013mf}
    have been expanded to include sub-leading colour $gg\to gg$ \cite{Currie:2013dwa} and $q\qb\to gg$ \cite{Currie:2014upa}. The $qg$ initiated
    processes are expected in the near future.
  \item[$t$/$\tb$:] Fully differential \NLOQ2~corrections have been completed for the dominant $t$-channel production process \cite{Brucherseifer:2014ama}.
\end{itemize}

\renewcommand{\arraystretch}{1}
\clearpage

\chapter{Parton distribution functions}
\label{cha:pdf}

\section{Construction and phenomenological applications of PDF4LHC parton distributions 
  \texorpdfstring{\protect\footnote{
    J.~Gao, T.-J.~Hou, J.~Huston, P.~M.~Nadolsky, B.~T.~Wang, K.~P.~Xie
  }}{}}

We revisit the construction and application of
combined PDF sets (PDF4LHC15) developed by the PDF4LHC group in
2015. Our focus is on the meta-analysis technique employed in the
construction of the 30-member PDF4LHC15 sets, and  
especially on aspects that were not fully described in the main
PDF4LHC recommendation document. These aspects include  
construction of the 30-member sets at NLO (in addition to NNLO),
extension of the NLO and NNLO sets to low QCD scales, and construction
of such sets for 4 active flavors. In addition, we clarify a point
regarding the calculation of parton luminosity uncertainties at low
mass. Finally, we present a website containing predictions based on
PDF4LHC15 PDFs for some crucial LHC processes. 

\subsection{Introduction}
To simplify applications of parton distribution functions (PDFs) in
several categories of LHC experimental simulations, 
the 2015  recommendations \cite{Butterworth:2015oua} of the PDF4LHC working group
introduce combinations of CT14~\cite{Dulat:2015mca}, MMHT2014~\cite{Harland-Lang:2014zoa}, and NNPDF3.0~\cite{Ball:2014uwa} PDF ensembles,
by utilizing the Monte Carlo (MC) replica
technique \cite{Watt:2012tq}. The central PDF and the uncertainties
of the combined set are derived from the
900 MC replicas of the error PDFs of the above three input
ensembles. As the 900 error PDFs are often too many to be manageable,
they are ``compressed'' into smaller PDF error sets using three reduction techniques
\cite{Gao:2013bia,Carrazza:2015hva,Carrazza:2015aoa}.  Consequently, the final combined PDFs come in
three versions, one with 30 error sets (PDF4LHC15\_30), and the other two
with 100 error sets (PDF4LHC15\_100 and PDF4LHC15\_MC). Two of these, 
PDF4LHC15\_30 and PDF4LHC\_100, are constructed in the form of Hessian
eigenvector sets \cite{Pumplin:2002vw}. The PDF4LHC15\_MC ensemble is
constructed from MC replicas. The central sets are the same in the
900-replica prior as well as in the  \_100, 
\_30, and \_MC ensembles. They are equal to the average of central
sets of CT14, MMHT2014, and NNPDF3.0 ensembles.  
The error sets of the three PDF4LHC15 ensembles are different,
reflecting the specifics of each reduction technique. They are available in
the LHAPDF library \cite{Buckley:2014ana} at NLO and NNLO in QCD coupling
strength $\alpha_s$, with the central value of $\alpha_{s}(M_{Z})$
equal to 0.118, and with additional sets corresponding to the
$\alpha_s$ variations by 0.0015 around the central value.  

The 30-member ensemble is constructed using the meta-parametrization
technique introduced in \cite{Gao:2013bia}.
This contribution describes additional developments in the
30-member ensemble that happened at the time, 
or immediately after, the release of
the original PDF4LHC recommendation document. They include
construction of the PDF4LHC15\_30 ensemble at NLO, extension of
PDF4LHC15\_30 to scales below 8 GeV, and the specialized ensemble with
4 active quark flavors. These features are already incorporated in the LHAPDF distributions. We provide comparisons of PDFs
and parton luminosities and introduce a website \cite{SMUgallery} 
illustrating essential
LHC cross sections computed with the PDF4LHC15 and other ensembles,
and using a variety of QCD programs.  

When deciding on which of the three PDF4LHC sets to use, it is
important to keep in mind that all of them reproduce well the uncertainties of the
900-replica ``prior'' PDF ensemble. 
This prior itself has some uncertainty both in its central value and
especially in the size of the PDF uncertainty itself, reflecting  differences between the
central values and the uncertainty bands of CT14, MMHT2014 and NNPDF3.0, 
which become especially pronounced at very low $x$ and high $x$.  
At moderate $x$ values, contributing to the bulk of
precision physics cross sections at the LHC, the agreement between the
three input PDF sets is often quite better, meaning that the combined
prior and the three reduced ensembles constructed from it are also known well.  
In general, the 30-member ensemble keeps the lowest,
best-known eigenvector sets, and thus provides a slightly lower estimate for
the uncertainty of the 900-replica prior, but one that is known
with higher confidence than the exact uncertainty of the prior set. 
We will demonstrate that, across many practical applications,  
the 30-member error estimates are typically close both to those of the
prior and of the Hessian 100 PDF error set. 

\subsection{QCD scale dependence of the 30-member NLO PDF4LHC ensemble}
The NLO meta-parametrizations 
are constructed in a slightly different manner compared
to the NNLO version. In Ref.~\cite{Gao:2013bia}, we have 
shown that the differences of the numerical
implementation of DGLAP evolution at NNLO in CT10~\cite{Gao:2013xoa}, 
MSTW2008~\cite{Martin:2009iq}, and
NNPDF2.3~\cite{Ball:2012cx} PDFs are
negligible compared to the intrinsic PDF uncertainties.\footnote{CT10 
PDFs use the $x$-space evolution provided by the program HOPPET~\cite{Salam:2008qg}.}   
However, at NLO, the NNPDF2.3 group uses evolution
that neglects some higher-order terms compared to HOPPET,  which can result 
in deviations by up to 1~\% in the small- and large-$x$ regions, 
compared to the evolution used by CT10 and MSTW2008. These
differences in NLO numerical DGLAP evolution, while formally allowed, 
also affect the most recent generation of NLO PDFs, i.e., 
CT14+MMHT2014 vs. NNPDF3.0. When the 900-replica prior ensemble
at NLO is constructed by taking 300 replicas from each of the input CT14,
MMHT14, and NNPDF3.0 ensembles, the implication is that $Q$-scale
dependence of these replicas is not strictly Markovian. 
Probability regions at the low $Q$ scale, as sampled by the MC
replicas, are not exactly preserved by DGLAP evolution to a higher
$Q$ scale. This is in contrast to the consistent DGLAP evolution of a
single input PDF set, which guarantees that the probability/confidence
value associated with a given error set is independent of the $Q$
scale.

\begin{figure}[t]
\includegraphics[width=0.47\textwidth]{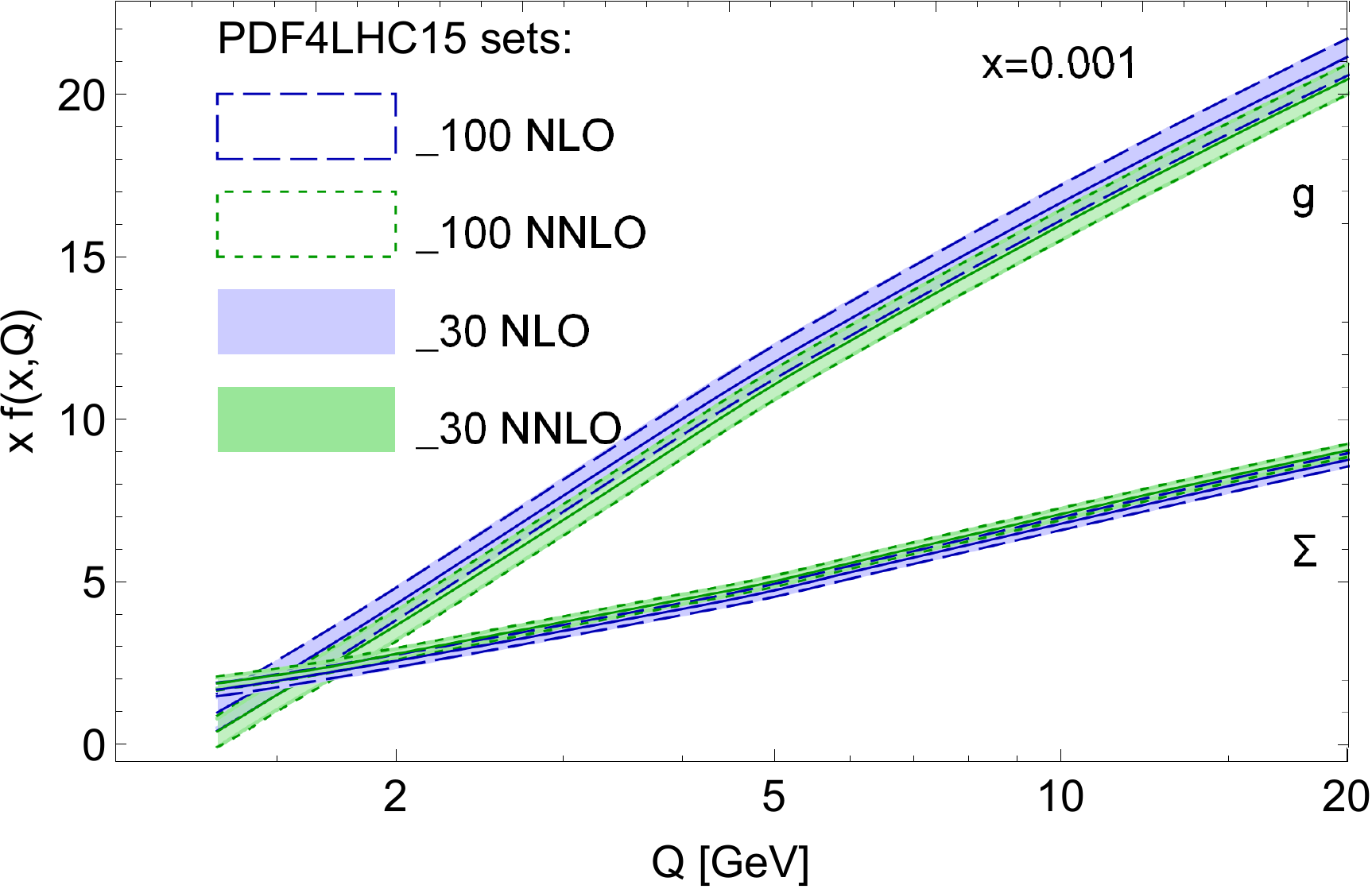}
\hfill
\includegraphics[width=0.47\textwidth]{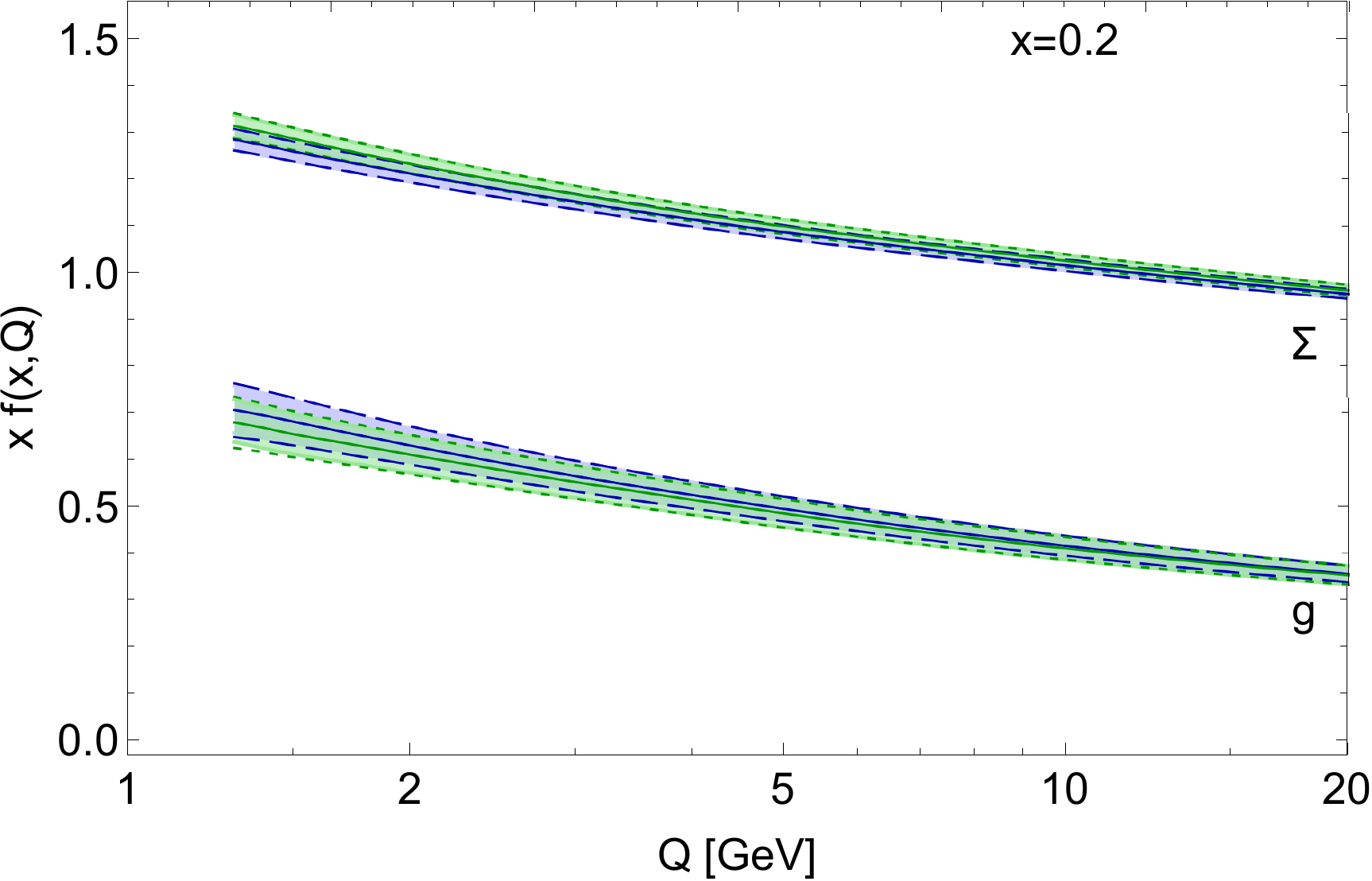}
\caption{The singlet and gluon PDFs, $\Sigma(x,Q)$ and $g(x,Q)$, from 100-
and 30-member PDF4LHC15 sets at NLO and NNLO, plotted vs. the QCD
scale $Q$ at $x=10^{-3}$ (left) and $0.2$ (right).\label{fig:SingletGluonVsQ}}
\end{figure}

Thus, the NLO prior ensemble is not inherently
consistent, even though the deviations in DGLAP evolution of
individual replicas are arguably small. One should apply 
a correction to restore the Markovian nature 
of the evolution. In the PDF4LHC15\_30 NLO set we do this by first
constructing the central PDF set at any $Q$ by averaging the CT14,
MMHT14, and NNPDF3.0 central sets that were evolved by their own
native programs. Then, we reduce the 900-member into the 30-member set
at scale $Q_0=8$ GeV and evolve all replicas to other $Q$ values using
HOPPET. Finally, we estimate the difference between the HOPPET
evolution and
native evolution of the central set, and subtract this difference at
every $Q$ from the HOPPET-evolved values of every error set. After such
universal shift, the $Q$ dependence of all error sets is
practically the same as the native evolution of the central PDF. The
probability regions are now independent of $Q$; this preserves
sum rules for momentum and quark quantum numbers. 

\subsection{PDF4LHC15\_30 PDFs at low $Q$}

\begin{figure}[t!]
\includegraphics[width=0.47\textwidth]{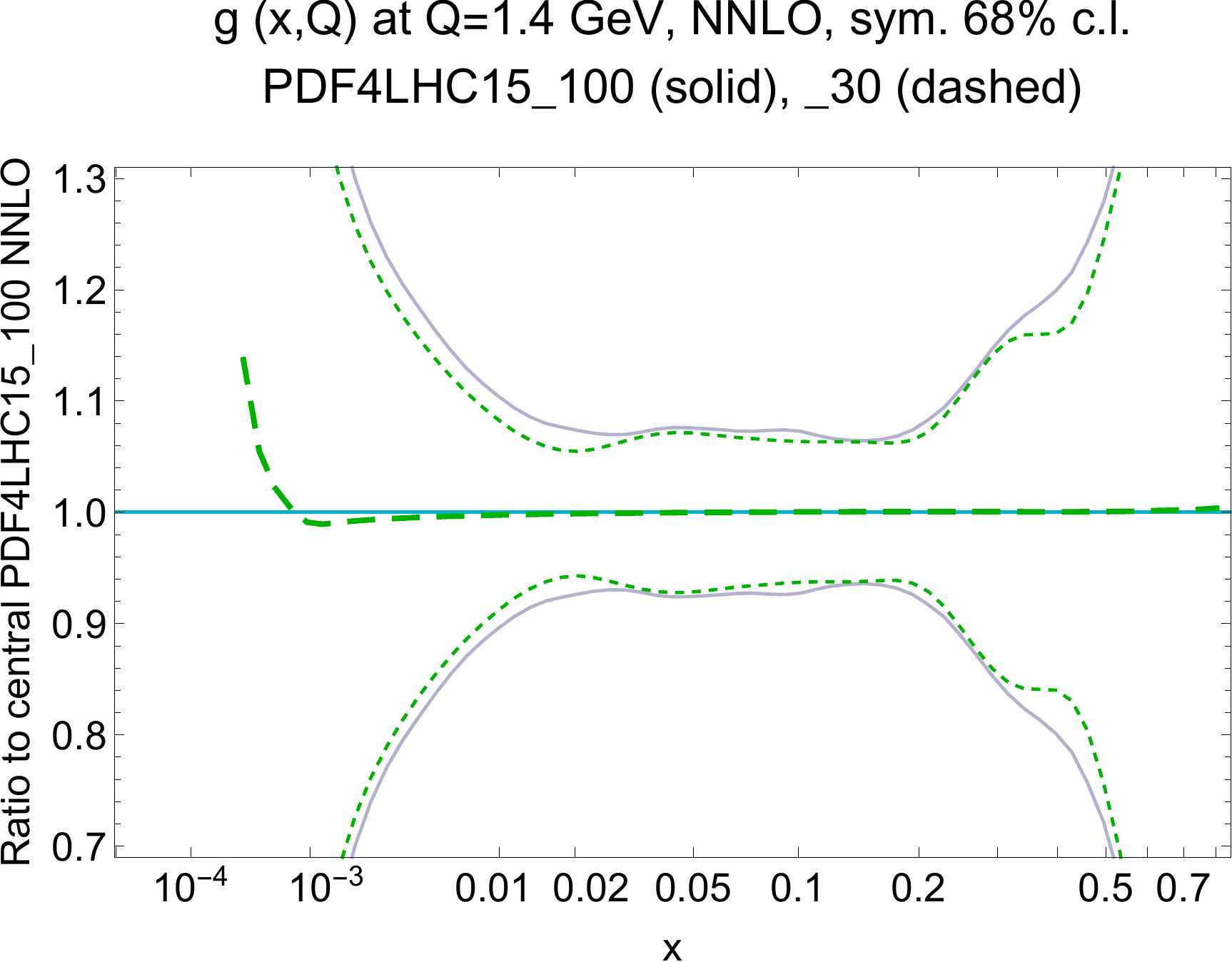}
\hfill
\includegraphics[width=0.47\textwidth]{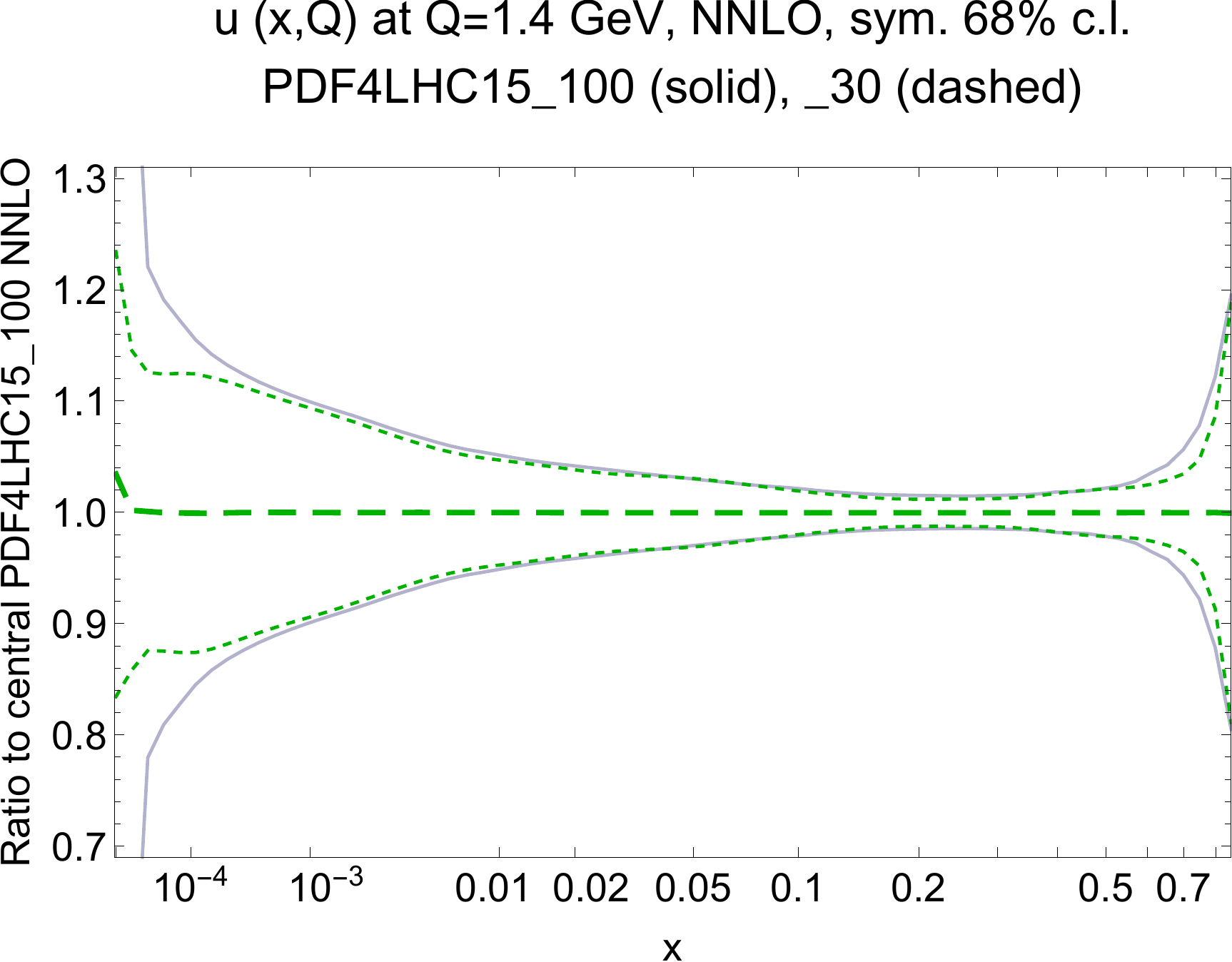}\\[4mm]
\includegraphics[width=0.47\textwidth]{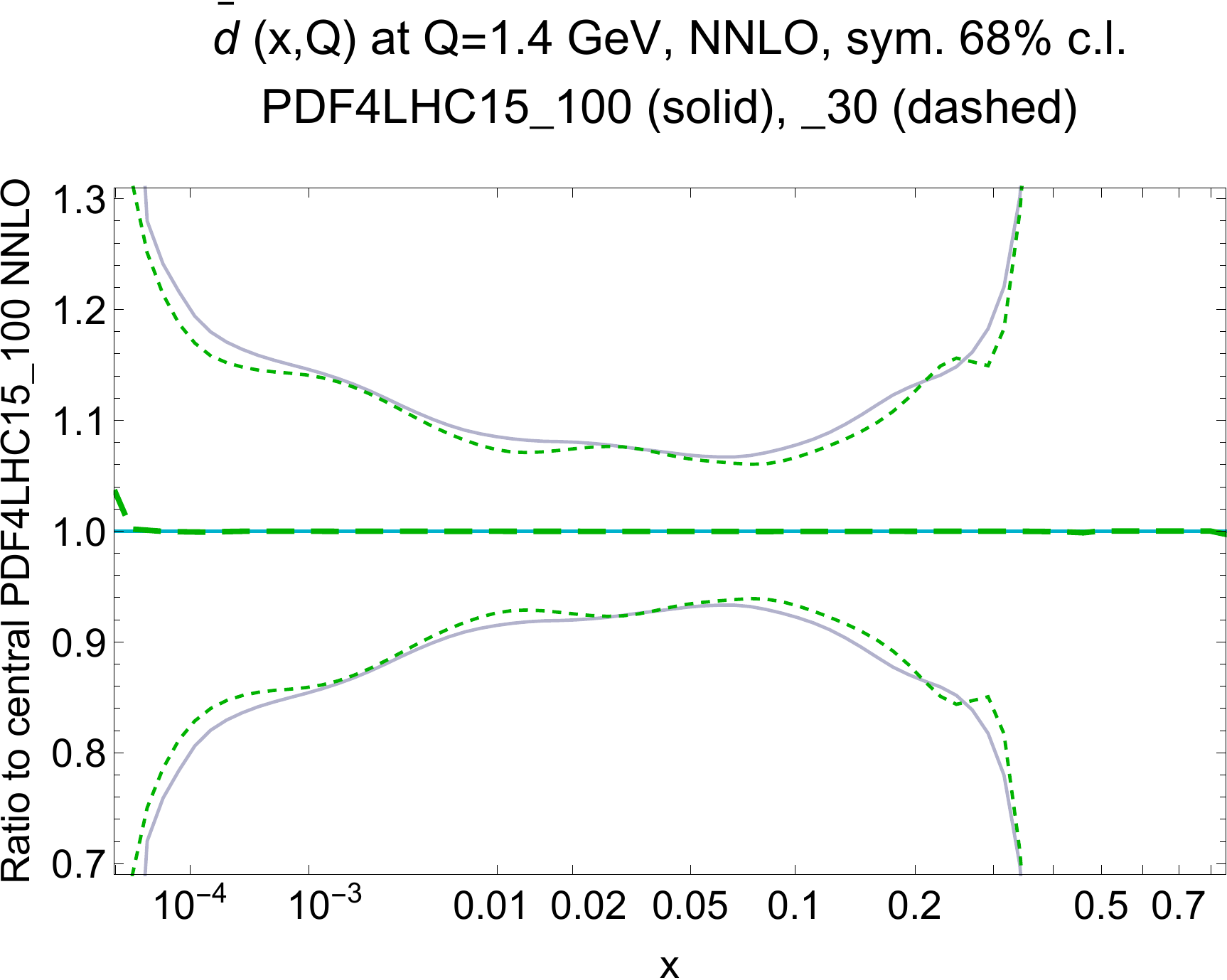}
\hfill
\includegraphics[width=0.47\textwidth]{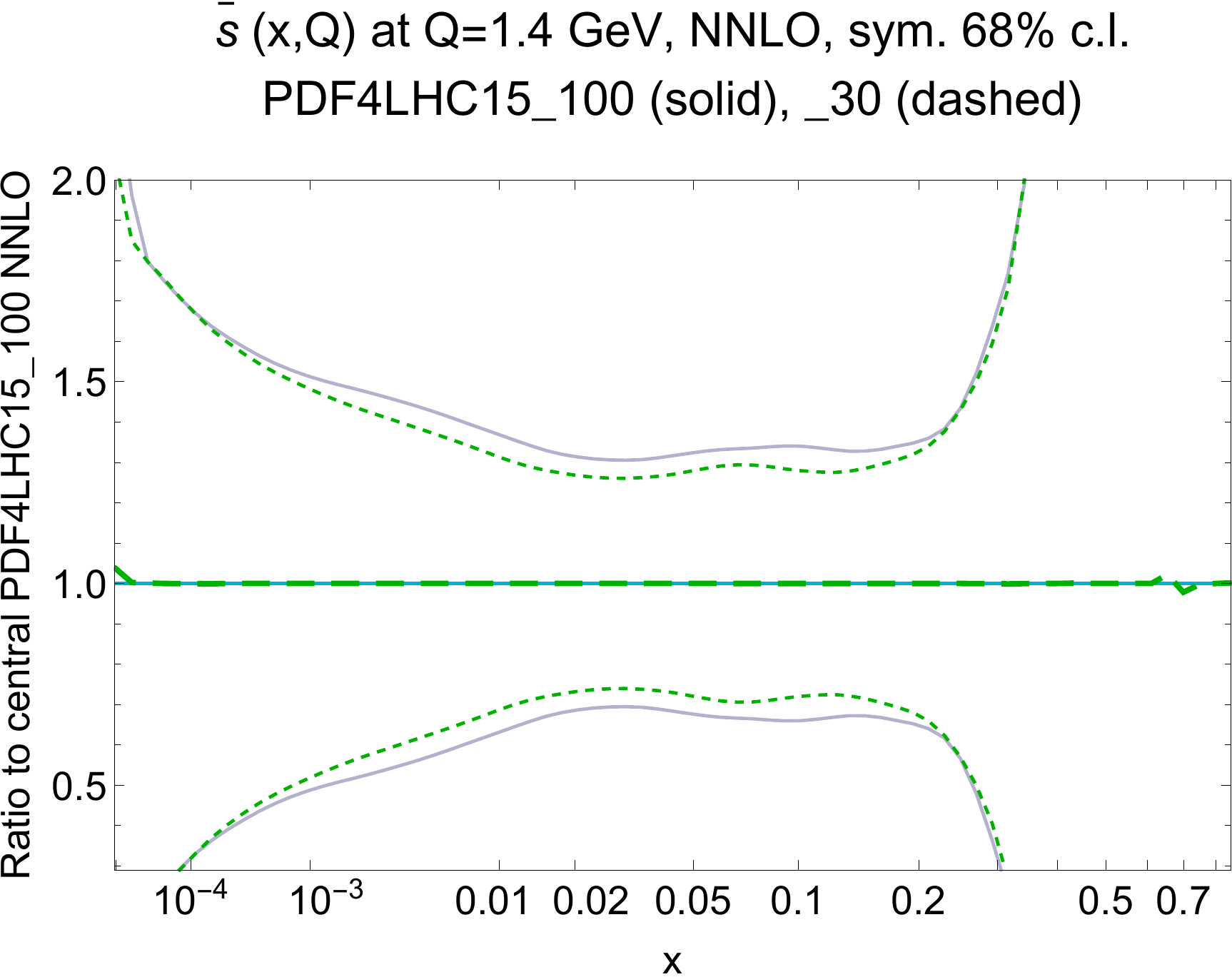}
\caption{
PDF central predictions and uncertainty bands for select parton flavors
from the 100- and 30-member NNLO PDF4LHC15 ensembles, plotted versus
$x$ at a QCD scale $Q=1.4$ GeV as ratios to the central PDF4LHC15\_100
distributions.\label{fig:NNLO100and30Vsx}}
\end{figure}

The original formulation of the meta-PDFs had a minimum $Q$ value of 8
GeV. The relatively high lower cutoff on $Q$ was introduced to justify 
the combination of PDFs obtained in different heavy-quark
schemes, and it is sufficient to describe all high-$Q^2$ physics at
the LHC. However, the extension of the \_30 PDFs
down to lower $Q$ values can be useful, too
as for
example in the simulatation of parton showers and the underlying event in
Monte-Carlo showering programs. The PDF4LHC15\_30 version on LHAPDF includes 
such an extension down to a $Q$ value of 1.4 GeV, obtained by backward
evolution from 8 GeV using HOPPET. It should be remembered
that the PDF4LHC15 combination is statistically consistent when
the factorization scale in the PDFs is much higher than the bottom
mass, as is typical in the bulk of LHC applications. The 
extension below $Q=8$ GeV should be used in less accurate aspects of the
calculation that are not sensitive to heavy-quark mass effects, 
such as inside the parton shower merged onto an (N)NLO fixed-order cross section.

\begin{figure}[t!]
\includegraphics[width=0.47\textwidth]{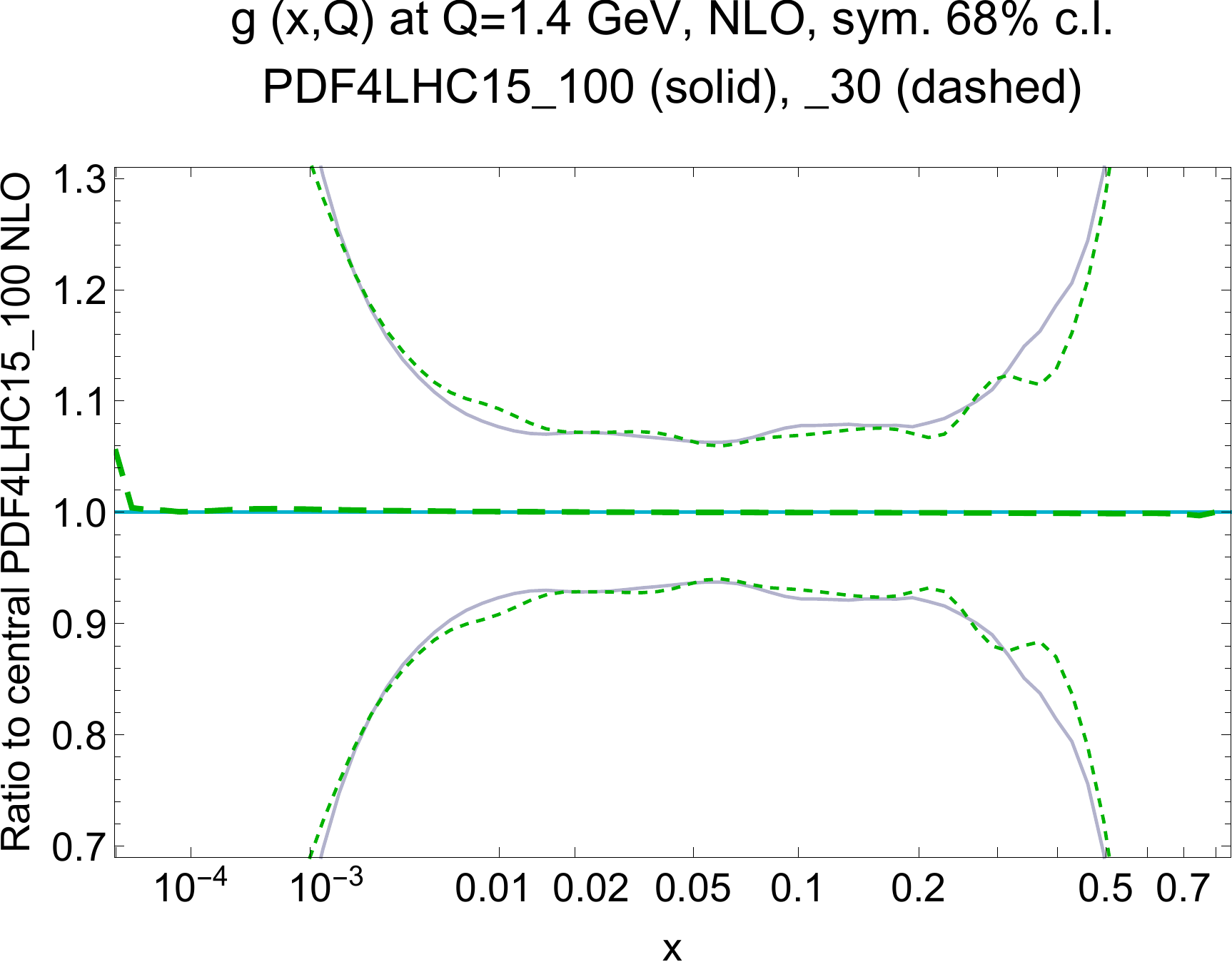}
\hfill
\includegraphics[width=0.47\textwidth]{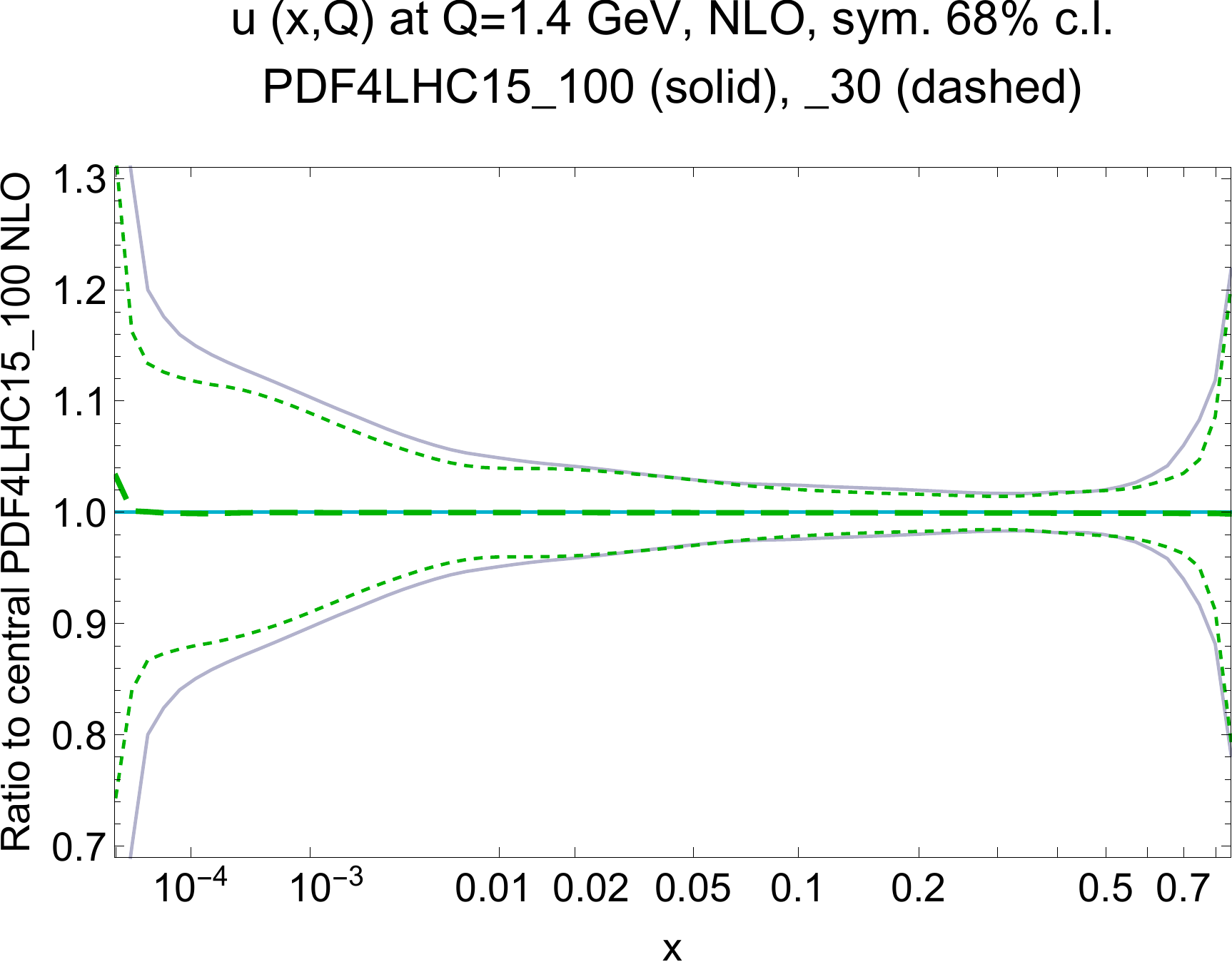}\\[4mm]
\includegraphics[width=0.47\textwidth]{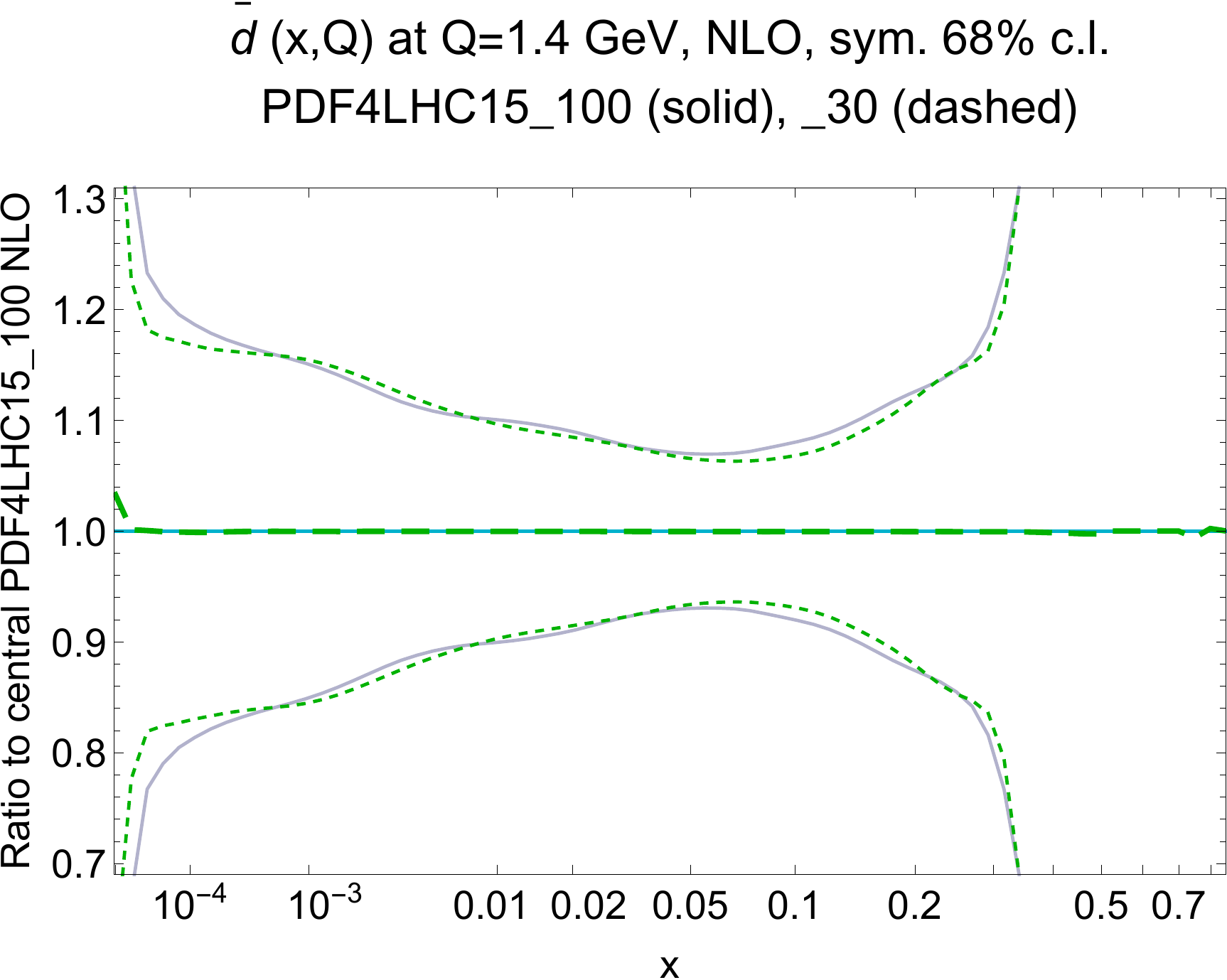}
\hfill
\includegraphics[width=0.47\textwidth]{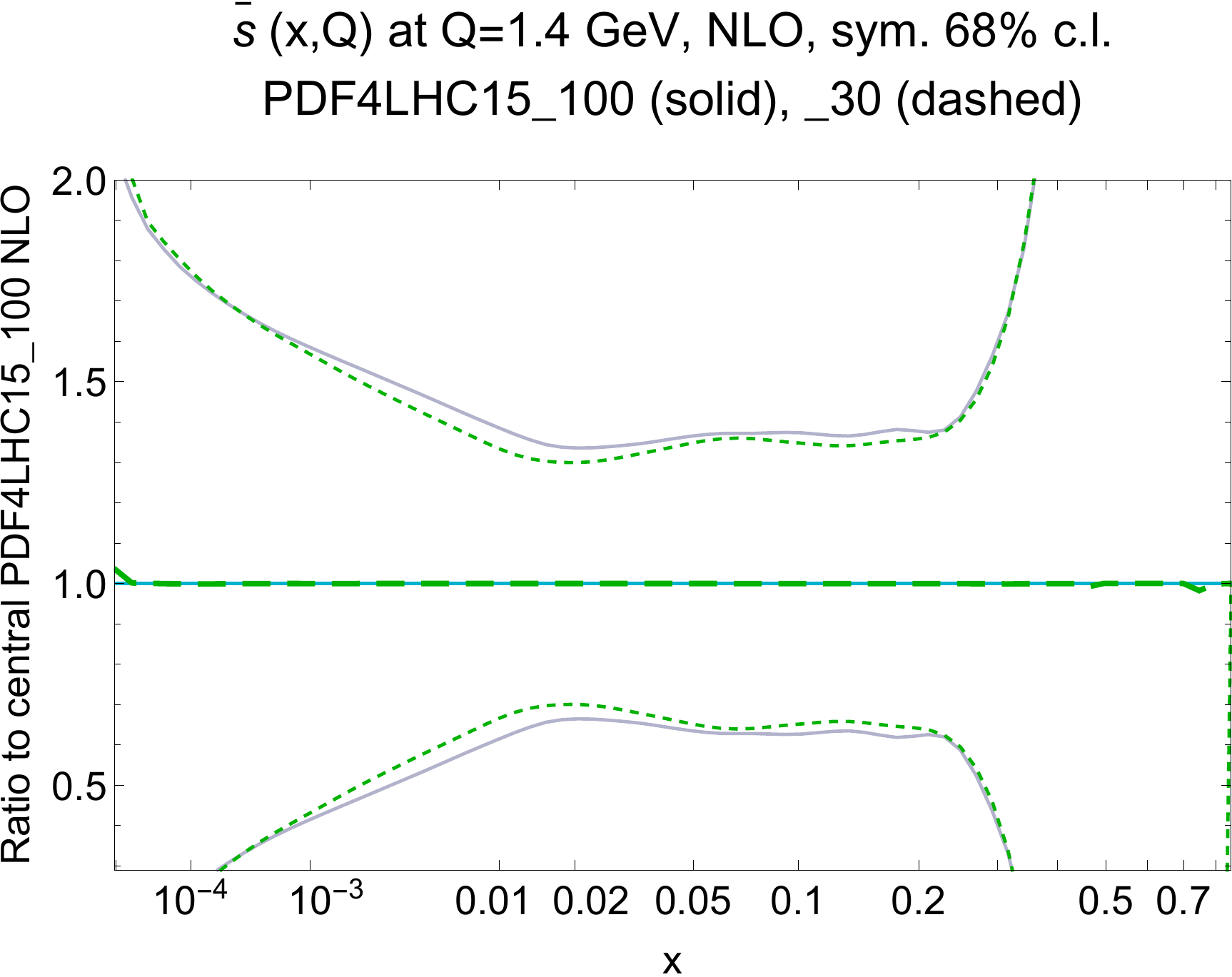}
\caption{Same as Fig.~\ref{fig:NNLO100and30Vsx}, for NLO PDF sets. \label{fig:NLO100and30Vsx}}
\end{figure}

Figure~\ref{fig:SingletGluonVsQ} illustrates the $Q$ dependence of singlet
and gluon PDFs of the \_30 and \_100 ensembles at NLO and NNLO, for
two select values of Bjorken $x$. Figs.~\ref{fig:NNLO100and30Vsx} and
\ref{fig:NLO100and30Vsx} compare the uncertainty bands 
for the $g,u,\bar{d}$ and $\bar{s}$ distributions
at a $Q$ value of 1.4 GeV, at NNLO and NLO, respectively, for the
PDF4LHC\_30 and PDF4LHC\_100 PDF sets. Good agreement between the two
sets is found in all cases; the backward evolution is smooth and
stable across the covered $Q$ range, with only minor deviations observed
below 2 GeV. [When examining the figures, recall
  that the \_30 error bands can be slightly narrower for unconstrained
  $x$ regions and PDF flavors at any $Q$].

\subsection{PDF4LHC15 parton luminosities at NLO and NNLO}

\begin{figure}[t!]
\includegraphics[width=0.47\textwidth]{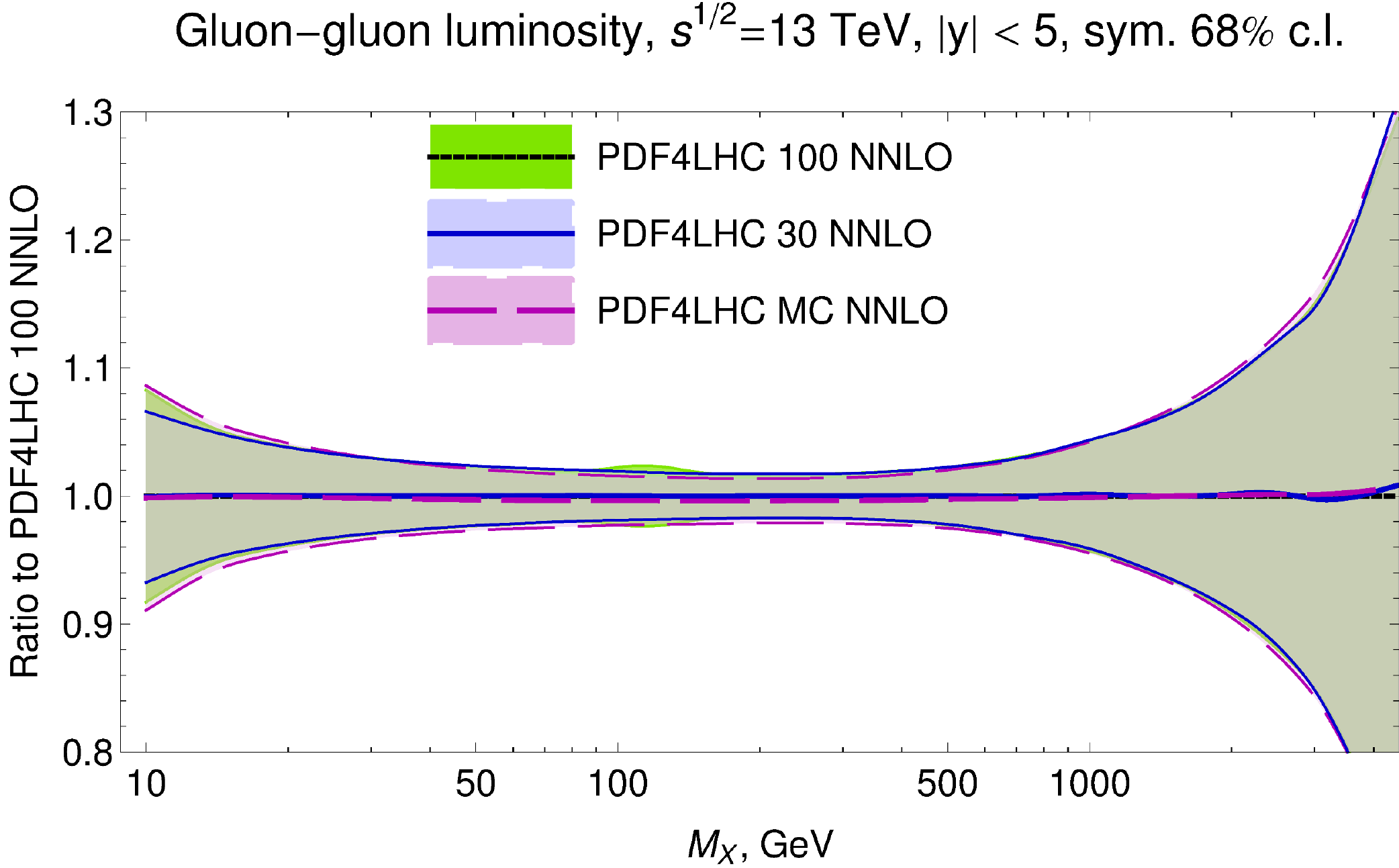}
\hfill
\includegraphics[width=0.47\textwidth]{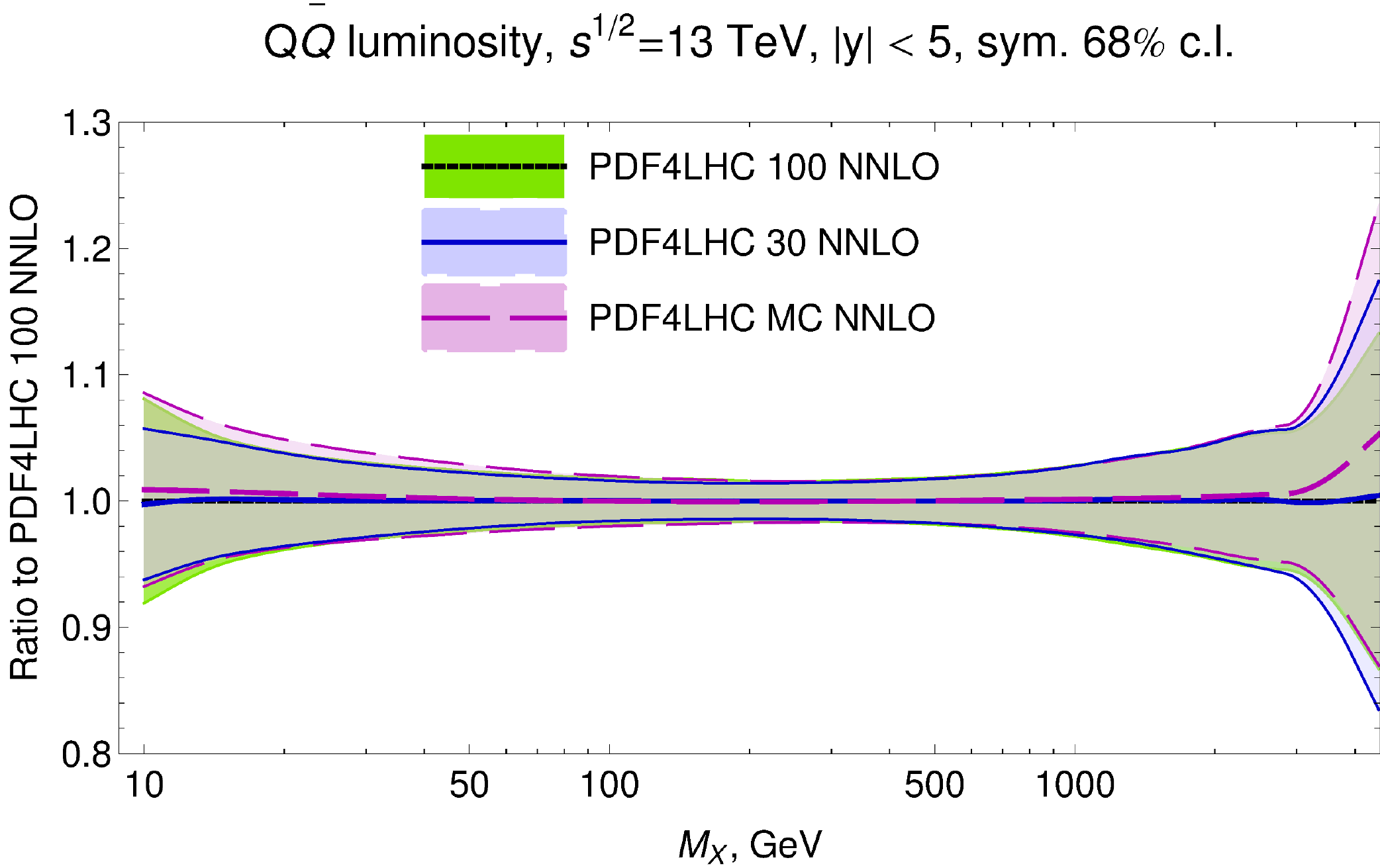}\\[4mm]
\includegraphics[width=0.47\textwidth]{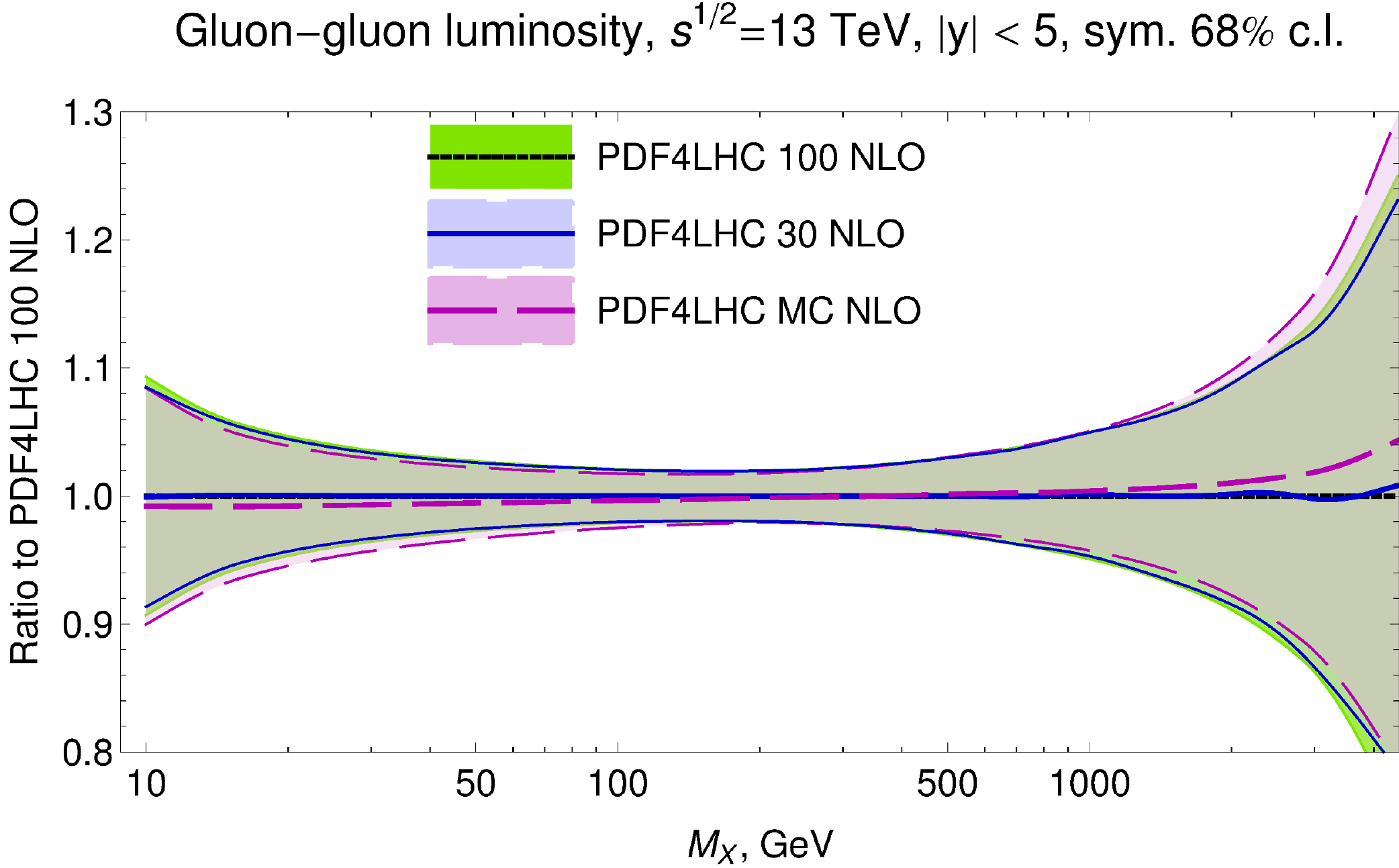}
\hfill
\includegraphics[width=0.47\textwidth]{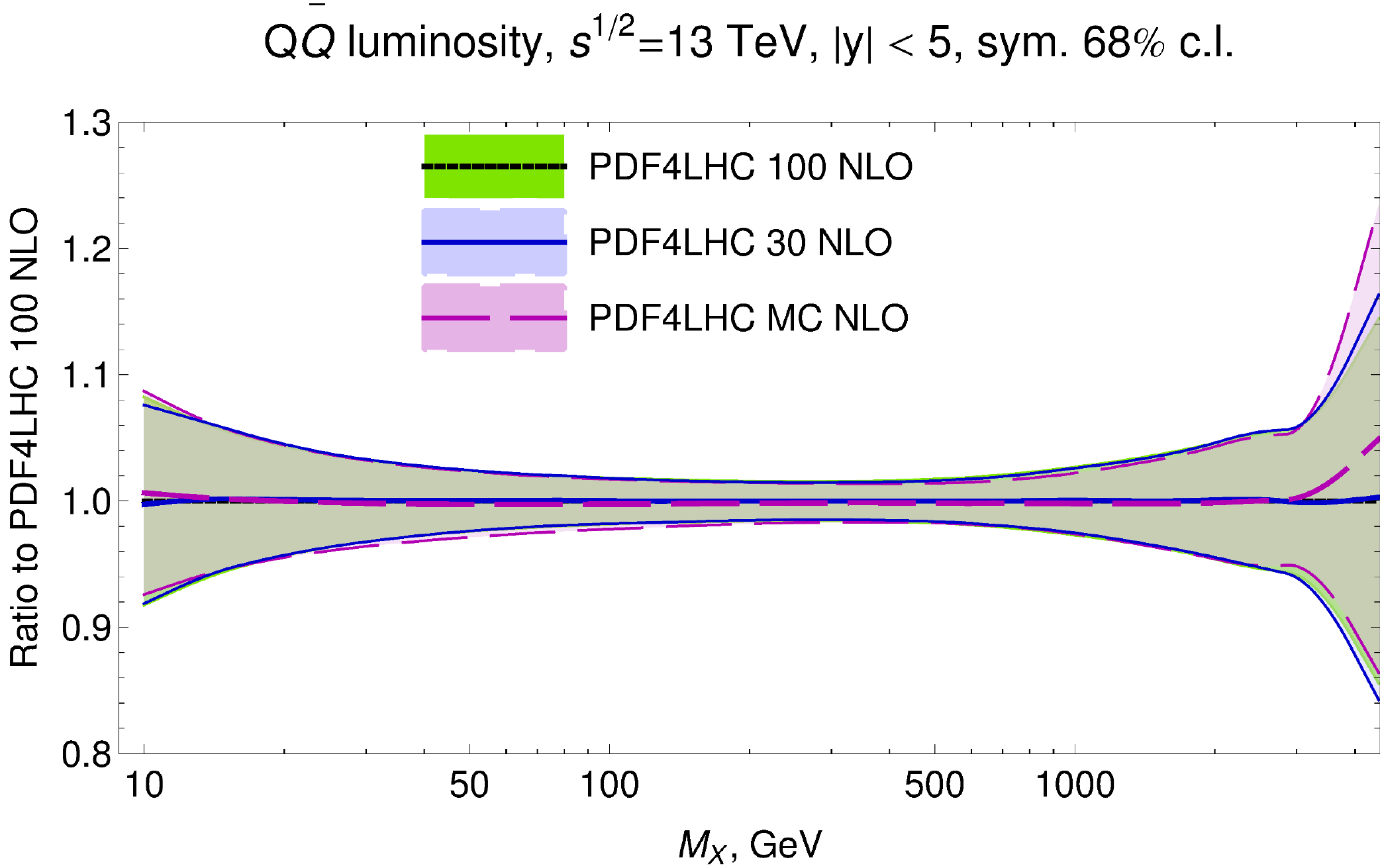}
\caption{PDF4LHC15 NNLO and NLO parton luminosities at $\sqrt{s}=13$ TeV in the experimentally accessible rapidity region $|y|<5$. \label{fig:lumi}}
\end{figure}

\begin{figure}[t!]
\includegraphics[width=0.47\textwidth]{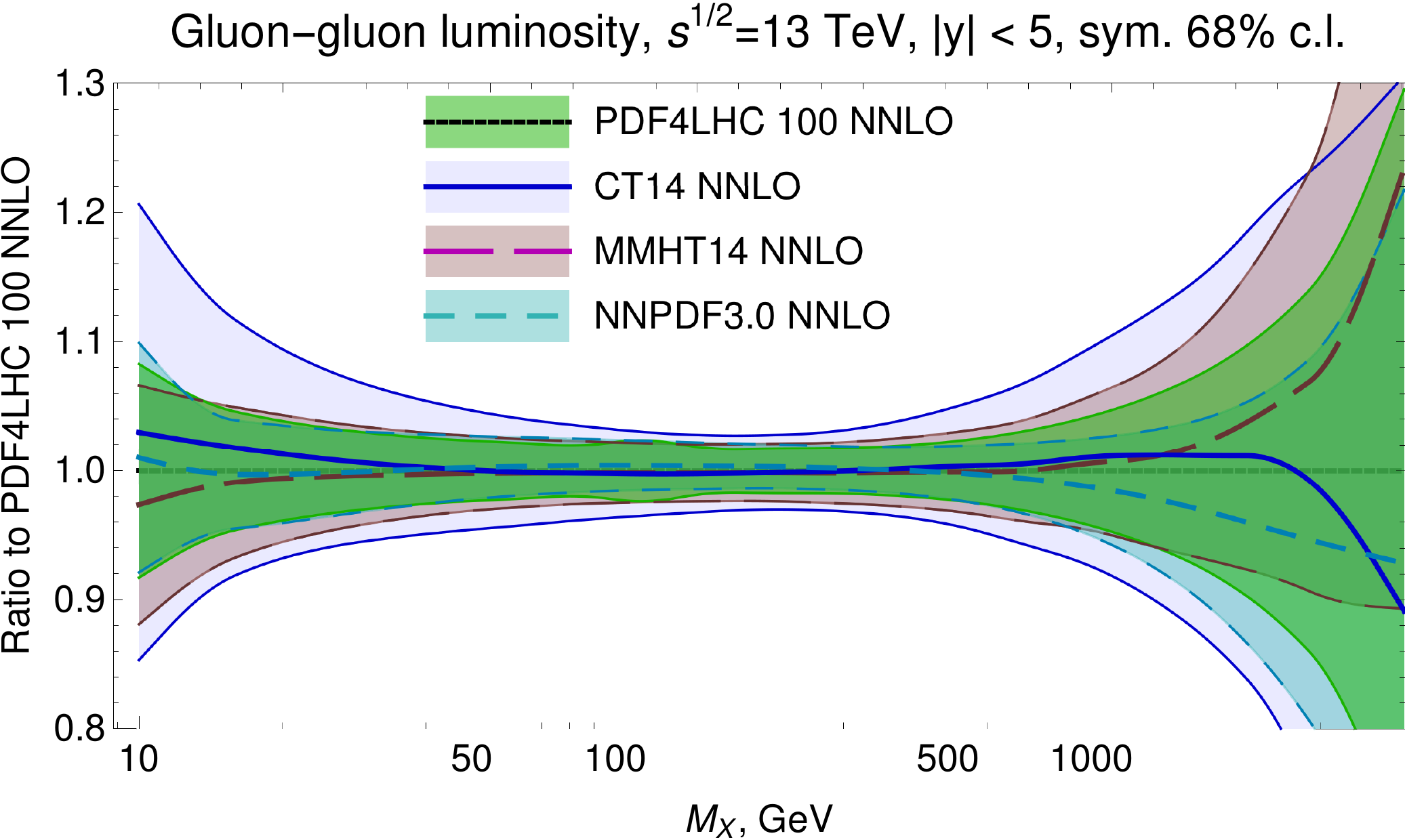}
\hfill
\includegraphics[width=0.47\textwidth]{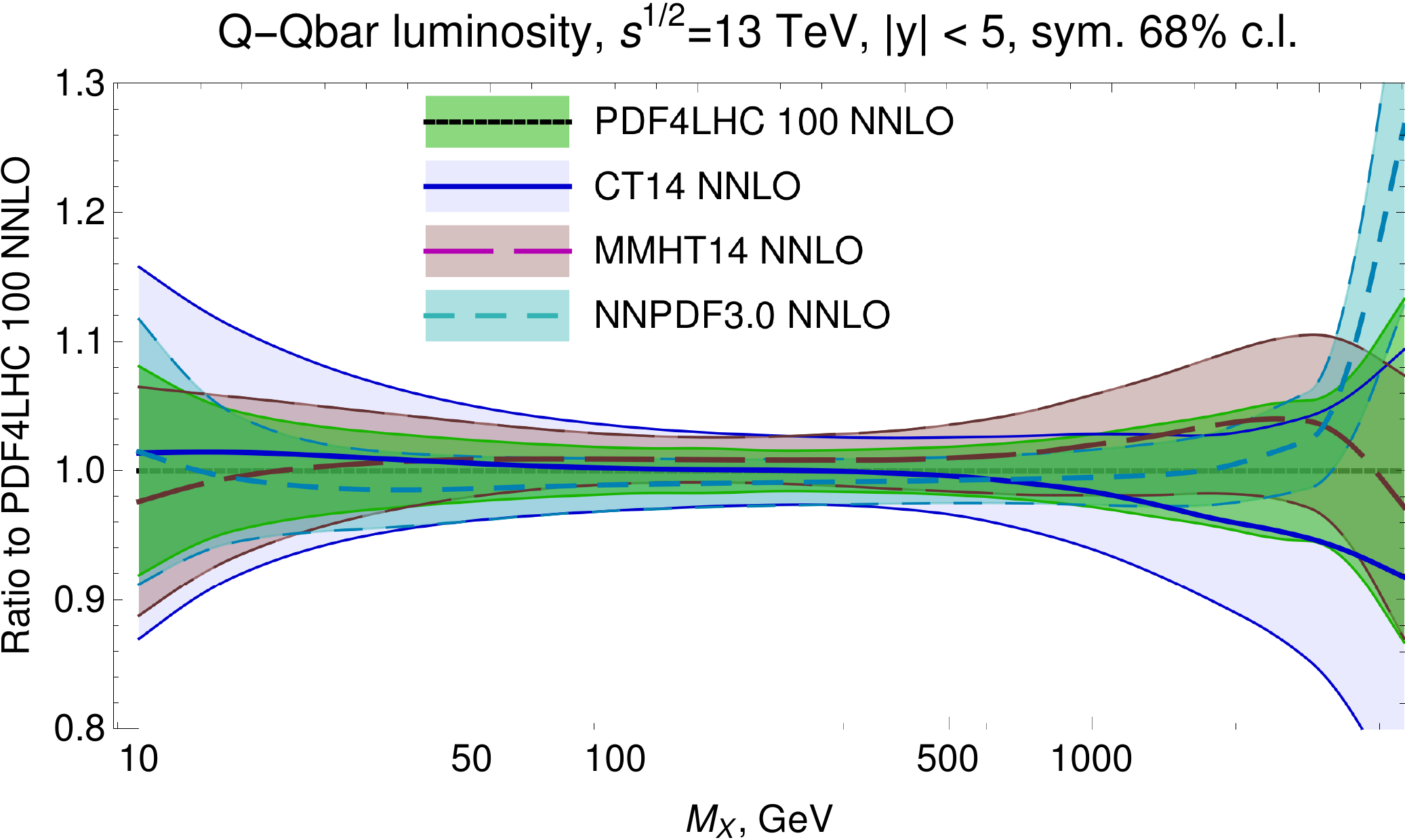}\\[4mm]
\includegraphics[width=0.47\textwidth]{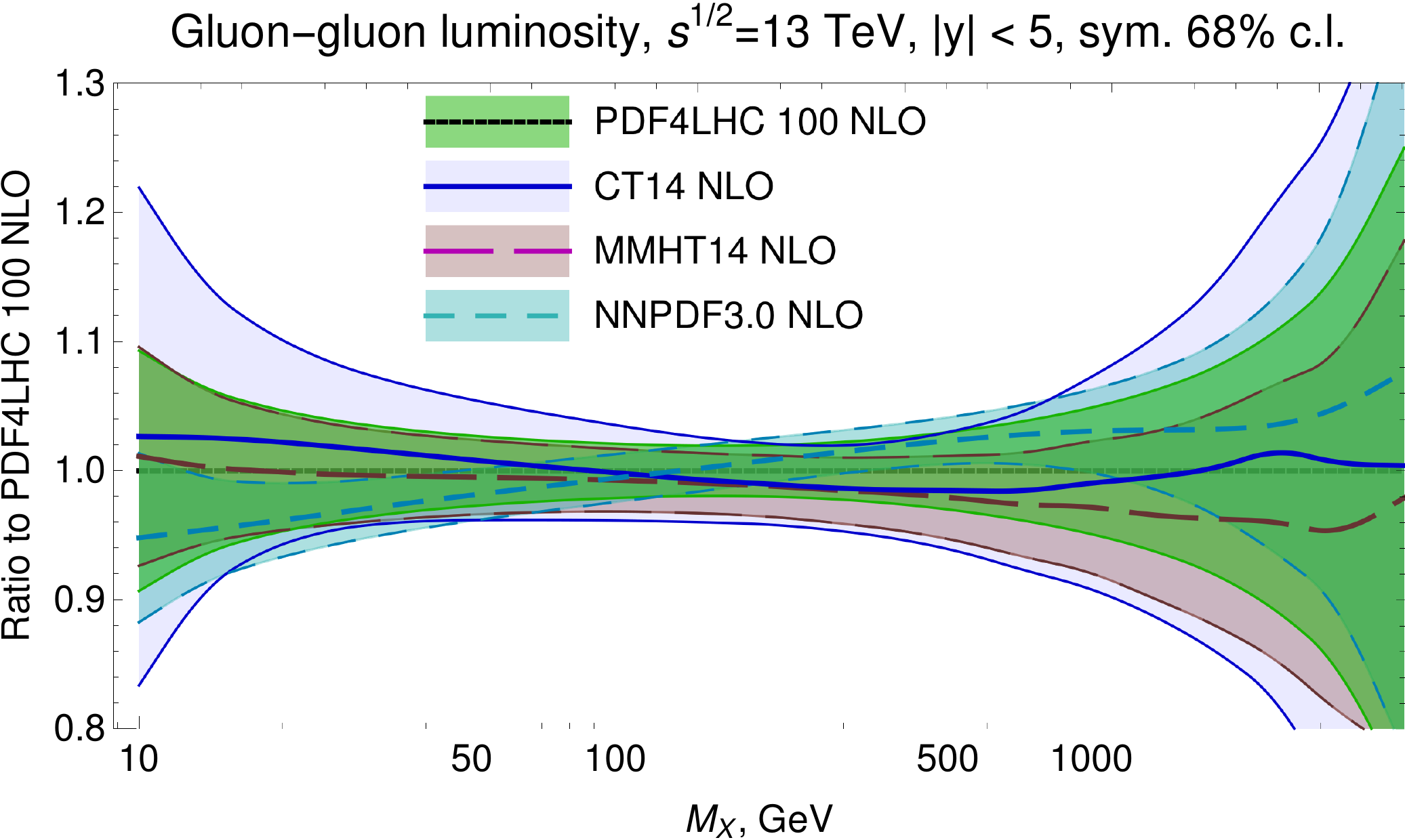}
\hfill
\includegraphics[width=0.47\textwidth]{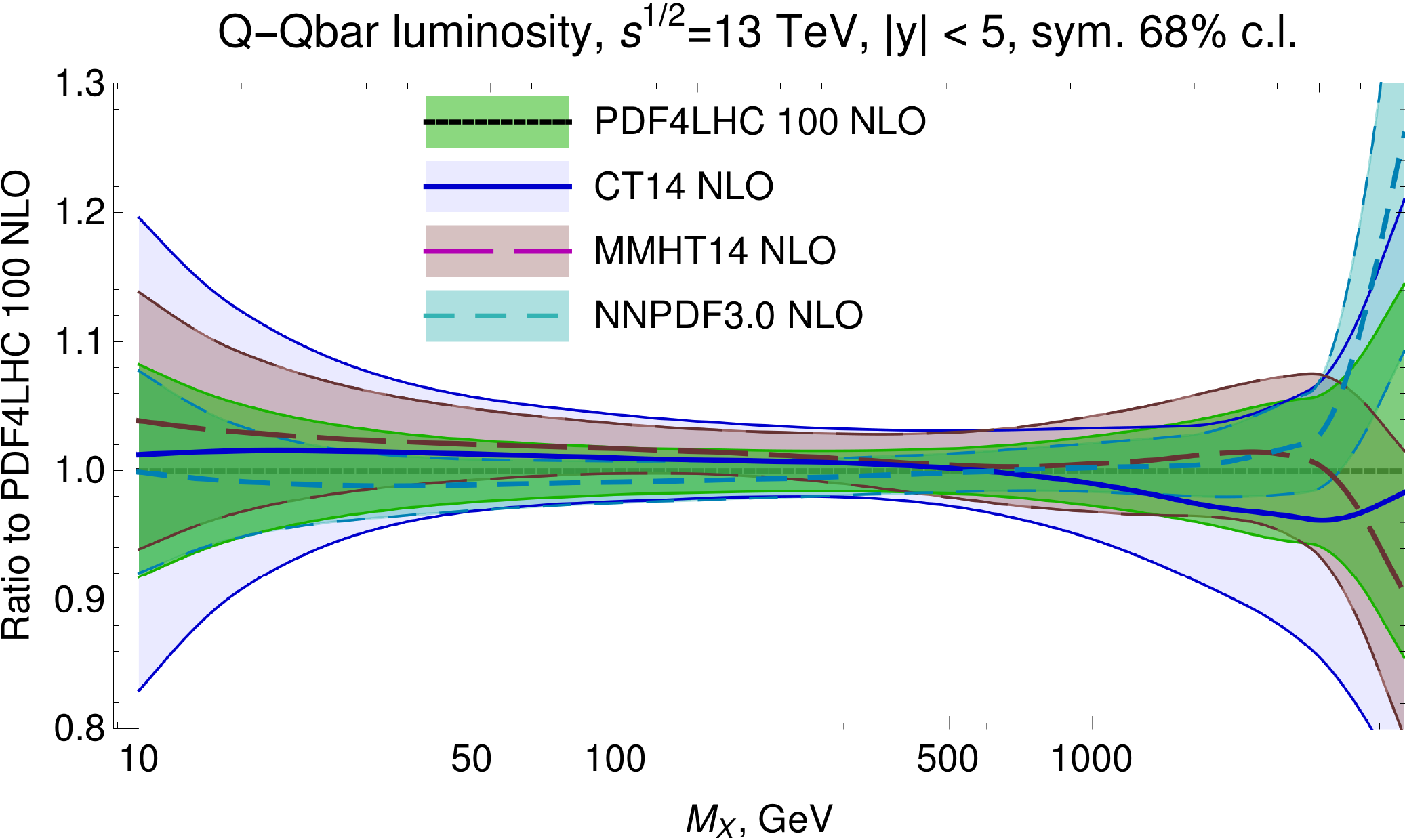}
\caption{NNLO and NLO parton luminosities for PDF4LHC15\_100, CT14, MMHT14, and NNPDF3.0 ensembles at $\sqrt{s}=13$ TeV in the experimentally accessible rapidity region $|y|<5$. \label{fig:lumi2}}
\end{figure}

Even more relevant for physics applications than the PDF error bands  are the parton luminosities. We have calculated the luminosities as a function of the mass of the final state, 
for a center-of-mass energy of
13 TeV. Comparisons of the $gg$ and $q\overline{q}$ PDF luminosities,
at NLO and NNLO, and defined as in \cite{Campbell:2006wx}, are
shown in Fig.~\ref{fig:lumi} for PDF4LHC15\_100, \_30, and \_MC sets, and 
in Fig.~\ref{fig:lumi} for PDF4LHC15\_100, CT14, MMHT14, and NNPDF3.0 sets. 
Note that the size of the uncertainties shown here,
and the level of agreement among the error bands, are 
different at low mass from those
shown in the PDF4LHC document \cite{Butterworth:2015oua}. That is
because, in our plots, a restriction has been 
applied on the $x$ values of the PDFs to
correspond to a rapidity cut of $|y| < 5$ on the produced state. 
Without such a cut, the luminosity integral at masses below 40 GeV receives
contributions from extremely low $x$ of less than $10^{-5}$, 
where (a) the uncertainties are
larger, (b) the LHAPDF grids provided for the 30 PDF sets are outside of
their tabulated range, and (c) the final state is produced in the
forward region outside of the experimental acceptance of the LHC
detectors. Without the constraints on the $x$ range, the
comparisons of parton luminosities at low mass 
are less relevant to LHC measurements. 

\subsection{4-flavor PDF4LHC15\_30 sets}

The nominal \_30 ensemble has been generated for a maximum number of
quark flavors of up to $N_f=5$. An alternative \_30 ensemble
have been now provided for a maximum quark flavors of $N_f=4$ at NLO,
based on the same prescription as for the $N_f=5$ sets, except that 
they are combined at an initial scale of 1.4 GeV in order to avoid 
backward evolution. We choose $\alpha_S(M_Z, N_f=4)=0.1126$ based on 
matching to  $\alpha_S(M_Z, N_f=5)=0.118$ with a
pole mass of 4.56 GeV for the bottom quark (equal to the average
of masses of 4.75 and 4.18 GeV from the CT14, MMHT14, and NNPDF3.0
ensembles, and consistent with the PDG pole mass value). 

\subsection{PDF4LHC15 predictions for QCD observables}
\begin{table}
\begin{center}
\begin{tabular}{|c|c|c|}\hline
 Process          &   Order  &   Type of calculation\\
 \hline\hline
 $p + p \to Z  + X $ &         NLO &    {\sc aMCFast/APPLgrid}\\
 $p + p \to W^+ + X $ &         NLO &     {\sc aMCFast/APPLgrid}    \\
 $p + p \to W^- + X $ &         NLO &     {\sc aMCFast/APPLgrid}    \\
 $p + p \to W + X $, $A_{ch,W}$ &         NLO &     {\sc aMCFast/APPLgrid}    \\
 CMS $p + p \to W(l\nu) + X $, $A_{ch,l}$ &         NLO &     {\sc aMCFast/APPLgrid}    \\
 $p + p \to W^+ \bar{c} + X $ &       NLO &     {\sc aMCFast/APPLgrid}    \\
 $p + p \to W^- c + X $ &      NLO & {\sc aMCFast/APPLgrid}    \\
 $p + p \to t\bar{t} + X $ &         NLO &     {\sc aMCFast/APPLgrid}    \\
 $p + p \to t\bar{t} \gamma\gamma +X $ &         NLO &     {\sc aMCFast/APPLgrid}    \\
 ATLAS inclusive jets   & NLO &     {\sc NLOJET++/APPLgrid}    \\
 ATLAS inclusive dijets & NLO &     {\sc NLOJET++/APPLgrid}    \\
 $p + p \to H(\gamma\gamma) + X $ &          LO, NLO & {\sc MCFM}    \\
 $p + p \to H(\gamma\gamma)+ jet + X $ &     LO, NLO &     {\sc MCFM} \\   
\hline
\end{tabular}
\end{center}
\caption{
Processes, QCD orders, and computer codes employed for comparisons of PDFs
in the online gallery \cite{SMUgallery}.}
\label{tab:Processes}
\end{table}

The PDF4LHC recommendation document \cite{Butterworth:2015oua} 
contains detailed guidelines to help decide which
individual or combined PDFs to use depending on the circumstances. 

To assist in this decision, predictions for 
typical LHC QCD observables have been calculated 
for an assortment of PDF sets. In Ref.~\cite{MC2Hgallery}, PDF4LHC15
predictions were made with the {\sc APPLgrid} fast interface
\cite{Carli:2010rw} for published LHC measurements within the fiducial
region. To provide a complementary perspective, 
at a gallery website \cite{SMUgallery}, 
we present LHC cross sections for processes listed 
in Table~\ref{tab:Processes} at 7, 8, and 13 TeV, and computed 
with no or minimal experimental cuts. 
The three (N)NLO ensembles of
the PDF4LHC15 family (\_100, \_30, \_MC \cite{Butterworth:2015oua}) 
are compared to those of 
ABM12 \cite{Alekhin:2013nda}, CT14 \cite{Dulat:2015mca}, 
HERA2.0 \cite{Abramowicz:2015mha},
MMHT14 \cite{Harland-Lang:2014zoa}, and NNPDF3.0 \cite{Ball:2014uwa}.
The cross sections are calculated using (N)LO hard matrix elements either by 
a fast convolution of the PDFs with the tabulated parton-level cross section 
in the {\sc APPLgrid} format \cite{Carli:2010rw}, or by direct Monte-Carlo
integration in {\sc MCFM} \cite{Campbell:2006xx}. 
Default $\alpha_s(M_Z)$ values are used with each PDF set. 
The {\sc APPLgrid} files used in the computations are linked 
to the website. In the {\sc
  APPLgrid} calculations, the hard cross sections are the same for all PDFs,
while the {\sc MCFM}-produced cross sections are sensitive
to Monte-Carlo integration fluctuations that vary depending
on the PDF ensemble, as will be discussed below.

The predictions were computed according to the following procedure. For production 
of $W^{\pm}$, $Z^0$, $t\bar{t}$,
$t\bar{t} \gamma\gamma$, $W^{+}\bar{c}\, (W^{-}c)$, we use
MadGraph\_aMC@NLO \cite{Alwall:2014hca}, 
combined with aMCfast \cite{Bertone:2014zva} to generate {\sc
  APPLgrid} files for different rapidities of the final-state particle. The  
renormalization and factorization scales are 
$\mu_{R}=\mu_{F}=M_W$, $M_Z$, $H_T/2$, $H_T/2$, $M_W$, respectively. $H_T$ is the
scalar sum of transverse masses $\sqrt{p_T^2+m^2}$ of final-state
particles. For $W^{+}\bar{c}\, (W^{-}c)$ production, we neglect small
contributions with initial-state $c$ or $b$ quarks. 
For NLO single-inclusive jet and dijet production, we use public
{\sc APPLgrid} 
files \cite{APPLgrid} in the bins of ATLAS measurements
\cite{Aad:2011fc}, created with the program {\sc NLOJET++}
\cite{Nagy:2001fj,Nagy:2003tz}. Similarly, the $W$ charge asymmetry in CMS experimental
bins \cite{Chatrchyan:2012xt,Chatrchyan:2013mza} is computed with {\sc APPLgrid} 
from \cite{Alekhin:2014irh}.

\begin{figure}[t]
\includegraphics[width=0.47\textwidth]{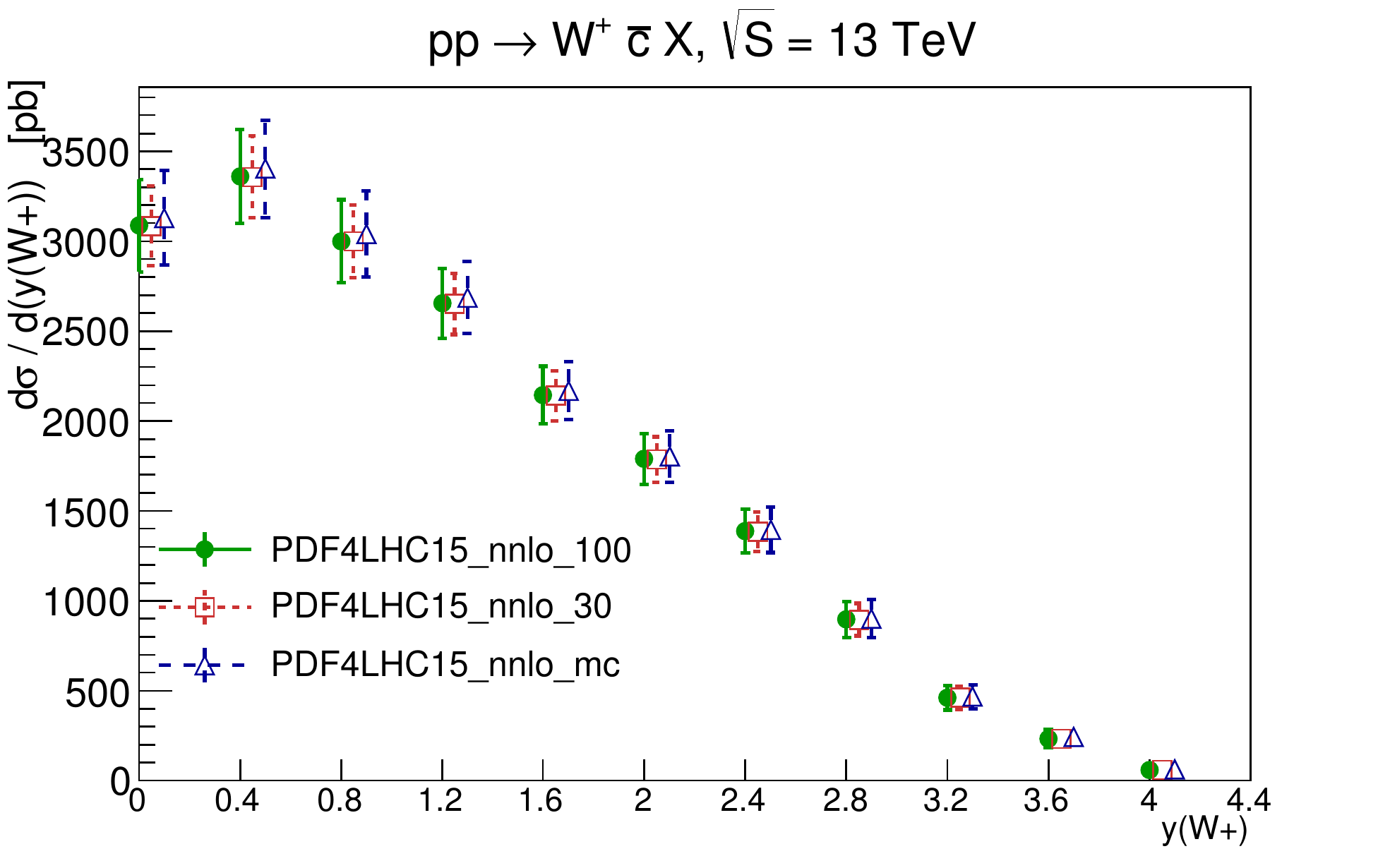}
\hfill
\includegraphics[width=0.47\textwidth]{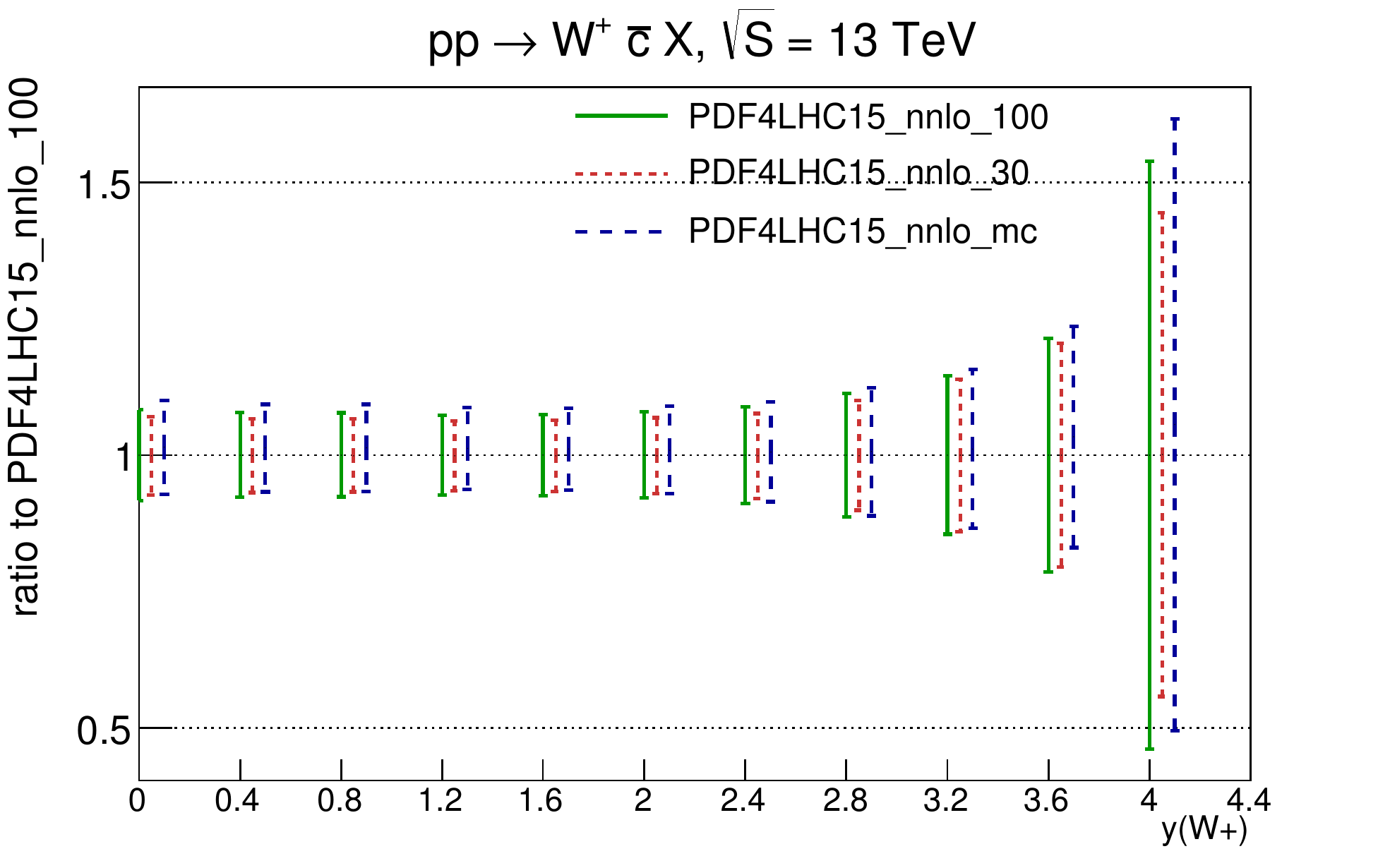}
\caption{NLO predictions of $d\sigma/dy(W^{+})$ in the process $pp \to W^{+}\bar{c}$ at the LHC 13 TeV, computed with {\sc APPLgrid}.  \label{fig:Wpc} }
\end{figure}

For cross sections of the Standard Model Higgs boson and Higgs boson+jet
production via gluon fusion, with subsequent decay to $\gamma\gamma$, 
we use {\sc MCFM} in the heavy-top quark approximation. 
Minimal cuts are imposed on the photons; the QCD scales are
$\mu_F=\mu_R=m_H$.

The PDF uncertainties shown are symmetric, computed according to the
prescriptions provided with each PDF ensemble, except for  the HERA2.0
predictions, which are shown with asymmetric uncertainties, including
contributions from both the eigenvector sets and the variation sets.

\begin{figure}[t]
\includegraphics[width=0.47\textwidth]{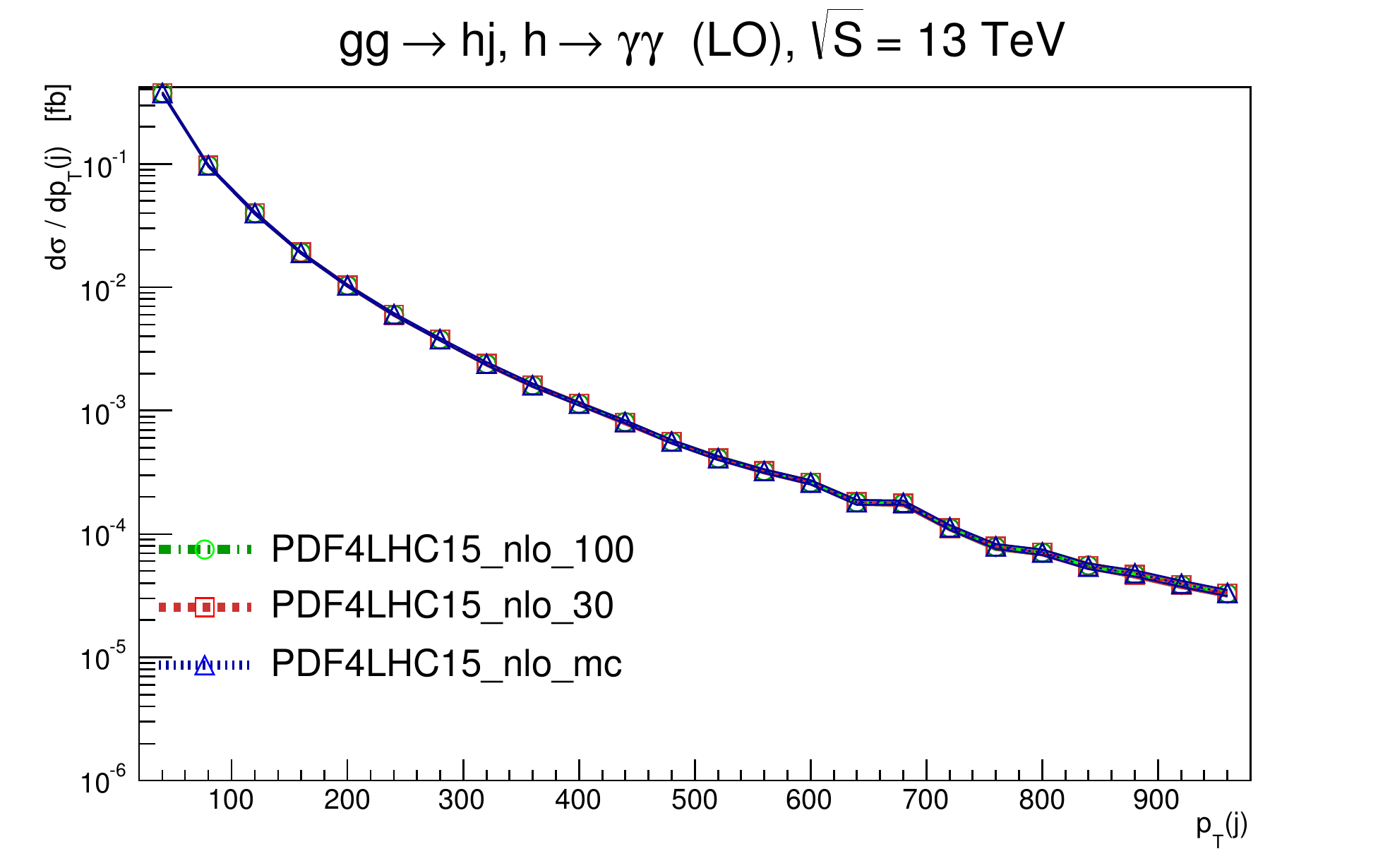}
\hfill
\includegraphics[width=0.47\textwidth]{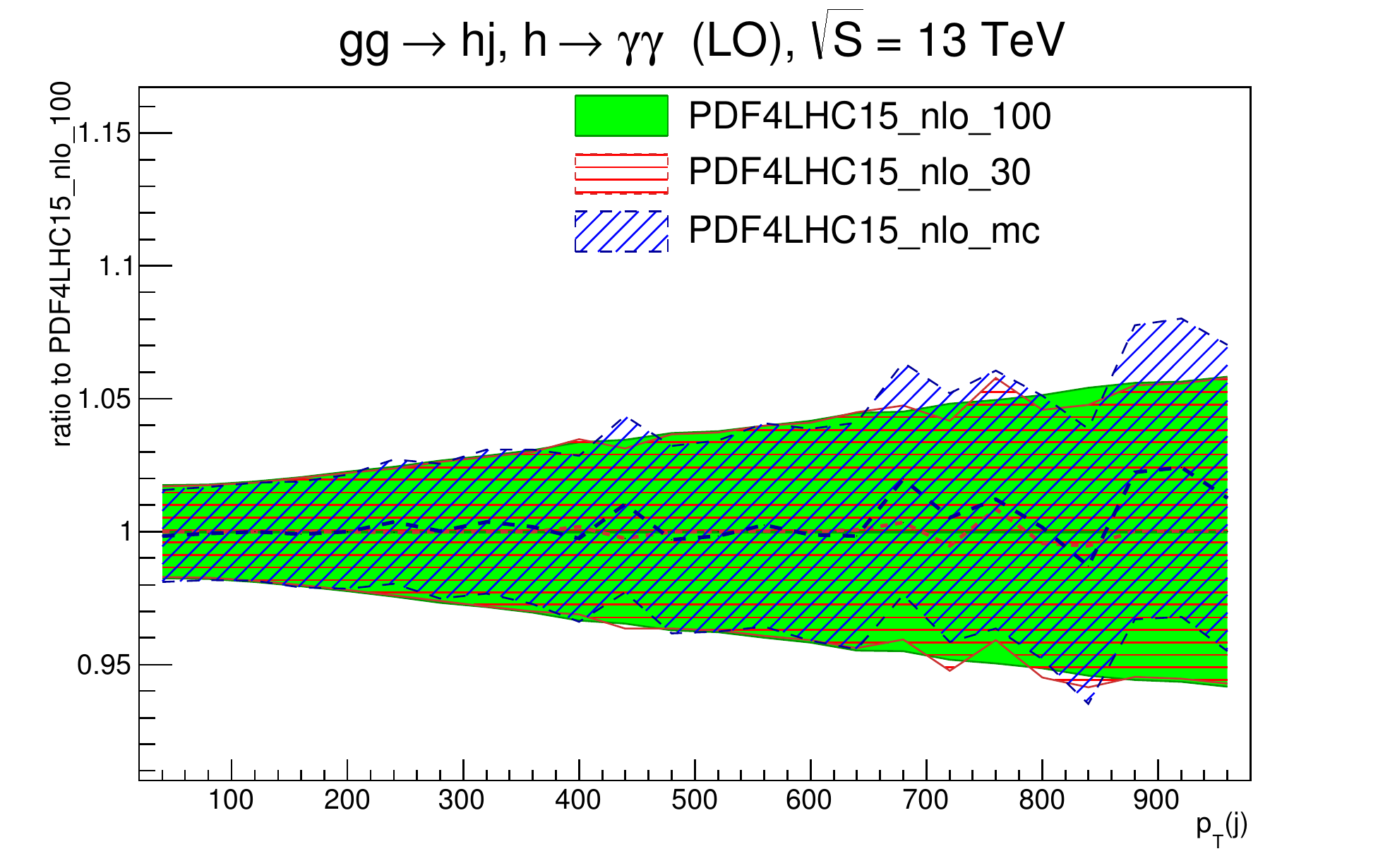}
\caption{$d\sigma/dp_{T}(j)$ in the process $gg \to
  H(\gamma \gamma)+jet$ at the LHC 13 TeV, and its relative PDF uncertainties.  \label{fig:gg2hjLO} }
\end{figure}

For each scattering process, our gallery shows plots of differential
cross sections and ratios of PDF uncertainties to the central
prediction based on  PDF4LHC15\_100.  
Figs.~\ref{fig:Wpc} and \ref{fig:gg2hjLO} provide two examples of
comparisons presented on the website. When computed with {\sc
  APPLgrid}, the cross sections reflect genuine differences in the
PDFs; the hard cross sections are the same with all PDF sets. Thus we
observe, for instance, in $W^+\bar c$ production in 
Fig.~\ref{fig:Wpc} that the uncertainties of 
\_100, \_30, and \_MC ensembles are very close across the
central-rapidity range for most processes, with the \_30 uncertainty
being only slightly smaller (as expected), and with the differences
that can be nearly always eliminated by slightly 
scaling the \_30 uncertainty up by a constant factor (e.g.,
by multiplying it by $\approx 1.05$ in Fig.~\ref{fig:Wpc}). 
The differences between the PDF4LHC15 ensembles grow at
rapidities above 2-3, where the cross sections also are rapidly decreasing. The PDF
uncertainties fluctuate more in the forward regions, reflecting
paucity of experimental constraints on the PDFs. 

Another perspective is glanced from 
$H$ and $H+\mbox{jet}$ production cross sections calculated by {\sc
  MCFM}, cf.~Fig.~\ref{fig:gg2hjLO}. [Additional comparisons can be
  viewed on the website.] These
illustrate that often the differences between the PDF4LHC15 reduced
ensembles will be washed out by Monte-Carlo integration errors, save
exceptionally precise calculations. To start, although the LHAPDF
grids for the \_100, \_30, and \_MC {\it central} sets are just independent
tabulations of the {\it same} prior central set
(they are equivalent up to roundoff errors), they will produce different
fluctuations during the Monte-Carlo integration in MCFM or alike program. 
This is exemplified in the right frame in
Fig.~\ref{fig:gg2hjLO}, where the Higgs boson production cross sections are
slightly different for the three LHAPDF tabulations of the central
set solely because of MC fluctuations. In this figure, the cross
sections were evaluated with $10^6$ Monte-Carlo samplings and with PDF
reweighting of events turned on. The events are exactly
the same for all PDF sets within a given ensemble,
and the event sequences are not the same among the ensembles because
of the different roundoffs of the central LHAPDF grids. Even when the event
reweighting is on, the PDF error bands fluctuate together with
their respective central predictions.

 The MCFM example touches on broader questions. The MC
 fluctuations can be suppressed by increasing the number of events or
 by using coarser binning for the cross sections. These adjustments 
  tend to either lengthen the calculations, especially with the \_100-replica
 ensembles, or to wash out the already small differences between the
 three PDF4LHC15 ensembles. There are several ways for ``averaging'' the input
 central PDF sets, e.g., because they use different evolution codes or
 round-offs. Each of these will lead to a different pattern of MC
 fluctuations.  Finally, if the MC integration is done without PDF event 
 reweighting, MC fluctuations will vary independently
 replica-by-replica. Using the combined PDF4LHC ensemble with fewer
 members may turn out to be preferable in such situations.
 
\subsection*{Acknowledgements}
Work at ANL is supported in part by the U.S. Department of Energy
under Contract No. DE-AC02-06CH11357. Work at SMU was supported
by the U.S. Department of Energy under Grant DE-SC0010129. Work at MSU was supported by
the National Science Foundation under Grant PHY-1410972.

\section{On the accuracy and Gaussianity of the PDF4LHC15
    combined sets of parton distributions
    \texorpdfstring{\protect\footnote{
      S.~Carrazza, S.~Forte, Z.~Kassabov, J.~Rojo
    }}{}}

We perform a systematic study of the combined PDF4LHC15 sets of parton
distributions which have been recently presented as a means to
implement the PDF4LHC prescriptions for the LHC Run II.
These combined sets
reproduce a prior large Monte Carlo (MC) sample in terms of
either a smaller  MC replica sample, or a Gaussian (Hessian)
representation with two different number of error sets, and obtained
using two different methodologies.
We study how well the three reduced sets reproduce the prior
for all the $N_{\sigma}\simeq 600$ hadronic
cross-sections used in the NNPDF3.0 fit.
We  then 
quantify deviations of the prior set from  Gaussianity, and
check that these deviations are well reproduced by the MC reduced set.
Our results indicate that generally the three reduced sets perform reasonably well,
and provide some guidance about which of these to use in specific applications.

\subsection{Introduction}
Recently, the PDF4LHC Working Group~\cite{PDF4LHC} has presented 
updated recommendations and guidelines~\cite{Butterworth:2015oua}
for the usage of Parton Distribution
Functions (PDFs) for LHC calculations at Run II.
These recommendations are specifically based on the usage of combined
PDF sets, which are obtained using
the  Monte Carlo (MC) method~\cite{Watt:2012tq,Forte:2013wc},
constructed
from the combination of $N_{\rm rep}=300$ MC replicas 
from the NNPDF3.0~\cite{Ball:2014uwa},
MMHT2014~\cite{Harland-Lang:2014zoa} and CT14~\cite{Dulat:2015mca} PDF sets,
for a total number of $N_{\rm rep}=900$ replicas.
The combination has been performed both at NLO and at NNLO, and
versions with $n_f=4$ and $n_f=5$ maximum number of active
quark flavors are available.
The impact of LHC measurements
from Run I on PDF determinations has been discussed in a companion
PDF4LHC report~\cite{Rojo:2015acz}. 

From this starting prior set,
three reduced sets, two Hessian and one MC, are delivered for general
usage.
The reduced sets, constructed using different compression
strategies, are supposed to reproduce as much as possible
the information
contained in the prior, but in terms of a substantially smaller number of error PDFs.
The reduced Monte Carlo set, {\tt PDF4LHC15\_mc}, is
constructed using the CMC-PDF method~\cite{Carrazza:2015hva},
and the two reduced Hessian sets,
{\tt PDF4LHC15\_100} and {\tt PDF4LHC15\_30}, are constructed using the
MC2H~\cite{Carrazza:2015aoa} and META-PDF~\cite{Gao:2013bia} techniques, respectively.
The PDF4LHC15 combined sets are available from {\tt LHAPDF6}~\cite{Buckley:2014ana},
and include additional PDF member sets to account for the uncertainty due to
the value of the strong coupling constant,
$\alpha_s(m_Z)=0.1180\pm 0.0015$.

The PDF4LHC 2015 report~\cite{Butterworth:2015oua} presented general
guidelines for the usage of the reduced sets, and some comparisons
 between them and the prior 
at the level of PDFs, parton luminosities, and LHC cross-sections,
while referring to a repository of cross-sections on the PDF4LHC
server~\cite{PDF4LHCrep} for a more detailed set of comparisons.
It is the purpose of this contribution to make these comparisons more
systematic and quantitative, in order to answer questions which have
been frequently asked on the usage of the reduced sets.
Specifically, we  will perform a systematic
 study of the accuracy of the PDF4LHC15 reduced sets, by assessing
 the relative accuracy of uncertainties determined using each of them
 instead of the prior, for all hadronic observables
 included in the NNPDF3.0
 PDF determination~\cite{Ball:2014uwa}. We will also compare the
 performance of the PDF4LHC15 reduced sets with that of the recently
 proposed SM-PDF sets~\cite{Carrazza:2016htc}: specialized PDF sets
 which strive to minimize the number of PDF error sets which are
 needed for the description of a particular class of processes.
 We will then
 address the issue of the validity of the Gaussian approximation to
 PDF uncertainties by testing for gaussianity of the distribution of
 results obtained using the prior PDF set for a very wide variety of
 observables, and then assessing the performance and accuracy of  
both the Monte Carlo sets (which
 allows for non-Gaussian behaviour) and the Hessian compressed sets
 (which do not, by construction).

\subsection{Validation of the PDF4LHC15 reduced PDF sets on a global dataset}
\label{sec:pdf2:validation}

We wish to compare the performance of the three
reduced NLO sets, the two Hessian
sets, {\tt PDF4LHC15\_nlo\_30} and {\tt PDF4LHC15\_nlo\_100}, and the Monte
Carlo set {\tt PDF4LHC15\_nlo\_mc}, for all the hadronic 
cross-sections included
in the NNPDF3.0 global analysis~\cite{Ball:2014uwa}.
These cross-sections have computed at $\sqrt{s}=7$ TeV using NLO theory with
{\tt MCFM}~\cite{Campbell:2004ch},
{\tt NLOjet++}~\cite{Nagy:2001fj}
and {\tt aMC@NLO}~\cite{Alwall:2014hca,Bertone:2014zva}
interfaced to {\tt APPLgrid}~\cite{Carli:2010rw}.
They include
 $N_{\sigma}\simeq 600$ independent observables for
a variety of hadron collider processes such as
 electroweak gauge boson, jet
production and top quark pair production, covering
a wide region in the $(x,Q)$ kinematical plane.
In this calculation, the PDF4LHC15 combined sets are
obtained from the {\tt LHAPDF6}
interface.

\begin{figure}[t]
\begin{center}
  \includegraphics[width=0.49\textwidth]{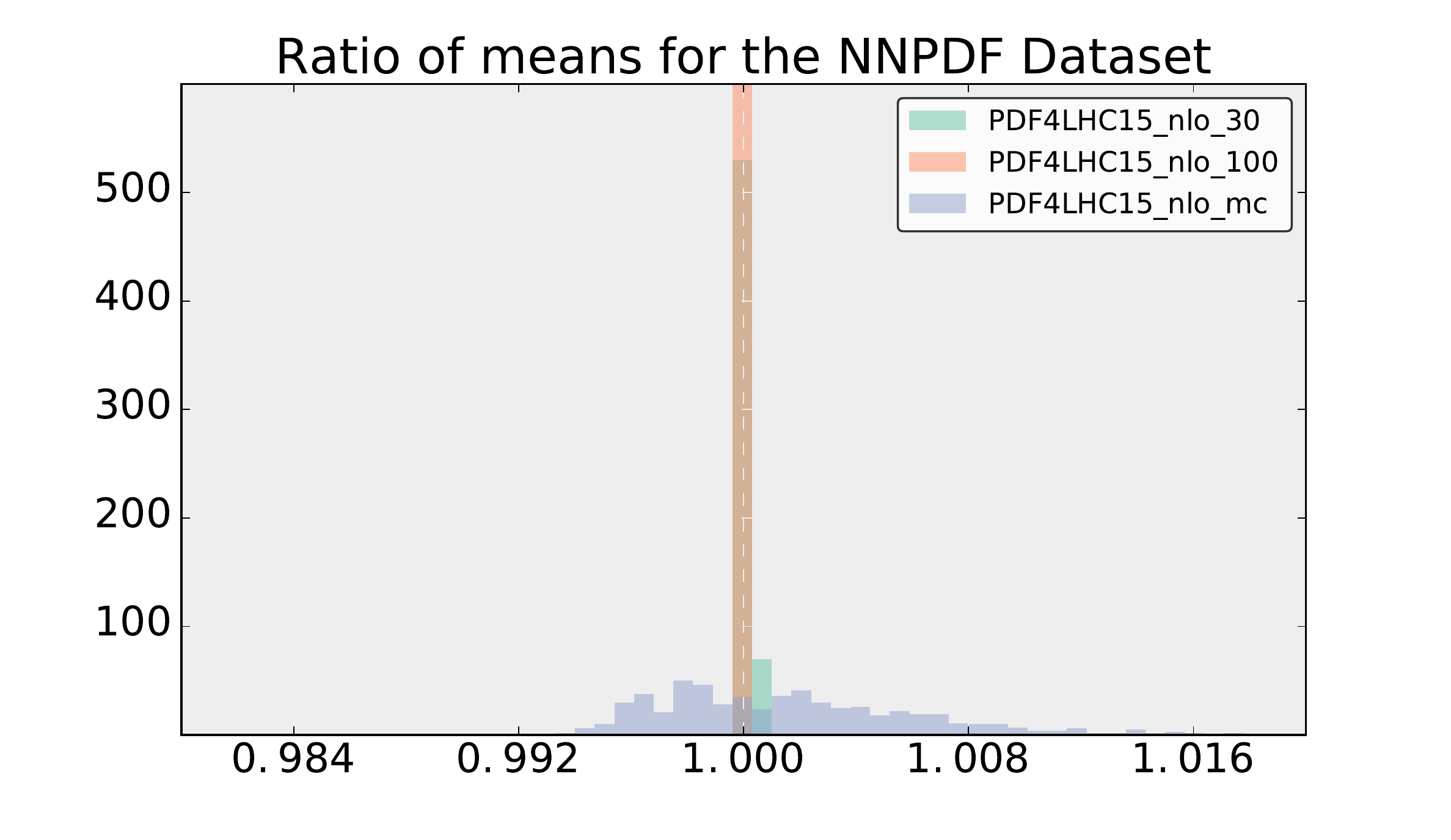}
  \includegraphics[width=0.49\textwidth]{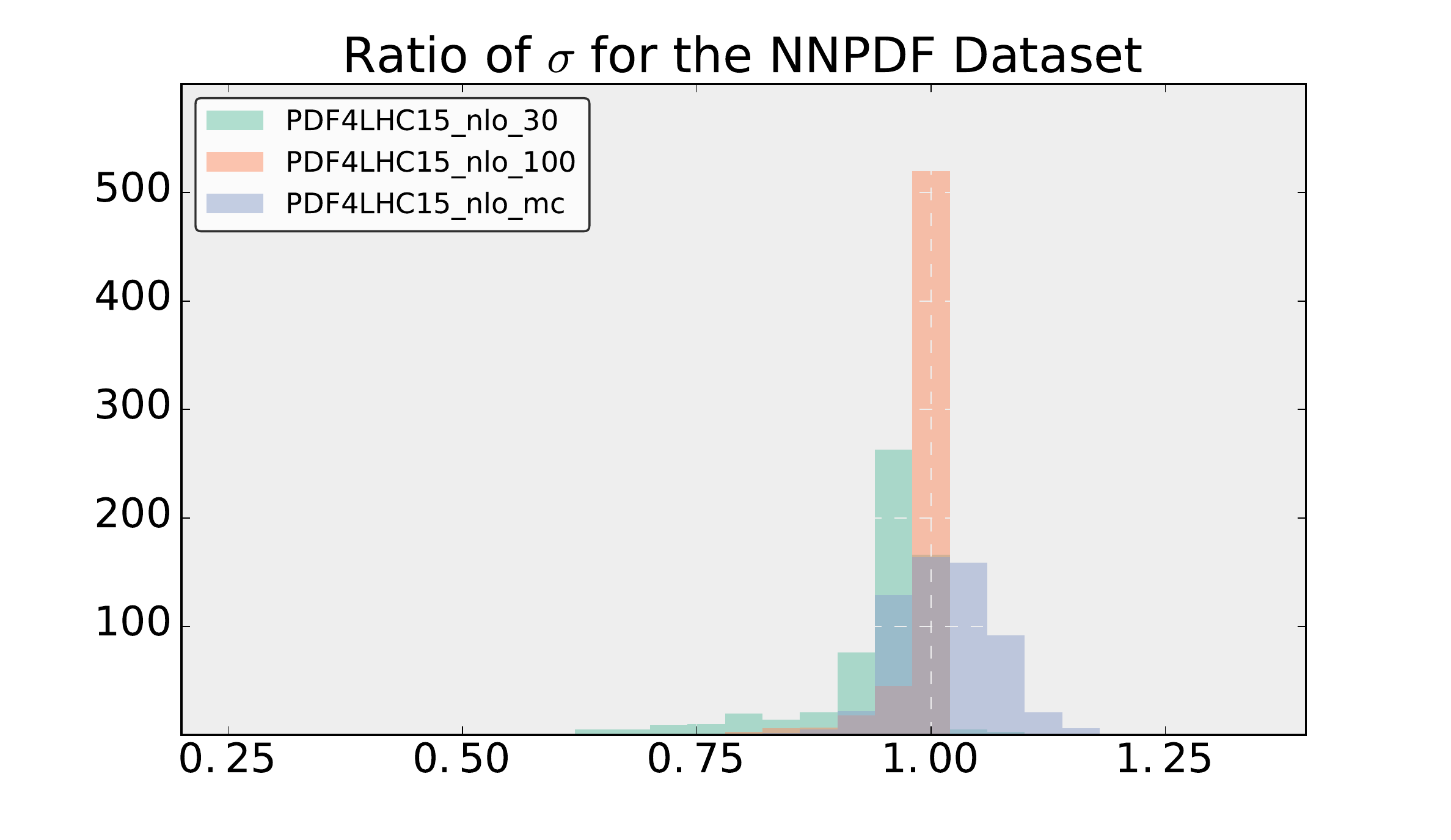}
\end{center}
\vspace{-0.5cm}
\caption{small 
  \label{fig:ratios}
  Distribution of the ratio to the prior of means Eq.~(\ref{eq:rat1}) (left) and of standard deviations
  Eq.~(\ref{eq:rat2}) (right) computed for each of the  $N_{\sigma}\simeq 600$
  hadronic cross-sections
included in the NNPDF3.0 global
analysis, using each of  the three reduced sets. 
}
\end{figure}

\begin{figure}[t]
\centering
  \includegraphics[width=0.45\textwidth]{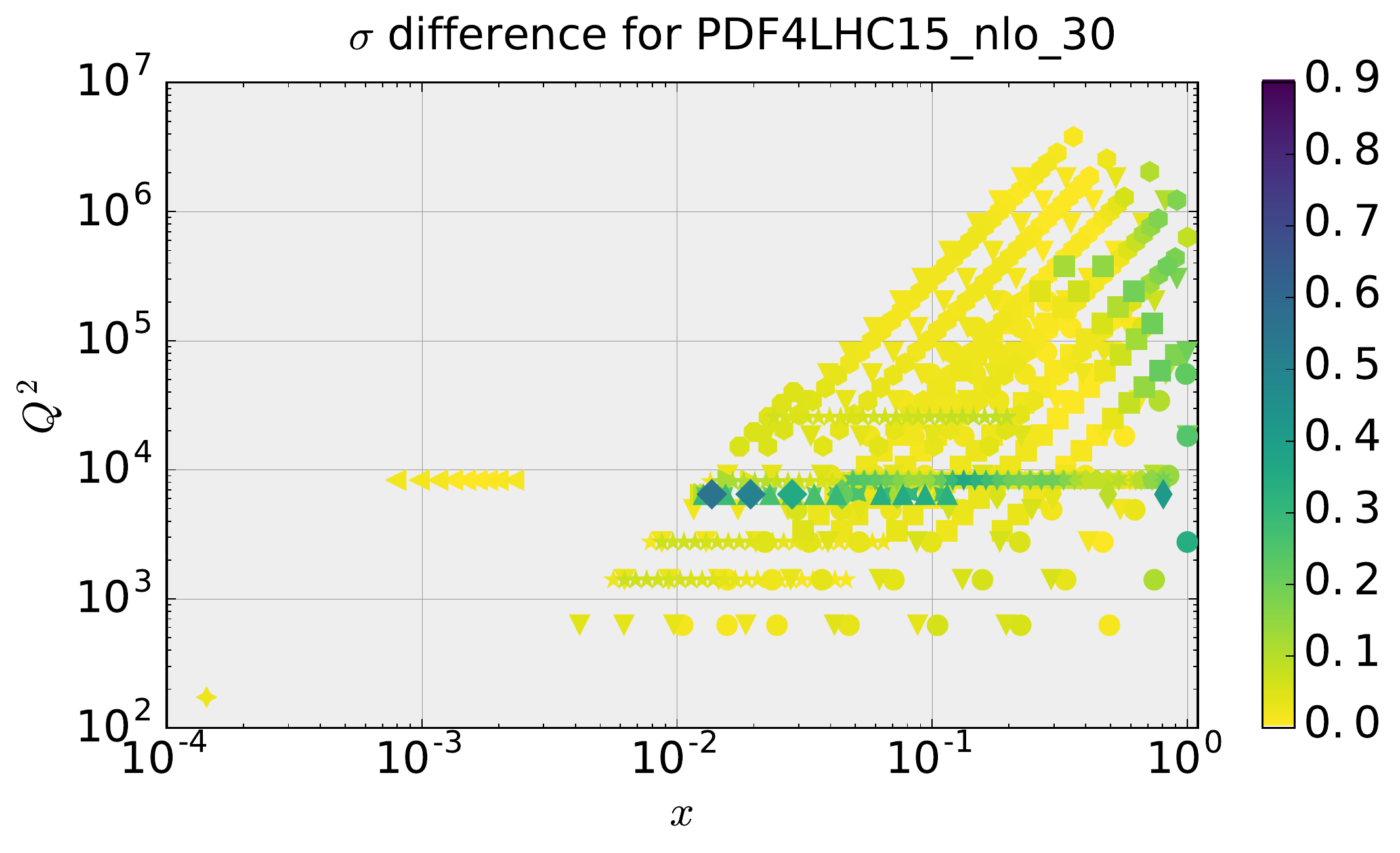}
  \includegraphics[width=0.45\textwidth]{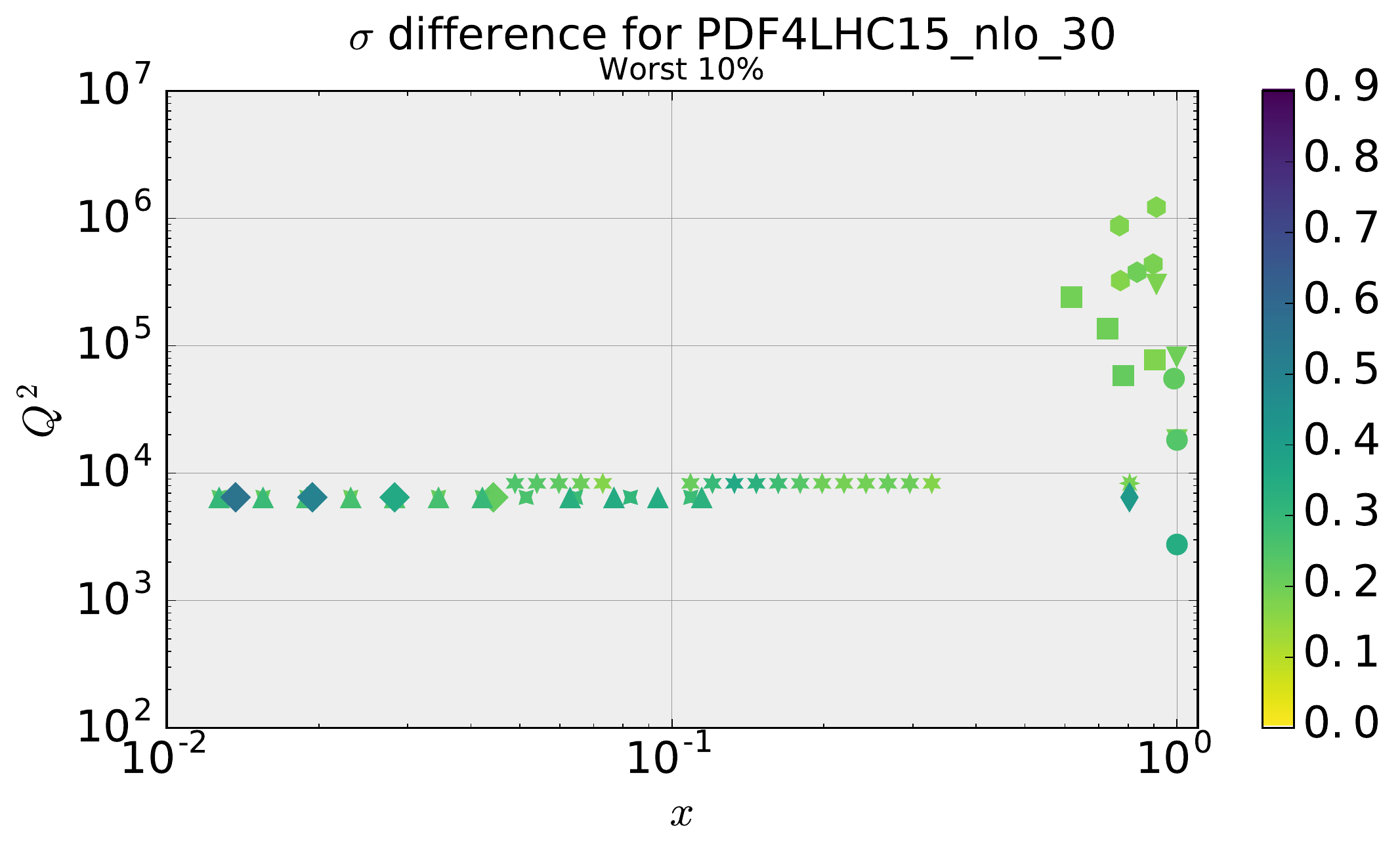}
  \includegraphics[width=0.45\textwidth]{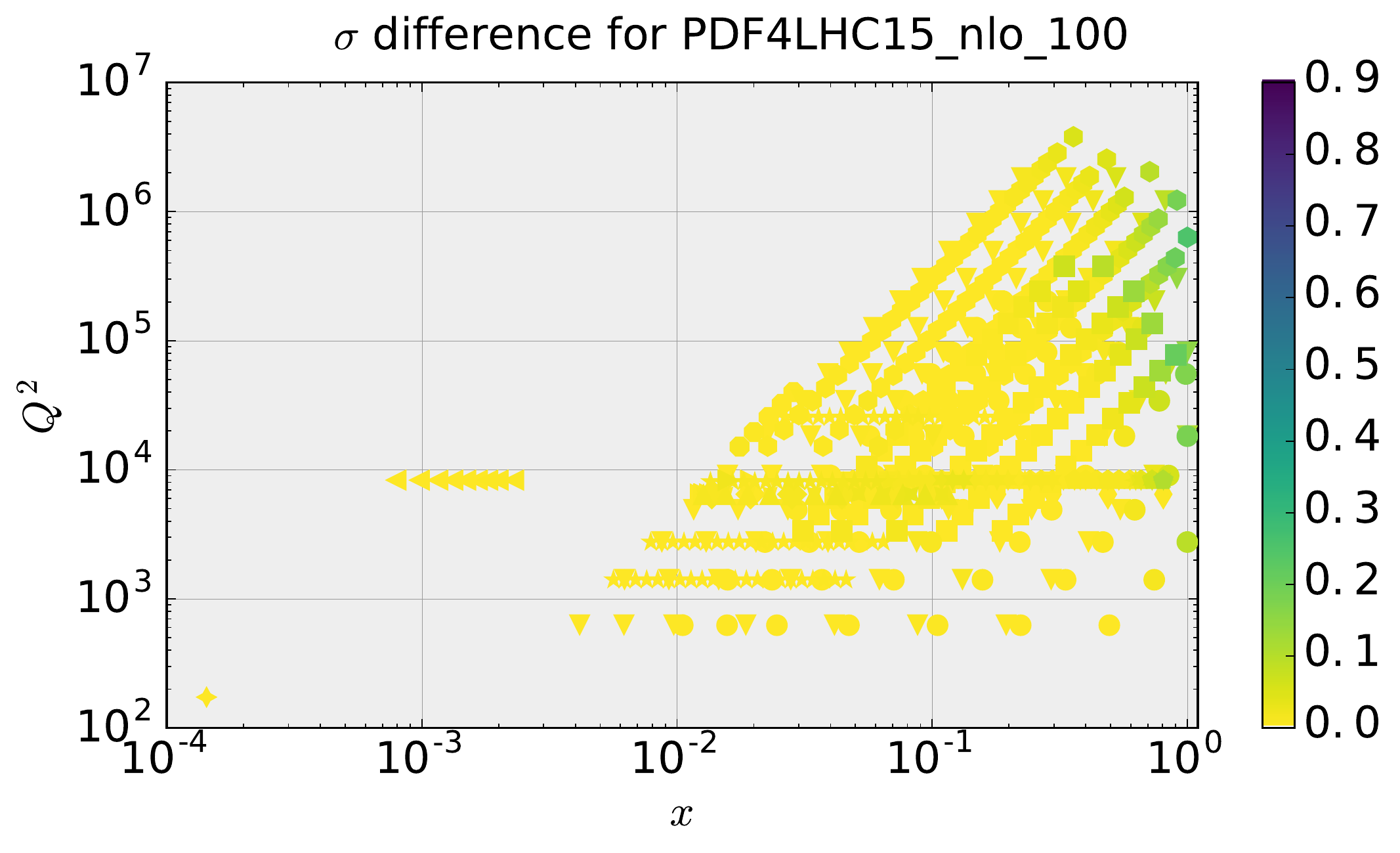}
  \includegraphics[width=0.45\textwidth]{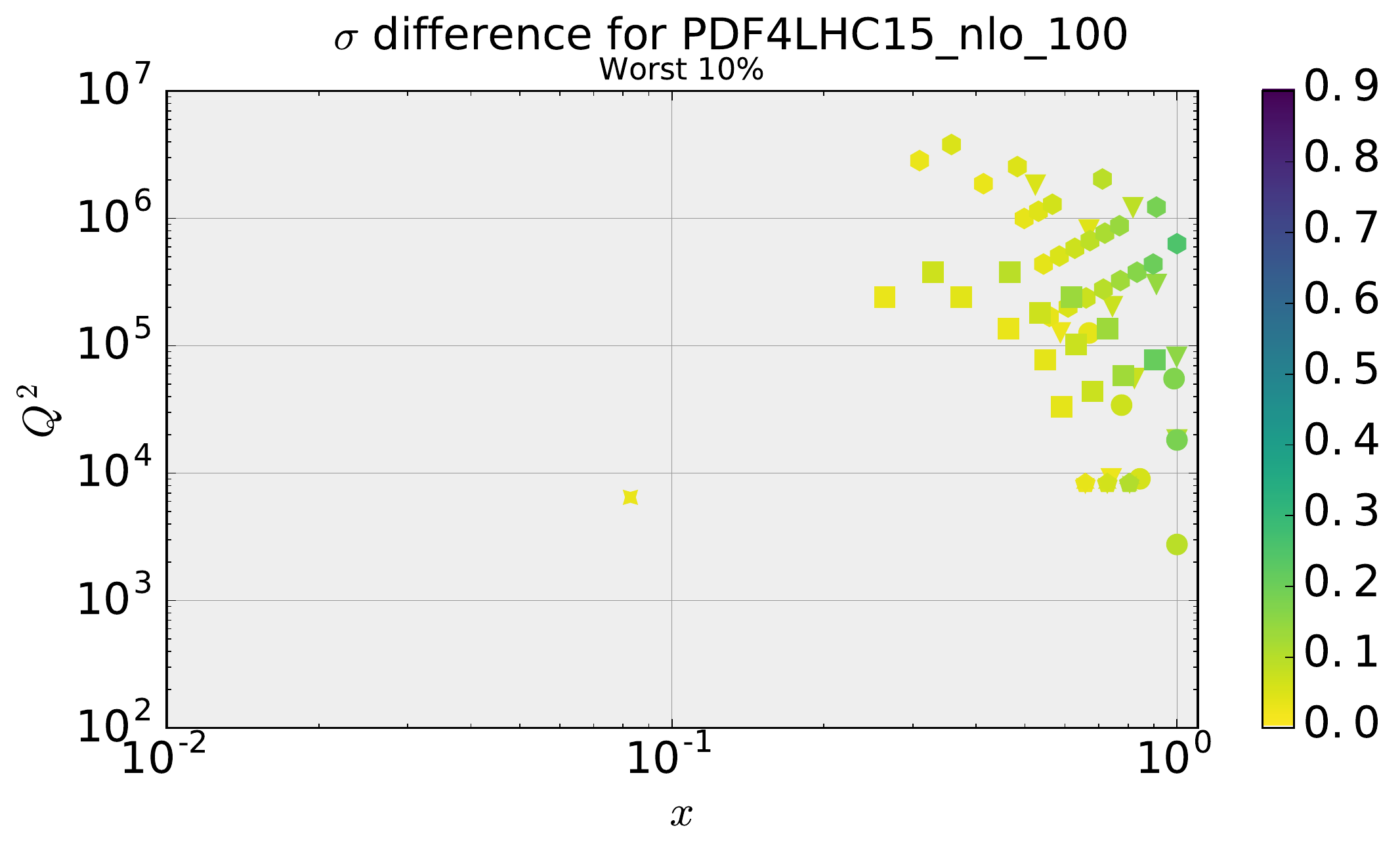}
  \includegraphics[width=0.45\textwidth]{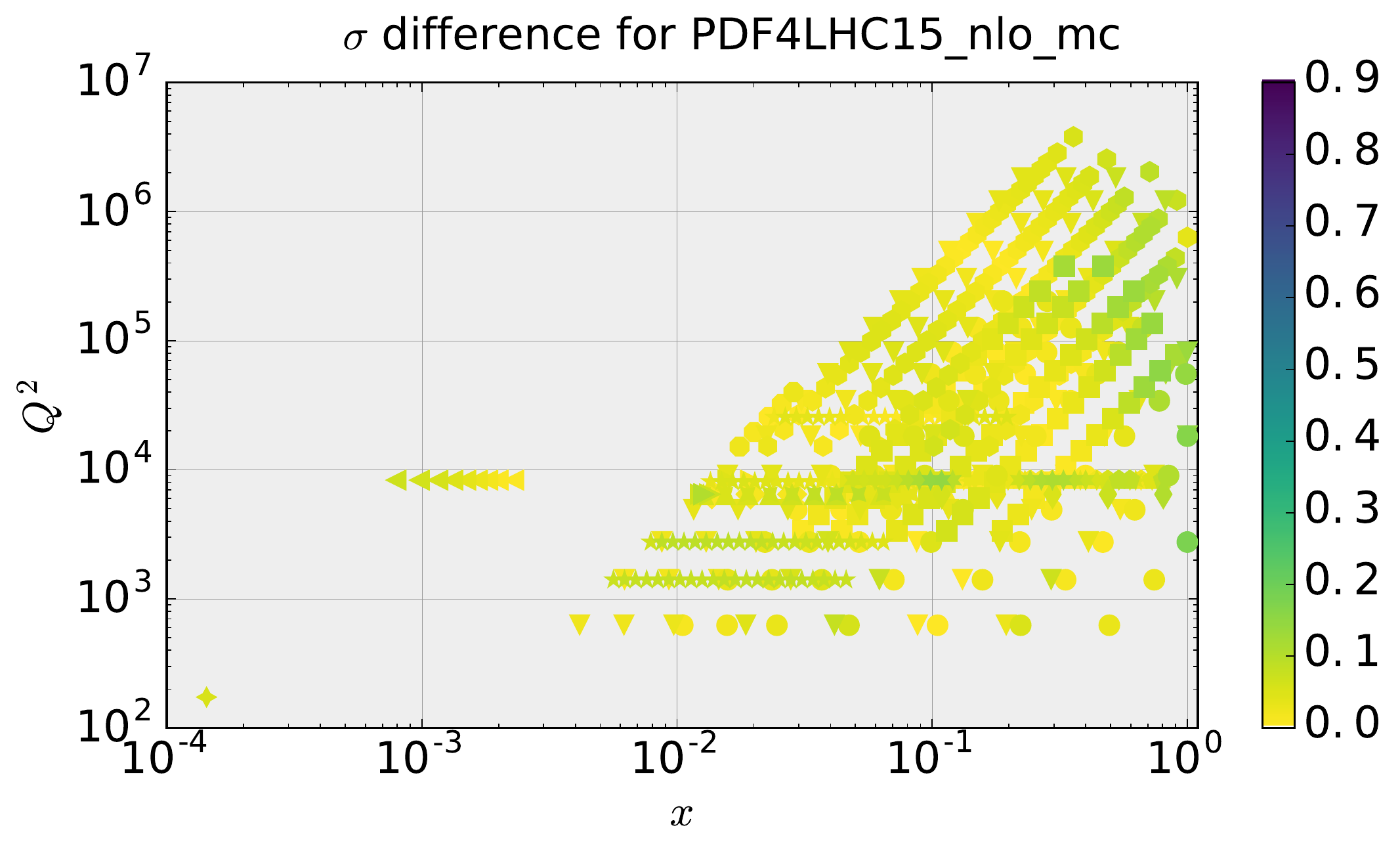}
  \includegraphics[width=0.45\textwidth]{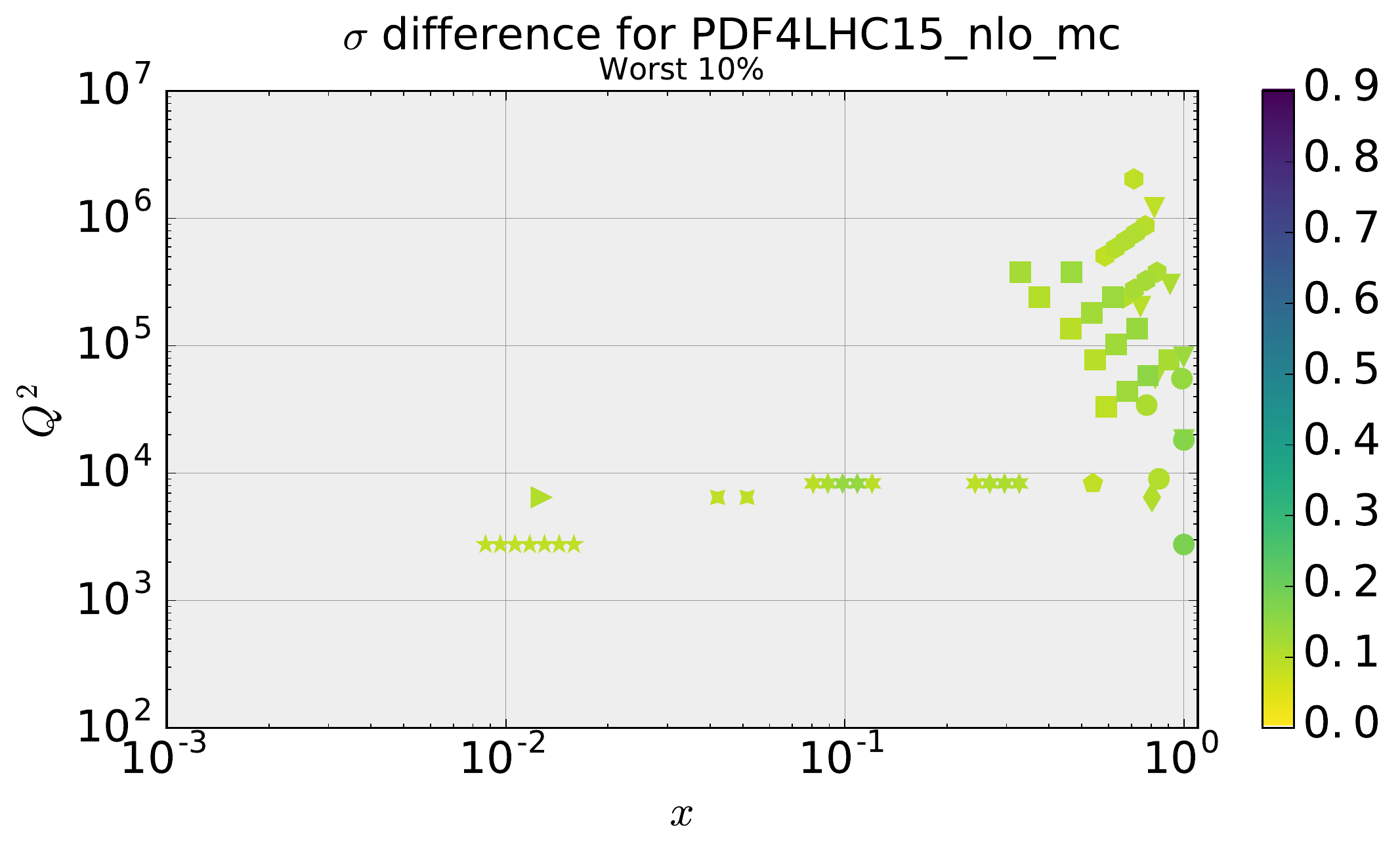}\\
  \includegraphics[width=.6\textwidth]{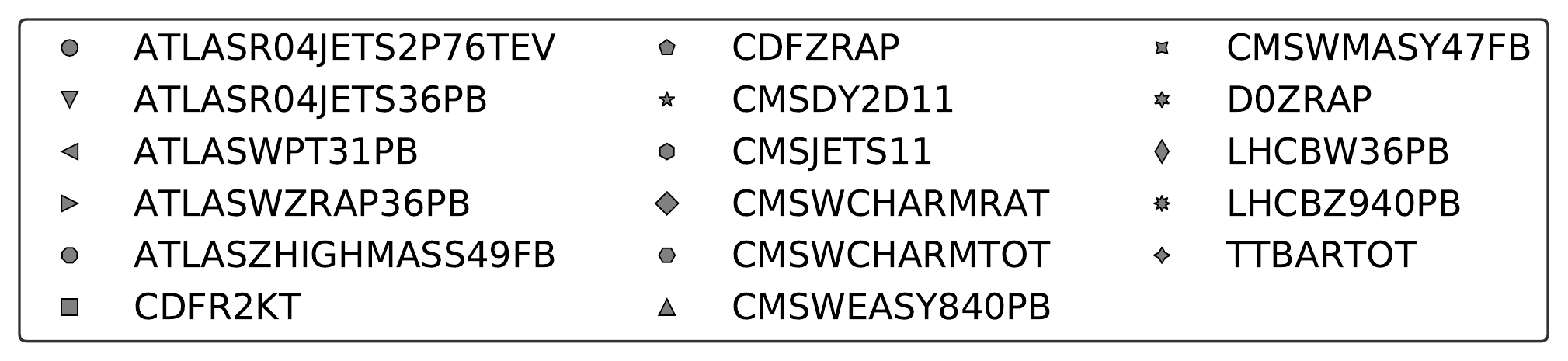}
  \caption{\small 
    Relative difference Eq.~(\ref{eq:reldiff}),
    between the PDF uncertainties computed using the reduced set and
    the prior
    computed for all hadronic observables included
    in the NNPDF3.0 fit, shown as a scatter plot in
    the $(x,Q^2)$ at the corresponding point, determined
    using leading-order kinematics. From top to bottom results for
    {\tt PDF4LHC\_nlo\_30}, 
    {\tt PDF4LHC\_nlo\_100} and {\tt PDF4LHC\_nlo\_mc} are shown. In
    the left plots, all points are shown, in the right plots only the
    10\% of points with maximal deviation.
   \label{fig:kinplots}}
\end{figure}

In Fig.~\ref{fig:ratios} we show the distribution of the ratios
\bea
\label{eq:rat1}
R_{\la \sigma_i\ra} &\equiv& \frac{\la \sigma_i\ra({\rm reduced})}{\la \sigma_i\ra({\rm prior})} \,,\quad i=1\,\ldots,N_{\sigma} \, , \\ 
R_{s_i} &\equiv& \frac{s_i({\rm reduced})}{s_i({\rm prior})} \, , \quad i=1\,\ldots,N_{\sigma} \, .
\label{eq:rat2}
\eea
between respectively the  
means $\la \sigma_i\ra$, 
and the  standard
deviations $s_i$  from each of the three reduced sets and the PDF4LHC15
prior, computed for all hadronic observables included in the NNPDF3.0
global analysis.
For the Hessian sets the central value coincides with that of the
prior, so the  ratio of means is supposed to equal one by
construction, with small deviations only due to rounding errors and
interpolation in the construction of the {\tt LHAPDF} grids, while for the
MC set the mean is optimized by the CMC construction to fluctuate due
to the finite size of the replica sample much less than expected on
purely statistical grounds.
Indeed, the histograms shows agreement of
central values at
the permille level.
For standard deviations  (i.e. PDF uncertainties)
Fig.~\ref{fig:ratios} shows that using 
the {\tt
  PDF4LHC15\_nlo\_100} set
 they are reproduced typically with better than  5\% accuracy. 
Differences are
somewhat larger for the {\tt PDF4LHC15\_nlo\_mc} and
{\tt PDF4LHC15\_nlo\_30} sets.

\begin{figure}[t]
\centering
  \includegraphics[width=0.45\textwidth]{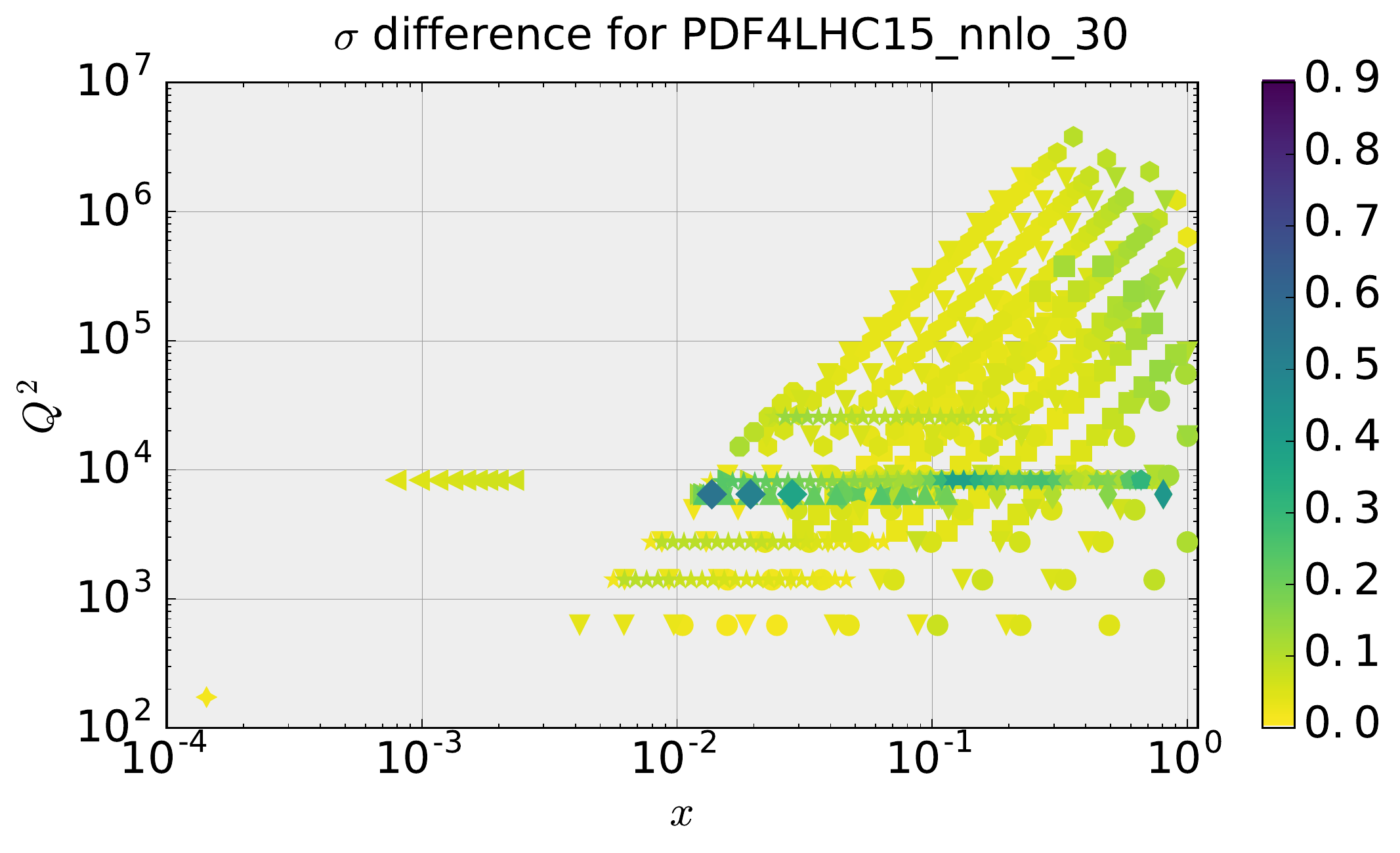}
  \includegraphics[width=0.45\textwidth]{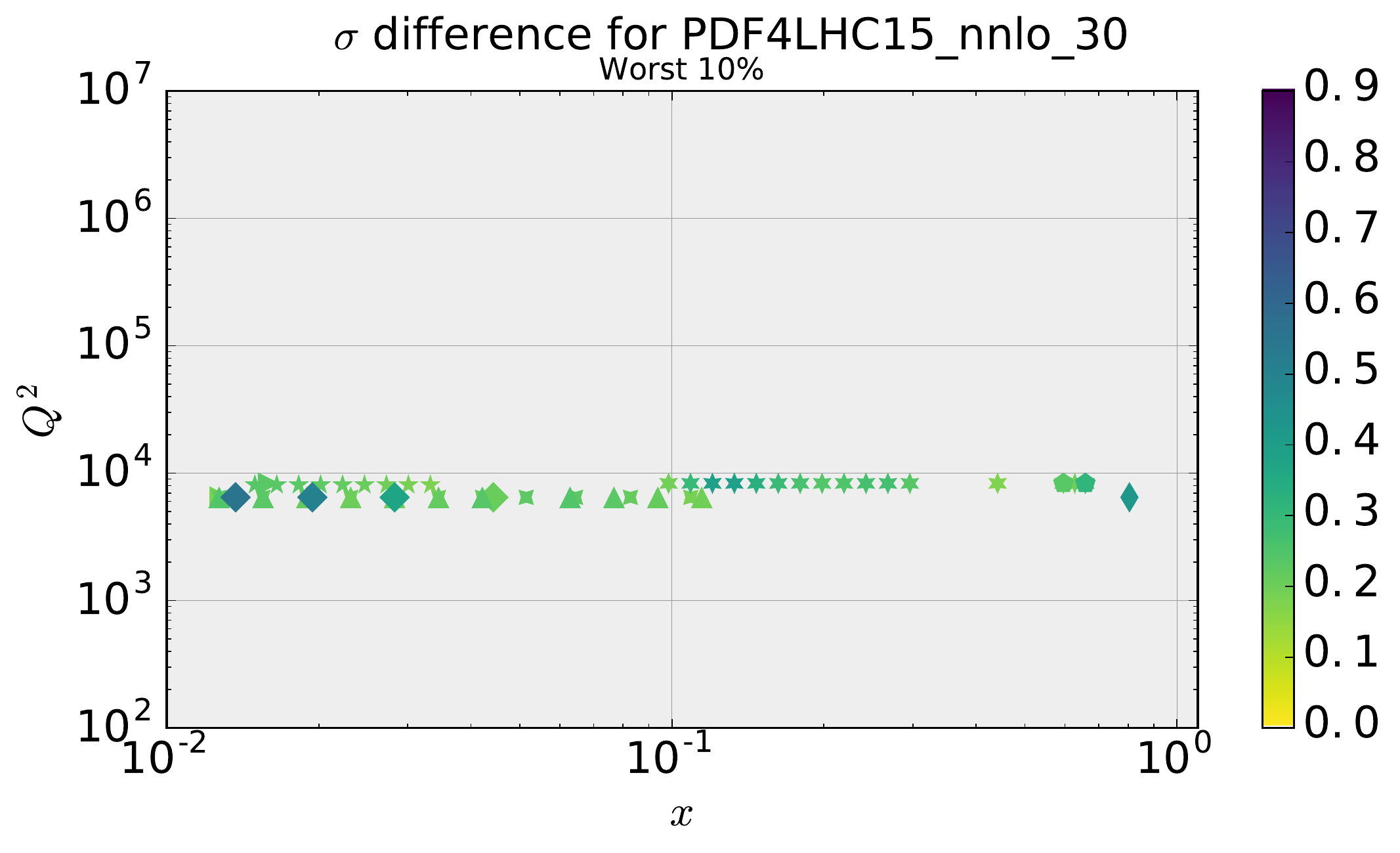}
  \includegraphics[width=0.45\textwidth]{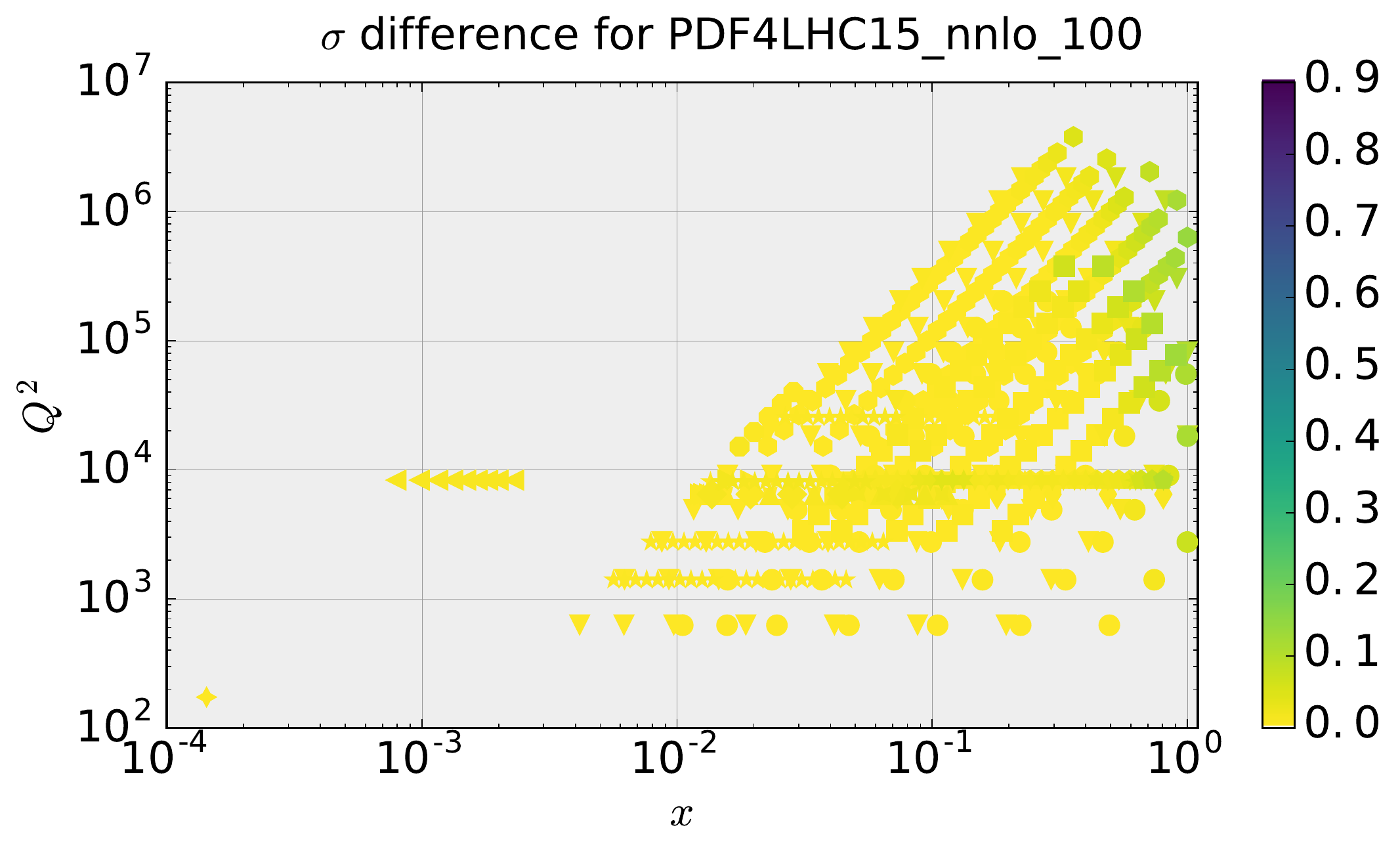}
  \includegraphics[width=0.45\textwidth]{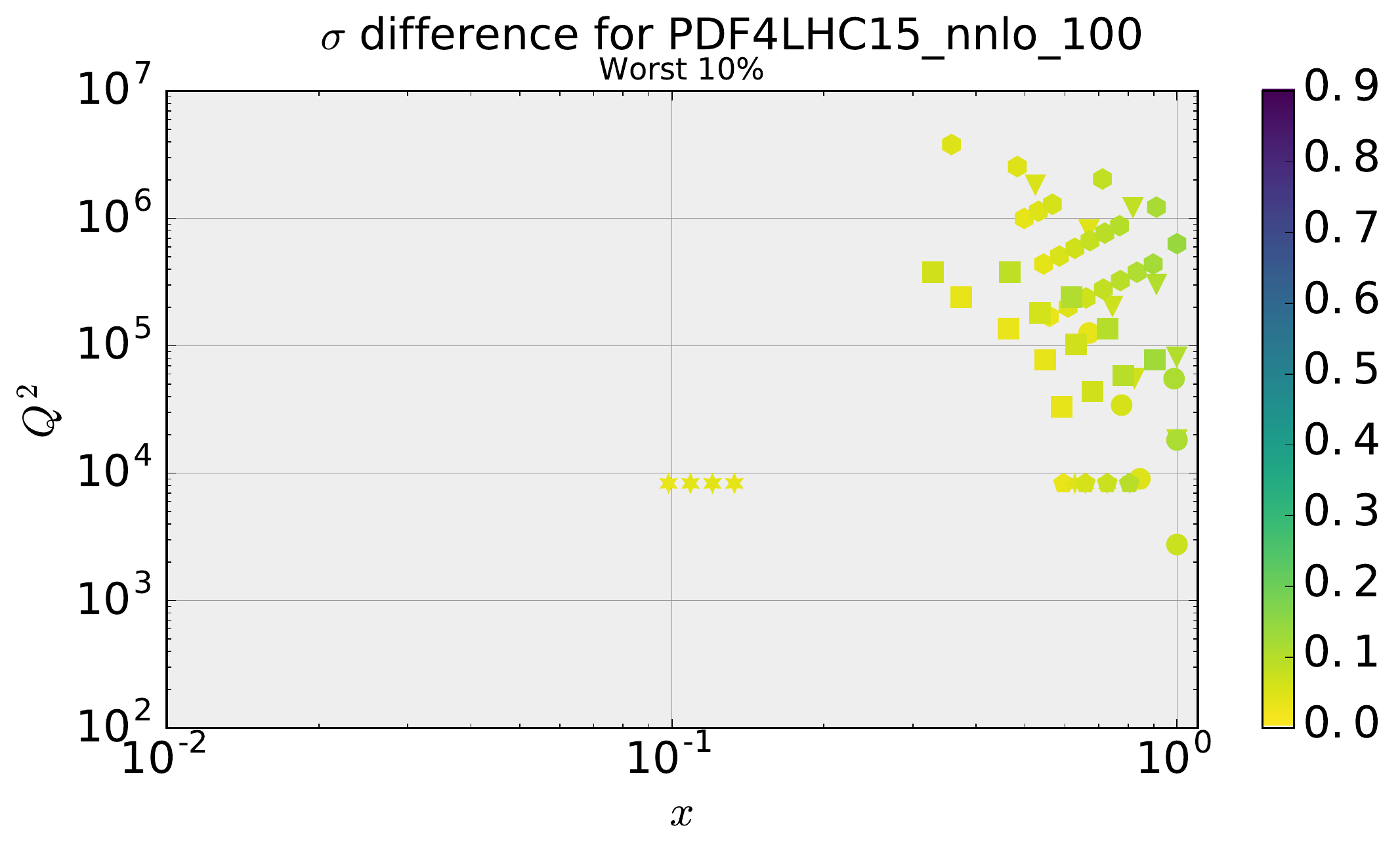}
  \includegraphics[width=0.45\textwidth]{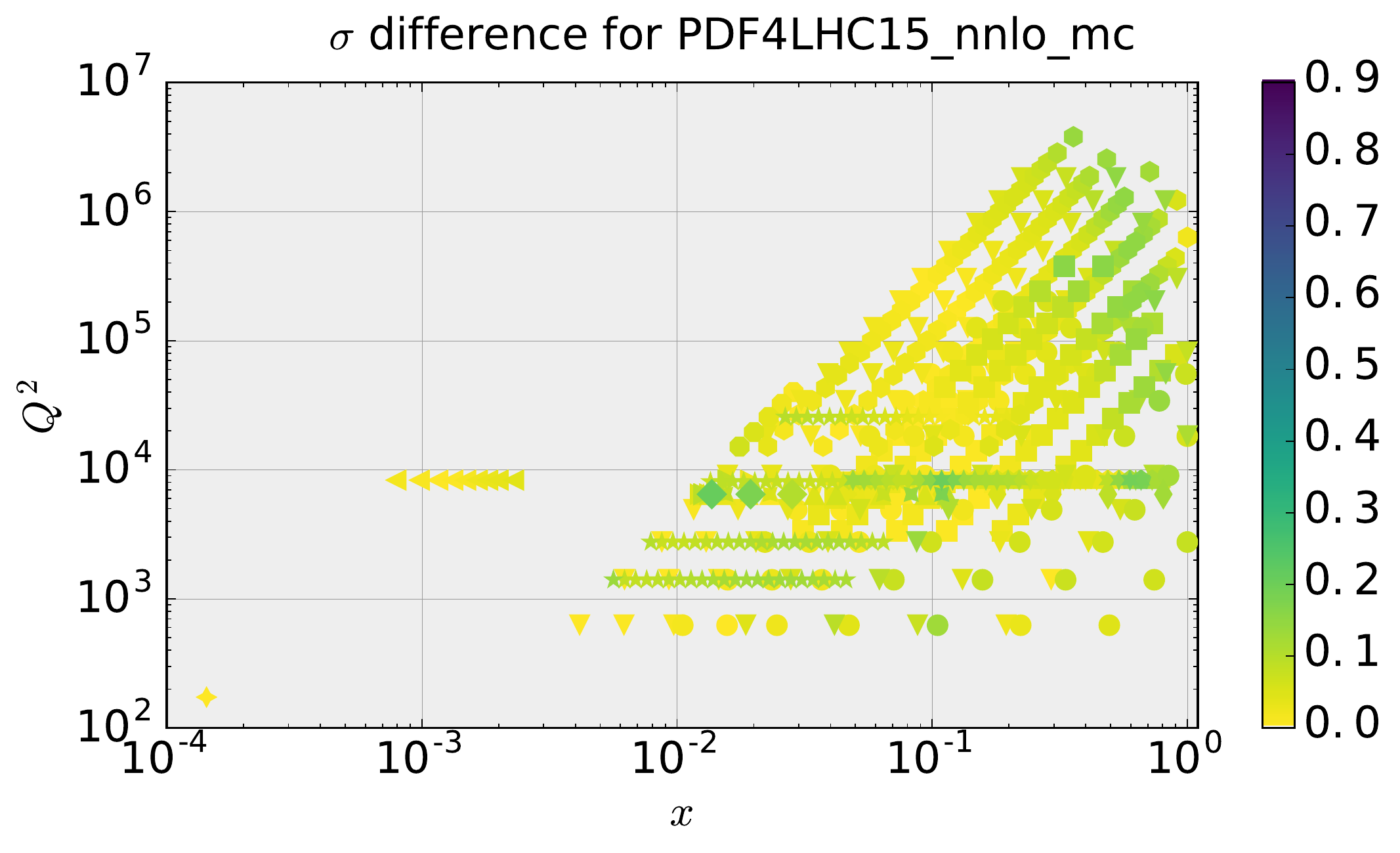}
  \includegraphics[width=0.45\textwidth]{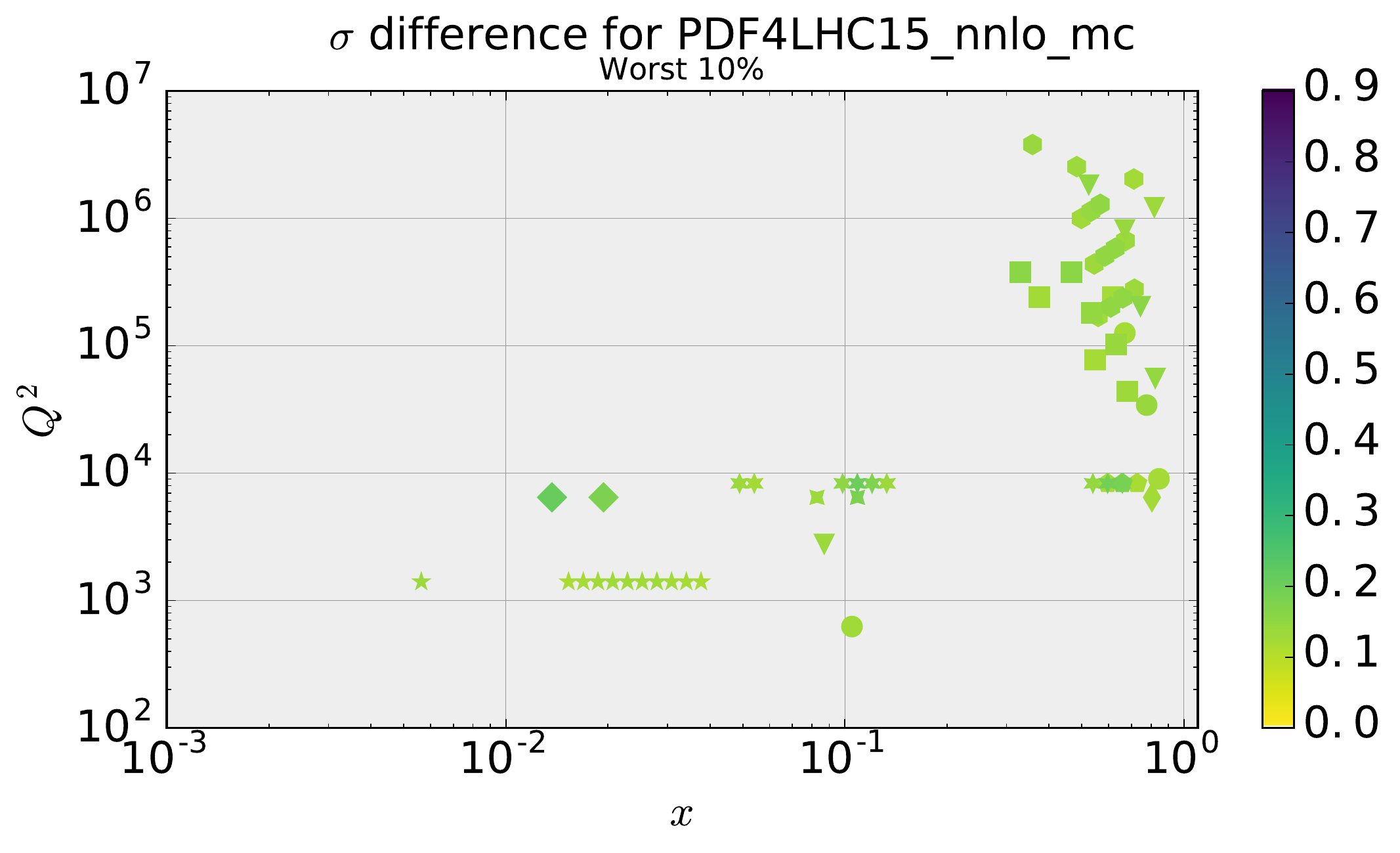}
  \caption{\small Same as Fig.~\ref{fig:kinplots}, but using the NNLO
    PDF sets.
   \label{fig:kinplots2}}
\end{figure}

In order to investigate the accuracy of PDF uncertainties in a more
detailed quantitative way, we define 
the relative
difference between the standard deviation, $s_i^{\rm (red)}$, of the cross-section $\sigma_i$
computed with the reduced sets, and  that of the prior,
$s^{\rm (prior)}_i$:
\begin{equation}
	\Delta_i \equiv \frac{\left| s^{\rm (prior)}_i  - s_i^{\rm (red)} \right|}{
	  s^{\rm (prior)}_i}
        \label{eq:reldiff}\, , \quad i=1,\ldots,N_{\sigma} \, .
\end{equation}
In Figs.~\ref{fig:kinplots} and~\ref{fig:kinplots2} 
the relative differences $\Delta_i$ are shown using NLO and NNLO PDFs 
for all hadronic observables which enter the NNPDF3.0 fit as a scatter plot in
the $(x,Q^2)$ kinematic plane, at the point corresponding to each
observable using leading order kinematics~\cite{Ball:2010de},
both for
all observables, and for the 10\% of observables
exhibiting the largest relative differences.
The $x$ value
    corresponding to the the parton with highest $x$ is always
    plotted, and for one-jet inclusive cross-sections, for which the
    $x$ values of the two partons are not fixed even at leading order,
    the largest accessible $x$ which corresponds to the rapidity range
    of each data point is plotted.
    Of course, these $x$ values should
    be only taken as indicative, and it should be born in mind that
    for most of the processes considered when one of the two partons
    involved is at large $x$, the other is at rather smaller $x$.
In this comparison, NLO theory is used
throughout.

From Figs.~\ref{fig:kinplots} and~\ref{fig:kinplots2}  we see that
    for {\tt PDF4LHC15\_nlo\_100} deviations are generally small, and
concentrated in regions in which  experimental information is scarce
and PDF uncertainties are largest,
such as the region of large $x$ and large $Q$.
For {\tt PDF4LHC15\_nlo\_mc} and {\tt PDF4LHC15\_nlo\_30} the
deviations are somewhat larger but still moderate in most
cases, with a few outliers. No significant difference is observed between NLO
and NNLO, consistent with the expectation that PDF uncertainties are
driven by data, not by theory, and thus are very similar at NLO and NNLO.

\begin{table}[t]
\footnotesize
\begin{centering}
  \begin{tabular}{|c|c|c|c|}
    \hline
process & distribution  &  $N_{{\rm bins}}$ & range  \tabularnewline
\hline 
\hline 
\multirow{2}{*}{$gg\to h$} &  $d\sigma/dp_t^h$  & 10 & {[}0,200{]} GeV\tabularnewline
 & $d\sigma/dy^h$ & 10 & {[}-2.5,2.5{]} \tabularnewline
\hline 
\multirow{2}{*}{VBF $hjj$} &  $d\sigma/dp_t^h$  & 5 & {[}0,200{]} GeV \tabularnewline
& $d\sigma/dy^h$ &  5 & {[}-2.5,2.5{]} \tabularnewline
\hline 
\multirow{2}{*}{$hW$}&  $d\sigma/dp_t^h$  & 10 & {[}0,200{]} GeV \tabularnewline
 & $d\sigma/dy^h$ &  10 & {[}-2.5,2.5{]} \tabularnewline
\hline 
\multirow{2}{*}{$hZ$} & $d\sigma/dp_t^h$  & 10 & {[}0,200{]} GeV \tabularnewline
& $d\sigma/dy^h$ &  10 & {[}-2.5,2.5{]} \tabularnewline
\hline
\multirow{2}{*}{$ht\bar{t}$} &  $d\sigma/dp_t^h$  & 10 & {[}0,200{]} GeV \tabularnewline
 & $d\sigma/dy^h$ &  10 & {[}-2.5,2.5{]} \tabularnewline
\hline
\end{tabular}$\quad $\begin{tabular}{|c|c|c|c|}
    \hline
process & distribution  &  $N_{{\rm bins}}$ & range  \tabularnewline
\hline 
\hline
\multirow{8}{*}{$Z$} & $d\sigma/dp_t^{l^-}$ &  10 & {[}0,200{]} GeV \tabularnewline
 & $d\sigma/dy^{l^-}$ &  10 & {[}-2.5,2.5{]} \tabularnewline
 & $d\sigma/dp_t^{l^+}$ &  10 & {[}0,200{]} GeV \tabularnewline
 & $d\sigma/dy^{l^-}$ &  10 & {[}-2.5,2.5{]} \tabularnewline
 & $d\sigma/dp_t^{Z}$ &  10 & {[}0,200{]} GeV \tabularnewline
 &  $d\sigma/dy^{Z}$&  5 & {[}-4,4{]} \tabularnewline
 & $d\sigma/dm^{ll}$&  10 & {[}50,130{]} GeV \tabularnewline
& $d\sigma/dp_t^{ll}$ &  10 & {[}0,200{]} GeV \tabularnewline
\hline
  \end{tabular}\\[0.2cm]
   \begin{tabular}{|c|c|c|c|}
    \hline
process & distribution  &  $N_{{\rm bins}}$ & range  \tabularnewline
\hline 
\hline 
\multirow{7}{*}{$t\bar{t}$}  &  $d\sigma/dp_t^{\bar{t}}$  &  10 & {[}40,400{]} GeV \tabularnewline
 & $d\sigma/dy^{\bar{t}}$ &  10 & {[}-2.5,2.5{]} \tabularnewline
 & $d\sigma/dp_t^{t}$ &  10 & {[}40,400{]} GeV \tabularnewline
 & $d\sigma/dy^{t}$&  10 & {[}-2.5,2.5{]} \tabularnewline
 & $d\sigma/dm^{t\bar{t}}$ &  10 & {[}300,1000{]} \tabularnewline
 & $d\sigma/dp_t^{t\bar{t}}$  & 10 & {[}20,200{]} \tabularnewline
 & $d\sigma/dy^{t\bar{t}}$ &  12 & {[}-3,3{]}  \tabularnewline
\hline
\end{tabular}$\quad $\begin{tabular}{|c|c|c|c|}
    \hline
process & distribution  &  $N_{{\rm bins}}$ & range  \tabularnewline
\hline 
\hline
\multirow{7}{*}{$W$} &   $d\sigma/d\phi$ &  10 & {[}0,200{]} GeV \tabularnewline
 & $d\sigma/dE_t^{\rm miss}$ &  10 & {[}-2.5,2.5{]} \tabularnewline
 & $d\sigma/dp_t^{l}$ &  10 & {[}0,200{]} GeV \tabularnewline
 & $d\sigma/dy^{l}$ &  10 & {[}-2.5,2.5{]} \tabularnewline
 & $d\sigma/dm_t$ &  10 & {[}0,200{]} GeV \tabularnewline
 &  $d\sigma/dp_T^{W}$&  5 & {[}-4,4{]} \tabularnewline
 & $d\sigma/y^{W}$&  10 & {[}50,130{]} GeV \tabularnewline
\hline
\end{tabular}
\par\end{centering}
\caption{\small LHC processes and the corresponding 
  differential distributions
  used as input in the construction of the
  {\tt SM-PDF-Ladder} set.
  In each case we indicate the range
  spanned by each distribution and the number of
  bins $N_{\rm bins}$.
  All processes have been computed for  $\sqrt{s}=13$ TeV.
  Higgs bosons and top quarks are stable, while weak gauge bosons are assumed
  to decay leptonically.
  No acceptance cuts are imposed with the exception of the leptons from
  the gauge boson decay, for which we require
  $p^l_{T}\geq10$ GeV and  $|\eta^{l}|\leq2.5$.
  \label{tab:processes_H}
}
\end{table}

This exercise shows that about $N_{\rm eig}=100$ Hessian eigenvectors are necessary
for a good accuracy general-purpose PDF set.
On the other hand, 
we have  recently argued~\cite{Carrazza:2016htc}  that
a much smaller set of Hessian eigenvectors is sufficient in order to
accurately reproduce a subset of cross-sections, and presented a
technique to construct such specialized minimal sets, dubbed SM-PDFs.
In order to test and
validate this claim, we have constructed two such SM-PDF sets, using
the methodology of Ref.~\cite{Carrazza:2016htc}, and starting
from the  PDF4LHC15 NLO prior:
\begin{itemize}
\item {\tt SM-PDF-ggh}: this SM-PDF set, with
  $N_{\rm eig}=4$
  symmetric eigenvectors, reproduces the inclusive cross-section
  and the $p_T$ and rapidity distributions of Higgs production
  in gluon fusion at $\sqrt{s}=13$ TeV.
\item {\tt SM-PDF-Ladder}: this SM-PDF set, with has $N_{\rm eig}=17$
  symmetric eigenvectors, reproduces all the observables listed in 
  Table~\ref{tab:processes_H}, which include a wide variety of LHC processes
  at $\sqrt{s}=13$ TeV.
\end{itemize}
The {\tt APPLgrid} grids
for the processes in Table~\ref{tab:processes_H}
have been  computed using {\tt aMC@NLO} interfaced
to {\tt aMCfast}.

\begin{figure}[t]
\begin{center}
  \includegraphics[width=0.49\textwidth]{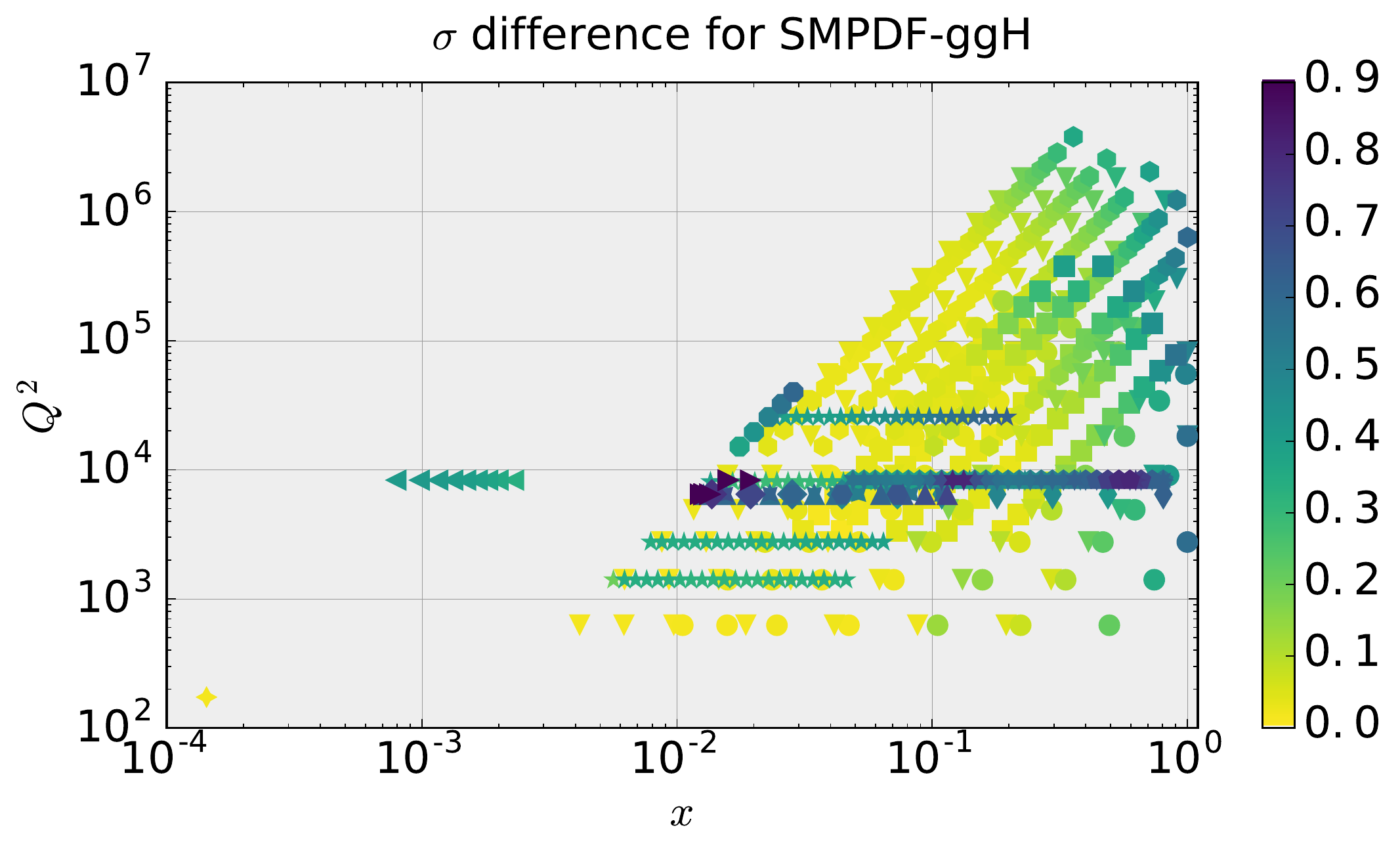}
  \includegraphics[width=0.49\textwidth]{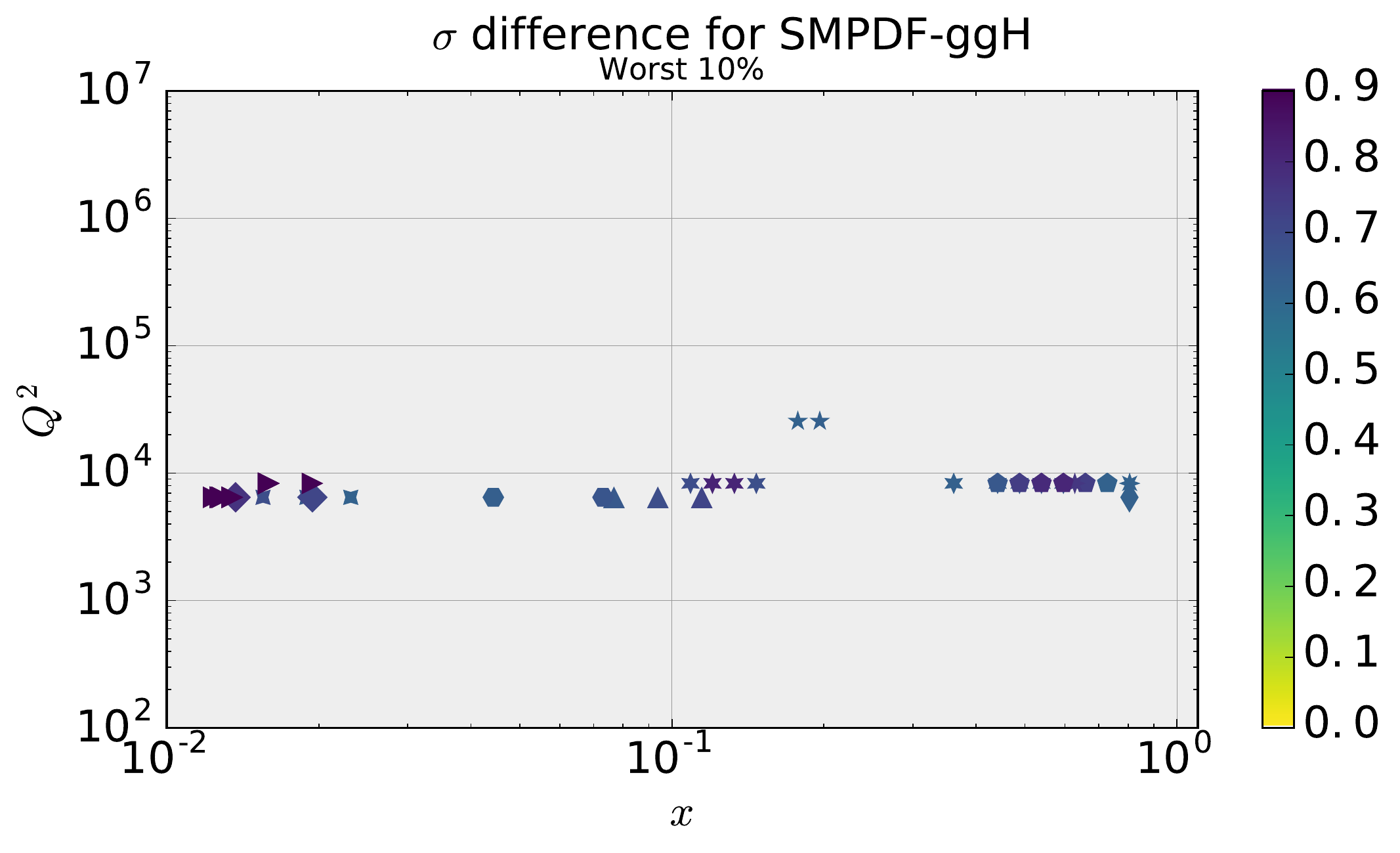}
  \includegraphics[width=0.49\textwidth]{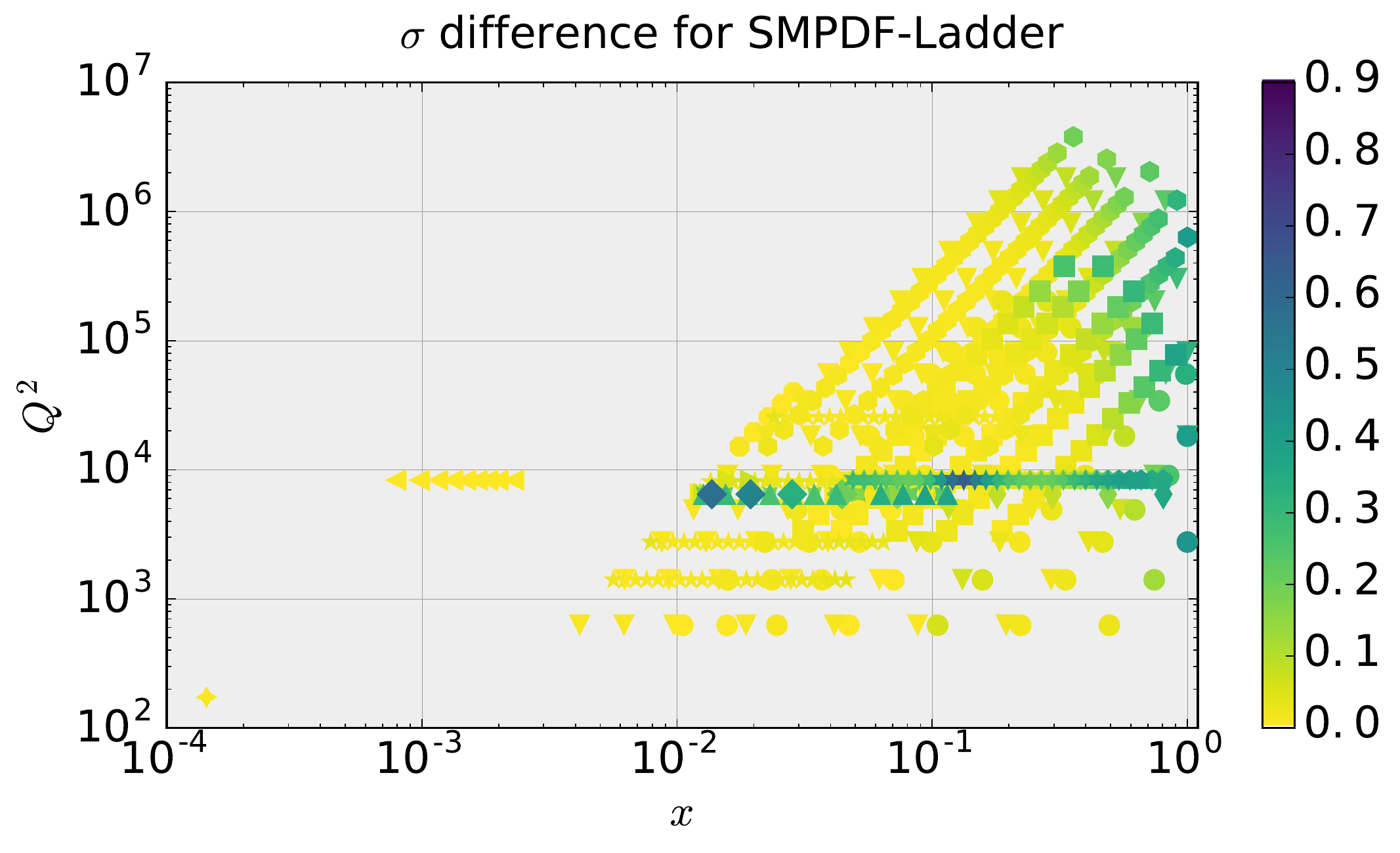}
  \includegraphics[width=0.49\textwidth]{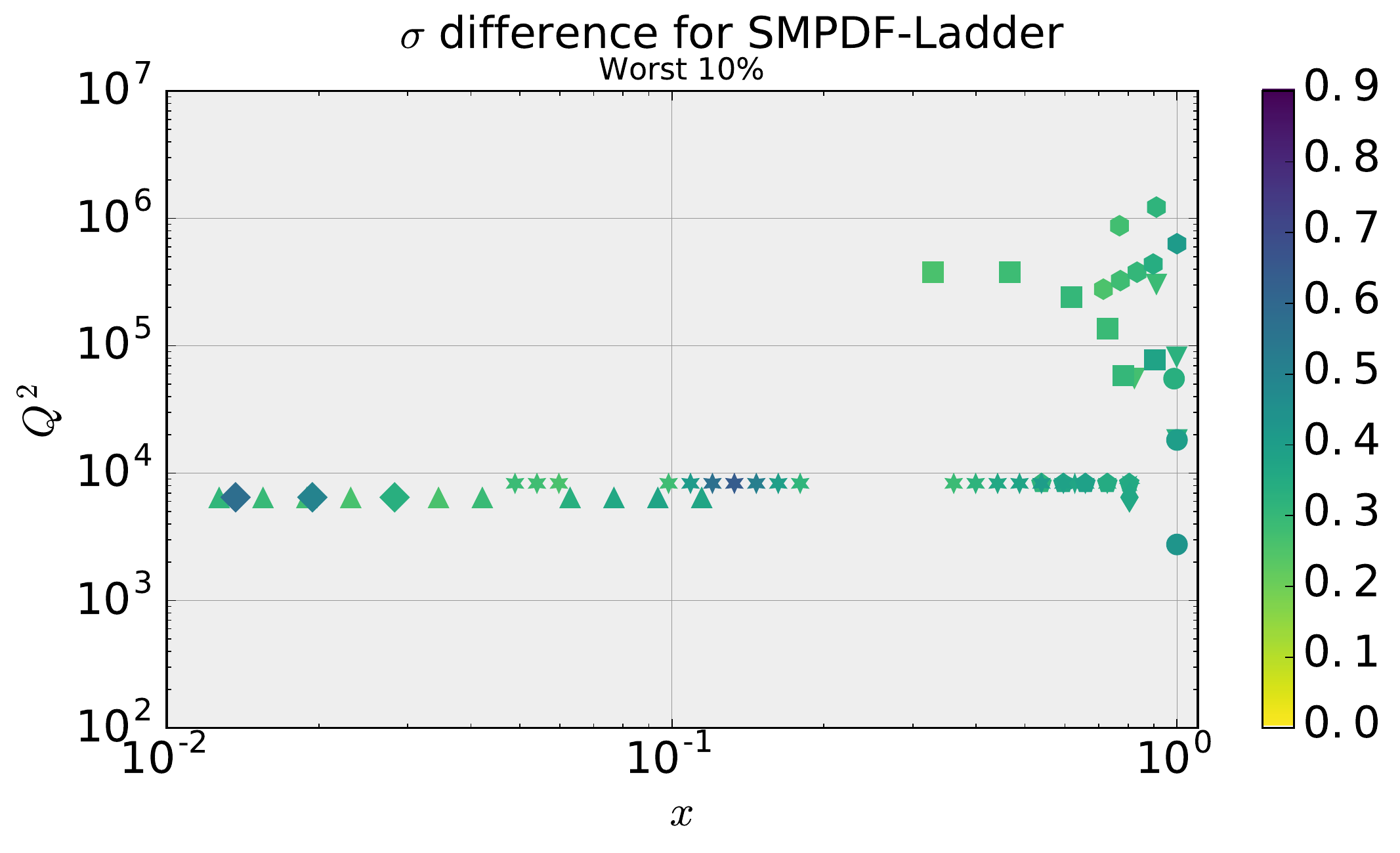}
\end{center}
\vspace{-0.6cm}
\caption{\small Same as Fig.~\ref{fig:kinplots}, this time for the two SM-PDF
  sets, {\tt SM-PDF-ggh} (upper plots) and {\tt SM-PDF-Ladder}
  (lower plots).
  \label{fig:kinplots3}}
\end{figure}

In Fig.~\ref{fig:kinplots3} we show again the relative difference
	$\Delta_i$ Eq.~(\ref{eq:reldiff}), which was shown in
Fig.~\ref{fig:kinplots}, but now comparing to the prior these two SM-PDF sets.
In the case of the {\tt SM-PDF-ggh} set,
we find good agreement with the prior
for all cross-sections on the region $x\simeq 0.01$ and
$Q\simeq 100$ GeV, relevant for Higgs production in gluon fusion.
On the other hand, as we move outside this region,
the accuracy rapidly deteriorates. This  exemplifies the virtues 
and limitations of the SM-PDF approach: a very small number of
eigenvectors is sufficient to reproduce a reasonably small set of
observables,  but if one tries to stretch results to too many
processes there is accuracy loss.
The {\tt SM-PDF-Ladder} set, on the other hand,
exhibits a similar performance
as the
{\tt
  PDF4LHC15\_nlo\_30} set.

\subsection{Non-Gaussianities in the PDF4LHC combination}

As discussed in the PDF4LHC15 report~\cite{Butterworth:2015oua},
the Monte Carlo combination of individual PDF set  in general is not
Gaussian. This is both because one of the three sets entering the
combination, NNPDF3.0, allows for non-Gaussian behaviour, and also
because in general the combination of Gaussian sets is not necessarily
Gaussian itself. 
We will now study in a more systematic way the degree of non-Gaussianity
of the prior set, and specifically correlate the comparison of the
reduced sets to the prior with  the degree of non-Gaussianity of the
prior. This has the threefold purpose of determining how much the
accuracy of the Hessian set deteriorates in the presence of
non-Gaussianities, of checking that the reduced MC set correctly
reproduces the non-Gaussianity of the prior, and of providing guidance
on when the MC set should be favored over the Hessian sets in order to
reproduce the non-Gaussianity.
\clearpage

In order to study non-Gaussianity, we proceed in two steps. First, we
turn the histogram, as
obtained from a Monte Carlo representation, into a continuous
probability distribution. Then, we compare this probability
distribution to a Gaussian with the same mean and standard deviation.
The first step is accomplished using the
Kernel
Density Estimate (KDE) method. The second, using the Kullback–Leibler
(KL) divergence as a measure of the difference between two probability
distributions (for a brief review of
both methods see e.g. Ref.~\cite{Bishop:1995}).

The KDE method consists of constructing the probability distribution
corresponding to a histogram as the
average of kernel
functions $K$ centered at the data from which the histogram would be constructed. In our case, given $k=1,\dots,N_{\rm
  rep}$ replicas of the $i$-th cross-section
 $\{\sigma_i^{(k)}\}$, the probability distribution is
\begin{equation}
  P(\sigma_i)=\frac{1}{N_{\rm rep}}\sum_{k=1}^{N_{\rm rep}}K\left(\sigma_i-\sigma_{i}^{(k)}\right)\,,\quad
  i=1,\ldots,N_{\sigma} \, .\label{eq:KDE}
\end{equation}
We specifically choose 
\begin{equation}
K(\sigma-\sigma_{i})\equiv \frac{1}{h\sqrt{2\pi}}\exp\lp -\frac{(\sigma-\sigma_{i})^{2}}{h}\rp\,,\label{eq:kdenormal}
\end{equation}
where the parameter $h$, known as  bandwidth, is 
 \begin{equation}
   h = \hat{s}_i \left( \frac{4}{3N_{\rm rep}}  \right)^\frac{1}{5} \, ,
 \end{equation}
 where $\hat{s}_i$ is the standard deviation of the given sample of
 replicas. This choice is known as {\it Silverman rule}, and, if the
 underlying probability distribution is Gaussian, it minimizes
 the integral of the square difference between the ensuing distribution and
 this underlying Gaussian~\cite{Silverman:1998}.
Once turned into continuous distributions via the KDE method, the
prior and reduced Monte Carlo sets can be compared to each other, to a
Gaussian, and to the Hessian sets. The comparison can be performed
using the  Kullback–Leibler (KL) divergence, which measures the
information loss when using
a probability
distribution  $Q(x)$ to approximate a  prior $P(x)$, and
is given by
\begin{equation}
  \label{eq:kldiv}
D_{\rm KL}^{(i)}(P|Q)=\int_{-\infty}^{+\infty}\lp P(x)\cdot \frac{\log
P(x)}{\log Q(x)} \rp dx. \ ,
\end{equation}

\begin{figure}[t]
\begin{center}
  \includegraphics[width=0.75\textwidth]{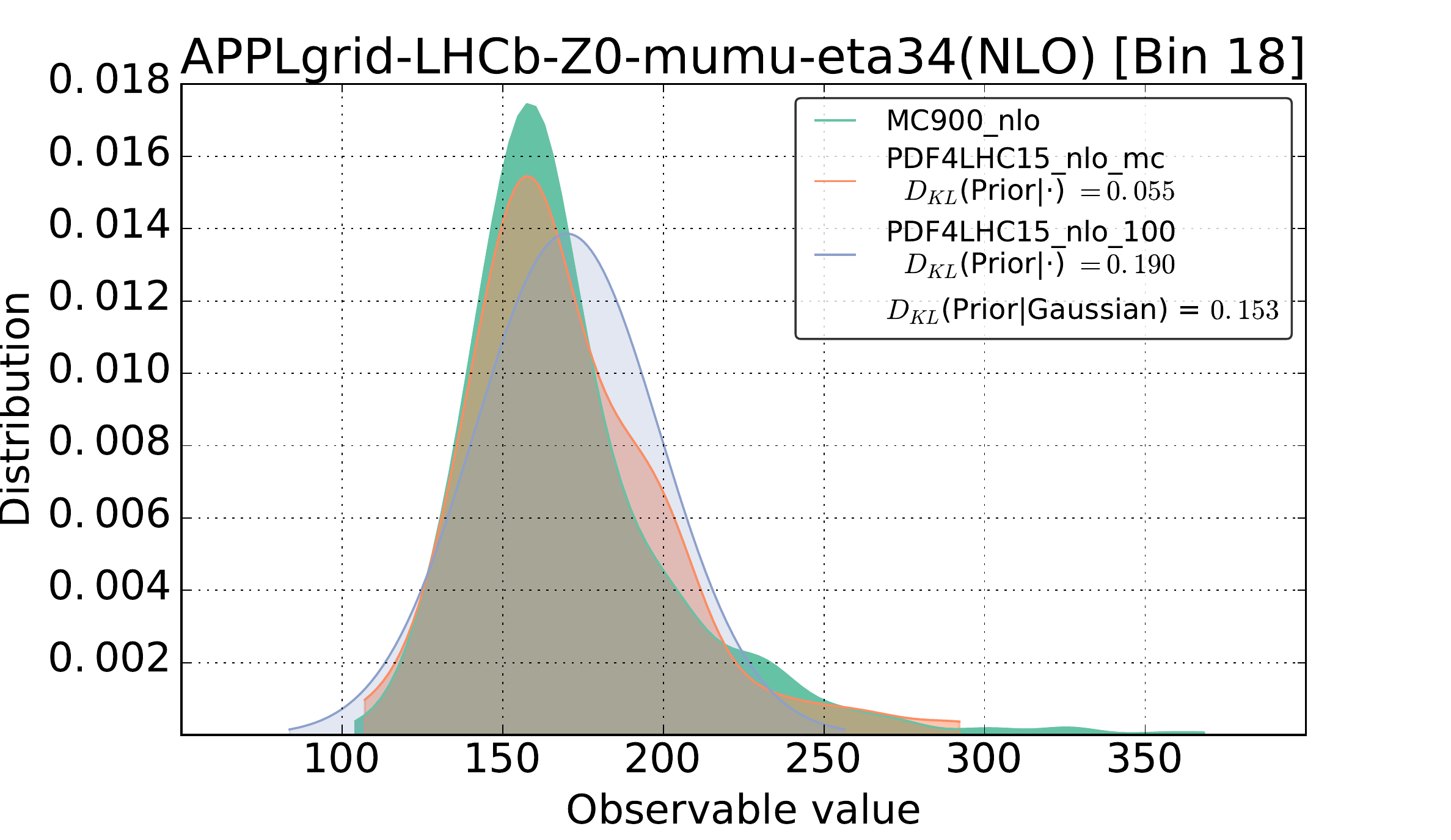}
\end{center}
\vspace{-0.3cm}
\caption{\small The probability distribution for 
the most forward bin
  in the LHCb $Z\to \mu\mu$ 8 TeV measurement obtained using the
  PDF4LHC15 prior and the  {\tt PDF4LHC15\_nlo\_mc} and
  {\tt PDF4LHC15\_nlo\_100} reduced sets.
 The value of the KL divergence $D_{\rm KL}$
  Eq.~(\ref{eq:kldiv}) between the prior and a Gaussian, and between 
each of the two reduced sets and the
  prior for this distribution, are also given.
\label{fig:example1}
}
\end{figure}

As a first example,
in Fig.~\ref{fig:example1} we select a data bin in which the
distribution of PDF replicas is clearly non-Gaussian,  namely the
most forward rapidity bin
  in the LHCb $Z\to \mu\mu$ 8 TeV measurement~\cite{Aaij:2015vua}, and
  we compare the distribution obtained  using the
  PDF4LHC15 prior to those found using 
the 
  {\tt PDF4LHC15\_nlo\_mc} and
  {\tt PDF4LHC15\_nlo\_100} reduced sets. The continuous distribution
  shown is obtained from the 
prior and reduced MC samples using the KDE method discussed above. 
For the {\tt PDF4LHC15\_nlo\_100} set the distribution shown is a
Gaussian with with and central value using the standard procedure,
based on linear error propagation, which is used to obtain predictions
from Hessian sets: namely, the central set provides the mean, and the
standard deviation is the sum in quadrature of the deviations obtained
using each of the error sets. 

\begin{figure}[t]
\begin{center}
 \includegraphics[width=0.75\textwidth]{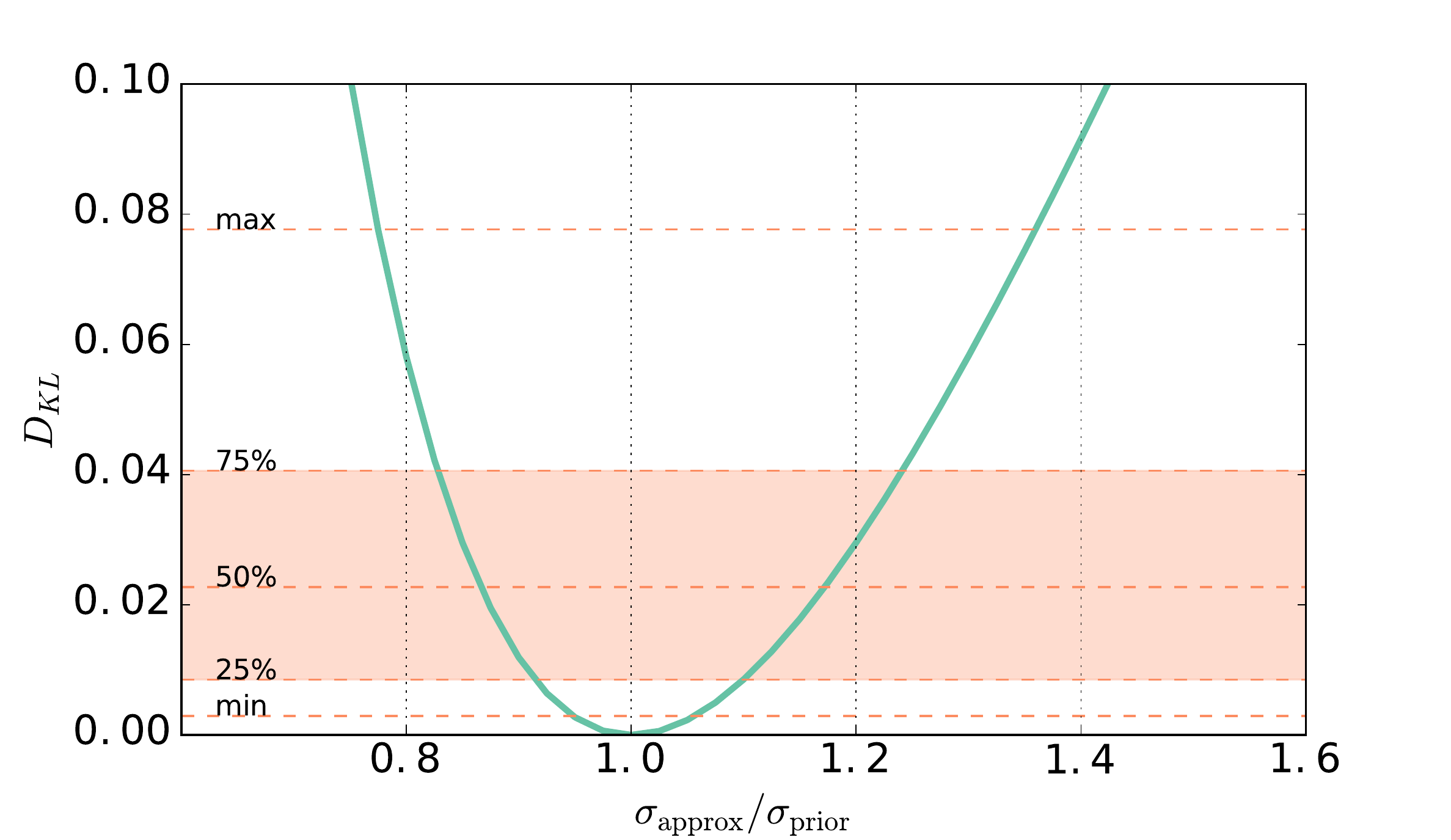}
\end{center}
\vspace{-0.3cm}
\caption{\small The
   KL divergence $D_{KL}$ Eq.~(\ref{eq:kldiv})  between two Gaussian distributions
  with the same mean but different widths, as a function of the ratio
  of their standard deviations.
We also show (horizontal lines) 
the highest value, lowest value, and the edges of the quartiles of
the distribution of $D_{KL}$ values between the prior and a Gaussian
approximation to it, for all observables listed in
Table~\ref{tab:processes_H}.
 \label{fig:example2}
}
\end{figure}

The KL divergence between the prior and a Gaussian is equal to
 $D_{KL}=0.153$, while the divergence between the prior
 and
its reduced MC representation is $D_{KL}=0.055$, and finally between the
prior and Hessian set it is
$D_{KL}=0.19$. This shows that the reduced MC representation of the
prior is much closer to it than the prior is to a Gaussian, while the
Hessian representation differs from it even more.
In order to facilitate the
interpretation these values of the KL
divergence, in Fig.~\ref{fig:example2} we plot the value of the KL
divergence between two Gaussian with different width, as a function
of the ratio of their width: the plot shows that  $D_{KL}\sim0.05$
corresponds to distorting the width of a Gaussian by about 20\%.
In this figure we also show as horizontal lines the minimum and
maximum values that we obtained, as well as the edges of the four
quartiles of the distribution of results.

We have extended the type of comparisons shown in
Fig.~\ref{fig:example1} into a systematic study
including
all the cross-sections listed in 
Table~\ref{tab:processes_H}.
As discussed in Sect.~\ref{sec:pdf2:validation},
this is a reasonably representative set of observables, since
it is possible to construct a PDF set, the {\tt SM-PDF-Ladder}, which is adequate to
describe  them and is also accurate to describe
all the hadronic cross-section from the NNPDF3.0 fit (see Fig.~\ref{fig:kinplots3}).
Specifically, for each cross-section
we have determined the probability distribution from the prior using
the KDE method Eq.~(\ref{eq:KDE}),
and also a Gaussian approximation to it, defined as
the Gaussian with the same mean and standard deviation as the
prior.
We have computed the KL divergence between the prior
distribution and this Gaussian approximation.
It is clear that the vast
majority of observables exhibits Gaussian behaviour to good
approximation, with extreme cases such as shown in
Fig.~\ref{fig:example1} happening in a small fraction of the first
quartile. 
\begin{figure}[t]
\begin{center}
\includegraphics[width=0.75\textwidth]{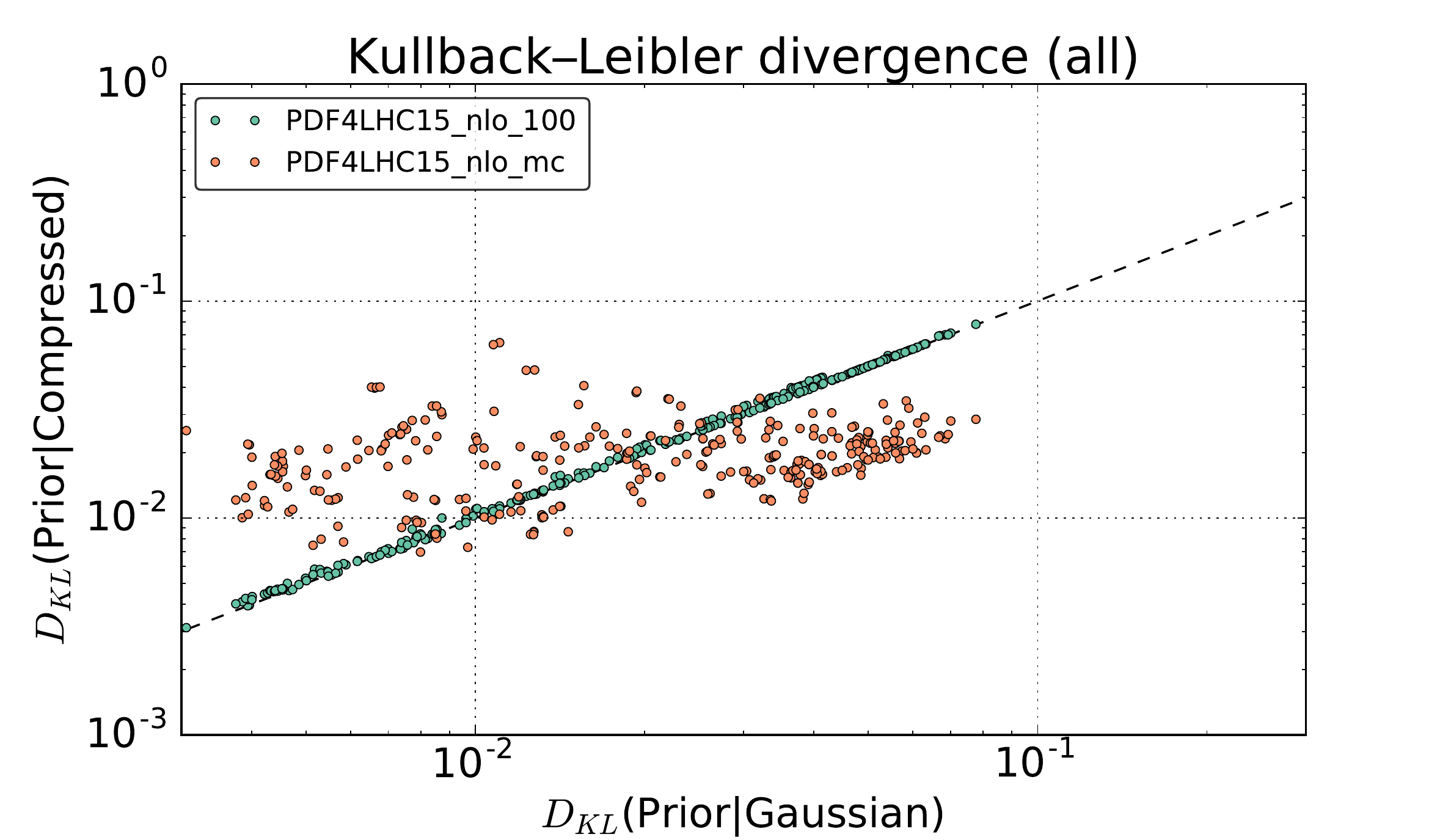}
\end{center}
\vspace{-0.3cm}
\caption{\small 
  \label{fig:kl_all} The KL divergence, Eq.~(\ref{eq:kldiv}) between
  the prior and each of its two reduced representations
  {\tt PDF4LHC15\_nlo\_prior} (Monte Carlo) and 
  {\tt PDF4LHC15\_nlo\_mc} (Hessian) vs. the divergence between the
  prior and its Gaussian approximation, computed for all observables
  listed  in Table~\ref{tab:processes_H}.
}
\end{figure}

\begin{figure}[t]
\begin{center}
\includegraphics[width=0.45\textwidth]{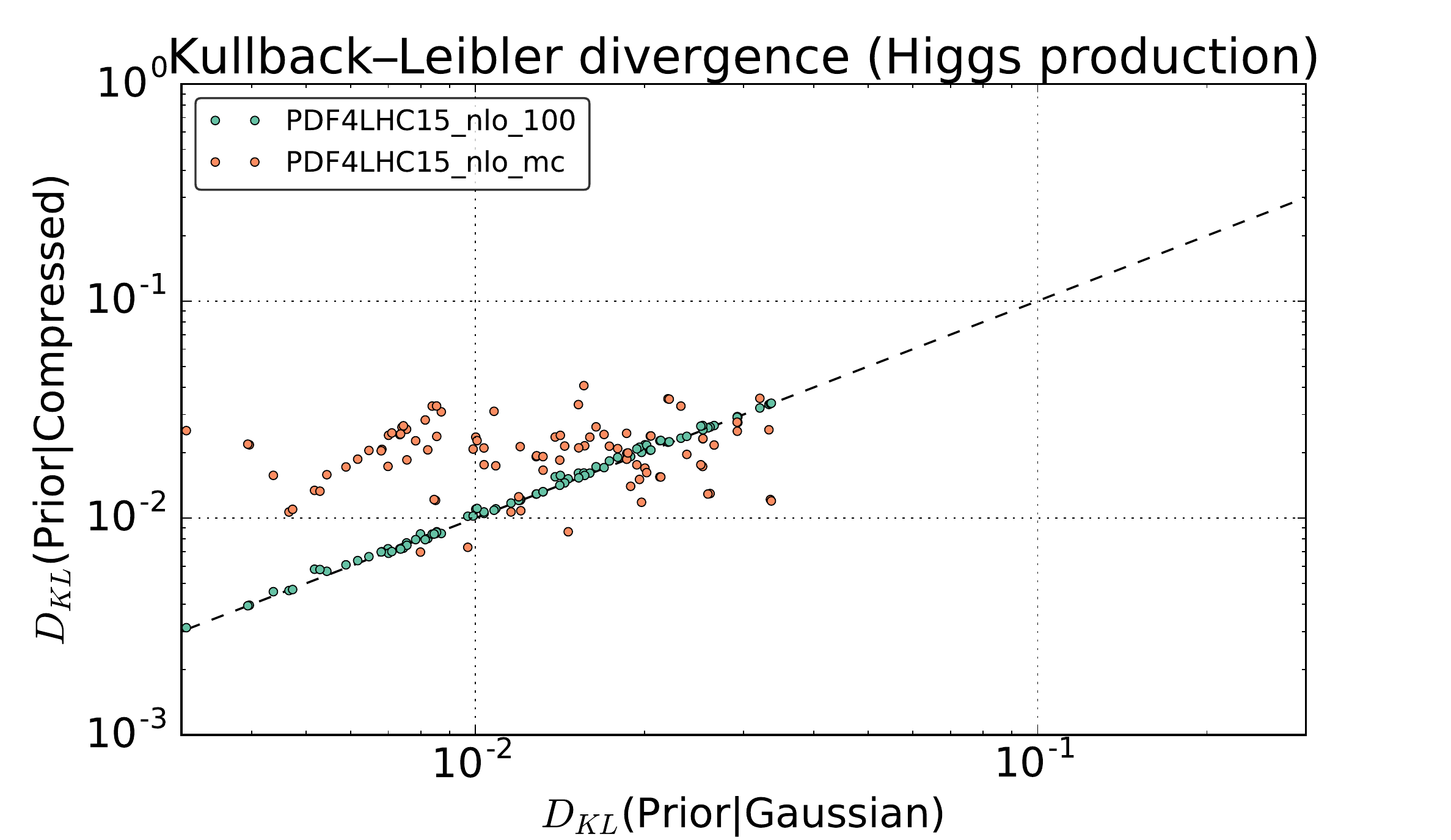}
\includegraphics[width=0.45\textwidth]{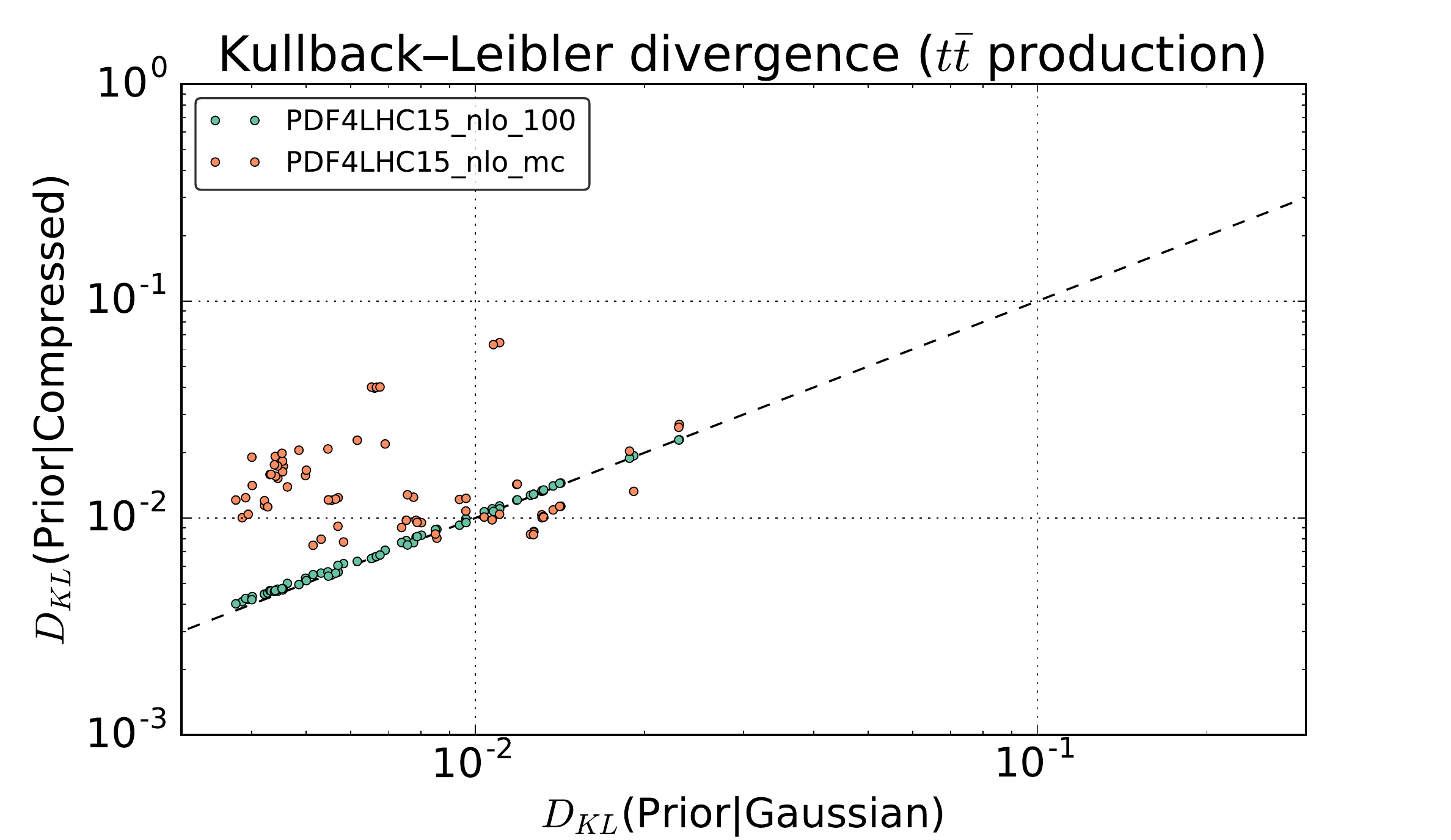}\\
\includegraphics[width=0.45\textwidth]{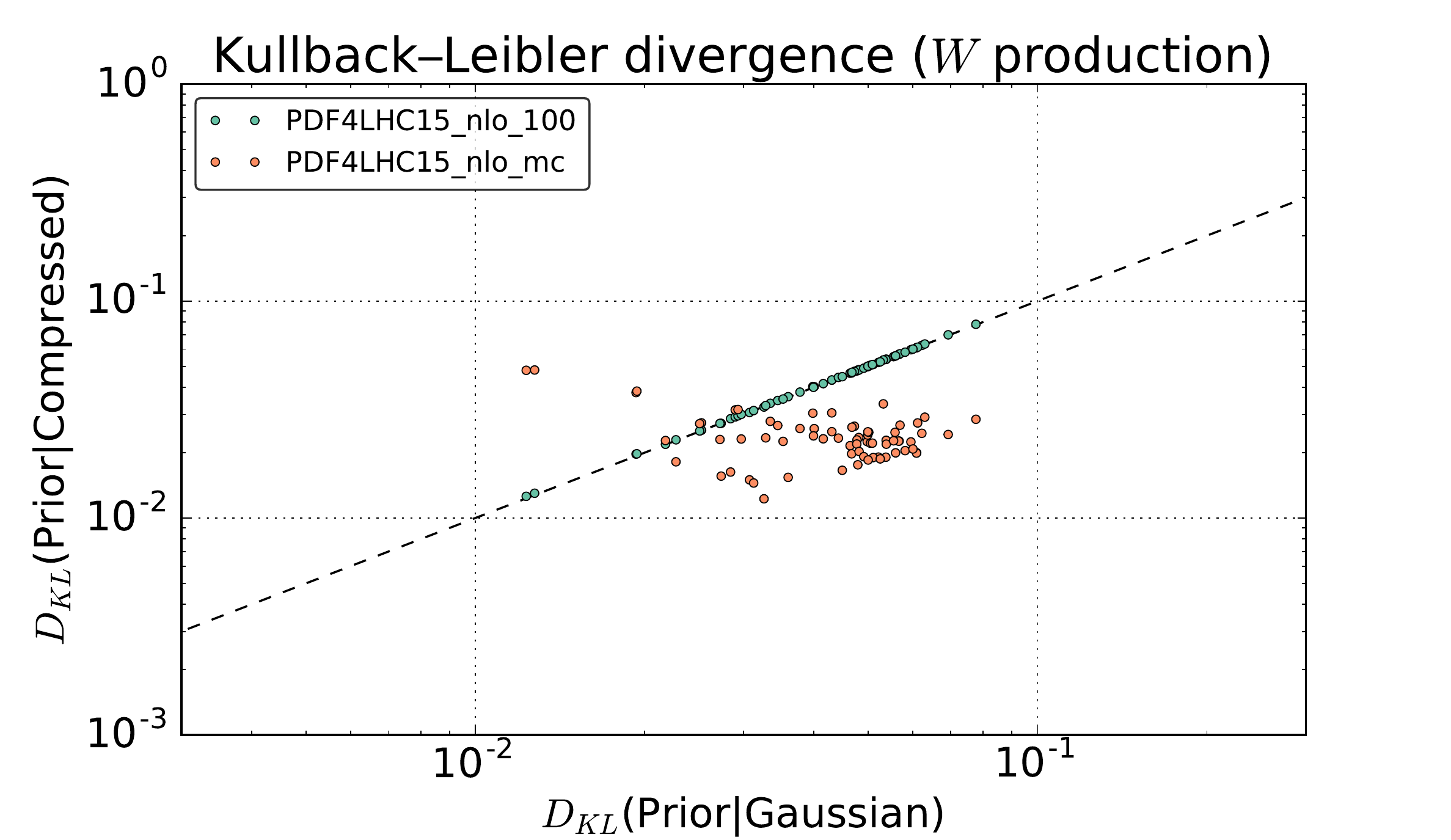}
\includegraphics[width=0.45\textwidth]{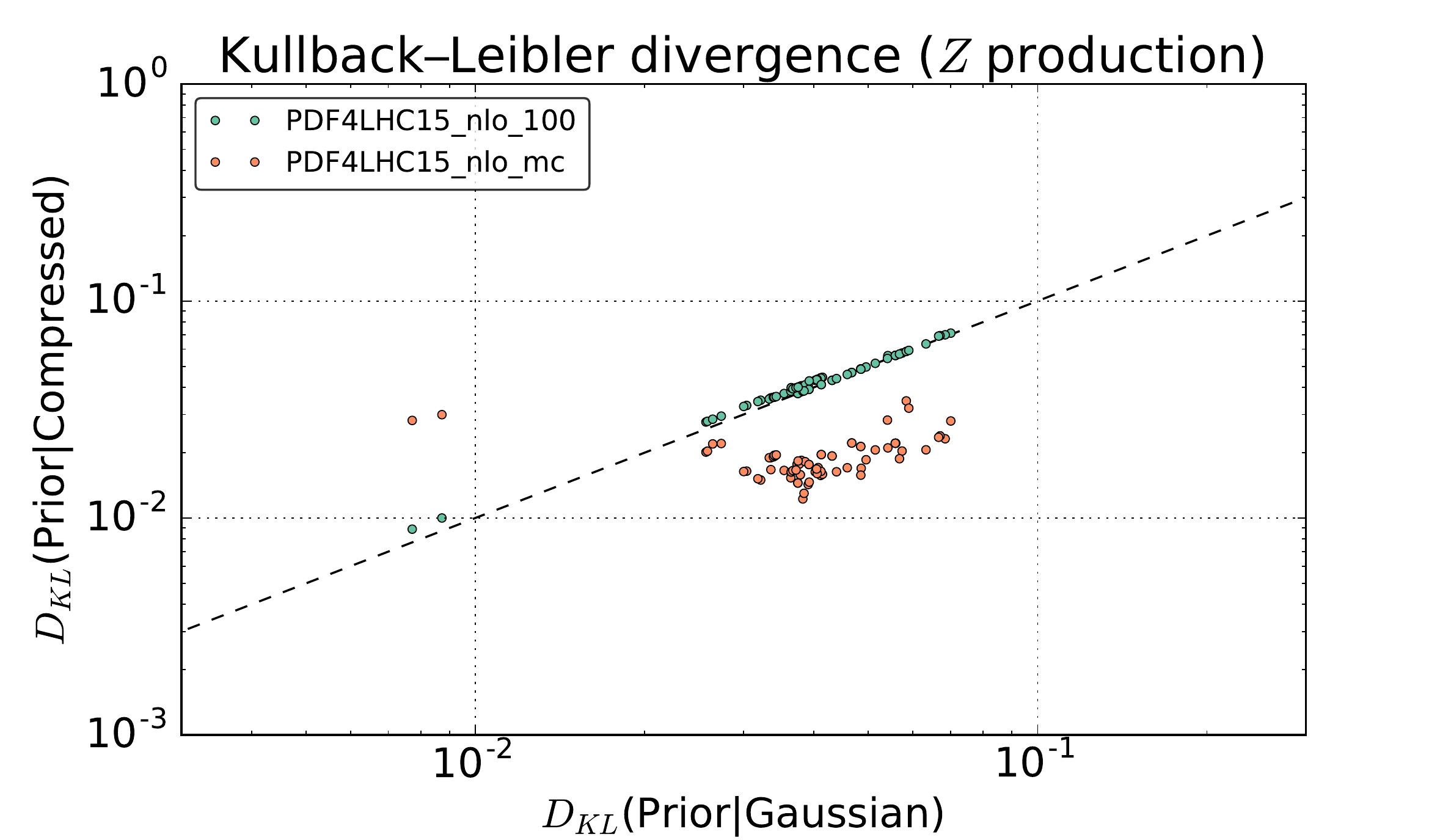}
\end{center}
\vspace{-0.3cm}
\caption{\small 
  \label{fig:kl_allbreak} Same as Fig.~\ref{fig:kl_all}, now separating
  the contributions of the different classes of processes of
  Table~\ref{tab:processes_H}: Higgs production (top left), top quark
  pair production (top right), $W$
  production (bottom left) and $Z$ production (bottom right).
}
\end{figure}

We have then computed for each observable the KL distance between
the prior and the  {\tt PDF4LHC15\_nlo\_mc} and  {\tt
  PDF4LHC15\_nlo\_100} combined sets.
Results are collected in
  Fig.~\ref{fig:kl_all} for all processes, while in
  Fig.~\ref{fig:kl_allbreak} we show a breakdown for the four classes
  of processes of Table~\ref{tab:processes_H}: Higgs, top, $W$ and $Z$ production.
For each cross-section there are two points on the plot, one
  corresponding to {\tt PDF4LHC15\_nlo\_mc}  and the other to {\tt
  PDF4LHC15\_nlo\_100}. The points are plotted with  on the $x$ axis the KL
  divergence between the prior and its gaussian approximation, and on
  the $y$ axis the same quantity now evaluated
  between the prior and the compressed set.
For the {\tt
  PDF4LHC15\_nlo\_100} all points cluster on the diagonal: this means that
the reduced Hessian set only deviates from the prior inasmuch as the
prior deviates from a Gaussian --- only for a more extreme deviation
from Gaussian such as shown in Fig.~\ref{fig:example1} does the
reduced Hessian deviate more. 
The  {\tt PDF4LHC15\_nlo\_mc} points
instead approximately fall within   a horizontal band: this means that the
quality of the approximation to the prior of the
reduced MC does not depend on the
degree of non-Gaussianity of the prior itself. 

Hence, the reduced MC
set does reproduce well the non-Gaussian features of the prior, when
they are present, and it will be advantageous to use it  
for points where the center
of the band corresponding to  {\tt PDF4LHC15\_nlo\_mc} is below the
diagonal. Figure~\ref{fig:kl_allbreak}   shows that this happens
for a significant fraction of the $W$ and $Z$ production
cross-sections, but not
for top and Higgs production.
This is consistent with the expectation that
non-Gaussian behaviour is mostly to be found in large $x$ PDFs, which
are probed by gauge boson production at high rapidity, but not by
Higgs and top production which are mostly sensitive to the gluon PDF at
medium and small $x$.
%

In order to further elucidate the dominant  non-Gaussian features, we
have  performed a comparison of 
the mean and the standard deviation of each probability
distribution with respectively the median and the 
minimum 68\% confidence interval $R$, defined as
\begin{equation}
R=\frac{1}{2}\min\{[x_{\min},x_{\max}];\qquad
\int_{x_{\min}}^{x_{\max}} P(x) =0.683 \, .
\end{equation}
The deviation of the median from the mean is a measure of the
asymmetry of the distribution, while the deviation  $R$ from the
standard deviation is a measure of the presence of outliers.
We then define two estimators, one for the deviation of the mean from the
median and for the deviation of the standard deviation $s$ from $R$:
\begin{equation}
	\label{eq:meanshift}
	\Delta_{\mu} = \frac{\textrm{median} - \mu}{s} \, ,
\end{equation}
\begin{equation}
	\label{eq:stdshift}
	\Delta_{s} = \frac{R - s}{s} \, .
\end{equation}
Both $\Delta_{\mu}$ and $\Delta_s$ would vanish for a Gaussian in the
limit of infinite sample size.

\begin{figure}[t]
\begin{center}
\includegraphics[width=0.49\textwidth]{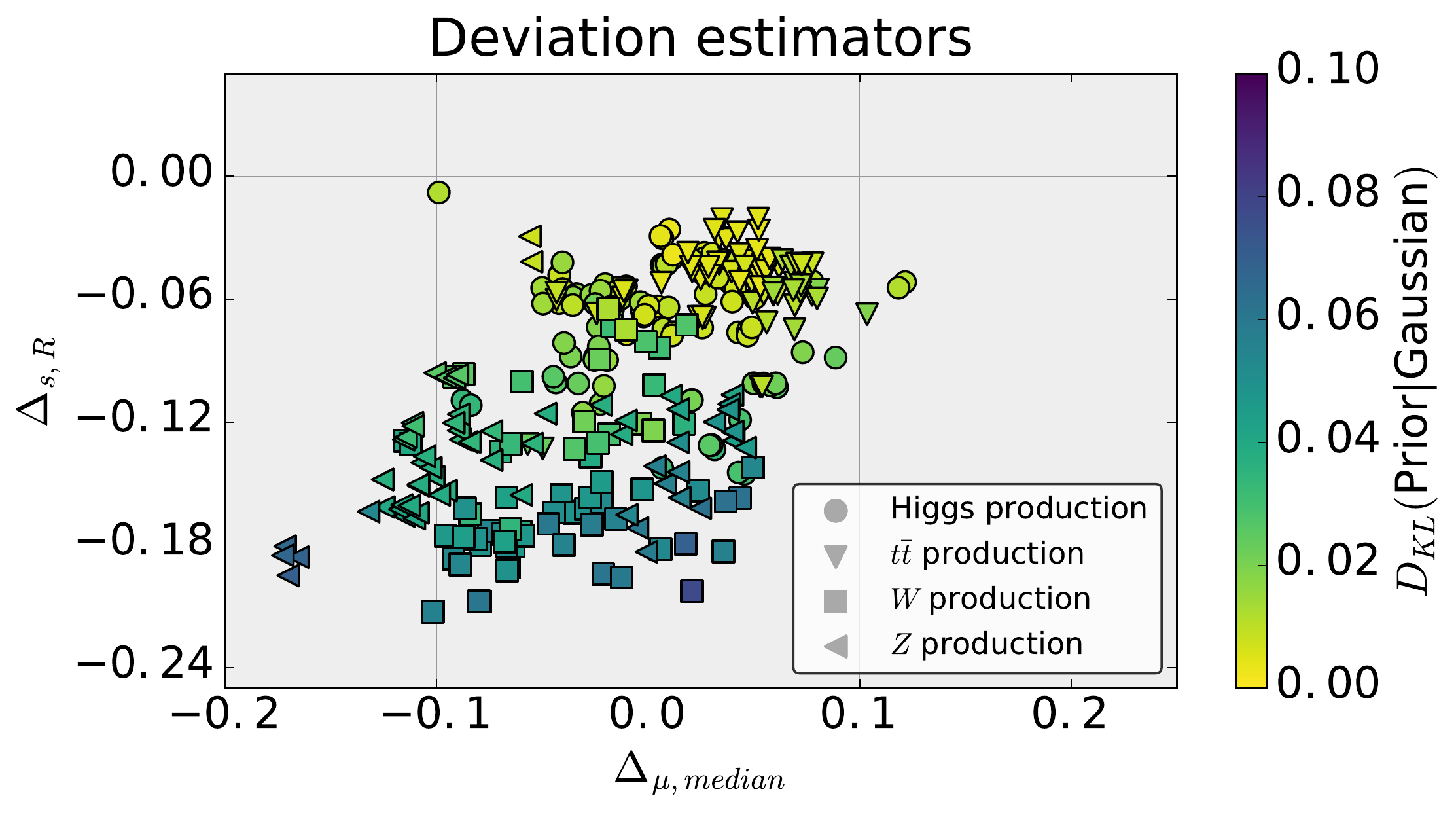}\\
\includegraphics[width=0.49\textwidth]{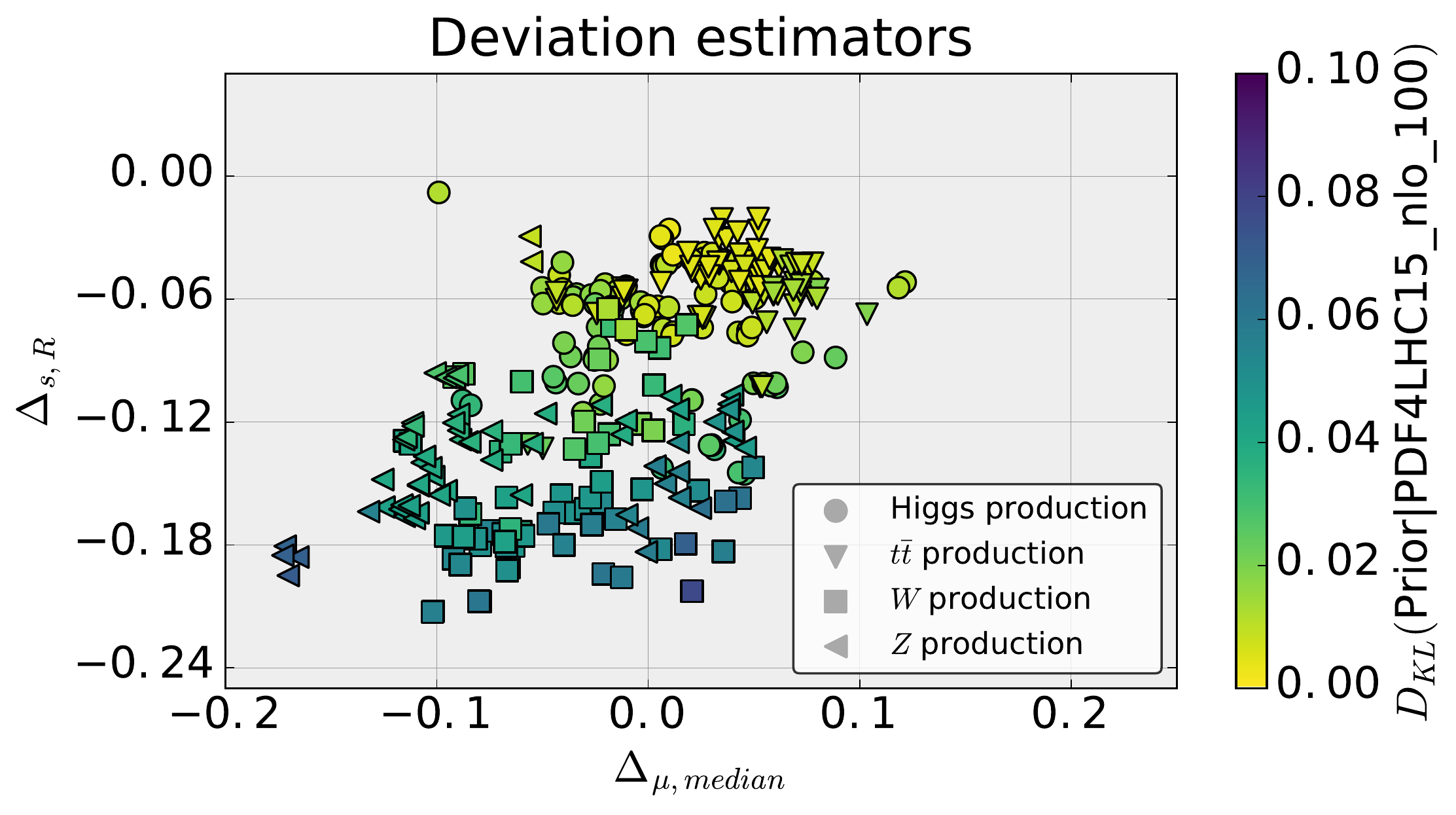}
\includegraphics[width=0.49\textwidth]{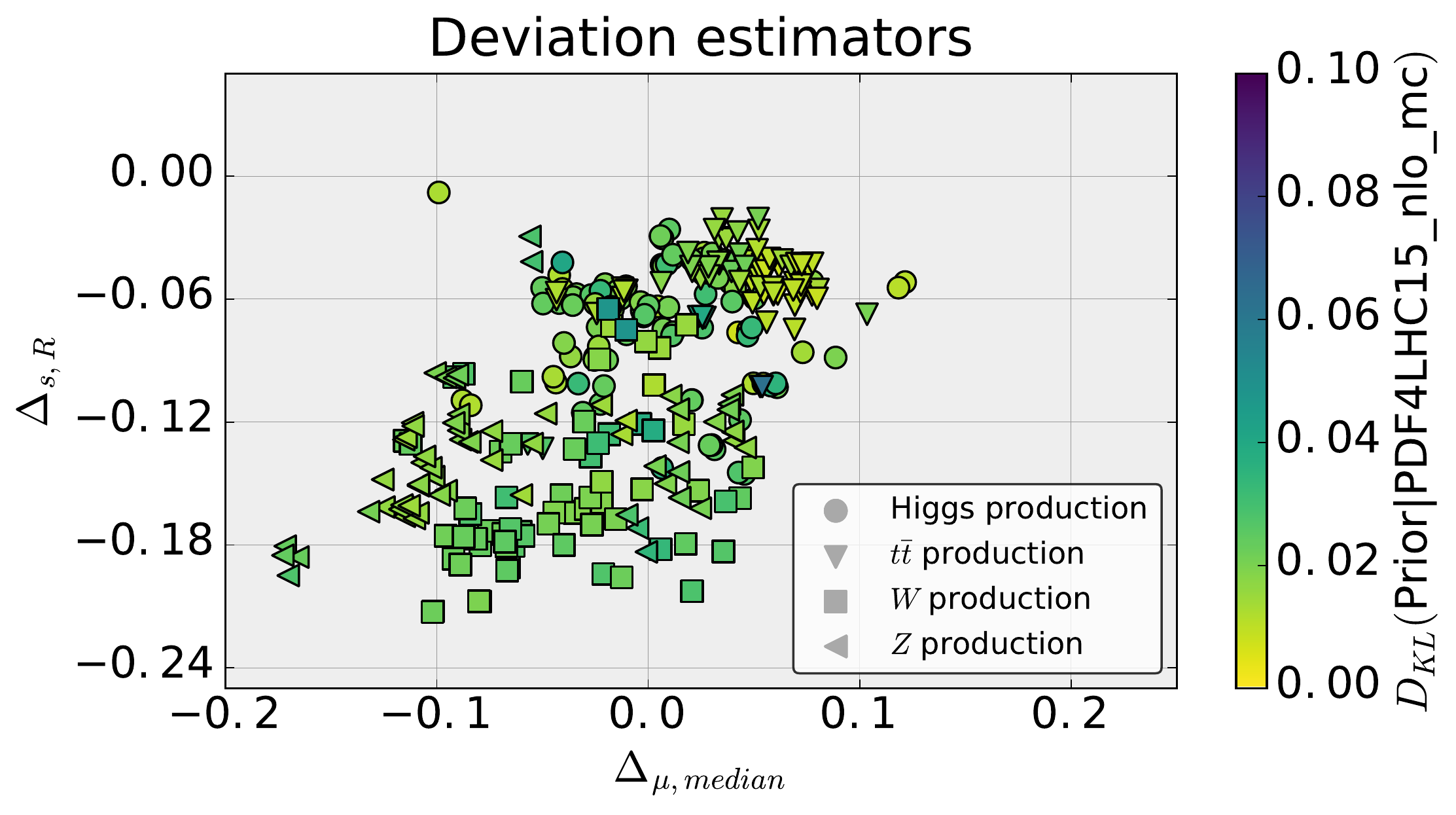}
\end{center}
\vspace{-0.3cm}
\caption{\small 
  \label{fig:estimators} Scatter plot of indicators of deviation from
  gaussianity for all the cross-sections of Table~\ref{tab:processes_H}.
  For each observable, the shift between the median and the mean
  Eq.~(\ref{eq:meanshift}) is shown
  in the horizontal axis, while the shift between standard deviation
  and the 68\% interval Eq.~(\ref{eq:stdshift}) is represented on the vertical
  axis.
  The color code shows the KL divergence between the prior and
  either a
  Gaussian (top) or the two reduced sets (bottom): Hessian
  (left) and MC (right).
  }
  
\end{figure}

In Fig.~\ref{fig:estimators} we represent these two
estimators in a scatter plot, with $\Delta_{\mu}$ and
$\Delta_s$ respectively on the $x$ and $y$ axis,
computed for all the cross-sections of Table~\ref{tab:processes_H}.
In addition,
we show a
color code with the KL
divergence between the prior and respectively its Gaussian
approximation and its two reduced MC and Hessian representations.
From this comparison, it is clear that  the shift in the median 
is  only weakly correlated to the degree of non-Gaussianity (top
plot), and also weakly correlated to the shift is standard deviation,
which instead is strongly correlated to non-Gaussianity.

In the presence of outliers, $R\le s$, and indeed $R$ is seen to be
always negative.
We expect asymmetries related to non-Gaussian
behaviour to be due to the fact that in some cases PDFs are bounded
from below by positivity, but not from above where outliers may be
present.
Indeed in the non-gaussian region $\Delta_{\mu}$ tends to be
negative, but with large fluctuations in its value.
The same correlations are seen with the KL divergence between prior
and Hessian, again showing that this is dominated by non-gaussian
behaviour.
On the other hand, no correlation is observed from the divergence between
prior and reduced MC, consistent with our conclusion that the
performance of the compressed MC set is independent of the degree of
non-Gaussianity.

\subsection{Summary and outlook}

In this contribution we have performed a systematic comparison of the three reduced
PDF4LHC15 PDF sets with the prior distribution they have been
constructed from, with particular regard to non-Gaussian features, by
comparing predictions for a wide variety of LHC
cross-sections.
Our
general conclusion is that the three sets all perform as expected. We
have specifically verified that the   {\tt PDF4LHC15\_nlo\_100}
Hessian set provides generally the most accurate representation of the
mean and standard deviation of the probability distribution, while the
{\tt PDF4LHC15\_nlo\_mc}  and {\tt PDF4LHC15\_nlo\_30} sets are
less accurate though still quite good.
We have also verified that specialized SM-PDF~\cite{Carrazza:2016htc}
sets can give an equally accurate representation, but with a smaller
number of error-sets, at the price of not being suited for all
possible processes, but with the option of combining them with other
more accurate sets.
We
have then verified that in the presence of substantial deviations from
Gaussianity, the  {\tt PDF4LHC15\_nlo\_mc} set
is the most accurate.
By
providing a breakdown of our comparisons by type of process, we have
verified that both deviations from Gaussianity and loss of accuracy of
the smaller Hessian set are more marked in regions which are sensitive
to poorly known PDFs, such as the anti-quarks at large $x$.

The results for  the $N_{\sigma}\simeq 600$
cross-sections used for the calculations in
Figs.~\ref{fig:kinplots}--\ref{fig:kinplots3} are
available from the link
\begin{center}
\url{http://pcteserver.mi.infn.it/~nnpdf/PDF4LHC15/gall}
\end{center}
from where they can be accessed in HTML, CSV and ODS formats.

\subsection*{Acknowledgements}
S.~C. and S.~F. are supported in part by an Italian PRIN2010 grant and
by a European Investment Bank EIBURS grant.
S.~C. is supported by the HICCUP ERC Consolidator grant (614577).
S.~F. and Z.~K. are
supported by the Executive Research Agency (REA) of the European
Commission under the Grant Agreement PITN-GA-2012-316704 (HiggsTools).
J.~R. is supported by an STFC Rutherford Fellowship
and Grant ST/K005227/1 and ST/M003787/1 and
by an European Research Council Starting Grant {\it ``PDF4BSM''}.


\clearpage

\chapter{SM Higgs working group report}
\label{cha:higgs}

\renewcommand{\arraystretch}{1.6}

A pillar of the physics program during the second Run of the LHC (Run II) is the detailed study of the 
Higgs boson. The Higgs discovery indeed represents the triumph of Run I, and now, in the second 
phase of exploration, the LHC will seek to constrain its properties in even greater detail. 
The Higgs boson is a special particle, both in the context of the Standard Model (SM) and in beyond the SM (BSM) scenarios. 
As the only fundamental scalar particle in the theory, the Higgs boson provides a unique window to extensions of the SM. 
For instance, the SM Higgs potential is uniquely determined once the Higgs mass is fixed.
New scalars from BSM physics would alter the Higgs potential,
resulting in modifications of the Higgs self couplings.  Another possibility 
arises through direct modification of the Higgs coupling to other SM states.
Theories such as Supersymmetry predict extended Higgs sectors 
which can naturally accommodate the SM-like properties of the Higgs boson observed in Run I.  

Clearly, in order to constrain BSM scenarios, the Higgs boson must be studied in much greater detail than ever before. This section of the Les Houches
proceedings details ongoing experimental and theoretical attempts to achieve this goal. The most obvious quantity for comparing data to the predictions of 
the SM is the inclusive Higgs cross section. Since the production of a Higgs boson through gluon fusion dominates this quantity, it is essential that 
accurate theoretical predictions are used to confront the LHC data, with a careful treatment of the uncertainties. This also allows for stringent constraints on Higgs coupling extractions. However, an additional 
advantage of the Run II program is the ability to study the Higgs in a differential setting. The limited amount of data in Run I resulted in rather crudely binned differential 
distributions for only the most accessible decay modes. The large data set anticipated in Run II, however, will allow for a much more varied set of observables 
and processes to be studied in ever greater detail. This will in turn allow us to constrain BSM contributions
which may have a too small impact at the inclusive level,
and thus may be only accessible through differential studies. At high values of the Higgs $p_T$ the effective theory approach, which works extremely well for the inclusive cross section,
begins to break down. If $p_T^H \sim m_t$ the top quark loop is resolved and theoretical predictions
must be modified to take the loop structure into account.
Addressing these various issues is the key goal of this report. 

This report proceeds as follows, in section~\ref{sec:N3LO} recent calculations of the N3LO cross section are summarized, with particular emphasis on the recent
attempts to estimate the remaining theoretical uncertainty.  In section~\ref{sec:Hmt} the effects of the top quark mass on differential distributions 
are studied in the Sherpa framework.
Section~\ref{sec:stxs} presents a proposal to link the new experimental measurements of differential properties 
to theoretical predictions in the SM and beyond using simplified template cross sections.

\section{Higgs boson production through gluon fusion at N3LO, and its theory uncertainty 
  \texorpdfstring{\protect\footnote{
    A.~Lazopoulos}}{}}
\label{sec:N3LO}

\subsection{Motivation}
Despite being a loop-induced process, Higgs production through gluon fusion is the dominant Higgs boson production mechanism at the LHC. The large leading-order cross-section, mainly due to the enhanced gluon luminosity at the LHC and the large value of the top Yukawa coupling, receives huge NLO corrections~\cite{Dawson:1990zj,Graudenz:1992pv,Djouadi:1991tka,Spira:1995rr,Harlander:2005rq,Aglietti:2006tp,Bonciani:2007ex,Anastasiou:2006hc,Anastasiou:2009kn} the sheer magnitude of which have cast doubts on the perturbativity of this observable in the past. The NNLO corrections~\cite{Anastasiou:2002yz,Harlander:2002wh,Ravindran:2003um} , computed in the heavy-top approximation, have been milder, and for the first time indications of reasonable perturbative behaviour appear. The N3LO corrections~\cite{
Anastasiou:2012kq, Hoschele:2012xc, Buehler:2013fha,
Moch:2004pa,Vogt:2004mw,Baikov:2009bg,Gehrmann:2010ue,Duhr:2014nda,Dulat:2014mda, Anastasiou:2013srw,Duhr:2013msa,Li:2013lsa, Anastasiou:2014vaa,Li:2014bfa,Li:2014afw, Anastasiou:2014lda, Anastasiou:2015yha, Anastasiou:2015ema}, also computed in the heavy-top limit, have recently been completed, and show that we now have very good control at the per cent level of this all-important quantity. The uncertainty from pure initial state QCD effects is now at the $2-3\%$ level. This brings to the foreground a number of other per-cent level uncertainties that have been hitherto ignored. An exhaustive phenomenological study of all these uncertainty sources is available in~\cite{Anastasiou:2016cez} where various approaches to assess these uncertainties to the best of our current knowledge is proposed. The purpose of this contribution is to clarify some salient features with respect to some of these estimates, which have been raised in discussions after the publication of~\cite{Anastasiou:2016cez}. The central value for the cross section, at $13$TeV and values for input parameters as discussed in~\cite{Anastasiou:2016cez} is 
\begin{equation}
\label{eq:finalresult_13TeV}
\sigma = 48.58\,{\rm pb} {}^{+2.22\, {\rm pb}\, (+4.56\%)}_{-3.27\, {\rm pb}\, (-6.72\%)} \mbox{ (theory)} 
\pm 1.56 \,{\rm pb}\, (3.20\%)  \mbox{ (PDF+$\alpha_s$)} \,.
\end{equation}

\subsection{Uncertainty sources}
There are currently three distinct sources of uncertainty on the best available estimate of the inclusive Higgs boson cross-section: 
\begin{itemize}
\item The value of $a_s(m_Z)$: this is the largest source of uncertainty since the inclusive cross-section, starting at ${\cal O}(a_s^2)$ is very sensitive to the precise value of the strong coupling constant. To the best of the community's current knowledge, a combination of various different extractions of $\alpha_s$ points to a value of $\alpha_s(m_Z)=0.118\pm0.0015$, which in turn induces an uncertainty of $2.6\%$ on the Higgs cross-section. We follow here the recommendation of the HXSWG~\cite{Denner:2047636}, the PDF4LHC working group~\cite{Butterworth:2015oua} and the upcoming PDG update. However it is worth pointing out there are approaches to $\alpha_s$ extraction that lead to values outside the quoted uncertainty range. If for example one adopts the best fit value for $\alpha_s$ from PDFs within the ABM collaboration, the Higgs cross section is $7.3\%$ smaller, a value that is clearly outside the recommended uncertainty range. In conclusion, the total Higgs cross-section uncertainty strongly depends on the reliability of the uncertainty on $\alpha_s$.
\item  The uncertainty due to the determination of PDFs from data: this is another intricate, global source of uncertainties. The recommendation of the PDF4LHC working group are followed here. This leads to an uncertainty of $1.86\%$. It is again worth noting that there are outlier PDFs, that have been excluded from the PDF4LHC combination for reasons explained in~\cite{Butterworth:2015oua}. Moreover the PDF sets that are included in the combination currently agree remarkably well in the region of Bjorken $x$ of the gluon density that affects the gluon fusion Higgs production, a situation that might change in the future, as more data are included in the individual fits. 
\item The total theory uncertainty due to missing higher order contributions to the total cross-section. These include uncertainties due to 
\vspace*{6pt}
	\begin{enumerate}
	\item QCD corrections beyond N3LO in the heavy-top approximation
	\item EW corrections beyond ${\cal O}(a_s^2a_{EW})$
	\item QCD corrections due to quark mass effects beyond ${\cal O}(a_s^3)$
	\item the truncation of the threshold expansion in which the N3LO result is derived
	\item missing higher orders in other processes used to extract the parton densities
	\item parametric uncertainties of quark mass effects
	\end{enumerate}
\end{itemize}
\vspace*{8pt}
The uncertainties for the Higgs cross-section through gluon fusion at the LHC with $\sqrt{S}=13$TeV are summarised in Table~\ref{table_from_paper}, from~\cite{Anastasiou:2016cez} that we reproduce here for convenience. 
\begin{center}
\begin{tabular}{|cccccc|}
\hline
\begin{tabular}{c} $\delta$(scale) \end{tabular} &
\begin{tabular}{c} $\delta$(trunc) \end{tabular} &
\begin{tabular}{c} $\delta$(PDF-TH) \end{tabular} &
\begin{tabular}{c} $\delta$(EW) \end{tabular} & 
\begin{tabular}{c} $\delta(t,b,c)$  \end{tabular} & 
\begin{tabular}{c} $\delta(1/m_t)$ \end{tabular}
\\ 
\hline
${}^{+0.10\textrm{ pb}}_{-1.15\textrm{ pb}} $ & $\pm$0.18 pb & $\pm$0.56 pb & $\pm$0.49 pb& $\pm$0.40 pb& $\pm$0.49 pb
\\ 
\hline
${}^{+0.21\%}_{-2.37\%}$ & $\pm 0.37\%$ & $\pm 1.16\%$ & $\pm 1\%$ & $\pm 0.83\%$ 
& $\pm 1\%$ 
\\ 
\hline
\end{tabular}
\label{table_from_paper}
\end{center}

\subsubsection{QCD corrections beyond N3LO in the EFT}
The uncertainty from missing higher order corrections is estimated by a typical scale variation procedure, around a central scale of $m_H/2$. It is asymmetric and amounts to $+0.21\% -2.37\%$. The asymmetry is typical of a scale variation that exhibits a maximum close to the central scale choice. In fact the only meaningful prediction from perturbation theory is the range in which the N3LO cross-section lies. The central value for the cross section is a reference point, but is not a priori preferred  to any other value in the uncertainty interval. So if one prefers, one can symmetrise the uncertainty by moving the nominal central value at the middle of the uncertainty range. 

A more interesting question is whether one should include resummation contributions beyond N3LO. The expected effect of resummation is to tame the renormalization scale dependence of the result and give an indication of the magnitude of missing higher order terms by probing its soft contributions. This relies on the validity of threshold dominance, as well as on cancelations due to sub-leading channels being negligible. It was shown in~\cite{Anastasiou:2016cez} that the effects of standard resummation approaches are actually negligible for the central choice of scale and within the range of the fixed order scale uncertainty. 

Resummation results are defined up to formally sub-leading terms in the soft expansion. The numerical impact of such sub-leading terms can be non-negligible, however, especially when the contributions of the leading logarithmic terms is very small, as is the case in Higgs production at N3LO. Such is the situation when constants are exponentiated as in~\cite{Bonvini:2014joa}. It is then impossible to maintain control, at the per cent level of precision, over the magnitude of the missing terms that might cancel, at N4LO, those that are arbitrarily included. We would therefore not consider such prescriptions as a probe of higher order corrections, and hence exclude them from the uncertainty estimate. 

\subsubsection{7-point scale variation}
In~\cite{Anastasiou:2016cez}, when estimating the scale uncertainty, we have kept the factorization and renormalization scales equal: the $\mu_F$ scale variation is remarkably flat, and it probes implicitly the parton density evolution, which, apart from being external to our computation, also interferes with our estimate of the PDF-TH uncertainty (due to the lack of N3LO parton densities). 

However, once could attempt to probe the higher order missing terms using the 7-point scale variation method, where 
$\mu_R\neq\mu_F$ and $\mu_R,\mu_F\in[m_H/4,m_H]$ with $\mu_R/\mu_F\neq 4,1/4$. In the present case this would fail to capture the maximum of the scale variation distribution, this maximum not being on one of the 7 points. We would need a scan over  the $\mu_R$-$\mu_F$ plane with the constraint $1/2\leq \mu_R/\mu_F\leq 2$. Such a study will take place within the scope of the Higgs Cross Section Working Group's YR4, and we refer the reader to that document, currently in preparation, for further details.

\subsection{Uncertainty due to light quark mass effects}
The gluon fusion cross section with the full quark mass dependence is only known at NNLO for the top quark and at NLO for the light quarks. The effects due to the top quark mass are quite different than those of the light quarks: the former largely factorise (deviation from factorization of the leading order cross section are at the $-0.6\%$ level at NLO and at $+1\%$ level at NNLO), and, for all we know, we expect them to stay at the level of $+5-7\%$ at all orders in perturbation theory. For the light quark  effects, including the top-light quark interference terms, on the contrary, there is no factorisation concept. The only available information, at LO and NLO, shows that the perturbative behaviour of the interference terms is actually quite good\footnote{In the sense that interference terms do not follow the pattern of the EFT or the top only terms: they do not exhibit as large an NLO K-factor ($1.56$ vs $2.28$)}, while the perturbative behaviour of the light quarks contributions squared is quite bad. Fortunately, for the parameters of the Standard Model, the light quark contributions squared are negligible, while the interference terms dominate. 

Note that we  only want to assign an uncertainty to the missing interference effects at NNLO. It is therefore appropriate to look at the progression of the interference contributions per order, as opposed to the progression of the total interference contribution that includes a large LO correction. We assume that the missing NNLO light quark coefficient cannot induce a relative shift from the top only cross section that is larger than the one induced at NLO. 
We therefore estimate the uncertainty due to light quarks by
\begin{equation}
\label{eq:tbc_uncertainty}
\delta(tbc)^{\overline{\rm MS}} = \pm\, \left| \,\frac{\delta\sigma_{ex;t}^{NLO}-\delta\sigma_{ex;t+b+c}^{NLO}}{\delta\sigma_{ex;t}^{NLO}}\,\right|\,
(R_{LO}\delta\sigma_{EFT}^{NNLO}+\delta_{t}\hat{\sigma}_{gg+qg,EFT}^{NNLO}) \simeq \pm 0.31\,\textrm{pb}\,,
\end{equation}
where 
\begin{equation}
\delta\sigma_{X}^{NLO} \equiv \sigma_{X}^{NLO}-\sigma_{X}^{LO}
{\rm~~and~~}
\delta\sigma_{X}^{NNLO} \equiv \sigma_{X}^{NNLO}-\sigma_{X}^{NLO}\,.
\end{equation}
In Table 7. of~\cite{Anastasiou:2016cez} the contributions due quark mass effects can be seen. The progression of the interference contributions per order is shown in the following table
\begin{center}
\begin{tabular}{|c|ccc|}
\hline
\begin{tabular}{c} $\overline{\rm MS}$ scheme  \end{tabular} &
\begin{tabular}{c} LO \end{tabular} &
\begin{tabular}{c} NLO \end{tabular} &
\begin{tabular}{c} NNLO \end{tabular} 
\\ 
\hline
$(\delta\sigma_{ex;t}-\delta\sigma_{ex;tbc} )/ \delta\sigma_{ex;t}$ &0.073 & 0.032& X? 
\\
\hline
\end{tabular}
\end{center}
The assumption on which the uncertainty estimate is based is that $X$ is not expected to be larger than the NLO value. 

Also note that there is an apparent cancellation between top quark effects and light quark effects at LO and NLO. We remark that this cancelation is not particularly motivated theoretically and it is also a scheme dependent statement\footnote{In the OS scheme at LO the light quark effects largely overwhelm the top mass corrections, while at NLO the level of the cancelation depends on whether the (sizeable) charm effects are included or not.} and a result of the particular ratios $m_b/m_H$ and $m_t/m_H$ in the Standard Model\footnote{For example the sign of the light quark effects would change for larger values of $m_b$. }. For the cancelation to persist at NNLO, the light quark interference terms themselves would have to induce a relative shift much larger than the one induced at NLO. 

The same progression for the OS scheme is shown in the table below.
\begin{center}
\begin{tabular}{|c|ccc|}
\hline
\begin{tabular}{c} $OS$ scheme  \end{tabular} &
\begin{tabular}{c} LO \end{tabular} &
\begin{tabular}{c} NLO \end{tabular} &
\begin{tabular}{c} NNLO \end{tabular} 
\\ 
\hline
$(\delta\sigma_{ex;t}-\delta\sigma_{ex;tbc} )/ \delta\sigma_{ex;t}$ &0.139 & 0.017& X? 
\\
\hline
\end{tabular}
\end{center}

The above estimated uncertainty accounts for the missing NNLO light quark effects but does not take into account the strong scheme dependence of the NLO cross section, as demonstrated in Table 8 of~\cite{Anastasiou:2016cez}. Although the OS scheme should be avoided in this context because of the bad convergence of the transition formula between the $\overline{MS}$ and OS bottom and charm quark masses, it is nevertheless prudent to enlarge the uncertainty due to light quarks to account for the difference between the two schemes. We use the factor 1.3 which is the ratio of the OS to $\overline{MS}$  contributions to the cross section from the top-bottom and top-charm interference terms, as explained in section 5 of~\cite{Anastasiou:2016cez}.

\subsubsection{Further theory uncertainties}
Various other theory uncertainty sources have been considered. In particular uncertainties due to electroweak effects and the values of the quark masses as external parameters were discussed extensively in the Les Houches 2015 meeting. They were extensively studied in~\cite{Anastasiou:2016cez} where we refer the reader for further details. 


\subsection{Summary}
With the completion of the computation of the Higgs cross section via gluon fusion at N3LO in the heavy-top effective theory, the scale uncertainty to the cross section has reached the $2-3\%$ level. This is comparable with other theory uncertainty sources the impact of which can no longer be neglected. In~\cite{Anastasiou:2016cez} we have proposed a quantitative assessment of these uncertainties. In this contribution I attempted  to clarify  salient features with respect to some of these estimates, which have been raised in discussions after the publication of~\cite{Anastasiou:2016cez}.


\section{Heavy quark mass effects in gluon fusion Higgs production
  \texorpdfstring{\protect\footnote{
    S.~Kuttimalai, F.~Krauss, P.~Maierh\"ofer, 
    M.~Sch\"onherr
  }}{}}
\label{sec:Hmt}

\subsection{Motivation}

The Higgs Effective Field Theory (HEFT) framework for perturbative
calculations of the gluon fusion Higgs production process is a well
established tool that allows a significant reduction of complexity
in higher-order QCD calculations.  In this approach, the
heavy-quark-loop induced Higgs-gluon coupling of the Standard Model
(SM) is approximated by taking into account only the top quark
contribution and by calculating production amplitudes in the limit of
an infinite top quark mass. This is typically achieved by deriving the
relevant amplitudes from the effective Lagrangian
\begin{eqnarray}
\mathcal{L}_\mathrm{HEFT} = -\frac{C_1}{4v} H G_{\mu\nu}G^{\mu\nu}\, ,
\end{eqnarray}
with the gluon field strength tensor $G_{\mu\nu}$, the Higgs field
$H$, and a perturbatively calculable Wilson coefficient $C_1$. This
Lagrangian gives rise to tree-level couplings that replace the
loop-induced SM couplings between gluons and the Higgs, effectively
reducing the number of loops in any calculation by one.

When considering the total inclusive Higgs production cross section,
finite top quark mass effects remain very moderate even at higher
orders in QCD~\cite{Marzani:2008az,Pak:2009bx,Pak:2009dg,
  Harlander:2009my,Harlander:2009mq,Harlander:2009bw}.
In the tail of the transverse momentum spectrum of the Higgs boson
or for heavy Higgs boson (virtual) masses, however, the corrections
can become very large, indicating a complete breakdown of the HEFT
approximation~\cite{Baur199038,Ellis1988221}. It has also
been known for a long time that the bottom quark loops, which are
entirely neglected in the HEFT, affect the spectrum in the
small-$p_\perp^H$ region \cite{Ellis1988221,Keung:2009bs}. In this
region, an all-order resummation of QCD corrections is
required. Standard techniques need to be adapted in order to achieve
this due to the bottom quark mass that introduces an additional scale
into the calculation \cite{Grazzini:2013mca}.

Several fully differential Monte Carlo codes have therefore been
developed that take into account the full heavy quark mass dependence
at NLO~\cite{Langenegger:2006wu,Anastasiou:2009kn,Bagnaschi:2011tu,
  Grazzini:2013mca}.
NLO results for Higgs production in association with a jet are not
available for finite heavy quark masses due to missing two-loop
amplitudes for this process.

In this note, we present an approximate treatment of finite top mass
effects at NLO based on one-loop amplitudes only. This allows us to
calculate Higgs plus $n$-jet processes at NLO, while retaining finite
top mass effects in an approximate way. Using this approximation, we
employ multijet merging techniques~\cite{Hoeche:2012yf} to
merge higher-multiplicity NLO processes matched to a parton shower
into one exclusive event sample, extending similar
approaches~\cite{Catani:2001cc,Krauss:2002up,Lonnblad:2001iq,Alwall:2011cy}
in terms of jet multiplicity and $\alpha_s$ accuracy.
Based on leading order merging, we also suggest a method to address
the issues raised in \cite{Grazzini:2013mca} concerning the inclusion
of bottom quark contributions in the low-$p_\perp^H$ region.

\subsection{Implementation of Quark Mass Corrections}

In order to take into account the full heavy quark mass effects in the
hard scattering at leading order, we replace the approximate HEFT
tree-level matrix elements provided by Sherpa's matrix element
generator Amegic++~\cite{Krauss:2001iv} with the exact one-loop matrix
elements provided by OpenLoops~\cite{Cascioli:2011va} in combination
with Collier~\cite{Denner:2014gla}. This allows the calculation of
processes with up to three additional jets in the final state at
leading order, with the full top and bottom quark mass dependency
taken into account.

At NLO, the cross section for the production of a Higgs accompanied by
a certain number $m-1$ of jets receives contributions from two
integrals of different phase space dimensionality.
\begin{eqnarray}
  \sigma = \int (B+V+I)\dif\phi_m + \int (R-S)\dif\phi_{m+1} \label{gghmq_nloxs}
\end{eqnarray}
The born term $B$ and the real emission term $R$ are present already
at leading order for processes of the respective jet multiplicity and
can be corrected as in the leading order calculation. $I(\phi_m)$ and
$S(\phi_{m+1})$ denote the integrated and differential Catani-Seymour
subtraction terms, respectively~\cite{Catani:1996vz}. They render both
integrals separately finite and are built up from leading-order
$m$-particle matrix elements dressed with appropriate splitting
kernels and can henceforth be corrected by using the full one-loop
matrix elements instead of the tree-level HEFT approximation. Note
that because we correct $R$ and $S$ with matrix elements of different
final state multiplicity, the mere convergence of the corresponding
integral already provides a crucial test for our implementation.

The IR-subtracted virtual correction $V$ receives contributions from
two-loop diagrams when taking into account the full heavy quark mass
dependencies. Since these amplitudes are available only for the Higgs
boson plus zero-jet final state, we employ an ad-hoc approximation that only
involves one-loop matrix elements (even for the Higgs boson plus zero-jet
final state). We assume a factorization of the $\alpha_s$-corrections
from the quark mass corrections and set
\begin{eqnarray}
V = V_\mathrm{HEFT}\frac{B}{B_\mathrm{HEFT}}\,.\label{gghmq_approx}
\end{eqnarray}
In this approximation, we can straightforwardly apply finite top mass
corrections in simulations employing CKKW multi jet merging at NLO in
the MEPS@NLO scheme~\cite{Hoeche:2012yf}.

We expect the approximation~\eqref{gghmq_approx} to give reasonable
results only if the HEFT-approxi\-mation is valid. For any contribution
involving the bottom Yukawa coupling $y_b$ , it cannot be used due to
the small bottom quark mass. This applies to the interference terms
proportional to $y_ty_b$ as well as the squared bottom contributions
proportional to $y_b^2$. We therefore calculate terms that involve
$y_b$ as separate processes at leading order. The NLO corrections to
the total inclusive cross sections for the $y_ty_b$ contributions and
the $y_b^2$ contributions are only of the order of \SI{1}{\percent}
and \SI{20}{\percent}, respectively~\cite{Bagnaschi:2015bop}.
Furthermore, the $y_b^2$ contributions featuring the slightly larger
NLO K-factor are significantly suppressed compared to the $y_ty_b$
terms~\cite{Bagnaschi:2015bop}. We therefore consider a treatment at
leading order sufficiently accurate. Any terms proportional to $y_t^2$
will however be calculated at NLO in the approximation described
above.

\subsection{Finite Top Mass Effects}

\begin{figure}
  \centering
  \begin{subfigure}[t]{.48\textwidth}
    \includegraphics[width=\textwidth]{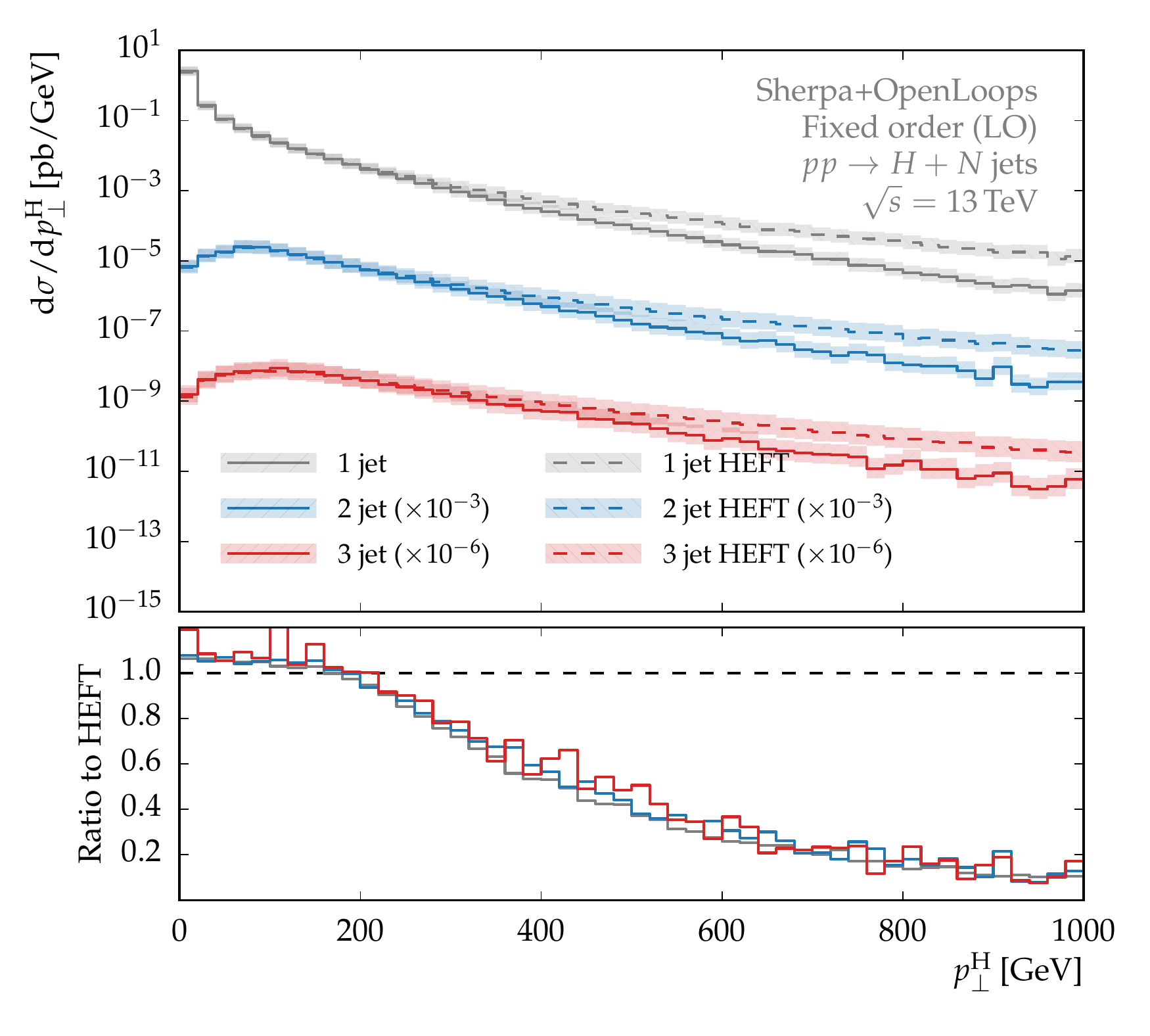}
    \caption{LO fixed order calculation for up to three jets. The
      error bands indicate the uncertainties obtained from variations
      of the factorization and renormalization scales.}
    \label{gghmq_fo_ratios}
  \end{subfigure}
  \hfill
  \begin{subfigure}[t]{.48\textwidth}
    \includegraphics[width=\textwidth]{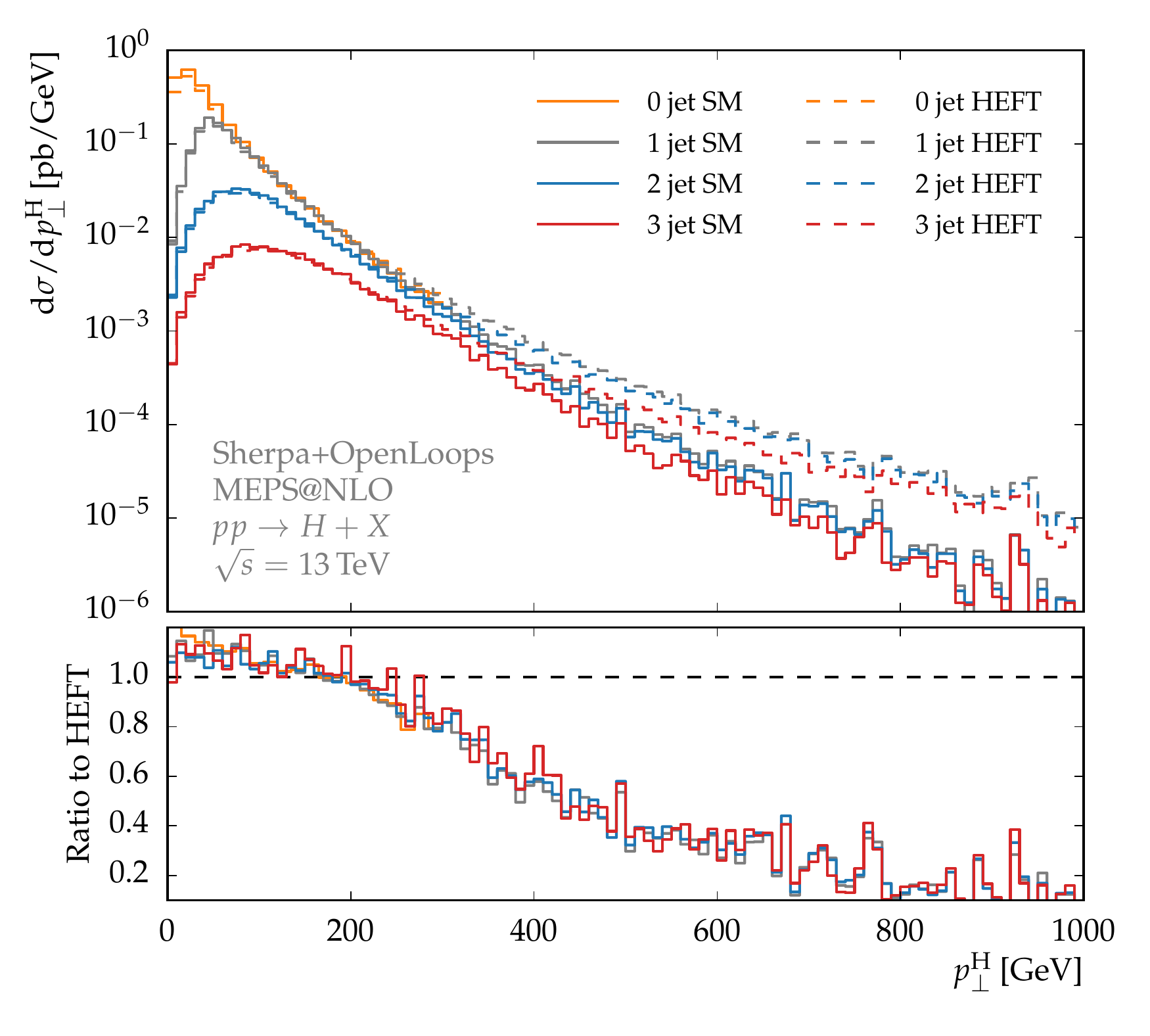}
    \caption{Multijet merged calculation. We include the zero and one
      jet final states at NLO as well as the two jet final state at
      leading order. The individual curves show inclusive $N$-jet
      contributions.}
    \label{gghmq_meps}
  \end{subfigure}
  \caption{The Higgs transverse momentum spectrum in gluon fusion. We
    show individual curves for the HEFT approximation (dashed) and the
    full SM result taking into account the mass dependence in the top
    quark loops. The lower panel shows the ratio of the SM results to
    the HEFT approximation.}
\end{figure}

As mentioned in the introduction, the total inclusive cross section is
only mildly affected by finite top mass effects. The low-$p^H_\perp$
region, where the bulk of the cross section is located, can therefore
be expected to exhibit only a moderate dependence on the top quark
mass. In kinematic regimes where any invariant significantly exceeds
$m_t$, however, we expect the HEFT approximation to break down. The
$p_\perp^H$ distributions in figure \ref{gghmq_fo_ratios} exemplify
this picture. We show Higgs boson transverse momentum distributions for
final states with one, two, and three jets calculated at fixed leading
order. Jets are reconstructed using the anti-$k_T$ algorithm with a
radius parameter of $R=0.4$ and a minimum jet $p_\perp$ of
\SI{30}{\giga\electronvolt} except in the 1-jet case, where we apply
only a small minimum $p_\perp$-cut of \SI{1}{\giga\electronvolt}. The
distributions for all three jet multiplicities exhibit a very similar
pattern when comparing the full SM result to the HEFT approximation.
Below $p_\perp^H\approx m_H$, we observe a flat excess of around
\SI{6}{\percent} that recovers the correction factor to the total
inclusive Higgs production cross section at leading order. The
deviations become very large when $p_\perp^H$ significantly exceeds
$m_t$, as expected. The similarity of the top mass dependency of the
$p_\perp^H$ spectrum for all jet final multiplicities confirms similar
findings for one- and two-jet configurations in
\cite{Buschmann:2014twa}.

In figure \ref{gghmq_meps}, we show analogous results obtained from
the MEPS@NLO simulation. We included NLO matrix elements for the zero-
and one-jet final states as well as leading order matrix elements for
the two-jet final state in the merged setup and set $Q_\mathrm{cut}$
to \SI{30}{\giga\electronvolt}. From the ratio plot in figure
\ref{gghmq_meps} it is evident that in our approximation we recover
the same suppression patterns as in the respective fixed leading order
calculations for all jet multiplicities. This is a nontrivial
observation as an $m$-jet configuration receives corrections from
$m$-jet matrix elements as well as from $m+1$-jet matrix elements
through the real emission corrections $R$ in \eqref{gghmq_approx}.

\subsection{Nonzero Bottom Mass Effects}

\begin{figure}
  \centering
  \includegraphics[width=.48\textwidth]{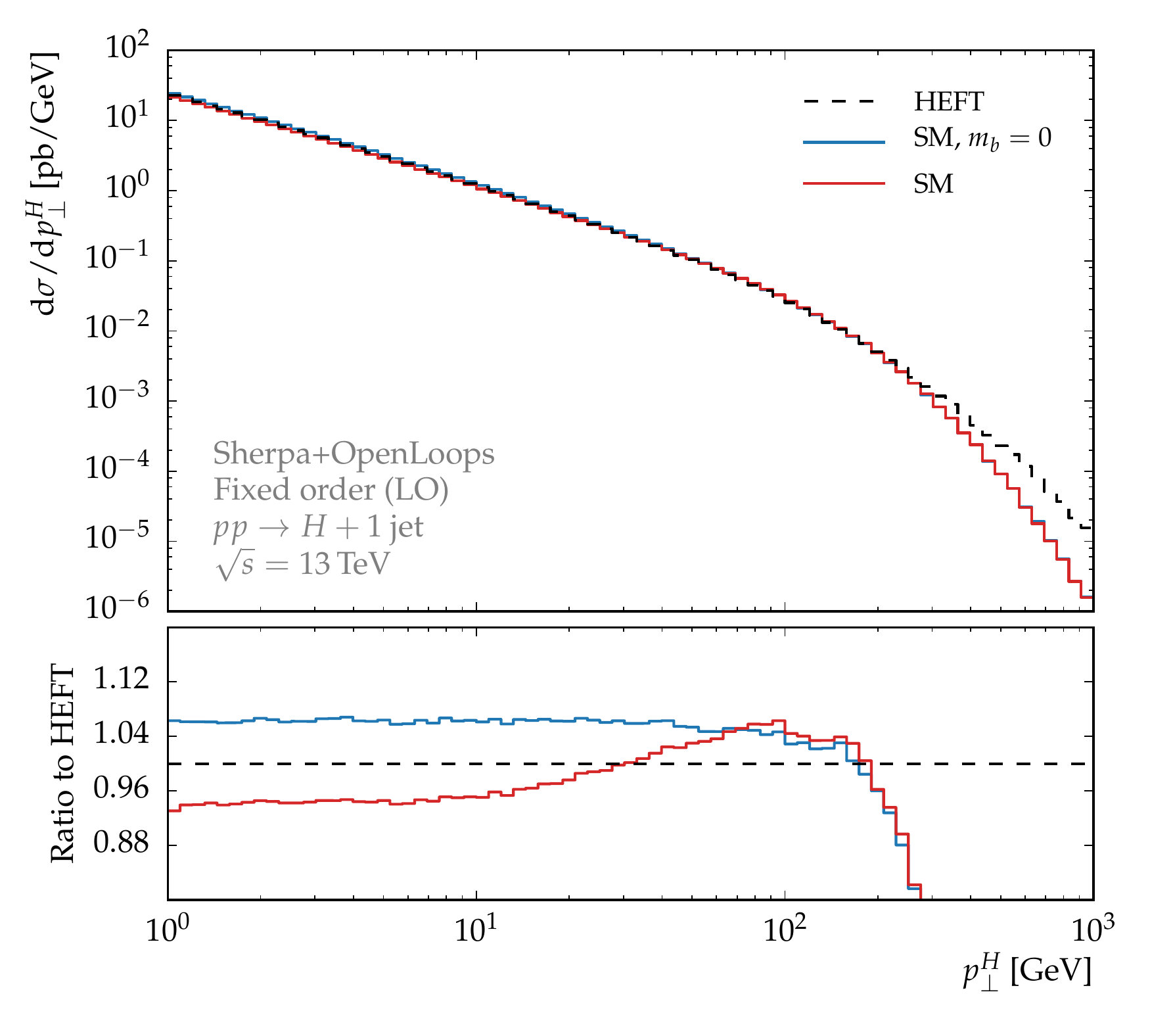}
  \caption{Bottom quark mass effects at fixed leading order. The
      minimum jet $p_\perp$ is set to \SI{1}{\giga\electronvolt} in
      order to display map out the low $p_\perp^H$ region as well.}
    \label{gghmq_tb_fo}
  \end{figure}
  
As pointed out already in~\cite{Ellis1988221,Keung:2009bs}, the
inclusion of the bottom quark in the loops affects the $p_\perp^H$
distribution only at small values of $p_\perp^H$ around $m_b$. In
figure \ref{gghmq_tb_fo} we reproduce these findings for the process
$pp\rightarrow H+j$ at fixed order. In the $p_\perp$ range around
$m_b$ where the bottom effects are large, a fixed order prediction is
of course unreliable due to the large hierarchy of scales between
$m_H$ and the transverse momentum. This large separation of scales
induces Sudakov logarithms $\ln(m_H/p_\perp)$ that spoil any fixed
order expansion and require resummation.

It was argued in~\cite{Grazzini:2013mca} that the resummation of these
logarithms is complicated by the presence of the bottom quark in loops
that couple to the Higgs boson. The bottom quark introduces $m_b$ as an
additional scale above which the matrix elements for additional QCD
emissions do not factorize. Since a factorization is essential for the
applicability of resummation techniques, it was proposed to use a
separate resummation scale of the order of $m_b$ for the contributions
involving $y_b$, thereby restricting the range of transverse momenta
where resummation is applied to the phase space where factorization is
guaranteed.  In this approach, the pure top quark contributions
proportional to $y_t^2$ are treated as usual, with the resummation
scale typically set to values of the order of $m_H$. 

\begin{figure}
  \centering
  \begin{subfigure}[t]{.48\textwidth}
    \includegraphics[width=\textwidth]{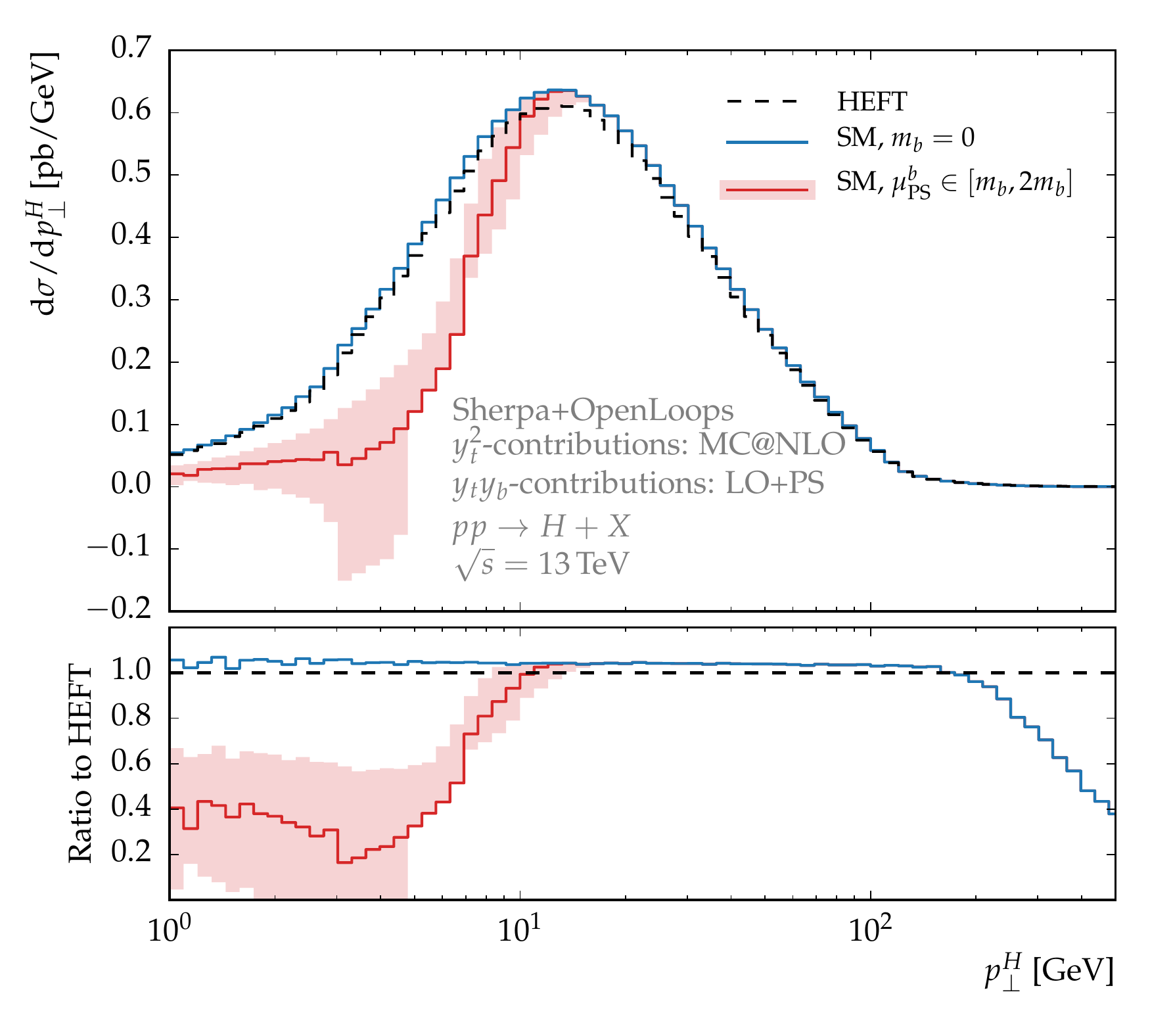}
    \caption{Bottom quark mass effects at LO+PS accuracy with a small
      parton shower starting scale of $\mu_\mathrm{PS}=\sqrt{2}m_b$.
      The red error band shows variations of this scale by factors of
      $\sqrt{2}$ up and down.}
    \label{gghmq_tb_lops}
  \end{subfigure}
  \hfill
  \begin{subfigure}[t]{.48\textwidth}
    \includegraphics[width=\textwidth]{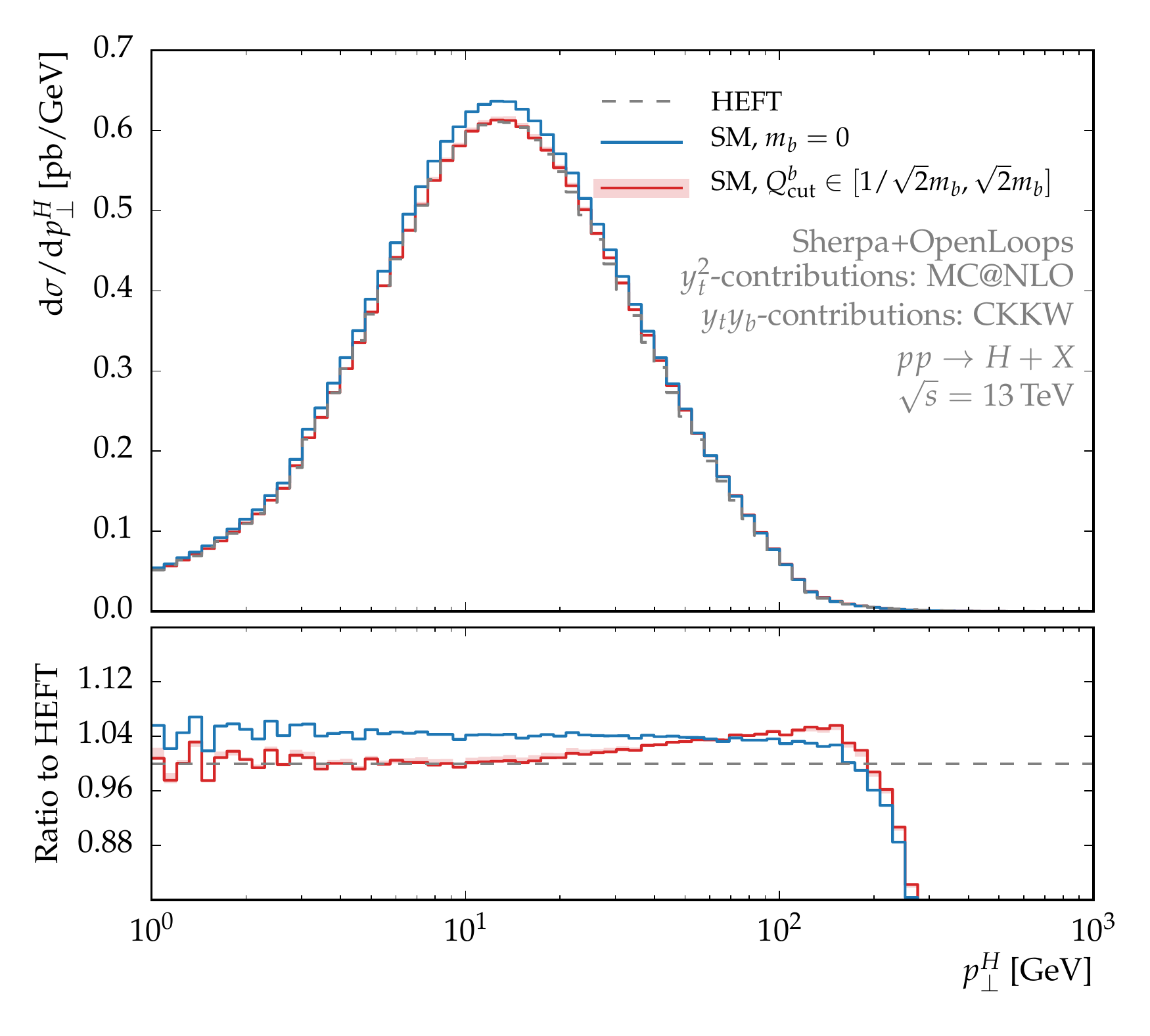}
    \caption{Bottom quark mass effects taken into account by means of
      CKKW merging with a small merging scale $Q^b_\mathrm{cut}=m_b$.
      The red error band (hardly visible in this plot) shows
      variations of this scale by factors of $\sqrt{2}$ up and down.}
    \label{gghmq_tb_meps}
  \end{subfigure}
  \caption{The Higgs transverse momentum spectrum in gluon fusion. We
     show individual curves for the HEFT approximation (dashed) and the
     full SM result taking into account the mass dependency in the top
     quark loops neglecting (blue) and accounting for (red) bottom mass
     effects. Uncertainties in the treatment of finite bottom mass effects
     are shown as a red band.}
\end{figure}

While reference~\cite{Grazzini:2013mca} was concerned with analytical
resummation techniques, similar approaches were studied in the context
of NLO-matched parton shower Monte Carlos
\cite{Bagnaschi:2015qta,Harlander:2014uea,Bagnaschi:2015bop}. Our
discussion will be restricted to the leading order as the
approximation used for the NLO calculation of the top quark
contributions \eqref{gghmq_approx} is invalid for the bottom quark
terms. The equivalent of the resummation scale in analytic
calculations is the parton shower starting scale $\mu_\mathrm{PS}$
because it restricts parton shower emissions to the phase space below
this scale and because this scale enters as the argument in the
Sudakov form factors. Using separate parton shower starting scales for
the top and the bottom contributions, respectively, requires to
generate and shower them separately as well. A corresponding
separation of terms in the one-loop matrix elements can be achieved
with OpenLoops. By means of this separation into terms proportional to
$y_t^2$ and the remainder, we can generate an MC@NLO sample for the
top quark contributions while calculating the terms involving $y_b$ at
leading order. Figure \ref{gghmq_tb_lops} shows the $p_\perp^H$
spectrum obtained this way. We set the parton shower starting scale
for the bottom contributions $\mu_\mathrm{PS}^b$ to $\sqrt{2}m_b$ and
show the error band obtained from variations of this scale by factors
of $\sqrt{2}$ up and down. The parton shower starting scale for the
top quark contributions will be $\mu_\mathrm{PS}^t=m_H$ throughout.
The suppression in the low $p_\perp$ region below $m_b$ is much more
pronounced than in the fixed order result in figure \ref{gghmq_tb_fo}.
This is because, pictorially speaking, without changing the cross
section of the individual contributions, the parton shower simulation
spreads the $y_t^2$ part over a much larger range, up to
$\mathcal{O}(m_H)$, than for the negative $y_ty_b$, up to
$\mathcal{O}(m_b)$ only.  The spectrum in this region is therefore
extremely sensitive to variations of $\mu_\mathrm{PS}^b$.  When varying
$\mu_\mathrm{PS}^b$ to sufficiently low values, the differential cross
section may even become negative, clearly an unphysical result.  We
stress, again, that this is not a physical effect but an artefact of
the unitary nature of the parton shower. Setting the value of
$\mu_\mathrm{PS}^b$ to a small value, the entire leading order bottom
cross section contributions will be distributed in a phase space with
Higgs transverse momenta not significantly exceeding
$\mu_\mathrm{PS}^b$. Since this cross section is negative, the
spectrum must become negative at some point when lowering
$\mu_\mathrm{PS}^b$.

We therefore suggest another approach at taking into account the
bottom quark contributions in a parton shower Monte Carlo simulation.
We account for the non-factorization of the real emission matrix
elements above some scale $Q^b_\mathrm{cut}$ by correcting parton
shower emissions harder than this scale with the appropriate fixed
order matrix elements. This can be done consistently in the CKKW
merging scheme~\cite{Catani:2001cc,Hoeche:2009rj}.  Setting the
merging scale for the bottom contributions $Q_\mathrm{cut}^b$ to $m_b$
allows the correction of the parton shower in the regime where the
matrix elements involving $m_b$ do not factorize (without restricting
all emissions to the phase space below).  Above $Q_\mathrm{cut}$, the
fixed-order accuracy of the real emission matrix
elements is thereby retained. Since any NLO prediction of the
inclusive process describes the $p_\perp$ spectrum only at leading
order, our approach retains the same parametric fixed order accuracy
when considering the $p_\perp^H$ distribution. Beyond fixed order, the
differences should be small since the NLO corrections to the inclusive
cross section are at the percent level for the $y_ty_b$ interference
terms.

In figure \ref{gghmq_tb_meps} we show the bottom quark effects on the
$p_\perp^H$ spectrum in this approach. We include matrix elements with
up to one jet in the merging such that a leading order accuracy in
$\alpha_s$ is guaranteed for both the top and the bottom contributions
to the $p_\perp^H$ spectrum. This allows a comparison to figure
\ref{gghmq_tb_fo}. The effects of the bottom quarks lead to a very
similar suppression pattern over the entire displayed range of
$p_\perp^H$. The large NLO K-factor that appears in the MC@NLO
calculations of the top contributions however affects the overall
relative normalization of the bottom quark effects. They are
correspondingly smaller by roughly \SI{50}{\percent} in figure
\ref{gghmq_tb_meps} when compared to figure \ref{gghmq_tb_fo}. The
sensitivity to variations of the scale in the calculation that
effectively accounts for the presence of the bottom mass in the
problem is drastically reduced. Figure \ref{gghmq_tb_meps} includes an
error band corresponding to variations of $Q_\mathrm{cut}^b$ by
factors of $\sqrt{2}$ up and down. On the displayed scale, these
variations are hardly visible.

\subsection{Summary}

We presented in this contribution an implementation of heavy quark mass
effects in gluon fusion Higgs production that allows to systematically
include finite top mass effects in an approximate way at NLO for in
principle arbitrary jet multiplicities in the final state. Based on
this approximation, we presented results for the Higgs boson transverse
momentum distributions obtained from NLO matched and merged samples.
When comparing the top quark mass dependence in one-, two-, and
three-jet final states, we observed a universal suppression pattern
that agrees very well with the corresponding leading order results.

Our treatment of contributions involving the bottom Yukawa coupling is
based on $m_b$- and $m_t$-exact leading order matrix elements in
combination with tree-level multijet merging techniques. We argued
that this approximation is appropriate since it allows to retain
leading order accuracy for the corresponding contributions in the
$p_\perp^H$-spectrum and it also allows to account for the
non-factorization of real emission amplitudes at scales above $m_b$.


\section{Simplified template cross sections
  \texorpdfstring{\protect\footnote{
    M.~Duehrssen-Debling, P.~Francavilla, 
    F.~J.~Tackmann, K.~Tackmann
  }}{}}
\label{sec:stxs} 

\subsection{Overview}
\label{PO_STCS:STCS_Overview}

After the successful Higgs coupling measurements during the LHC Run1, which had
as their main results measured signal strength and multiplicative coupling modifiers,
it is important to discuss in which way the experiments should present and perform
Higgs coupling measurements in the future. 
Simplified template cross sections were developed to provide a natural way to
evolve the signal strength measurements used during Run1. Compared to the Run1
measurements, the simplified template cross section framework allows one to
reduce in a systematic fashion the theory dependences that must be directly folded into
the measurements. This includes both the dependence on the theoretical uncertainties in the SM
predictions as well as the dependence on the underlying physics model (i.e.\ the SM or BSM models).
In addition, they provide more finely-grained measurements (and hence more information for theoretical
interpretations), while at the same time allowing and benefitting from the global combination of
the measurements in all decay channels.

The primary goals of the simplified template cross section framework are to maximize the sensitivity
of the measurements while at the same time to minimize their theory dependence. This means in
particular
\begin{itemize}
\item combination of all decay channels
\item measurement of cross sections instead of signal strengths, in mutually exclusive regions of
  phase space
\item cross sections are measured for specific production modes (with the SM production serving as kinematic template)
\item measurements are performed in abstracted/simplified fiducial volumes
\item allow the use of advanced analysis techniques such as event categorization, multivariate techniques, etc.
\end{itemize}

The measured exclusive regions of phase space, called ``bins'' for simplicity, are specific to
the different production modes. Their definitions are motivated by
\begin{itemize}
\item minimizing the dependence on theoretical uncertainties that are directly folded into the measurements
\item maximizing experimental sensitivity
\item isolation of possible BSM effects
\item minimizing the number of bins without loss of experimental sensitivity
\end{itemize}
These will of course be competing requirements in some cases and some compromise has to be achieved.
The implementation of these basic design principles is discussed in more detail below.

\begin{figure}
\begin{center}
\includegraphics[width=\textwidth]{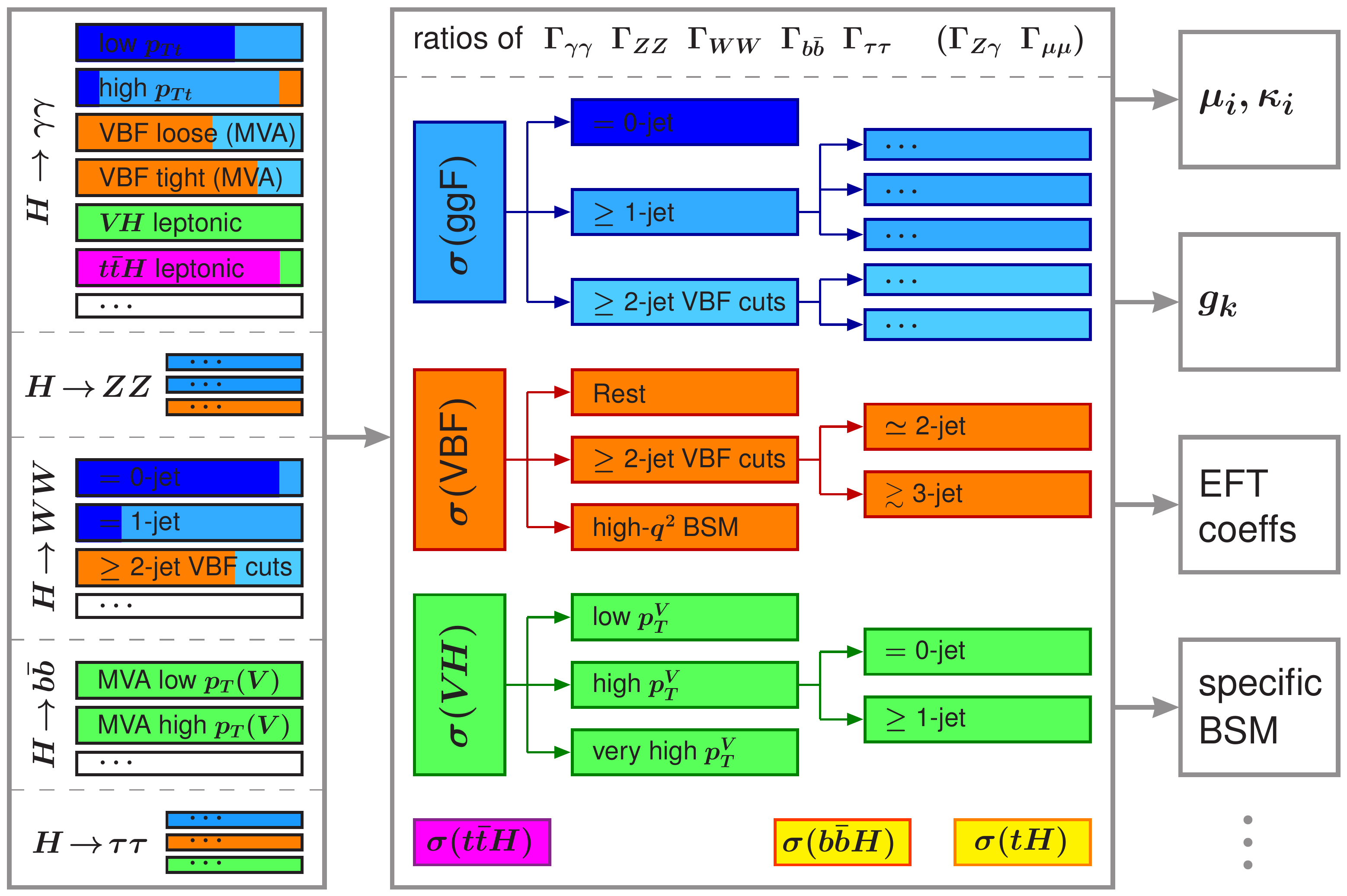}
\end{center}
\caption{Schematic overview of the simplified template cross section framework.}
\label{fig:STCS:sketch}
\end{figure}

A schematic overview of the simplified template cross section framework is shown in Fig.~\ref{fig:STCS:sketch}.
The experimental analyses shown on the left are very similar to the Run1 coupling measurements. For each decay channel, the
events are categorized in the analyses, and there are several motivations
for the precise form of the categorization. Typically, a subset of the experimental event categories
is designed to enrich events of a given Higgs production mode, usually making use of specific
event topologies. This is what eventually allows the splitting of the production modes in the global fit. Another subset
of event categories is defined to increase the sensitivity of the analysis by splitting events
according to their expected signal-to-background ratio and/or invariant-mass resolution. In other cases,
the categories are motivated by the analysis itself, e.g. as a consequence of the backgrounds being estimated
specifically for certain classes of events. While these are some of the primary motivations, in the future the details of the event
categorization can also be optimized in order to give good sensitivity to the
simplified template cross sections to be measured.

The center of Fig.~\ref{fig:STCS:sketch} shows a sketch of the simplified template cross sections, which are determined from
the experimental categories by a global fit that combines all decay channels and which represent the main results of
the experimental measurements. They are cross sections per production mode, split into mutually exclusive kinematic bins for each of
the main production modes.
In addition, the different Higgs decays are treated by fitting the partial decay widths.
Note that as usual, without additional assumptions on the total width, only ratios of partial widths
and ratios of simplified template cross sections are experimentally accessible.

The measured simplified template cross sections together with the partial decay widths then serve as input for
subsequent interpretations, as illustrated on the right of Fig.~\ref{fig:STCS:sketch}.
Such interpretations could for example be the determination of signal strength modifiers or
coupling scale factors $\kappa$ (providing compatibility with earlier results), EFT coefficients, tests
of specific BSM models, and so forth. For this purpose, the experimental results should quote the full covariance among the different bins.
By aiming to minimize the theory dependence that is folded into the first step of determining the simplified
template cross sections from the event categories, this theory dependence is shifted into the second interpretation step, making the measurements more long-term useful. For example, the treatment of theoretical uncertainties can be decoupled from the measurements and can be dealt with at the interpretation stage. In this way, propagating improvements in theoretical predictions and their uncertainties into the measurements itself, which is a very time-consuming procedure and unlikely to be feasible for older datasets, becomes much
less important. Propagating future theoretical advances into the interpretation, on the other hand, is generally much easier.

To increase the sensitivity to BSM effects, the simplified template cross sections can be
interpreted together with e.g. POs in Higgs boson decays. To make this possible, the experimental
and theoretical correlations between the simplified template cross sections and the decay POs would need
to be evaluated and taken into account in the interpretation. This point will not be expanded on further
in this section, but would be interesting to investigate in the future.

While the simplified template cross section bins have some similarity to a differential cross section measurement,
they aim to combine the advantages of the signal strength measurements and fiducial
and differential measurements. In particular, they are complementary to full-fledged fiducial and
differential measurements and are neither designed nor meant to replace these. Fully fiducial differential
measurements are of course essential but can only be carried out in a subset of decay channels in
the foreseeable future. They are explicitly optimized for maximal theory independence. In practice, this
means that in the measurements acceptance corrections are minimized, typically,
simple selection cuts are used, and the measurements are unfolded to a fiducial volume that is as close as possible to the fiducial
volume measured for a particular Higgs decay channel. In contrast,
simplified template cross sections are optimized for sensitivity while reducing the dominant theory
dependence in the measurement. In practice, this means that simplified fiducial volumes are used and larger acceptance corrections are allowed in order to maximally
benefit from the use of event categories and multivariate techniques. They are also
inclusive in the Higgs decay to allow for the combination of the different decay channels. The
fiducial and differential measurements are designed to be agnostic to the production modes as much
as possible. On the other hand, the separation into the production modes is an essential aspect of
the simplified template cross sections to reduce their model dependence.

\subsection{Guiding principles in the definition of simplified template cross section bins}
\label{PO_STCS:STCS_GuidingPrinciple}

As outlined above, several considerations have been taken into account in the definition
of the simplified template cross section bins.

One important design goal is to reduce the dependence of the measurements on theoretical uncertainties
in SM predictions. This has several aspects. First, this requires
avoiding that the measurements have to extrapolate from a certain region in phase space to the full
(or a larger region of) phase space whenever this extrapolation carries nontrivial or sizeable theoretical
uncertainties. A example is the case where an event category selects an exclusive
region of phase space, such as an exclusive jet bin. In this case, the associated theoretical uncertainties
can be largely avoided in the measurement by defining a corresponding truth jet bin.
The definition of the bins is preferably in terms of quantities that are directly measured by the experiments
to reduce the needed extrapolation.

There will of course always be residual theoretical uncertainties due to the experimental acceptances for each
truth bin. Reducing the theory dependence thus also requires to avoid cases with large variation in the
experimental acceptance within one truth bin, as this would introduce a direct dependence on the
underlying theoretical distribution in the simulation. If this becomes an issue, the bin can be further split into two
or more smaller bins, which reduces this dependence in the measurement and moves it to the interpretation step.

To maximize the experimental sensitivity, the analyses should continue to use event categories primarily optimized for sensitivity, while
the definition of the truth bins should take into consideration the experimental requirements. However, in cases where multivariate analyses are used in the analyses, it has to be carefully checked and balanced against the requirement to not introduce theory dependence, e.g., by selecting specific regions of phase space.

Another design goal is to isolate regions of phase space, typically at large kinematic scales, where BSM effects could be potentially large and visible above the SM background. Explicitly separating these also reduces the dependence of the measurements on the assumed SM kinematic distribution.

In addition, the experimental sensitivity is maximized by allowing the combination of all decay channels, which requires the framework
to be used by all analyses. To facilitate the experimental implementation, the bins should be mutually exclusive to avoid introducing statistical correlations between different bins. In addition, the number of bins should be kept
minimal to avoid technical complications in the individual analyses as well as the global fit, e.g. in the evaluation of the full covariance matrix. For example, each bin should typically have some sensitivity from at least one event category in order to avoid the need to statistically combine many poorly constrained or unconstrained measurements. On the other hand, in BSM sensitive bins experimental limits are already very useful for the theoretical interpretation.

\subsubsection{Splitting of production modes}
\label{PO_STCS:STCS_GuidingPrincipleSplitting}

The definition of the production modes has some notable differences compared to Run1 to deal
with the fact that the naive distinction between the $q\bar{q}\to VH$ and VBF processes, and
similarly between $gg\to VH$ and gluon-fusion production, becomes ambiguous at higher order
when the $V$ decays hadronically. For this reason, the $VH$ production mode is explicitly
defined as Higgs production in association with a leptonically decaying $V$ boson.
The $q\bar{q}\to VH$ process with a hadronically decaying $V$ boson is considered to be part of
what is called ``VBF production'', which is defined as electroweak $qqH$ production. Similarly,
the $gg\to ZH$ process with hadronically decaying $Z$ boson is included in what is called
``gluon-fusion production''.

In principle, also the separation of $ZH$ production with a leptonic $Z$ into $q\bar{q}$ or $gg$ initial states
becomes ambiguous at higher order. For present practical purposes, on the experimental side the split
can be defined according to the separate MC samples for $q\bar{q}\to ZH$ and $gg\to ZH$ used in the analyses.

\subsubsection{Staging}
\label{PO_STCS:STCS_GuidingPrincipleStaging}

In practice, it will be impossible to define a set of bins that satisfies all of the above requirements for every analysis.
Some analyses will only be able to constrain a subset of all bins or only constrain the sum of a set of bins.
In addition, the number of bins that will be possible to measure increases with increasing amount of
available data. For this reason, several stages with an increasing number of bins are defined.
The evolution from one stage to the next can take place independently for each production mode.

\begin{figure}
\begin{center}
\includegraphics[width=\textwidth]{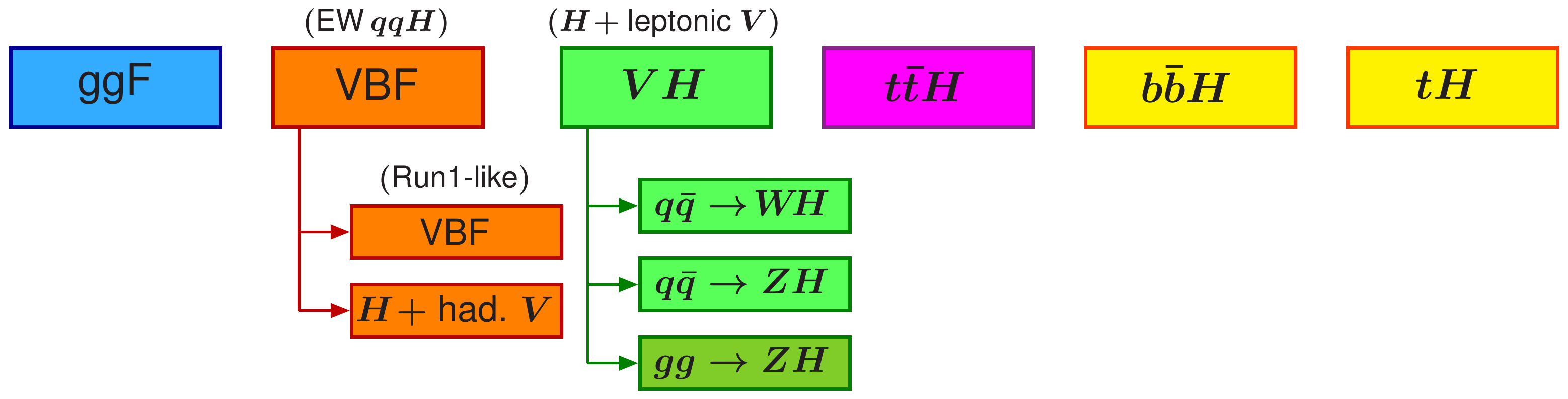}
\end{center}
\caption{Stage 0 bins.}
\label{fig:STCS:stage0}
\end{figure}

\paragraph{Stage 0} Stage 0 is summarized in Fig.~\ref{fig:STCS:stage0} and corresponds most closely to the
measurement of the production mode $\mu$ in Run1. At this stage, each main production mode has a single inclusive bin,
with associated Higgs production separated into $q\bar q\to WH$, $q\bar q\to ZH$ and $gg\to ZH$ channels.

As discussed in Sec.~\ref{PO_STCS:STCS_GuidingPrincipleSplitting}, VBF production is
defined as electroweak $qqH$ production. For better compatibility with Run1 measurements,
the VBF production is split into a Run1-like VBF and
Run1-like $V(\to jj)H$ bin, where the splitting is defined by the conventional Feynman diagrams included in the
simulations. In practice, most decay channels will only provide a measurement for
the Run1-like VBF bin.

\paragraph{Stage 1} Stage 1 defines a binning that is targeted to be used by all analyses on an intermediate
time scale. In principle, all analyses should aim to eventually implement the full stage 1 binning.
If necessary, intermediate stages to reach the full stage 1 binning can be implemented by a given analysis by
merging bins that cannot be split. In this case, the analysis should ensure that the merged bins have similar acceptances,
such that the individual bins can still be determined in an unbiased way in the global combination of all channels.
In the diagrams presented below, the possibilities for merging bins are indicated by ``(+)''.

\paragraph{Stage 2} Defining the stage 2 binning in full detail is very difficult before having gained experience with the
practical implementation of the framework with the stage 1 binning. Therefore, instead of giving a detailed
proposal for the stage 2 binning, we only give indications of interesting further separation of bins that should be considered
for the stage 2 binning.

\subsection{Definition of leptons and jets}
\label{PO_STCS:STCS_ObjectsAndJets}

The measured event categories in all decay channels are unfolded by the global fit to the
simplified template cross sections bins. For this purpose, and for the
comparison between the measured bins and theoretical predictions from either analytic
calculations or MC simulations, the truth final state particles need to be defined unambiguously.
The definition of the final state particles, leptons, jets, and in particular also the Higgs
boson are explicitly kept simpler and more idealized than in the fiducial cross section measurements.
Treating the Higgs boson as a final state particle is what allows the combination of the
different decay channels.

For the moment, the definitions are adapted to the current scope of the measurements. Once a
finer binning is introduced for $t\bar{t}H$ or processes such as VBF with the emission of a hard
photon are added, some of the definitions below might have to be adapted or refined.

\subsubsection{Higgs boson}
\label{PO_STCS:STCS_ObjectsAndJets_Higgs}

The simplified template cross sections are defined for the production of an on-shell Higgs boson,
and the unfolding should be done accordingly. A global cut on the Higgs rapidity at $|Y_H| < 2.5$
is included in all bins. As the current measurements have no sensitivity beyond this rapidity range, this
part of phase space would only be extrapolated by the MC simulation. On the other
hand, it is in principle possible to use forward electrons (up to $|\eta|$ of 4.9) in
$H\to ZZ^*\to 4\ell$ and extend the accessible rapidity range. For this purpose, an 
additional otherwise
inclusive bin for $|Y_H| > 2.5$ can be included.

\subsubsection{Leptons}
\label{PO_STCS:STCS_ObjectsAndJets_leptons}

Electrons and muons from decays of signal vector bosons, e.g. from $VH$ production, are defined 
as dressed, i.e. FSR photons should be added back to the
electron or muon. $\tau$ leptons are defined from the sum of their decay products (for any $\tau$ decay
mode). There should be
no restriction on the transverse momentum or the rapidity of the leptons. 
That is, for a leptonically decaying vector boson
the full decay phase space is included.

\subsubsection{Jets}
\label{PO_STCS:STCS_ObjectsAndJets_Jets}

Truth jets are defined as anti-$k_t$ jets with a jet radius of $R = 0.4$, and are built from all
stable particles (exceptions are given below), including neutrinos, photons and leptons 
from hadron decays or produced in the shower.
Stable particles here have the usual definition, having a lifetime greater than 10~ps, i.e. those
particles that are passed to GEANT in the experimental simulation chain. All decay products from
the Higgs boson decay are removed as they are accounted for by the truth Higgs boson.
Similary, leptons (as defined above) and neutrinos from decays of the signal $V$ bosons are 
removed as they are
treated separately, while decay products from hadronically decaying signal $V$ bosons are included in
the inputs to the truth jet building.

By default, truth jets are defined without restriction on their rapidity. A possible rapidity cut can 
be included
in the bin definition. A common $p_\mathrm{T}$ threshold for jets should be used for all truth jets.
A lower threshold would in principle have the advantage to split the events more evenly between the 
different jet bins.
Experimentally, a higher threshold at 30~GeV is favored due to pile up and is therefore used for the
jet definition to limit the amount of phase-space extrapolation in the measurements.

\subsection{Bin definitions for the different production modes}
\label{PO_STCS:STCS_bins}

In the following, the bin definitions for the different production modes in each stage are given.
The bins are easily visualized through cut flow diagrams. In the diagrams, the bins on each branch
are defined to be mutually exclusive and sum up to the preceding parent bin. For simplicity, sometimes
not all cuts are explicitly written out in the diagrams, in which case the complete set of cuts are specified in the text.
In case of ambiguities, a more specific bin is excluded from a more generic bin. As already mentioned, for the stage 1 binning
the allowed possibilities for merging bins at intermediate stages are indicated by a ``(+)'' between two bins.

\subsubsection{Bins for $gg \to H$ production}
\label{PO_STCS:STCS_ggH}

\paragraph{Stage 0} Inclusive gluon fusion cross section within $|Y_H|<2.5$. Should the
measurements start to have acceptance beyond 2.5, an additional bin for $|Y_H| > 2.5$ can be 
included.

\begin{figure}
\begin{center}
\includegraphics[width=\textwidth]{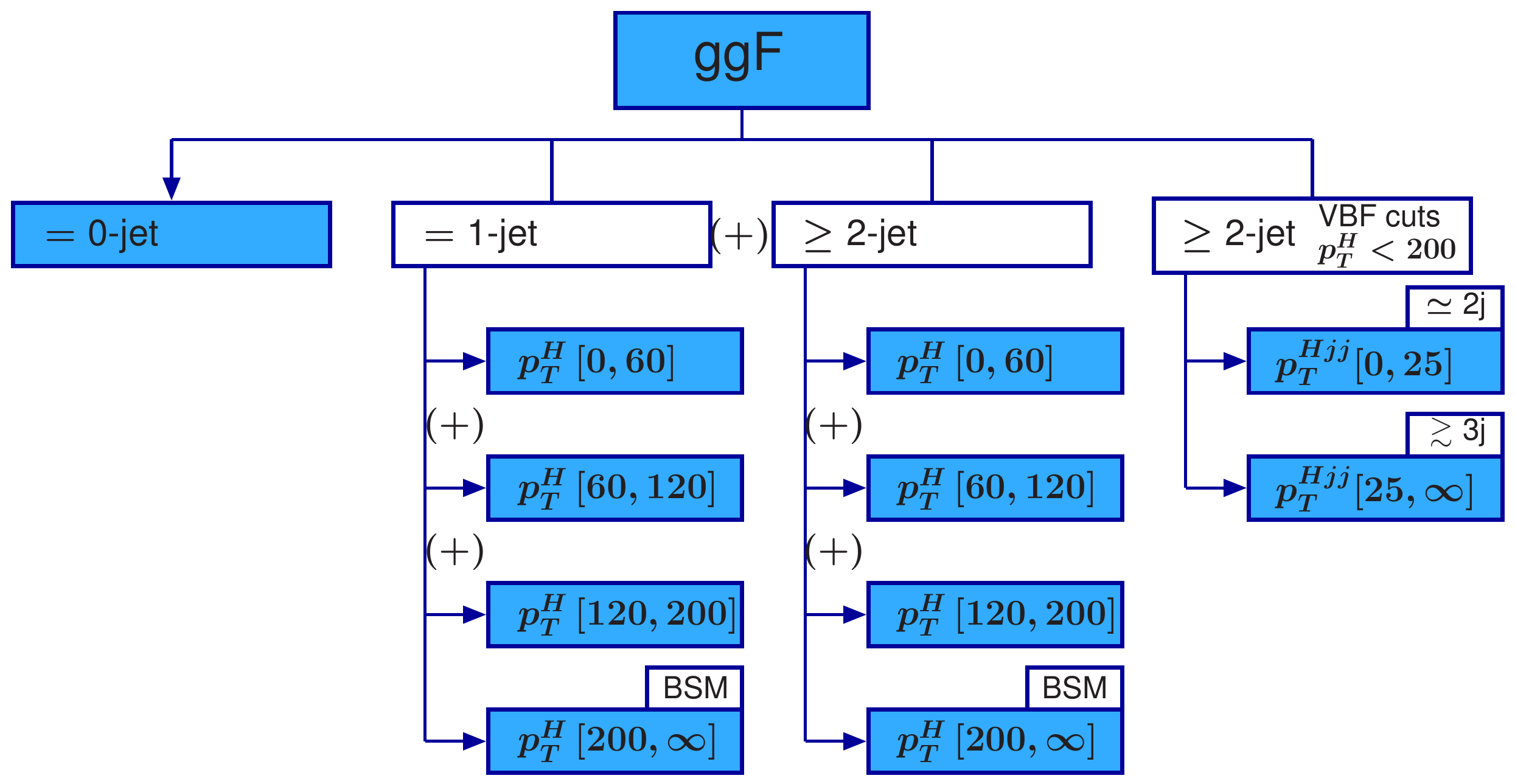}
\end{center}
\caption{Stage 1 binning for gluon fusion production.}
\label{fig:STCS:ggH}
\end{figure}

\paragraph{Stage 1}

Stage 1 refines the binning for $|Y_H|<2.5$.
The stage 1 binning is depicted in Fig.~\ref{fig:STCS:ggH} and summarized as follows:
\begin{itemize}
\item  Split into jet bins: $N_j=0$, $N_j=1$, $N_j\geq2$, $N_j\geq2$ with VBF topology cuts
  (defined with the same cuts as the corresponding bin in VBF production). For the 
  $N_j\geq2$ with VBF topology cuts, $p_\mathrm{T}^H <$ 200~GeV is required, which gives priority
  to the $p_\mathrm{T}^H >$ 200~GeV bin for $N_j\geq2$. Otherwise, the $N_j \geq 2$ with VBF topology cuts
  is excluded from the $N_j \geq 2$ bins. The jet bins are
  motivated by the use of jet bins in the experimental analyses. Introducing them also for the
  simplified template cross sections avoids folding the associated theoretical uncertainties into
  the measurement. The separation of the $N_j\geq 2$ with VBF topology cuts is motivated by the wish to
  separately measure the gluon fusion contamination in the VBF selection. If the fit has no
  sensitivity to determine the gluon fusion and the VBF contributions with this topology, the sum
 of the two contributions can be quoted as result.
\item The $N_j\geq2$ with VBF topology bin is split further into an exclusive $2$-jet-like and inclusive $3$-jet-like
  bin. The split is implemented by a cut on $p_\mathrm{T}^{Hjj} = |\vec p_\mathrm{T}^H + \vec p_\mathrm{T}^{j1} + \vec p_\mathrm{T}^{j2}|$ at 25~GeV.
  See the corresponding discussion for VBF for more details.
  This split is explicitly included here since it induces
  nontrivial theory uncertainties in the gluon-fusion contribution. 
\item The $N_j=1$ and $N_j\geq2$ bins are further split into $p_\mathrm{T}^H$ bins.
  \begin{itemize}
  \item  \vspace{0cm} 0~GeV $< p_\mathrm{T}^H<$ 60~GeV: The boson channels have most sensitivity in the low
    $p_\mathrm{T}^H$ region. The upper cut is chosen as low as possible to give a more even split
    of events but at the same time high enough that no resummation effects are expected. The cut
    should also be sufficiently high that the jet $p_{\rm T}$ cut introduces a negligible bias.
  \item 60~GeV $< p_\mathrm{T}^H<$ 120~GeV: This is the resulting intermediate bin between the low and high
    $p_\mathrm{T}^H$ regions. The lower cut here is high enough that this bin can be safely treated
    as a hard $H+j$ system in the theoretical description.
  \item 120~GeV $< p_\mathrm{T}^H<$ 200~GeV: The boosted selection in $H\to\tau\tau$ contributes to
    the high $p_\mathrm{T}^H$ region. Defining a separate bin avoids large extrapolations for the
    $H\to\tau\tau$ contribution. For $N_j = 2$, this bin likely provides a substantial part of the gluon-fusion
    contribution in the hadronic $VH$ selection.
  \item $p_\mathrm{T}^H>$ 200~GeV: Beyond the top-quark mass, the top-quark loop gets resolved and top-quark mass effects become relevant.
    Splitting off the high-$p_\mathrm{T}^H$ region ensures the usability of the heavy-top expansion
    for the lower-$p_\mathrm{T}^H$ bins. At the same time, the high $p_\mathrm{T}^H$ bin in principle
    offers the possibility to distinguish a pointlike $ggH$ vertex induced by heavier BSM particles in the loop from
    the resolved top-quark loop.
\end{itemize}
\end{itemize}

At intermediate stages, all lower three $p_\mathrm{T}^H$ bins, or any two adjacent bins, can be merged. Alternatively
or in addition the $N_j=1$ and $N_j\geq2$ bins can be merged by individual analyses as needed, and potentially
also when the combination is performed at an intermediate stage.

\paragraph{Stage 2}

In stage 2, the high $p_\mathrm{T}^H$ bin should be split further, in particular if evidence for new heavy particles 
arises.
In addition, the low $p_\mathrm{T}^H$ region can be split further to reduce any theory dependence there. If desired by
the analyses, another possible option is to further split the $N_j \geq 2$ bin into $N_j=2$ and $N_j \geq 3$.

\subsubsection{Bins for VBF production}
\label{PO_STCS:STCS_VBF}

At higher order, VBF production and $VH$ production with hadronically decaying $V$ become ambiguous.
Hence, what we refer to as VBF in this section, is defined as as electroweak $qq'H$ production, 
which includes both VBF and $VH$ with hadronic $V$ decays.

\paragraph{Stage 0} Inclusive vector boson fusion cross section within $|Y_H|<2.5$. Should the
measurements start to have acceptance beyond 2.5, an additional bin for $|Y_H| > 2.5$ can be 
included.

\begin{figure}
\begin{center}
\includegraphics[width=\textwidth]{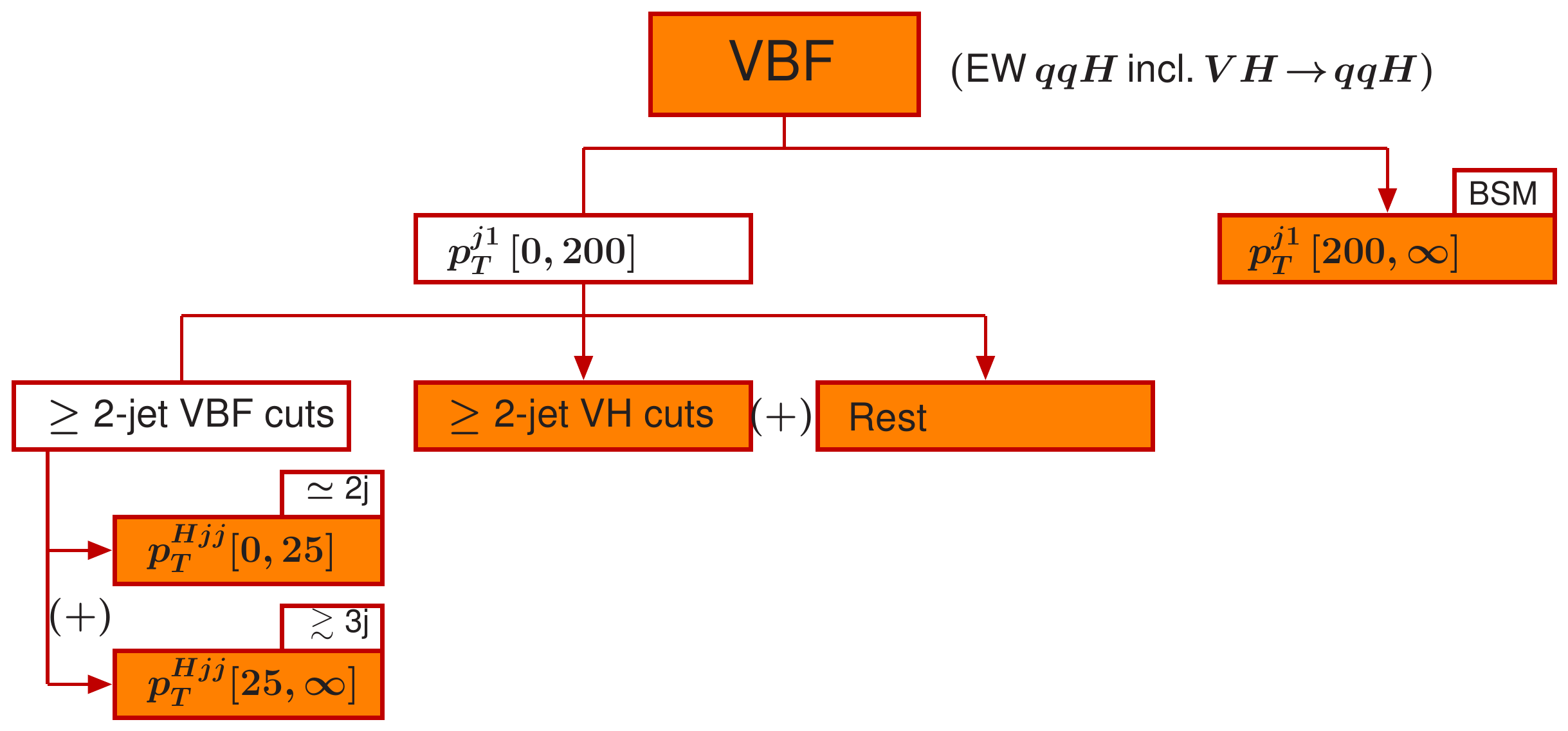}
\end{center}
\caption{Stage 1 binning for vector boson fusion production.}
\label{fig:STCS:VBF}
\end{figure}

\paragraph{Stage 1}

Stage 1 refines the binning for $|Y_H|<2.5$.
The stage 1 binning is depicted in Fig.~\ref{fig:STCS:VBF} and summarized as follows:
\begin{itemize}
\item VBF events are split by $p_\mathrm{T}^{j1}$, the transverse momentum of the highest-$p_\mathrm{T}$
jet. The lower $p_\mathrm{T}^{j1}$ region is expected to be dominated by SM-like events, while the
high-$p_\mathrm{T}^{j1}$ region is sensitive to potential BSM contributions, including events with
typical VBF topology as well as boosted $V(\to jj)H$ events where the $V$ is reconstructed as one jet.
The suggested cut is
at 200~GeV, to keep the fraction of SM events in the BSM bin small. Note that events with $N_j = 0$, 
corresponding to $p_\mathrm{T}^{j1}<$ 30~GeV, is included in the $p_\mathrm{T}^{j1}<$ 200~GeV bin.
\item The $p_\mathrm{T}^{j1}<$ 200~GeV bin is split further:
  \begin{itemize}
  \item \vspace{0cm} Typical VBF topology: The adopted VBF topology cuts are $m_{jj} >$ 400~GeV, 
    $\Delta\eta_{jj} > 2.8$ (and without any additional rapidity cuts on the signal jets). 
    This should provide a good intermediate compromise among the various VBF selection cuts
    employed by different channels.
  \begin{itemize}
  \item The bin with typical VBF topology is split into an exclusive $2$-jet-like and inclusive $3$-jet-like
  bin using a cut on $p_\mathrm{T}^{Hjj}$ at 25~GeV, where the cut value is a compromise between providing
  a good separation of gluon fusion and VBF and the selections used in the measurements.
  $p_\mathrm{T}^{Hjj}$ as quantity to define this split is chosen as a compromise between 
  the different kinematic
  variables used by different channels to enrich VBF production. (In particular the kinematic 
  variables $\Delta\phi_{H-jj}$
  and $p_\mathrm{T}^{j3}$ are both correlated with $p_\mathrm{T}^{Hjj}$).
\end{itemize}
\item \vspace{0cm} Typical $V(\to jj)H$ topology: events with at least two jets and 60~GeV $< m_{jj} <$ 120~GeV.
\item  Rest: all remaining events, including events with zero or one jet. 
  The ``rest'' bin can be sensitive to certain BSM
  contributions that do not follow the typical SM VBF signature with two forward jets.
\end{itemize}
\end{itemize}

\begin{figure}
\begin{center}
\includegraphics[width=\textwidth]{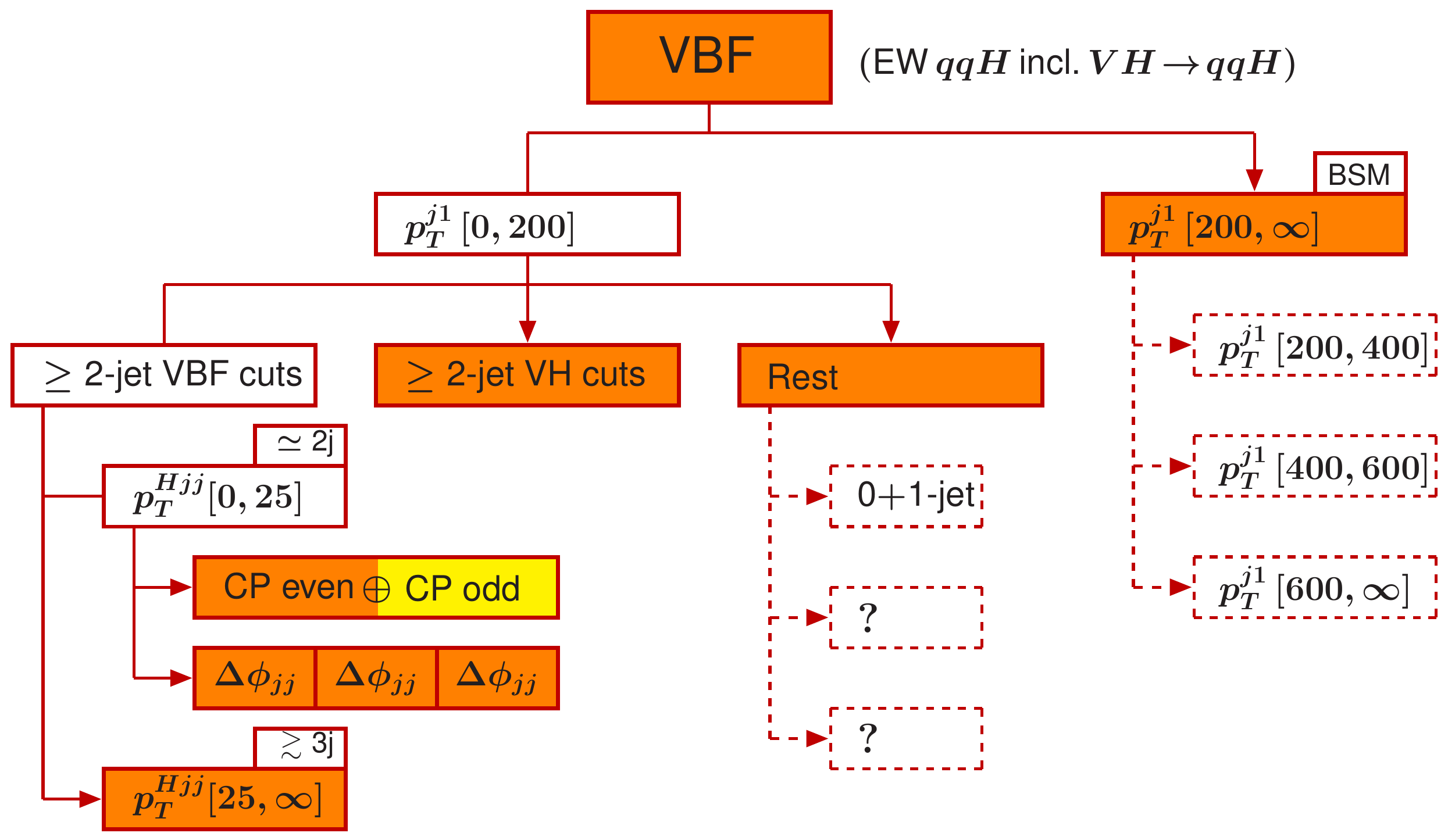}
\end{center}
\caption{Possible stage 2 binning for vector boson fusion production.}
\label{fig:STCS:VBF2}
\end{figure}

\paragraph{Stage 2}

More splits are introduced at stage 2 as illustrated in Fig.~\ref{fig:STCS:VBF2}. While the details require more discussion and
cannot be finalized at the present, this could include

\begin{itemize}
\item The high-$p_\mathrm{T}^{j1}$ bin can be split further by separating out very high-$p_\mathrm{T}^{j1}$
  events for example with additional cuts at 400~GeV and 600~GeV.
\item The ``rest'' bin can be split further, e.g., by explicitly separating out a looser VBF selection,
  and/or by separating out events with zero or one jets.
\item The $N_j\simeq 2$ VBF topology bin can be split further to gain sensitivity to CP odd
  contributions, e.g. by splitting it into subbins of $\Delta\phi_{jj}$ or alternatively by measuring a
  continuous parameter.
\end{itemize}

\subsubsection{Bins for $VH$ production}
\label{PO_STCS:STCS_VH}

In this section, $VH$ is defined as Higgs production in association with a leptonically
decaying $V$ boson.

Note that $q\bar{q}\to VH$ production with a hadronically decaying $V$ boson is considered part of 
VBF production. Similarly, $gg\to VH$ production with hadronically decaying $V$ boson is
considered part of gluon fusion production.

\paragraph{Stage 0} Inclusive associated production with vector bosons cross section within
 $|Y_H|<2.5$. Should the
measurements start to have acceptance beyond 2.5, an additional bin for $|Y_H| > 2.5$ can be 
included.

\begin{figure}
\begin{center}
\includegraphics[width=\textwidth]{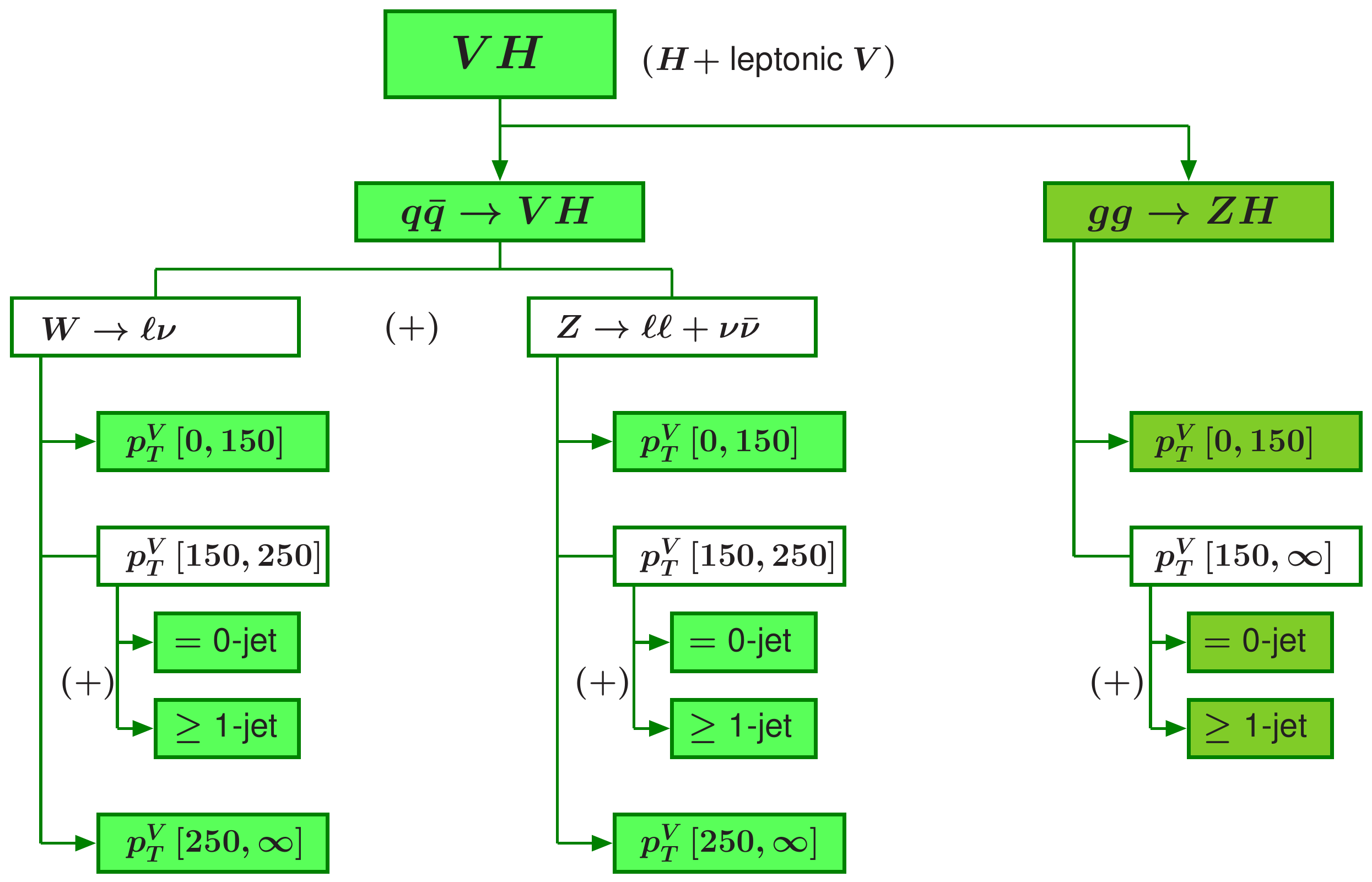}
\end{center}
\caption{Stage 1 binning for associated production with vector bosons.}
\label{fig:STCS:VH}
\end{figure}

\paragraph{Stage 1}

Stage 1 refines the binning for $|Y_H|<2.5$.
The stage 1 binning is depicted in Fig.~\ref{fig:STCS:VH} and summarized as follows:
\begin{itemize}
\item   \vspace{0cm}$VH$ production is first split into the production via a $q\bar{q}$ or $gg$ initial state.
This split becomes ambiguous at higher order. For practical
purposes, on the experimental side the split can be defined according to the MC samples used in
the analyses, which are split by $q\bar{q}$ and $gg$.
\begin{itemize}
\item   \vspace{0cm} The production via $q\bar{q}\to VH$ is split according to the vector boson: $W\to\ell\nu$ and
  $Z\to\ell\ell + \nu\bar{\nu}$.
\item $W\to\ell\nu$ and $Z\to\ell\ell + \nu\bar{\nu}$ are split further into bins of $p_\mathrm{T}^V$,
  aligned with the quantity used in the $H\to b\bar{b}$ analysis, which is one of the main 
  contributors to the $VH$ bins.
  \begin{itemize}
  \item  \vspace{0cm}
  $p_\mathrm{T}^V<$ 150~GeV receives contributions from the bosonic decay channels and from
    $H\to b\bar{b}$ with $W\to\ell\nu$ and $Z\to\ell\ell$, which do not rely on $E_\mathrm{T}^\mathrm{miss}$
    triggers.
  \item   \vspace{0cm}
  150~GeV $<p_\mathrm{T}^V<$ 250~GeV receives contributions from $H\to b\bar{b}$ with
    $Z\to\nu\bar{\nu}$ due to the high threshold of the $E_\mathrm{T}^\mathrm{miss}$ trigger, as
    well as from $H\to b\bar{b}$ with $W\to\ell\nu$ and $Z\to\ell\ell$.
    \begin{itemize}
    \item  \vspace{0cm}
     This bin is split further into a $N_j=0$ and a $N_j \geq 1$ bin, reflecting the different
      experimental sensitivity and to avoid the corresponding theory dependence.
    \end{itemize}
  \item   \vspace{0cm} $p_\mathrm{T}^V>$ 250~GeV is sensitive to BSM contributions.
  \end{itemize} 
\item   \vspace{0cm} The production via $gg \to ZH$ is split in analogy to production from the $q\bar{q}$
  initial state, apart from the $p_\mathrm{T}^V>$ 250~GeV bin, which is not split out.
\end{itemize}
\end{itemize}

\begin{figure}
\begin{center}
\includegraphics[width=\textwidth]{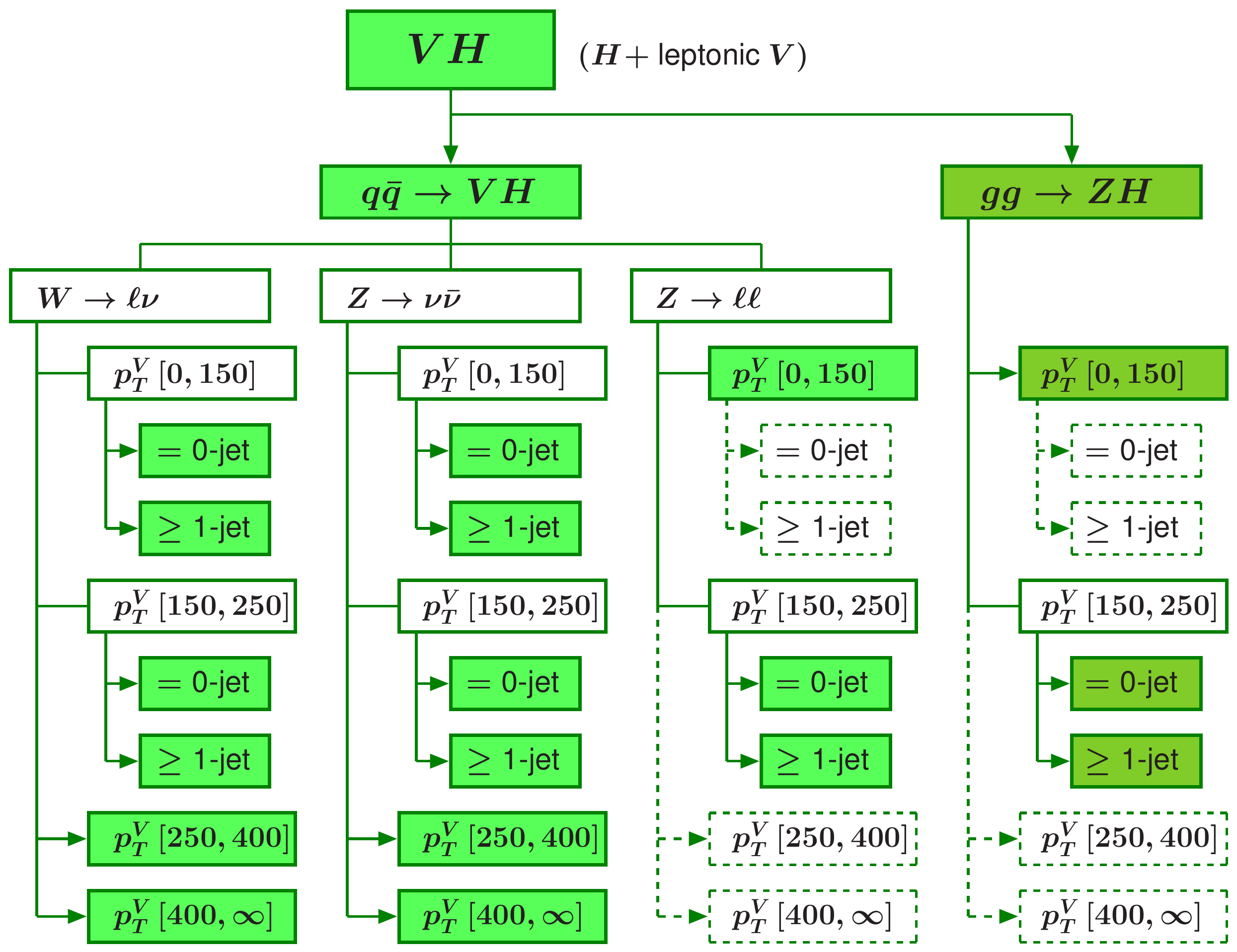}
\end{center}
\caption{Possible Stage 2 binning for associated production with vector bosons.}
\label{fig:STCS:VH2}
\end{figure}

\paragraph{Stage 2}

More splits are introduced at stage 2 as illustrated in Fig.~\ref{fig:STCS:VBF2}. While the details need more discussion, this could include

\begin{itemize}
\item Split of the $Z\to\ell\ell + \nu\bar{\nu}$ into $Z\to\ell\ell$ and $Z\to\nu\bar{\nu}$.
\item Split of the $p_\mathrm{T}^V<$ 150~GeV into a $N_j=0$ and a $N_j \geq 1$ bin, except maybe for
  the $Z\to\ell\ell$ channel, which will suffer from the low $Z\to\ell\ell$ branching ratio.
\item Split of the $p_\mathrm{T}^V>$ 250~GeV bin into $p_\mathrm{T}^V<$ 400~GeV and $p_\mathrm{T}^V>$ 400~GeV,
  to increase the sensitivity to BSM contributions with very high $p_\mathrm{T}^V$, potentially apart
  from the $Z\to\ell\ell$.
\item Potentially analoguous splits for $gg\to ZH$ production.
\end{itemize}

\subsubsection{Treatment of $t\bar tH$ production}
\label{PO_STCS:STCS_ttH}

\paragraph{Stage 0} Inclusive $t\bar t H$ production with $|Y_H|<2.5$. Should the
measurements start to have acceptance beyond 2.5, an additional bin for $|Y_H| > 2.5$ can be 
included.

\paragraph{Stage 1}

Currently no additional splits beyond stage 0 are foreseen. One option might be to
separate different top decay channels for $|Y_H|<2.5$.

\paragraph{Stage 2}

In the long term it could be useful to split into bins with 0 and $\geq 1$ additional jets or
one or more bins tailored for BSM sensitivity.

\subsubsection{Treatment of $b\bar b H$ and $tH$ production}
\label{PO_STCS:STCS_bbH}

In the foreseeable future, there will only be one inclusive bin for $b\bar{b}H$ production and
only one inclusive bin for $tH$ production for $|Y_H|<2.5$. Should the
measurements start to have acceptance beyond 2.5, an additional bin for $|Y_H| > 2.5$ can be 
included.

\subsection{Practical considerations}
\label{PO_STCS:practicalities}

To facilitate the combination of the results from ATLAS and CMS, the same bin definitions
need to be used by the two collaborations.
As for the Run1 Higgs coupling measurements, a combination of results from ATLAS and CMS will
also require that the two collaborations estimate systematic and residual theoretical uncertainties
in a compatible way.
This might be facilitated for example by the use of the same Monte Carlo generators in the measurements.

After first experience with the measurement and
interpretation has been collected, the stage 1 bin definitions should be reviewed. This
should in particular include the definition of the VBF topology cuts as well as the $p_\mathrm{T}^{Hjj}$ split.
In cases where the bin definitions are clearly inadequate, they should be improved for future
measurements. The stage 2 bins will be defined in detail taking into account the
experience gained during the measurements based on the stage 1 definitions.

An implementation of the bin definitions in Rivet is in development. This will ensure a consistent
implementation used by both experiments as well as for theoretical studies.

\subsection{Summary}
\label{PO_STCS:summary}

Simplified template cross sections provide a way to evolve the signal strength measurements that were
performed during LHC Run1, by reducing the theoretical uncertainties that are directly folded into
the measurements and by providing more finely-grained measurements, while at the same time 
allowing and benefitting from the combination of measurements in many decay channels. Several stages
are proposed: stage 0 essentially corresponds to the production mode measurements of Run1 and stage 1
defines a first complete setup, with indications for potential bin merging when a given channel
cannot yet afford the full stage 1 granularity. A complete proposal for the stage 2 binning will need to be based on
experience of using the simplified template cross section framework in real life, but some indications
of what could be interesting are already given here.

\subsection*{Acknowledgements}
We acknowledge discussions in Les Houches 2015 and WG2 of the LHC Higgs cross section working 
group, and contributions and feedback from Aaron Armbruster, 
Josh Bendavid, Fawzi Boudjema, Andr\'e David, Marco Delmastro, Dag Gillberg, Admir Greljo, Thibault Guillemin, 
Chris Hays, Gino Isidori, Sabine Kraml, James Lacey, Kirtimaan Mohan, Carlo Pandini, Elisabetta Pianori, Michael Rauch, 
Chris White, and many others. The work of PF was partially supported by the ILP LABEX (under reference 
ANR-10-LABX-63 and ANR-11-IDEX-0004-02). The work of FT was supported by the 
DFG Emmy-Noether Grant No. TA 867/1-1. KT acknowledges partial support from 
the Collaborative Research Center SFB676 of the DFG "Particles, Strings and 
the early Universe".

\renewcommand{\arraystretch}{1}

\chapter{Phenomenological studies}
\label{cha:pheno}

\section{Electroweak corrections in Drell--Yan production
  \texorpdfstring{\protect\footnote{A.\ Huss, M.\ Sch\"onherr}}{}}

\subsection{Introduction}
\label{sec:dyew:intro}

The Drell--Yan-like production of electroweak gauge bosons represents one of the key Standard Model processes at hadron colliders whose detailed understanding is crucial in order to exploit the full potential of the measurements performed at the LHC.
The neutral-current process $\mathrm{p}\mathrm{p}\to\mathrm{Z}/\gamma\to\ell^+\ell^-$, in particular, has a very clean experimental signature owing to the two charged leptons in the final state which further allows for the full reconstruction of the kinematics of the intermediate gauge boson.
Not only does this process constitute a powerful tool for detector calibration, but it also delivers important constraints in the fit of PDFs and allows for precision measurements such as the extraction of the weak mixing angle $\sin^2(\theta^{\ell}_\mathrm{eff})$.

On the theory side, Drell--Yan production belongs to one of the most precisely predicted processes:
QCD corrections are known up to NNLO~~\cite{Hamberg:1990np,Harlander:2002wh,Anastasiou:2003ds,Melnikov:2006di,Melnikov:2006kv,Catani:2009sm,Gavin:2010az,Gavin:2012sy}, the electroweak~(EW) corrections up to NLO~\cite{Baur:1997wa,Zykunov:2001mn,Baur:2001ze,Dittmaier:2001ay,Baur:2004ig,Arbuzov:2005dd,CarloniCalame:2006zq,Zykunov:2005tc,CarloniCalame:2007cd,Arbuzov:2007db,Brensing:2007qm,Dittmaier:2009cr}, and many further improvements beyond fixed-order predictions (see, e.g., references in Ref.~\cite{Dittmaier:2015rxo}).
Until recently, the largest missing piece in terms of fixed-order predictions were given by the NNLO mixed QCD--EW corrections.
Different approaches of combining QCD and EW corrections revealed that the missing $\mathcal{O}(\alpha_\mathrm{s}\alpha)$ corrections could have an impact of a few per cent in the resonance region, i.e.\ at the level that is relevant for precision phenomenology.
In a series of papers~\cite{Dittmaier:2014qza,Dittmaier:2015rxo}, the calculation of these corrections have been tackled using the so-called pole approximation~(PA).
This approach is suitable to describe observables that are dominated by resonances with sufficient accuracy.

In this work, we investigate how the $\mathcal{O}(\alpha_\mathrm{s}\alpha)$ corrections 
generated through a universal process-independent resummation approach 
compares to the results of Ref.~\cite{Dittmaier:2015rxo}.
Section~\ref{sec:dyew:comp} gives a brief overview of the computation 
employing the PA and the implementation of the QED shower in the 
\textsc{Sherpa} Monte Carlo program~\cite{Gleisberg:2008ta}.
In Sect.~\ref{sec:dyew:results} we present our numerical results of the 
comparison before we conclude in Sect.~\ref{sec:dyew:conclusions}.

\subsection{Computational setup}
\label{sec:dyew:comp}

\paragraph*{The Pole approximation}
The calculation of the $\mathcal{O}(\alpha_\mathrm{s}\alpha)$ corrections presented in Refs.~\cite{Dittmaier:2014qza,Dittmaier:2015rxo} was performed in the framework of a pole expansion, which is based on a systematic expansion of the cross section around the gauge-boson resonance $p_V^2\sim\mu_V^2$, with $\mu_V^2=M_V^2-\mathrm{i} M_V\Gamma_V$ denoting the gauge-invariant location of the propagator pole in the complex plane. 
Only retaining the leading contributions that are enhanced by a resonant propagator, we obtain the so-called pole approximation~(PA).
As a result of applying the PA, the calculation is split into separate well-defined parts that can be classified into the non-factorizable and factorizable corrections: 
The non-factorizable corrections involve soft-photon exchange between the production and decay stages of the process and constitute the conceptually most difficult part of the calculation.
They have been computed in Ref.~\cite{Dittmaier:2014qza} and were found to be negligible for all phenomenological purposes.
The factorizable contributions, on the other hand, involve corrections where the production of the intermediate gauge boson and its decay proceed independently. 
Here, the factorizable corrections of ``initial--final'' type were identified as the numerically dominant contribution---combining the sizeable QCD corrections to the production sub-process with the large EW corrections of the gauge boson decay---and were computed in Ref.~\cite{Dittmaier:2015rxo}.
The remaining factorizable corrections are given by the ``initial--initial'' and ``finial--final'' types.
The latter were found to be numerically negligible~\cite{Dittmaier:2015rxo}, while the former are not expected to deliver a sizeable correction, in particular for observables that are less sensitive to initial-state radiation effects.
In the remainder of this work, we will thus focus our attention to the initial--final factorizable corrections which we will often simply refer to as the $\mathcal{O}(\alpha_\mathrm{s}\alpha)$ corrections in the PA.
More details on this calculation can be found in Refs.~\cite{Dittmaier:2014qza,Dittmaier:2015rxo}.

\paragraph*{The QED resummation in Sherpa}
Another approach to higher order QED or electroweak corrections is 
presented in the soft-photon resummation of Yennie, Frautschi and 
Suura (YFS)~\cite{Yennie:1961ad}. Therein the universal structure 
of real and virtual soft photon emissions is exploited to construct 
an all-order approximation to the process at hand which can be 
systematically supplemented with process-dependent finite hard 
real and virtual emission corrections. The implementation presented 
in Ref.~\cite{Schonherr:2008av} focusses on higher-order QED corrections 
to particle decays and is used since as the default mechanism for such 
corrections in \textsc{Sherpa}~\cite{Gleisberg:2008ta}, both for 
elementary particle (e.g.\ $\mathrm{W}^{\pm}$, $\mathrm{Z}$, $\tau^\pm$) as well as 
hadron decays. 

In the present context of lepton pair production the higher-order 
QED corrections are effected in a factorised approach. The complete 
process $\mathrm{p}\mathrm{p}\to\ell^+\ell^-$ is calculated at LO or NLO in the strong 
coupling constant keeping all off-shell effects. Then, an intermediate 
resonance $X$ is reconstructed from the lepton pair and assigned its 
invariant mass. Its decay width is then corrected for higher order 
QED corrections, including YFS resummation, to 
\begin{equation}
  \begin{split}\label{eq:dyew:comp:yfs}
    \Gamma
    \,=\;&
      \frac{1}{2m_X}\sum\limits_{n_\gamma=0}^\infty\frac{1}{n_\gamma !}
      \int\mathrm{d}\Phi\;\mathrm{e}^{Y(\Omega)}
      \prod\limits_{i=1}^{n_\gamma}\mathrm{d}\Phi_i\,\tilde{S}(k_i)\,\Theta(k_i,\Omega)
      \left[
        \tilde\beta_0^0
        +\tilde\beta_0^1
        +\sum\limits_{i=1}^{n_\gamma}\frac{\tilde\beta_1^1(k_i)}{\tilde{S}(k_i)}
        +\mathcal{O}(\alpha^2)
      \right]\;.
  \end{split}
\end{equation}
Therein, $m_X$ is the mass of the decaying resonance and $\mathrm{d}\Phi$ 
is the phase element of the leading order decay, and $\tilde\beta_0^0$ is 
the leading order decay squared matrix element. The $Y(\Omega)$ then 
is the sum of the eikonal approximations to virtual photon exchange and 
unresolved soft real photon emission, $\Omega$ denoting the region in which 
soft photons are not resolvable. The YFS form factor, $\mathrm{e}^{Y(\Omega)}$, then 
resums these leading logarithmic universal corrections to all orders. 
Resolved photons are then described explicitly, emission by emission, 
by the eikonal $\tilde{S}$ depending on the individual photon momentum 
$k_i$. $\mathrm{d}\Phi_i$ is the corresponding phase space element. The 
eikonal approximations used in both the YFS form factor and for 
resolved real emissions can then, order-by-order, be corrected by 
supplementing the corresponding infrared-subtracted squared matrix 
elements $\tilde\beta_i^{i+j}$ of $\mathcal{O}(\alpha^{i+j})$ relative to 
the Born decay and containing $i$ resolved photons. Since all charged 
particles are considered massive in the context of YFS resummation, 
all $\tilde\beta_i$ are free of any infrared singularity. Finally, it is 
interesting to note that in the case of multi-photon emission 
each emitted photon receives the hard emission correction $\tilde\beta_1^1$ 
in the respective one-photon emission projected phase space.

The implementation used here, as we restrict the $\gamma^*/Z$ propagator 
virtuality to be near the $Z$ mass, always identifies the resonance $X$ with 
the $Z$ boson. The calculation thus contains the $\mathcal{O}(\alpha)$ virtual 
corrections $\tilde\beta_0^1$ and real emission corrections $\tilde\beta_1^1$ 
resulting in an NLO QED accurate description. As NLO weak corrections 
are finite they can in principle be incorporated in the $\tilde\beta_0^1$. 
This is left to a future work.

\subsection{Results}
\label{sec:dyew:results}

The numerical results presented in this section are obtained using 
the same input parameters and event selection cuts as in 
Ref.~\cite{Dittmaier:2015rxo}. The electroweak coupling constant 
$\alpha$ is defined in the $G_\mu$-scheme, with the exception of 
the photonic corrections which use $\alpha(0)$ as their coupling. 
For the parton distribution functions we use the NNPDF2.3QED NLO PDF 
set~\cite{Ball:2012cx}, in particular also the LO predictions shown in 
the following were evaluated using this choice. For the charged 
leptons in the final state we consider two different reconstruction 
strategies: In the case of ``dressed'' electrons, we apply a photon 
recombination procedure in order to treat all collinear 
lepton--photon configurations inclusively, whereas in the ``bare'' 
muon setup no such recombination is performed. Further details on 
the calculational setup and the event reconstruction are given in 
Ref.~\cite{Dittmaier:2015rxo}.

In order to establish the setup of the two computations, we first 
consider the relative $\mathcal{O}(\alpha)$ corrections by applying 
the YFS resummation in \textsc{Sherpa} to the LO prediction, 
$\sigma^{\mathrm{LO}\otimes\mathrm{YFS}}$, and compare it to the full 
EW corrections denoted as $\sigma^{\mathrm{NLO}[\mathrm{EW}]}$%
\footnote{
  A comparison of the PA at $\mathcal{O}(\alpha)$ against the full 
  NLO EW corrections has been performed in Ref.~\cite{Dittmaier:2014qza}.
}.
The respective relative correction factors, normalised to the LO 
prediction, are then given by
\begin{align}
  \label{eq:dyew:delta:nlo}
  \delta_\alpha^\mathrm{YFS} &= 
  \frac{\sigma^{\mathrm{LO}\otimes\mathrm{YFS}} - \sigma^\mathrm{LO}}{\sigma^\mathrm{LO}}
  , &
  \delta_\alpha &= 
  \frac{\sigma^{\mathrm{NLO}[\mathrm{EW}]} - \sigma^\mathrm{LO}}{\sigma^\mathrm{LO}}
  .
\end{align}
Please note, the thus defined $\delta_\alpha^\mathrm{YFS}$ retains also 
higher orders of $\alpha$ contrary to $\delta_\alpha$. For the 
mixed QCD--EW corrections, we generate terms of $\mathcal{O}(\alpha_\mathrm{s}\alpha)$ 
by applying the YFS resummation on top of the fixed-order NLO QCD 
prediction, which we denote by $\sigma^{\mathrm{NLO}[\mathrm{QCD}]\otimes\mathrm{YFS}}$.
These results are compared to the best prediction of 
Ref.~\cite{Dittmaier:2015rxo}, $\sigma^{\mathrm{NNLO}[\mathrm{QCD}\times\mathrm{EW}]}_{\mathrm{PA}}$, 
which includes the full NLO QCD and EW corrections, supplemented by 
the dominant $\mathcal{O}(\alpha_\mathrm{s}\alpha)$ corrections in the PA. In order 
to extract the genuine $\mathcal{O}(\alpha_\mathrm{s}\alpha)$ contribution from the 
prediction based on the YFS resummation, we define the relative 
correction factor as follows,
\begin{align}
  \label{eq:dyew:delta:nnlo:yfs}
  \delta_{\alpha_\mathrm{s}\alpha}^\mathrm{YFS} &= 
  \frac{ (\sigma^{\mathrm{NLO}[\mathrm{QCD}]\otimes\mathrm{YFS}} - \sigma^{\mathrm{NLO}[\mathrm{QCD}]}) 
        -(\sigma^{\mathrm{LO}\otimes\mathrm{YFS}} - \sigma^\mathrm{LO})}
       {\sigma^\mathrm{LO}} .
\end{align}
For the fixed-order prediction in the PA, the corresponding correction 
factor is given by
\begin{align}
  \label{eq:dyew:delta:nnlo:pa}
  \delta_{\alpha_\mathrm{s}\alpha}^\mathrm{PA} &= 
    \frac{\sigma^{\mathrm{NNLO}[\mathrm{QCD}\times\mathrm{EW}]}_{\mathrm{PA}} - \sigma^{\mathrm{NLO}[\mathrm{QCD}+\mathrm{EW}]}}
       {\sigma^\mathrm{LO}} .
\end{align}
Again, as in Eq.~\ref{eq:dyew:delta:nlo}, $\delta_{\alpha_\mathrm{s}\alpha}^\mathrm{YFS}$ 
also contains higher orders in $\alpha$, contrary to 
$\delta_{\alpha_\mathrm{s}\alpha}^\mathrm{PA}$.

\begin{figure}
  \includegraphics[width=.48\linewidth]{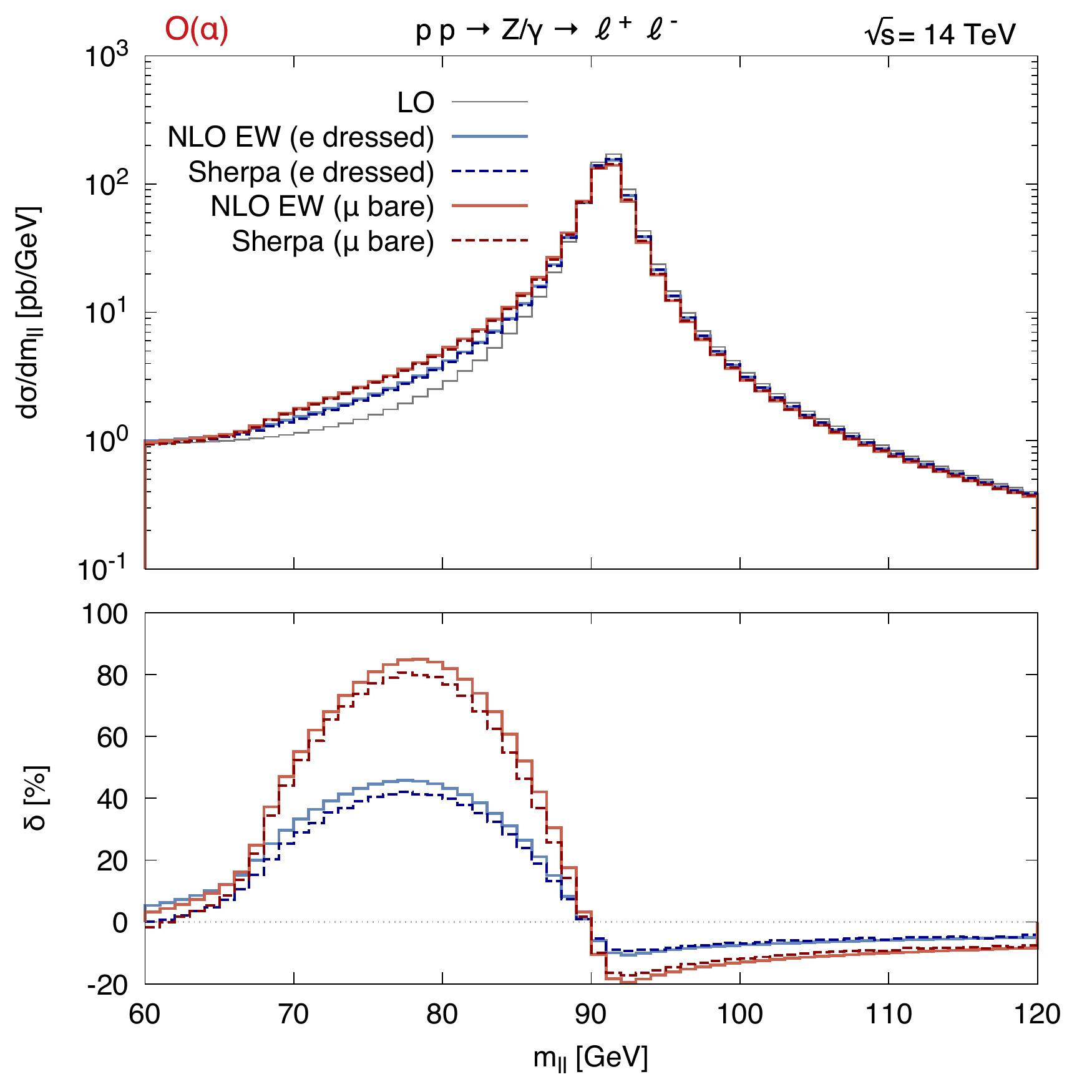} \hfill
  \includegraphics[width=.48\linewidth]{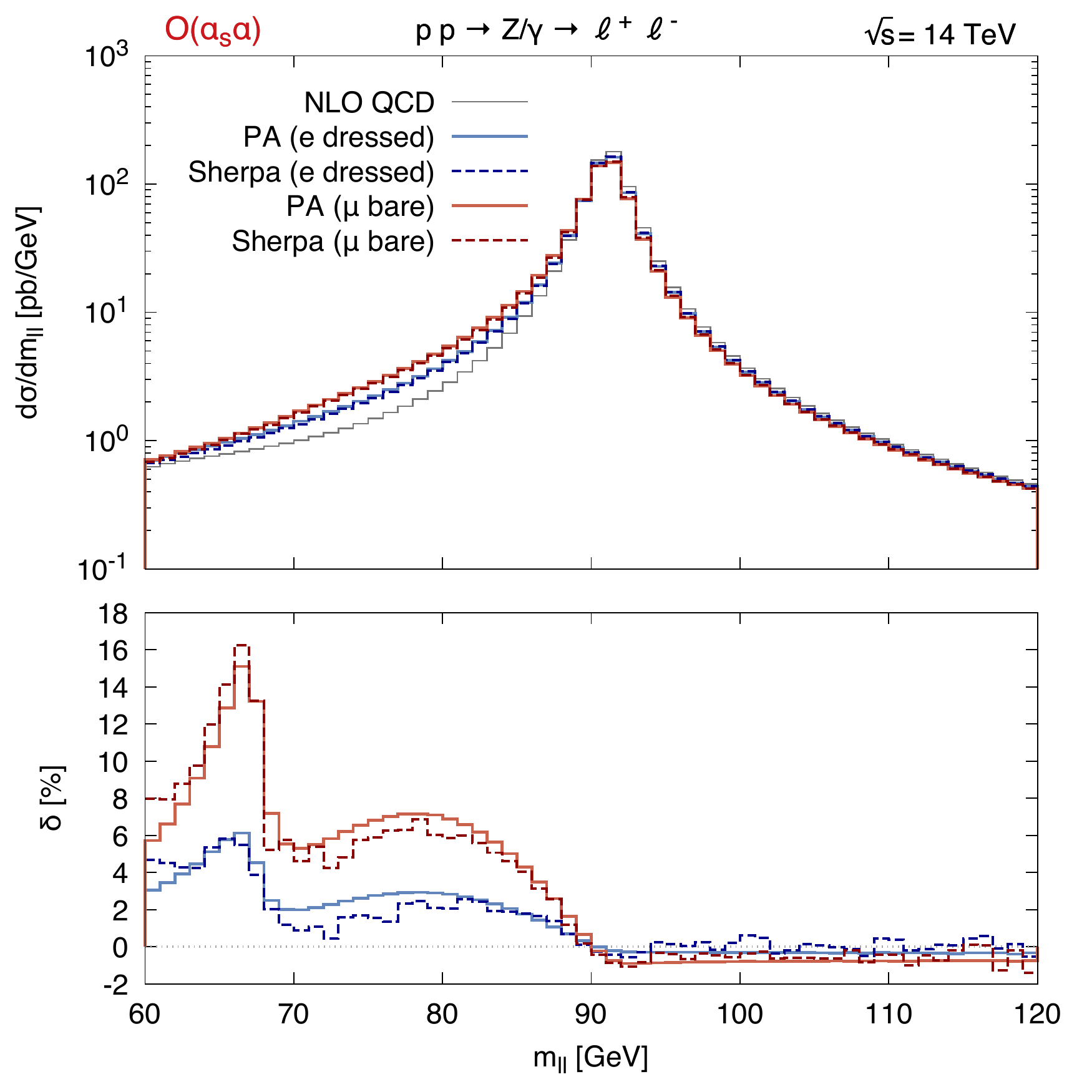} 
  \caption{
    Comparison of the $\mathcal{O}(\alpha)$~(left) and $\mathcal{O}(\alpha_\mathrm{s}\alpha)$ 
    (right) corrections to the invariant-mass distribution of the lepton 
    pair $m_{\ell\ell}$ between Ref.~\cite{Dittmaier:2015rxo} and Sherpa. 
    The absolute distributions and the relative corrections at the 
    respective order are shown in the top and bottom panels, respectively. 
    Collinear lepton--photon configurations are treated both inclusively 
    with a recombination procedure resulting in the ``e dressed'' setup 
    (blue) or exclusively in the case of muons labelled as ``$\mu$ bare'' 
    (red).
  }
  \label{fig:dyew:mll}
\end{figure}
\begin{figure}
  \includegraphics[width=.48\linewidth]{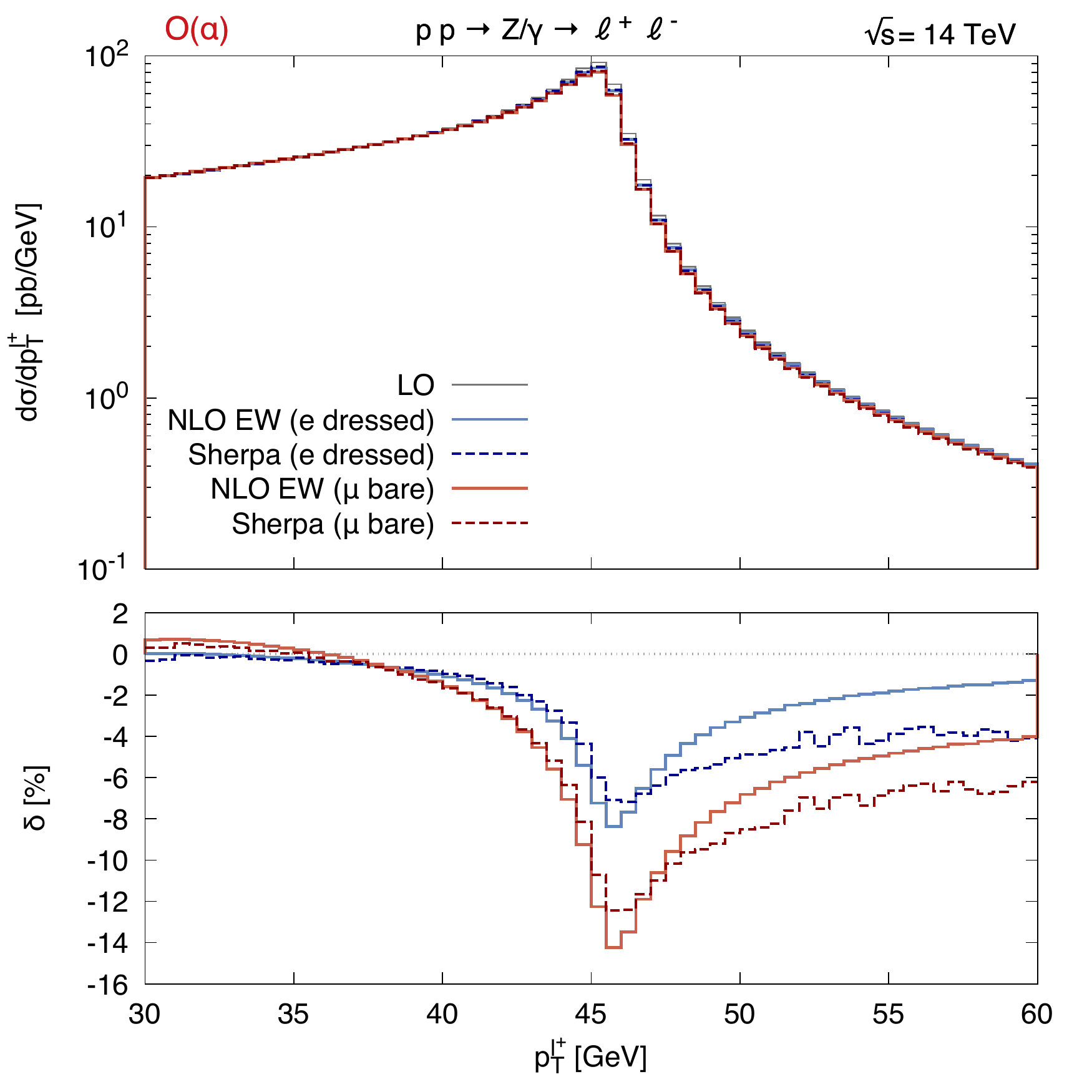} \hfill
  \includegraphics[width=.48\linewidth]{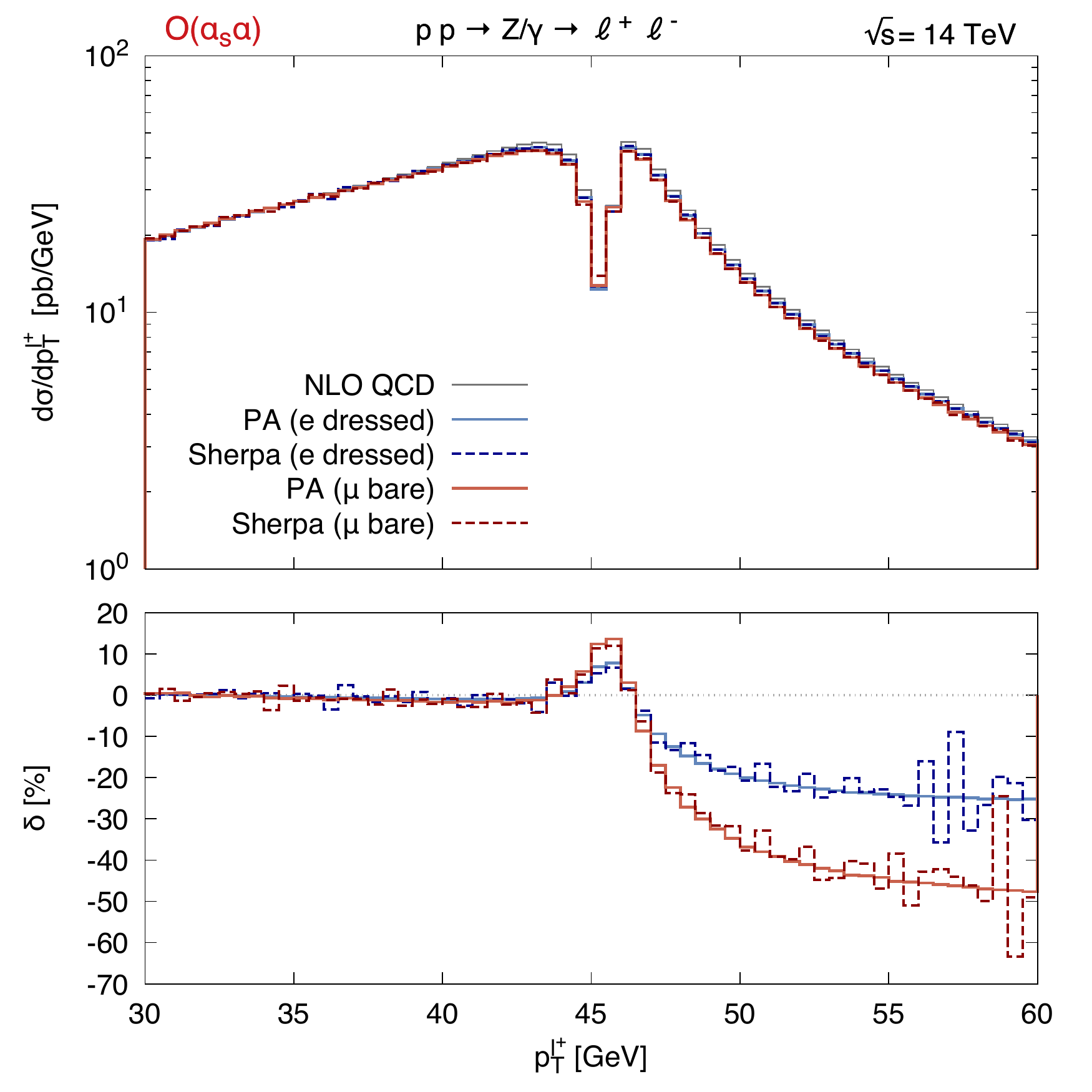} 
  \caption{
    Comparison of the $\mathcal{O}(\alpha)$~(left) and $\mathcal{O}(\alpha_\mathrm{s}\alpha)$ 
    (right) corrections to the transverse-momentum distribution of the 
    positively charged lepton $p_\mathrm{T}^{\ell^+}$ between 
    Ref.~\cite{Dittmaier:2015rxo} and Sherpa. The absolute distributions 
    and the relative corrections at the respective order are shown in the 
    top and bottom panels, respectively. Collinear lepton--photon 
    configurations are treated both inclusively with a recombination 
    procedure resulting in the ``e dressed'' setup~(blue) or exclusively in 
    the case of muons labelled as ``$\mu$ bare''~(red).
  }
  \label{fig:dyew:ptl}
\end{figure}
\begin{figure}
  \includegraphics[width=.48\linewidth]{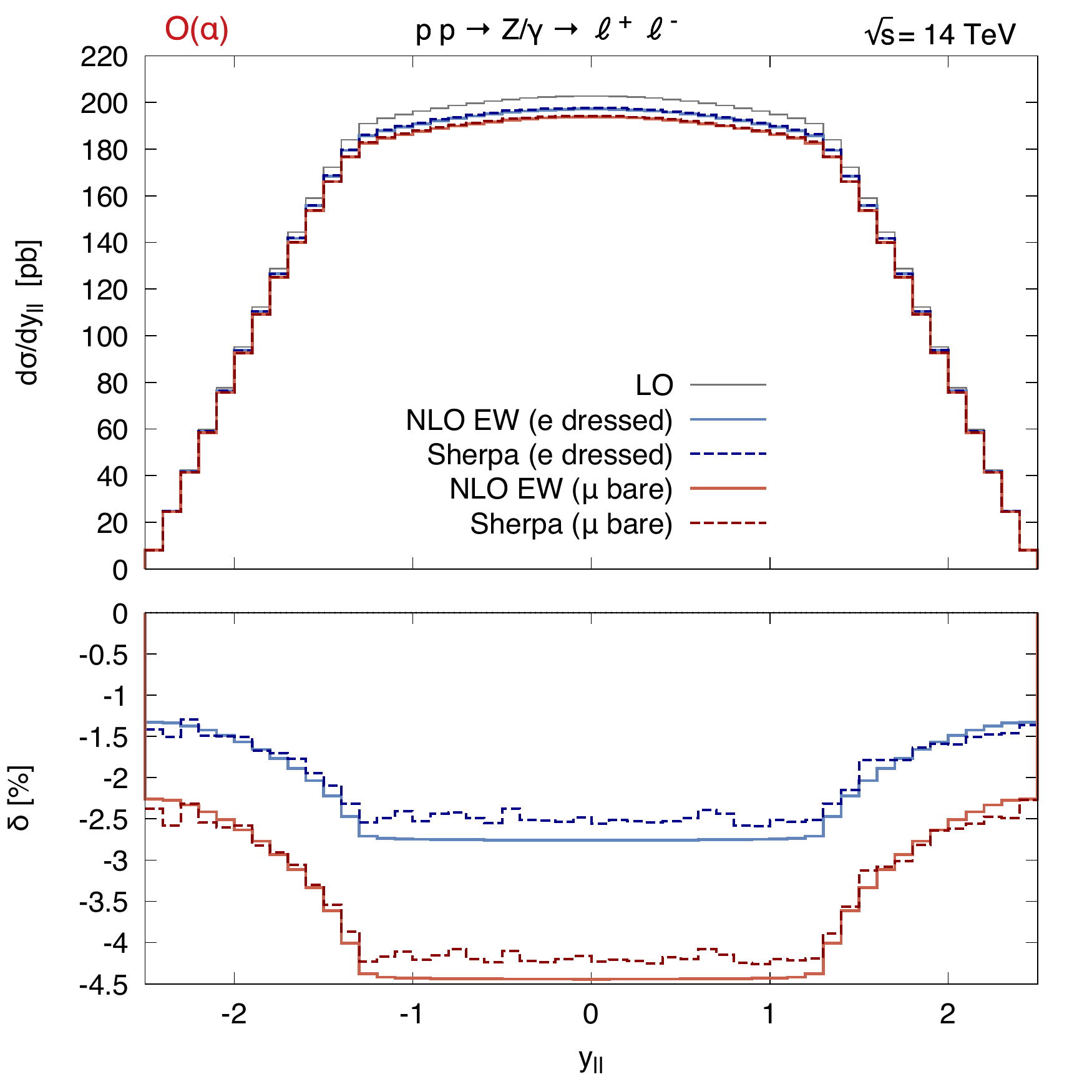} \hfill
  \includegraphics[width=.48\linewidth]{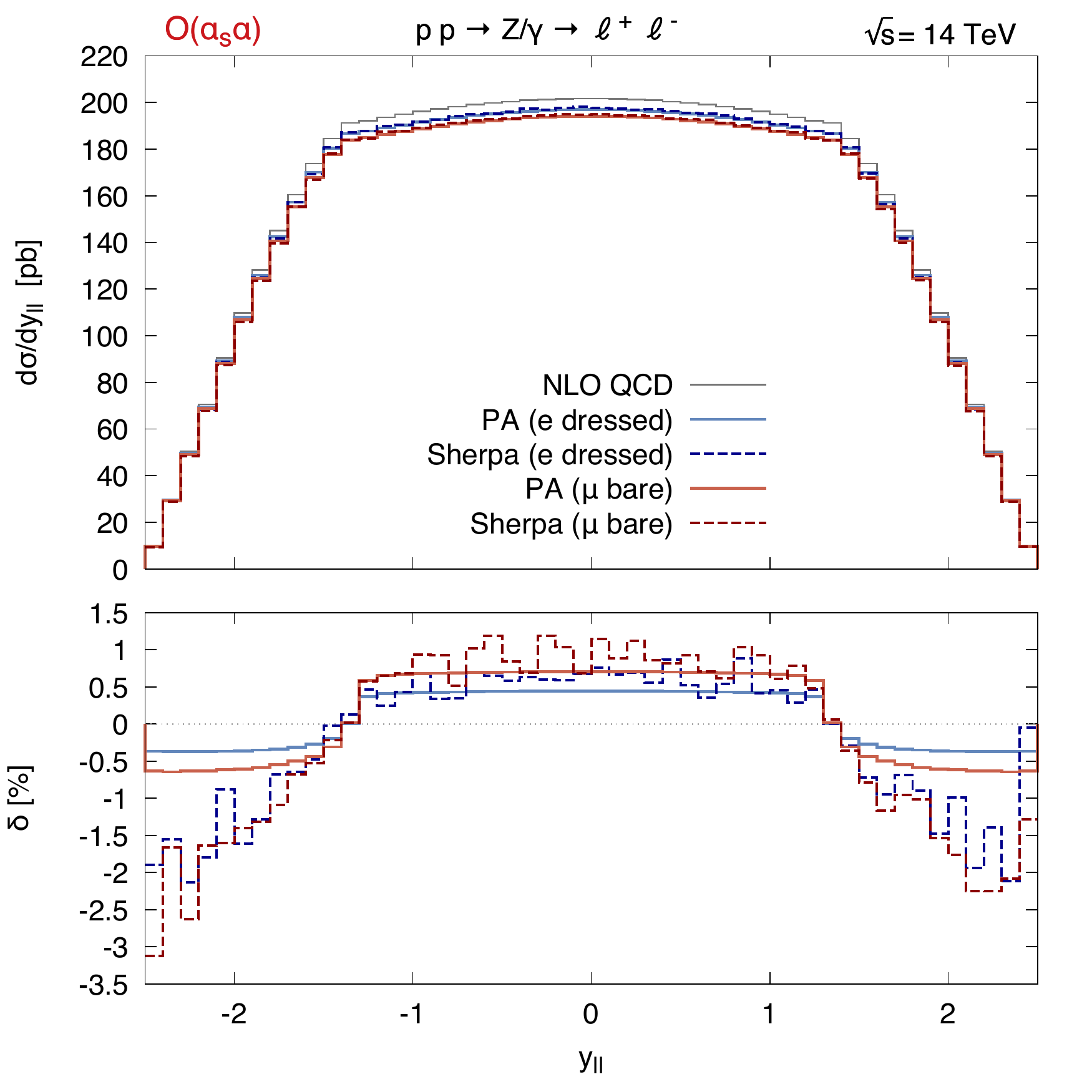}
  \caption{
    Comparison of the $\mathcal{O}(\alpha)$~(left) and $\mathcal{O}(\alpha_\mathrm{s}\alpha)$ 
    (right) corrections to the rapidity distribution of the lepton pair 
    $y_{\ell\ell}$ between Ref.~\cite{Dittmaier:2015rxo} and Sherpa. The 
    absolute distributions and the relative corrections at the respective 
    order are shown in the top and bottom panels, respectively. Collinear 
    lepton--photon configurations are treated both inclusively with a 
    recombination procedure resulting in the ``e dressed'' setup~(blue) or 
    exclusively in the case of muons labelled as ``$\mu$ bare''~(red).
  }
  \label{fig:dyew:yll}
\end{figure}

The numerical results comprise differential distributions in the 
lepton invariant mass $m_{\ell\ell}$, the transverse momentum of the 
positively charged lepton $p_\mathrm{T}^{\ell^+}$, and the rapidity of the 
lepton pair $y_{\ell\ell}$, which are shown in 
Figs.~\ref{fig:dyew:mll}--\ref{fig:dyew:yll}, respectively. The left plot 
in the figures shows a comparison of the $\mathcal{O}(\alpha)$ corrections, 
while the right-hand plot compares the corresponding $\mathcal{O}(\alpha_\mathrm{s}\alpha)$ 
corrections. In each plot we show the absolute distributions in 
the top frame and the relative correction factors in the bottom 
panel as defined in Eq.~\eqref{eq:dyew:delta:nlo} for the $\mathcal{O}(\alpha)$ 
and Eqs.~\eqref{eq:dyew:delta:nnlo:yfs} and \eqref{eq:dyew:delta:nnlo:pa} 
for the mixed QCD--EW corrections.

For the invariant mass distribution shown in Fig.~\ref{fig:dyew:mll} we 
observe an overall good agreement between the YFS resummation and 
the fixed-order result. This reflects the property of this observable 
whose corrections are known to be dominated by final-state photon 
emission. Both at $\mathcal{O}(\alpha)$ and $\mathcal{O}(\alpha_\mathrm{s}\alpha)$ we 
observe a small offset between the two predictions where the resummed 
approach leads to slightly smaller corrections below the $\mathrm{Z}$ 
resonance. This difference originates from multi-photon emissions, 
which is included in the YFS formalism, cf.\ Eq.~\eqref{eq:dyew:comp:yfs}, 
whereas the fixed-order prediction is restricted to at most one 
photon emission.
Figure~\ref{fig:dyew:ptl} shows the corrections to the transverse momentum 
distribution of the positively charged lepton, $p_\mathrm{T}^{\ell^+}$. Although 
qualitatively displaying a similar shape in the $\mathcal{O}(\alpha)$ corrections, 
we observe larger differences between the two computations around the 
resonance and in the higher transverse momentum tails. This difference 
can be understood from the fact that this observable is sensitive to 
recoil effects from initial-state radiation which are not accounted for 
in the YFS approach as used here.  Indeed, comparing the final-state 
factorizable $\mathcal{O}(\alpha)$ corrections in PA to the full NLO EW 
corrections, as it is done in Fig.~9 in Ref.~\cite{Dittmaier:2014qza},
a very similar behaviour can be observed.
For the mixed QCD--EW corrections, on 
the other hand, the EW corrections contained in the 
$\mathcal{O}(\alpha_\mathrm{s}\alpha)$ PA prediction are confined to the decay 
sub-process similarly to the case of the YFS resummation. As a result, 
we see a much better agreement between the two computations here.
Lastly, the numerical results for the rapidity distribution of the 
lepton pair, $y_{\ell\ell}$, are shown in Fig.~\ref{fig:dyew:yll}. At 
$\mathcal{O}(\alpha)$, the resummed prediction is able to reproduce the 
exact fixed-order result to a large extent up to a small offset in 
the normalisation. This shift can be attributed to the finite weak 
corrections which are missing in the YFS resummed prediction here. 
The purely weak corrections amount to a flat correction of 
approximately $-0.5\%$ in this distribution as can be read off from
Fig.~14 in Ref.~\cite{Dittmaier:2009cr} and matches well with 
the observed offset. For the mixed QCD--EW corrections we obtain 
corrections from the YFS resummation that are similar in shape to 
those of the fixed-order prediction in the PA, however, with larger 
negative corrections in the forward regime which possibly stem from 
multi-photon effects on the event acceptance in this region.

\subsection{Conclusions and outlook}
\label{sec:dyew:conclusions}

The Drell--Yan process is one of the most important ``standard candle'' 
processes at the LHC and, as such, has a wide range of applications.
In this work, we have explored the possibility of generating mixed 
QCD--EW corrections of $\mathcal{O}(\alpha_\mathrm{s}\alpha)$ to this process using 
the YFS resummation available in the \textsc{Sherpa} Monte Carlo and 
performed a comparison to the fixed-order calculation of 
Ref.~\cite{Dittmaier:2015rxo}. To this end, we have considered both 
$\mathcal{O}(\alpha)$ and $\mathcal{O}(\alpha_\mathrm{s}\alpha)$ corrections to various 
differential distributions given by the invariant mass of the leptons, 
the lepton transverse momentum, and the rapidity of the lepton pair.
We find that the QED resummation is able to capture the electroweak 
corrections to these observables remarkably well.
Furthermore, we were able to identify various sources for the differences 
that were observed between the two predictions.

Building on the insights that were gained from this study, it will 
be interesting to investigate further observables and also repeating 
this comparison for the charged-current process. Potential 
improvements were identified in both computations which should be 
explored in order to gain further insights into the numerical impact of 
the various ingredients that enter the two predictions. Such 
improvements include the finite weak corrections that can be 
incorporated to the YFS approach through the $\tilde{\beta}^1$ 
coefficient in Eq.~\eqref{eq:dyew:comp:yfs} on the one side, and 
supplementing the fixed-order calculation with multi-photon emission 
effects.

\subsection*{Acknowledgements}

We would like to thank the Les Houches workshop for hospitality offered 
during which some of the work contained herein was performed. 
A.H. acknowledges support from the ERC Advanced Grant MC@NNLO (340983). 
MS acknowledges support by the Swiss National Science Foundation (SNF) 
under contract PP00P2-128552.

\renewcommand{\arraystretch}{1.6}
\section{NLO EW technical and physics comparisons 
  \texorpdfstring{\protect\footnote{
     A.~Denner, V.~Ciulli,
      M. Chiesa, 
      R. Frederix,  
      L.~Hofer,
      S.~Kallweit,
      J.~M.~Lindert,
      P.~Maierh\"ofer,
      A.~Marini, 
      G.~Montagna, 
      M.~Moretti, 
      O.~Nicrosini, 
      D.~Pagani,
      F.~Piccinini,
      S.~Pozzorini,
      M.~Sch\"onherr, 
      E.~Takasugi,
      S.~Uccirati, 
      M.~A. Weber,
      M.~Zaro}}{}}
\label{sec:SMewcomparison}

Adequate predictions for scattering processes at particle colliders
such as the LHC require the inclusion of next-to-leading order (NLO)
perturbative corrections of the strong and electroweak (EW) interactions.
For the calculation of NLO QCD corrections the automation has been
established, and several tools exist
\cite{Hahn:1998yk,Arnold:2008rz,Berger:2008sj,Becker:2010ng,
  Badger:2010nx,Hirschi:2011pa,Bevilacqua:2011xh,Cullen:2011ac,Cascioli:2011va}.
On the other hand, the automation of NLO EW corrections has just
started, and only a few tools are available
\cite{Actis:2012qn,Kallweit:2014xda,Frixione:2014qaa,Chiesa:2015mya}.
In this section we review the status of the existing tools for the
automated calculation of EW one-loop amplitudes and provide a
comparison of some of them for specific processes. In addition,
we also show a comparison of the Sudakov approximation for the EW
corrections as implemented in {\sc Alpgen} \cite{Mangano:2002ea}
with complete NLO calculations. 
Finally, we present a comparison of theoretical predictions from
various codes with experimental results from CMS for the ratio of the
associated production of a $\PZ/\gamma^*$ or an on-shell photon with
additional jets as a function of the transverse momentum of the
vector boson.

\subsection{Codes for the automated calculation of electroweak NLO corrections}

\subsubsection*{\sc Recola}

{\sc Recola} is a Fortran95 computer program for the automated
generation and numerical computation of scattering amplitudes in the
full Standard Model (SM) (including QCD and the EW sector) at tree
and one-loop level which has become publicly available very recently 
\cite{Actis:2016mpe}. It is based on a one-loop generalization of
Dyson--Schwinger recursion relations \cite{vanHameren:2009vq} and
allows to generate one-loop amplitudes for (in principle) arbitrary
decay and scattering processes in the SM with 
particular emphasis on high particle multiplicities. Counterterms
\cite{Denner:1991kt} and rational terms \cite{Garzelli:2009is} are
included via dedicated tree-level Feynman rules. {\sc Recola} was the
first automated tool to calculate EW NLO corrections
\cite{Actis:2012qn}.

{\sc Recola} provides numerical results for scattering amplitudes in
the 't~Hooft--Feynman gauge. 
Dimensional regularization is used for ultraviolet singularities,
while collinear and soft singularities can be treated either in
dimensional or in mass regularization. {\sc Recola} is interfaced to
{\sc Collier} \cite{Denner:2016kdg,Denner:2014gla,collier-hepforge}, 
a recently published library for
the fast and numerically stable calculation of one-loop tensor
integrals. For renormalization, apart from
the traditional on-shell scheme~\cite{Denner:1991kt}, {\sc Recola} in
particular features its generalization to the complex-mass scheme~\cite{Denner:2005fg}.
{\sc Recola} further supports the $G_\mu$, the $\alpha(0)$ and the
$\alpha(M_{\PZ})$ scheme for the renormalization of the
electromagnetic coupling constant, and a dynamical
$N_{\mathrm{f}}$-flavour scheme for the strong coupling constant. 
Internal resonant particles can be
treated in the complex-mass scheme. Moreover,
resonant contributions can be consistently isolated such that matrix
elements involving specific intermediate resonances can be calculated. The
calculation of squared amplitudes, summed over spin and colour, is
supported at leading order (LO), NLO, and for loop-induced processes.
Besides the calculation of complete tree-level and one-loop results,
{\sc Recola} allows to select or discard specific orders of $\alpha_s$
(and thus also of $\alpha$) in all computed objects, such as in the
amplitudes, in the square of the Born amplitude or in the interference
of the Born with the one-loop amplitude.  Moreover, the code allows
the computation of colour- and spin-correlated LO squared amplitudes
required for dipole subtraction \cite{Catani:1996vz,Catani:2002hc}.

For the calculation of physical cross sections {\sc Recola} has to be
interfaced to a Monte Carlo code or a multi-purpose event generator;
such interfaces are presently being developed.  Together with private
Monte Carlo integrators, {\sc Recola} has been used for the computation
of EW corrections to $\Pp\Pp \to \Pl^+ \Pl^-{\rm j j},
\nu_{\Pl}\bar\nu_{\Pl}{\rm j j} $ \cite{Denner:2014ina} and the QCD
corrections to $\Pp\Pp\to\Pe^+\nu_{\Pe}
\mu^-\bar{\nu}_\mu\PQb\bar{\PQb}\PH$ \cite{Denner:2015yca}.

\subsubsection*{{\sc Sherpa/Munich+OpenLoops}}

The frameworks {\sc Munich + OpenLoops} and {\sc Sherpa + OpenLoops}
automate the full chain of operations -- from process definition to
collider observables -- that enter NLO QCD+EW simulations at parton
level.
The relevant scattering amplitudes at NLO QCD are publicly available
in the form of an automatically generated library
\cite{openloops-hepforge} that supports all interesting LHC processes
(more than one hundred processes), can be easily extended upon user
request and will be extended to NLO EW soon.
The recently achieved automation of EW
corrections~\cite{Kallweit:2014xda,Kallweit:2015dum} is based on the
well established QCD implementations and allows for NLO QCD+EW
simulations for a vast range of SM processes, up to high particle
multiplicity, at current and future colliders.
To be precise, the new implementations allow for NLO calculations at
any given order $\alpha_s^n\alpha^m$, including all relevant QCD--EW
interference effects.  Full NLO SM calculations that include all
possible $\mathcal{O}(\alpha_s^{n+k} \alpha^{m-k})$ contributions to a
certain process are also supported.

In these frameworks virtual amplitudes are provided by the {\sc OpenLoops}
program~\cite{openloops-hepforge}, which is based on the Open Loops algorithm~\cite{Cascioli:2011va} --
a fast numerical recursion for the evaluation of one-loop scattering 
amplitudes.
The extension to NLO EW corrections required the
implementation of all $\mathcal{O}(\alpha)$ EW Feynman rules in the
framework of the numerical Open Loops recursion including counterterms
associated with so-called $R_2$ rational parts~\cite{Garzelli:2009is}
and with the on-shell renormalization of UV
singularities~\cite{Denner:1991kt}.  Additionally, for the treatment
of heavy unstable particles the complex-mass
scheme~\cite{Denner:2005fg} has been implemented in a fully general
way.
Combined with the {\sc Collier} tensor-reduction library
\cite{Denner:2014gla}, which implements the Denner--Dittmaier
reduction techniques~\cite{Denner:2002ii,Denner:2005nn} and the scalar
integrals of Ref.~\cite{Denner:2010tr}, or with {\sc
  CutTools}~\cite{Ossola:2007ax}, which implements the OPP
method~\cite{Ossola:2006us}, together with the {\sc OneLOop}
library~\cite{vanHameren:2010cp}, the employed recursion permits to
achieve very high CPU performance and a high degree of numerical
stability.

All remaining tasks, i.e.\ the bookkeeping of partonic subprocesses,
phase-space integration and the subtraction of QCD and QED
bremsstrahlung are supported by the two independent and fully
automated Monte Carlo generators {\sc Munich}~\cite{munich} and {\sc
  Sherpa}~\cite{Gleisberg:2007md,Gleisberg:2008ta}.  The first one,
{\sc Munich}, is a fully generic and very fast parton-level Monte
Carlo integrator. {\sc Sherpa} is a particle-level Monte Carlo
generator providing all stages of hadron-collider simulations,
including parton showering, NLO matching, multi-jet merging,
hadronization and underlying event simulations.  For tree amplitudes,
with all relevant colour and helicity correlations, {\sc Munich}
relies on {\sc OpenLoops}, while {\sc Sherpa} generates them
internally with {\sc Amegic++} \cite{Krauss:2001iv} and {\sc Comix}
\cite{Gleisberg:2008fv}.
For the cancellation of infrared singularities both Monte Carlo tools,
{\sc Munich} and {\sc Sherpa}, employ the dipole-subtraction
scheme~\cite{Catani:1996vz,Catani:2002hc} and its extension to EW
corrections~\cite{Dittmaier:1999mb,Dittmaier:2008md}.  Both codes were
extensively checked against each other, and sub-permille level
agreement was found. The implementation of parton-shower matching and
multi-jet merging including NLO EW effects is available in an
approximate way~\cite{Kallweit:2015dum}, while a fully consistent
implementation is under way.

Employing the described frameworks in Ref.~\cite{Kallweit:2014xda} the
simulation of $\PW+1,2,3$ jet production at NLO EW+QCD was presented.
In order to make the calculation of the process with the highest jet
multiplicity feasible, it was important to factorize these processes
into a production part and into a decay part. At the NLO EW level this
required a careful implementation of the narrow-width approximation in
order to control numerical stability given the appearance of
pseudo-resonances for two or more associated jets. It was found that
$V$ + multijet final states feature genuinely different EW effects as
compared to the case of $V + 1$ jet. Subsequently in
Ref.~\cite{Kallweit:2015dum} NLO QCD+EW simulations were presented for
$\PZ+{}$jets and $\PW^{\pm}+{}$jets production including off-shell leptonic
decays and multijet merging with up to two jets within the MEPS@NLO
framework of the {\sc Sherpa} Monte Carlo program.

\subsubsection*{{\sc\small MadGraph5\_\-aMC@NLO}}


The results presented in Section~\ref{sec:comp;ew} have been obtained
via an extension of the public code {\sc\small
  MadGraph5\_\-aMC@NLO}~\cite{Alwall:2014hca} that allows to
automatically calculate also NLO EW corrections. This version of the
code is, at the moment, private and has already been used for the
calculation of NLO QCD and EW corrections to the hadroproduction of a
top-quark pair in association with a heavy boson
\cite{Frixione:2014qaa, Frixione:2015zaa}.

The automation of the NLO QCD and EW corrections for a generic SM
process has required major improvements for all the building blocks of
the {\sc\small MadGraph5\_\-aMC@NLO} code
\cite{Frederix:2009yq,Hirschi:2011pa}.  The generation of the
amplitudes and matrix elements at LO and NLO accuracy for the combined
expansion in powers of $\alpha_s$ and $\alpha$ has been lengthly
discussed in Refs.~\cite{Alwall:2014hca, Frixione:2014qaa}. The code
at the moment is able to automatically calculate all the possible
perturbative orders stemming from tree-level amplitudes and their
interference with one-loop amplitudes, for any SM process.
Moreover, it is possible to select any combination of perturbative
orders. Thus, all the necessary $R_2$ and UV counterterms have been
calculated both in the complex-mass scheme and with real masses
and on-shell renormalization conditions for unstable particles. These
counterterms are part of the so-called NLO UFO models
\cite{Degrande:2011ua}, which is the general format used in {\sc\small
  MadGraph5\_\-aMC@NLO} for importing Feynman rules of a given
Lagrangian. In the case of the SM at NLO, two different UFO models
have been created in order to calculate EW corrections
respectively in the $\alpha(m_{\PZ})$ or $G_{\mu}$~scheme.

The subtraction of infrared divergences and the integration over the
Born-like and real-emission phase space is automatically performed for
any process by extending the framework described in
Refs.~\cite{Alwall:2014hca,Frederix:2009yq}, taking into account all
the EW-QCD IR divergences appearing at any perturbative order that
enter into NLO predictions.

At the moment, NLO corrections for massive-only final states can be
automatically calculated by running a code that can be generated in
{\sc\small MadGraph5\_\-aMC@NLO} via few commands of the form:%
\footnote{These are representative commands, the future public version
  of the code may have to be used with different syntax.}
\begin{verbatim}
    import model QCDandEW_renormalization-scheme
    generate process QCD=m QED=n [QCD QED]
    output process_at_orders_m_n_with_QCD_and_QED_corrections
\end{verbatim}
The indices $m$ and $n$ and the flags QCD and QED can be used to
select the desired perturbative orders.  In the case of massless final
states, code-wise no additional improvement is required. However, with
QCD and EW corrections a generic approach for the treatment of IR
divergences and a IR-safe definition of final-state products is not
straightforward. Thus, besides the generation of the code on the same
line of the case of massive-only final states, with massless particle
specific solutions are necessary to obtain IR-safe definition of the
final state, as done, e.g.\ in
Refs.~\cite{Kallweit:2014xda,Denner:2014ina}, which are considered in
these proceedings.

At the moment results can be obtained only at fixed order, without
matching to shower effects. The matching with QED or in general
EW shower will be also included in the future, by extending
the matching procedure with QCD parton showers that is already
available in the public version of the code.

\subsection{Technical comparison of EW NLO cross sections}
\label{sec:comp;ew}

In order to assess the status of the automated tools for the
calculation of EW corrections a technical comparison has been
performed.  Given the complexity of this enterprise, we have chosen
the recent calculations of EW corrections to $\Pp\Pp\to
\mathrm{\Pl^+\Pl^-}+2{}$\,jets, $\Pp\Pp\to \mathrm{\Pl^+\nu+2}$\,jets, and
$\Pp\Pp\to \PQt\bar\PQt\PH$ obtained with the automated tools {\sc
  Recola}, {\sc Sherpa/Munich+Openloops}, and {\sc MadGraph5\_aMC@NLO},
respectively and published in
Refs.~\cite{Denner:2014ina,Kallweit:2015dum,Frixione:2015zaa}.  We
invited all other groups to provide numbers for comparison with
specific results in these publications and in addition some cumulative
histograms using the setups of the respective publications.  While the
{\sc  Sherpa/Munich+OpenLoops} collaboration provided numbers for all three
processes, none of the other groups delivered new results for
comparison.

The input parameters and set-ups of the calculations in
Refs.~\cite{Denner:2014ina,Kallweit:2015dum,Frixione:2015zaa} are
summarized in Table~\ref{tab:input;ew}. The large number of parameters
and settings that have to be adapted reflects the complexity of the
calculations. Note that $\hat{H}_{\rm T}$ is
calculated from the sum of transverse energies of all final-state
particles, i.e.\ 
\begin{equation}
\hat{H}_{\rm T} = \sum_i E_{\rm T,i}= \sum_i \sqrt{p_{\rm T,i}^2+m^2_i},
\end{equation} 
while $H_{\rm T}$ from the sum of all transverse momenta,
\begin{equation}
{H}_{\rm T} =  \sum_i p_{\rm T,i}.
\end{equation} 
The calculation of the factorization and renormalization scale for $\Pp\Pp\to \mathrm{\Pl^+\nu+2}$\,jets,
is performed using the transverse energy of the lepton--neutrino system,
\begin{equation}
\label{eq:hprime;ew}
\hat{H}'_{\rm T} = p_{\rm T,j_1}+p_{\rm T,j_2}+ p_{\rm T,g}+p_{\rm
  T,\gamma}+E_{\rm T,\Pl\nu}, \qquad E_{\rm T,\Pl\nu} = \sqrt{p_{\rm T,\Pl\nu}^2+m^2_{\rm \Pl\nu}}.
\end{equation} 
Radiated partons and photons are included at NLO.
 The cuts for the processes $\Pp\Pp\to
\mathrm{\Pl^+\Pl^-}+2{}$\,jets and  $\Pp\Pp\to
\mathrm{\Pl^+\nu+2}$\,jets are listed in Table~\ref{tab:cuts;ew}, while
no cuts have been applied for $\Pp\Pp\to \PQt\bar\PQt\PH$.
 The jet with highest
transverse momentum is denoted $\rm j_1$ in the following.
\begin{table}
\small
\begin{tabular}{l|p{9em}|p{9em}|p{9em}}
parameter/setting &  $\Pp\Pp\to \mathrm{\Pl^+\Pl^-+2}$\,jets &
$\Pp\Pp\to \mathrm{\Pl^+\nu+2}$\,jets & $\Pp\Pp\to \PQt\bar{\PQt}\PH$ \\
\hline
order of LO contribution &      all & ${\cal O}(\alpha_s^2\alpha^2)$  & ${\cal
  O}(\alpha_s^2\alpha)$
\\
order of NLO corrections & ${\cal O}(\alpha_s^2\alpha^3)$& 
${\cal O} ( \alpha_s^3\alpha^2), {\cal O}( \alpha_s^2\alpha^3)$ &
${\cal O}(\alpha_s^3 \alpha),  {\cal O}(\alpha_s^2 \alpha^2)$ \\
renormalization scheme &  $G_\mu$ scheme & $G_\mu$ scheme & $
\alpha(M_{\PZ})$ scheme \\
complex/real masses &   complex-mass scheme &   complex-mass scheme &
real masses \\
jet algorithm & anti-$k_{\rm T}$, $R=0.4$  &   anti-$k_{\rm T}$, $R=0.4$ &   \qquad --  \\
partons clustered for & $|y| < 5$ &         $|y| < \infty $
& \qquad--   \\
photon/jet separation & democratic clustering + fragmentation &
fermion--photon  \mbox{recombination~and} democratic clustering &    \qquad--  \\
$\max\frac{E_\gamma}{E_\gamma+E_{j}} $ &       0.7 &   0.5 &   \qquad--    \\
PDF set &  MSTW2008LO &        NNPDF2.3QED &       NNPDF2.3QED\\
factorization scale &   $M_{\rm \PZ,pole}$ &    $\hat{H}'_{\rm T} / 2$
&      $\hat{H}_{\rm T} / 2$  \\
renormalization scale & $M_{\rm \PZ,pole}$ &    $\hat{H}'_{\rm T} / 2$ 
 &     $\hat{H}_{\rm T} / 2$ \\
partons at LO &  g,u,c,d,s,b &  g,u,c,d,s,b & g,u,c,d,s,b,$\gamma$ \\
partons at NLO &        g,u,c,d,s & g,u,c,d,s,b &g,u,c,d,s,b,$\gamma$ \\
$\gamma$-induced contributions &  none &  only at LO &    all/none\\
collider energy  &  $13\UTeV$ &    $13\UTeV$ &   $ 13\UTeV$ \\
$\alpha_s$ (from PDF)  & $0.139395\ldots$       & $0.118$ & $0.118$  \\
$G_\mu$ [GeV$^{-2}]$&        $1.1663787\cdot10^{-5}$ &
$1.16637\cdot10^{-5} $ &  calculated      from $\alpha$   \\
$\alpha$ &      calculated from $G_\mu$ & calculated from $G_\mu$ &
$1/128.93$ \\
$M_{\rm \PZ,on-shell}$         &$91.1876\UGeV$ &  $91.1876\UGeV$&    $91.188\UGeV$ \\
$\Gamma_{\rm \PZ,on-shell}$   & $2.4952\UGeV$&    $2.4955\UGeV$&     $0\UGeV$\\
$M_{\rm \PZ,pole}$, $\Gamma_{\rm \PZ,pole}$ &        calculated &    \qquad-- &     \qquad-- \\
$M_{\rm \PW,on-shell}$ & $80.385 \UGeV$ &    $80.385 \UGeV$ &    $80.385 \UGeV$ \\
$\Gamma_{\rm \PW,on-shell}$&         $2.085\UGeV$ &     $2.0897\UGeV$ &    $0\UGeV$ \\
$M_{\rm \PW,pole}$, $\Gamma_{\rm \PW,pole}$ &       calculated &    \qquad-- &     \qquad-- \\
complex masses &  from pole masses & from on-shell masses  &        \qquad-- \\
$m_{\PQb}$ & $0\UGeV$ & $0\UGeV$ & $0\UGeV$ \\
$m_{\PQt}$ & $173.2 \UGeV$ &     $173.2 \UGeV$ &    $ 173.3\UGeV$ \\
$\Gamma_{\PQt}$ &  $0\UGeV$ & $1.339\UGeV$ & $0 \UGeV$ \\
$M_{\PH}$ & $125 \UGeV$ &       $125 \UGeV$ &        $125 \UGeV$ \\
$\Gamma_{\PH}$ &   $0\UGeV$ & $4.07 \UMeV$ &      $0\UGeV$\\
\end{tabular}
\caption{\label{tab:input;ew} Settings and input parameters used for technical comparisons.}
\end{table}

\begin{table}
\small
\centering
\tabcolsep 15pt
\begin{tabular}{c|c}
  $\Pp\Pp\to \mathrm{\Pl^+\Pl^-+2}$\,jets &
$\Pp\Pp\to \mathrm{\Pl^+\nu+2}$\,jets \\
\hline
   $p_{\rm T,j} > 30 \UGeV$  &   $p_{\rm T,j} > 30 \UGeV$      \\
       $|y_{\rm j}| < 4.5 $    & $|\eta_{\rm j}| < 4.5 $   \\
       $p_{T,l} > 20 \UGeV$    & $p_{\rm T,\Pl} > 25 \UGeV$      \\
       $|y_{\Pl}| < 2.5 $    & $|\eta_{\Pl}| < 2.5 $   \\
       $\Delta R_{\Pl\Pl} > 0.2 $& $\Delta R_{\rm j\Pl} > 0.5$   \\
       $\Delta R_{\rm j\Pl} > 0.5 $& $ M_{\rm T,\PW} > 40 \UGeV$     \\
       $ 66 \UGeV < M_{\Pl\Pl} < 116 \UGeV$&$ E_{\rm T,miss} > 25 \UGeV$\\   
\end{tabular}
\caption{\label{tab:cuts;ew} Cuts used for technical comparisons.}
\end{table}


\begin{table}
\begin{center}
\begin{tabular}{cr|rrr|rrr}
\hline 
\multicolumn{2}{c}{$\Pp\Pp\to\Pl^+\Pl^-+2\,\mathrm{j}$}   & \multicolumn{3}{|c}{\sc Recola}  &
\multicolumn{3}{|c}{\sc Sherpa/Munich+OpenLoops}  \\ 
\multicolumn{2}{c}{$G_{\mu}$ scheme}   & \multicolumn{3}{|c}{[arXiv:1411.0916]}  &  \multicolumn{3}{|c}{}  \\ 
\multicolumn{2}{c|}{} &  $\sigma^{\rm LO}$ & $\delta^{\textrm{NLO}}_{\textrm{EW}}$ & &  $\sigma^{\rm LO}$  & $\delta^{\textrm{NLO}}_{\textrm{EW}}$ &\\
&&[fb]& [\%] &   & [fb] & [\%] &  \\  \hline\hline
$p_{\mathrm{T,j}_1}>$   & 0.0\UTeV         & $5.120\cdot 10^{4}$  & $-2.5$  && $5.122\cdot 10^{4}$  & $-3.4$  &\\
                        & 0.25\UTeV      & $2.071\cdot 10^{3}$  & $-7.6$  && $2.072\cdot 10^{3}$  & $-8.5$  &\\
                        & 0.5\UTeV       & $2.060\cdot 10^{2}$  & $-13.1$ && $2.061\cdot 10^{2}$  & $-13.7$ &\\
                        & 0.75\UTeV      & $3.603\cdot 10^{1}$  & $-17.5$ && $3.603\cdot 10^{1}$  & $-17.4$ &\\
                        & 1.0\UTeV       & $7.806\cdot 10^{0}$  & $-21.5$ && $7.801\cdot 10^{0}$  & $-20.5$ &\\ \hline
$M_{\mathrm{j}_1\mathrm{j}_2}>$ & 0.5\UTeV & $4.203\cdot 10^{3}$& $-4.3$  && $4.202\cdot 10^{4}$  & $-5.1$  &\\
                        & 1.0\UTeV       & $8.085\cdot 10^{2}$  & $-5.8$  && $8.088\cdot 10^{2}$  & $-6.5$  &\\
                        & 2.0\UTeV       & $8.377\cdot 10^{1}$  & $-7.3$  && $8.368\cdot 10^{1}$  & $-7.9$  &\\
                        & 4.0\UTeV       & $2.485\cdot 10^{0}$  & $-8.2$  && $2.476\cdot 10^{0}$  & $-8.4$  &\\ \hline
$p_{\mathrm{T,\ell}^-}>$& 0.25\UTeV      & $3.176\cdot 10^{2}$  & $-12.9$ && $3.177\cdot 10^{2}$  & $-14.4$ &\\
                        & 0.5\UTeV       & $2.099\cdot 10^{1}$  & $-20.5$ && $2.097\cdot 10^{1}$  & $-20.3$ &\\
                        & 0.75\UTeV      & $2.673\cdot 10^{0}$  & $-26.3$ && $2.676\cdot 10^{0}$  & $-27.3$ &\\
                        & 1\UTeV         & $4.552\cdot 10^{-1}$ & $-30.9$ && $4.532\cdot 10^{-1}$ & $-31.3$ &\\ \hline
$p_{\mathrm{T,\ell\ell}}>$& 0.25\UTeV    & $1.356\cdot 10^{3}$  & $-9.3$  && $1.356\cdot 10^{3}$  & $-10.9$ &\\
                        & 0.5\UTeV       & $1.094\cdot 10^{2}$  & $-16.4$ && $1.093\cdot 10^{2}$  & $-17.5$ &\\
                        & 0.75\UTeV      & $1.528\cdot 10^{1}$  & $-22.0$ && $1.526\cdot 10^{1}$  & $-21.1$ &\\
                        & 1.0\UTeV         & $1.879\cdot 10^{0}$  & $-27.1$ && $1.873\cdot 10^{0}$  & $-27.7$ &\\ \hline
$H_{\mathrm{T}}>$       & 0.5\UTeV       & $3.293\cdot 10^{3}$  & $-7.1$  && $3.294\cdot 10^{3}$  & $-8.0$ &\\
                        & 1.0\UTeV         & $3.012\cdot 10^{2}$  & $-12.8$ && $3.012\cdot 10^{2}$  & $-13.2$ &\\
                        & 1.5\UTeV       & $5.166\cdot 10^{1}$  & $-17.6$ && $5.165\cdot 10^{1}$  & $-16.8$ &\\
                        & 2.0\UTeV         & $1.087\cdot 10^{1}$  &
                        $-21.8$ && $1.087\cdot 10^{1}$  & $-19.0$ &\\
                        \hline           \end{tabular}
\caption{\label{tab:Zjj_comp;ew} Comparison of results from {\sc
    Sherpa/Munich+OpenLoops} for $\Pp\Pp\to \Pl^+\Pl^-+2$\,jets
  in the setup defined in Tables \ref{tab:input;ew} and \ref{tab:cuts;ew}.}
\end{center}
\end{table}
In Table \ref{tab:Zjj_comp;ew} we present a comparison of results from
{\sc Recola} and {\sc Sherpa/Munich+Open\-Loops} for the process
$\Pp\Pp\to \Pl^+\Pl^-+2$\,jets in the setup defined in Tables
\ref{tab:input;ew} and \ref{tab:cuts;ew}. The results for the LO cross
sections $\sigma^{\rm LO}$ including all SM tree-level contributions
of order ${\cal O}(\alpha_s^2\alpha^2)$, ${\cal O}(\alpha_s\alpha^3)$,
and ${\cal O}(\alpha^4)$, agree within 0.5\% or better. The relative
EW corrections $\delta^{\rm NLO}_{\rm EW} = \delta\sigma^{\rm
  NLO}_{\rm EW}/\sigma^{\rm LO}$, where $\delta\sigma^{\rm NLO}_{\rm
  EW}$ contains the complete NLO corrections of order ${\cal
  O}(\alpha^2_s\alpha^3)$, agree at the level of $\mathcal{O}(1\%)$ or
better.  The difference in the NLO EW correction factors can be
attributed to a different treatment of b-quark-initiated processes and
of final-state photon radiation. While both calculations use
democratic clustering for jets and photons, in
Ref.~\cite{Denner:2014ina} the quark--photon fragmentation function
has been used for a consistent photon--jet separation; in the {\sc
  Sherpa/Munich+OpenLoops} approach the cancellation of collinear
singularities is enforced by recombining (anti)quark-photon pairs in a
tiny cone around the singular region as described in
Ref.~\cite{Kallweit:2014xda}.  While in the calculation of
Ref.~\cite{Denner:2014ina} the contributions of bottom quarks have
been neglected at NLO, these are included in the calculation based on
{\sc Sherpa/Munich+OpenLoops}.  Note that the agreement for
$\delta^{\rm NLO}_{\rm EW}$ is better for large transverse momenta,
where the contributions of b-quark-initiated processes are small.



\begin{table}
\begin{center}
\begin{tabular}{cr|rrr}
\hline
\multicolumn{2}{c}{$\Pp\Pp\to\Pl^+\nu+2\,\mathrm{j}$}   & \multicolumn{3}{|c}{\sc
  Sherpa/Munich+OpenLoops}  
\\ 
\multicolumn{2}{c}{$G_{\mu}$ scheme}   &
\multicolumn{3}{|c}{[arXiv:1511.08692]}  
\\ 
\multicolumn{2}{c|}{} & $\sigma^{\rm LO}_{\textrm{QCD}}$ & $\delta^{\textrm{NLO}}_{\textrm{QCD}}$ &
$\delta^{\textrm{NLO}}_{\textrm{EW}}$  
\\ 
&&[pb]& [\%] & [\%]  
\\
\hline\hline
$p_{\mathrm{T,j}_1}>$   & 0\UTeV         & $1.114\cdot 10^{2}$  &
$15.2$ & $-2.7$  
\\
                        & 0.5\UTeV       & $1.551\cdot 10^{-1}$ & $2.3$
                        & $-12.5$ 
\\
                        & 1\UTeV         & $4.092\cdot 10^{-3}$ & $8.7$
                        & $-19.5$ 
\\ \hline
$E_{\mathrm{T,miss}}>$  & 0.5\UTeV       & $7.863\cdot 10^{-3}$ &
$-4.6$ & $-22.4$ 
\\
                        & 1\UTeV         & $7.863\cdot 10^{-3}$ &
                        $-2.9$ & $-27.9$ 
\\ \hline
$p_{\mathrm{T,\Pl}^+}>$& 0.5\UTeV       & $1.647\cdot 10^{-2}$ & $0.1$
& $-21.1$ 
\\
                        & 1\UTeV         & $2.912\cdot 10^{-4}$ & $0.6$
                        & $-30.4$ 
\\ \hline    
$H_{\mathrm{T}}>$       & 0.5\UTeV       & $1.2635\cdot 10^{0}$ &
$12.3$ & $-8.0$  
\\
                        & 1\UTeV         & $2.304\cdot 10^{-1}$ &
                        $58.6$ & $-12.3$ 
\\
                        & 2\UTeV         & $5.749\cdot 10^{-3}$ &
                        $61.9$ & $-18.0$ 
\\ \hline    
\end{tabular}
\caption{\label{tab:Wjj_comp;ew} Results from {\sc
    Sherpa/Munich+OpenLoops} for $\Pp\Pp\to \mathrm{\Pl^+\nu+2}$\,jets
  in the setup defined in Tables \ref{tab:input;ew} and \ref{tab:cuts;ew}.}
\end{center}
\end{table}
In Table \ref{tab:Wjj_comp;ew} we present results from {\sc
  Sherpa/Munich+OpenLoops} for the process $\Pp\Pp\to
\Pl^+\nu+2$\,jets in the setup defined in Tables~\ref{tab:input;ew}
and \ref{tab:cuts;ew}.  The results for the LO cross sections
$\sigma^{\rm LO}_{\rm QCD}$ include the tree-level contributions of order
${\cal O}(\alpha_s^2\alpha^2)$; the relative QCD corrections, $\delta^{\rm
  NLO}_{\rm QCD} = \delta\sigma^{\rm NLO}_{\rm QCD}/\sigma^{\rm LO}_{\rm QCD}$,
and EW corrections, $\delta^{\rm NLO}_{\rm EW} = \delta\sigma^{\rm
  NLO}_{\rm EW}/\sigma^{\rm LO}_{\rm QCD}$, involve the complete NLO
contributions of order ${\cal O}(\alpha^3_s\alpha^2)$ and ${\cal
  O}(\alpha^2_s\alpha^3)$, respectively.  While none of the other
groups provided results for this process, we nevertheless show these
numbers as benchmarks for future comparisons. Moreover, in Section
\ref{sec:sudakov;ew} these results are compared with a calculation in
the Sudakov approximation for the EW corrections.



\begin{table}
\begin{center}
\scalebox{0.897}{
\begin{tabular}{r|llll|llll}
\hline
\multicolumn{1}{c}{$\Pp\Pp\to\PQt\PAQt\PH$} & \multicolumn{4}{|c}{\sc MadGraph5\_aMC@NLO} &
\multicolumn{4}{|c}{\sc Sherpa/Munich+OpenLoops}  \\ 
\multicolumn{1}{c}{$\alpha(M_{\PZ})$  scheme} & \multicolumn{4}{|c}{[arXiv:1504.03446]} &  \multicolumn{4}{|c}{}  \\ 
\multicolumn{1}{c|}{} & 
$\sigma^{\textrm{LO}}_{\textrm{QCD}}$  & $\delta^{\textrm{NLO}}_{\textrm{QCD}}$ &  $\delta^{\textrm{NLO}}_{\textrm{EW}}$ &  $\delta^{\textrm{NLO}}_{\textrm{EW, no}~\gamma}$  & 
$\sigma^{\textrm{LO}}_{\textrm{QCD}}$  & $\delta^{\textrm{NLO}}_{\textrm{QCD}}$ & $\delta^{\textrm{NLO}}_{\textrm{EW}}$ &  $\delta^{\textrm{NLO}}_{\textrm{EW, no}~\gamma}$ \\ 
&[pb]& [\%] & [\%] & [\%] & [pb] & [\%] & [\%] & [\%] \\
\hline\hline
incl.                                   & $3.617\cdot 10^{-1}$ & $28.9$ & $-1.2$  & $-1.4$  & $3.617\cdot 10^{-1}$ & $28.3$ & $-1.3$  & $-1.4$\\
$p_{\mathrm{T,H/t/\bar t}}> 200\UGeV$    & $1.338\cdot 10^{-2}$ & $23.4$ & $-8.2$  & $-8.5$  & $1.338\cdot 10^{-2}$ & $22.5$ & $-8.2$  & $-8.4$\\
$p_{\mathrm{T,H/t/\bar t}}> 400\UGeV$    & $3.977\cdot 10^{-4}$ & $9.6$ & $-13.8$ & $-13.9$ & $3.995\cdot 10^{-4}$ & $10.4$ & $-13.9$ & $-14.0$\\
$p_{\mathrm{T,H}}> 500\UGeV$             & $2.013\cdot 10^{-3}$ & $37.8$ & $-10.6$ & $-11.6$ & $2.014\cdot 10^{-3}$ & $37.3$ & $-10.8$ & $-11.7$\\
$|y_{\PQt}|>2.5$                           & $4.961\cdot 10^{-3}$ & $37.5$ & $~~~0.2$   & $~~~0.5$   & $5.006\cdot 10^{-3}$ & $36.9$ & $~~~0.2$   & $~~~0.5$\\ \hline
\end{tabular}
}
\caption{\label{tab:tth_comp;ew} Comparison of results from {\sc MadGraph5\_aMC@NLO} and {\sc
    Sherpa/Munich+OpenLoops} for $\Pp\Pp\to \PQt\bar{\PQt}\PH$ in the
  setup defined in Table \ref{tab:input;ew}.}
\end{center}
\end{table}

In Table \ref{tab:tth_comp;ew} we present a comparison of results from
{\sc MadGraph5\_aMC@NLO} and {\sc Sherpa/Munich+OpenLoops} for the
process $\Pp\Pp\to \PQt\bar{\PQt}\PH$ in the inclusive setup defined
in Table \ref{tab:input;ew}. The relative corrections are normalized
to the LO QCD cross section, $\delta^{\rm NLO}_X =\delta\sigma^{\rm
  NLO}_X/\sigma^{\rm LO}_{\rm QCD}$.  The absolute QCD corrections, $
\delta\sigma^{\rm NLO}_{\rm QCD}$ and EW corrections
$\delta\sigma^{\rm NLO}_{\rm EW}$ comprise the complete NLO
contributions of order ${\cal O}(\alpha^3_s\alpha)$ and ${\cal
  O}(\alpha^2_s\alpha^2)$, respectively.  For the entries with ``no
$\gamma$'' the PDF of the photon has been artificially set to zero in
order to gauge the impact of photons in the initial state.  All the
results are at per mille accuracy w.r.t.~the corresponding LO QCD
predictions. In the {\sc Sherpa/Munich+Openloops} calculation, a
six-flavour scheme is employed, consistently with the NNPDF2.3\_QED
PDF distribution. The gluon splitting into top-quark pairs in the PDF
evolution as well as the six-flavour running and renormalization of
$\alpha_s$ are taken into account.  However, contributions from
initial-state top quarks and top-quark bremsstrahlung are not
included. This is justified by the fact that the former are formally
of order $\alpha_s^2\log^2(\mu_{\rm R}/m_{\PQt})$, while
top-bremsstrahlung gives rise to Higgs+multi-top signatures.
Conversely, the results from {\sc MadGraph5\_aMC@NLO} are obtained
within the five-flavour scheme, renormalizing top-quark loops in the
decoupling scheme. This means that top-quark contributions to the
initial state and to the Bremsstrahlung should not be included in the
hard cross section, and a running $\alpha_s$ with five active flavours
should be employed.  Therefore, one has to compensate the inputs from
the NNPDF2.3\_QED PDF distribution, i.e.\ the luminosities and
$\alpha_s$ for given Bjorken $x$'s and scales, which are in the
six-flavour scheme, to be consistent with the five-flavour scheme
approach used in the matrix elements.\footnote{While {\sc
    NNPDF2.3\_QED} PDF distributions are in the variable-flavour
  scheme, with our choice of the scale, which is always larger than
  $m_{\PQt}$, the variable-flavour scheme is equivalent to the
  six-flavour scheme.  In order to remove the impact of the sixth
  flavour, one has to compensate its contribution to the running of
  $\alpha_s$ and to the DGLAP equation for the PDF evolution. At NLO
  QCD, this corresponds to
$$
\sigma^{\rm NLO}_{\rm QCD~(6f-PDFs)} = \\ \sigma^{\rm NLO}_{\rm QCD} + \nonumber
\alpha_s \frac{2 T_{\rm F}}{3 \pi}\left[\log\left(\frac{m_{\PQt}^2}{\mu_{\rm R}^2}\right)
  \sigma^{\rm LO}_{\rm QCD,q\bar{q}}+ \log\left(\frac{\mu_{\rm F}^2}{\mu_{\rm R}^2}\right)
  \sigma^{\rm LO}_{\rm QCD,gg}\right]
\label{eq:6fto5f}
$$
as explained in Ref.~\cite{Cacciari:1998it}, where the same issue
has been addressed for the five- and the four-flavour scheme.  The
quantities $\sigma^{\rm LO}_{\rm QCD,q\bar{q}}$ and $ \sigma^{\rm
  LO}_{\rm QCD,gg}$ correspond to the LO QCD cross sections from
respectively only the $q\bar{q}$ and $\Pg\Pg$ initial states. By
setting $\mu_{\rm F}=\mu_{\rm R}$, as done here, the term proportional
to $ \sigma^{\rm LO}_{\rm QCD,gg}$ in the r.h.s.\ of
Eq.~\eqref{eq:6fto5f} is equal to zero. In
Ref.~\cite{Frixione:2015zaa} the $N_{\rm f}=5$ scheme was used in
combination with {\sc NNPDF2.3\_QED} without including such a
subtraction term.}  In practice, both approaches are internally
consistent, and the different treatments of the $\alpha_s$ evolution
are exactly equivalent, as $\log(\mu_{\rm R}/m_{\PQt})$ terms turn out
to be accounted for to all orders in both calculations. However, the
fact that top-quark contributions to the evolution of the gluon
density are taken into account (through the NNPDF2.3\_QED PDFs) in
{\sc Sherpa/Munich+Openloops} and are subtracted in {\sc
  MadGraph5\_aMC@NLO} gives rise to a difference of
$\mathcal{O}(\alpha_s\log(\mu_{\rm F}/m_{\PQt}))$, which manifests
itself as one-percent level deviations in the numerical predictions.

\subsection{Comparison of Sudakov approximation with EW NLO corrections for distributions}
\label{sec:sudakov;ew}

In Refs.~\cite{Denner:2000jv,Denner:2001gw} a process-independent
algorithm for the computation of one-loop EW corrections has
been developed. According to the algorithm, the $\mathcal{O}(\alpha)$
corrections to a generic process involving $N$ external particles of
flavour $i_1, \cdots, i_N$ factorize in the high-energy limit as
follows:
\begin{equation}
\label{eq:dlsl}
\delta \mathcal{M}_{i_{1}\cdots i_{n}}^{\rm NLL} \Bigg|_{\rm Sudakov}=
\sum_{k=1}^{N}\sum_{l>k}\delta_{kl}^{\rm DL}\mathcal{M}_{i_{1}\cdots j_{k}\cdots j_{l}\cdots i_{n}}^{\rm LO} +
\sum_{k=1}^{N}\delta_{k}^{\rm SL}\mathcal{M}_{i_{1}\cdots j_{k}\cdots i_{n}}^{\rm LO}+
\delta^{\rm PR}\mathcal{M}_{i_{1}\cdots i_{n}}^{\rm NLL}. 
\end{equation}
In Eq.~(\ref{eq:dlsl}), the functions $\delta_{kl}^{\rm DL}$ and
$\delta_{k}^{\rm SL}$ contain the Sudakov double and single logarithmic
contributions, respectively. They depend only on the flavour and on
the kinematics of the external particles. These terms multiply
LO matrix elements that are obtained from the one of the
original process $\mathcal{M}_{i_{1}\cdots i_{n}}^{\rm LO}$ via $\rm SU(2)$
transformations of pairs or single external legs, $j_k$ being in
Eq.~(\ref{eq:dlsl}) the $\rm SU(2)$ transformed of the particle $i_k$. The
last term in Eq.~(\ref{eq:dlsl}) comes from parameter renormalization:
\begin{equation}
\label{eq:slpr}
\delta^{\rm PR}\mathcal{M}_{i_{1}\cdots i_{n}}^{\rm NLL}=
\delta e\frac{\delta\mathcal{M}_{i_{1}\cdots i_{n}}^{\rm LO}}{\delta e}+
\delta c_{\rm w}\frac{\delta\mathcal{M}_{i_{1}\cdots
    i_{n}}^{\rm LO}}{\delta c_{\rm w}}+
\delta h_{\PQt}\frac{\delta\mathcal{M}_{i_{1}\cdots i_{n}}^{\rm LO}}{\delta h_{\PQt}}+
\delta h_{\PH}\frac{\delta\mathcal{M}_{i_{1}\cdots i_{n}}^{\rm LO}}{\delta h_{\PH}},
\end{equation}
where $h_t=m_t/M_{\PW}$, $h_{\PH}=M_H^2/M_{\PW}^2$ and
$c_{\rm w}=M_{\PW}/M_{\PZ} $. In
Ref.~\cite{Chiesa:2013yma}, the algorithm of
Refs.~\cite{Denner:2000jv,Denner:2001gw} has been implemented in the
{\sc Alpgen v2.1.4}~\cite{Mangano:2002ea} event generator for the
vector-boson + multi-jet production and applied to study the phenomenological
impact of the one-loop weak corrections to New Physics searches in missing transverse energy plus multi-jets
production~\cite{Chiesa:2013yma,Mishra:2013una,Campbell:2013qaa,Butterworth:2014efa}.
Following Eq.~(\ref{eq:dlsl}), the analytic expressions of the
process-independent corrections factors have been coded and all the
required LO matrix elements are computed numerically by
means of the {\sc ALPHA} algorithm~\cite{Caravaglios:1995cd}.

\begin{table}[t!]
\centering
\small{
\begin{tabular}{l|c|c}
\hline
$\Pp\Pp\to\PW+2\,\rm j$ & {\sc Sherpa/Munich}  & {\sc Alpgen}  \\[-1ex]
                          & {\sc +OpenLoops}  &  \\
\hline
\hline
$H_{\rm T}>0.5\UTeV$ & $-8.09(2)\%$  & $-4.7(2)\%$       \\
$H_{\rm T}>1\UTeV$   & $-12.37(4)\%$ & $-9.6(2)\%$      \\
$H_{\rm T}>2\UTeV$   & $-17.8(2)\%$ & $-16.6(3)\%$       \\
\hline
$p_{\rm T,j_1}>0.5\UTeV$  & $-12.56(5)\%$ & $-9.4(2)\%$      \\
$p_{\rm T,j_1}>1\UTeV$    & $-19.1(2)\%$  & $-16.0(3)\%$       \\
\hline
$p_{\rm T,\Pl}>0.5\UTeV$  & $-21.0(3)\%$   &  $-20.1(2)\%$   \\
$p_{\rm T,\Pl}>1\UTeV$    & $-31(1)\%$     &  $-31.9(5)\%$     \\
\hline
$E_{\rm T}^{\rm miss}>0.5\UTeV$  & $-22.0(3)\%$  &  $-20.2(2)\%$   \\
$E_{\rm T}^{\rm miss}>1\UTeV$    & $-30(1)\%$    &  $-31.7(4)\%$     \\
\hline
\end{tabular}
}
\caption{ \label{tabelwtotj} Relative corrections $\frac{d \sigma^{\rm NLO}}{d \sigma^{\rm LO}}-1$ to the combined 
$\PW^++2$~jets and $\PW^-+2$~jets production processes. Comparison between the full one-loop results ({\sc Sherpa/Munich+OpenLoops}) and the 
predictions of the logarithmic approximation ({\sc Alpgen}). }
\end{table}

\begin{table}[t!]
\centering
\small{
\begin{tabular}{l|c|c}
\hline
$\Pp\Pp\to\PW^-+2\,\rm j$ & {\sc Sherpa/Munich}  & {\sc Alpgen} \\[-1ex]
                          & {\sc +OpenLoops}  &  \\
\hline
\hline
$H_{\rm T}>0.5\UTeV$ & $-8.12(2)\%$  & $-4.3(2)\%$       \\
$H_{\rm T}>1\UTeV$   & $-12.40(6)\%$ & $-9.4(2)\% $      \\
$H_{\rm T}>2\UTeV$   & $-17.7(2)\%$ & $-16.6(2)\%$     \\
\hline
$p_{\rm T,j_1}>0.5\UTeV$  & $-12.57(6)\%$ & $-9.3(2)\%$     \\
$p_{\rm T,j_1}>1\UTeV$    & $-18.9(2)\%$  & $-15.3(3)\%$      \\
\hline
$p_{\rm T,\Pl^-}>0.5\UTeV$  & $-20.8(5)\%$   &  $-20.0(2)\%$       \\
$p_{\rm T,\Pl^-}>1\UTeV$    & $-32(2)\%$     &  $-32.1(3)\%$       \\
\hline
$E_{\rm T}^{\rm miss}>0.5\UTeV$  & $-21.9(4)\%$  &  $-20.0(3)\%$      \\
$E_{\rm T}^{\rm miss}>1\UTeV$    & $-31(1)\%$    &  $-32.1(3)\%$     \\
\hline
\end{tabular}
}
\caption{ \label{tabelwmj} Relative corrections $\frac{d \sigma^{\rm NLO}}{d \sigma^{\rm LO}}-1$ to  
$\PW^-+2$~jets production. Comparison between the full one-loop results ({\sc Sherpa/Munich+OpenLoops}) and the 
predictions of the logarithmic approximation ({\sc Alpgen}).}
\end{table}

\begin{table}[t!]
\centering
\small{
\begin{tabular}{l|c|c}
\hline
$\Pp\Pp\to\PW^++2\,\rm j$ & {\sc Sherpa/Munich}  & {\sc Alpgen}  \\[-1ex]
                          & {\sc +OpenLoops}  &  \\
\hline
\hline
$H_{\rm T}>0.5\UTeV$ & $-8.03(2)\%$  & $-4.5(3)\%$     \\
$H_{\rm T}>1\UTeV$   & $-12.33(6)\%$ & $-9.9(2)\%$       \\
$H_{\rm T}>2\UTeV$   & $-18.0(3)\%$ & $-17.3(2)\%$       \\
\hline
$p_{\rm T,j_1}>0.5\UTeV$  & $-12.56(7)\%$ & $-9.5(3)\%$      \\
$p_{\rm T,j_1}>1\UTeV$    & $-19.5(3)\%$  & $-15.9(4)\%$       \\
\hline
$p_{\rm T,\Pl^+}>0.5\UTeV$  & $-21.1(3)\%$   &  $-20.1(3)\%$   \\
$p_{\rm T,\Pl^+}>1\UTeV$    & $-30(2)\%$     &  $-31.9(3)\%$     \\
\hline
$E_{\rm T}^{\rm miss}>0.5\UTeV$  & $-22.4(6)\%$  &  $-20.2(3)\%$    \\
$E_{\rm T}^{\rm miss}>1\UTeV$    & $-28(3)\%$    &  $-31.7(4)\%$    \\
\hline
\end{tabular}
}
\caption{ \label{tabelwpj} Relative corrections $\frac{d \sigma^{\rm NLO}}{d \sigma^{\rm LO}}-1$ to 
$\PW^++2$~jets production. Comparison between the full one-loop results ({\sc Sherpa/Munich+OpenLoops}) and the 
predictions of the logarithmic approximation ({\sc Alpgen}).}
\end{table}

The {\sc Alpgen} results shown in
Tables~\ref{tabelwtotj}--\ref{tabelz2j} have been computed by using
the {\tt vbjet} package for the production of
$n\PW+m\PZ+j\gamma+l\PH+k$~jets with $n+m+j+l+k \leq 8$ and $k \leq 3$. In
order to compare the results of the different codes, it is worth
recalling the following features of the {\tt vbjet} package: it
includes only the first two generations of quarks, the external
massive vector bosons are produced on-shell and the matrix elements
are computed including the effect of both QCD and EW interactions.
Within the {\tt vbjet} package, the Sudakov corrections are computed
using on-shell external vector bosons (the $\PZ$ and $\PW$ bosons are
allowed to decay including spin-correlation effects only at the
analyses level in order to apply the cuts listed in Table
\ref{tab:cuts;ew}), they include the full logarithmic dependence in
Eq.~(\ref{eq:dlsl}) for the leading $\mathcal{O}(\alpha_s^{2} \alpha)$
LO contributions, while they only have double logarithmic accuracy for
the subleading $\mathcal{O}( \alpha^3 )$ Born processes. In
Refs.~\cite{Denner:2000jv,Denner:2001gw} photonic contributions to
virtual one-loop EW corrections are split into purely weak
and purely electromagnetic terms by introducing a photon \emph{mass}
of the order of $M_{\PW}$: in Tables~\ref{tabelwtotj}--\ref{tabelz2j}
only the weak part of the photonic contribution has been included. 
No real corrections have been included in the results from {\sc Alpgen}.
The input parameters used for the simulation are the ones defined in
Table~\ref{tab:input;ew}, with the following exceptions for
$\PW+2$~jets: as we are dealing with on-shell $\PW$ bosons, the
factorization and renormalization scale for QCD in
Eq.~(\ref{eq:hprime;ew}) depend on $M_{\PW}^2$, rather than $m_{\Pl
  \nu}^2$. As the {\sc NNPDF2.3 QED} PDF set is not available in {\sc
  ALPGEN v2.1.4}, we used the {\sc MSTW2008LO} PDF set for both the
$\PZ+2\,$jets and $\PW+2\,$jets calculations: even though this implies
that a comparison at the level of cross sections is not possible for
$\PW+2\,$jets, the PDF choice does not affect the relative corrections
coming from virtual weak contributions.

In Tables~\ref{tabelwtotj}--\ref{tabelz2j} we compare the results for
the full relative EW corrections, $\delta^{\rm NLO}_{\rm EW}$, from
Section \ref{sec:comp;ew} with the relative EW corrections calculated
in the Sudakov approximation. Note that in the latter case no real
corrections are included, and the
virtual photonic corrections are regularized by a photon mass equal to
$M_{\PW}$.

As reported in Tables~\ref{tabelwtotj}--\ref{tabelwpj} the Sudakov
approximation implemented in {\sc ALPGEN} and the full one-loop
results are in good agreement (at the percent level) for $\PW+2$~jets
production for the distributions in the lepton transverse momentum
$p_{\rm T,\Pl}$ and the missing energy $E_{\rm T}^{\rm miss}$. For
the distributions in the transverse momentum of the leading jet
$p_{\rm T,j_1}$ and the variable $H_{\rm T}=p_{\rm T,j_1}+p_{\rm T,j_2}+p_{\rm T,\Pl}+E_{\rm T}^{\rm miss}$, the Sudakov
approximation tends to deviate by up to $4\%$ from the exact
fixed-order calculations in those regions of phase space where the
vector boson $p_{\rm T}$ tends to be soft compared to the transverse
momentum of the jets.

\begin{table}
\centering
\small{
\begin{tabular}{l|c|c|c}
\hline
$\Pp\Pp\to\PZ+2\,\rm j$ & {\sc Recola} full & {\sc Recola}$^{*}$ & {\sc Alpgen} 
\\
\hline
\hline
inclusive  & $-2.47(2)\%$ & $-2.69(2)\%$ & -- 
\\
\hline
$H_{\rm T}>0.5\UTeV$ & $-7.05(3)\%$  & $-7.38(3)\%$  & $-5.2(3)\% $    \\
$H_{\rm T}>1\UTeV$   & $-12.79(4)\%$ & $-13.41(5)\%$ & $-10.4(1)\%$    \\
$H_{\rm T}>1.5\UTeV$ & $-17.59(8)\%$ & $-18.70(9)\%$ & $-13.7(1)\%$    \\
$H_{\rm T}>2\UTeV$   & $-21.7(2)\%$  & $-23.8(2)\%$  & $-14.9(2)\%$    \\
\hline
$m_{\rm jj}>0.5\UTeV$ & $-4.33(5)\%$  & $-4.56(5)\%$ & $-1.4(3)\%$     \\
$m_{\rm jj}>1\UTeV$   & $-5.76(8)\%$  & $-6.03(8)\%$ & $-2(1)\%$       \\
$m_{\rm jj}>2\UTeV$   & $-7.3(1)\%$   & $-7.6(1)\%$  & $-6(1)\%$    \\
$m_{\rm jj}>4\UTeV$   & $-8.2(2)\%$   & $-8.5(2)\%$  & $-15(2)\%$    \\
\hline
$p_{\rm T,j_1}>0.25\UTeV$ & $-7.64(3)\%$  & $-7.99(3)\%$  & $-5.8(1)\% $    \\
$p_{\rm T,j_1}>0.5\UTeV$  & $-13.09(6)\%$ & $-13.83(6)\%$ & $-10.4(1)\%$    \\
$p_{\rm T,j_1}>0.75\UTeV$ & $-17.5(1)\%$  & $-19.0(1)\%$  & $-13.0(1)\%$    \\
$p_{\rm T,j_1}>1\UTeV$    & $-21.4(2)\%$  & $-24.1(3)\%$  & $-14.0(2)\%$    \\
\hline
$p_{\rm T,\Pl^-}>0.25\UTeV$ & $-12.93(8)\%$ & $-13.47(9)\%$ & $-11.0(1)\%$    \\
$p_{\rm T,\Pl^-}>0.5\UTeV$  & $-20.5(2)\%$  & $-21.1(2)\%$  & $-18.7(1)\%$    \\
$p_{\rm T,\Pl^-}>0.75\UTeV$ & $-26.3(4)\%$  & $-26.9(4)\%$  & $-23.9(1)\%$    \\
$p_{\rm T,\Pl^-}>1\UTeV$    & $-30.9(8)\%$  & $-31.3(9)\%$  & $-28.4(2)\%$    \\
\hline
$p_{\rm T,\PZ}>0.25\UTeV$ & $-9.31(4)\%$   & $-9.77(4)\%$   & $ -8.5(1)\%$  \\
$p_{\rm T,\PZ}>0.5\UTeV$  & $-16.37(8)\%$  & $-16.9(8)\%$   & $-16.3(1)\%$  \\
$p_{\rm T,\PZ}>0.75\UTeV$ & $-22.0(1)\%$   & $-22.5(1)\%$   & $-21.7(1)\%$  \\
$p_{\rm T,\PZ}>1\UTeV$    & $-27.1(4)\%$   & $-27.6(4)\%$   & $-26.8(2)\%$  \\
\hline
\hline
\end{tabular}
}
\caption{ \label{tabelz2j} Relative corrections $\frac{d \sigma^{\rm
      NLO}}{d \sigma^{\rm LO}}-1$ to $\PZ+2\,$jets production. Comparison
  between the full one-loop results ({\sc Recola}) and the predictions
  of  the logarithmic approximation ({\sc Alpgen}, {\sc Sherpa}). 
{\sc Recola}$^{*}$ is the full one-loop result where the contribution 
of $b$-quarks has been removed from  both LO and NLO and the
contribution of gluon radiation has been removed from the NLO. }
\end{table}
For the $\PZ+2$\,jets case shown in Table \ref{tabelz2j} this
behaviour is emphasized. While the Sudakov approximation works fine
for the distribution in the transverse momentum of the $\PZ$~boson,
deviations between the Sudakov approximation
and the exact fixed-order calculation amount to $3\%$ for the
distribution in the  transverse momentum of the lepton, and reach even
$10\%$ for the distributions in the transverse momentum of the leading jet
$p_{\rm T, j_1}$ and the variable $H_{\rm T}=p_{\rm T, j_1}+p_{\rm T,j_2}+p_{\rm T,\Pl^+}+p_{\rm T,\Pl^-}$.
It is worth noticing, however, that for
$\PZ+2\,$jets the difference between the {\sc ALPGEN} predictions and
the full one-loop results also originates from the on-shell
approximation, as in this approximation the contribution of the
diagrams containing a photon connecting the lepton pair to a quark
line in the process $\Pp\Pp \to \Pl^+\Pl^-+2\,$jets is not included.

\subsection{Ratio of the Z to gamma transverse momentum differential
  cross-section}

The ratio of the associated production of a $\mathrm{Z}/\gamma^*$ or a
$\gamma$ with one or more jets has been recently measured in
proton--proton collisions at $8\UTeV$ center-of-mass energy by the CMS
Collaboration at the CERN LHC~\cite{Khachatryan:2015ira}. In the limit
of high transverse momentum of the vector boson $p_{\mathrm{T}}^{V}$
($V=\PZ,\gamma$) and at LO in perturbative quantum chromodynamics
(QCD), effects due to the mass of the Z boson ($M_\mathrm{Z}$) are
small, and the cross section ratio of Z${}+{}$jets to $\gamma+{}$jets
as a function of $p_{\mathrm{T}}^{V}$ is expected to become constant,
reaching a plateau for $p_{\mathrm{T}}^{V} \geq
300\UGeV$~\cite{Ask:2011xf}. (Hereafter, production of
$\PZ/\gamma^*+{}$jets is denoted by $\PZ+{}$jets.) A QCD calculation at
NLO for pp$\to\PZ+{}$jets and pp$\to\gamma+{}$jets
was provided by the {\sc BLACKHAT} Collaboration~\cite{Bern:2012vx}.  The
NLO QCD corrections tend to decrease the value of the cross section
ratio. At higher energies, EW corrections can also
introduce a dependence of the cross section on logarithmic terms of
the form $\ln(p_{\mathrm{T}}^{\mathrm{Z}}/m_{\mathrm{Z}})$ that become
large and pose a challenge for perturbative calculations.

Searches for new particles involving final states characterized by the
presence of large missing transverse energy and hard jets, use the
$\gamma+{}$jets process to model the invisible Z decays,
Z$\to\nu\bar{\nu}$, since the $\gamma+{}$jets cross section is larger
than the $\PZ+{}$jets process where the Z decays to leptons. A precise
estimate of EW corrections on the cross section ratio for $\PZ+{}$jets and
$\gamma+{}$jets is therefore crucial to reduce uncertainties related to
the Z$\to\nu\bar{\nu}$ background estimation in these searches.

In the CMS measurement~\cite{Khachatryan:2015ira}, results are
unfolded into a fiducial region defined at particle level. For
$\PZ+{}$jets events, the leading leptons are required to have
$p_{\mathrm{T}}>20\UGeV$ and $|\eta|<2.4$, while jets are required to
have $p_{\mathrm{T}}>30\UGeV$ within the region of $|\eta|<2.4$.
Electrons and muons have different energy losses due to final-state
radiation at particle level. In order to compensate for these
differences, a ``dressed'' level is defined to make the electron and
muon channels compatible to within 1\%. This is achieved by defining
in simulation a particle momentum vector by adding the momentum of the
stable lepton and the momenta of all photons with a radius of $\Delta
R=0.1$ around the stable lepton. All jets are required to be separated
from each lepton by $\Delta R>0.5$. At the particle level, a true
isolated photon is defined as a prompt photon, around which the scalar
sum of the $p_{\mathrm{T}}$ of all stable particles in a cone of
radius $\Delta R=0.4$ is less than $5\UGeV$. A true isolated photon is
defined as a prompt photon (not generated by a hadron decay), around
which the scalar sum of the $p_{\mathrm{T}}$ of all stable particles
in a cone of radius $\Delta R=0.4$ is less than $5\UGeV$. When comparing
the cross sections for $\PZ+{}$jets and $\gamma+{}$jets, the rapidity
range of the bosons is restricted to $|{y^{V}}|<1.4$ because this is
the selected kinematic region for the photons.  The ratio of the
differential cross sections as a function of $p_{\mathrm{T}}$ is
measured in the phase-space regions: $N_{\mathrm{jets}}\geq
1,\,2,\,3$ and $H_{\mathrm{T}}>300\UGeV$, $N_{\mathrm{jets}}\geq1$.

\begin{figure}[t!]
  \includegraphics[width=0.47\textwidth]{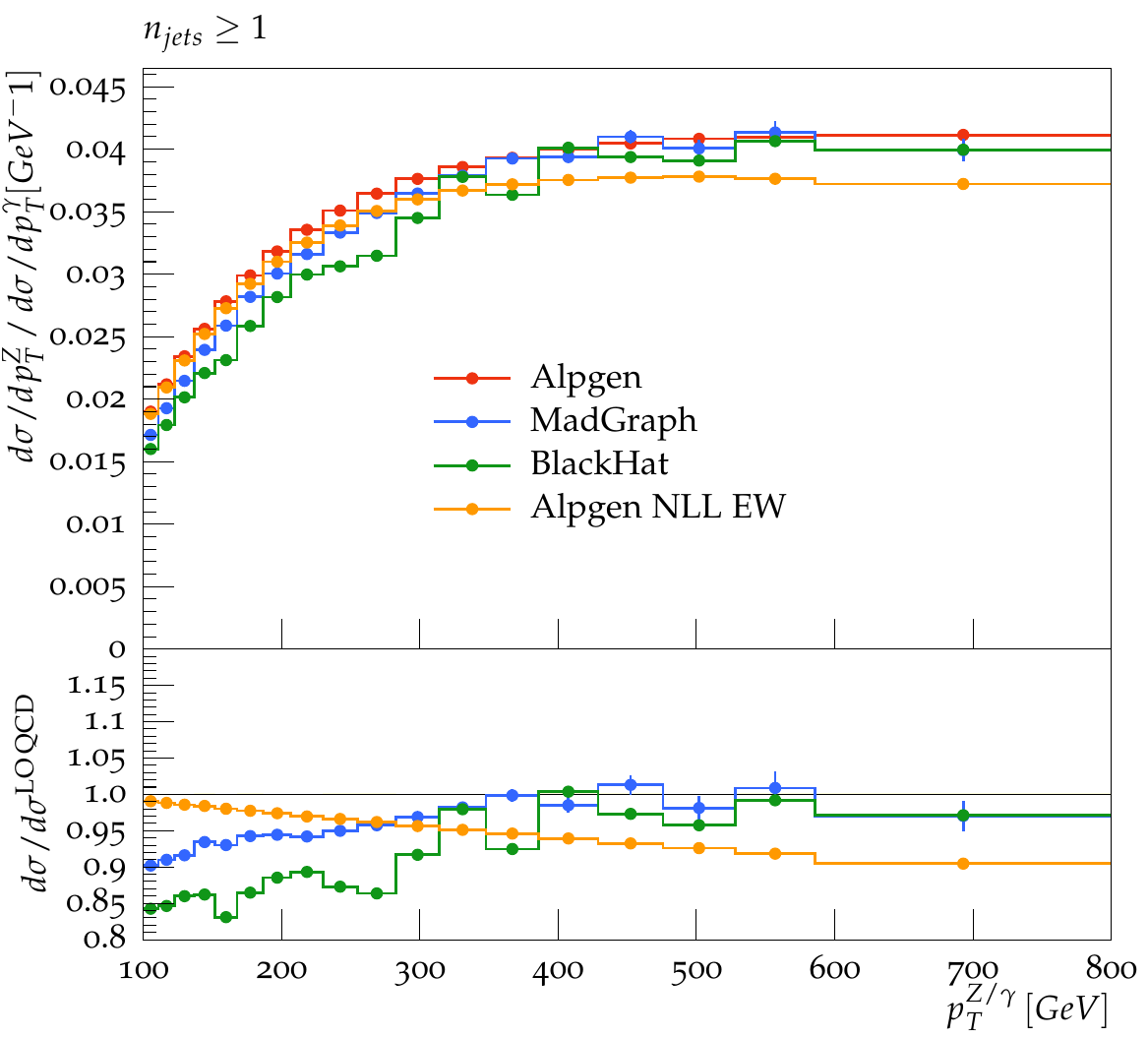}
  \hfill
  \includegraphics[width=0.47\textwidth]{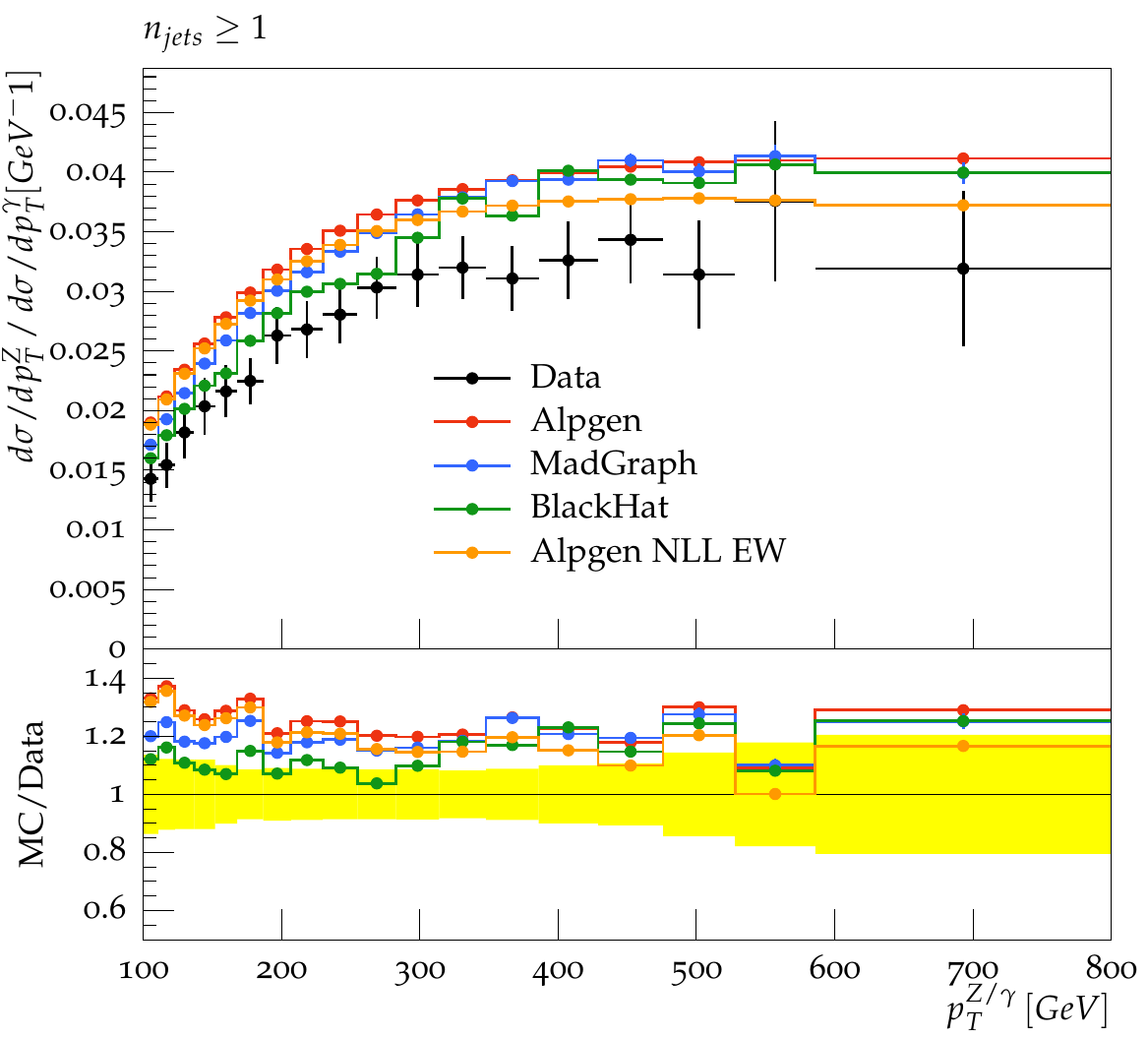} \\
  \includegraphics[width=0.47\textwidth]{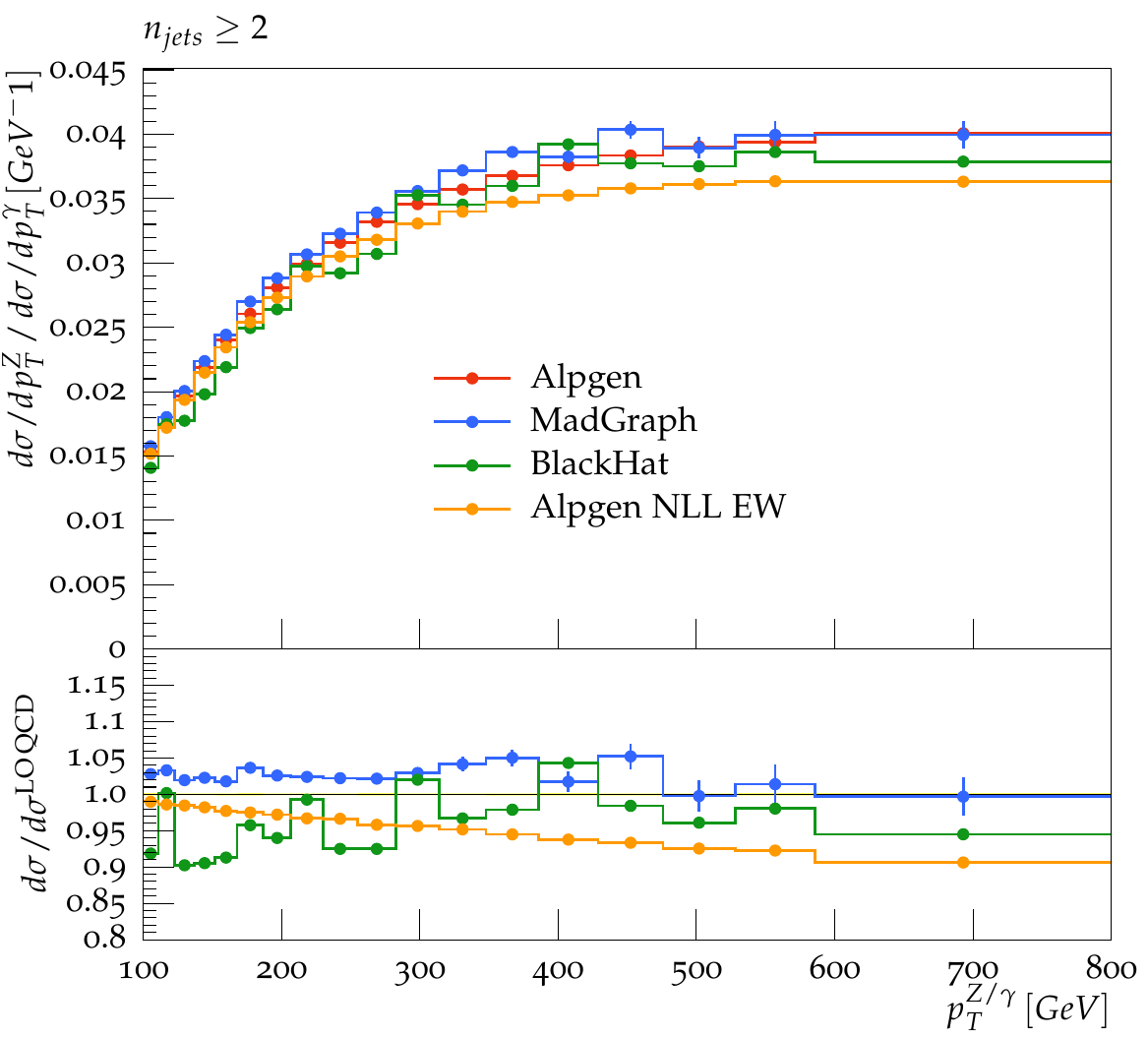}
  \hfill
  \includegraphics[width=0.47\textwidth]{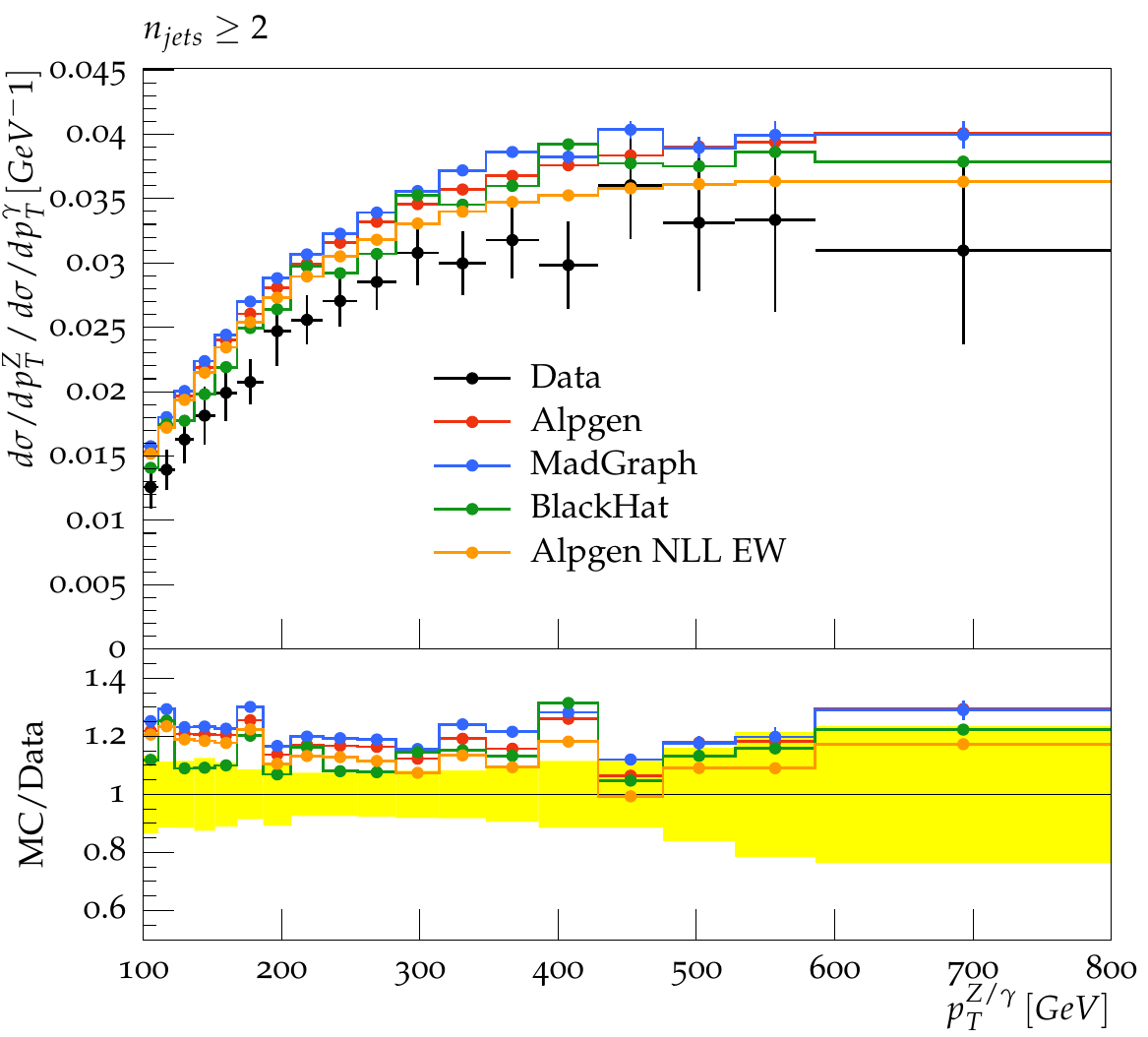} \\
  \includegraphics[width=0.47\textwidth]{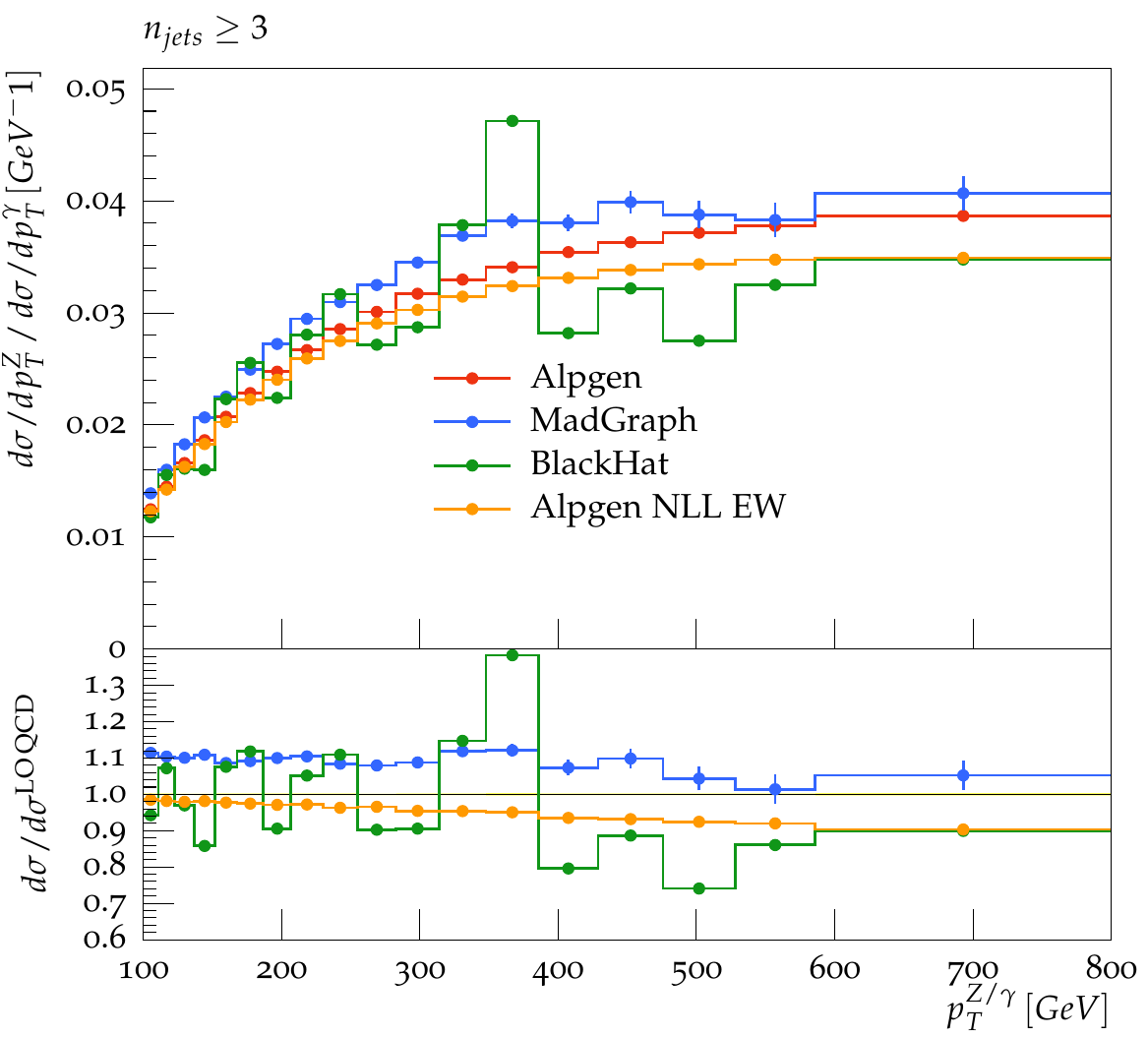}
  \hfill
  \includegraphics[width=0.47\textwidth]{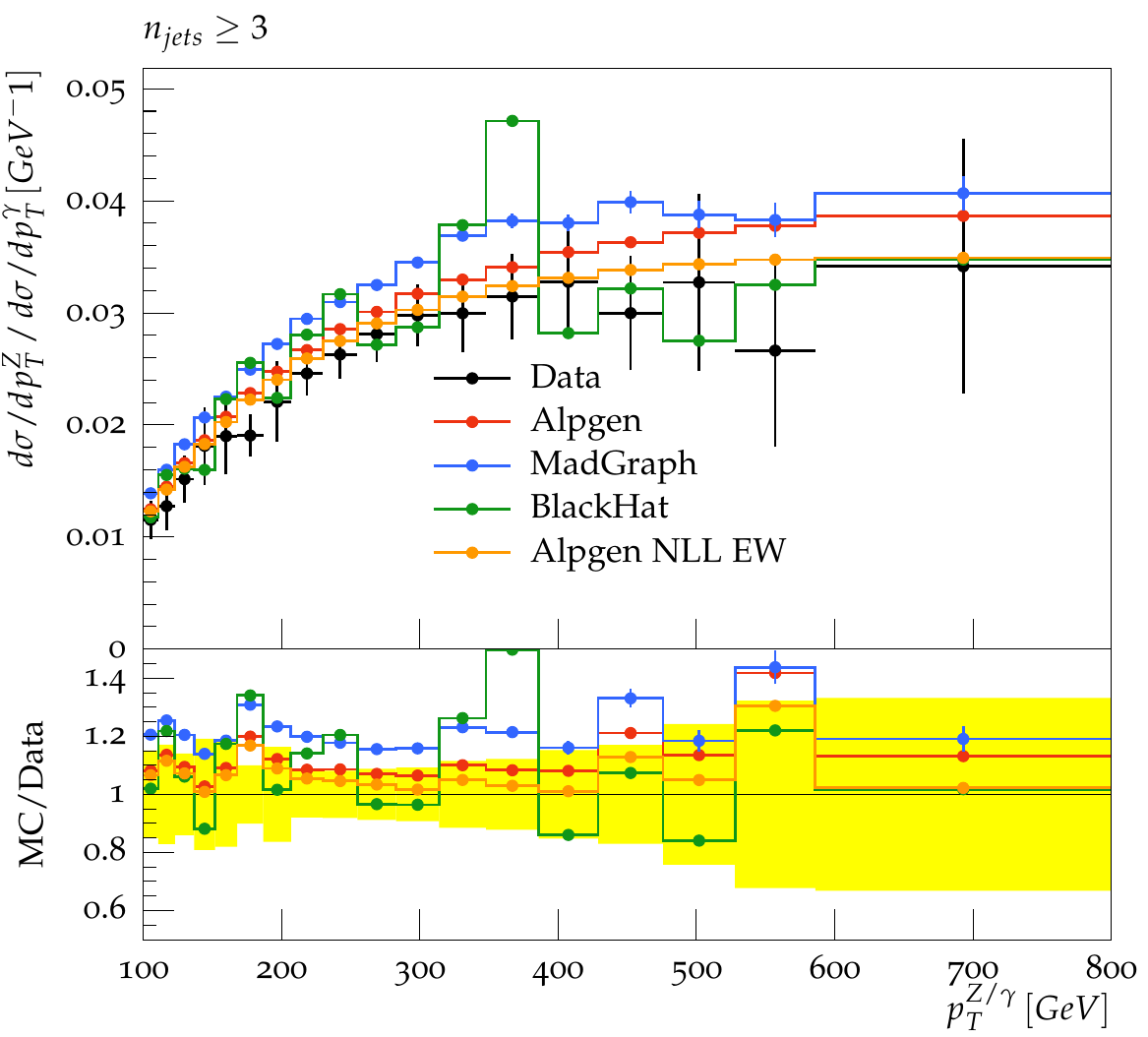}
  \caption{Comparison of different predictions, among them (left) and
    to CMS data (right), for the ratio of
    $\PZ+{}$jets over $\gamma+{}$jets at $8\UTeV$ pp 
    center-of-mass energy. From top to bottom, results are shown 
    for events with at least 1, 2 and 3 jets accompanying the
    boson. For fixed-order predictions this value corresponds to 
    the number of partons in the final state at the lowest order.
    \label{zgr}
  }

\end{figure}

Figure~\ref{zgr} compares several predictions for the ratio within
the fiducial regions as defined above\footnote{We dropped the
  comparison for $H_{\mathrm{T}}>300\UGeV$ because fixed-order
  predictions are known to fail describing high jet activity with a
  comparatively low vector boson $p_{\mathrm{T}}$.}. The fixed-order
partonic predictions computed with the {\tt ALPGEN} generator are
shown at LO and at approximated NLO accuracy~\cite{Chiesa:2013yma},
i.e.\ including the effect of virtual weak corrections in the Sudakov
approximation obtained by means of the algorithm of
Refs.~\cite{Denner:2000jv,Denner:2001gw}, as described in Section
\ref{sec:sudakov;ew}. The predictions for $\PZ+{}$jets and
$\gamma+{}$jets are computed in the $G_{\mu}$ and $\alpha(0)$ schemes,
respectively, with the set of parameters listed above for the
calculation of the $\mathcal{O}(\alpha)$ corrections to the process
$\PW+2\,$jets. The factorization scale is set to $\sum_j
p_{\mathrm{T}}^j +\sqrt{M^2_V +p^2_{\mathrm{T},V}}$. CTEQ5L is used as
PDF set: it is worth noticing, however, that PDFs largely cancel in
the $\PZ/\gamma$ ratio as pointed out in Ref.~\cite{Ask:2011xf}.  More
precisely, the predictions for $\gamma+{}$jets and $\PZ+{}$jets are
computed by using the {\tt phjet} and {\tt zjet} packages,
respectively: at variance with the {\tt vbjet} package, where the
external vector bosons are produced on-shell, in {\tt zjet} the $\PZ$
boson decays in a fermion--antifermion pair including all the off-shell
and spin-correlation effects. These packages include only the QCD
contributions of order $\alpha_s^{n \, {\rm jets}} \alpha$ to the LO
predictions. Though the LO results for $\PZ+{}$jets include exactly
off-shell and spin-correlation effects, the Sudakov corrections are
obtained in the on-shell approximation by using the phase-space
mapping described in Ref.~\cite{Denner:2014ina}.

The other predictions are also shown in the CMS paper, where a
detailed description of the configuration used can be found. For
$\PZ+{}$jets and $\gamma+{}$jets generated with the {\sc
  Madgraph5}~\cite{Alwall:2011uj} program, the LO multiparton
matrix-element calculation includes up to four partons in the final
state. The showering and hadronization, as well as the underlying
event, are modelled by {\sc Pythia6}~\cite{Sjostrand:2006za}. The events are
generated with the CTEQ6L1~\cite{Pumplin:2002vw} parton distribution functions,
and the ktMLM matching scheme~\cite{Alwall:2007fs} with a matching
parameter of $20\UGeV$ is applied.  In addition to these Monte Carlo
signal data sets, a NLO perturbative QCD calculation from the {\sc
  BLACKHAT} Collaboration~\cite{Bern:2012vx} is available for a boson
accompanied by up to three jets. These calculations use
MSTW2008nlo68cl~\cite{Martin:2009iq} with $\alpha_{s}=0.119$ as the PDF set,
and the renormalization and factorization scales are set to
$\mu_{\mathrm{R}}=\mu_{\mathrm{F}}=H_{\mathrm{T}}+E_{\mathrm{T}}^{\mathrm{V}}$, where
$H_{\mathrm{T}}$ is the scalar $p_{\mathrm{T}}$ sum of all outgoing
partons with $p_{\mathrm{T}}>20\UGeV$ and
$E_{\mathrm{T}}^{\mathrm{V}}$ is defined as
$\sqrt{m_{\mathrm{Z}}^{2}+\left(p_{\mathrm{T}}^{\mathrm{Z}}\right)^{2}}$
and $p_{\mathrm{T}}^{\gamma}$, respectively, for $\PZ+{}$jets and
$\gamma+{}$jets.  In addition, the photons must satisfy the Frixione
cone isolation condition~\cite{Frixione:1998jh}.

From the plot it is clear that both NLO QCD and EW corrections are
negative with respect to the fixed-order LO predictions. The NLO QCD
corrections are larger for lower transverse momentum of the bosons,
reaching a 15\% effect for $N_{\mathrm{jets}} \geq 1$. A fraction of
this effect seems to be included by {\sc Madgraph5} predictions, which
include higher-order real-parton emissions in the matrix-element
calculation. The EW corrections increase with the boson transverse
momentum, up to about 10\% for $p_{\mathrm{T}} > 600\UGeV$ in events
with at least one jet. Both QCD and EW corrections decrease for
larger jet multiplicities. It can be also noticed that the {\tt Madgraph}
prediction overshoots the NLO QCD ones for larger multiplicities.

In Figure~\ref{zgr}, these predictions are also compared to CMS
results, showing that the agreement improves when NLO corrections are
included, both in the case of QCD and EW ones. In particular,
including the EW corrections, results are in better agreement in the
region of high boson transverse-momenta, especially for larger jet
multiplicities.

\begin{figure}[t!]
  \includegraphics[width=0.47\textwidth]{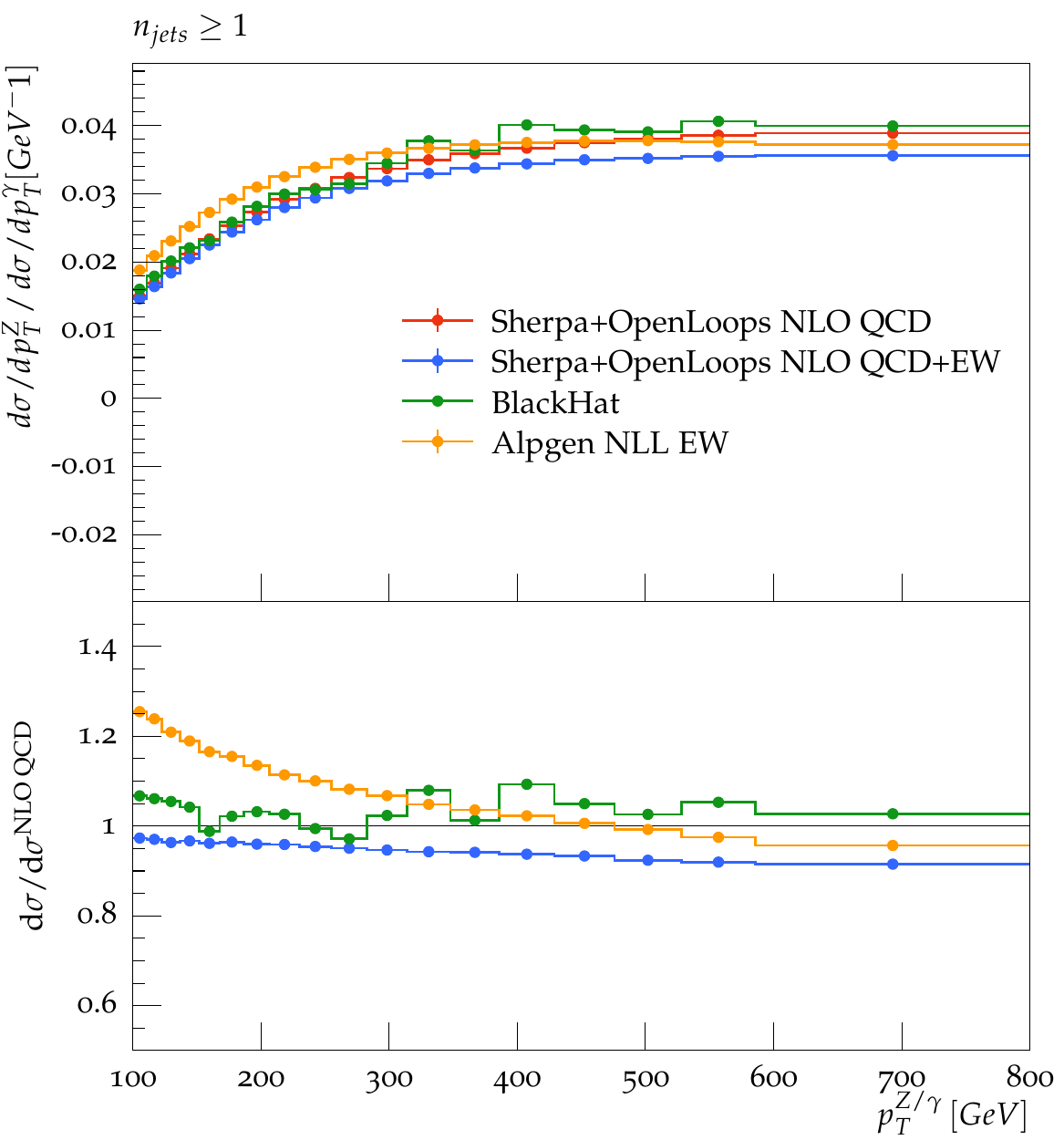}
  \hfill
  \includegraphics[width=0.47\textwidth]{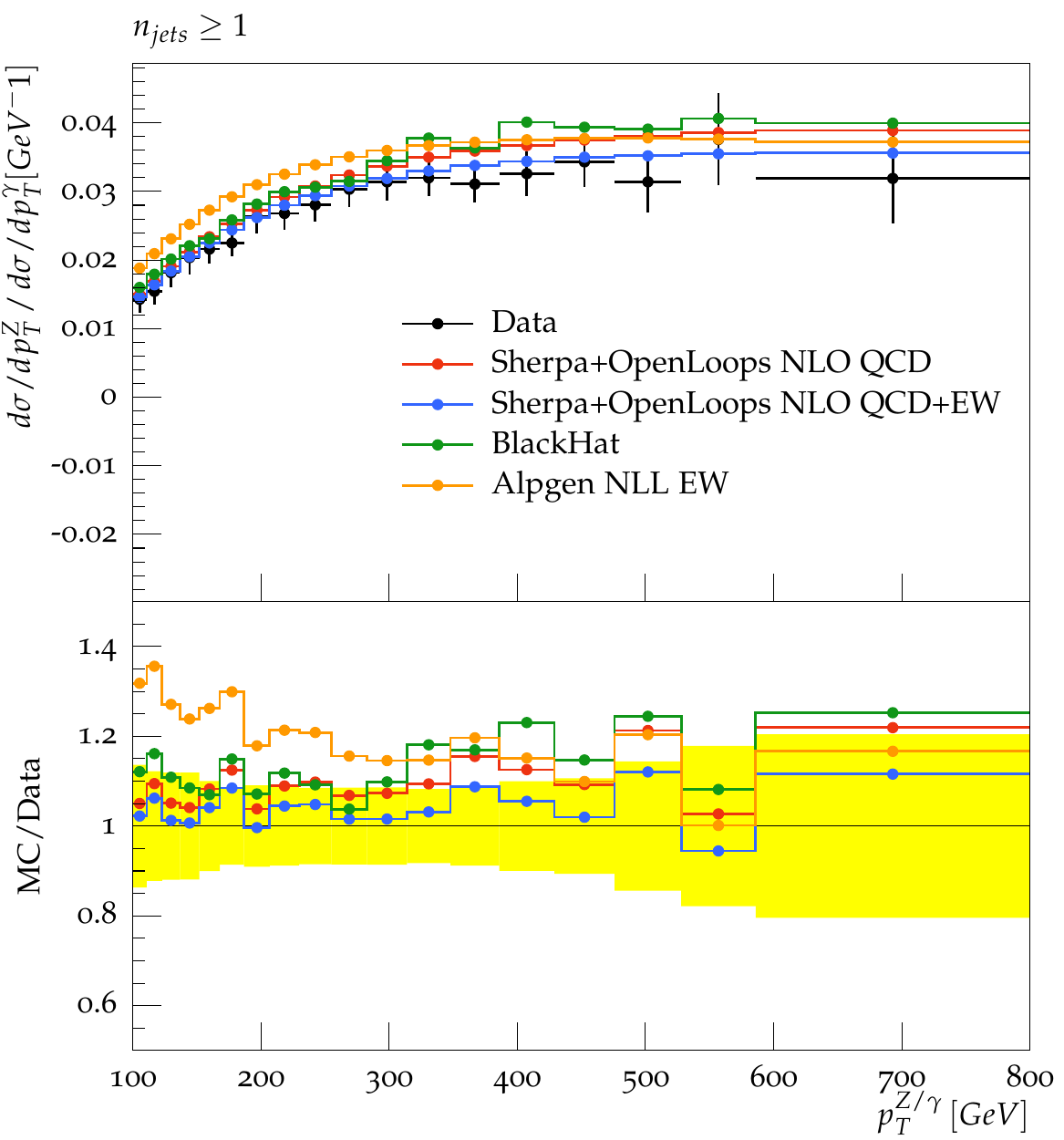}
  \caption{Comparison of different predictions at NLO QCD,  NLL EW and
    NLO QCD+EW order for the  
    ratio of $\PZ+{}$jets over $\gamma+{}$jets at $8\UTeV$ pp
    center-of-mass energy, in events with at least 1 jet accompanying
    the boson. For fixed-order predictions this value corresponds to
    the number of partons in the final state at the lowest order. The
    left plot shows a comparison among the predictions, the right plot
    a comparison to CMS data. \label{zgrNLO}
  }
\end{figure}
Finally, Figure~\ref{zgrNLO} shows in addition, for events with a
vector boson and at least one jet, fixed-order predictions from {\tt
  Sherpa+OpenLoops}. The $\PZ+{}$jets prediction is obtained from an
off-shell calculation for $\Pl^+\Pl^-+{}$jets including all
$\PZ/\gamma^*$ interference effects.  The presented predictions are
based on the recently achieved automation of NLO QCD+EW
calculations~\cite{Kallweit:2014xda,Kallweit:2015dum}, as described in
Section~\ref{sec:comp;ew}. Related predictions for the
$\PZ+{}$jets/$\gamma+{}$jets ratio (with an on-shell $\PZ$ boson) from {\tt
  Munich+OpenLoops} have already been presented
in Ref.~\cite{Kallweit:2015fta} and have for example been employed for
background predictions in Ref.~\cite{CMS:2015jdt}. Here, we employ
NNPDF2.3QED \cite{Ball:2013hta} parton distributions with
$\alpha_s=0.118$, and all input parameters and scale choices are as
detailed in Ref.~\cite{Kallweit:2015dum}. In particular, all unstable
particles are treated in the complex-mass scheme~\cite{Denner:2005fg},
and renormalization and factorization scales are set to
$\mu_{\mathrm{R,F}}=\hat{H}_{\mathrm{T}}'/2$, where $\hat{H}_{\mathrm{T}}'$ is
the scalar sum of the transverse energy of all parton-level
final-state objects, $\hat{H}_{\mathrm{T}}' = \sum_{i\in
  \{\mathrm{quarks,gluons}\}} p_{\mathrm{T},i} + p_{\mathrm{T},\gamma}
+ E_{\mathrm{T}, V}$. QCD partons and photons that are radiated at NLO
are included in $\hat{H}_{\mathrm{T}}'$, and the vector-boson
transverse energy, $E_{\mathrm{T},V}$, is computed using the total
(off-shell) four-momentum of the corresponding (dressed) decay
products, i.e.\
$E^2_{\mathrm{T},\PZ}=p^2_{\mathrm{T},\Pl\Pl}+m_{\Pl\Pl}^2$ and
$E^2_{\mathrm{T},\gamma}=p_{\mathrm{T},\gamma}^2$. The weak coupling
constant $\alpha$ is renormalized in the $G_{\mu}$ scheme for the
$\Pl^+\Pl^-+{}$jets prediction, while the $\alpha(0)$~scheme is used
for the $\gamma+{}$jets prediction. Results are presented at the NLO QCD
level and combining QCD and EW corrections via an additive
prescription, i.e.\ $\sigma^{\mathrm{NLO}}_{\mathrm{QCD+EW}} =
\sigma^{\mathrm{LO}}+\delta\sigma^{\mathrm{NLO}}_{\mathrm{QCD}} +
\delta\sigma^{\mathrm{NLO}}_{\mathrm{EW}}$. Isolated photons in the
$\gamma+{}$jets predictions are required to satisfy Frixione cone
isolation~\cite{Frixione:1998jh} with parameters as specified in
Ref.~\cite{Khachatryan:2015ira}.

The agreement of the combined NLO QCD+EW prediction with the CMS data is
remarkable over the whole spectrum. As already noted, at low transverse momentum
NLO QCD corrections to the ratio are relevant due to mass effects, but sizable
EW corrections due to EW Sudakov logarithms (of different size for the two
processes) alter the shape of the ratio prediction already much below $1\UTeV$.
These results show the importance of combining NLO QCD and EW corrections in a
unified framework.

\subsection{Conclusions}
In this contribution we have performed a comparison of calculations of
EW corrections between different automated codes. While it is
a non-trivial task to precisely adjust the settings of the different
calculations such that they are consistent with each other,  we found a
typical agreement at the level of one per cent. We also compared the
Sudakov approximation to exact NLO calculations. Depending on the
considered process and the considered observable the Sudakov
approximation can describe the complete EW NLO corrections at the
level of one to ten per cent.
When comparing CMS data for the ratio of the associated production of a
$\PZ/\gamma^*$ or an on-shell photon with one or more jets to theoretical
predictions, the inclusion of EW corrections results in a better
agreement for high boson transverse momenta.

\subsection*{Acknowledgements}
The work of A. Denner was supported by the Deutsche Forschungsgemeinschaft
(DFG) under reference number DE~623/2-1. 
The work of D. Pagani is supported  by the ERC grant 291377 ``LHCtheory: Theoretical predictions and analyses of LHC physics:
advancing the precision frontier''.
M. Zaro is supported by the European Union's Horizon 2020 research and
innovation programme under the Marie Sklodovska-Curie grant agreement
No 660171 and in part by the ILP LABEX (ANR-10-LABX-63), in turn
supported by French state funds managed by the ANR within the ``Investissements d'Avenir'' programme under reference
ANR-11-IDEX-0004-02.

\renewcommand{\arraystretch}{1}
\section{A comparative study of Higgs boson plus jets production in
  gluon fusion
  \texorpdfstring{\protect\footnote{
    J.~R.~Andersen, S.~Badger, K.~Becker, J.~Bellm, R.~Boughezal,
    G.~Falmagne, R.~Frederix, M.~Grazzini, N.~Greiner,
    S.~H\"oche, J.~Huston, J.~Isaacson, Y.~Li, X.~Liu, G.~Luisoni,
    F.~Petriello, S.~Pl\"atzer, S.~Prestel, I.~Pogrebnyak,
    P.~Schichtel, M.~Sch\"onherr, P.~Sun, F.~J.~Tackmann, E.~Vryonidou,
    J.~Winter, C.-P.~Yuan, F.~Yuan}}{}}
\label{sec:hjetscomp}

\subsection{Introduction}
\label{sec:hjetscomp:intro}

There has been a great deal of progress in
the attack on the Les Houches precision wish list. Such higher order
calculations are needed for the full exploitation of precision LHC
measurements. However, measurements by the ATLAS and CMS experiments are 
often compared 
to predictions involving parton shower Monte Carlos, often
supplemented with matrix element information at leading order (LO) and
next-to-leading order (NLO). Such frameworks allow for the generation
of fully exclusive final states, often more amenable to comparisons
with experimental data. There are a number of such matrix-element plus
parton-shower (ME+PS) frameworks used by the LHC collider experiments,
as can be seen from the predictions in this section. But the higher
(fixed) order calculations often provide the highest precision. It is
thus important to understand: (1) the degree to which the various
ME+PS predictions agree with each other, (2) how well the ME+PS
predictions agree with fixed-order predictions and (3) the impact of
Sudakov regions\,\footnote{By Sudakov region, we refer to kinematic
situations where there is a severe restriction on phase space for
gluon emission, such as the Higgs boson transverse momentum
distribution at low $p_\perp$.} and/or the imposition of jet
vetoes/binning on both fixed-order and ME+PS predictions.  We come to
these comparisons with several expectations: outside of Sudakov regions, the
influence of parton showering/resummation should be mild, and cross
sections that are fairly inclusive should not be subject to
large jet veto logs. This means that for observables like the leading 
jet's transverse momentum distribution for $h$~+~$\ge1$ jet or the 
inclusive $n$-jet cross sections we do not expect any 
significant resummation correction, originating in either the parton 
showers or the dedicated resummed expressions. On the other hand, 
the more exclusive the cross section, the more different scales are 
involved, the larger the impact of such corrections should be, 
stabilizing the result. Besides the jet vetoed cross sections, 
exclusive $p_\perp$ spectra, both of the Higgs and any jet, should 
be highly dependent on the type of resummation included.

The production of a Higgs boson through gluon--gluon fusion is an
excellent testing ground for such comparisons, because of the
importance of the process, and the enhanced rate for additional jet
production associated with gluon--gluon fusion. First comparisons of 
Higgs production in gluon fusion were done in Les Houches
2013~\cite{AlcarazMaestre:2012vp}, with comparisons of various
predictions for $h$~+~$\ge2$ jets from gluon--gluon fusion, a critical
background for the measurement of vector boson fusion. In the present 
contribution to the Les Houches 2015 proceedings, we extend that study
to more observables, for a variety of jet multiplicities, and with
comparisons to fixed-order predictions as well as to ME+PS frameworks.

To allow for as standardized a comparison as possible, a group of
generation parameters were agreed upon. MMHT2014 NLO PDFs (and the
NNLO version for the NNLO calculations) were to be used, with a
central value of $\alpha_\mathrm{s}(m_Z)=0.118$.  
Variations of scale choice are allowed; however, they
should reproduce a scale of $\tfrac{1}{2}m_h$ in the zero-jet limit. 
All computations were done in the Higgs effective field theory approach 
in the strict $m_\text{top}\to\infty$ limit. Although this does not constitute 
a best-of setup for most contributions these common parameters do not 
alter their underlying methods' properties, capabilites and limits.

Alas, in some cases, there are deviations from these standards. These
will be noted where present, and the impact on the comparisons will be
discussed.  The Higgs boson was left undecayed. Jets were reconstructed 
with the anti-$k_T$ jet clustering algorithm \cite{Cacciari:2008gp} using
$R=0.4$, and a transverse momentum constraint of $30\hjetscompgev$ was imposed,
along with a rapidity cut of $|\eta(j)|<4.4$.  To provide a common
framework for the display of the results, a Rivet
routine~\cite{Buckley:2010ar,webpage} was created and distributed to each group
providing a prediction.

In this contribution, predictions have been made with fixed-order
calculations at NLO for $pp\to h+1,2,3j$ with standard scale choices 
and the \hjetscompMinlo approach including Sudakov factors (using \hjetscompGoSam in 
comnbination with \hjetscompSherpa, cf.\ Secs.\ 
\ref{sec:hjetscomp:tools:fo:hnj}-\ref{sec:hjetscomp:tools:fo:hnjminlo}), 
at approximate NNLO (using the \hjetscompLoopsim approach, cf. Sec.\ 
\ref{sec:hjetscomp:tools:fo:hnjloopsim}) and full NNLO for $pp\to h$ 
(using \hjetscompSherpa, cf.\ Sec.\ \ref{sec:hjetscomp:tools:fo:sherpa}) and 
$pp\to h+j$ (using the results of the BFGLP group, cf.\ 
\ref{sec:hjetscomp:tools:fo:BFGLP}). These are compared to explicit 
high-precision resummation calculations for observables of interest, 
i.e.\ the inclusive Higgs boson transverse momentum (using \hjetscompHqT and \hjetscompResbos, cf.\ 
Secs.\ \ref{sec:hjetscomp:tools:ares:hqt} and 
\ref{sec:hjetscomp:tools:ares:resbos}), and the leading-jet $p_\perp$ spectrum 
and jet-vetoed zero jet 
cross section (using the results of the STWZ group, cf.\ Sec.\ 
\ref{sec:hjetscomp:tools:ares:jvres}), as well as to the generic 
parton shower matched predictions of inclusive Higgs boson production 
at NNLO (using \hjetscompPowheg and \hjetscompSherpa, cf.\ Secs.\ 
\ref{sec:hjetscomp:tools:nnlops:powheg} and 
\ref{sec:hjetscomp:tools:nnlops:sherpa}) and the multijet merged 
predictions (provided by \hjetscompMGaMC, \hjetscompHerwig and \hjetscompSherpa, cf.\ Secs.\ 
\ref{sec:hjetscomp:tools:mc:mgamc}-\ref{sec:hjetscomp:tools:mc:sherpa}), 
using NLO matrix element information for up to two (three) jets (\hjetscompSherpa). 
For observables requiring the presence of at least two jets, results 
obtained resumming BFKL-type logarithms (using \hjetscompHej, cf.\ Sec.\ 
\ref{sec:hjetscomp:tools:bfkl:hej}) are added. 
Sec.\ \ref{sec:hjetscomp:results} then presents the results in detail 
for a multitude of relevant observables.

\subsection{Tools}
\label{sec:hjetscomp:tools}

\label{sec:hjetscomp:tools:fo}
\subsubsection{NLO calculation of $pp\to h+1,2,3\,\text{jets}$}
\label{sec:hjetscomp:tools:fo:hnj}

We compute $h+1$ jet, $h+2$ jets and
$h+3$ jets at NLO in QCD in the infinite top mass limit 
\cite{vanDeurzen:2013rv,Cullen:2013saa,Greiner:2015jha}
using \hjetscompSherpa~\cite{Gleisberg:2008ta} and
\hjetscompGoSam~\cite{Cullen:2011ac,Cullen:2014yla}, linked via the
interface defined in the Binoth Les Houches
Accord~\cite{Binoth:2010xt,Alioli:2013nda}.
The one-loop amplitudes are generated with \hjetscompGoSam employing
\textsc{QGraf}~\cite{Nogueira:1991ex},
\textsc{Form}~\cite{Vermaseren:2000nd,Kuipers:2012rf} and
\textsc{Spinney}~\cite{Cullen:2010jv}, The reduction of the loop
integrals is performed using
\textsc{Ninja}~\cite{Mastrolia:2012bu,vanDeurzen:2013saa,Peraro:2014cba},
\textsc{Golem95}~\cite{Heinrich:2010ax,Binoth:2008uq,Cullen:2011kv}
and \textsc{OneLoop}~\cite{vanHameren:2010cp} for the evaluation of
the scalar integrals.
The calculation of tree-level matrix elements for the Born and the
real emission contribution as well as the subtraction terms in the
Catani-Seymour approach~\cite{Catani:1996vz} have been done within
\hjetscompSherpa using the matrix element generator
\textsc{Comix}~\cite{Gleisberg:2008fv}.

The computation is performed for a Higgs boson with mass
$m_h=125\,\hjetscompgev$, which is left undecayed. We used the 
\texttt{MMHT2014nlo68clas118} PDF set. We present results 
obtained by processing events pre-generated and
stored in form of \textsc{Root} Ntuples as described
in~\cite{Bern:2013zja}. Theoretical
uncertainties are estimated by varying renormalization and
factorization scales by factors of $\tfrac{1}{2}$ and $2$ 
around the central scale
\begin{equation}
  \mu_0 \;=\; \tfrac{1}{2}\,\sqrt{m_{h}^2+\sum p_{T,j_i}^2}\;,
\end{equation}
where $i$ runs over all identified jets.
This scale was chosen to facilitate comparison with the $h+1$ jet NNLO
calculation of Sec.\ \ref{sec:hjetscomp:tools:fo:BFGLP}.

\subsubsection{\hjetscompMinlo calculation of $pp\to h+1,2,3\,\text{jets}$}
\label{sec:hjetscomp:tools:fo:hnjminlo}

Reprocessing the \textsc{Root} Ntuples of Sec.\ 
\ref{sec:hjetscomp:tools:fo:hnj}, we present fixed order NLO results evaluated 
with a \hjetscompMinlo~\cite{Hamilton:2012np} scale chioce, as implemented in
\hjetscompSherpa, for $pp\to h+1,2\,\text{jets}$ and, for the first time, 
for $pp\to h+3\,\text{jets}$. Events are read-in by \hjetscompSherpa, which applies
the \hjetscompMinlo prescription event by event. As a \hjetscompMinlo
core scale we choose
\begin{equation} \label{eq:hthatprime}
  \mu_\text{core}^\text{\textsc{MiNLO}}
  \;=\;\tfrac{1}{2}\,\hat{H}^\prime_T
  \;=\;\tfrac{1}{2}\left(\sqrt{m_h^{2}+p_{T,h}^{2}}
       +\sum_{i}p_{T,i}^{}\right)\,
\end{equation}
where $i$ runs over all partons of the identified core.

\subsubsection{\hjetscompLoopsim merged nNLO calculation of 
               $pp\to h+(1,2)\,\text{jets}$ and $pp\to h+(2,3)\,\text{jets}$}
\label{sec:hjetscomp:tools:fo:hnjloopsim}

The fixed order Ntuples produced by \hjetscompGoSam{}+\hjetscompSherpa 
detailed in Sec.\ \ref{sec:hjetscomp:tools:fo:hnj} can be combined 
using the \hjetscompLoopsim \cite{Rubin:2010xp} procedure to make an approximate 
NNLO prediction, labelled nNLO in the following, which is missing the 
double virtual corrections but captures much of the double unresolved 
radiation contributions. There is a cut-off dependence on the additional 
real radiation since the fixed order Ntuples where generated with a jet 
$p_T>25$ GeV. The \hjetscompLoopsim procedure uses a flavour sensitive 
$k_T$-algorithm where a jet radius of $R=1$ was used. All other 
parameters and scales were chosen the same as in the fixed order case 
with the exception of the PDF where the NNLO PDF set 
\texttt{MMHT2014nnlo68cl} was used.

\subsubsection{NNLO calculation of $pp\to h$}
\label{sec:hjetscomp:tools:fo:sherpa}

For the fully differential calculation of inclusive Higgs production 
at NNLO accuracy we use the implementation in the \hjetscompSherpa Monte Carlo 
event generator presented in \cite{Gleisberg:2008ta,Hoche:2014dla}, in 
this case without matching it to the parton shower. This calculation is 
performed using the $q_\text{T}$-slicing technique and was extensively 
checked against \hjetscompHNNLO \cite{Catani:2007vq}. For this study here we set 
$\mu_R=\mu_F=\tfrac{1}{2}\,m_h$ and use the \texttt{MMHT2014nnlo68cl} 
PDF set \cite{Harland-Lang:2014zoa}.

\subsubsection{NNLO calculation of $pp\to h+j$}
\label{sec:hjetscomp:tools:fo:BFGLP}

The NNLO calculation for Higgs production in association with one or 
more jets is obtained using the $N$-jettiness subtraction 
scheme~\cite{Boughezal:2015dva,Gaunt:2015pea}.  The application of 
this technique to obtain the full NNLO result for Higgs+jet, and its 
validation against another calculation using sector-improved residue 
subtraction~\cite{Boughezal:2015dra}, was described in 
\cite{Boughezal:2015aha}.  In this study we adapt the dynamical scale choice
\begin{equation}\label{eq:bfglpScale}
  \mu_0\;=\;\tfrac{1}{2}\,\sqrt{m_{h}^2+\sum p_{T,j_i}^2}
\end{equation}
for both the renormalization and factorization scales, where the sum 
runs over all reconstructed jets.  This dynamical scale correctly 
captures the characteristic energy throughout the entire kinematic 
range.  To estimate the theoretical uncertainties we equate $\mu_R$ 
and $\mu_F$ and vary them in the range $\tfrac{1}{2}\mu_0 \leq \mu_{R,F} \leq 2 \mu_0$.  
We use \texttt{CT14nnlo} parton distribution functions 
\cite{Dulat:2015mca}. The effects of the use of this PDF, as compared to the nominal PDF, should be 
minimal for most of the kinematic regions studied in this contribution. Jets are reconstructed using the anti-$k_T$ algorithm 
with $R=0.4$, subject to the requirements $p_T>30$ GeV and $|\eta|<4.4$.

\label{sec:hjetscomp:tools:ares}
\subsubsection{\hjetscompHqT}
\label{sec:hjetscomp:tools:ares:hqt}

\hjetscompHqT \cite{Bozzi:2005wk,deFlorian:2011xf} is a numerical program which 
combines the exact fixed order calculation of the transverse momentum 
spectrum valid at large $q_\text{T}$ at ${\cal O}(\alpha_\mathrm{s}^4)$ with the 
resummation of the logarithmically enhanced contributions at small 
transverse momenta at next-to-next-to-leading logarithmic accuracy.
The calculation is performed according to the formalism of 
\cite{Catani:2000vq,Bozzi:2005wk} and it implements a unitarity 
constraint such that, upon integration over $q_\text{T}$, the inclusive 
NNLO cross section is recovered. The results are valid in the large 
$m_{top}$ approximation. As any perturbative QCD computation in hadron 
collisions, the results depend on the factorization ($\mu_F$) and 
renormalization ($\mu_R$) scales. In addition, the resummation procedure 
introduces an additional scale, dubbed \lq\lq resummation scale\rq\rq\ 
($Q$). The three scales must be chosen of the order of the hard scale 
of the process, $m_h$. The numerical results presented here are obtained 
by using \hjetscompHqT-2.0 with $\mu_F=\mu_R=Q=\tfrac{1}{2}\,m_h$ GeV as central scale choice. 
The procedure to estimate perturbative uncertainties is to perform 
independent variations of $\mu_F$, $\mu_R$ and $Q$ around the central 
value by a factor of 2 with the constraints $\tfrac{1}{2} < \mu_F/\mu_R < 2$ and 
$\tfrac{1}{2} <Q/\mu_R<2$.
\subsubsection{\hjetscompResbos}
\label{sec:hjetscomp:tools:ares:resbos}

\hjetscompResbos is the updated version of \textsc{ResBos} \cite{Balazs:1997xd}, 
which resums the effect of multiple soft gluon radiation to all orders 
in the strong coupling via the $q_\text{T}$-resummation formalism proposed by 
Collins-Soper-Sterman \cite{Collins:1984kg}. The \hjetscompResbos prediction 
for inclusive Higgs boson production via the gluon fusion process 
includes the full NNLO QCD corrections in the total rate (similar to 
the setup of a \hjetscompHNNLO calculation) and 
the NLO contribution to the Higgs boson distribution at large transverse 
momentum ($q_\text{T}$) \cite{Glosser:2002gm}. In this calculation, we include 
NNLL accuracy in the Sudakov factor and NNLO accuracy in the Wilson coefficient functions 
of the resummed piece, which is matched to the perturbative piece, 
evaluated at the NLO, in the high $q_\text{T}$ region \cite{Wang:2012xs}. The 
renormalization and resummation scales have been varied by a factor 
of 2 around the central value of $\tfrac{1}{2}m_h$, with the 
non-perturbative parameters needed for the $q_\text{T}$ resummation calculation 
taken to be the same values as those in \cite{Wang:2012xs}. The 
\texttt{MMHT2014nnlo68cl} PDF set has been used.

The \hjetscompResbos prediction of inclusive Higgs plus jet production via 
gluon fusion is based on the transverse momentum dependent (TMD) 
resummation formalism, in the Collins 2011 scheme \cite{Collins:2011zzd}.
The prediction includes the NLL Sudakov factor and the NLO inclusive rate in the 
resummed piece which is matched to NLO QCD prediction in the large 
$q_\text{T}$ region \cite{Sun:2016kkh}. Here, $q_\text{T}$ denotes the transverse 
momentum of the Higgs boson and jet system. Thus, the total 
inclusive rate of Higgs plus jet production, applying a minimum cut on the jet transverse 
momentum, will agree well with that given by the NLO calculation, such 
as produced by \textsc{MCFM} \cite{Campbell:2010ff}. 
The resummation calculation provides a better description for kinematic distributions 
of Higgs boson and jet when they are almost back-to-back. In this 
calculation, we have fixed the resummation scale to be the jet 
transverse momentum ($p_\perp(j)$), as suggested in \cite{Sun:2016kkh}, 
and have varied the renormalization scale by a factor of 2 around its 
central value $\tfrac{1}{2}m_h$, with the non-perturbative parameters 
needed for the TMD resummation calculation taken to be the same values 
as those in \cite{Sun:2016kkh}. The jets are defined based on the \hjetscompantikt 
algorithm using $R = 0.4$; they furthermore have to obey the criteria that 
$p_\perp(j) > 30$ GeV and its rapidity $|\eta(j)| < 4.4$. The 
\texttt{MMHT2014nnlo68cl} PDF set has been used.

\subsubsection{Jet veto resummation}
\label{sec:hjetscomp:tools:ares:jvres}

To resum the $p_\perp$ spectrum of the leading jet as well as the exclusive $0$-jet
(jet-vetoed) cross section, the STWZ predictions~\cite{Stewart:2013faa} utilize
the soft-collinear effective theory (SCET) formalism for jet-veto resummation at
hadron colliders as developed in \cite{Stewart:2009yx, Berger:2010xi,
Tackmann:2012bt, Stewart:2013faa}. The results are obtained in the HEFT 
limit, taking $m_\text{top}\to\infty$.
Jets are defined with a jet radius of $R = 0.4$ and without any cut on the jet
rapidity. The \texttt{MMHT2014nnlo68cl} PDFs are used. The calculation is carried out to
NNLL$'+$NNLO order. The resummation is performed by renormalization group
evolution in virtuality and rapidity space in SCET. The NNLL$'$ resummation
includes the RG evolution at next-to-next-to-leading logarithmic order together
with the full NNLO singular matching correction, thus incorporating all two-loop
virtual and singular real-emission contributions in the resummation. This allows
to perform the matching to the full NNLO result by adding purely nonsingular
corrections to the resummed contributions, and to achieve a smooth transition to
the fixed-order result simply by turning off the RG evolution using profile
scales~\cite{Ligeti:2008ac, Abbate:2010xh}. The perturbative uncertainties are
estimated by varying the relevant virtuality and rapidity renormalization scales
using profile scale variations, which has been established as a reliable method
to assess perturbative uncertainties in resummed predictions. We evaluate
separate fixed-order and resummation uncertainties which are added in
quadrature~\cite{Berger:2010xi, Stewart:2011cf, Stewart:2013faa}.
The predictions use a complex value for the hard scale $\mu_H = -i \mu_{FO}$ where
$\mu_{FO} = m_h$ is the fixed-order scale, which allows to resum large virtual
corrections in the $gg\to h$ form factor in both the 0-jet limit and the
inclusive cross section. This scale choice results in a similar 
inclusive cross section compared to a standard NNLO calculation with 
$\mu_R=\mu_F=\tfrac{1}{2}m_h$. The uncertainty related to this resummation is
estimated by varying the phase of $\mu_H$ and is also added in quadrature.

\label{sec:hjetscomp:tools:nnlops}
\subsubsection{\hjetscompPowheg \hjetscompNNLOPS}
\label{sec:hjetscomp:tools:nnlops:powheg}

One of the generators under investigation is the \hjetscompNNLOPS prediction 
of Higgs boson production via gluon fusion~\cite{Hamilton:2013fea}.
To produce this sample, the Higgs-boson-plus-jet \hjetscompMinlo~\cite{Hamilton:2012np} 
predictions generated with \hjetscompPowhegBox{}\texttt{-v2}~\cite{Campbell:2012am} 
are combined with an NNLO accurate analytical calculation by the 
program \hjetscompHNNLO (v.2.0)~\cite{Catani:2007vq,Grazzini:2008tf,Grazzini:2013mca} 
following the instructions in \cite{Hamilton:2013fea}. Here, the rapidity 
spectrum of the Higgs boson of the \hjetscompPowheg \texttt{HJ} \hjetscompMinlo prediction 
is reweighted with the spectrum from \hjetscompHNNLO, while ensuring that NLO 
precision is kept for the kinematic properties of the first additional jet. 
In the generation and the \hjetscompHNNLO calculation, the default parameters are 
set as follows, in particular the renormalization and factorization scales 
have been set to $\mu_R = \mu_F = \tfrac{1}{2}\,m_h$ and the 
\texttt{MMHT2014nnlo68cl} \cite{Harland-Lang:2014zoa} has been used. 
To evaluate the uncertainties in the choice of the renormalization and 
factorization scales, the 21-point scale variations described in detail 
in \cite{Hamilton:2013fea} are evaluated.
The \hjetscompNNLOPS sample is showered with \hjetscompPythia~8 (v.8.253) \cite{Sjostrand:2014zea} 
using the \texttt{NNPDF} \texttt{2.3} leading-order PDF set~\cite{Ball:2012cx} 
and the A14 tune~\cite{ATL-PHYS-PUB-2014-021}. 
The variation of the renormalization scale also varies consistently the 
resummation scale in the Sudakov form factor of the \hjetscompMinlo 
procedure~\cite{Hamilton:2012rf}. Other resummation properties are not 
varied in this simulation.

\subsubsection{\hjetscompSherpa \hjetscompNNLOPS}
\label{sec:hjetscomp:tools:nnlops:sherpa}

A second NNLO prediction matched to a parton shower is investigated 
in this study. The computation is performed in the \textsc{uN$^2$loPs} 
method~\cite{Hoeche:2014aia,Hoche:2014dla}, which is based on the 
\textsc{uNloPs} algorithm for merging inclusive NLO matched calculations 
of varying jet multiplicity, while leaving their total cross sections 
unchanged~\cite{Lonnblad:2012ix,Lonnblad:2012ng}. The event samples 
used in this comparison were generated with an implementation of the 
\textsc{uN$^2$loPs} matching scheme in the event generator \hjetscompSherpa. The 
NNLO prediction needed for the matching is computed within \hjetscompSherpa itself, 
using a $q_\text{T}$ cutoff method~\cite{Gao:2012ja}, for details see 
\cite{Hoeche:2014aia,Hoche:2014dla}. We used a parton shower based 
on Catani-Seymour dipole factorization~\cite{Schumann:2007mg,Hoeche:2009xc},
which has been matched to the Higgs plus jet NLO calculation using 
the \hjetscompSMCatNLO method~\cite{Hoeche:2011fd,Hoeche:2012ft}. Renormalization 
and factorization scales for the fixed-order calculation have been 
set to $\mu_R = \mu_F = \tfrac{1}{2}\,m_h$, and the the parton shower 
starting scale is set to the same value. The \texttt{MMHT2014nnlo68cl} 
PDF set \cite{Harland-Lang:2014zoa} has been used throughout.

\label{sec:hjetscomp:tools:mc}
\subsubsection{\hjetscompMGaMC}
\label{sec:hjetscomp:tools:mc:mgamc}

Higgs production in gluon fusion, in association with multiple jets,
is generated with \hjetscompMGaMC \cite{Alwall:2014hca}~at NLO
accuracy using the following commands
\begin{verbatim}
  import model HC_NLO_X0_UFO-heft
  generate p p > x0 /t [QCD] @0
  add process p p > x0 j /t [QCD] @1
  add process p p > x0 j j /t [QCD] @2
  output MG5aMC_FxFx_Hjets
\end{verbatim}
The first command loads the model that includes the Higgs boson
effective coupling to gluons in the $m_t\to\infty$ limit. This model
can be found on the
FeynRules~\cite{Alloul:2013bka}~website\footnote{\texttt{http://feynrules.irmp.ucl.ac.be}}
and was originally created for the studies in
\cite{Demartin:2014fia}. In this model, the Standard Model Higgs
boson is called `\texttt{x0}', and therefore the second to fourth
commands generate this boson in association with 0, 1 and 2 jets,
respectively. The model, and therefore also the definitions of
\texttt{p} and \texttt{j}, are in the 5 flavour scheme. The
`\texttt{/t}' syntax is needed to remove the explicit top quark
contributions from the loops (they are already integrated out in the
effective Higgs-gluon vertices). By setting the parameter
\texttt{ickkw} to \texttt{3} in the \texttt{run\_card.dat}, the FxFx
merging~\cite{Frederix:2012ps}~is turned on. The LHE events are
matched to the \hjetscompPythia 8 (v.210) parton
shower~\cite{Sjostrand:2014zea}, using the FxFx interface also used in
\cite{Frederix:2015eii}. As central choices for the factorisation
and renormalisation scales we use the default value in the
\hjetscompMGaMC code, which, in the context of FxFx merging, is
roughly given by the transverse energy of the Higgs boson, after the
partons entering the matrix elements have been clustered to a $pp \to h
j$ configuration. $pp\to h$ configurations thus are calculated using 
$m_h$ as the scale. The central merging scale is taken to be $35\,\hjetscompgev$,
while the variations include $25\,\hjetscompgev$ and $50\,\hjetscompgev$. These scales include 
values both below as well as above the default minimal jet transverse
momentum used in the analysis and should therefore cover the complete
range of uncertainties coming from the merging.  The uncertainty band
is computed by varying the factorisation and renormalisation scales by
a factor of 2 up and down from the central value, using the reweighting
technique as described in \cite{Frederix:2011ss}, for each of the
three choices of merging scales. It is therefore obtained
as the bin-by-bin envelope of $3 \times 3 \times 3 = 27$ individual
values. This is similar to what is done in
\cite{Frederix:2015eii}---apart from the uncertainties coming
from the parton distribution function, which are not taking into
account here. Throughout the \texttt{MMHT2014nlo68cl} PDF set with 
$\alpha_\mathrm{s}(m_Z)=0.12$ has been used, as compared to the canonical choice
of $0.118$. There is a partial cancellation between the effects of the larger scale choice
and the larger  value of $\alpha_\mathrm{s}(m_Z)$, but for most of the comparisons, 
when there are notable differences, they can be traced to the larger scale choice of \hjetscompMGaMC. 

\subsubsection{\hjetscompHerwig}
\label{sec:hjetscomp:tools:mc:herwig}

We provide predictions from NLO merging of $h+0,1,2$ jets at NLO and $h+3,4$
jets at LO in the Higgs effective theory. The merging is carried out with the
\hjetscompHerwig \ \cite{Bellm:2015jjp} dipole shower module based on
\cite{Platzer:2009jq,Platzer:2011bc}, in a modified version of the algorithms
set out in \cite{Platzer:2012bs,Lonnblad:2012ix}, and implemented in
\cite{Bellm:thesis,Bellm:2016xxx}. The merging implementation will become
publicly available in \hjetscompHerwig. We use \hjetscompMGaMC
\cite{Alwall:2014hca} generated amplitudes together with \textsf{ColorFull}
\cite{Sjodahl:2014opa} for the point-by-point evaluation of tree-level type
objects (tree level matrix elements squared, colour- and spin-correlated
matrix elements), and \textsf{OpenLoops} \cite{Cascioli:2011va} for the
evaluation of Born/one-loop interferences.  Subtraction terms and their
integrated counter-parts, phase space generation, integration and process
bookkeeping is handled by the \textsf{Matchbox} module as outlined in
\cite{Bellm:2015jjp}.

The algorithm we use is a modified, unitarized merging algorithm. We allow
finite, higher-order cross section corrections in higher multiplicity jet
bins, but still choose a unitarization procedure to remove potentially
dangerous, logarithmic enhanced terms in inclusive quantities. Below the
merging scale of $30\,\hjetscompgev$, NLO accuracy of the first additional
emission off each contribution is reached by the standard subtractive
matching. Scales are determined through clustering and the core scale 
is defined as $\mu_R=\mu_F=\tfrac{1}{2}\,m_h$. The shower starting scale 
is set to the same value. The uncertainty band is obtained by variation of 
the renormalization and factorization scales of the hard input processes, 
and is covering all other uncertainties present in the algorithm 
(specifically, merging and shower scale variations).
The \texttt{MMHT2014nlo68clas0118} PDF set \cite{Harland-Lang:2014zoa} 
is used.

\subsubsection{\hjetscompSherpa \hjetscompMEPSatNLO}
\label{sec:hjetscomp:tools:mc:sherpa}

We provide a multijet merged sample of Higgs boson production in 
gluon fusion in assiciation jets wherein the $h+0,1,2,3\,\text{jet}$ 
final states are calculated at NLO accuracy \cite{Gleisberg:2008ta,
  Hoeche:2012yf,Gehrmann:2012yg,Hoeche:2014lxa}. Therein, NLO QCD 
computations are matched to the parton shower using a variant of the 
\hjetscompSMCatNLO method \cite{Hoeche:2011fd,Hoeche:2012ft,Hoche:2012wh}.
The one-loop matrix elements are provided by \hjetscompGoSam \cite{Cullen:2011ac,
  Cullen:2013saa,Greiner:2015jha} and are interfaced using the BLHA 
\cite{Alioli:2013nda} standard.

The individual jet multiplicity contributions are merged using a merging 
cut of $Q_\text{cut}=20\,\hjetscompgev$. The $n$-parton matrix elements are then 
clustered in the scheme of \cite{Hoeche:2009rj,Hoeche:2012yf,Gehrmann:2012yg} 
to find a suitable core configuration and defining $\mu_\text{core}
=\tfrac{1}{2}\,m_h$ for the present study. The factorisation 
and parton shower starting scales are then set directly to 
$\mu_\text{core}$ while the renormalisation scale is determined by 
through CKKW scale setting using the identified emission scales and 
the above defined core scale.
The uncertainties are assessed varying $Q_\text{cut}\in[15,30]\,\hjetscompgev$, 
ensuring that the description of all analysis jets always proceeds 
at NLO accuracy. Renormalisation and factorisation scales are varied 
indepently by a factor of two keeping $\tfrac{1}{2} < \mu_F/\mu_R < 2$. 
The shower starting starting scale is varied by a factor $\sqrt{2}$. 
The \texttt{MMHT2014nlo68clas0118} PDF set \cite{Harland-Lang:2014zoa} 
is used throughout.

\label{sec:hjetscomp:tools:bfkl}
\subsubsection{\hjetscompHej}
\label{sec:hjetscomp:tools:bfkl:hej}

High Energy Jets (\hjetscompHej) describes hard, wide angle (high energy-)
emissions to all orders and to all multiplicities. The predictions are based
on events generated according to an all-order resummation, merged with
high-multiplicity full tree-level matrix-elements. The explicit all-order summation
is built on an approximation to the n-parton hard scattering matrix element
\cite{Andersen:2009nu,Andersen:2009he,Andersen:2011hs} which becomes exact in
the limit of wide-angle emissions, ensuring leading logarithmic accuracy for
both real and virtual corrections. A first set of sub-leading logarithmic
terms are included by allowing one un-ordered gluon emission from quarks. All
of these logarithmic terms are important when
the partonic invariant mass is large compared to the typical transverse
momentum in the event. This is precisely the situation which arises in
typical \lq\lq VBF\rq\rq\ cuts, including those used in this study.  Matching to the
full tree level accuracy for up to three jets is obtained by supplementing
the resummation with a merging
procedure~\cite{Andersen:2008ue,Andersen:2008gc}.

The implementation of this framework in a fully-flexible Monte Carlo event
generator is available at \texttt{http://hej.web.cern.ch}, and produces
exclusive samples for events with at least two jets.  The predictions include
resummation also for events with up to two un-ordered emissions,
i.e.~contributions from the first sub-leading configurations.

The factorisation and renormalisation scales can be chosen arbitrarily, just
as in a standard fixed-order calculation. Here, we have chosen to evaluate
two powers of the strong coupling at a scale given by the Higgs mass, and for
the central predictions the remaining scales are evaluated at $\mu_R=\tfrac{1}{2}H_T$. Thus,
for the $n$-jet tree-level evaluation, 
\begin{equation}
  \alpha_\mathrm{s}^{n+2}=\alpha_\mathrm{s}^2(m_h)\cdot \alpha_\mathrm{s}^n(\mu_R).
\end{equation}
The scale variation bands shown in the plots here correspond to varying
$\mu_F,\mu_R\in \{\tfrac{1}{2}\mu_c,\,\tfrac{1}{\sqrt{2}}\mu_c,\,\mu_c,$ $\,\sqrt{2}\mu_c,\,2\mu_c\}$ with 
$\mu_c=\tfrac{1}{2}H_T$, but discarding evaluations where any ratio $\mu_F/\mu_R$ or 
$\mu_R/\mu_F$ is bigger than two (which results in a total of 18 variations
around the central scale). The \texttt{CT10nlo} \cite{Lai:2010vv,Gao:2013xoa} parton 
distribution functions were used in the predictions.

\subsection{Results}
\label{sec:hjetscomp:results}

In the following comparisons, we show the central
values of each prediction on the left (both as absolute predictions
and in ratio to a reference prediction), and a similar comparison of
the predictions with uncertainty bands on the right. The reason for
the former is that with the overlapping uncertainty bands, it can be
difficult to discriminate the behavior of the central predictions. But
it is also useful to compare the uncertainty bands from each
prediction given similar prescriptions for scale variation.  Note that
it is not enough to say that different predictions agree within their
scale uncertainty bands. In most cases, the predictions should be held
to a higher standard, as the scale logs are common to all of the
calculations that are being compared.

In general, the \hjetscompPowheg \hjetscompNNLOPS calculation has been chosen as the 
reference for the ratios presented. For some observables, other calculations
have been chosen, especially if that reference calculation has the highest
precision. Predictions of similar precision are typically grouped together 
to improve readability. The top ratio plot always compares the \hjetscompNNLOPS 
predictions to the reference. 
The second ratio plot then contains the multijet merged predictions, while 
the third ratio plot has the predictions at pure fixed order. Additional ratios are added as needed. The BFKL resummation 
of \hjetscompHej is inserted into the top ratio plot where applicable.

\subsubsection{Inclusive observables}
\label{sec:hjetscomp:results:inclobs}

This section contains observables which characterize the $h+{}$jets
phase space in the most inclusive way. Only the presence of a Higgs
boson is required, with no restrictions on its momentum. Two important
predictions in this category then are  the rapidity and the transverse momentum
distribution of the Higgs boson. The latter is differentiated into
two cases: the inclusive spectrum of the Higgs boson and the exclusive spectrum, requiring
zero  accompanying jets. Also, to get an overview on the amount
of QCD activity accompanying Higgs boson production, we examine the predictions for 
the inclusive as well as exclusive cross sections for various jet
multiplicities as obtained by the different approaches.

\begin{figure}[t!]
  \centering
  \includegraphics[width=0.47\textwidth]{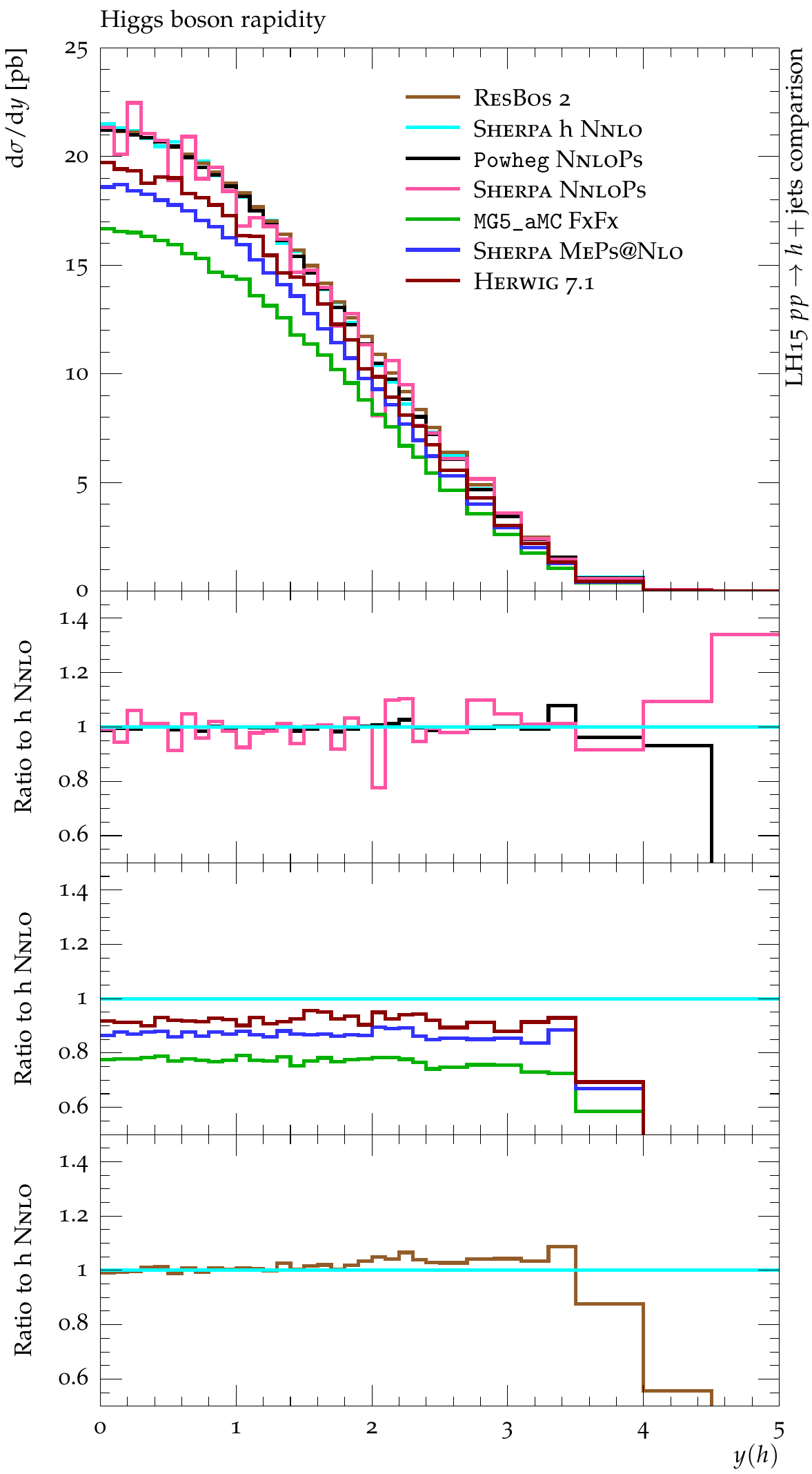}
  \hfill
  \includegraphics[width=0.47\textwidth]{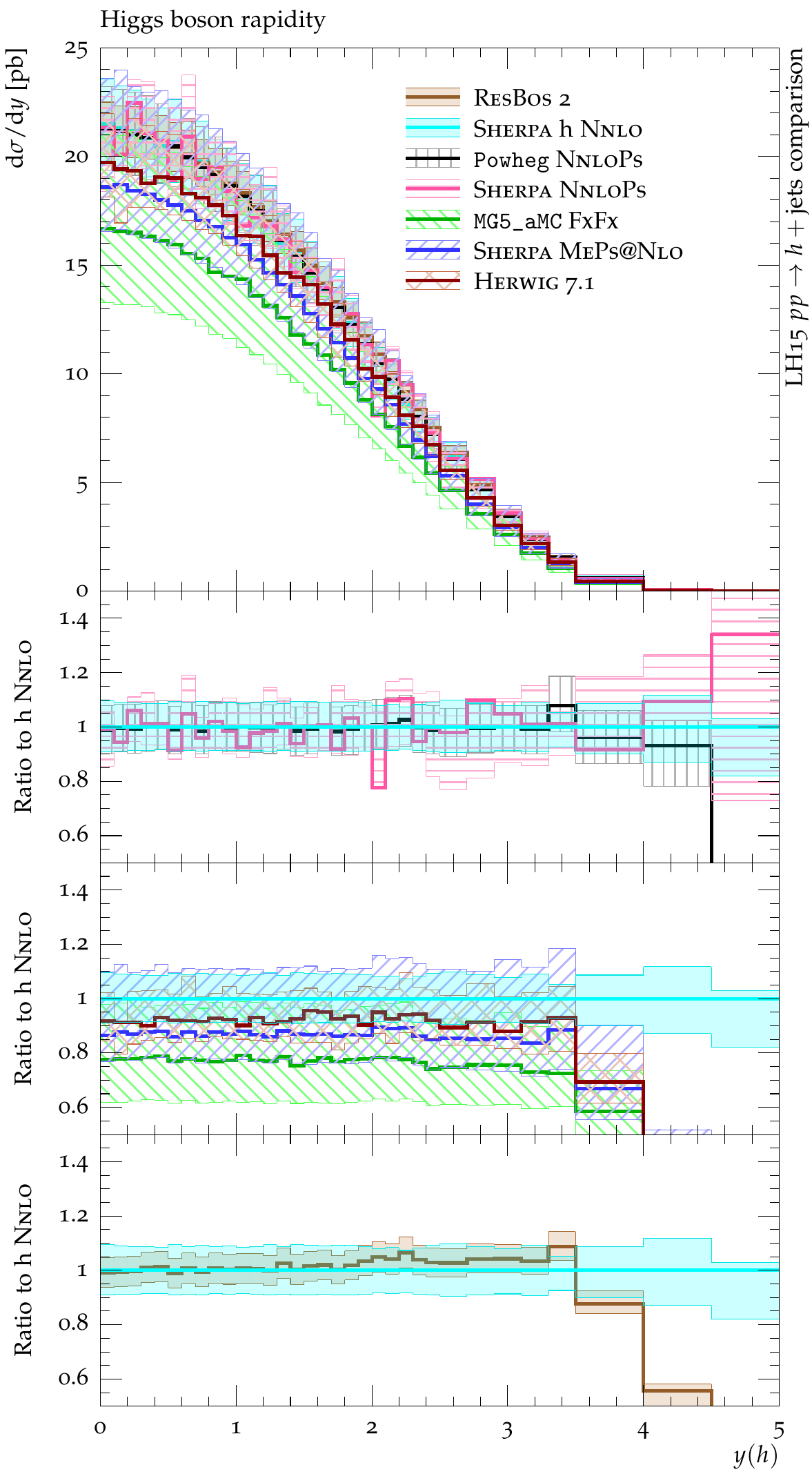}
  \caption{\label{fig:hjetscomp:results:inclobs:hy}%
    The inclusive Higgs boson rapidity without (left) and with (right)
    uncertainties. To enhance visibility, the \hjetscompNNLOPS, multijet merged and
    analytic $q_\text{T}$-resummation predictions are grouped together and
    shown with respect to the same reference curve in the upper, lower
    and middle ratio plots, respectively. The reference prediction is
    taken from the NNLO-accurate description of inclusive $h$ production.}
\end{figure}

We start by examining the inclusive Higgs boson rapidity distribution in
Figure~\ref{fig:hjetscomp:results:inclobs:hy}. While the absolute
predictions are given in the top panel, the plots in the bottom panel
depict the respective ratios to the NNLO prediction. For better
visibility, we have divided the predictions into three groups based on
their simulation type and/or claimed accuracy. The upper ratio plot
contains the NNLO predictions, while the middle ratio plot shows predictions obtained
from different strategies for  merging matrix elements plus parton showers
at NLO. The lower ratio plot displays the comparison between the pure
NNLO prediction and the matched result from \hjetscompResbos that includes the
effects from the $q_\text{T}$-resummation. Overall, we find very good
agreement in the description of the shape of the Higgs boson rapidity
distribution. The main source of deviations stems from the different
normalizations given at NNLO or NLO and the different (core) scale
choices. As expected, the \hjetscompSherpa \hjetscompNNLOPS and
\hjetscompPowheg \hjetscompNNLOPS results agree well with the fixed-order NNLO prediction. 
The larger fluctuations found for \hjetscompSherpa \hjetscompNNLOPS wrt.~\hjetscompPowheg \hjetscompNNLOPS
stem from the fact that the former is computed directly rather than
reweighted from an NLO computation for $hj$ final states.
\hjetscompResbos predicts a slightly smaller rate for $y(h)<2$ and a
slightly larger rate for $2<y(h)<3.5$ than the one given by the NNLO
calculation. The \hjetscompMGaMC, \hjetscompSherpa \hjetscompMEPSatNLO and \hjetscompHerwig have slightly
lower (NLO) normalizations. Here, the \hjetscompMGaMC scale choice reducing to
$m_h$, rather than $\tfrac{1}{2}m_h$, is clearly noticeable. The upper edge of
the \hjetscompMGaMC uncertainty band (equal to a scale that reduces to $\tfrac{1}{2}m_h$)
agrees with the central value of the other NLO ME+PS
predictions. There are no major differences in the size of the
uncertainty envelopes, although to some extent, the NNLO scale
uncertainty bands (including \hjetscompResbos) are smaller than those at NLO,
as expected. Note
that the NLO-based predictions fall off more rapidly at higher $y$
than the NNLO-based predictions do. This is expected because of the
influence of additional $\ln(1-x)$ corrections present in the
determination of NNLO PDFs. Similar effects can be observed in going
from LO to NLO.

\begin{figure}[t!]
  \centering
  \includegraphics[width=0.47\textwidth]{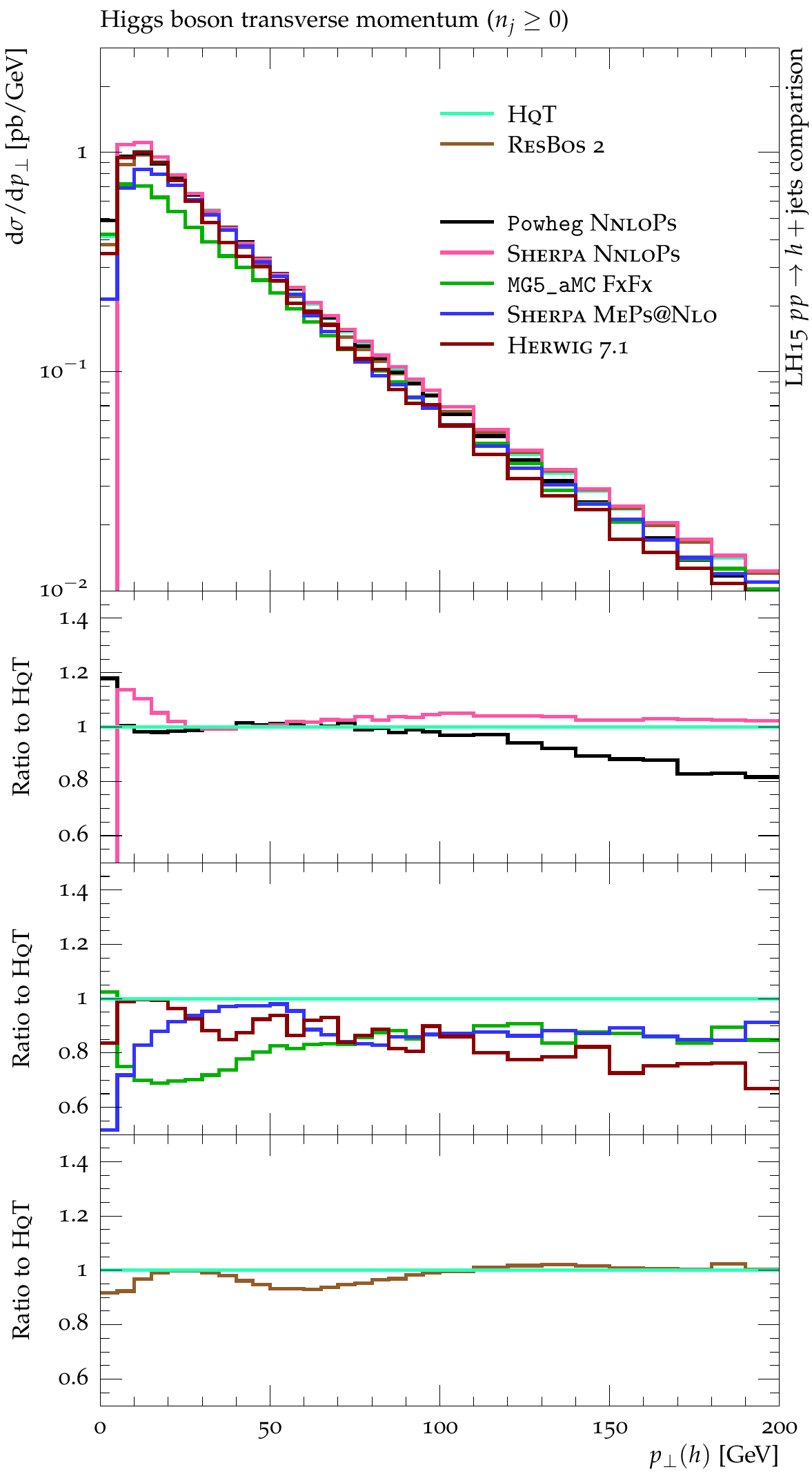}
  \hfill
  \includegraphics[width=0.47\textwidth]{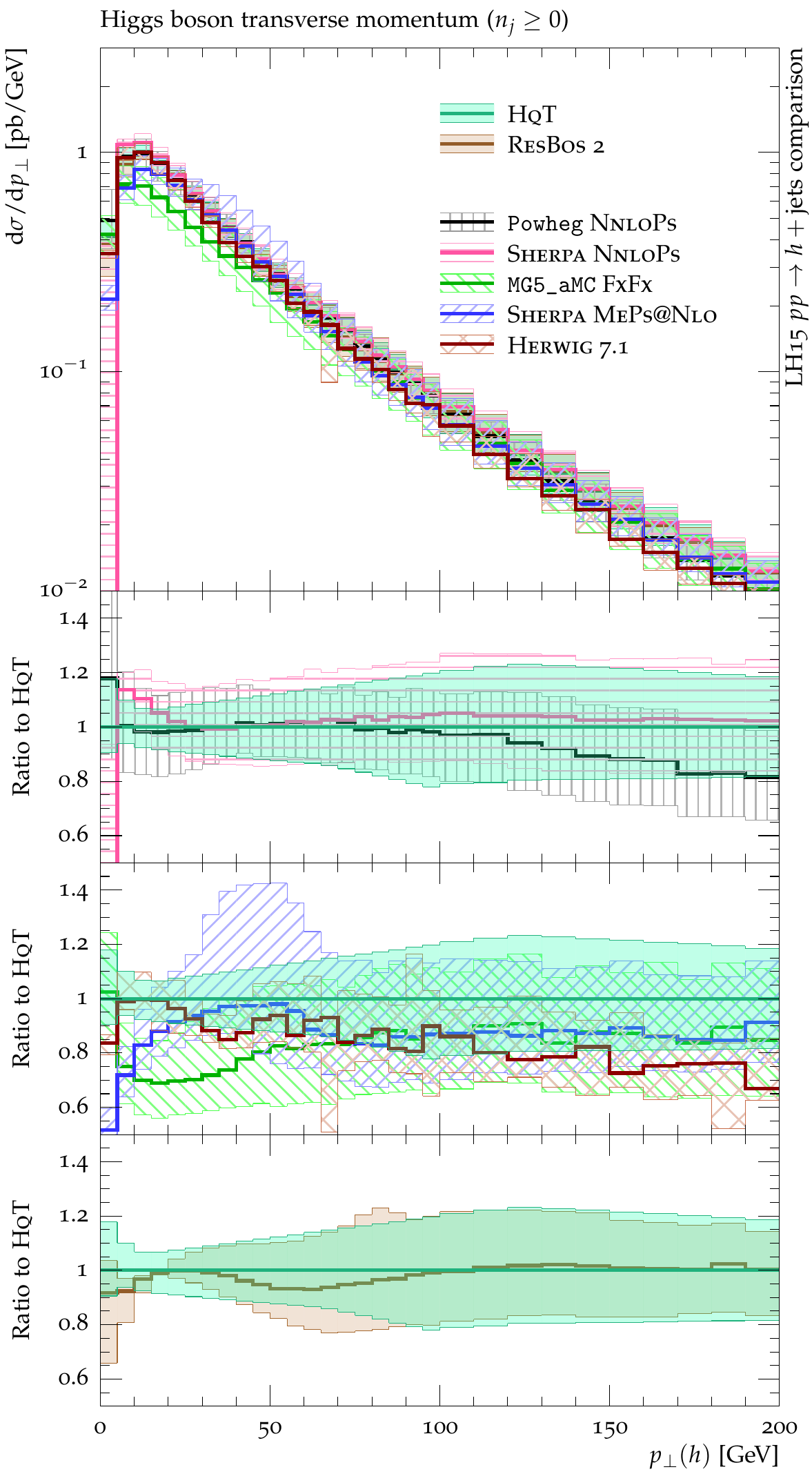}
  \caption{\label{fig:hjetscomp:results:inclobs:hpt}%
    The Higgs boson transverse momentum in the inclusive event
    selection without (left) and with (right) uncertainties. For the
    ratios in the bottom panel, the same grouping strategy has been
    used as in Figure~\ref{fig:hjetscomp:results:inclobs:hy}, while
    the reference prediction has been changed from that of pure NNLO
    to the one as given by \hjetscompHqT.}
\end{figure}

\begin{figure}[t!]
  \centering
  \includegraphics[width=0.47\textwidth]{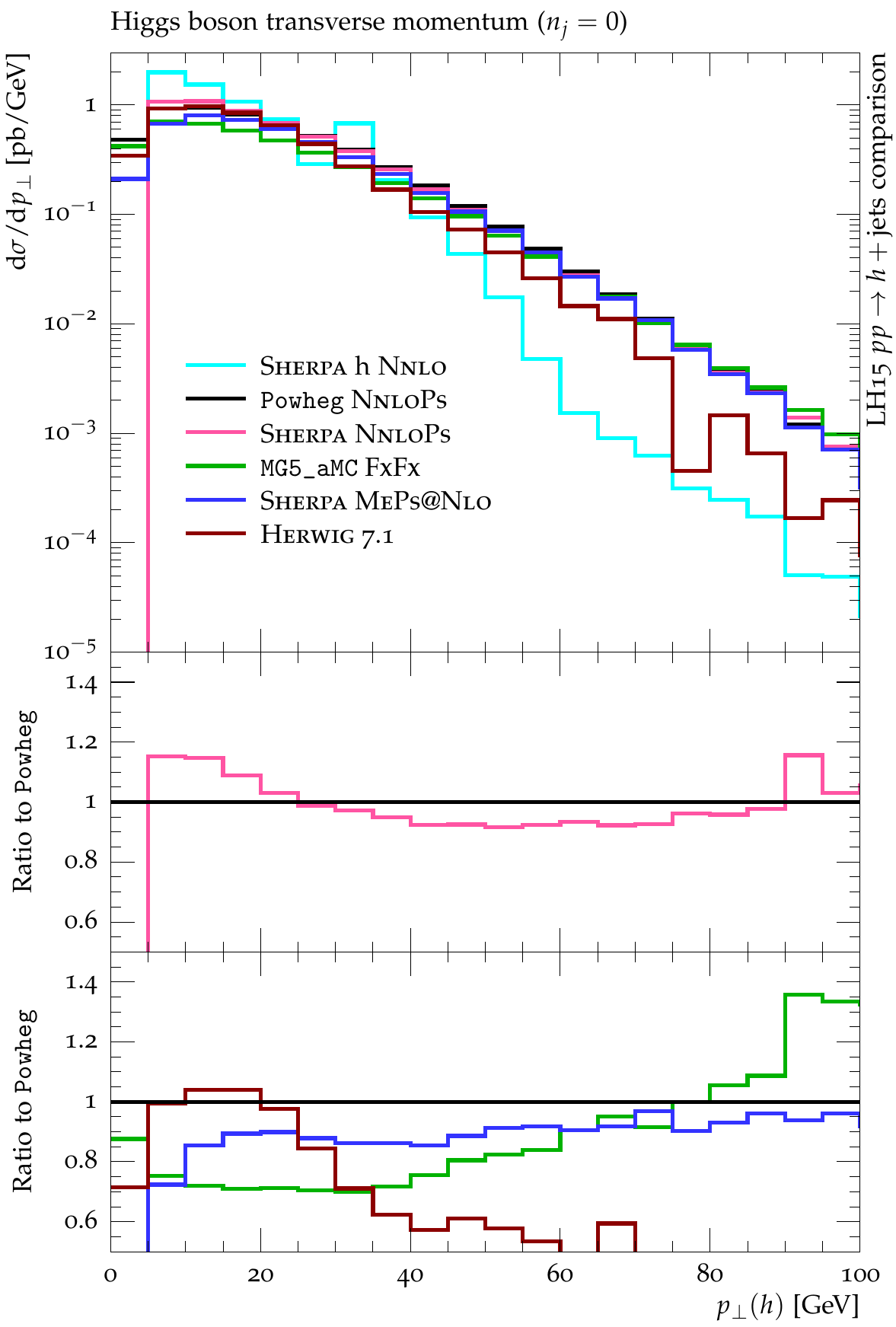}
  \hfill
  \includegraphics[width=0.47\textwidth]{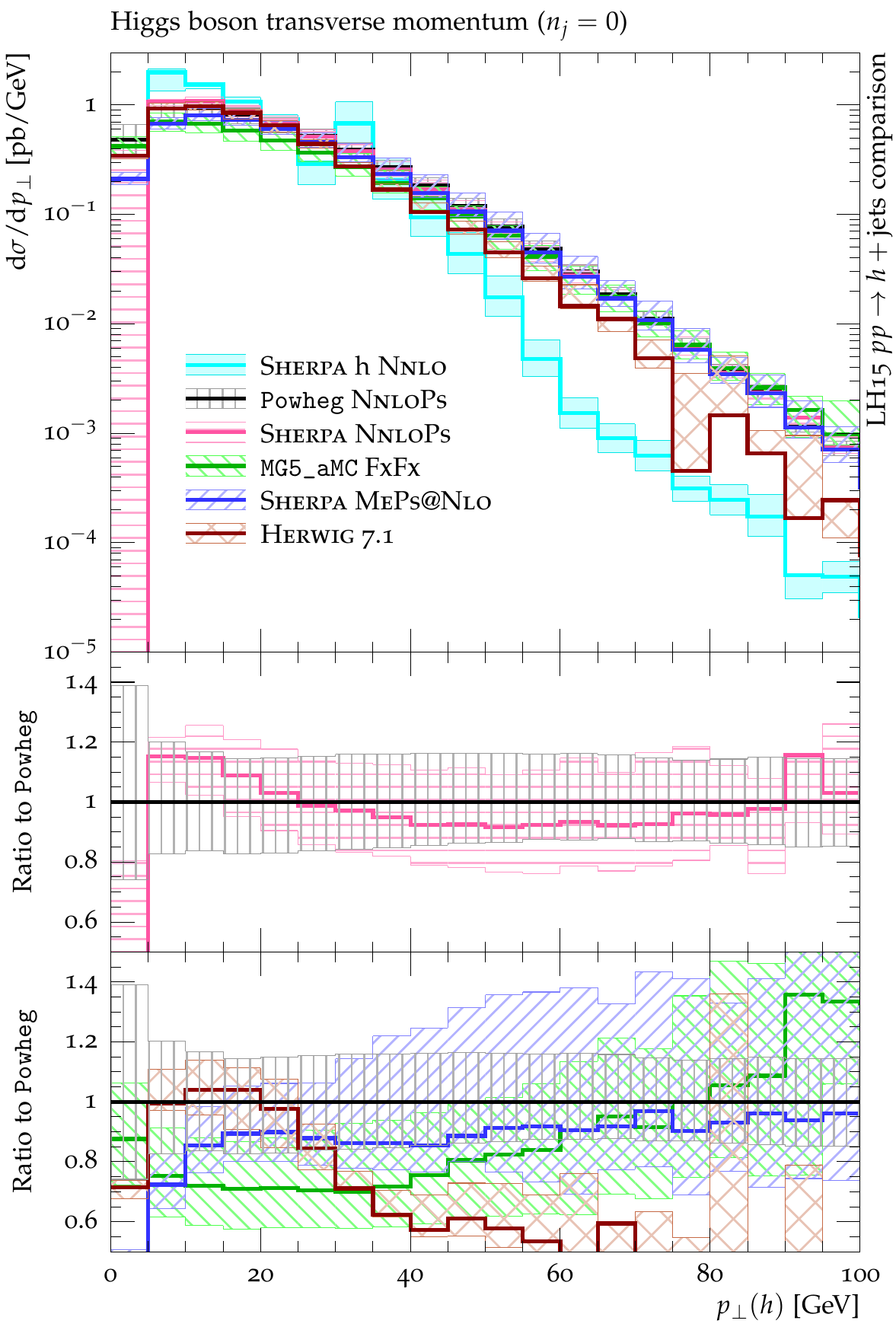}
  \caption{\label{fig:hjetscomp:results:exclobs:hpt}%
    The Higgs boson transverse momentum in the exclusive event
    selection (i.e.~in the absence of any jet) without (left) and with
    (right) uncertainties. The  panels have been arranged as in
    the previous figure, apart from dropping the last panel,  and
    switching to a new reference curve  obtained from
    \hjetscompPowheg \hjetscompNNLOPS.}
\end{figure}

In Figures~\ref{fig:hjetscomp:results:inclobs:hpt} and
\ref{fig:hjetscomp:results:exclobs:hpt}, the inclusive and exclusive
(i.e.~vetoing events with jets above $30\,\hjetscompgev$) Higgs boson transverse momentum
distributions are shown, respectively.  For the former, the ratios in
the bottom panels are taken with respect to the \hjetscompHqT result, while
\hjetscompPowheg \hjetscompNNLOPS serves as the reference for the latter case. In
general, good agreement is found, with differences being somewhat more
pronounced in the exclusive case. For the inclusive version of the
$p_\perp(h)$ observable,  good agreement is observed with \hjetscompHqT,
with some larger deviations evident at very low $p_\perp$. Here, the
resummation properties of the different parton showers dominate the
spectra of the matched and merged predictions. While at low $p_\perp$
the \hjetscompSherpa \hjetscompNNLOPS curve starts about 15\% higher than both \hjetscompHqT and
\hjetscompPowheg \hjetscompNNLOPS, it approaches the \hjetscompHqT results at higher $p_\perp$.
The \hjetscompPowheg \hjetscompNNLOPS  prediction follows 
\hjetscompHqT closely up to $p_\perp$ values of $\sim m_h$. The differences 
observed beyond that value are due to the 
dynamical scale choice employed by \hjetscompPowheg \hjetscompNNLOPS. The multijet merged calculations, due to their
similar scale choices, follow the pattern of the \hjetscompPowheg \hjetscompNNLOPS
prediction. Note that the differences in \hjetscompMGaMC's central scale
choice becomes less significant as the Higgs boson transverse momentum
increases. \hjetscompHerwig clearly provides the softest spectrum and \hjetscompSherpa
as well as \hjetscompMGaMC predict a noticeably different shape for the Sudakov
suppression at low $p_\perp$, which is not covered by the \hjetscompHqT
uncertainty envelope. For \hjetscompSherpa, the more significant suppression of
the lowest $p_\perp$ values can be traced back to the performance of the
\textsc{Csshower} which tends to radiate more strongly in this region.
The third ratio plot finally presents the direct comparison between
the two analytic resummation approaches, \hjetscompHqT and \hjetscompResbos,
which are in good agreement with each other. The leftover
$\mathcal{O}(10\%)$ deviations between the two approaches can be
attributed to non-perturbative effects, still included in the latter, and the different handling of
the process-dependent pieces in the CFG and CSS schemes. The
\hjetscompResbos scale variation band features a cross-over point at
$p_\perp(h)\approx20\,\hjetscompgev$ but this does not indicate or imply a
vanishing uncertainty at this point.
Lastly, we refrain from showing any fixed-order prediction here
because they are neither stable nor reliable at low $p_\perp$ in the
Sudakov region where resummation effects play a dominant role.

As shown in Figure~\ref{fig:hjetscomp:results:exclobs:hpt}, the
exclusive version of  the Higgs boson $p_\perp(h)$ distribution exhibits deviations among the
predictions that are more sizable. The $p_\perp(h)$ distribution 
declines much faster, easily spanning three orders of magnitude 
between zero and $100\,\hjetscompgev$. This observable is less straightforward 
than the inclusive $p_\perp$-spectrum, as not only do Sudakov effects 
dominate the low-$p_\perp$ region, but resummation effects are also 
entering through the veto on any jet activity. A reliable description
of the observable therefore necessitates both a proper understanding
of the small $p_\perp(h)$ region and of jet production. Thus,
this is a stringent test of all predictions combining matrix elements
and parton showers (ME+PS), as
the high transverse momentum of the Higgs boson is produced by a
combination of soft jets (those that are below the $30\,\hjetscompgev$ threshold)
and soft gluon radiation. Note that the comparison is now taken with respect to
\hjetscompPowheg \hjetscompNNLOPS. The inclusive 
NNLO calculation is shown for this case in order  to demonstrate the failure of a fixed-order 
calculation on both accounts; thus, only the parton showered 
predictions are included in the study of the respective differences. 
Among them, apart from the differences already seen in the inclusive 
spectrum, both \hjetscompNNLOPS calculations agree  well with one another, 
remaining within the 20\% uncertainty bands throughout the spectrum. 
While the \hjetscompSherpa \hjetscompMEPSatNLO prediction remains mostly flat with respect to the \hjetscompNNLOPS 
predictions, both \hjetscompMGaMC and \hjetscompHerwig exhibit shape differences  at
larger transverse momenta. In the case of \hjetscompHerwig, the deviations  grow
larger than 40\%. From the resummation (i.e.~parton-shower)
point-of-view, all predictions are at a similar level here, although
formally the \hjetscompNNLOPS techniques should lead to a more accurate description of
the exact zero-jet bin. The NLO merging approaches reduce to an \hjetscompNLOPS
treatment in this zero-jet bin. It is however hard to infer this
formal difference from the behavior of the scale variation bands as
they are very comparable in size among all predictions. We conclude
that the deviation of the predictions probably provides us with a
better reflection of the true uncertainty.

\begin{table}[t!]
  \centering
  \begin{tabular}{c||c|c|c}
    Order \vphantom{$\int\limits_a^b$} &
    $\mu_\mathrm{R}=\mu_\mathrm{F}=\tfrac{1}{2}\,m_h$ &
    $\mu_\mathrm{R}=\mu_\mathrm{F}=\tfrac{1}{2}\,\sqrt{\vphantom{\big[}\Sigma_T}$ &
    $\mu_\mathrm{R}=\mu_\mathrm{F}=\tfrac{1}{2}\,\hat{H}_T'$ \\
    \hline\hline
    NLO \vphantom{$\int\limits_a^b$} &
    $17.0^{+3.0}_{-2.9}~\mathrm{pb}$ &
    $16.2^{+3.1}_{-2.8}~\mathrm{pb}$ &
    $13.5^{+2.0}_{-2.1}~\mathrm{pb}$ \\\hline
    NNLO \vphantom{$\int\limits_a^b$} & -- & $16.4^{+0.0}_{-0.9}~\mathrm{pb}$ & -- \\
    \hline
  \end{tabular}
  \caption{\label{tab:H1jXS}%
    The total cross section for the inclusive production of a Higgs
    boson and one additional jet using different core scale choices.
    The two dynamical scales are $\Sigma_T=m_h^2+\sum_\mathrm{jets}p_T^2$, see also
    Eq.~\eqref{eq:bfglpScale}, and
    $\hat{H}_T'=m_{T,h}+\sum_\mathrm{partons}p_T$,
    see also Eq.~\eqref{eq:hthatprime}. We note that though the NNLO figures
    are not available for $\tfrac{1}{2}\,\hat{H}_T'$ or $\tfrac{1}{2}m_h$, the variations
    will be very small.}
\end{table}

\begin{figure}[t!]
  \centering
  \includegraphics[width=0.47\textwidth]{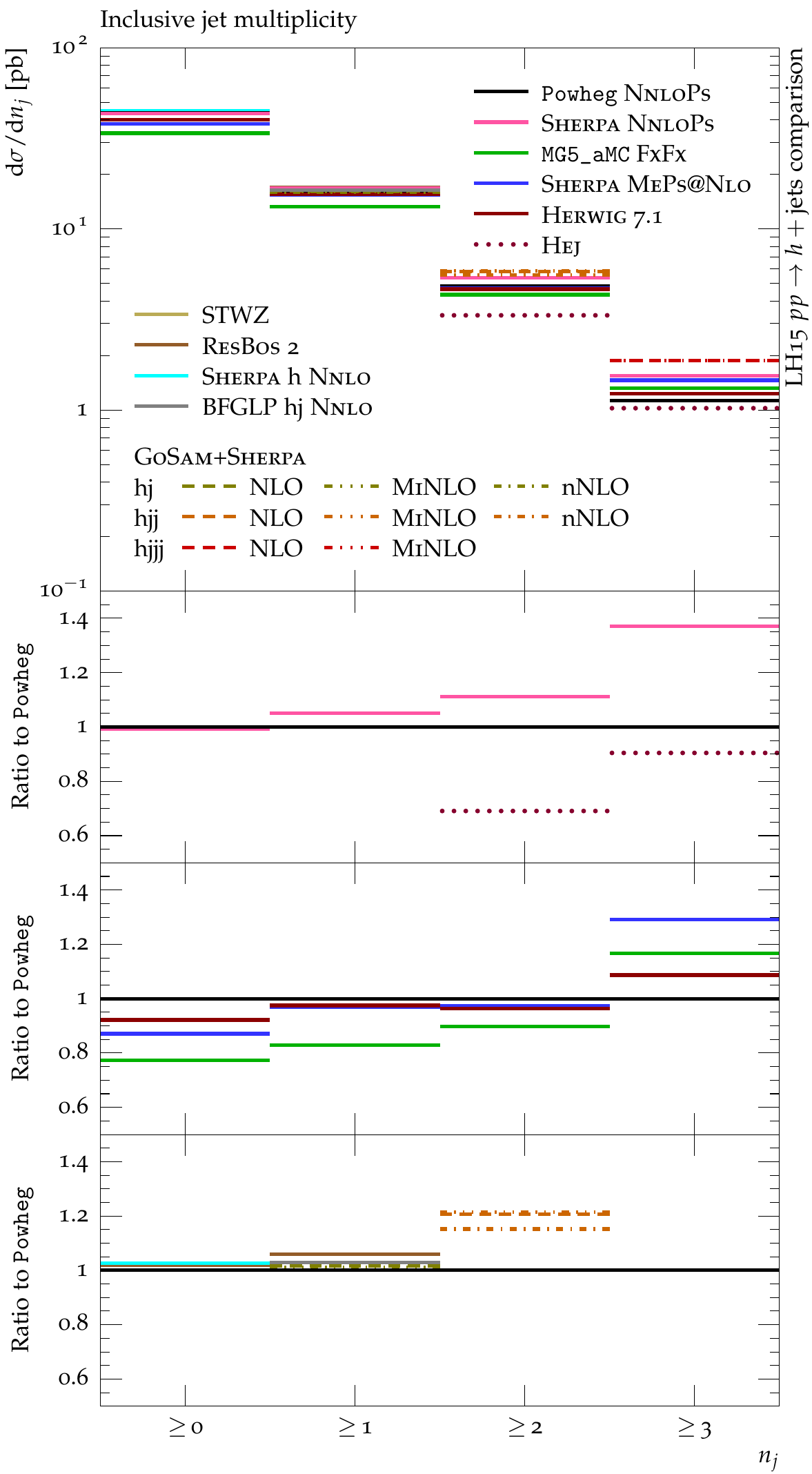}
  \hfill
  \includegraphics[width=0.47\textwidth]{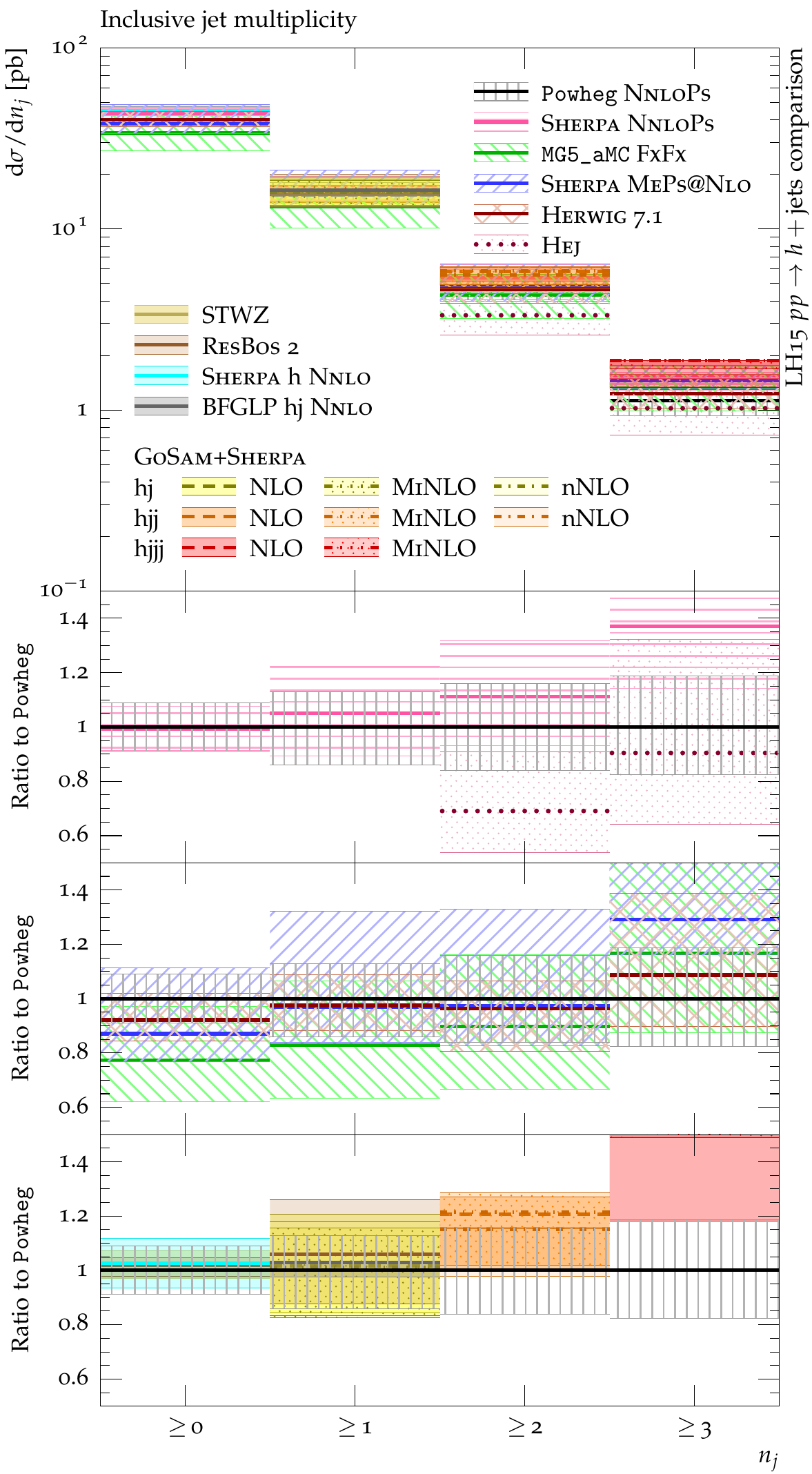}
  \caption{\label{fig:hjetscomp:results:inclobs:njets}%
    The central predictions on their own (left panel) and including their
    theoretical uncertainties (right panel) for the inclusive jet
    multiplicities as predicted by fixed-order calculations, resummed
    calculations, NNLO and NLO Monte Carlos. The bottom panel is
    divided up into three subplots all showing the ratios with respect
    to the \hjetscompPowheg \hjetscompNNLOPS prediction. The upper of these plots
    contains the \hjetscompHej and \hjetscompSherpa \hjetscompNNLOPS ratios, while the middle one
    includes all NLO merged predictions (\hjetscompMGaMC, \hjetscompHerwig and \hjetscompSherpa)
    and the lower one shows all those listed in the bottom left legend
    of the main panel.}
\end{figure}

\begin{figure}[t!]
  \centering
  \includegraphics[width=0.47\textwidth]{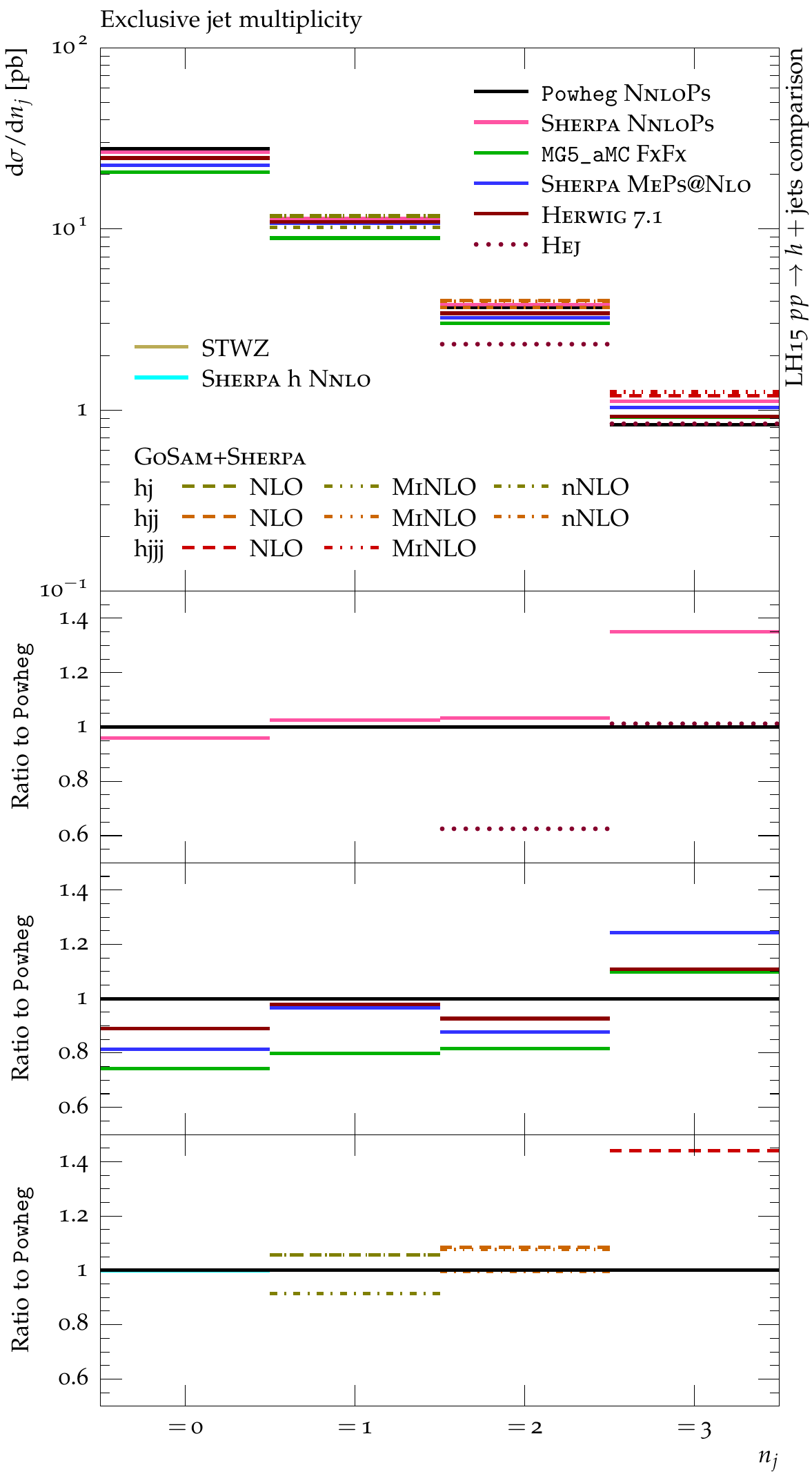}
  \hfill
  \includegraphics[width=0.47\textwidth]{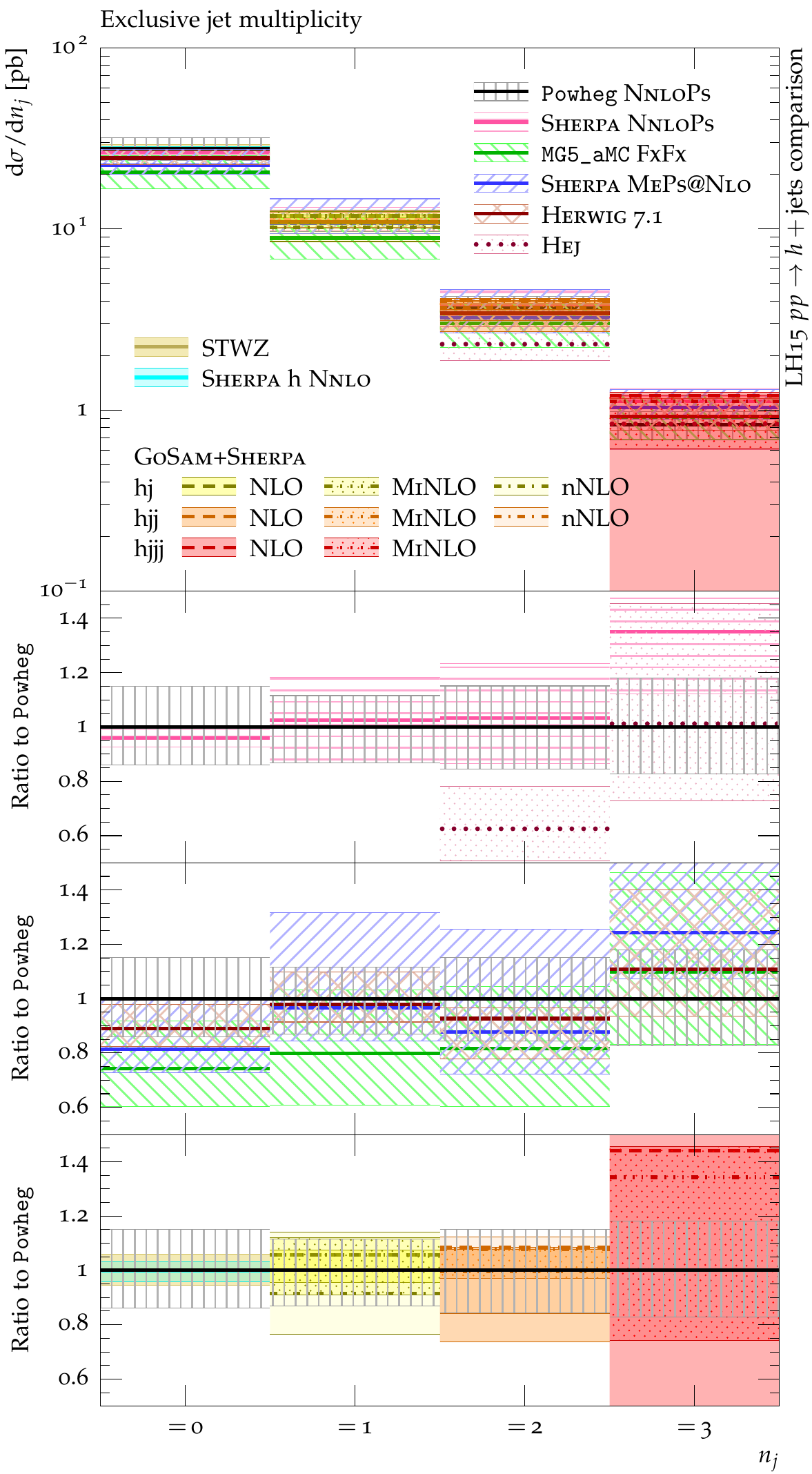}
  \caption{\label{fig:hjetscomp:results:inclobs:njets_excl}%
    The central predictions on their own (left panel) and including their
    theoretical uncertainties (right panel) for the exclusive jet multiplicities
    as predicted by fixed-order calculations, resummed calculations,
    NNLO and NLO Monte Carlos. The bottom panel is divided up into
    three subplots all showing the ratios with respect to the \hjetscompPowheg
    \hjetscompNNLOPS prediction. The upper of these plots contains the \hjetscompHej and
    \hjetscompSherpa \hjetscompNNLOPS ratios, while the middle one includes all NLO
    merged predictions (\hjetscompMGaMC, \hjetscompHerwig and \hjetscompSherpa) and the lower one
    shows all those listed in the bottom left legend of the main panel.}
\end{figure}

To obtain a first impression of how the different predictions 
compare beyond the zero-jet bins ($n_j\ge0$ and $n_j=0$), we now
examine the various inclusive and exclusive $n_j$ cross sections.
Accordingly, Figures~\ref{fig:hjetscomp:results:inclobs:njets} and
\ref{fig:hjetscomp:results:inclobs:njets_excl}, respectively, show the
inclusive ($n_j\ge N$) and the exclusive ($n_j=N$) jet multiplicity
distributions up to $N=3$, requiring \hjetscompantikt jets
with $p_\perp>30\,\hjetscompgev$. Two statements can be made before discussing
the individual results in more detail:
first of all, the agreement between all results is basically very good, and
second, the level of agreement is barely altered for exclusive jet
predictions compared to the inclusive jet predictions. 
Although not
shown here, when  the minimum jet-$p_\perp$ threshold is increased to
$50\,\hjetscompgev$ the picture does not change significantly.
Again, the bottom panels are split up into several ratio plots with
the common reference provided by the \hjetscompPowheg \hjetscompNNLOPS result. The
upper ratio plot depicts the \hjetscompNNLOPS methods together
with \hjetscompHej, which provides predictions only starting at $N=2$. Due to the similar core scale
choices in either \hjetscompNNLOPS calculation, they share a common fully
inclusive cross section, with \hjetscompSherpa being greater than \hjetscompPowheg for higher
jet multiplicities. Conversely, \hjetscompHej undershoots by 30\% in the
two-jet bin; however, this bin is described with an accuracy no higher
than LO by all predictions in this upper panel. In the three-jet bin
($N=3$), \hjetscompHej retains  LO accuracy, while both \hjetscompNNLOPS calculations
produce this cross section by their respective parton showers
only. It is thus natural to find the largest differences between the
\hjetscompNNLOPS predictions for $N=3$. Please note that the respective parton
shower uncertainties are partially incorporated in the \hjetscompSherpa \hjetscompNNLOPS
uncertainty estimate, while they are not assessed for \hjetscompPowheg,
resulting in a rather slowly increasing uncertainty band for rising
$N\le n_j$. And, more surprisingly, the \hjetscompPowheg band remains very flat
in the $N=n_j$ case. The central ratio plot
compares the NLO matched and merged predictions with each other and
against the common reference \hjetscompPowheg \hjetscompNNLOPS. All these predictions
claim NLO accuracy for $N=0,1,2$, and thus overlap well within
uncertainties, where the lower value for \hjetscompMGaMC can again be attributed
to the different scale choice. Again, with the same scale choice, all of
the calculations should agree much better. For $N=3$, only the \hjetscompSherpa \hjetscompMEPSatNLO
prediction retains NLO accuracy, while \hjetscompMGaMC and \hjetscompHerwig revert to LO
accuracy, which is nicely reflected in the sizes of the uncertainty estimates. 
Unsurprisingly, in the $N=3$ case, all three NLO merged calculations
predict larger cross sections with respect to the reference whose rate
is only given by the parton shower, i.e.~\hjetscompPythia.

Lastly, fixed-order predictions are shown for all jet multiplicities
at NLO (provided by \hjetscompGoSam{}+ \hjetscompSherpa) for the $1$-jet, $2$-jet and
$3$-jet bins and at approximate NNLO (labeled nNLO, provided by \hjetscompLoopsim) 
for the $1$-jet and $2$-jet bins.
Complete NNLO predictions are shown for the $0$-jet inclusive and
exclusive bins using \hjetscompSherpa (without PS), and for the $1$-jet inclusive
bin using the prediction of Boughezal et al.~(BFGLP).
The zero-jet bin comparison
also contains the resummation prediction of Stewart et
al.~(STWZ); the comparisons for the $n_j\ge 1$, $2$, $3$ jet bins also
include the \hjetscompMinlo enhanced NLO calculations.
Figure~\ref{fig:hjetscomp:results:inclobs:njets} also
contains the \hjetscompResbos $q_\text{T}$-resummed predictions in the zero-jet bin
(with precision corresponding to NNLO+NNLL) as well as in the
one-jet bin (with precision corresponding to NLO+NLL). All of the
above are grouped together in the lower ratio plot.
For the zeroth bin, good agreement can be found between \hjetscompPowheg and
\hjetscompSherpa $h$ NNLO as well as with the STWZ approach and \hjetscompResbos; the
uncertainties also are of comparable size, except for the
significantly wider $n_j=0$ envelope of \hjetscompPowheg.
In the $1$-jet case, \hjetscompPowheg (being NLO-accurate in this bin)
sits less than 3\% below the pure NLO prediction, with the small difference
being due to slightly different scale choices. Unlike the inclusive Higgs
boson production case, the NNLO corrections for inclusive $1$-jet
production are small (cf.~Table~\ref{tab:H1jXS})-- slightly positive
for the central scale choice as given in Eq.~\eqref{eq:bfglpScale}.
There is a notable decrease in the scale uncertainty with respect to
the NLO band given by \hjetscompGoSam{}+\hjetscompSherpa. As expected, the uncertainty
envelope of the \hjetscompResbos prediction is of similar size while its
central value is about 5\% higher than the reference as a result of
its scale choice being $\mu=\tfrac{1}{2}m_h$.
The $2$-jet bin shows the \hjetscompGoSam{}+\hjetscompSherpa NLO prediction 10-20\%
above \hjetscompPowheg, which gives a LO prediction in this case. For the same
reason, we assume that the \hjetscompPowheg uncertainties in the higher jet
multiplicities are probably underestimated.
In the $3$-jet bin (both inclusive and exclusive), the
\hjetscompGoSam{}+\hjetscompSherpa and \hjetscompSherpa \hjetscompMEPSatNLO predictions clearly indicate the
presence of  NLO corrections in the third jet bin. The \hjetscompLoopsim
inclusive results for $hj$ and $hjj$ are always somewhat below the
respective \hjetscompGoSam{}+\hjetscompSherpa results, and in the exclusive case this
difference gets more pronounced. However, the relatively large Monte
Carlo generation cut of $25\,\hjetscompgev$ means that the total rates predicted
by \hjetscompLoopsim should be interpreted with care.
Furthermore, compared to the NLO benchmark, the \hjetscompMinlo approach
predicts 10-20\% larger cross sections for all $n_j\ge 1$, $2$, $3$ jet bins.
Note that the \hjetscompMinlo ratio for inclusive $hjjj$ turns out to be
outside the plot range appearing at around $1.65$, with the lower edge
of the uncertainty band at a ratio value of $1.5$.
In the cases where NNLO precision is available, the reduction in scale
uncertainty is clear. For $n_j\ge1$ the variation around
$\mu_\mathrm{R}=\sqrt{\Sigma_T}/2$ is about $5.5\%$ while for
$n_j\ge0$ it is about $10.0\%$ around $\mu_\mathrm{R}=\tfrac{1}{2}m_h$.
The latter result can be improved to only a few percent using the
N${}^3$LO prediction of Anastasiou et al.~\cite{Anastasiou:2015ema}.
In order to compare more easily with results presented previously in
the literature, we give numerical values at NLO and NNLO for different
scale choices in Table~\ref{tab:H1jXS}. As observed previously
\cite{Boughezal:2015dra}, the convergence of the total cross section predictions
is improved for scales that limit to $\tfrac{1}{2}m_h$. The dynamical
scale of $\sqrt{\Sigma_T}$ defined in Eq.~\eqref{eq:bfglpScale} is
slightly harder than the fixed scale given the minimum jet
$p_\perp(j)>30\,\hjetscompgev$. On the other hand it is softer than $\hat{H}_T'$
defined in Eq.~\eqref{eq:hthatprime} which explains the differences observed at
NLO.


\begin{figure}[p!]
  \centering
  \includegraphics[width=0.47\textwidth]{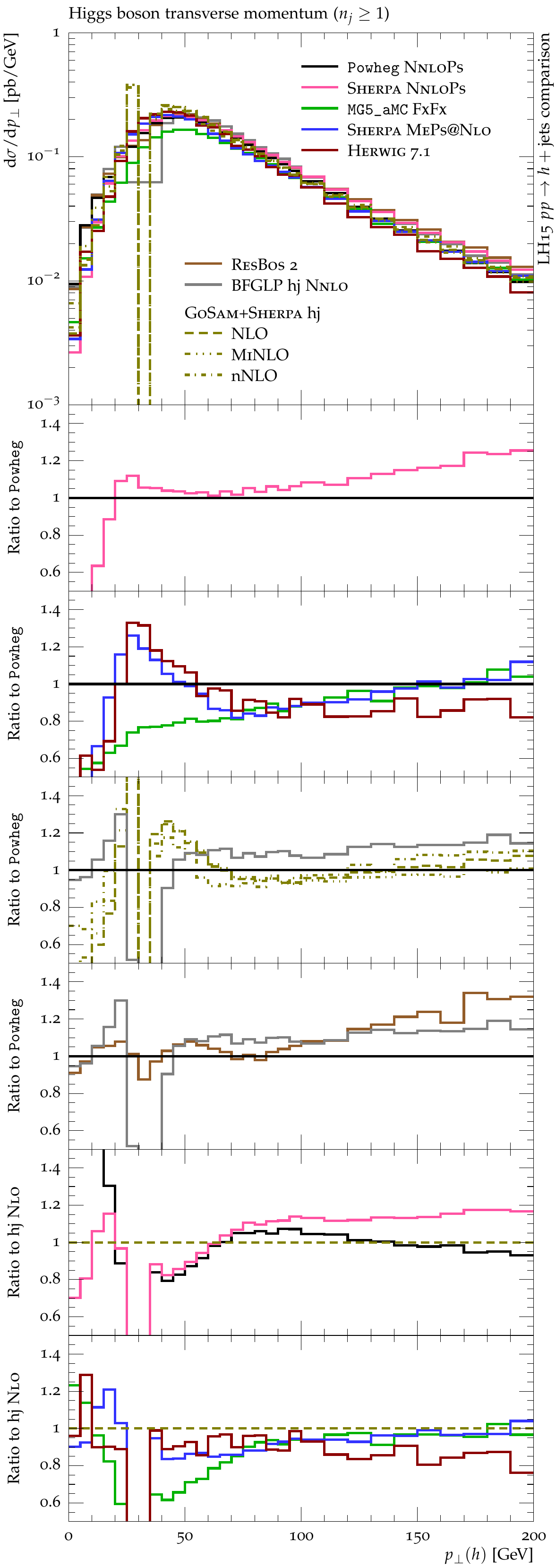}
  \hfill
  \includegraphics[width=0.47\textwidth]{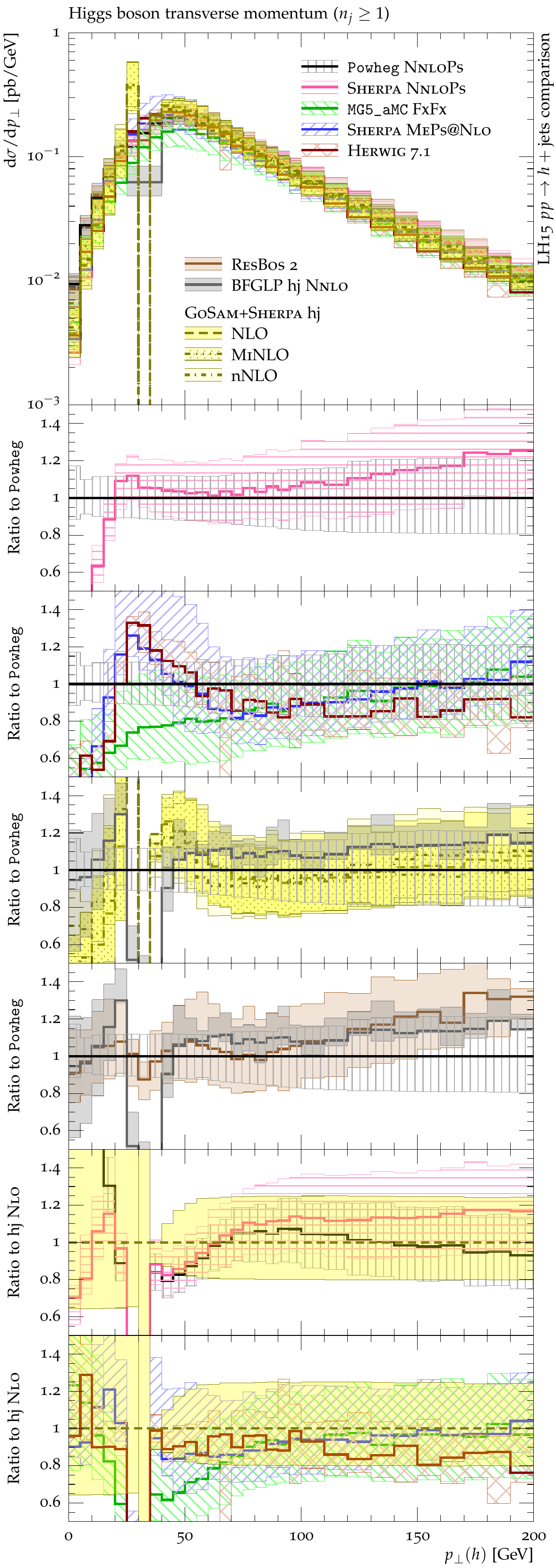}
  \caption{\label{fig:hjetscomp:results:1obs:hpt}%
    The Higgs boson transverse momentum in the presence of at least
    one jet without (left) and with (right) uncertainty bands. The
    ratio plot panel is divided into six parts where the upper four
    exhibit the ratios wrt.~the \hjetscompPowheg \hjetscompNNLOPS result while the lower
    two show them wrt.~the NLO calculation for $h+1$~jet as
    provided by \hjetscompGoSam{}+\hjetscompSherpa. The grouping in the ratio plots has
    been arranged to separately compare with each other the \hjetscompNNLOPS
    predictions (first and fifth subplot), the NLO merging
    predictions (second and last subplot) and the fixed-order
    predictions (third subplot in the middle).}
\end{figure}

\begin{figure}[t!]
  \centering
  \includegraphics[width=0.47\textwidth]{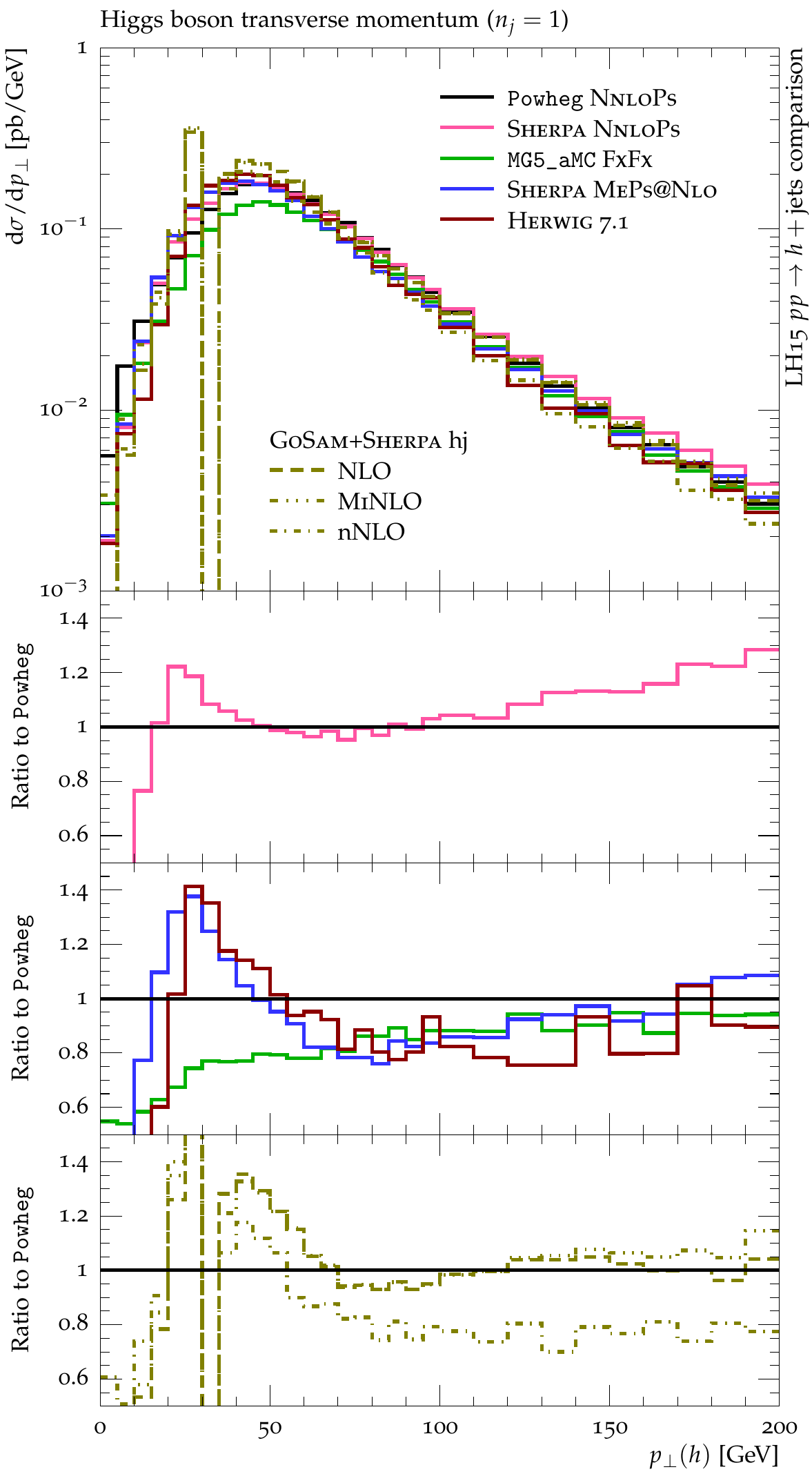}
  \hfill
  \includegraphics[width=0.47\textwidth]{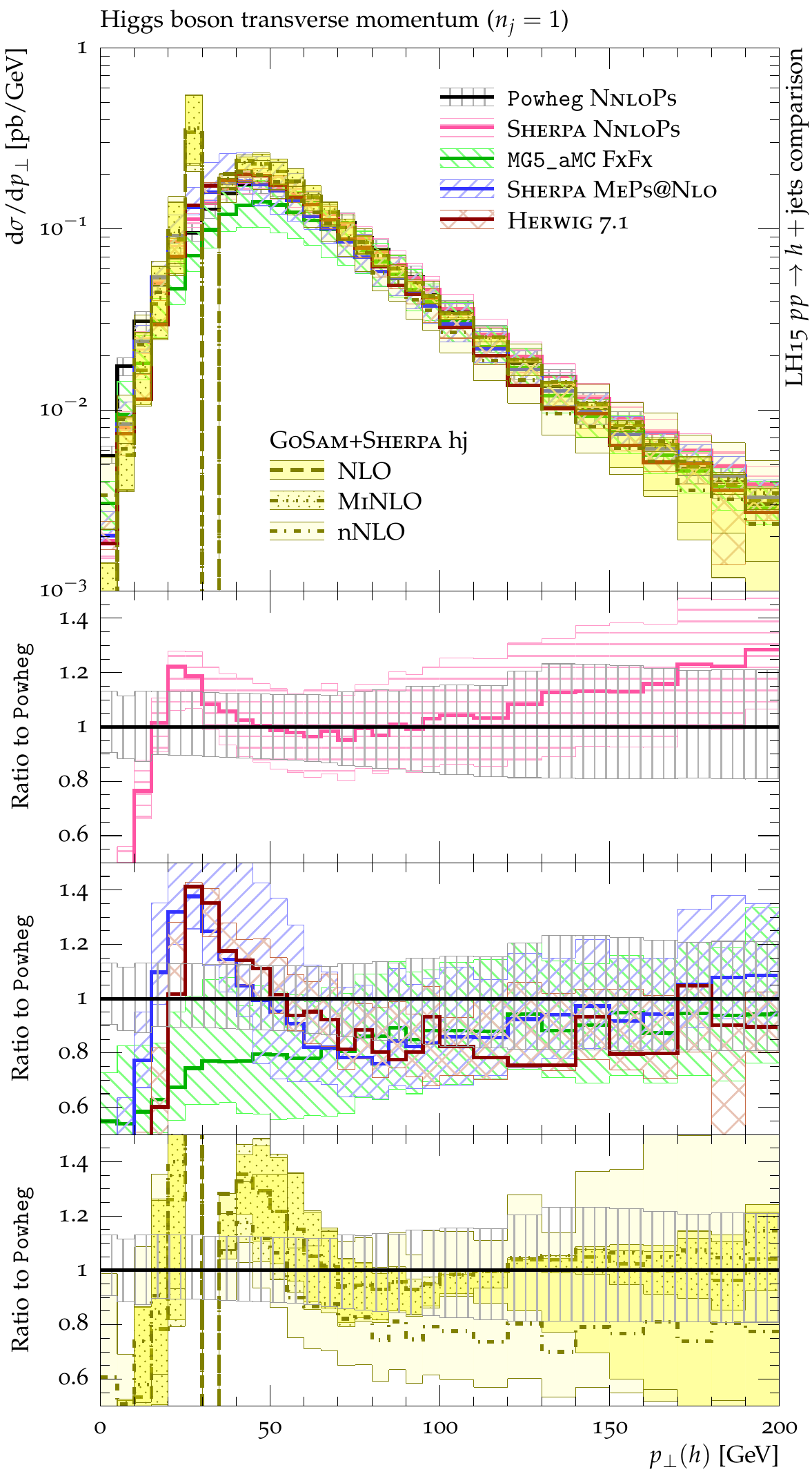}
  \caption{\label{fig:hjetscomp:results:1obs:hpt_excl}%
    The Higgs boson transverse momentum in the presence of exactly one
    jet without (left) and with (right) uncertainty bands. The ratio
    plot panel is divided into three parts all of which depict the
    corresponding ratios wrt.~the \hjetscompPowheg \hjetscompNNLOPS result. From top to
    bottom, the predictions are grouped such that the \hjetscompNNLOPS results,
    the ME+PS results at NLO and the fixed-order results are compared
    directly in the first, second and third ratio plot, respectively.}
\end{figure}

\subsubsection{One-jet observables}
\label{sec:hjetscomp:results:1jobs}

In this section, we move away from the fully inclusive picture and
require the presence of at least one or exactly one jet associated with 
the Higgs boson. Recall that the jets are defined based on the
\hjetscompantikt algorithm using $R=0.4$; they furthermore have to fulfil the selection
criteria that $p_\perp(j)>30\,\hjetscompgev$ and $|\eta(j)|<4.4$. The set of
observables presented here includes the transverse momentum
distributions of the Higgs boson $h$, the leading jet $j_1$ and the
$hj_1$ two-body system as well as the rapidity spectrum of the leading
jet.

The Higgs boson transverse momentum distribution in the presence of at
least one jet is shown in
Figure~\ref{fig:hjetscomp:results:1obs:hpt}. The exclusive version of
this plot, i.e.~where one requires the Higgs boson and the jet to be 
the only resolved final state objects,
is presented in Figure~\ref{fig:hjetscomp:results:1obs:hpt_excl}. As
for the zero-jet cases discussed earlier in
Figures~\ref{fig:hjetscomp:results:inclobs:hpt}~and~\ref{fig:hjetscomp:results:exclobs:hpt},
the one-jet $p_\perp(h)$ variables are prone to large Sudakov effects that arise at low $p_\perp$, 
but are also present beyond
this region, in particular for the exclusive final states. Moreover, the
Sudakov shoulder effect can be observed for all fixed-order
predictions shown here. The jet-$p_\perp$ threshold leads to a
non-smooth behavior of the $p_\perp(h)$ observable at LO, and
therefore to the existence of a critical point at $30\,\hjetscompgev$, for which
the cancellations between real and virtual soft-gluon singularities
will be imperfect at any given fixed, higher order in perturbation
theory~\cite{Catani:1997xc}. For the BFGLP $hj$ NNLO  prediction, an
averaging procedure has been used to dampen the effect around the jet
threshold, while for the NLO predictions, the large oscillations are a
clear indication of the instability emerging at the jet threshold. The
\hjetscompNNLOPS, NLO ME+PS and \hjetscompResbos predictions do not suffer from
the Sudakov shoulder effect since they include the necessary all-order
resummation corrections.

Comparing the different fixed-order predictions, which are detailed in 
the third ratio plot, noticeable differences only
occur between the NNLO prediction and the four NLO predictions as
obtained from \hjetscompPowheg and the three variations on \hjetscompGoSam{}+\hjetscompSherpa (pure
NLO, \hjetscompMinlo and \hjetscompLoopsim). The NNLO tail is harder by about 15\% which
is expected since the $p_\perp$ tail is affected by multijet
contributions. The NNLO treatment includes these contributions to a
larger extent, as it includes $h+2$-jet and $h+3$-jet contributions at
 NLO and LO, respectively. For $p_\perp(h)<30\,\hjetscompgev$, the (N)NLO
description is degraded to (N)LO. Here, the presence of the
$\mathcal{O}(\alpha_\mathrm{s})$ term of the Sudakov resummation, also
included in the NLL resummation of \hjetscompResbos, affects the BFGLP $hj$
NNLO calculation. However, it can also be noticed (see fourth ratio plot in
Figure~\ref{fig:hjetscomp:results:1obs:hpt}) that apart from the BFGLP
$hj$ NNLO calculation, \hjetscompResbos and also \hjetscompPowheg, all other approaches
predict a more steeply falling shoulder resulting in a significantly
lower cross section as $p_\perp(h)\to0$.
For larger $p_\perp(h)$ values, $p_\perp(h)\gtrsim\tfrac{1}{2}m_h$, in
general there is good agreement between \hjetscompPowheg, \hjetscompMGaMC, \hjetscompSherpa
\hjetscompMEPSatNLO and the NLO curves; this can be expected as these
predictions are all NLO-accurate. As before, \hjetscompHerwig tends to be
softer, whereas \hjetscompMGaMC, using the nominal $\tfrac{1}{2}m_h$ core scale,
turns out to be harder by almost 40\%, as indicated by the upper edge
of its corresponding uncertainty band (see the second or last subplot
to the right in Figure~\ref{fig:hjetscomp:results:1obs:hpt}). \hjetscompSherpa's
\hjetscompNNLOPS prediction also features a harder tail than \hjetscompPowheg owing to the
different scale setting procedures employed by the two approaches.
While in \hjetscompPowheg the scale setting is accomplished through the \hjetscompMinlo
procedure, \hjetscompSherpa's \hjetscompNNLOPS uses the fixed scale choice of $\tfrac{1}{2}m_h$
and therefore enhances the $p_\perp$ tail with respect to \hjetscompPowheg's
result. In this region, the \hjetscompResbos prediction closely resembles the
one given by \hjetscompSherpa \hjetscompNNLOPS, since it is dominated by the fixed-order
contribution evaluated with the same scale choice as used in \hjetscompSherpa.

We furthermore observe that apart from the BFGLP $hj$ NNLO computation,
all uncertainty envelopes are of similar size. This does not come as
a surprise because all of these predictions are effectively given at
NLO. The NNLO uncertainty band (shown in grey) is found to be
significantly smaller. Comparing the second and last ratio plots with
each other, we also notice that the ME+PS predictions are in better
overall agreement with the pure NLO prediction given by
\hjetscompGoSam{}+\hjetscompSherpa than they are in agreement with \hjetscompPowheg, with the
exception of \hjetscompMGaMC below $p_\perp(h)\lesssim\tfrac{1}{2}m_h$. This is
surprising since all parton shower matched calculations are of the
same intrinsic accuracy (NLO), use similar local scale definitions
along the lines of CKKW and include Sudakov factors of NLL accuracy.
It may, however, be related to the way the resummation is controlled
in \hjetscompPowheg and the \hjetscompMCatNLO-type matchings used in \hjetscompMGaMC, \hjetscompHerwig and
\hjetscompSherpa. The different parton shower starting scales employed in
\hjetscompHerwig and \hjetscompSherpa ($\sim\tfrac{1}{2}m_h$), \hjetscompMGaMC ($\sim m_h$) and
\hjetscompPowheg ($E_\mathrm{cm}$) do also play a role.

For the exclusive one-jet case, we reduce the number of ratio plots
and show only those that display the ratios to \hjetscompPowheg in the same way
as before. Note that for the exclusive version of the observable, the
NNLO result is not available, turning this comparison into one between
NLO-accurate predictions, except for the NNLO-approximate result given
by \hjetscompLoopsim, labelled \hjetscompGoSam{}+\hjetscompSherpa $hj$ nNLO.
Figure~\ref{fig:hjetscomp:results:1obs:hpt_excl} clearly shows that
the differences among the results are very similar to those discussed
in the inclusive case; they are however pronounced such that the
deviations wrt.~the \hjetscompPowheg \hjetscompNNLOPS prediction for $h$ production
become larger for the full tranverse momentum range. There is one
exception to this: \hjetscompLoopsim predicts a softer tail of the $p_\perp(h)$
distribution by about 20\%, which in the end is a restatement of the
fact that there is a 20\% difference in the one-jet rate between the
pure NLO and the \hjetscompLoopsim result, as already shown in
Figure~\ref{fig:hjetscomp:results:inclobs:njets_excl}.

\begin{figure}[t!]
  \centering
  \includegraphics[width=0.47\textwidth]{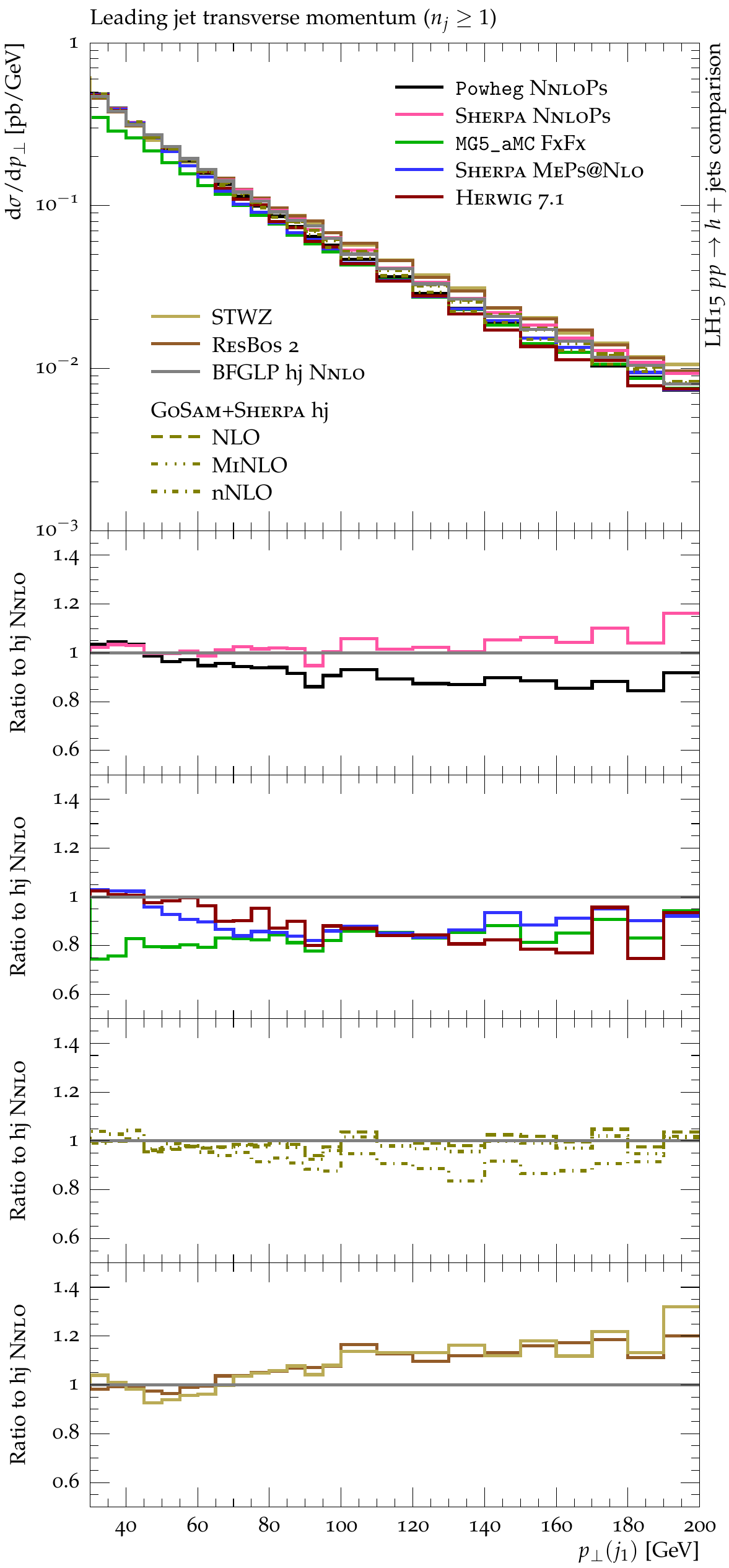}
  \hfill
  \includegraphics[width=0.47\textwidth]{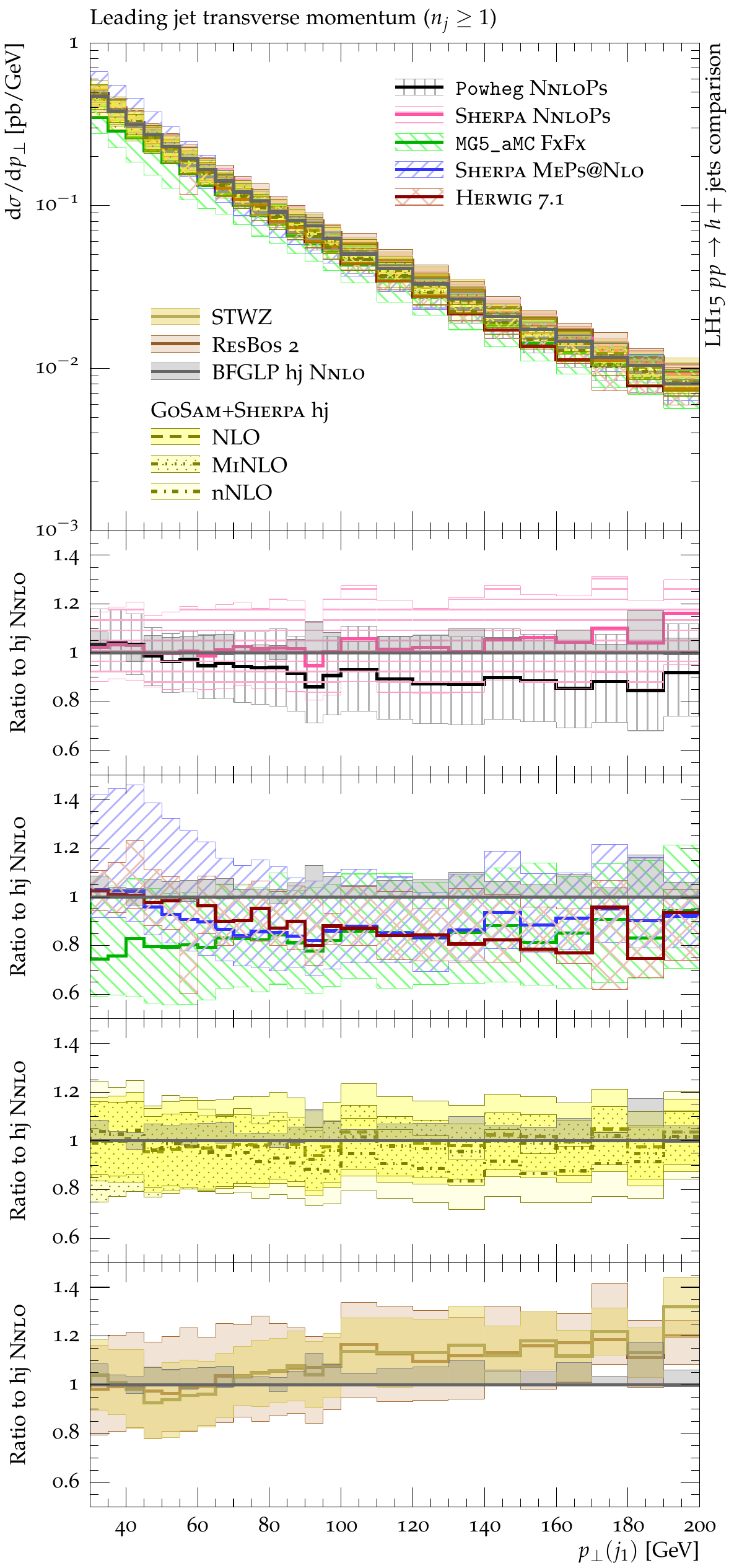}
  \caption{\label{fig:hjetscomp:results:1obs:j1pt}%
    The leading jet transverse momentum distribution for
    $h\,+\!\ge\!1$-jet production, to the right (left) shown with
    (without) the uncertainty bands provided by the various
    calculations. The part below the main plot contains four ratio
    plots taken wrt.~the NNLO result of the BFGLP group following the
    same strategy for grouping the predictions as before (\hjetscompNNLOPS
    versus NLO ME+PS versus fixed-order and resummation results).}
\end{figure}

\begin{figure}[t!]
  \centering
  \includegraphics[width=0.47\textwidth]{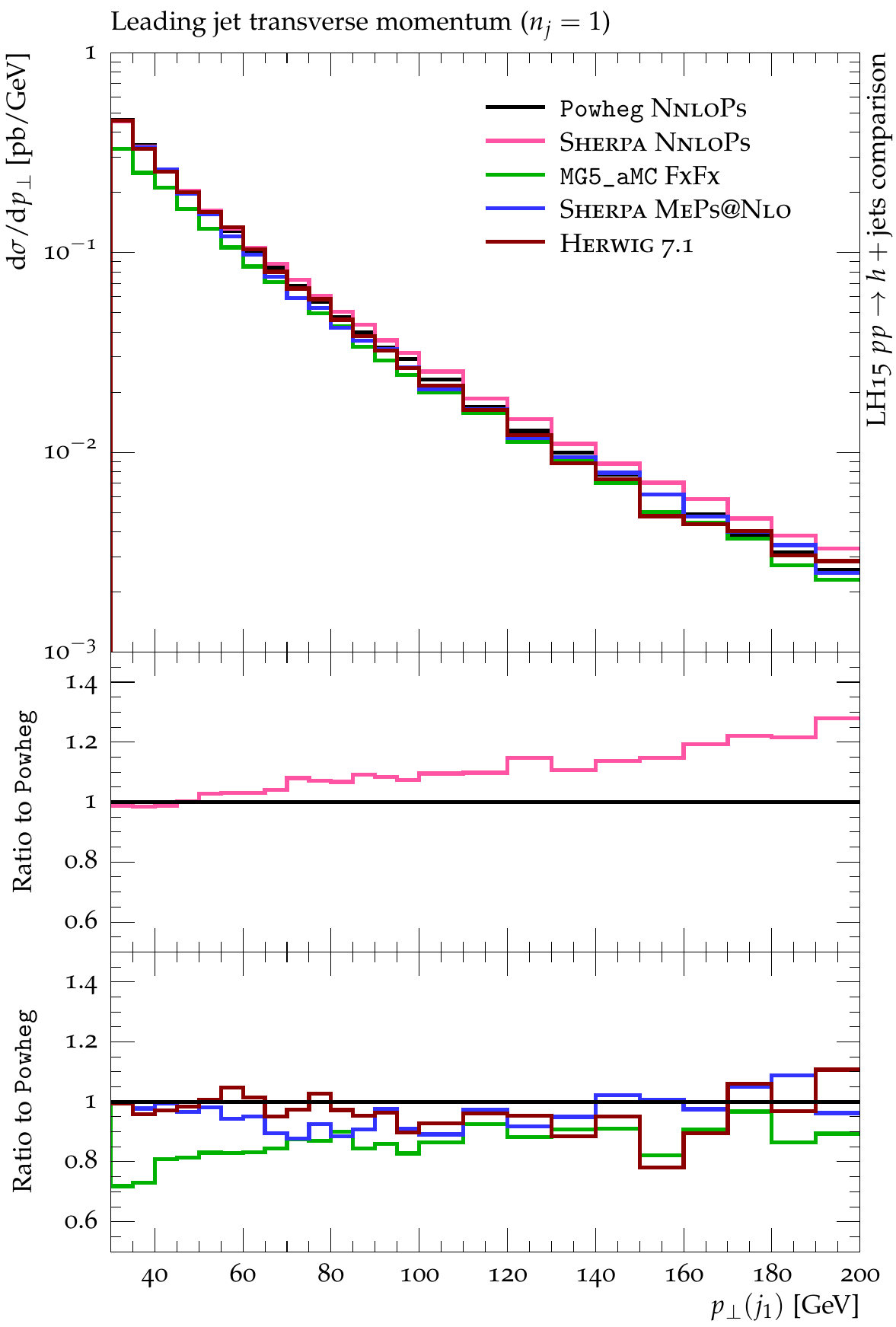}
  \hfill
  \includegraphics[width=0.47\textwidth]{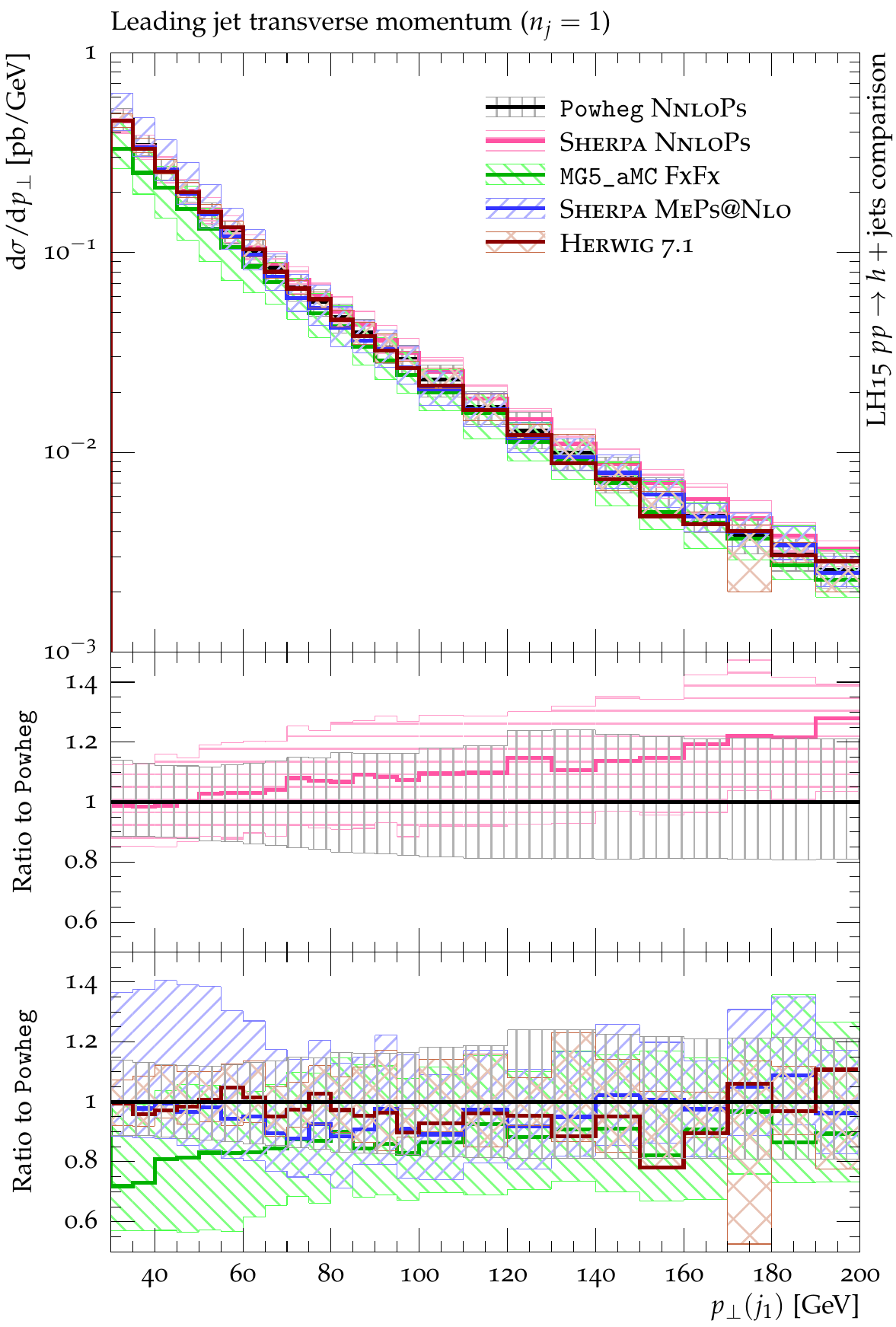}
  \caption{\label{fig:hjetscomp:results:1obs:j1pt_excl}%
    The leading jet transverse momentum distribution for exclusive
    $h+1$-jet production, to the right (left) shown with (without)
    the uncertainty bands provided by the various calculations. Ratio
    plots are displayed in the lower part of the plot using the
    \hjetscompPowheg \hjetscompNNLOPS result for Higgs boson production as their
    reference. Predictions are grouped in similar fashion to the
    previous plots.}
\end{figure}

Next we discuss the leading jet transverse momentum distribution for
$h\,+\!\ge\!1$-jet final states. For this type of observable, we do
not expect large Sudakov effects (i.e.~shifts owing to parton
showering/resummation). The impact of jet veto logarithms (owing to
the restriction that all jets be greater than $30\,\hjetscompgev$) has been
examined and found to be reasonably small at NLO and
NNLO~\cite{Banfi:2012jm,Banfi:2015pju}.
In the exclusive jet case, the $p_\perp(j_1)$ variable however is
prone to larger resummation effects, and we note that the scale
uncertainties shown will not reflect the true uncertainty. The
inclusive jet results for all approaches are shown in
Figure~\ref{fig:hjetscomp:results:1obs:j1pt} including the NNLO
prediction of the BFGLP group, the prediction of Stewart, Tackmann
et al.~and the prediction provided by \hjetscompResbos.
Figure~\ref{fig:hjetscomp:results:1obs:j1pt_excl} depicts the
exclusive one-jet case presenting the results obtained by the Monte
Carlo tools only; no fixed-order/resummation predictions are shown.
Accordingly, different reference predictions (NNLO and \hjetscompPowheg) are
used in the ratio plots associated with
Figures~\ref{fig:hjetscomp:results:1obs:j1pt} and
\ref{fig:hjetscomp:results:1obs:j1pt_excl}. 
Overall we find a rather remarkable agreement
between all results where the largest deviations rarely exceed the
20\% mark. For the exclusive lead-jet transverse momentum distribution
of Figure~\ref{fig:hjetscomp:results:1obs:j1pt_excl}, this means that
all predictions are in reasonably good agreement with \hjetscompPowheg. The
\hjetscompMGaMC prediction is lower than \hjetscompPowheg for almost the entire
transverse momentum range, again because of the central scale choice
being higher than in the other approaches. For the inclusive lead-jet
transverse momentum spectrum (see
Figure~\ref{fig:hjetscomp:results:1obs:j1pt}), the remarkable
agreement implies that all predictions indeed lie within each other's
uncertainty bands (as they should). The quoted uncertainties are similar in size, with
values at and around the 20\% level, and only those of the BFGLP $hj$
NNLO calculation are significantly smaller. 

\begin{figure}[t!]
  \centering
  \includegraphics[width=0.47\textwidth]{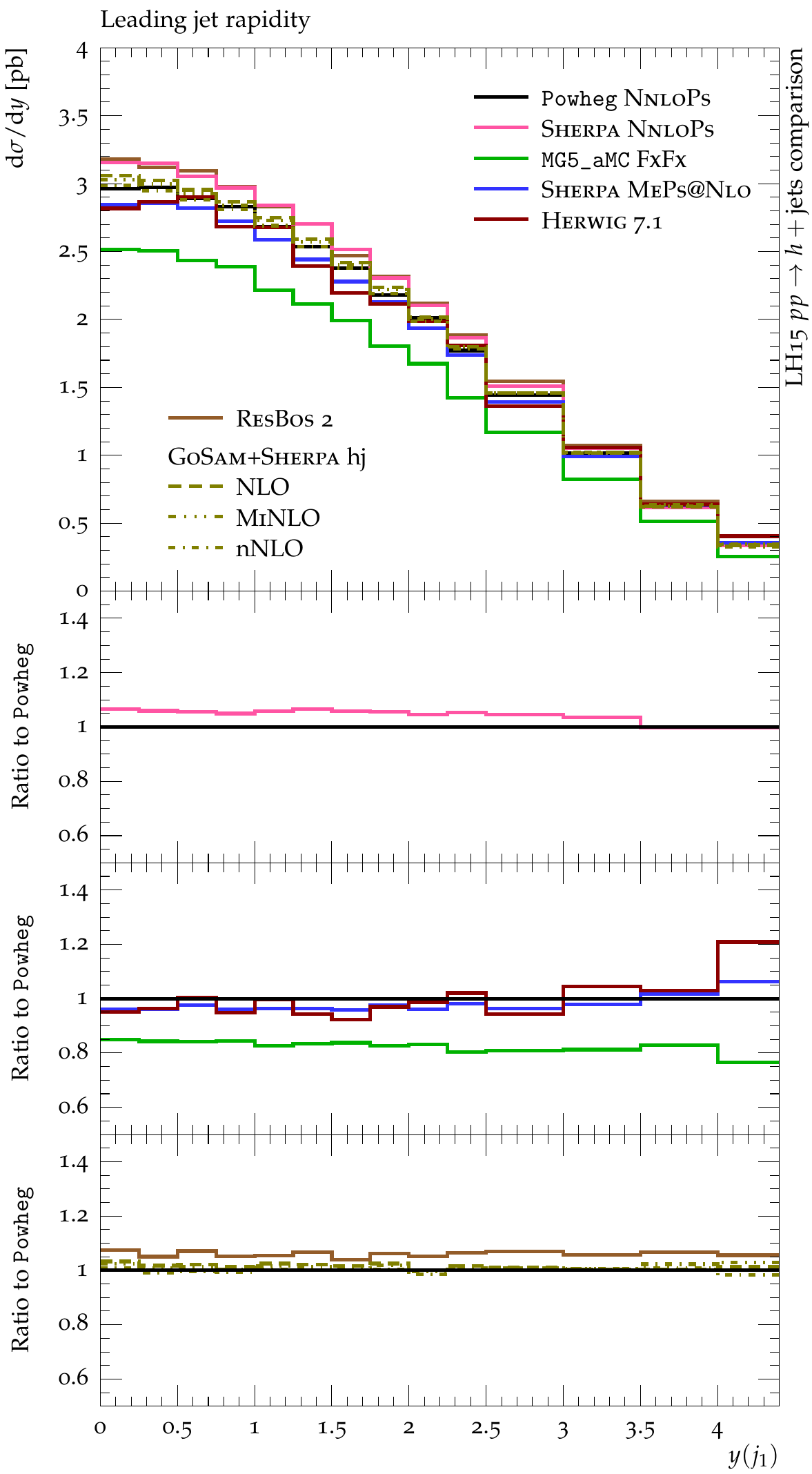}
  \hfill
  \includegraphics[width=0.47\textwidth]{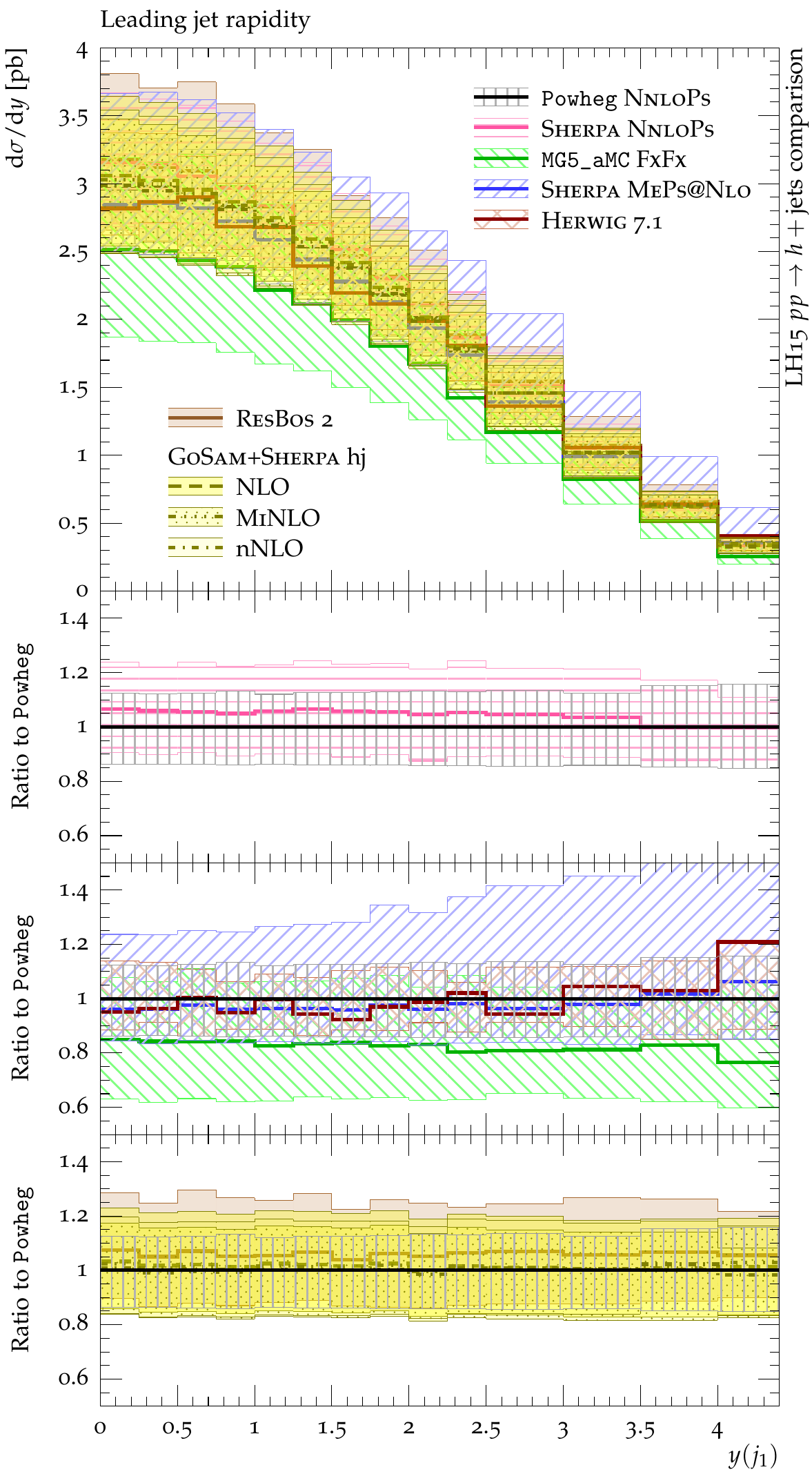}
  \caption{\label{fig:hjetscomp:results:1obs:j1y}%
    The rapidity distribution for the leading jet in $h\,+\!\ge\!1$-jet
    production, shown without (left) and with (right) theoretical
    uncertainties. Ratio plots are displayed in the lower part of the
    plot using the \hjetscompPowheg \hjetscompNNLOPS result for Higgs boson production
    as their reference. Predictions are grouped, from top to bottom,
    according to the categories \hjetscompNNLOPS, ME+PS at NLO and NLO fixed
    order as well as resummation.}
\end{figure}

Despite the good agreement seen among all predictions in
Figure~\ref{fig:hjetscomp:results:1obs:j1pt}, it is worthwhile to go
through the ratio plots and discuss some of the interesting features.
In the top ratio panel, the two \hjetscompNNLOPS predictions are compared to the
NNLO $h\,+\!\ge\!1$-jet prediction. The agreement among all three is
good at low transverse momentum, but at higher $p_\perp(j_1)$ there is
a tendency for \hjetscompSherpa \hjetscompNNLOPS to move to the upper edge of the NNLO 
uncertainty band and \hjetscompPowheg \hjetscompNNLOPS to move slightly below, resulting 
in a 20\% net difference between the two. Again, this is a result of using
$\mu=\tfrac{1}{2}m_h$ within \hjetscompSherpa versus using \hjetscompMinlo/CKKW scales within the
\hjetscompPowheg approach. In the second ratio panel, \hjetscompHerwig, \hjetscompSherpa and \hjetscompMGaMC
(taking into account its larger limit scale) agree reasonably well
with each other over the entire transverse momentum range, but fall
about 15\% low wrt.~the BFGLP prediction in the mid-range of the
$p_\perp$ distribution. The third ratio panel shows that there is almost no
difference in normalization nor shape between the NNLO and the NLO
$h\,+\!\ge\!1$-jet predictions using the given scale choice,
cf.~Eq.~\eqref{eq:bfglpScale}. This extends to the \hjetscompMinlo reweighted
NLO result and the nNLO prediction provided by the \hjetscompLoopsim approach,
although the latter is somewhat softer. Nevertheless we should bear in
mind that if we were to include in this comparison a fixed-order
calculation based on the scale choice $\tfrac{1}{2}\hat H_T^\prime$,
the resulting prediction would fall close to the multijet merged
predictions, showcasing the still largish scale dependence of the $hj$
NLO calculation. In the bottom ratio panel, the STWZ and \hjetscompResbos
predictions agree very well with one another, partly because in their
calculation they both rely on the same fixed-order piece dominating
this observable and coincide in their use of fixed scale. Both
approaches provide a resummation improved NLO calculation for this
observable, agreeing with the NNLO prediction at low $p_\perp$. At the
largest $p_\perp$ values, where no resummation effects are present,
deviations rise up to 30\% owing to their fixed scale choice, which
furthermore brings them into agreement with \hjetscompSherpa's \hjetscompNNLOPS result.
As a matter of fact, for sufficiently large values of $p_\perp(j_1)$,
all three predictions converge to an NLO prediction employing a fixed
scale of $\mu=\tfrac{1}{2}m_h$.

In summary, the fixed-order NNLO prediction of the BFGLP group, which
has the best theoretical uncertainties available, is in good agreement
with the \hjetscompSherpa \hjetscompNNLOPS prediction, and to a
somewhat lesser extent, with the \hjetscompPowheg \hjetscompNNLOPS predictions. The level of
moderate disagreements observed with the multijet merged calculations 
may be due to the different scale choices that are still important 
at NLO. The largest deviation seen with \hjetscompMGaMC can mainly be traced
back to its different choice of central scales. There is no sign of
any serious impact of the merging of the fixed-order predictions with
parton showers, as expected for such an inclusive cross section.

Even better agreement is observed  for predictions for the rapidity distribution of the lead jet,
$y(j_1)$, as demonstrated by Figure~\ref{fig:hjetscomp:results:1obs:j1y}.
This level of agreement seems to be even slightly better than the one
found for $y(h)$ in Figure~\ref{fig:hjetscomp:results:inclobs:hy}.
We do not observe any shape differences, and the rate differences 
follow the already established pattern where most noticeably the
\hjetscompMGaMC cross section is reduced owing to their choice of using a
higher central scale. Similarly, \hjetscompSherpa \hjetscompNNLOPS and \hjetscompResbos are
higher by about 6-7\% as a result of using the fixed scale
$\mu=\tfrac{1}{2}m_h$. Again, uncertainty envelopes are similar in
size and do not point to any shape changes when varying the scales;
\hjetscompHerwig's band is slightly narrower while \hjetscompSherpa \hjetscompMEPSatNLO's and
\hjetscompMGaMC's band are somewhat wider compared to all other NLO-accurate
predictions.

\begin{figure}[t!]
  \centering
  \includegraphics[width=0.47\textwidth]{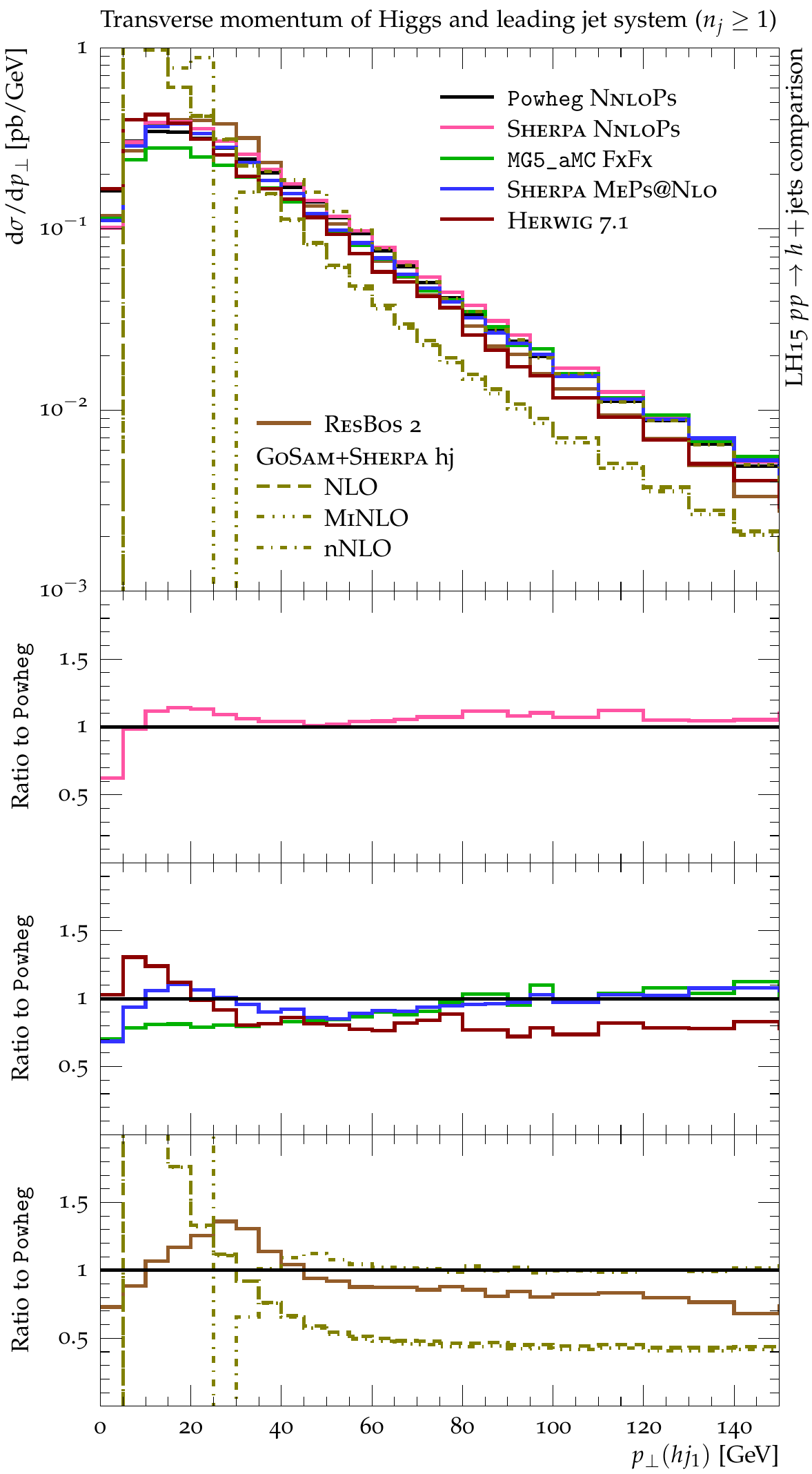}
  \hfill
  \includegraphics[width=0.47\textwidth]{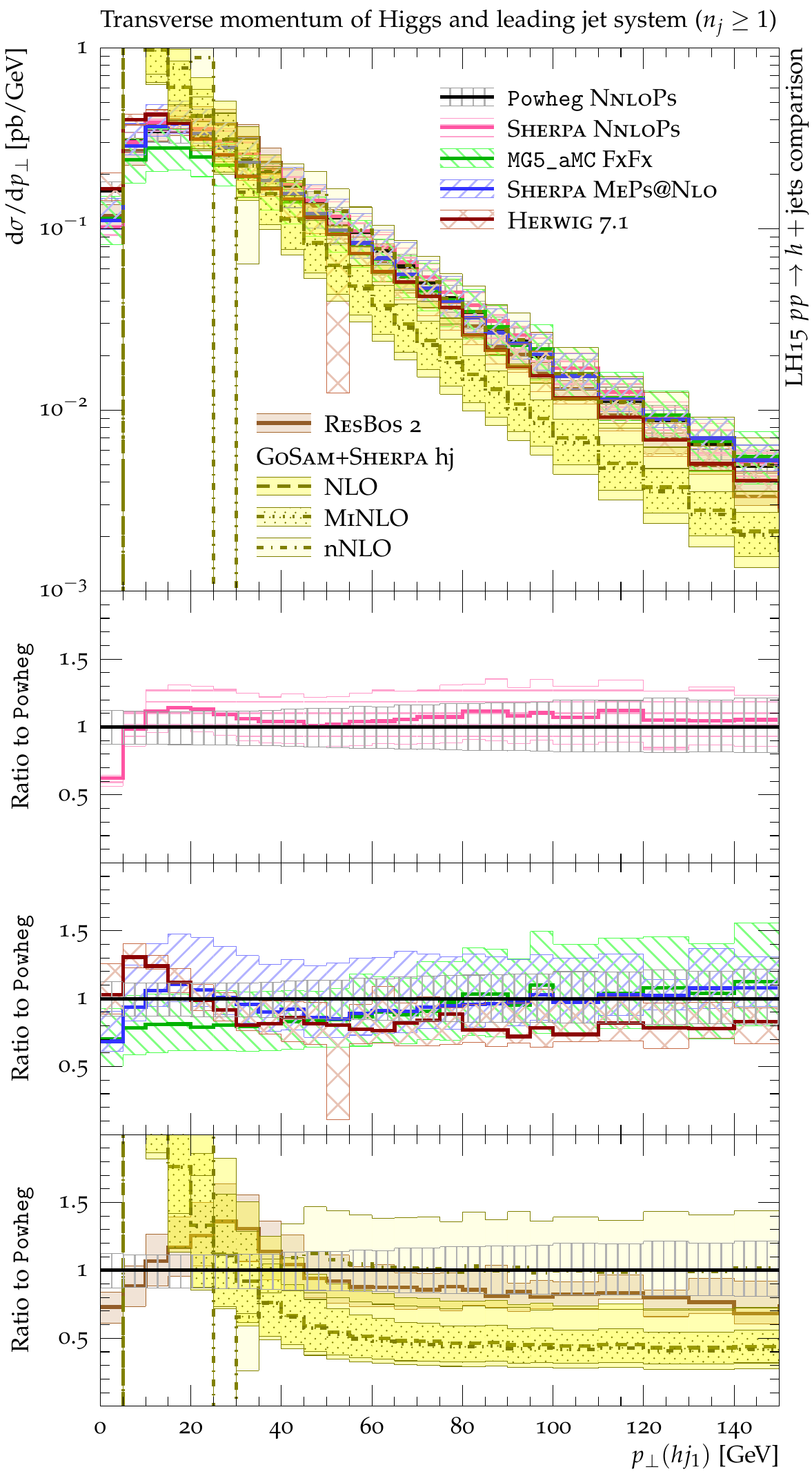}
  \caption{\label{fig:hjetscomp:results:1obs:hj_pt}%
    The transverse momentum of the Higgs-boson-leading-jet system in
    the presence of at least one jet. For better visibility, results
    are shown without (left) and with (right) theoretical
    uncertainties. The plot layout exactly corresponds to that of
    Figure~\ref{fig:hjetscomp:results:1obs:j1y}, except for the
    extended $\hat y$-axis range in the ratio plots.}
\end{figure}

\begin{figure}[t!]
  \centering
  \includegraphics[width=0.47\textwidth]{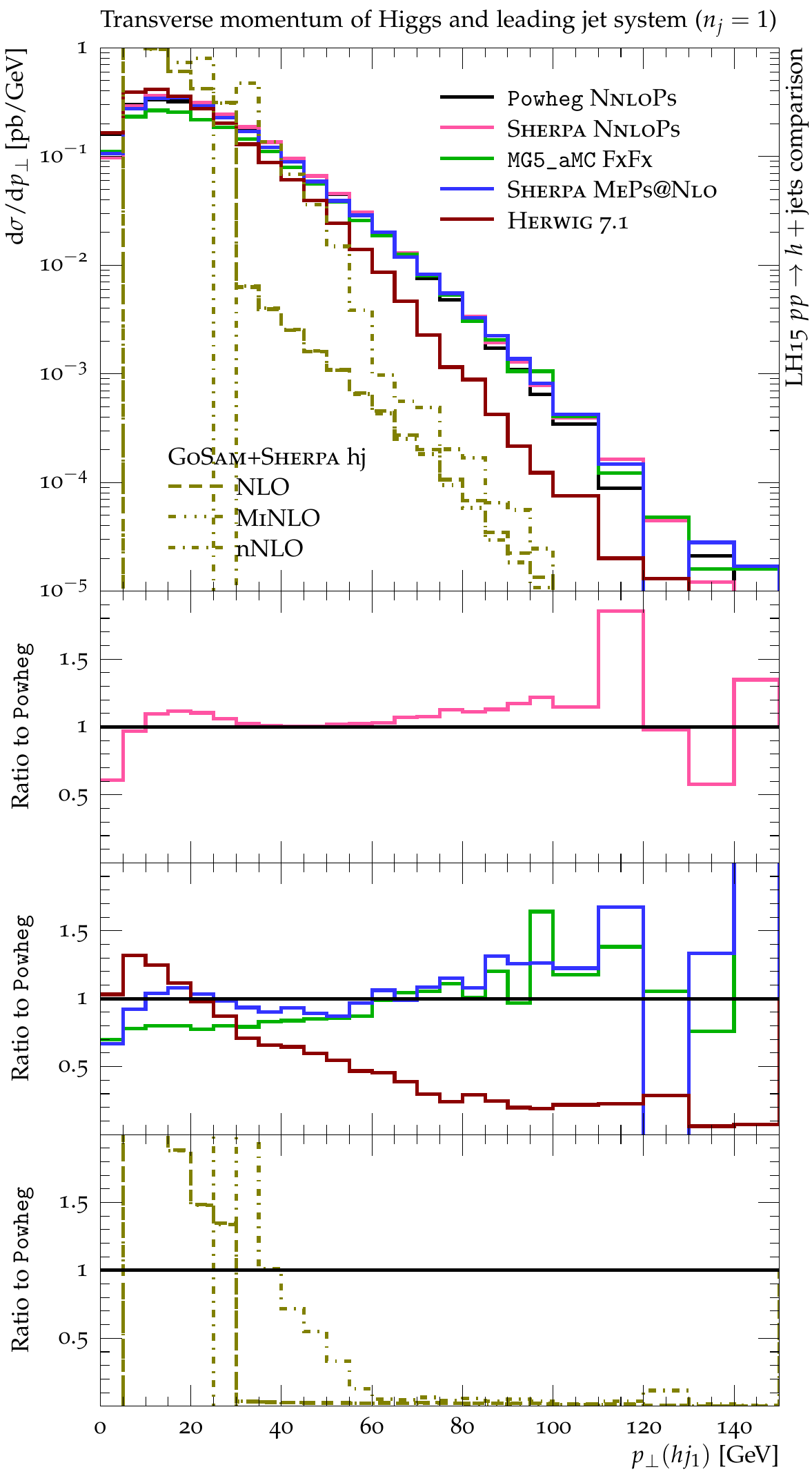}
  \hfill
  \includegraphics[width=0.47\textwidth]{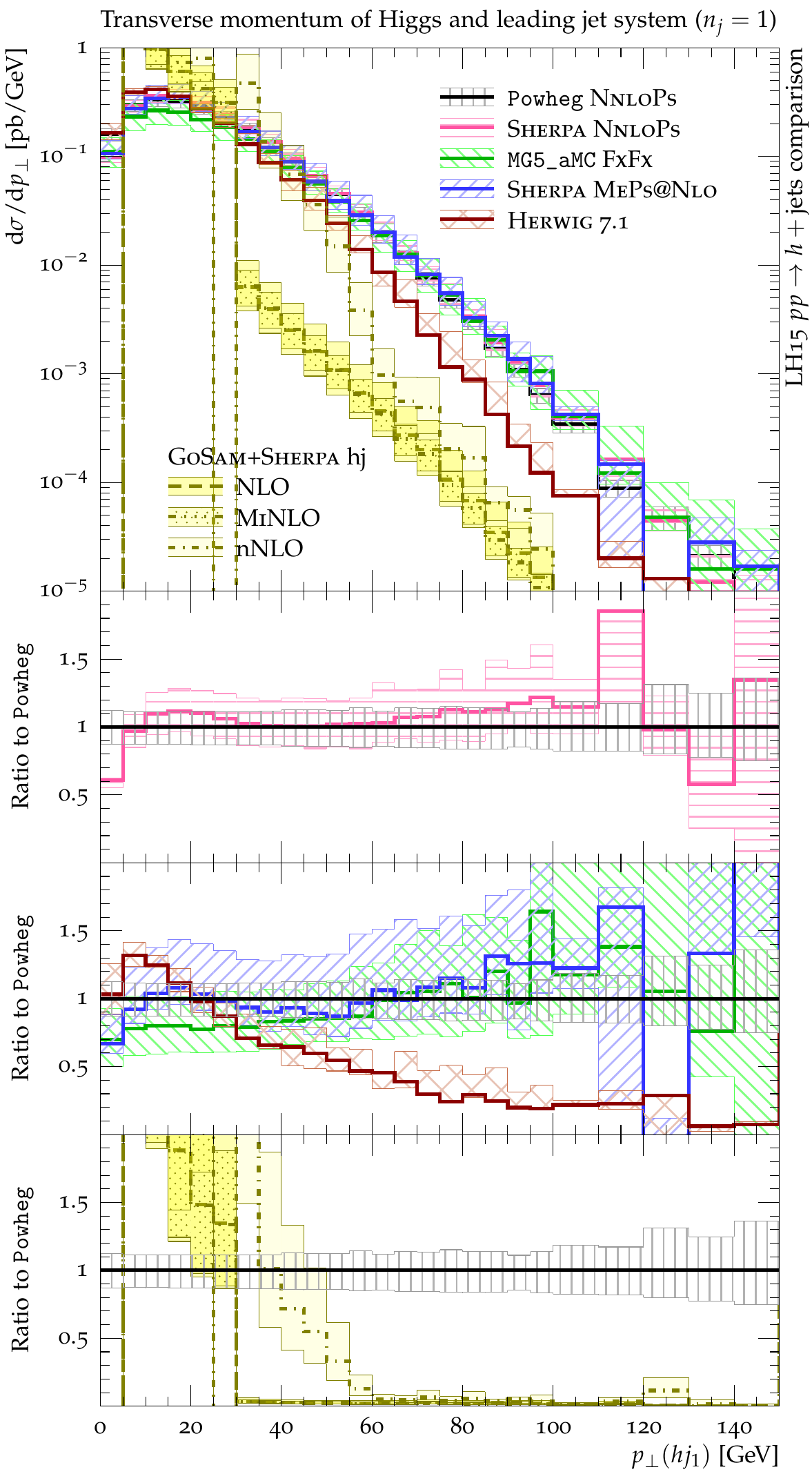}
  \caption{\label{fig:hjetscomp:results:1obs:hj_pt_excl}%
    The transverse momentum of the Higgs-boson-leading-jet system in
    the presence of exactly one jet. Again, results are shown without
    (left) and with (right) theoretical uncertainties as given by the
    different groups. Note that the plot layout corresponds exactly to
    that of Figure~\ref{fig:hjetscomp:results:1obs:hpt_excl}, except
    for the extended $\hat y$-axis range used in the ratio plots.}
\end{figure}

We finish this section by examining the results for the transverse
momentum of the Higgs-boson plus leading-jet system. In other words,
we are interested in studying the different descriptions of the
recoil of the $hj_1$ system. In the inclusive one-jet case depicted in
Figure~\ref{fig:hjetscomp:results:1obs:hj_pt}, the system may recoil
against a second jet (or secondary jets) plus soft radiation, while
for the exclusive jet scenario shown in
Figure~\ref{fig:hjetscomp:results:1obs:hj_pt_excl}, it is only soft
radiation that recoils against the $hj_1$ system. The latter case will
therefore be strongly affected by the level at which resummation is
taken into account. Formally, for the observable in question, the
predictions of highest accuracy are facilitated by the NLO merging
approaches as they provide an NLO description of the second jet; all
other predictions are only LO accurate in the second jet, i.e.~larger
differences can be expected. In the exclusive ($=1$-jet) case, all
predictions that include parton showering operate at the same level of
precision while the fixed-order calculations cannot do anything but
fail in describing the $p_\perp(hj_1)$ distribution.

For the $h\,+\!\ge\!1$-jet events, differences of $\mathcal{O}(30\%)$ are
observed among the ME+PS predictions below the jet threshold, while
there is better agreement at higher $p_\perp$ values, where again
\hjetscompHerwig turns out to be on the softer side. The \hjetscompPowheg and \hjetscompSherpa \hjetscompNNLOPS 
curves surprisingly fit right in with the ME+PS results throughout 
the spectrum. The mostly comparable
behavior of the \hjetscompNNLOPS results to those obtained from NLO merging is
not what one would expect a priori, however the NLO matching (\hjetscompPowheg- 
and \hjetscompSMCatNLO-type, respectively) of the $hj$ configurations transfers 
a differential $K$-factor to their second parton emission such that 
the \hjetscompLOPS treatment of the $h\,+\!\ge\!2$-jet rate
obtains a more appropriate normalization. The $hj$ NLO calculations
(i.e.~the pure and \hjetscompMinlo reweighted \hjetscompGoSam{}+\hjetscompSherpa results) cannot
compete with this performance since they miss an adequate description 
of the second jet giving recoil to the $hj$ system.
Correspondingly, the $p_\perp(hj_1)$ observable is described poorly
with values clearly overshooting below the jet threshold due to the
missing Sudakov suppression and undershooting by 60\% beyond
$p_\perp=50\,\hjetscompgev$ due to missing higher (than two-) jet multiplicity
contributions. It is interesting to see that the \hjetscompLoopsim procedure
lifts this large discrepancy in the $p_\perp$ tail. This is evidence
that an adequate description of a second and third jet is sufficient
to describe this observable in this regime. Thus, the good agreement
with the ME+PS results is largely driven by the $hjj$ NLO component
used to build the nNLO prediction for the $h\,+\!\ge\!1$-jet
process. However, as a result of the cut-off dependence of the
procedure nothing can be said about the $p_\perp<25\,\hjetscompgev$ region.
On the contrary, \hjetscompResbos predicts this region with NLL precision but
reverts to a LO description in the tail of the distribution leveling
off about 30\% above the \hjetscompGoSam{}+\hjetscompSherpa result as a consequence of
employing a lower scale (that is $\mu=\tfrac{1}{2}m_h$). Note that for
the jet associated Higgs boson production, the transverse momentum of
the $hj_1$ system constitutes exactly the $q_\text{T}$ quantity that is to be
resummed by \hjetscompResbos imposing the constraint $p_\perp(hj_1)<p_\perp(j_1)$,
which is to ensure that $j_1$ is indeed the leading jet (as any other
jet is integrated out in the resummation formalism used by \hjetscompResbos).
In addition, terms of the form $\ln(1/R^2)$ are also taken into
account by the resummation carried out in \hjetscompResbos and lead to a
somewhat broader, upward shifted Sudakov peak.

Lastly, we comment on the uncertainties quoted by the different
calculations: the \hjetscompResbos, \hjetscompGoSam{}+\hjetscompSherpa NLO and \hjetscompMinlo envelopes
have an appropriate size reflecting the underlying LO nature of the
$p_\perp(hj_1)$ prediction above the jet threshold. The \hjetscompLoopsim
procedure leaves us with a somewhat wider band as it involves two
real-emission (LO-like) contributions ($hjj$ and $hjjj$), which impact
the $p_\perp(hj_1)$ observable.
\hjetscompMGaMC and \hjetscompSherpa \hjetscompMEPSatNLO on the one side and \hjetscompHerwig on the other
side produce NLO variations that are fairly different in size.
However, the \hjetscompHerwig as well as the \hjetscompPowheg and \hjetscompSherpa \hjetscompNNLOPS envelopes 
are most likely underestimated; in particular the \hjetscompPowheg and \hjetscompSherpa \hjetscompNNLOPS 
bands do not behave as
expected from a LO variation for above-jet-threshold $p_\perp(hj)$. 

The exclusive (exactly one jet) case for the $p_\perp(hj_1)$
observable is shown in Figure~\ref{fig:hjetscomp:results:1obs:hj_pt_excl}.
Apart from the \hjetscompNNLOPS outcomes, there is a much greater divergence of
the predictions for exactly one jet, especially at high $p_\perp$.
Recall that for this situation, the recoil is generated only from soft emissions, and
it is clear that a highly exclusive distribution such as the one in
question serves as a stress test for the ME+PS as well as the \hjetscompNNLOPS
predictions (in fact any parton shower or resummed prediction). For
the same reason, caution has to be taken in interpreting the 
uncertainties. The current case is similar to the case for the Higgs
boson $p_\perp$ distribution with no jets and it is no surprise that
different approaches can lead to different answers. Most notably, we
observe \hjetscompNNLOPS predictions that are in slightly worse agreement as
compared to the inclusive case, and the expected complete failure of the
\hjetscompGoSam{}+\hjetscompSherpa results (including the \hjetscompLoopsim result where the
one-jet requirement removes the effect that yielded the
improvement in the inclusive case),%
\footnote{Owing to the kinematic constraints on the jets, harder
  radiation that goes forward does not get identified as a jet;
  these contributions actually form the (naively unexpected)
  $p_\perp>30\,\hjetscompgev$ tail of the \hjetscompGoSam{}+\hjetscompSherpa results though the
  mechanism is highly suppressed.}
and the severe decline of the \hjetscompHerwig differential cross section
dropping to about 25\% wrt.~the \hjetscompPowheg result at $p_\perp\sim100\,\hjetscompgev$.

\subsubsection{Dijet observables}
\label{sec:hjetscomp:results:2jobs}

\begin{figure}[t!]
  \centering
  \includegraphics[width=0.47\textwidth]{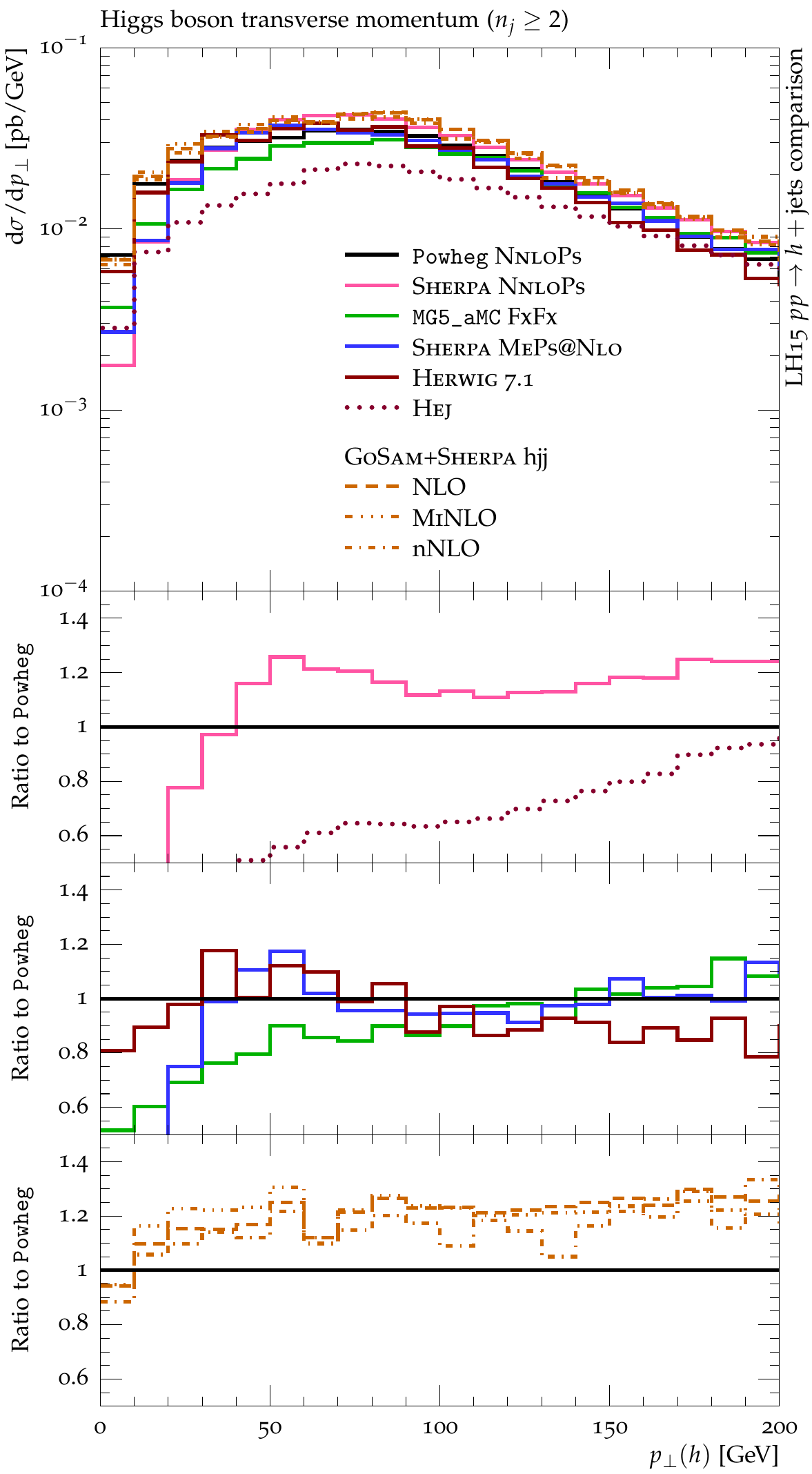}
  \hfill
  \includegraphics[width=0.47\textwidth]{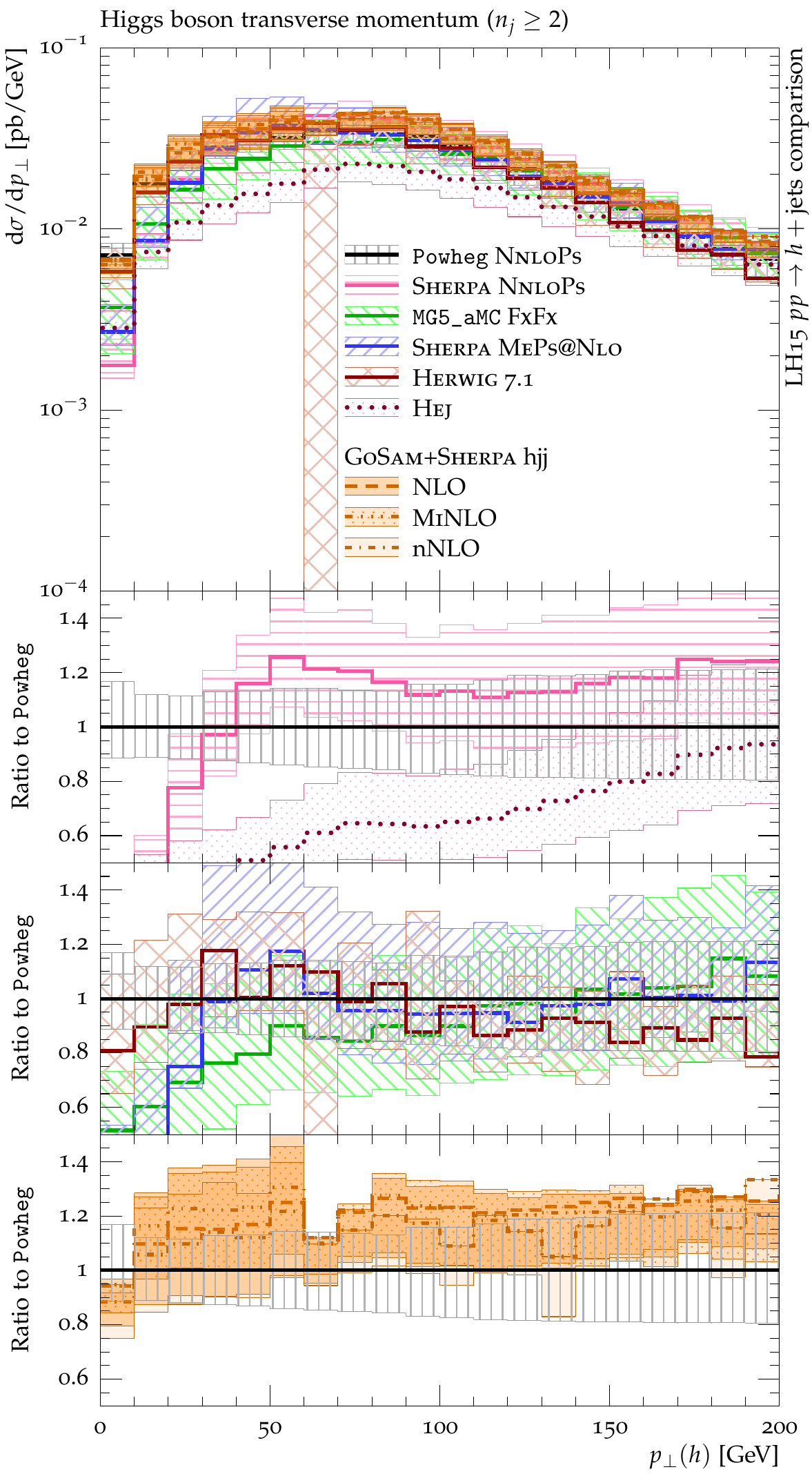}
  \caption{\label{fig:hjetscomp:results:2obs:hpt}%
    The transverse momentum of the Higgs boson in the presence of at
    least two jets without (left) and with (right) uncertainties,
    supplemented by three ratio plots using the reference result as
    obtained from \hjetscompPowheg's \hjetscompNNLOPS calculation for $h$ production.
    The predictions are grouped -- from top to bottom -- according to
    the categories \hjetscompNNLOPS $h$ production, ME+PS merging at NLO (at
    least up to two jets) and NLO fixed-order $hjj$ production. \hjetscompHej's
    prediction is added to the first, the \hjetscompNNLOPS subpanel.}
\end{figure}

Moving to topologies with one more jet in the final state, we discuss a 
number of $h+2$-jet observables in this section. It is important
to understand, in detail, the level of (dis-)agreement among the
available predictions since the $h+2$-jet gluon fusion contribution
constitutes the major irreducible background in any LHC Run-II analysis
targeting the prominent vector boson fusion channel. Following the layout 
of the previous sections, we compare predictions from three
different categories with each other: \hjetscompNNLOPS $h$ production,
\hjetscompMEPSatNLO $h+\text{jets}$ production and fixed-order (and related)
$hjj$ production at $\mathcal{O}(\alpha_\mathrm{s}^5)$. As before we have
provided one ratio plot for each category in order to enhance the
readability of our plots; common to all figures is the use of \hjetscompPowheg's
\hjetscompNNLOPS prediction to serve as the reference result. Note that for
each category, the accuracy with which the $h+2$-jet events are
described is different: the \hjetscompNNLOPS predictions are at the \hjetscompLOPS level
while the multijet merged results incorporate the precision given by \hjetscompNLOPS
matching. \hjetscompHej generates predictions derived from the behavior of QCD
in the high-energy limit, starting from a LO accurate $h+2$-jet configuration.
With two or more jets in the final state,
\hjetscompHej can provide meaningful predictions for the first time. As
for earlier cases, the last category contains NLO-accurate predictions
solely, free of any parton showering.

\begin{figure}[t!]
  \centering
  \includegraphics[width=0.47\textwidth]{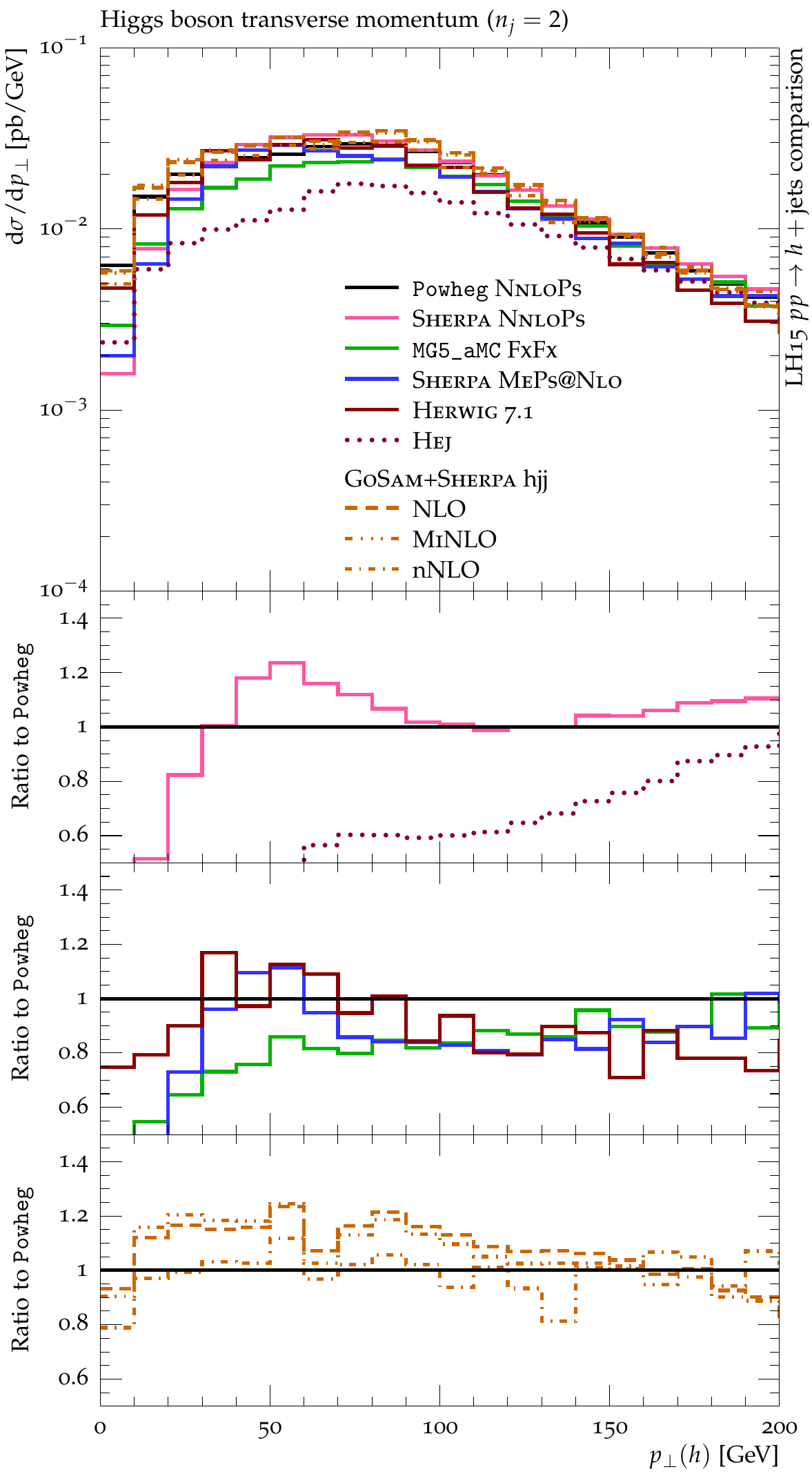}
  \hfill
  \includegraphics[width=0.47\textwidth]{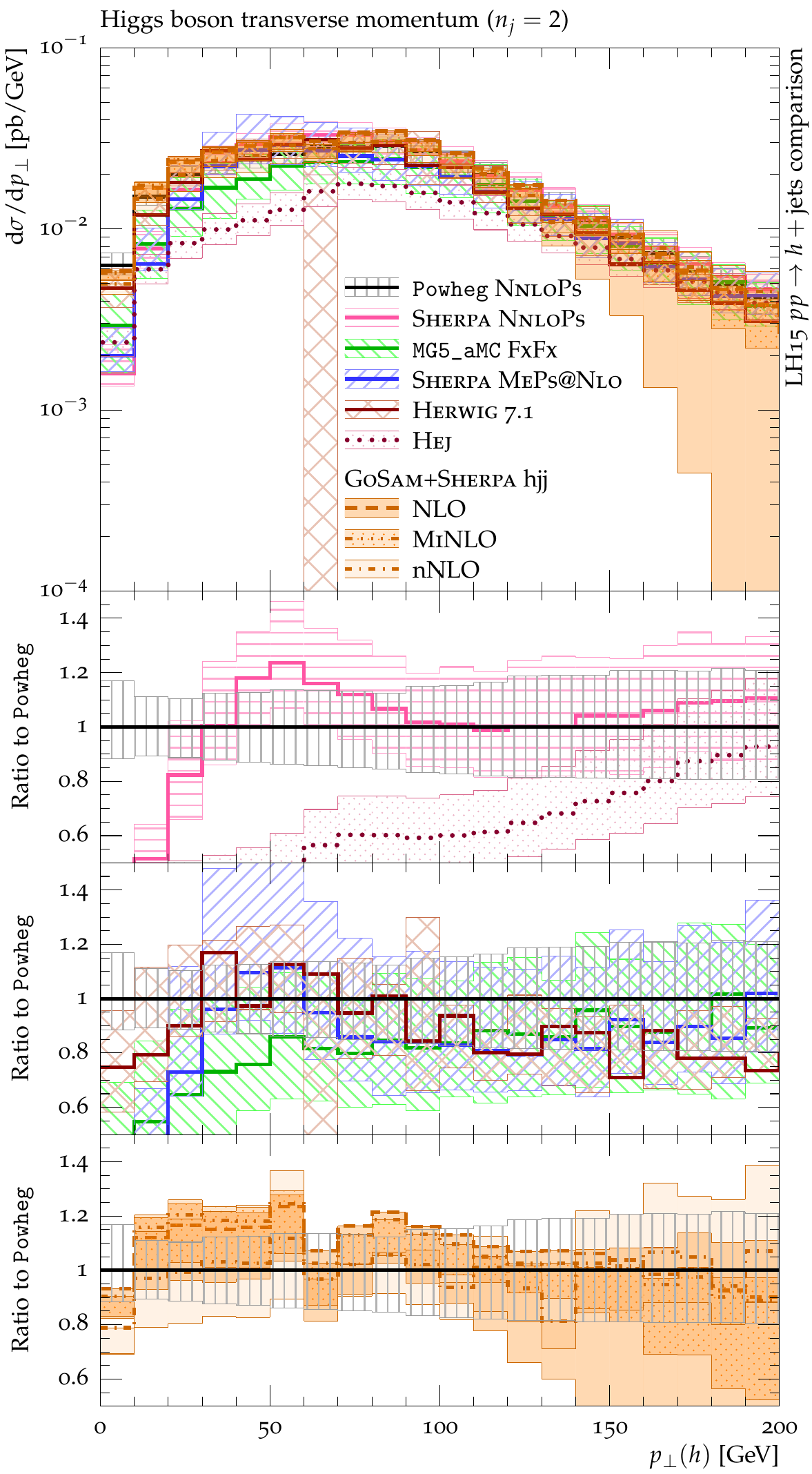}
  \caption{\label{fig:hjetscomp:results:2obs:hpt_excl}%
    The transverse momentum of the Higgs boson in the presence of 
    exactly two jets without (left) and with (right) uncertainties,
    supplemented by three ratio plots using the reference result as
    obtained from \hjetscompPowheg's \hjetscompNNLOPS calculation for $h$ production.
    The predictions are grouped -- from top to bottom -- according to
    the categories \hjetscompNNLOPS $h$ production, ME+PS merging at NLO (at
    least up to two jets) and NLO fixed-order $hjj$ production. \hjetscompHej's
    prediction is added to the first, the \hjetscompNNLOPS subpanel.}
\end{figure}

The Higgs boson $p_\perp$ spectrum, in the presence of at least two
jets, is shown in Figure~\ref{fig:hjetscomp:results:2obs:hpt}.
Although all calculations agree on the position of the maximum
of the distribution (situated at $p_\perp\gtrsim 60\,\hjetscompgev$, i.e.~twice the 
jet threshold),\footnote{Note that around this two-jet threshold, we
  again find remnants of the Sudakov shoulder effect affecting the
  fixed-order predictions.}
varying behavior is observed for both larger and smaller 
transverse momenta. Examining first the region where $p_\perp\gtrsim 
60\,\hjetscompgev$, good agreement between all multijet merged calculations and 
\hjetscompPowheg is found. Only \hjetscompHerwig predicts a somewhat more rapidly falling 
distribution. \hjetscompSherpa \hjetscompNNLOPS, due to its scale choice, exhibits a 
more or less constant $\mathcal{O}(+20\%)$ offset with respect to the
\hjetscompPowheg \hjetscompNNLOPS, but remains well within their respective uncertainties.
In comparison to all the NLO uncertainty bands, the \hjetscompNNLOPS envelopes
are expected to be larger reflecting the loss of one order in accuracy.
A clear difference however is not found indicating that the \hjetscompNNLOPS
estimates are too optimistic. From this point of view, the larger
difference seen between the two \hjetscompNNLOPS predictions is no surprise at
all -- rather typical for comparisons at the \hjetscompLOPS level, and two
different parton showers at work. The fixed-order
predictions (NLO, \hjetscompMinlo and \hjetscompLoopsim) agree well in this region with
one another, but surpass the \hjetscompPowheg reference and, more importantly,
the multijet merged calculations with the same accuracy, by about 
20\%. This deviation is just covered by the edge of the uncertainty
bands associated with either of the NLO calculations. Finally, \hjetscompHej
clearly produces the hardest spectrum in this region, exhibiting a
considerably different slope with respect to all other calculations,
albeit it starts out from an approximately 40\% lower cross section at
$p_\perp\approx 60\,\hjetscompgev$.

In the region below the peak, $p_\perp\lesssim 60\,\hjetscompgev$, the various
calculations are more widely spread. This is expected as effects from
parton showering have a larger impact here, but none of the considered
shower Monte Carlos in Figure~\ref{fig:hjetscomp:results:2obs:hpt}
works at a higher level regarding the resummation precision. We can
readily distinguish between two topologies if we assume the jets to be
produced near their $p_\perp$ threshold: while for
$p_\perp(h)>30\,\hjetscompgev$, both the leading and subleading jet have to be in
the same hemisphere opposite the Higgs boson, for
$p_\perp(h)<30\,\hjetscompgev$, the subleading jet has to cross over into the
Higgs boson's hemisphere opposite the leading jet.
Because of  parton shower effects, deviations among the predictions
are greater in the former region, but it is the latter region that
receives the larger resummation corrections, as both jets recoil mainly
against each other (to form a jet-balanced configuration) and soft
radiation off them has a large influence on the small Higgs boson
transverse momentum. An important role is also played by the
assignment of scales in the context of multijet merging.
To assign a meaningful history in this regime, in particular
for sufficiently hard jets, the clustering algorithm that is needed to
define the local CKKW or \hjetscompMinlo scales must allow for the possibility
of Higgs boson radiation off a dijet process. In consequence, the value
of the scales increases and the cross section is reduced, explaining
the identical behavior of both the \hjetscompSherpa \hjetscompMEPSatNLO and \hjetscompSherpa
\hjetscompNNLOPS predictions.
The \hjetscompHerwig, \hjetscompMGaMC and \hjetscompHej predictions are largely similar in shape,
but offset as a result of the larger central scale choice in \hjetscompMGaMC
and the LO accuracy of the total cross section in \hjetscompHej. \hjetscompPowheg
\hjetscompNNLOPS gives the slowest decline of the differential cross section as
$p_\perp\to0$.
The various fixed-order predictions are somewhat more widely spread than
for large $p_\perp$ (as indicated by the wider uncertainty bands), and
start to exhibit a slope wrt.~the \hjetscompPowheg reference, retaining however
a more or less constant ratio wrt.~\hjetscompHerwig and \hjetscompMGaMC. In the
$p_\perp(h)$ region below $30\,\hjetscompgev$, the global scale setting of the
fixed-order calculations is certainly not flexible enough to deal with
the jet-balanced configurations in a similar way as done by \hjetscompSherpa.

Figure \ref{fig:hjetscomp:results:2obs:hpt_excl} displays the 
Higgs boson's transverse momentum in the presence of exactly two jets. 
Apart from the overall reduction of the cross section, the general 
features as observed in Figure~\ref{fig:hjetscomp:results:2obs:hpt}
for each generator are reproduced. There are, however, two notable
exceptions. The decrease in the \hjetscompPowheg \hjetscompNNLOPS prediction at large $p_\perp(h)$ 
is less, as compared to
the other generators. On the one hand, it now overshoots the multijet
merged predictions by about 10-20\%, but on the other hand it achieves
a much better agreement with \hjetscompSherpa \hjetscompNNLOPS. This, however, is not
very surprising given that the jet veto (on the third jet) in the
\hjetscompNNLOPS calculations is only described at parton shower level, lacking
any matrix element input. The latter is mandatory for a good
description of the rejected $30\,\hjetscompgev$ emission. As seen before, the
fixed-order predictions start to lose their perturbative stability when
the $p_\perp$ ratio between the observed Higgs boson and the rejected
jet turns out to be large. This effect is more pronounced for the
\hjetscompMinlo procedure because it generates lower scale values on
average. In the \hjetscompLoopsim approach, perturbative stability is
improved as a result of the inclusion of $hjjj$ information.

\begin{figure}[t!]
  \centering
  \includegraphics[width=0.47\textwidth]{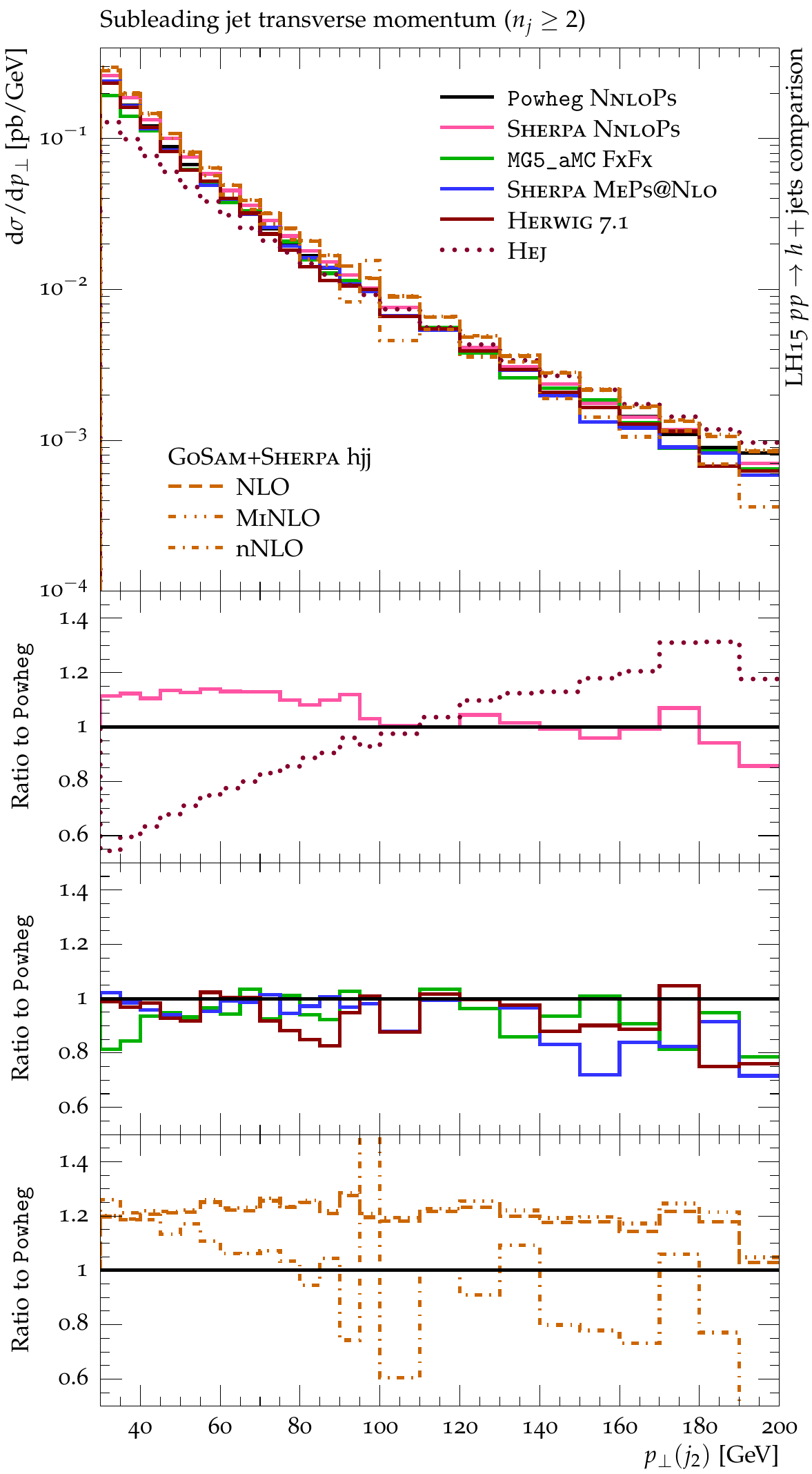}
  \hfill
  \includegraphics[width=0.47\textwidth]{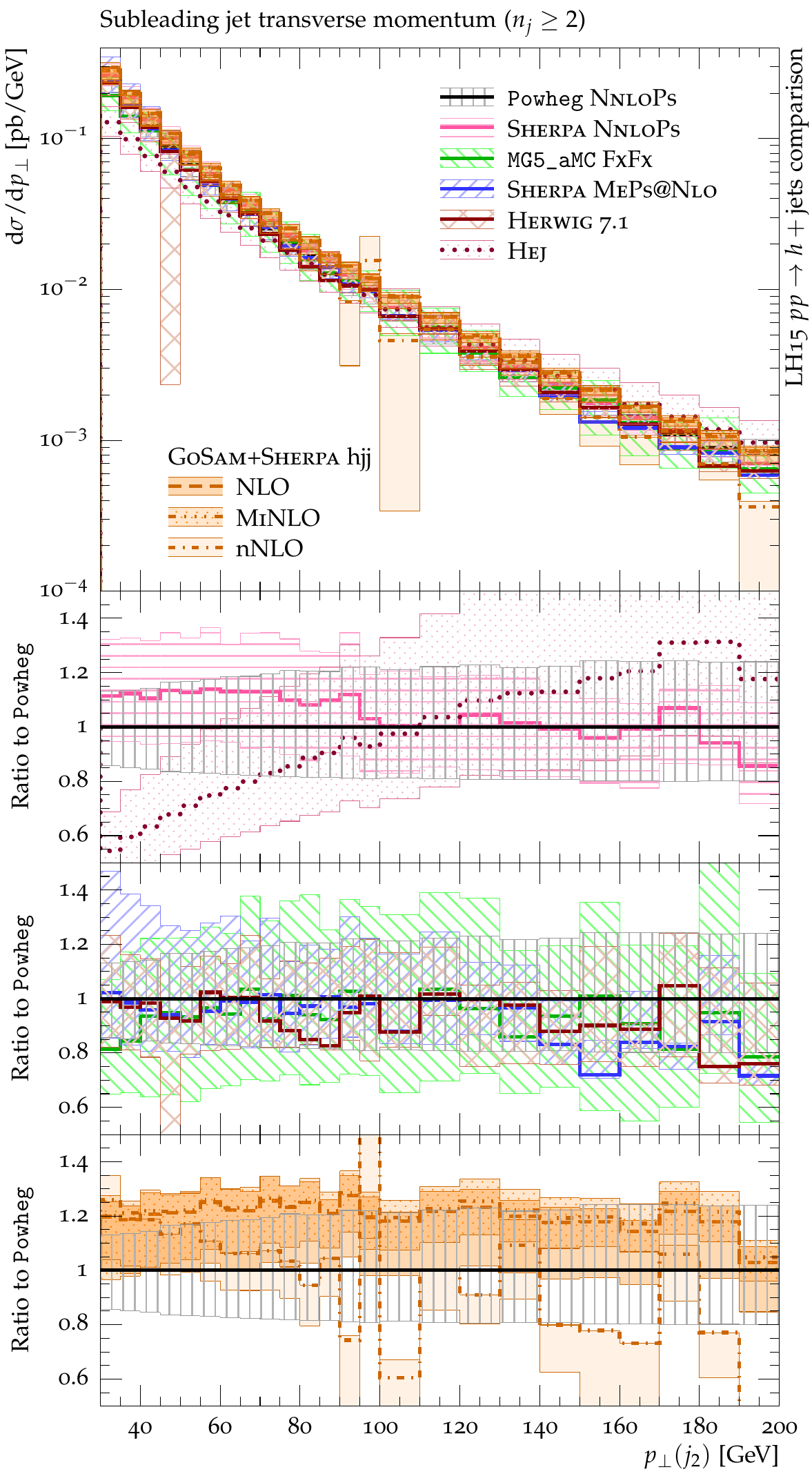}
  \caption{\label{fig:hjetscomp:results:2obs:jet2_pt}%
    The subleading jet $p_\perp$ for $h\,+\!\ge\!2$-jets production shown
    without (left) and with (right) theoretical uncertainties. The
    plot layout is the same as in the previous figure.}
\end{figure}

\begin{figure}[t!]
  \centering
  \includegraphics[width=0.47\textwidth]{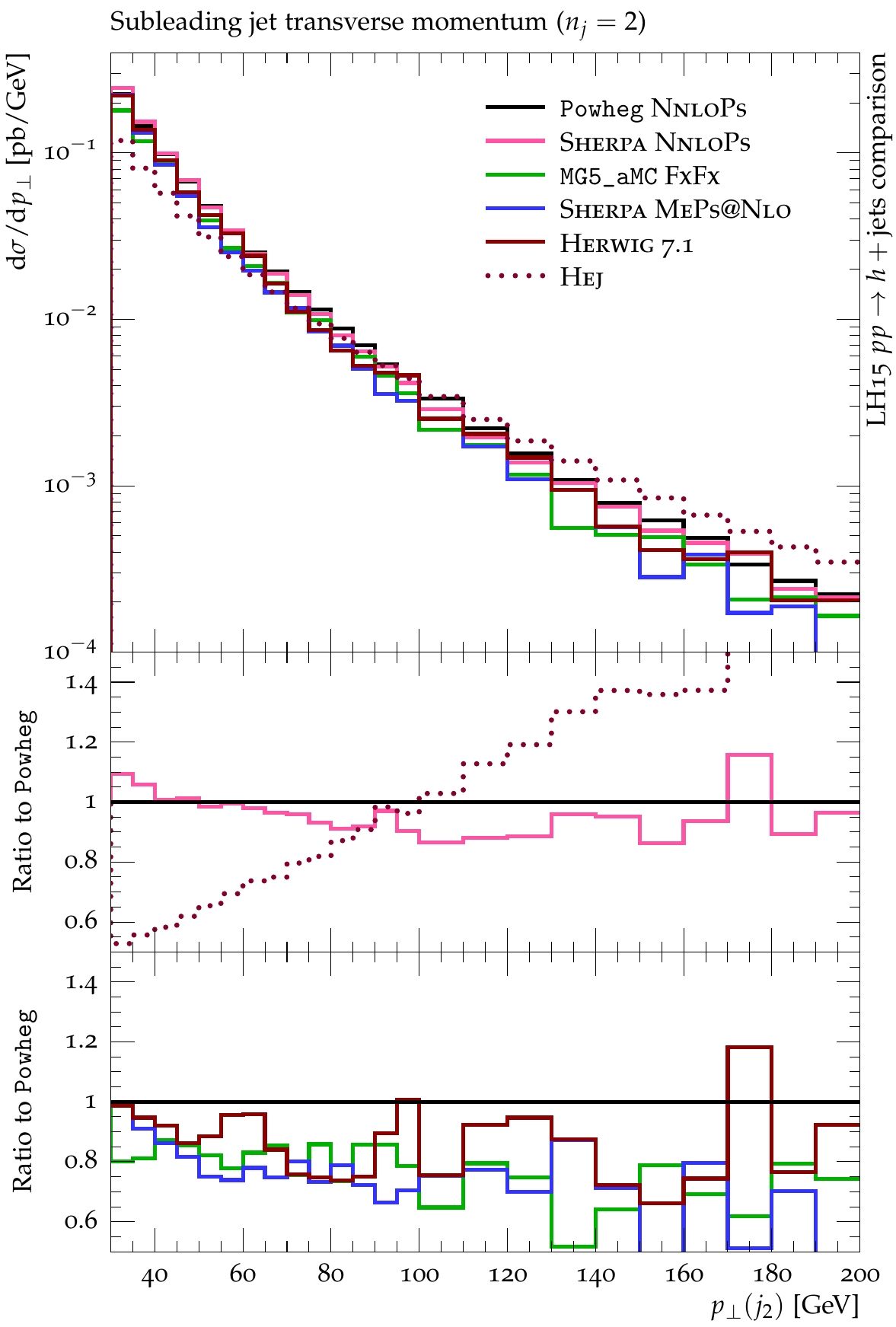}
  \hfill
  \includegraphics[width=0.47\textwidth]{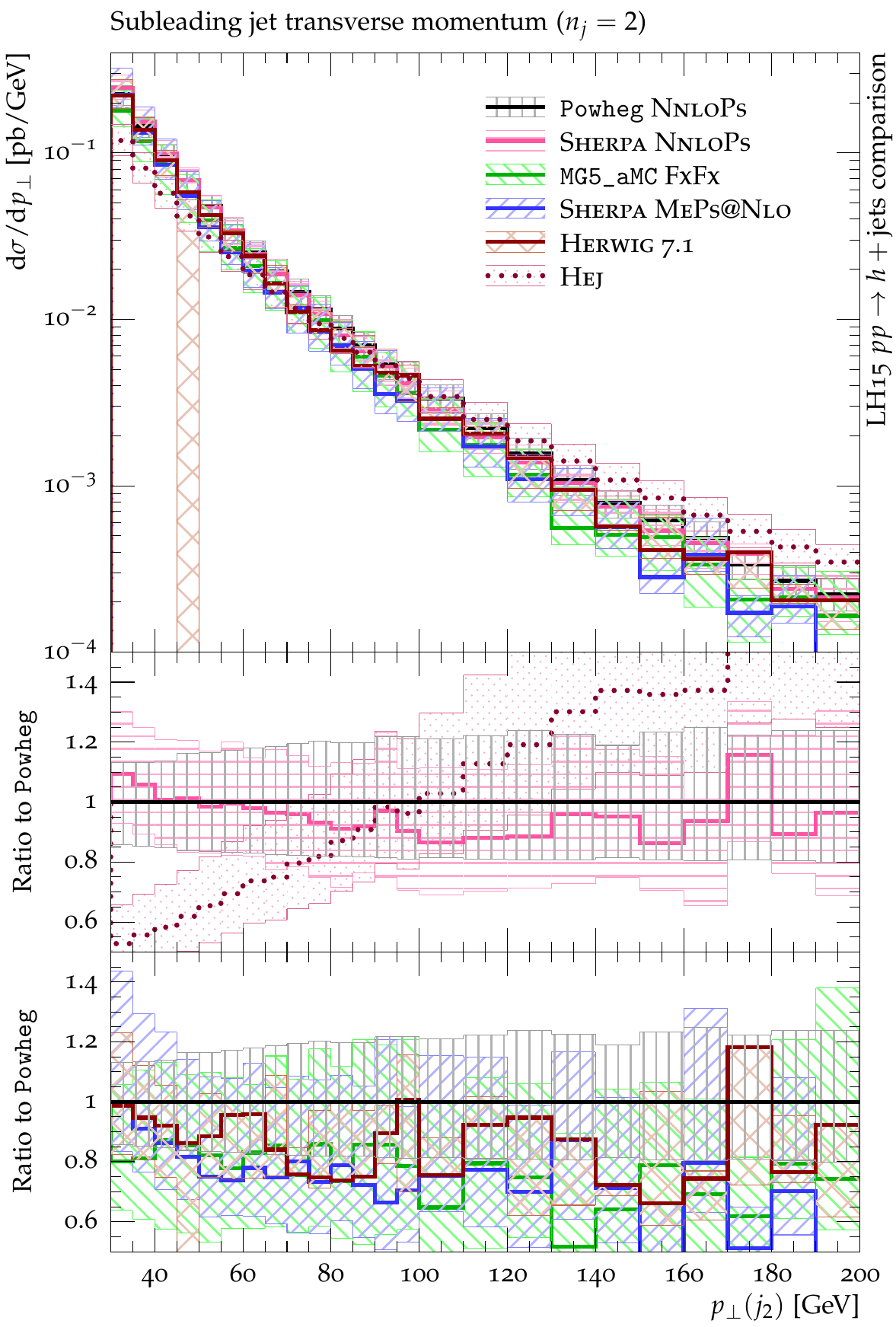}
  \caption{\label{fig:hjetscomp:results:2obs:jet2_pt_excl}%
    The subleading jet $p_\perp$ for exclusive $h+2$-jets production
    shown without (left) and with (right) theoretical uncertainties.
    Again, the layout is similar to that of
    Figure~\ref{fig:hjetscomp:results:2obs:hpt} except for
    dropping the results of the NLO fixed-order $hjj$ production and
    the associated ratio subpanel.}
\end{figure}

The subleading jet $p_\perp$ spectra for $h+\ge2$ and $h$ plus exactly
two jets are shown in Figures~\ref{fig:hjetscomp:results:2obs:jet2_pt}
and \ref{fig:hjetscomp:results:2obs:jet2_pt_excl}, respectively. A
comparison between the second jet results (of the
$n_j\ge2$ and $n_j=2$ case) with the first jet results (of the
$n_j\ge1$ and $n_j=1$ case, depicted in
Figures~\ref{fig:hjetscomp:results:1obs:j1pt} and
\ref{fig:hjetscomp:results:1obs:j1pt_excl}, respectively) reveals a
rather consistent picture. The agreement among the ME+PS
predictions, and between the ME+PS and the \hjetscompPowheg predictions, is
slightly better than in the case of the leading jet. For $n_j=2$
(cf.~Figure~\ref{fig:hjetscomp:results:2obs:jet2_pt_excl}), \hjetscompPowheg's
\hjetscompNNLOPS approach predicts harder subleading jets than the others do,
apart from \hjetscompHej. \hjetscompMGaMC, \hjetscompSherpa and \hjetscompHerwig lie in reasonable
agreement with each other, but systematically lower than \hjetscompPowheg by
about 10-20\%. The agreement in the inclusive (i.e.~$n_j\ge2$) case
eventually results from a compensating effect, since all the different
methods for NLO merging give larger three-jet rates wrt.~\hjetscompPowheg. The
\hjetscompGoSam{}+\hjetscompSherpa $hjj$ NLO predictions do not have a different shape,
but their rate is higher, which likely comes from choosing
$\tfrac{1}{2}\sum_T^{1/2}$ as the central scale; recall that in the
one-jet case this choice already produced large NLO cross sections, as
large as the NNLO one, and larger than the ME+PS ones. With two jets
in the final state, this effect is easily seen to be enhanced. The
\hjetscompLoopsim NNLO estimate for the $p_\perp(j_2)$ spectrum gradually
decreases wrt.~the other \hjetscompGoSam results, in a similar, slightly more
pronounced way to what we found for $p_\perp(j_1)$ in the $n_j\ge1$
case. 
\hjetscompLoopsim combines the NLO-accurate $hjj$ and $hjjj$ bins, and in doing
so  seems closer to the NLO-merged results, and therefore reproduces
some of their characteristics. The hardest spectrum is
delivered by \hjetscompHej even though its total rate is significantly lower. 
In the plot range, \hjetscompHej's prediction is consistent with the others due 
to overlapping uncertainties (LO for \hjetscompHej), but this statement gets
stretched for the exclusive two-jet scenario. Moreover, for $n_j=2$,
the   uncertainties reported here have to be considered with caution,
owing to the neglect of important resummation effects. However, the
statements made regarding error estimates when discussing 
$p_\perp(h)$ for $n_j\ge2$, carry over to the current case, as well as
for all other observables in this section.

\begin{figure}[t!]
  \centering
  \includegraphics[width=0.47\textwidth]{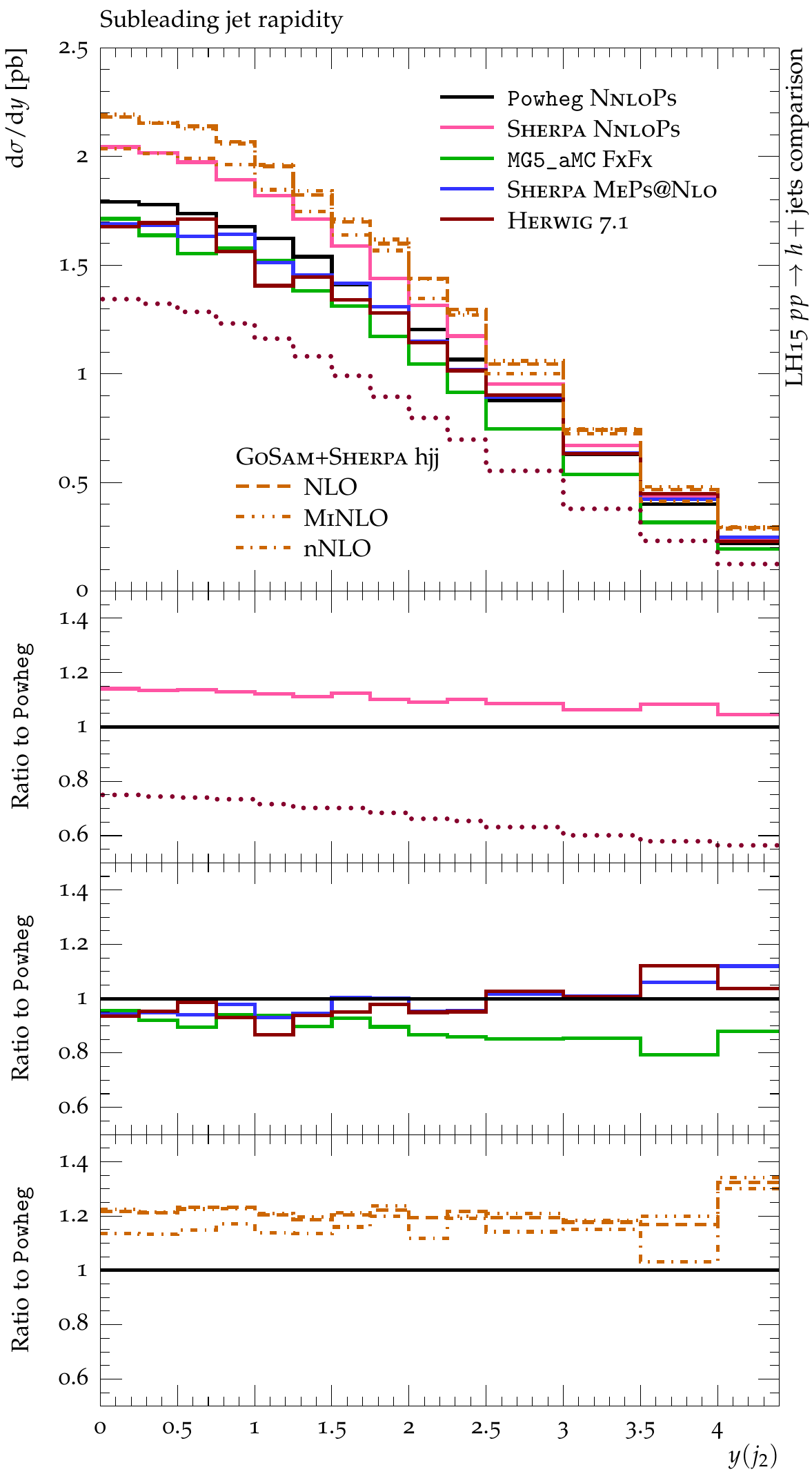}
  \hfill
  \includegraphics[width=0.47\textwidth]{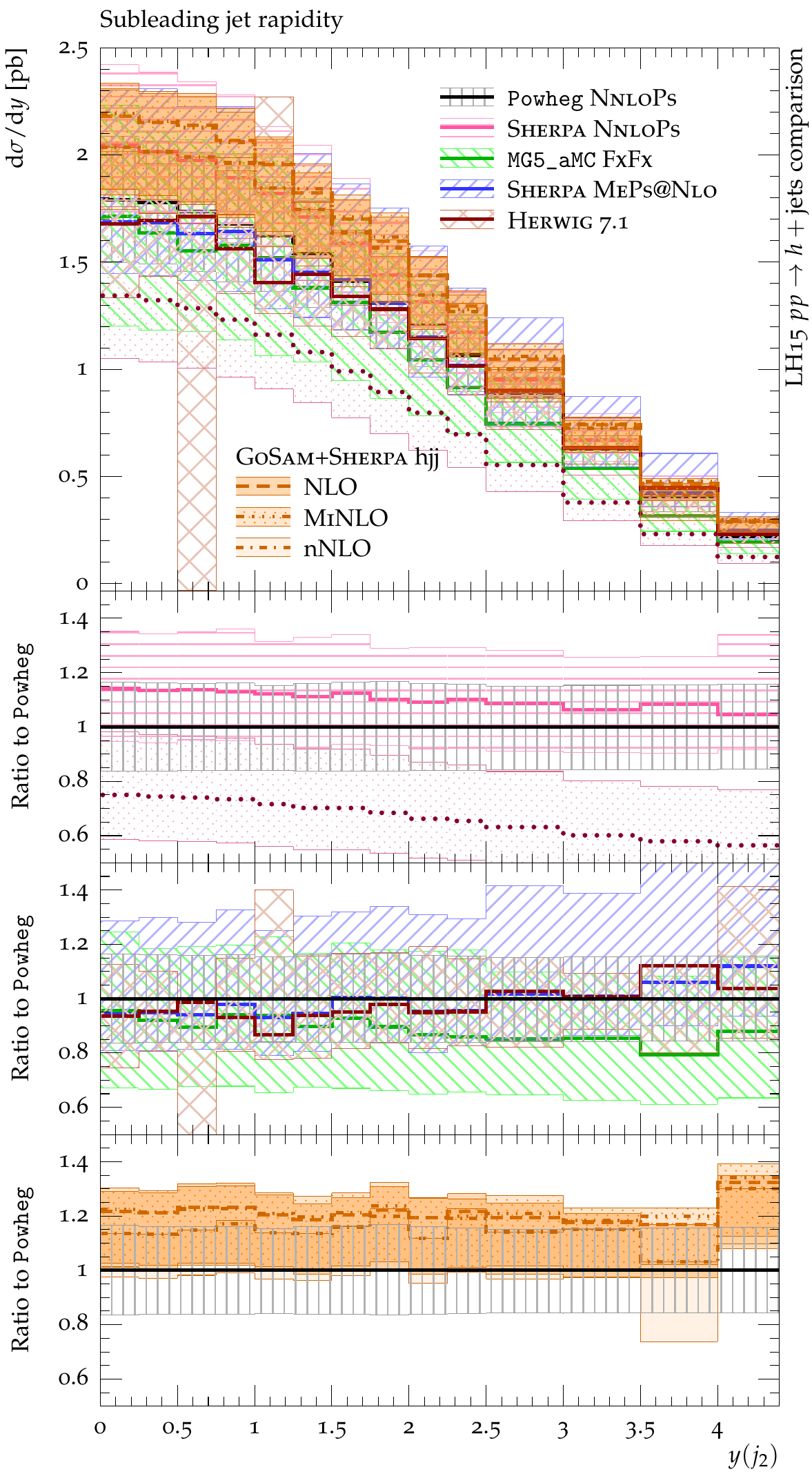}
  \caption{\label{fig:hjetscomp:results:2obs:j2y}%
    The rapidity separation between the leading and subleading jets
    for $h\,+\!\ge\!2$-jets production, shown without (left) and with
    (right) theoretical uncertainties. The plot layout is the same as
    the one used in Figure~\ref{fig:hjetscomp:results:2obs:hpt}.}
\end{figure}

Figure~\ref{fig:hjetscomp:results:2obs:j2y} depicts the rapidity
distribution of the subleading jet. The conclusion of this comparison is
more or less the same as for the leading jet, see
Figure~\ref{fig:hjetscomp:results:1obs:j1y}. Only minor differences 
between the computations are encountered, in particular concerning the
shape of the rapidity spectrum -- the larger effects are driven by the
different rate predictions. While \hjetscompSherpa \hjetscompNNLOPS and \hjetscompMGaMC prefer a
slightly more central production than the other calculations, the
rate difference previously observed between \hjetscompPowheg \hjetscompNNLOPS, \hjetscompMGaMC,
\hjetscompHerwig and \hjetscompSherpa \hjetscompMEPSatNLO on the one side, and \hjetscompSherpa \hjetscompNNLOPS and
the various \hjetscompGoSam{}+\hjetscompSherpa NLO calculations on the other side, is the
feature that most stands out when plotting the subleading jet rapidity.
The largest deviations are again delivered by \hjetscompHej; apart from the
lower total rate, \hjetscompHej predicts fewer subleading jets at large
rapidities than the other approaches. It again becomes apparent that
the quoted uncertainties of the \hjetscompNNLOPS calculations are likely
underestimated.

\begin{figure}[t!]
  \centering
  \includegraphics[width=0.47\textwidth]{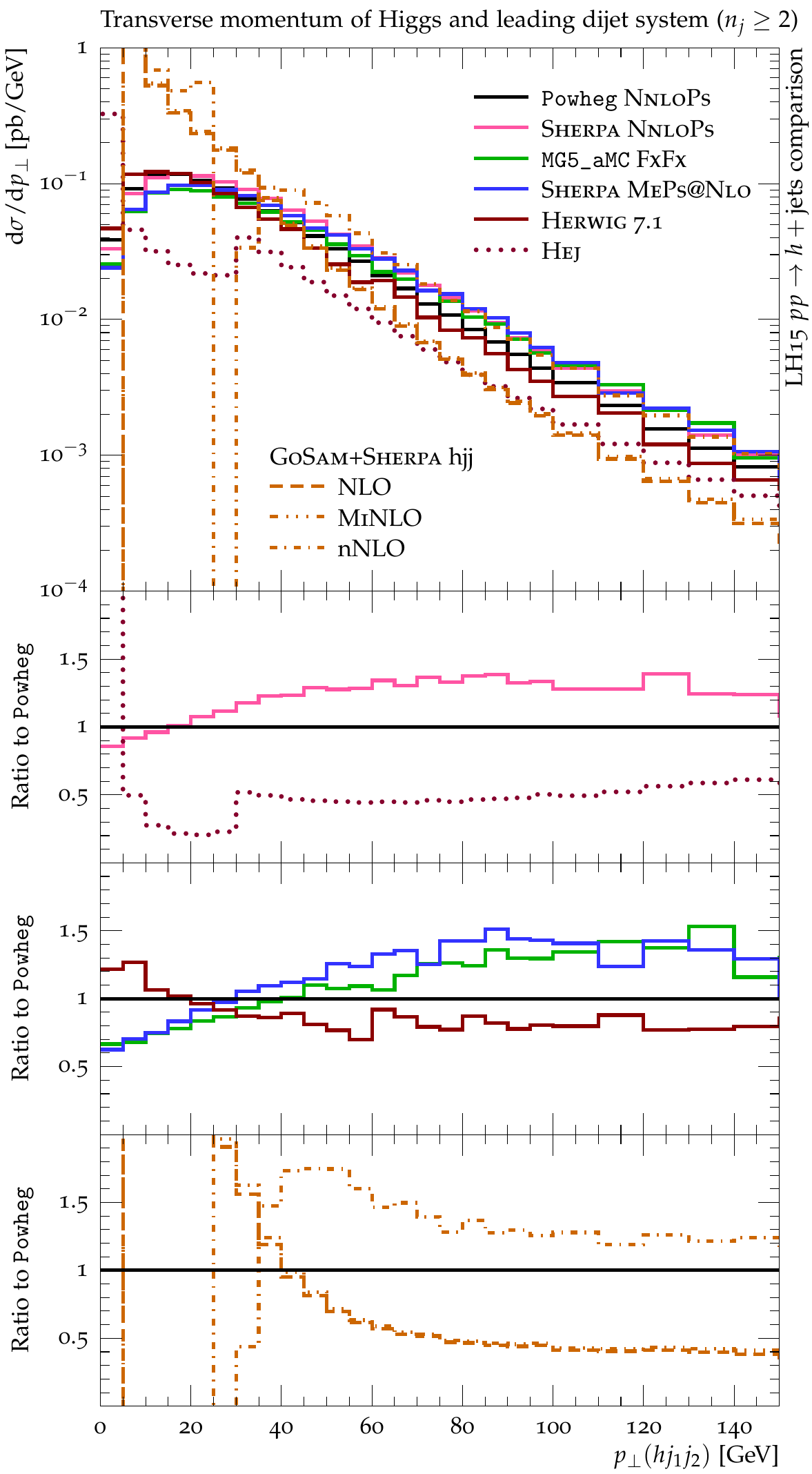}
  \hfill
  \includegraphics[width=0.47\textwidth]{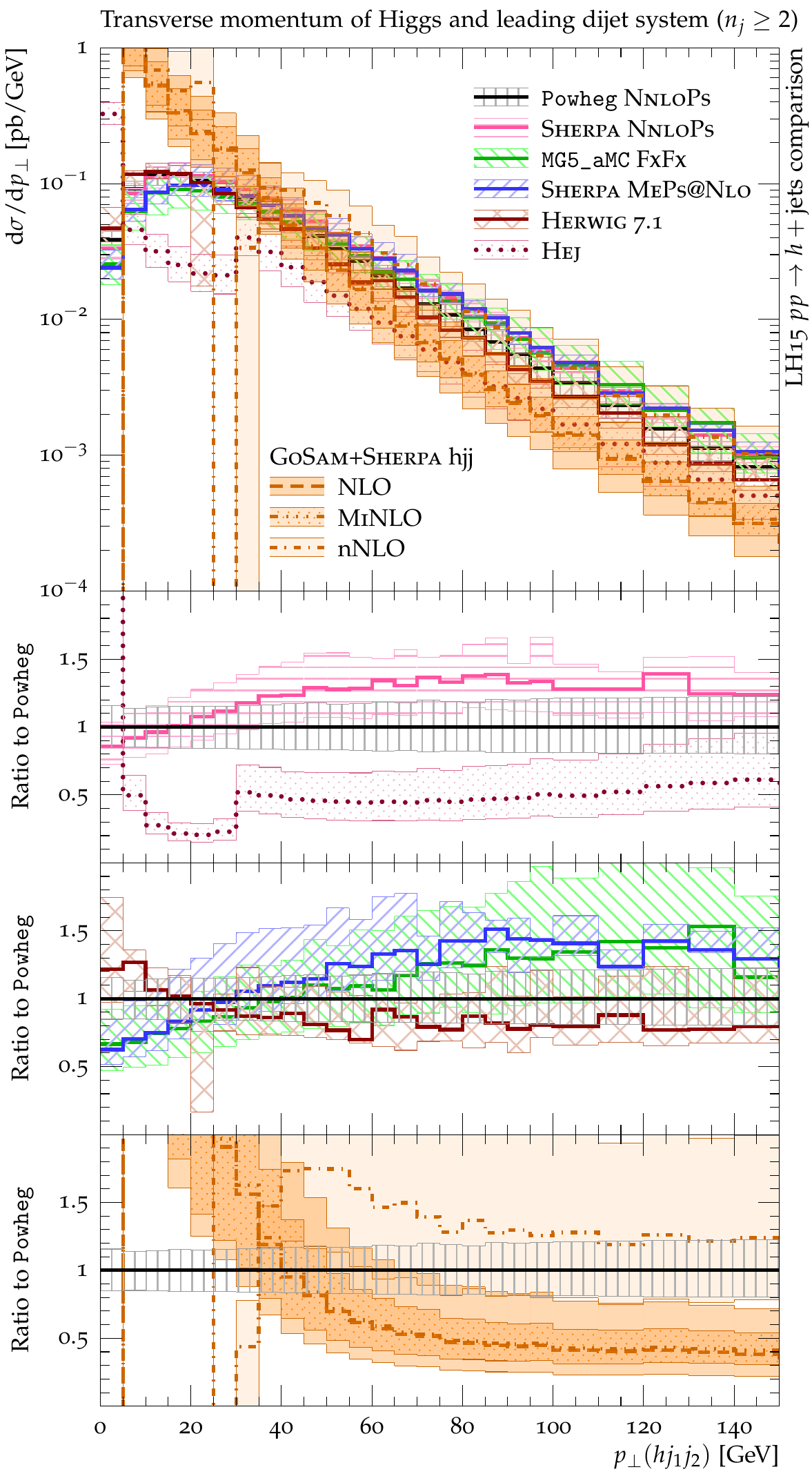}
  \caption{\label{fig:hjetscomp:results:2obs:hjj_pt}%
    The transverse momentum distribution of the Higgs boson plus two
    leading jet system, shown without (left) and with (right)
    uncertainties. The plot layout is the same as in
    Figure~\ref{fig:hjetscomp:results:2obs:hpt}.}
\end{figure}

As before, we are also interested in multiparticle or system
observables. As an example we show the transverse momentum
distribution of the Higgs boson plus two leading jet system in
Figure~\ref{fig:hjetscomp:results:2obs:hjj_pt}. The $p_\perp(hj_1j_2)$
variable is the two-jet analog of the $p_\perp(hj_1)$ variable for
inclusive one-jet events. The physics of this type of recoil
observable has been already discussed in detail for 
Figure~\ref{fig:hjetscomp:results:1obs:hj_pt}. The variations between 
the different approaches are qualitatively similar to those of 
$p_\perp(hj_1)$, cf.\ Figure~\ref{fig:hjetscomp:results:1obs:hj_pt}, 
only their absolute size is increased. \hjetscompMGaMC and \hjetscompSherpa have a slope with respect
to \hjetscompPowheg at low $p_\perp$ and overshoot by about 25\% at higher
$p_\perp$, while \hjetscompHerwig is again somewhat lower than \hjetscompPowheg at high
$p_\perp$. In the soft domain, the \hjetscompSherpa \hjetscompNNLOPS result resembles
the corresponding ME+PS one as a result of using the same parton
shower. While the latter prediction levels off above \hjetscompPowheg (due to
the ME description of a third jet at NLO), the former falls back to
the same level slightly above the plotted range. Recall that for 
\hjetscompSherpa \hjetscompNNLOPS a third, fourth and so
forth jet is described by the parton shower only, the same as for \hjetscompPowheg.
The behavior of \hjetscompHej is again largely affected by the lower rate. The
shape difference once more puts emphasis on the fact that \hjetscompHej
generates harder transverse momentum spectra in general while possessing 
a discontinuity where the events start to possess a resolved third jet. 
In the \hjetscompGoSam{}+\hjetscompSherpa NLO predictions, the recoil is solely
described by the real emission contribution, hence the divergent
behavior in the soft domain and the significantly lower rate but
constant LO shape wrt.~\hjetscompPowheg at higher $p_\perp$. For the same
reason, the associated uncertainty bands are found to be larger as
well. \hjetscompLoopsim, as before, benefits from the fact that the third
jet enters at NLO accuracy, but cannot describe the soft region due to
the procedure's cut-off dependence. Also, \hjetscompLoopsim's uncertainty band
is relatively wide due to the strong impact of the four-jet
contributions on the $p_\perp(hj_1j_2)$ distribution. On the positive
side, these contributions are only included because of \hjetscompLoopsim's
combination of jet bins, while on the negative side they are described
with  LO accuracy. Note that this peculiar behavior of the
$p_\perp(hj_1j_2)$ observable has been already pointed out and
discussed in more detail in Ref.~\cite{Greiner:2015jha}.

\begin{figure}[t!]
  \centering
  \includegraphics[width=0.47\textwidth]{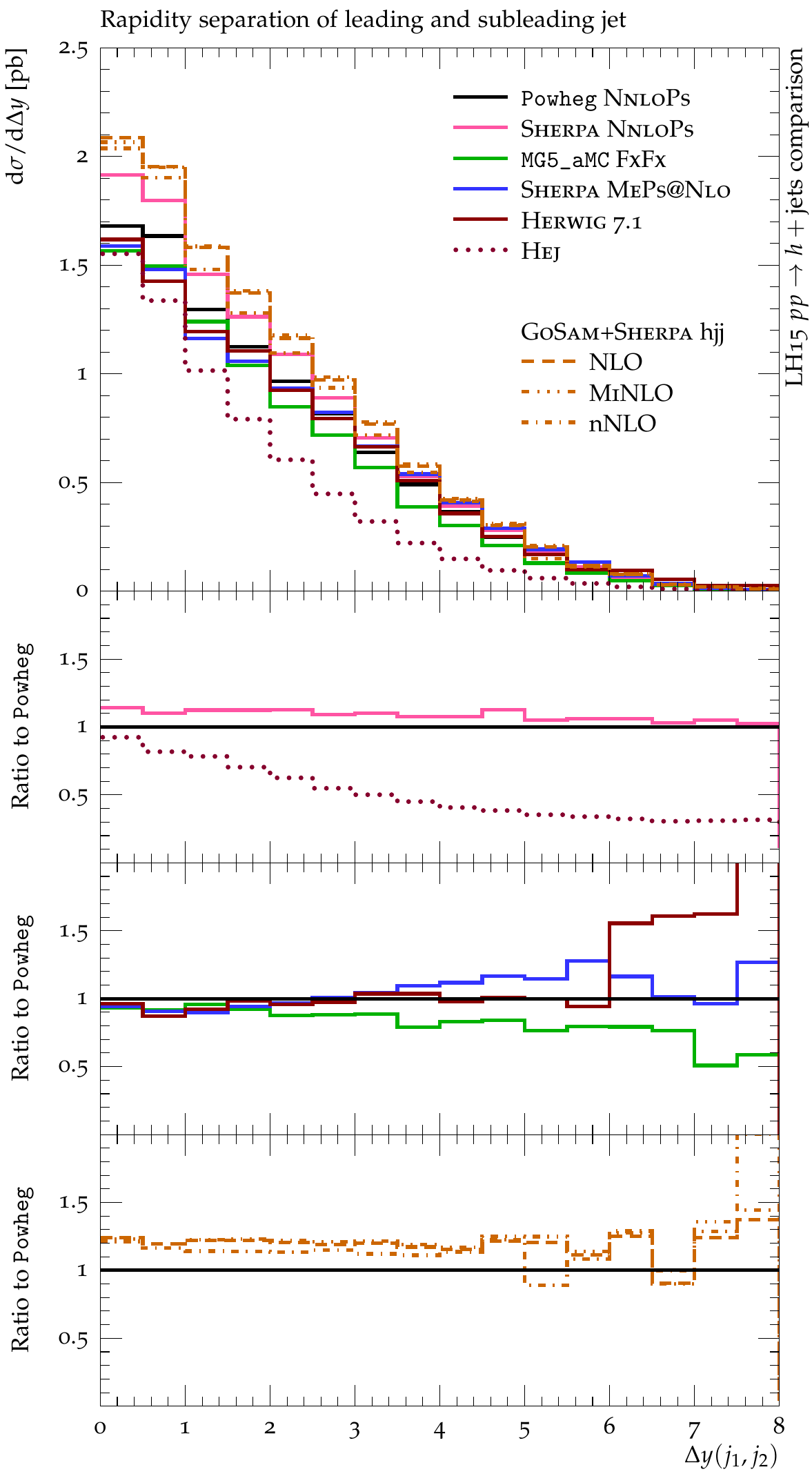}
  \hfill
  \includegraphics[width=0.47\textwidth]{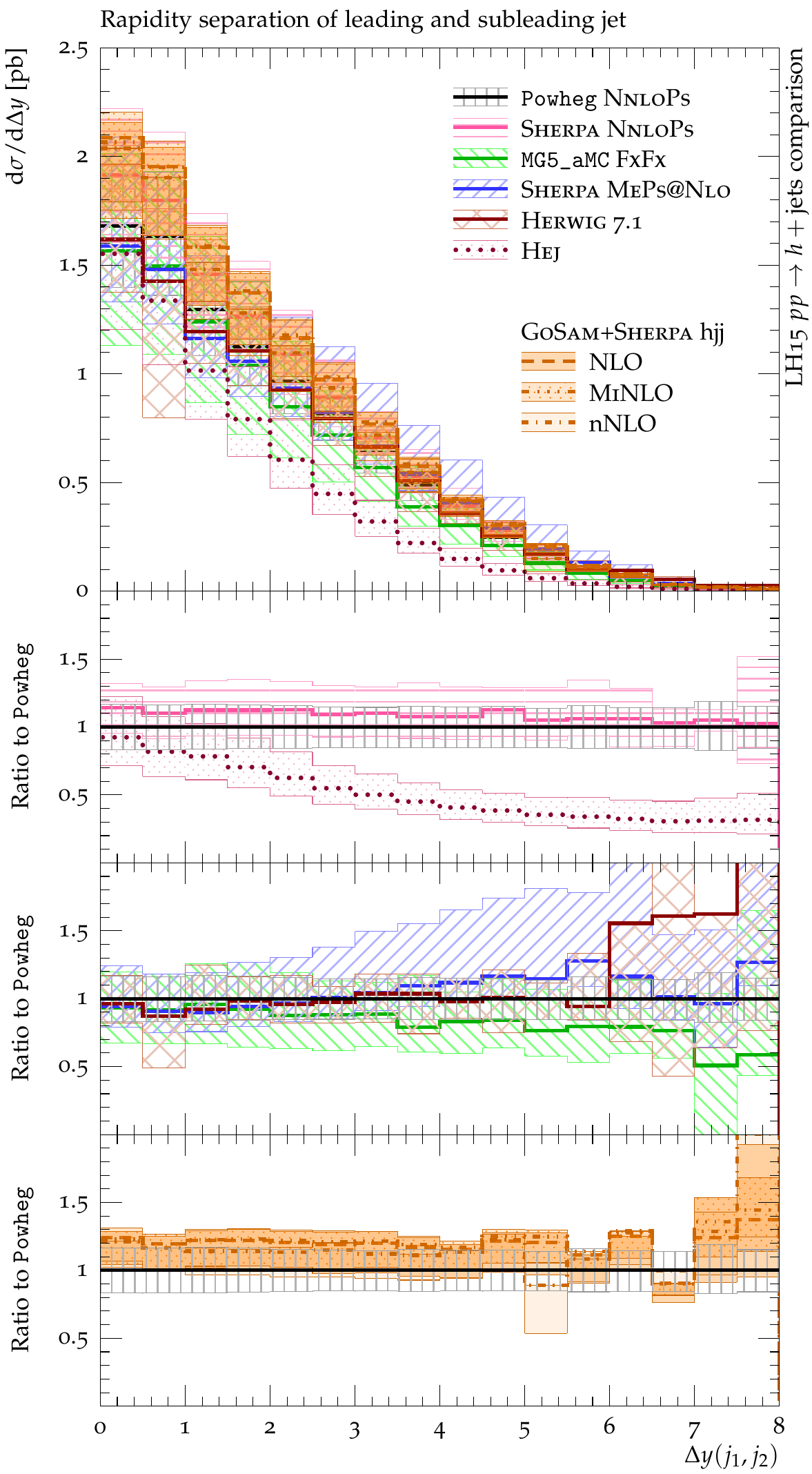}
  \caption{\label{fig:hjetscomp:results:2obs:dyjj}%
    The rapidity separation between the leading and subleading jets
    for $h\,+\!\ge\!2$-jets production, shown without (left) and with
    (right) theoretical uncertainties. The plot layout is the same as
    the one used in Figure~\ref{fig:hjetscomp:results:2obs:hpt}.}
\end{figure}

We now turn to the discussion of jet--jet correlations.
The rapidity separation between the leading and subleading jets is
shown in Figure~\ref{fig:hjetscomp:results:2obs:dyjj}, considering
$h\,+\!\ge\!2$-jet final states. Taking the findings regarding the
individual jet rapidity spectra into account,
cf.~Figures~\ref{fig:hjetscomp:results:1obs:j1y} and
\ref{fig:hjetscomp:results:2obs:j2y}, the $\Delta y(j_1,j_2)$ results
are found to behave very similarly. While \hjetscompSherpa \hjetscompNNLOPS and \hjetscompMGaMC
tend to be slightly more central than \hjetscompPowheg, the \hjetscompSherpa \hjetscompMEPSatNLO
implementation predicts the leading jets to have a somewhat larger
rapidity separation. Surprisingly, \hjetscompHerwig shows a large increase 
beyond $\Delta y>6$. However, it is not clear whether this is simply
due to a lack of statistics. The various \hjetscompGoSam{}+\hjetscompSherpa NLO results
are in good agreement with each other and, as seen in the subleading
jet's spectrum, larger than  \hjetscompPowheg  by about 20\%. For the NLO-based
calculations, the uncertainties are as expected or observed previously, except
for the fact that \hjetscompSherpa's quoted uncertainties for the \hjetscompMEPSatNLO
calculation rise at larger rapidity separation, likely due to
identified scales of low value in that region. The \hjetscompNNLOPS predictions
formally possess LO accuracy only, but as mentioned several times this is
not reflected by the given uncertainty estimates. \hjetscompHej again
predicts a more rapid decrease of the $\Delta y(j_1,j_2)$
distribution. As this observable is used to identify VBF topologies,
it is clear that the cross section computed by \hjetscompHej after imposing VBF
selection criteria will turn out to be substantially different.

\begin{figure}[t!]
  \centering
  \includegraphics[width=0.47\textwidth]{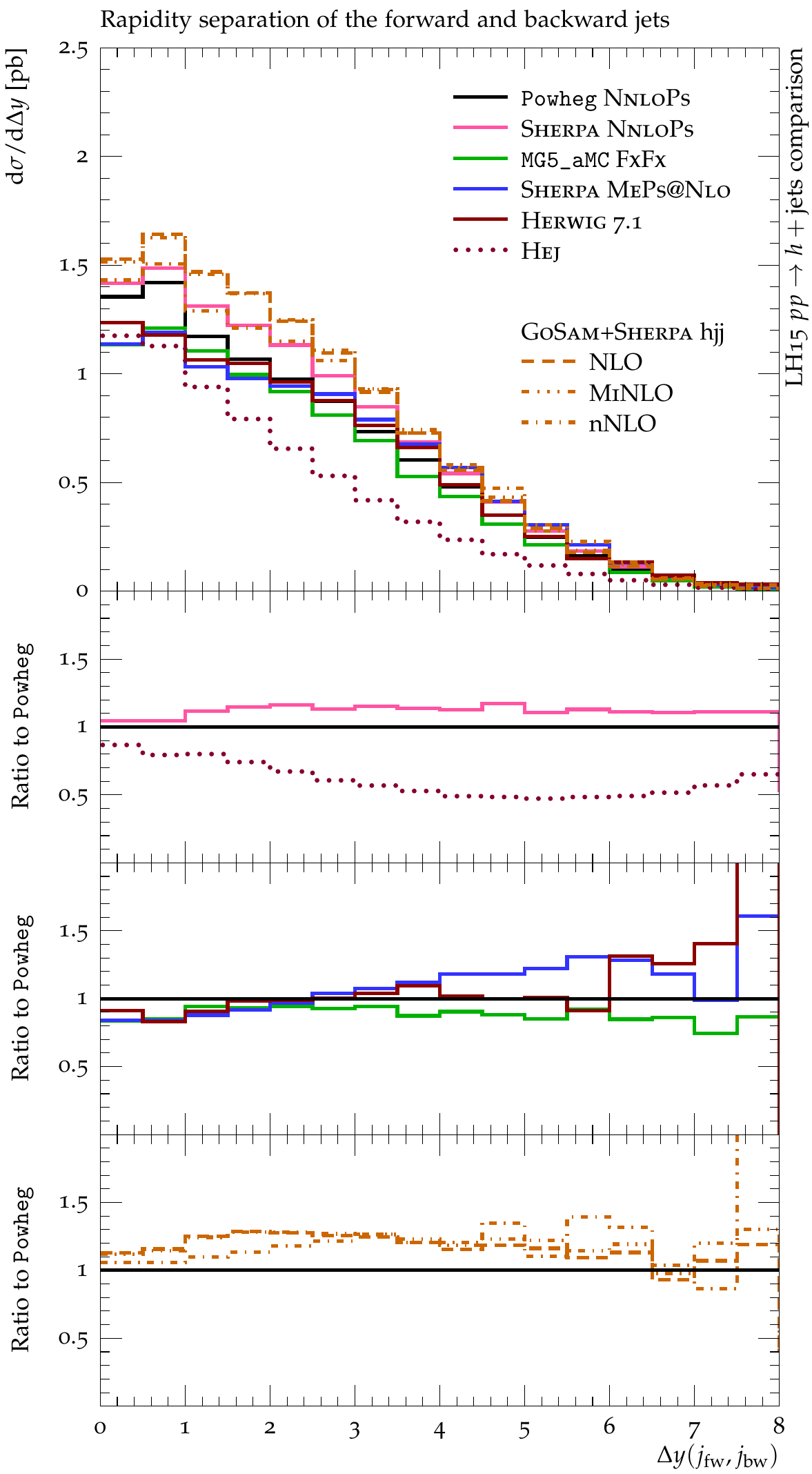}
  \hfill
  \includegraphics[width=0.47\textwidth]{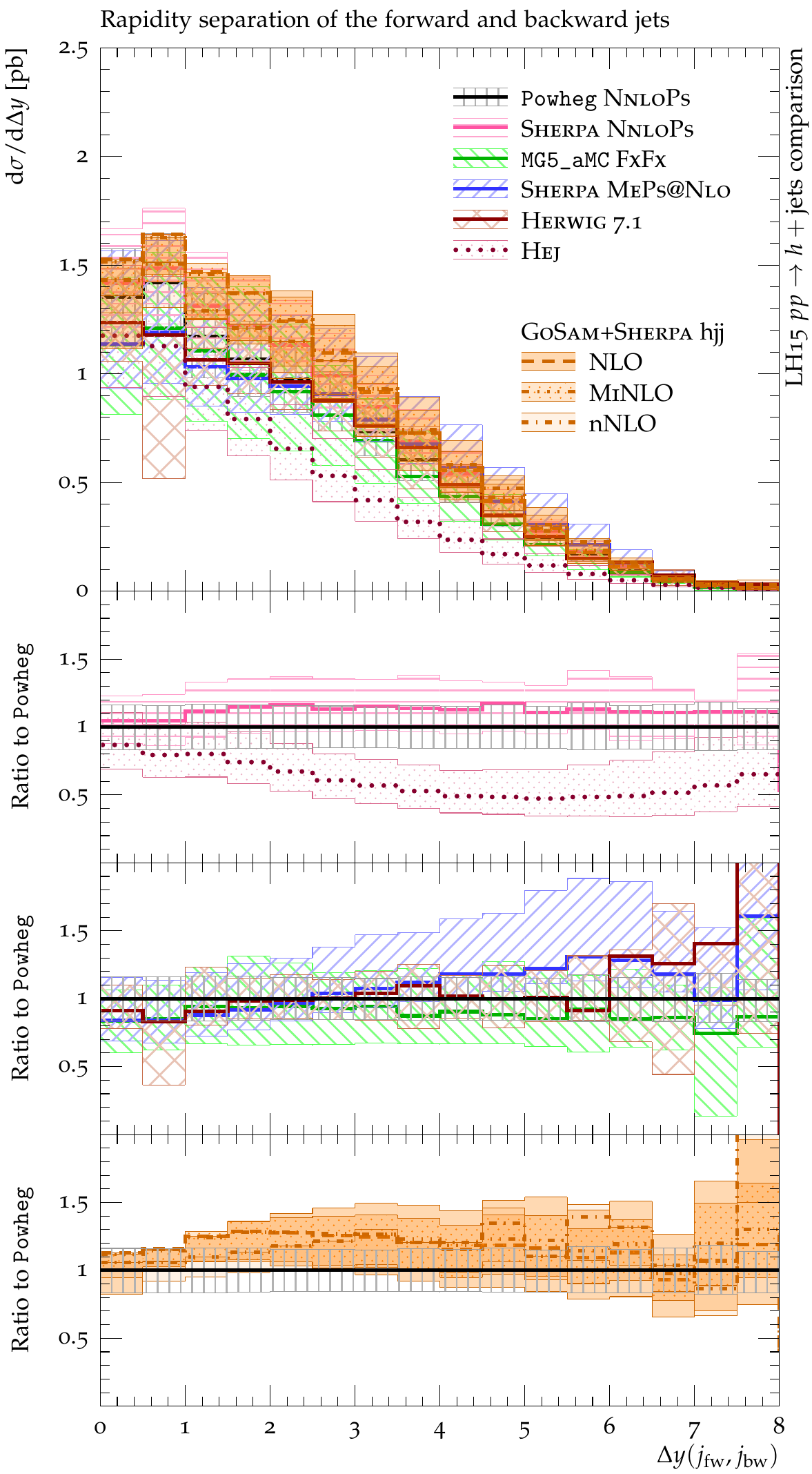}
  \caption{\label{fig:hjetscomp:results:2obs:dyjj_fb}%
    The rapidity separation between the two jets most widely separated
    in rapidity, i.e.~the two most forward/backward jets for
    $h\,+\!\ge\!2$-jets production, shown without (left) and with
    (right) uncertainties. The plot layout is the same as the one used
    in Figure~\ref{fig:hjetscomp:results:2obs:hpt}.}
\end{figure}

As a short digression, it is interesting to consider the rapidity 
separation of the most forward and backward jets, instead of the two
leading jets. Of course, only that fraction of dijet events which are accompanied by 
a third (or more) jet (i.e.~three-jet events) result in 
a different value for this observable, as compared to $\Delta
y(j_1,j_2)$ for the $p_\perp$ ordered case. It is argued that
additional jet production in such rapidity ordered states is well
described by calculations incorporating BFKL effects.
The overall pattern of results between the $p_\perp$ and rapidity
ordered case is the same; certainly, the latter selection leads to a
wider $\Delta y$ distribution. In more detail,
Figure~\ref{fig:hjetscomp:results:2obs:dyjj_fb} shows a somewhat
different relative and absolute behavior of the various calculations.
\hjetscompPowheg \hjetscompNNLOPS here tends to generate slightly more 
central jets than the other calculations, while the opposite is true 
for \hjetscompSherpa \hjetscompMEPSatNLO, which exhibits a clear slope wrt.~\hjetscompPowheg. Again,
for \hjetscompSherpa \hjetscompMEPSatNLO, this behavior can be traced back to \hjetscompSherpa \hjetscompMEPSatNLO using its parton
shower in the scale setting procedure for forward jet production. 
The fixed order calculations of \hjetscompGoSam{}+\hjetscompSherpa now possess a small
shape difference compared to the \hjetscompPowheg reference, which was not
present in the leading jets version of this observable. On the
other hand, \hjetscompHej produces the same relative shape that it showed in the 
leading jets case, up to rapidity separations of around $4$. For
larger rapidity differences, it moves closer to
the reference prediction, being 30\% below the reference at
$\Delta y=8$. Taking the assumption that  \hjetscompHej provides the best description in
this kinematic regime  at face value, all DGLAP based parton shower 
resummations, as well as fixed order calculations, predict 
cross sections that are too large at large rapidity intervals.

\begin{figure}[t!]
  \centering
  \includegraphics[width=0.47\textwidth]{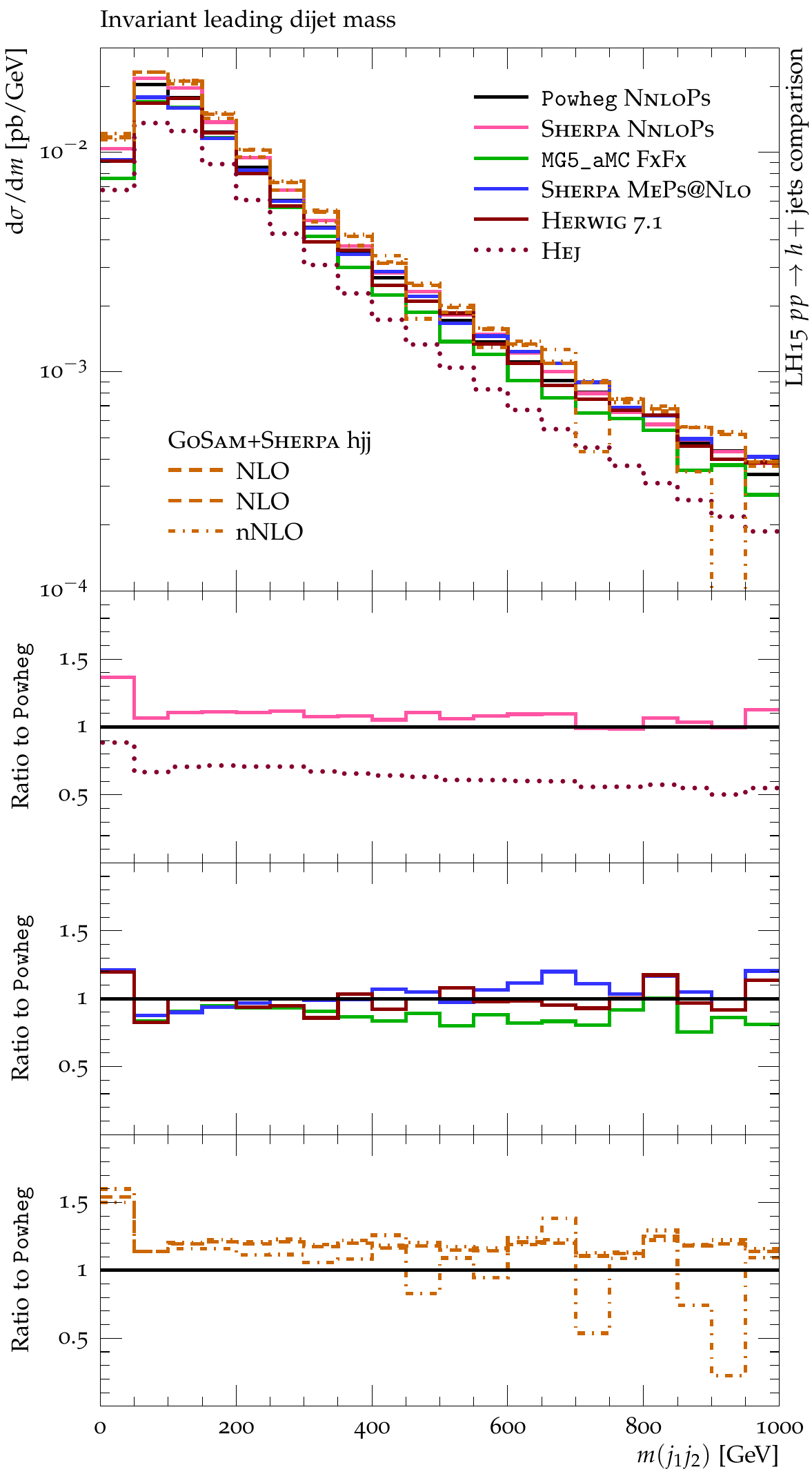}
  \hfill
  \includegraphics[width=0.47\textwidth]{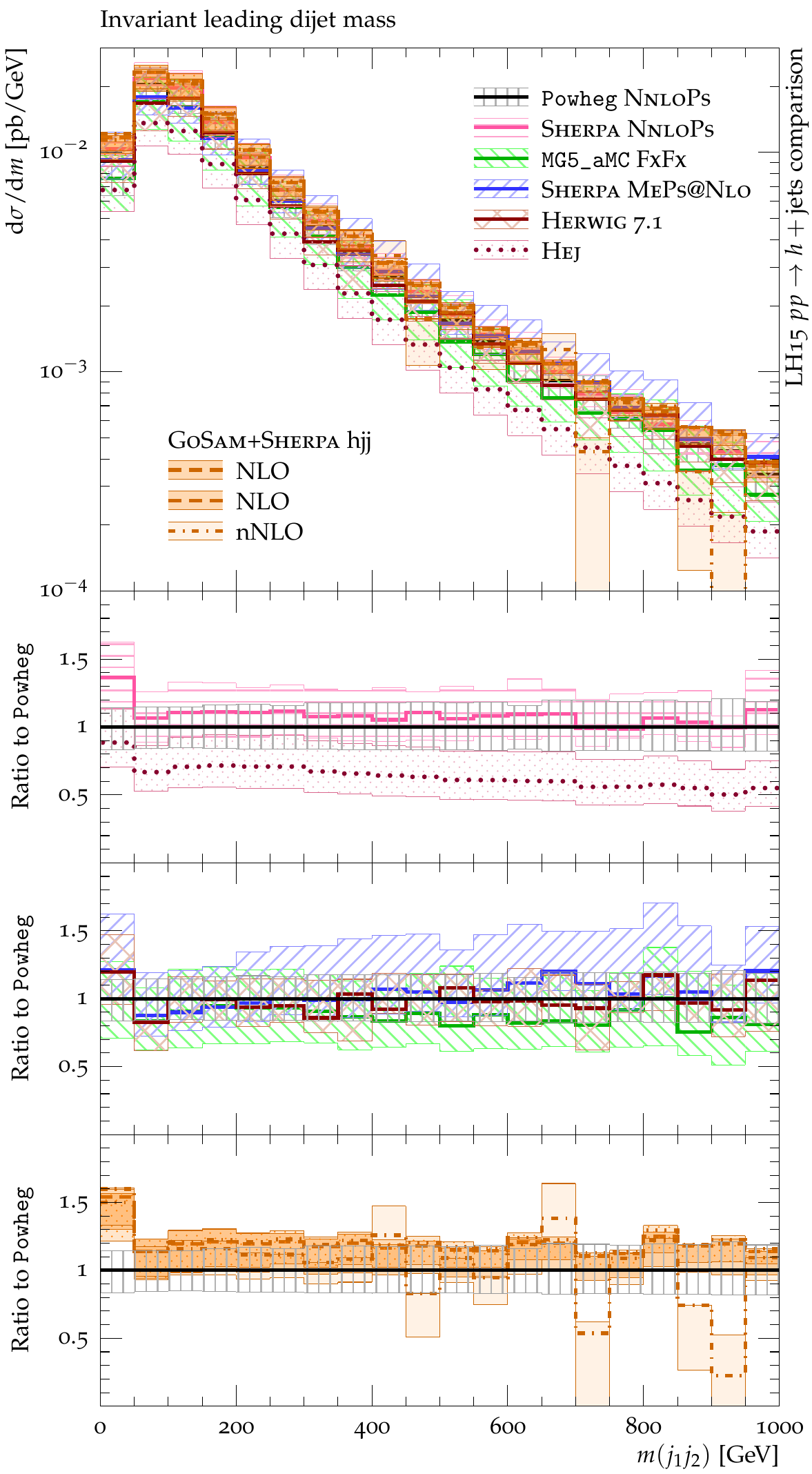}
  \caption{\label{fig:hjetscomp:results:2obs:mjj}%
    The invariant mass distribution of the leading dijet system in
    $h\,+\!\ge\!2$-jets production, shown without (left) and with (right)
    theoretical uncertainties. The plot layout is the same as the one
    used in Figure~\ref{fig:hjetscomp:results:2obs:hpt}.}
\end{figure}

Similar conclusions as for Figure~\ref{fig:hjetscomp:results:2obs:dyjj} 
hold for for the dijet invariant mass distribution formed by the 
leading jets, although the differences among predictions are smaller.
In fact, Figure~\ref{fig:hjetscomp:results:2obs:mjj} shows that there
are almost no shape deviations among the various predictions, except
for the first-to-second bin transition, which \hjetscompPowheg and \hjetscompMGaMC
predict to rise more strongly. All other deviations are mainly driven
by rate differences. \hjetscompSherpa \hjetscompNNLOPS is larger than \hjetscompPowheg \hjetscompNNLOPS
by about 10\%, and maintains this excess almost throughout the whole
range. The multijet merged calculations are also in good agreement
with the \hjetscompPowheg reference and, more importantly, with one another. Only
\hjetscompMGaMC falls below \hjetscompHerwig and \hjetscompSherpa \hjetscompMEPSatNLO, being about 20\%
lower at $1\,\hjetscomptev$, but well within the uncertainties of the other
calculations. The \hjetscompGoSam{}+\hjetscompSherpa NLO calculations display the same
increase of the cross section wrt.~the reference as seen before -- a
constant 20\% increase. The scale variations in all three methods,
pure NLO, \hjetscompMinlo and \hjetscompLoopsim, are also in agreement for this
observable. Only \hjetscompLoopsim suffers somewhat from the limited statistics
of its NLO $hjjj$ component. Again, \hjetscompHej deviates more strongly,
but this time mainly as a result of its LO normalization, being a
near constant $\sim 40\%$ below the other predictions.

\begin{figure}[t!]
  \centering
  \includegraphics[width=0.47\textwidth]{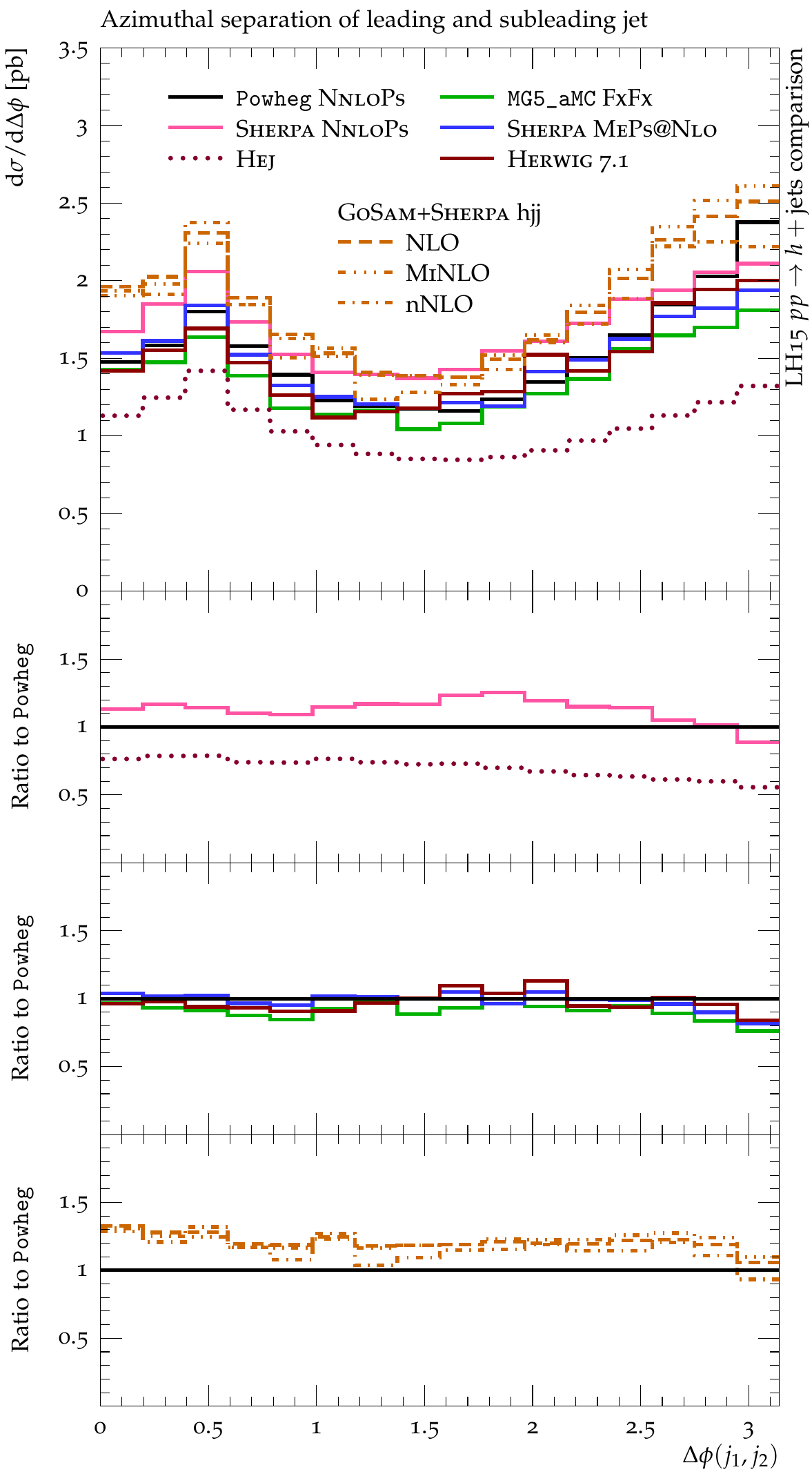}
  \hfill
  \includegraphics[width=0.47\textwidth]{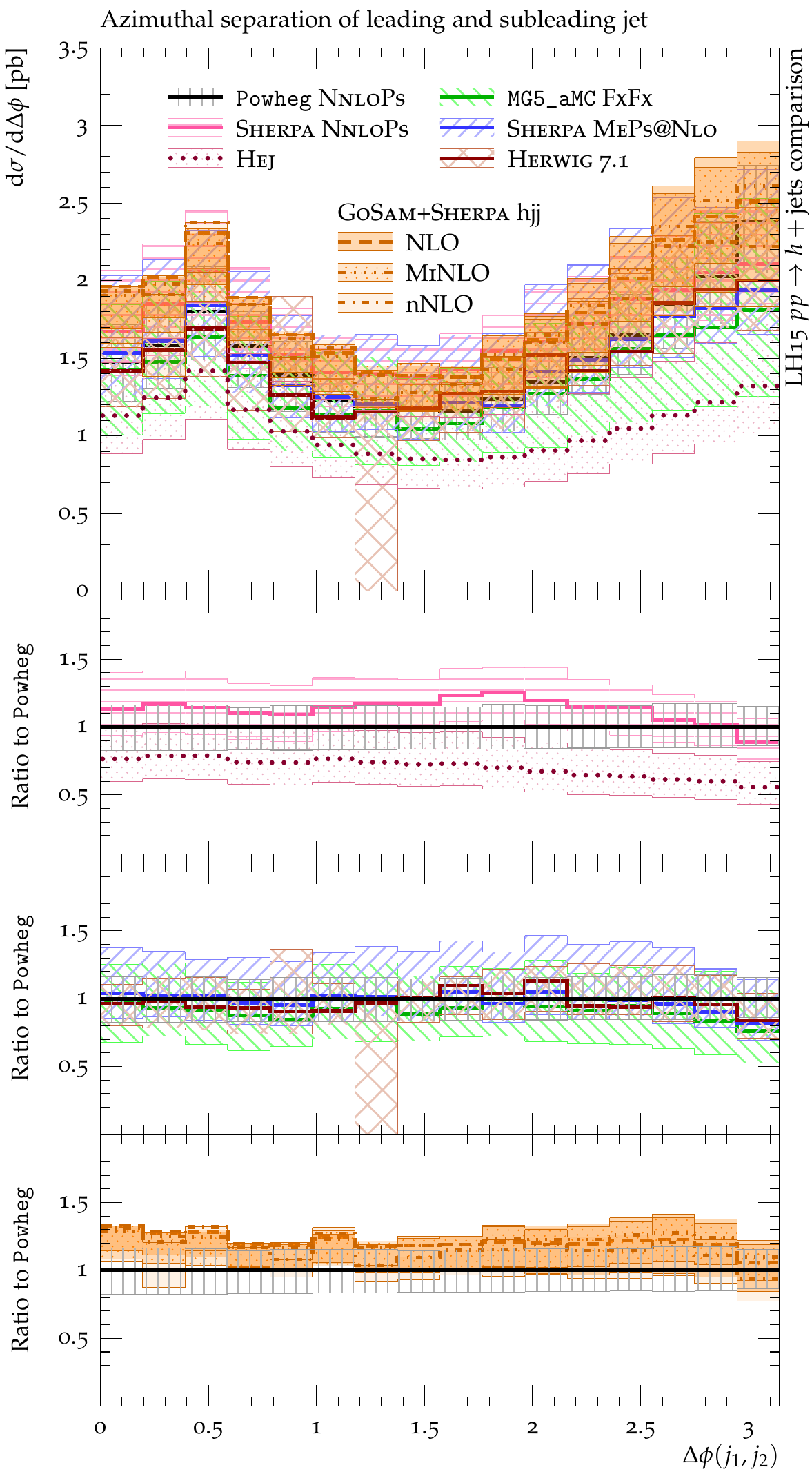}
  \caption{\label{fig:hjetscomp:results:2obs:dphijj}%
    The azimuthal angle separation, $\Delta\phi(j_1,j_2)$, between the
    two leading jets for $h\,+\!\ge\!2$-jets production, presented
    without (left) and with (right) the associated theoretical
    uncertainties. Again, the plot layout is the same as the one used in
    Figure~\ref{fig:hjetscomp:results:2obs:hpt}.}
\end{figure}

As a last multi-object observable, we examine the azimuthal separation of 
the leading dijet pair. Although this observable plays a crucial role 
only after VBF selection cuts are made,  it is interesting to analyze this
observable for the inclusive dijet selection in order to judge which features
are impressed upon it by the VBF event selection.
Inherent differences in the description on the inclusive level will
invariably feed through to the VBF analysis level, and will then be
overlayed with the cut efficiency effects of the various calculations.
The observable possesses two topologically distinct regions. In the
first at $\Delta\phi\gtrsim0$, both jets are in the same azimuthal
hemisphere and recoil against the Higgs boson plus softer radiation
(including further jets if present). 
For this configuration, larger
parton shower effects are expected.
Conversely, for $\Delta\phi\lesssim\pi$, both jets are back-to-back,
recoiling against each other -- but they do not necessarily have to be
$p_\perp$ balanced. In this situation, the Higgs boson will
mostly have small-sized to medium-sized transverse momentum.
While the Higgs boson's transverse momentum is affected by parton
showering off these topologies, the $\Delta\phi$ distribution is more
robust regarding these corrections.
The various predictions for the $\Delta\phi$ separation between the
two leading jets are shown in Figure~\ref{fig:hjetscomp:results:2obs:dphijj},
and are found to be in reasonable agreement with each other,
neglecting for a moment the lower rate of \hjetscompHej due to the LO
normalization. \hjetscompHej
features a roughly 20\% smaller cross section for small $\Delta\phi$,
which increases to about 40\% in the dijet back-to-back region.
\hjetscompSherpa \hjetscompNNLOPS exhibits again a larger cross section,
$\mathcal{O}(15\%)$ for $\Delta\phi\lesssim2.5$, when compared to the
\hjetscompPowheg \hjetscompNNLOPS result. It agrees with \hjetscompPowheg \hjetscompNNLOPS  in the dijet back-to-back
region. All multijet merged distributions agree very well with each
other and the reference. The same applies to the various
\hjetscompGoSam{}+\hjetscompSherpa NLO calculations; they agree very well among
themselves, despite their different scales and inputs, but exhibit a
20\% higher rate wrt.~the reference for $\Delta\phi>0.5$, which
increases to 30\% as $\Delta\phi\to0$. 
There is one feature of the \hjetscompPowheg reference that stands out from the
crowd of other predictions, which is that it rises more strongly
than the others towards the $\Delta\phi=\pi$ limit. The shape change
wrt.~\hjetscompHej is however fairly mild. Recalling that the second jet is
only described at LO in \hjetscompPowheg, it is more than plausible that any
NLO treatment will depopulate this area to some extent. As for the
case of \hjetscompSherpa \hjetscompNNLOPS, it is more subtle, but the reason for the
different behavior, as before, lies in the choice of 
scales. Again, the example of the $p_\perp$ balanced jet topologies and
their scale setting, which we discussed for the $p_\perp(h)$
distribution in the $n_j\ge2$-jet bin, can be used to understand why
\hjetscompSherpa's \hjetscompNNLOPS generates this different behavior in this region.

\subsubsection{VBF observables}
\label{sec:hjetscomp:results:VBFobs}

A situation where Higgs boson production through gluon fusion primarily
serves as a background is found in analyses intended to measure its
couplings to weak vector bosons. To enhance the relative contribution
of processes where the Higgs boson production proceeds through weak
vector boson fusion (VBF), additional cuts are placed on the so-called
tagging jets. The tagging jets themselves can now be defined in
multiple ways. In the study performed during the Les Houches 2013
workshop \cite{AlcarazMaestre:2012vp}, the standard leading
($p_\perp$-ordered) jet selection was supplemented with the
forward-backward selection defining the pair of jets with the largest
rapidity separation as tagging jets. Another strategy using the
highest invariant mass jet pair was also studied more recently
\cite{Greiner:2015jha}.

In this analysis, the leading and subleading jets are required to have
a mass greater than $400\,\hjetscompgev$ and a rapidity separation greater than
$2.8$. This set of cuts is referred to as VBF cuts. An alternative set
of cuts (VBF2) requires that any two jets satisfy the above
requirements. For many observables, the distributions are similar for
the two cases, and only the VBF cut scenario will be presented.
Results for the alternative choice can be found at the project's
webpage \cite{webpage}.

\begin{figure}[t!]
  \centering
  \includegraphics[width=0.47\textwidth]{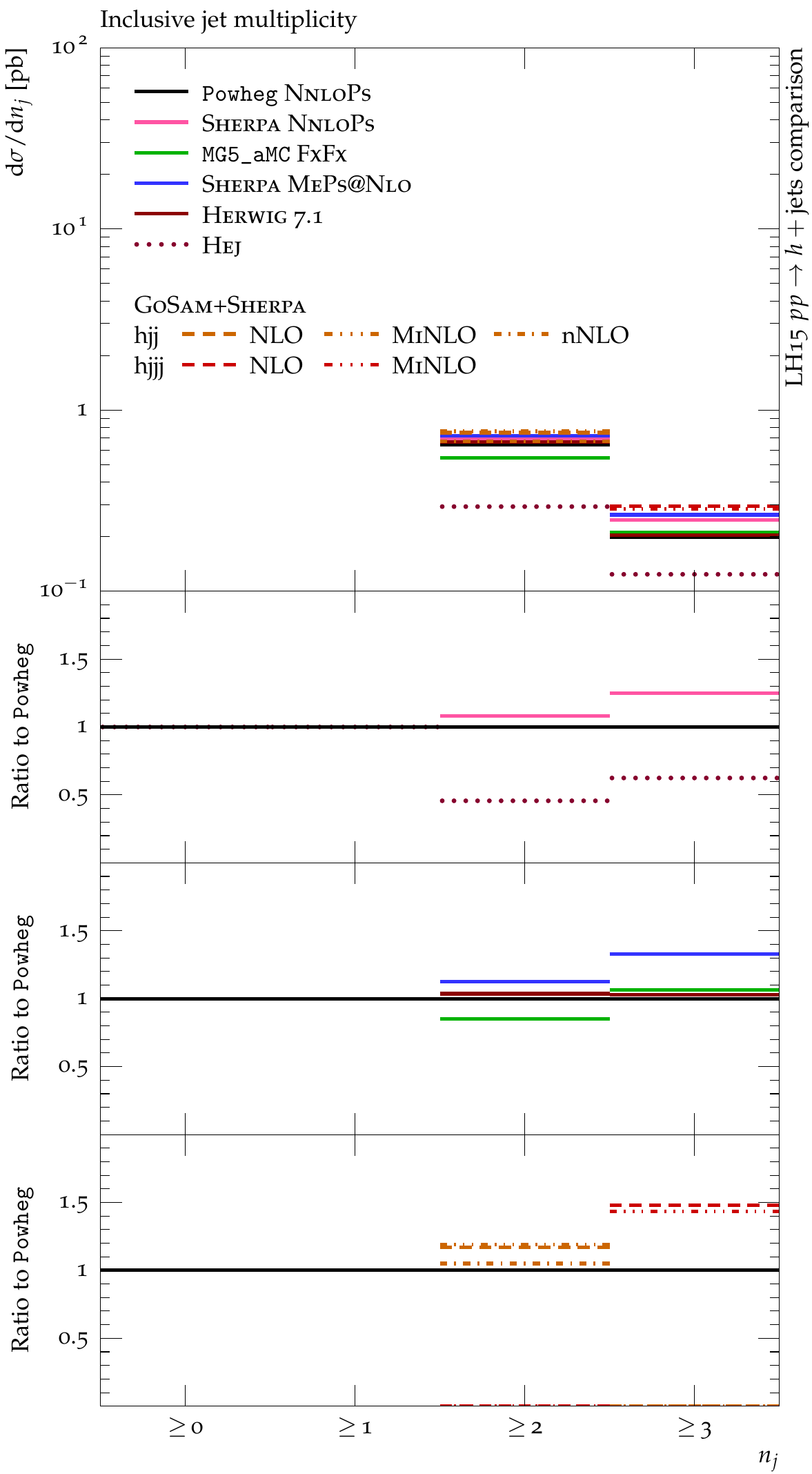}
  \hfill
  \includegraphics[width=0.47\textwidth]{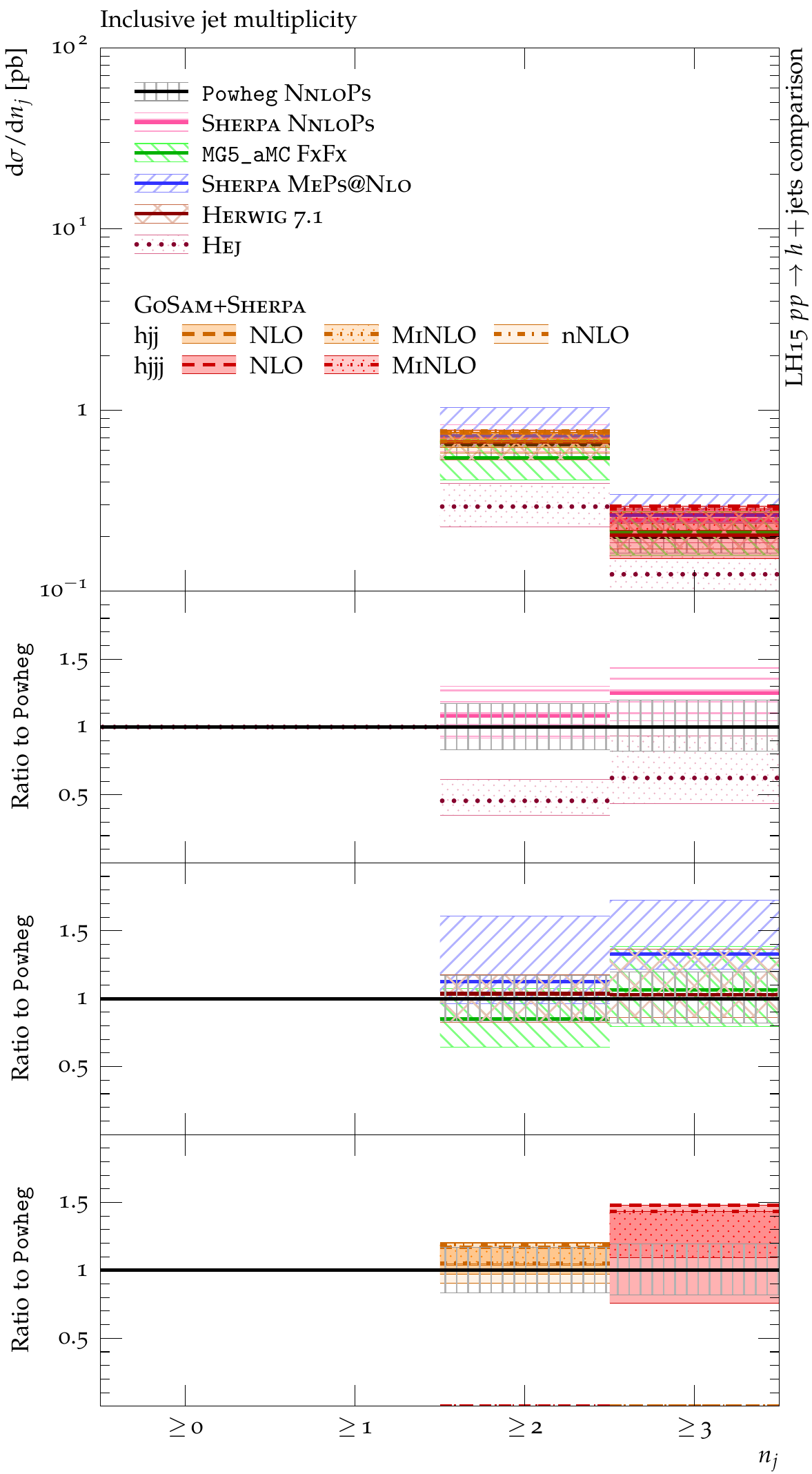}
  \caption{\label{fig:hjetscomp:results:inclobs:njets_VBF}%
    The inclusive jet multiplicities after imposing the leading
    tag-jets definition (see text for the details of the `VBF'
    selection), shown without (left) and with (right) uncertainties.}
\end{figure}

\begin{figure}[t!]
  \centering
  \includegraphics[width=0.47\textwidth]{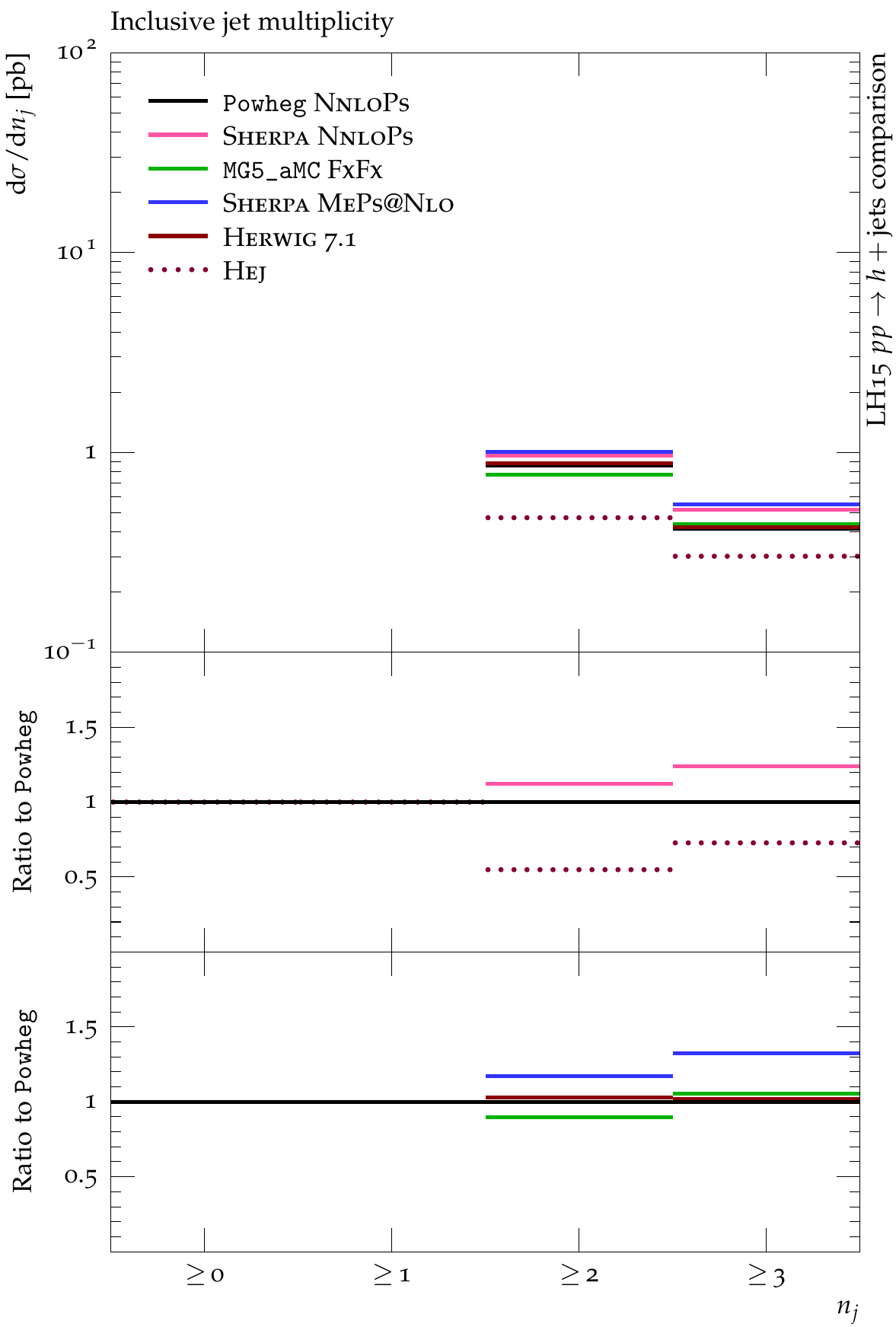}
  \hfill
  \includegraphics[width=0.47\textwidth]{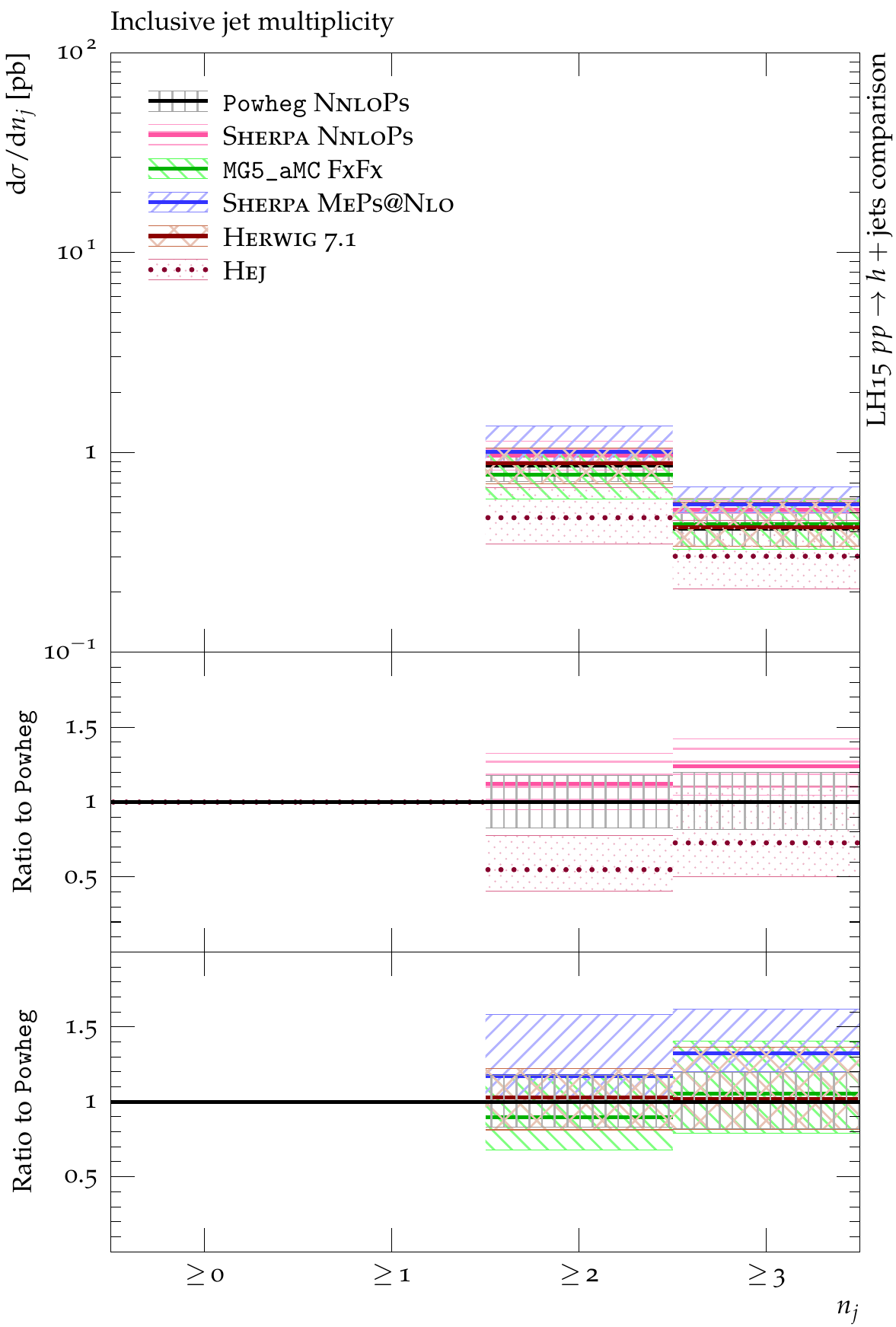}
  \caption{\label{fig:hjetscomp:results:inclobs:njets_VBF2}%
    The inclusive jet multiplicities after the application of the `VBF2'
    tag-jets cuts (see text for the definition), shown without (left)
    and with (right) uncertainties.}
\end{figure}

The inclusive jet multiplicity distributions after the application of
the VBF (VBF2) cuts are shown in
Figure~\ref{fig:hjetscomp:results:inclobs:njets_VBF}
(Figure~\ref{fig:hjetscomp:results:inclobs:njets_VBF2}).  The
hierarchy observed is essentially the same as for the inclusive jet
multiplicity distribution without any cuts. The differences among the
predictions are slightly smaller with the VBF2 cuts than with the VBF
cuts.  In both cases, the \hjetscompNNLOPS calculations are again in good 
agreement, with \hjetscompSherpa \hjetscompNNLOPS predicting slightly larger cross sections.
\hjetscompHej predicts only about 50\% of the cross section of \hjetscompNNLOPS.
For the $\ge2$-jet bin, the NLO merged predictions vary from a 20\%
smaller cross section (\hjetscompMGaMC) to a 20\% larger cross section (\hjetscompSherpa
\hjetscompMEPSatNLO); both are at the edge of the \hjetscompNNLOPS uncertainty
band. Interestingly, their uncertainty is of similar size as \hjetscompNNLOPS,
$\sim$20\%, despite being of NLO accuracy for these observables (only
LO for \hjetscompNNLOPS). This hints at underestimated uncertainties of the
\hjetscompNNLOPS calculations for this observable. The NLO and \hjetscompMinlo
\hjetscompGoSam{}+\hjetscompSherpa predictions are also 20\% larger than the \hjetscompNNLOPS
result in the $\ge2$ jet bin, close to the \hjetscompSherpa \hjetscompMEPSatNLO
prediction, while the \hjetscompLoopsim prediction is about 5\% higher than the
\hjetscompNNLOPS predictions.
More differences are apparent after requiring a third jet. The NLO and
\hjetscompMinlo \hjetscompGoSam{}+\hjetscompSherpa predictions are substantially larger than the
\hjetscompNNLOPS predictions and the prediction from \hjetscompMGaMC and \hjetscompHerwig, but in better
agreement with that from \hjetscompSherpa \hjetscompMEPSatNLO. This is expected given
the NLO normalization present in those calculations (NLO,\hjetscompMinlo
\hjetscompGoSam{}+\hjetscompSherpa, \hjetscompSherpa \hjetscompMEPSatNLO). The third jet arises from parton
showering in the \hjetscompNNLOPS calculations and from the parton shower matched 
leading order matrix elements in \hjetscompMGaMC and \hjetscompHerwig. Owing to the  
matching techniques employed and the  resummation uncertainties that have not been evaluated, 
the uncertainties are underestimated in both calculations. The same 
therefore holds for \hjetscompPowheg \hjetscompNNLOPS and (partially) for \hjetscompSherpa 
\hjetscompNNLOPS. 

\begin{figure}[t!]
  \centering
  \includegraphics[width=0.47\textwidth]{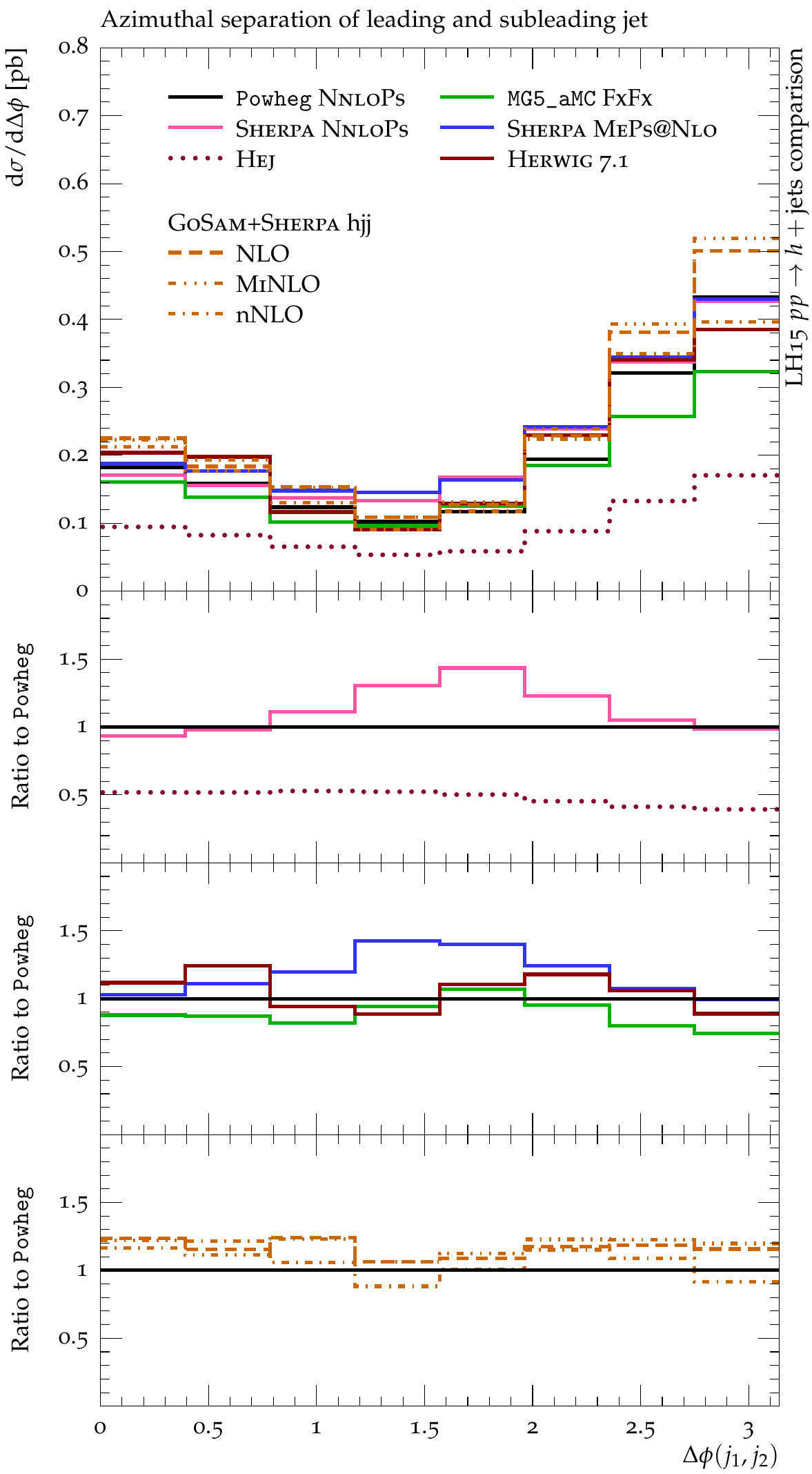}
  \hfill
  \includegraphics[width=0.47\textwidth]{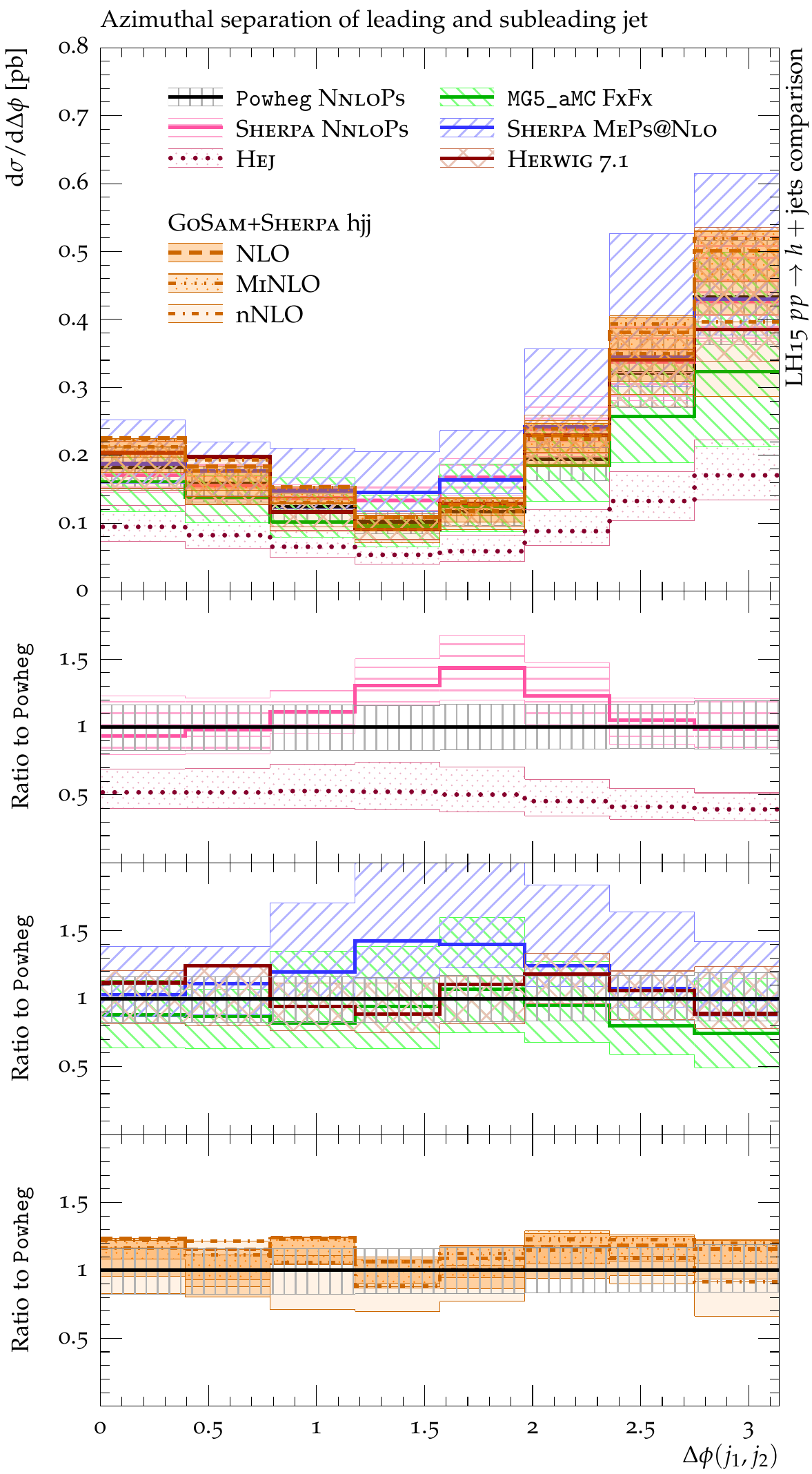}
  \caption{\label{fig:hjetscomp:results:VBFobs:dphijj}%
    The azimuthal separation of the leading jet pair (VBF cuts) shown
    without (left) and with (right) uncertainities in $h\,+\!\ge\!\!2$-jet
    production.}
\end{figure}

The azimuthal separation between the two tagging jets is a crucial
observable in VBF measurements. In general, there is good agreement 
among the various predictions for this observable, with the caveat
that \hjetscompSherpa, in both its \hjetscompNNLOPS and its \hjetscompMEPSatNLO forms, predicts a
shallower dip at $\Delta\phi(j_1,j_2)\approx\tfrac{\pi}{2}$, something
that can be traced to its parton shower. Similar distributions (and
agreement) are observed if the generalised tagging jet definition
(VBF2) is used.

\subsubsection{Multijet observables}
\label{sec:hjetscomp:results:mjobs}

In this section we consider observables sensitive to more than two jets.
The most accurate predictions here come from the NLO computations of 
\hjetscompGoSam{}+\hjetscompSherpa and \hjetscompSherpa \hjetscompMEPSatNLO both of which include the 3rd 
jet at NLO accuracy and the 4th jet at LO accuracy. \hjetscompMGaMC and \hjetscompHerwig 
both include the 3rd jet at LO while the 4th jet is LO in \hjetscompHerwig but 
showered in \hjetscompMGaMC. \hjetscompHej also includes LO matrix elements for the production 
of the thirf jet. The \hjetscompNNLOPS codes only have parton shower accuracy 
throughout the observables of this section.

\begin{figure}[t!]
  \centering
  \includegraphics[width=0.47\textwidth]{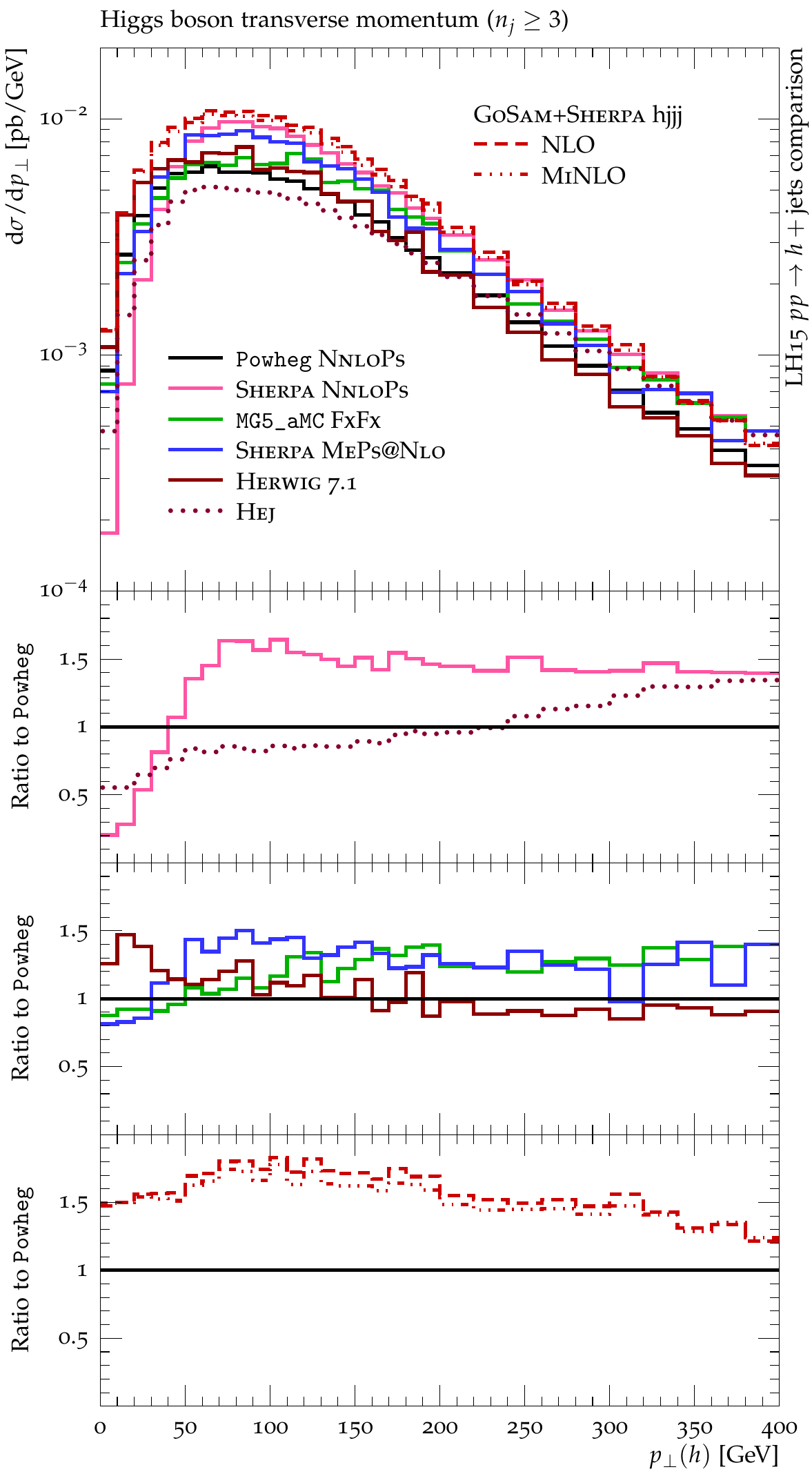}
  \hfill
  \includegraphics[width=0.47\textwidth]{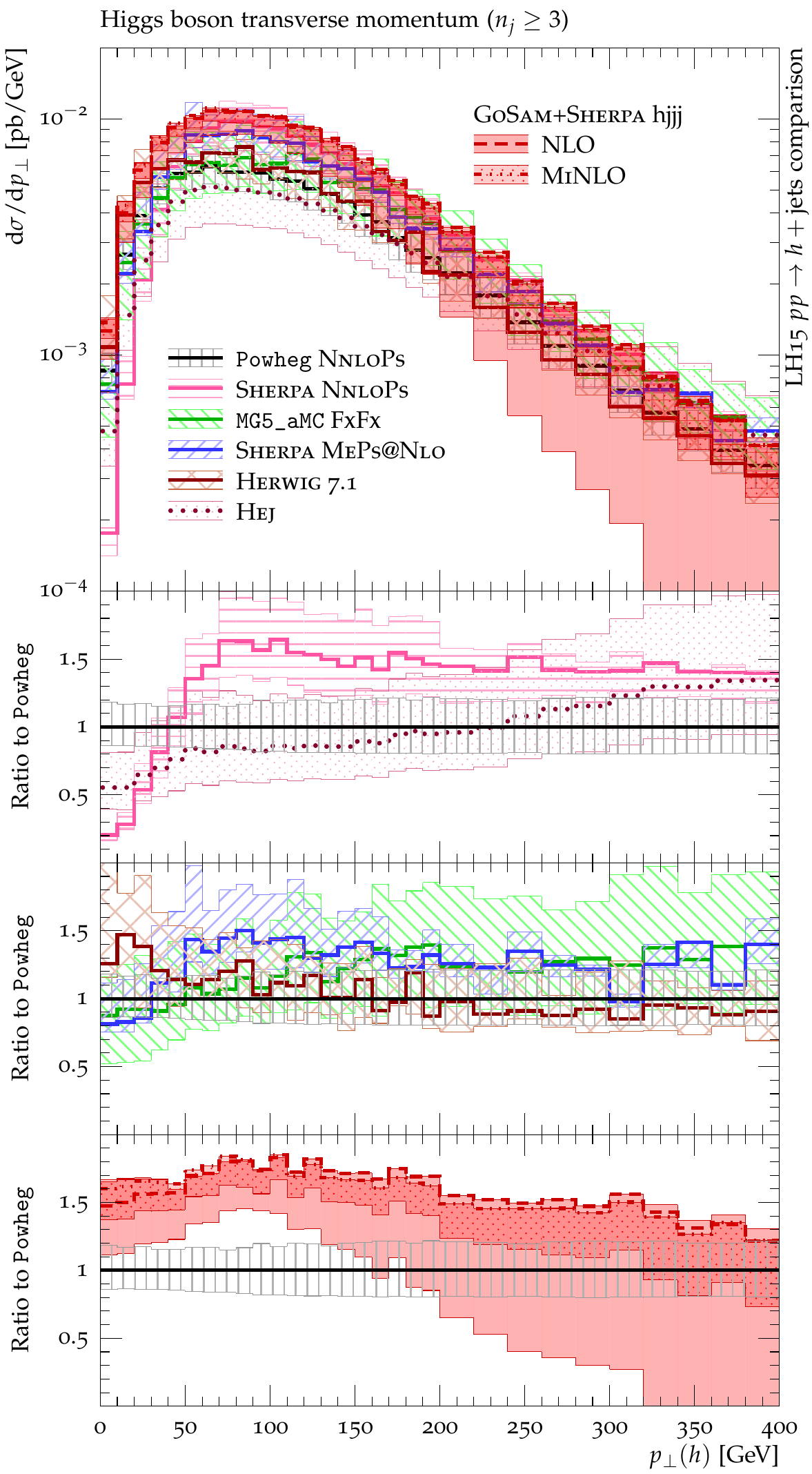}
  \caption{
    The Higgs boson transverse momentum in the presence of at least three 
    jets, shown without (left) and with (right) theoretical uncertainties.
    \label{fig:hjetscomp:results:mobs:hpt_j3}
  }
\end{figure}

The Higgs boson transverse momentum distribution for $h\,+\!\ge\!\!3$ jets is shown in
Figure~\ref{fig:hjetscomp:results:mobs:hpt_j3}. In general we see larger
results from predictions with at least LO accuracy over \hjetscompPowheg, which includes
the 3rd jet via the parton shower. The \hjetscompSherpa \hjetscompNNLOPS result is also
significantly larger than \hjetscompPowheg which shows that the \hjetscompSherpa shower generates
more radiation than \hjetscompPythia~8. The \hjetscompHej prediction begins to increase over
\hjetscompPowheg at high $p_T$ and is closer there to the NLO results of \hjetscompGoSam{}+\hjetscompSherpa and
\hjetscompSherpa \hjetscompMEPSatNLO. The multijet merged calculations on the whole agree well
within scale uncertainties, though these are rather large for \hjetscompMGaMC, and the
central value for \hjetscompHerwig shows a clear deviation below the jet $p_T$ cut. The
benefit of NLO accuracy is clearly seen in the \hjetscompGoSam{}+\hjetscompSherpa result, which has large
corrections with respect to  \hjetscompPowheg. The \hjetscompMinlo scale choice used in the fixed order
computation has a central value that is almost identical to the dynamical scale
of $\tfrac{1}{2}\,\sqrt{m_{h}^2+\sum p_{T,j_i}^2}$. The scale variations
however are much smaller than the fixed order NLO.

\begin{figure}[t!]
  \centering
  \includegraphics[width=0.47\textwidth]{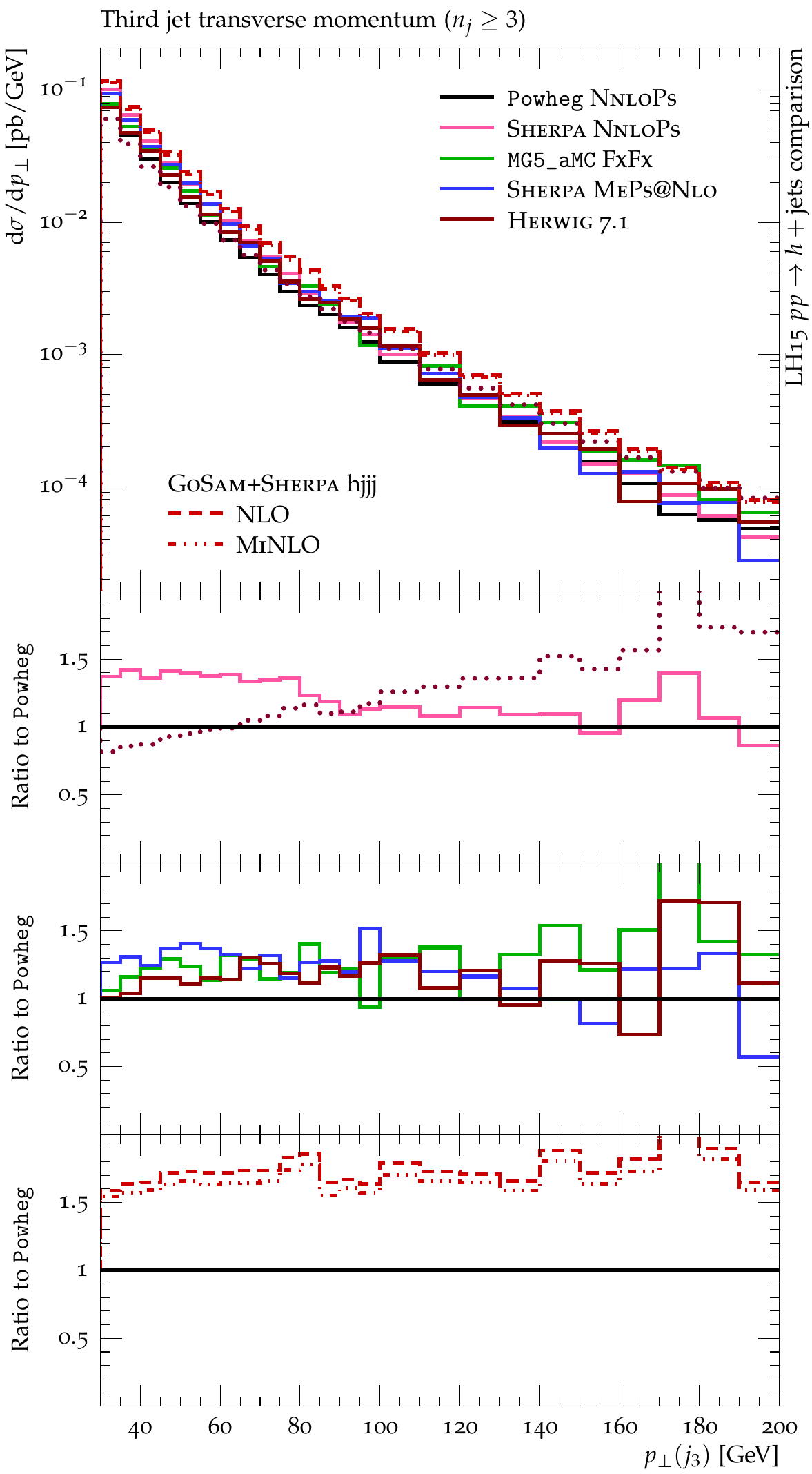}
  \hfill
  \includegraphics[width=0.47\textwidth]{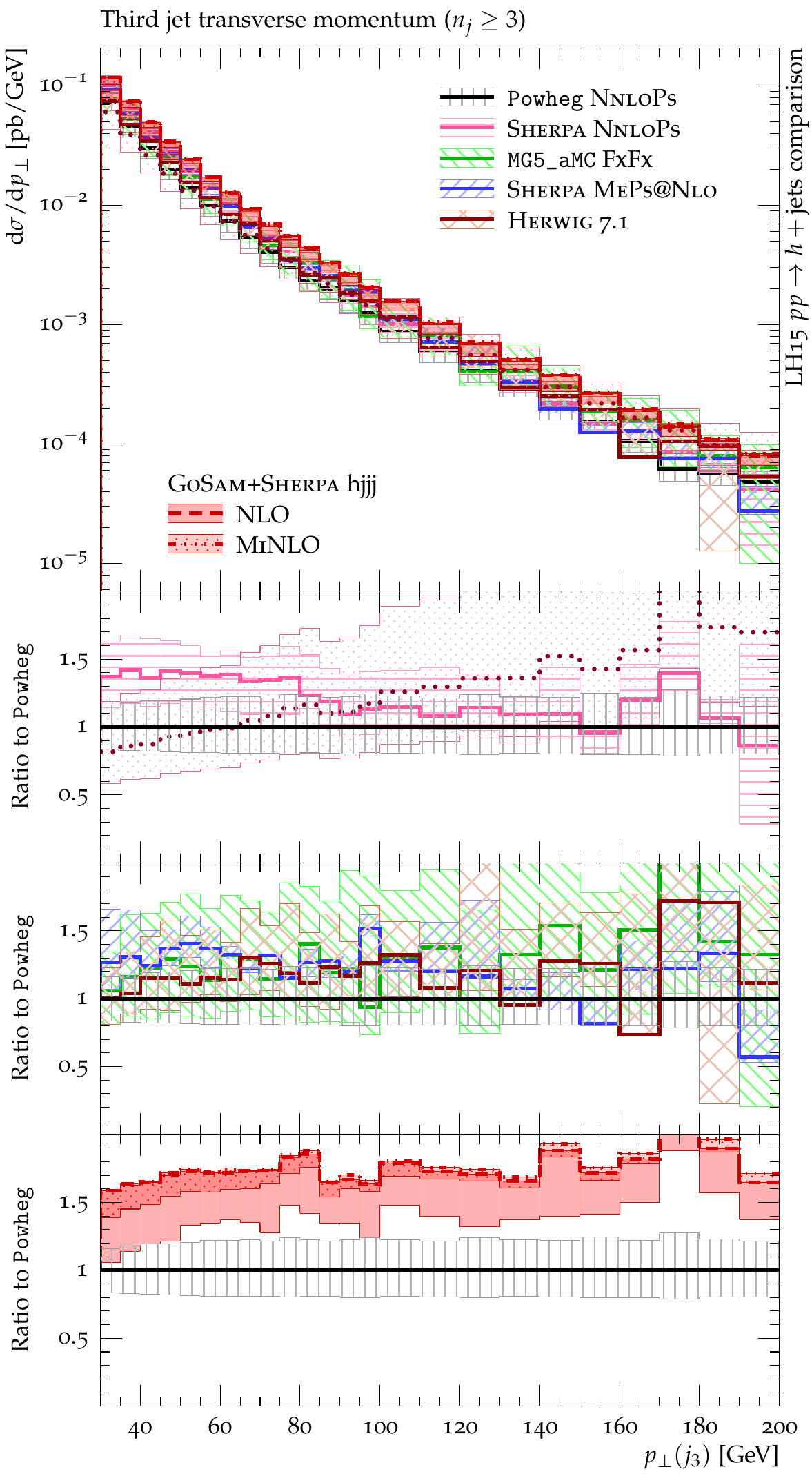}
  \caption{
    The third jet transverse momentum distribution for $H+\ge3$ jets
    shown without (left) and with (right) theoretical uncertainty bands.
    \label{fig:hjetscomp:results:mobs:j3pt}
  }
\end{figure}

The third jet $p_T$ for $h\,+\!\ge\!\!3$ jets is shown in
Figure~\ref{fig:hjetscomp:results:mobs:j3pt}. Overall, a similar pattern to the
previous Higgs boson transverse momentum distribution is followed. \hjetscompSherpa
\hjetscompMEPSatNLO has a much smaller scale uncertainty than \hjetscompMGaMC, which mirrors 
the small scale uncertainty of the NLO \hjetscompGoSam{}+\hjetscompSherpa calculation. The fixed
order \hjetscompGoSam{}+\hjetscompSherpa results show more than a 50\% increase over \hjetscompPowheg at large
$p_T$.  The close agreement between central values of \hjetscompGoSam{}+\hjetscompSherpa using the
dynamical scale of $\tfrac{1}{2}\,\sqrt{m_{h}^2+\sum p_{T,j_i}^2}$, and \hjetscompMinlo
is somewhat suprising. The \hjetscompMinlo results again show a significant decrease in
scale variations - particularly at high $p_T$. Care
should be taken in interpreting the high $p_T$ region since due to the
complexity of the phase space, the MC statistical error is beginning to come into
play.

\begin{figure}[t!]
  \centering
  \includegraphics[width=0.47\textwidth]{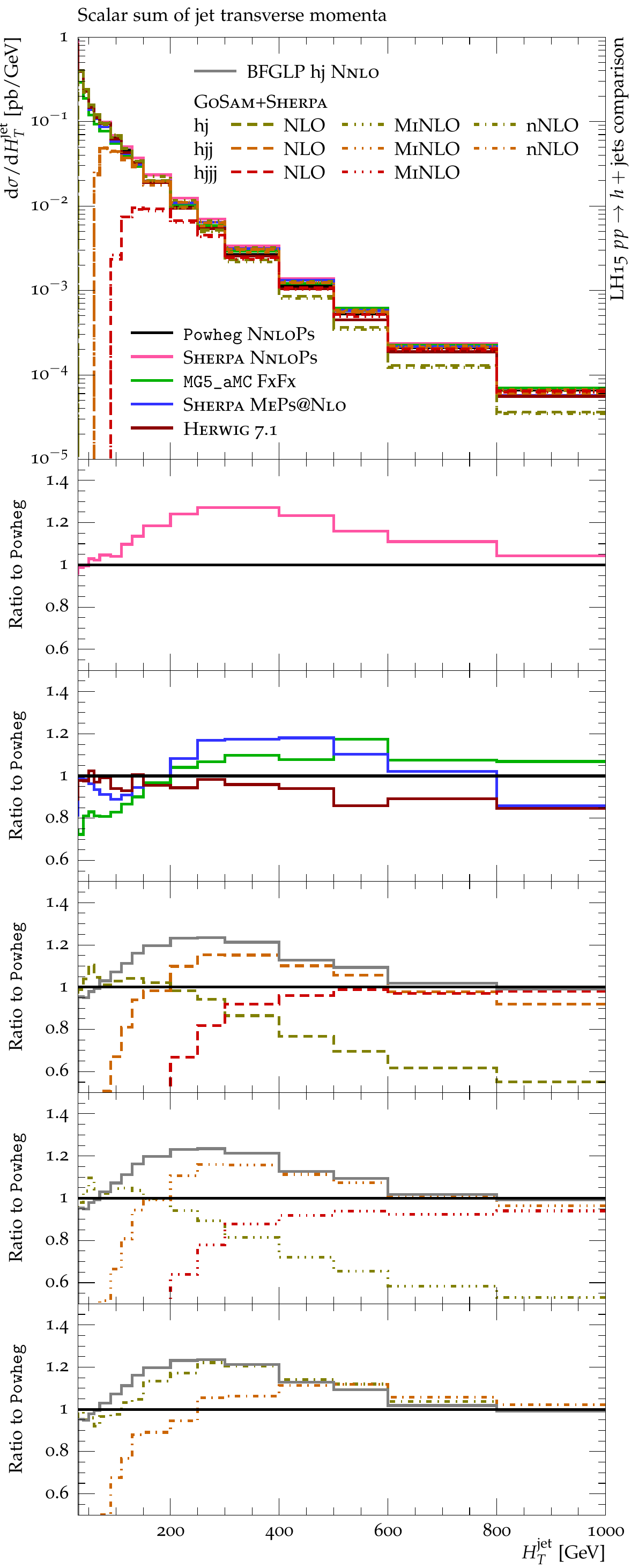}
  \hfill
  \includegraphics[width=0.47\textwidth]{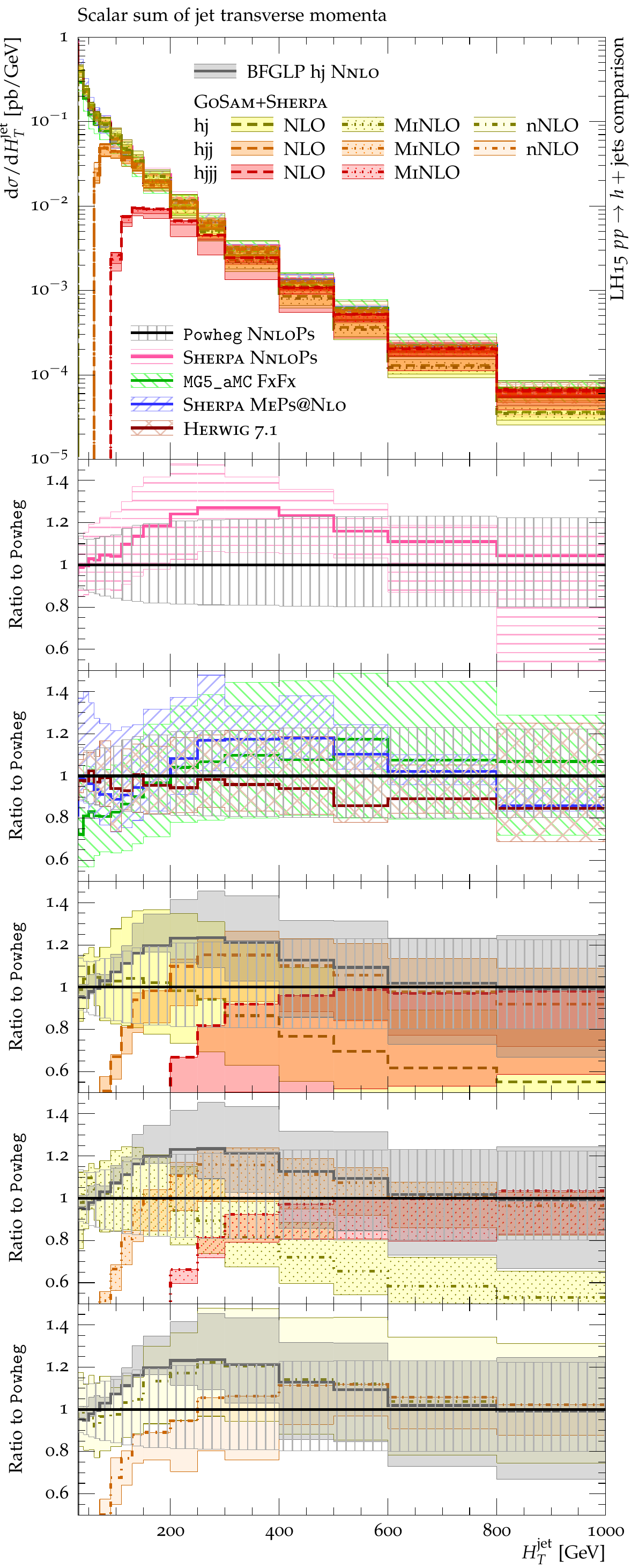}
  \caption{
    The $H_{T,\mathrm{jets}}$ distribution for $H+\ge1$ jet final
    states, without (left) and with (right) uncertainties.
    \label{fig:hjetscomp:results:mobs:HT_jets}
  }
\end{figure}

The $H_T^\text{jet}$ distribution, defined as sum scalar sum of all jet 
transverse momenta, is shown in Figure~\ref{fig:hjetscomp:results:mobs:HT_jets}. 
Requiring $h\,+\!\ge\!\!1$-jet final states this observable is
extremly sensitive to additional radiation and is a useful case to observe
the impact of different jet multiplicities. While the \hjetscompNNLOPS 
computations of \hjetscompSherpa and \hjetscompPowheg deviate significantly for 
$150\,\hjetscompgev < H_T^\text{jet} < 600\,\hjetscompgev$, \hjetscompSherpa \hjetscompNNLOPS agrees better with 
the BFGLP $hj$ NNLO fixed order prediction and, to a certain extent, 
the \hjetscompSherpa \hjetscompMEPSatNLO computation. \hjetscompHerwig and \hjetscompMGaMC give results 
consistent with \hjetscompPowheg except for deviations at low energies but with 
relatively large errors.

For very high $H_T^\text{jet}$ all tools including some approximation to the 3rd
jet converge, while NLO predictions for $h+1$ jet quickly begin to fall away
from the other predictions. The \hjetscompLoopsim $h+(1,2)$ jet prediction at nNLO does a good
job of matching the complete NNLO result here, as it is designed to do. One also
clearly sees the benefit of the 3rd jet at NLO accuracy.  The scale variations
for the NLO $h+3$ jet cross section,  either at fixed order with \hjetscompLoopsim or  with the \hjetscompMinlo
procedure or with the fully merged prediction with \hjetscompSherpa \hjetscompMEPSatNLO, have
significantly smaller errors above $400\,\hjetscompgev$ than the other predictions.

\clearpage
\subsubsection{Jet veto cross sections}
\label{sec:hjetscomp:results:jvobs}

\begin{figure}[t!]
  \centering
  \includegraphics[width=0.47\textwidth]{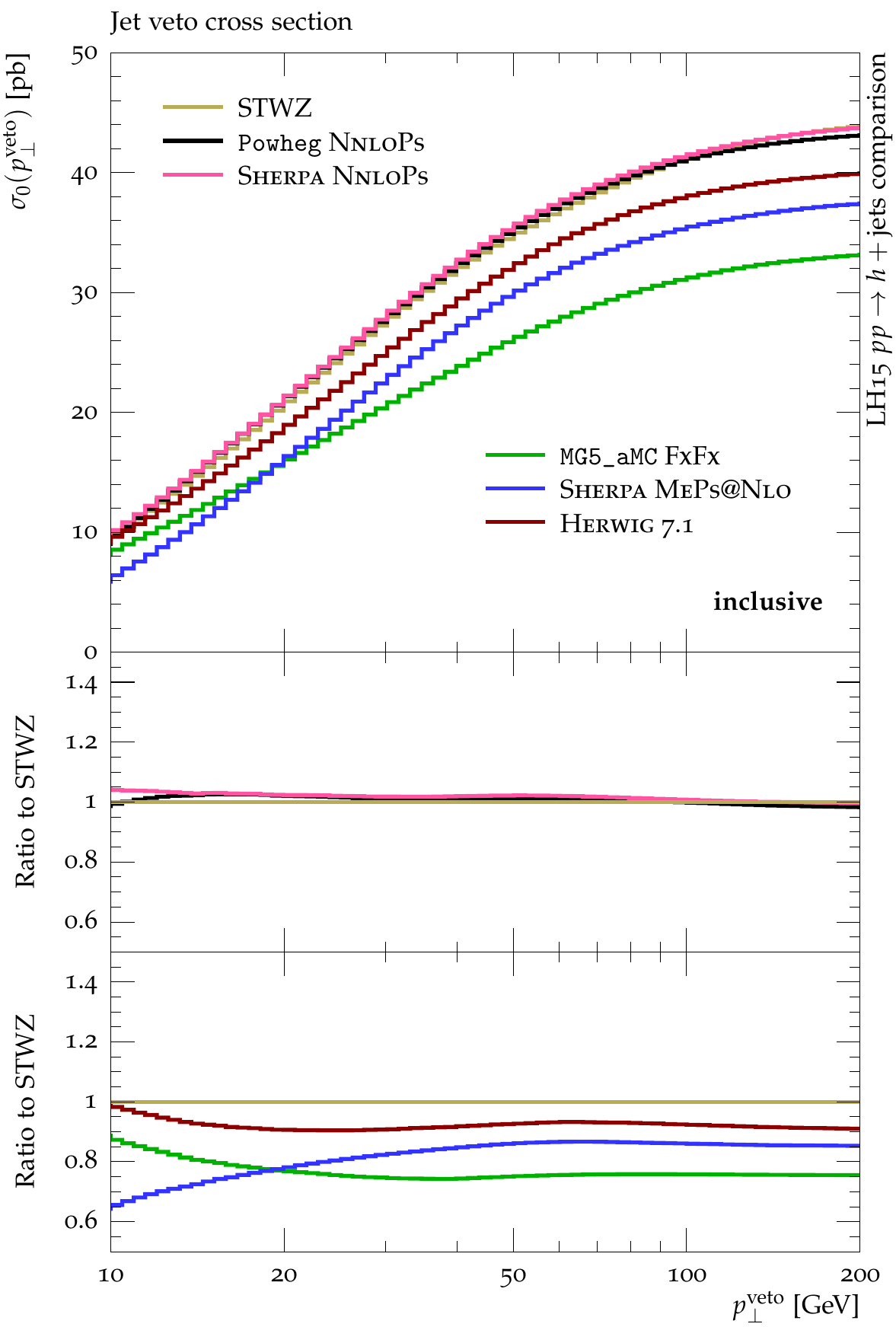}
  \hfill
  \includegraphics[width=0.47\textwidth]{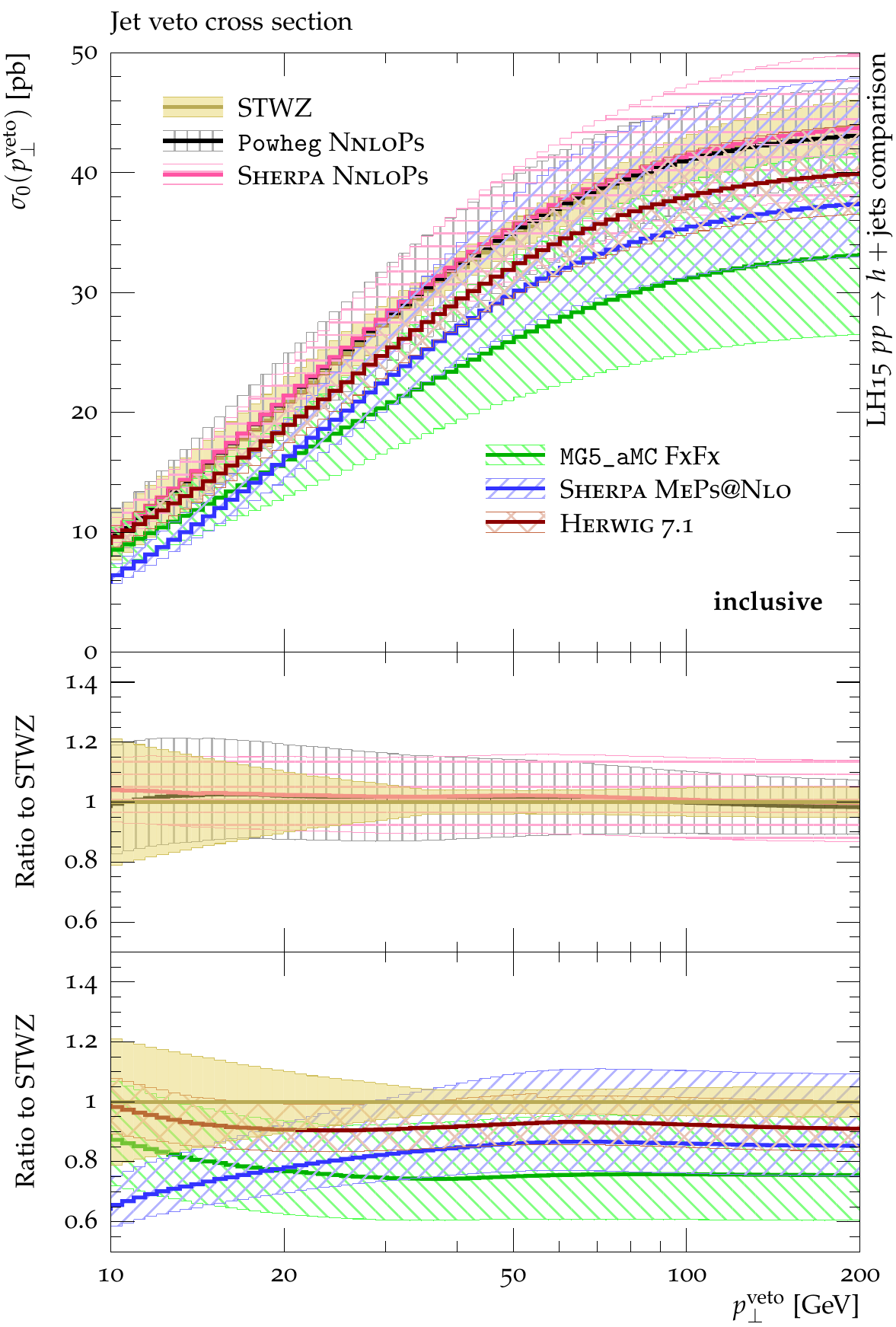}
  \caption{
    The exclusive zero jet cross section as a function of 
    the vetoed minimal leading jet transverse momentum,
    without (left) and with (right) uncertainties.
    \label{fig:hjetscomp:results:jvobs:jvxs0}
  }
\end{figure}

\begin{figure}[t!]
  \centering
  \includegraphics[width=0.47\textwidth]{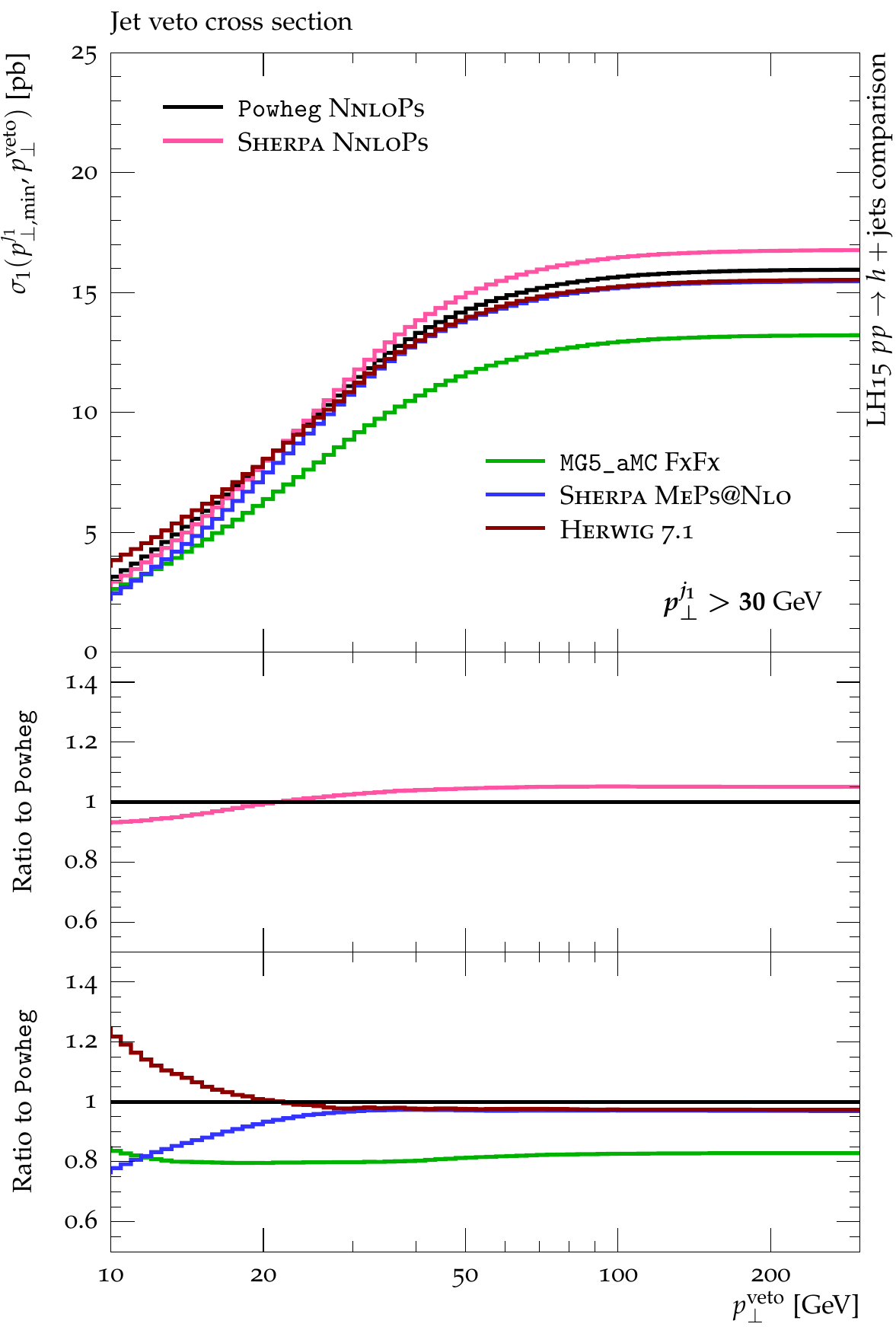}
  \hfill
  \includegraphics[width=0.47\textwidth]{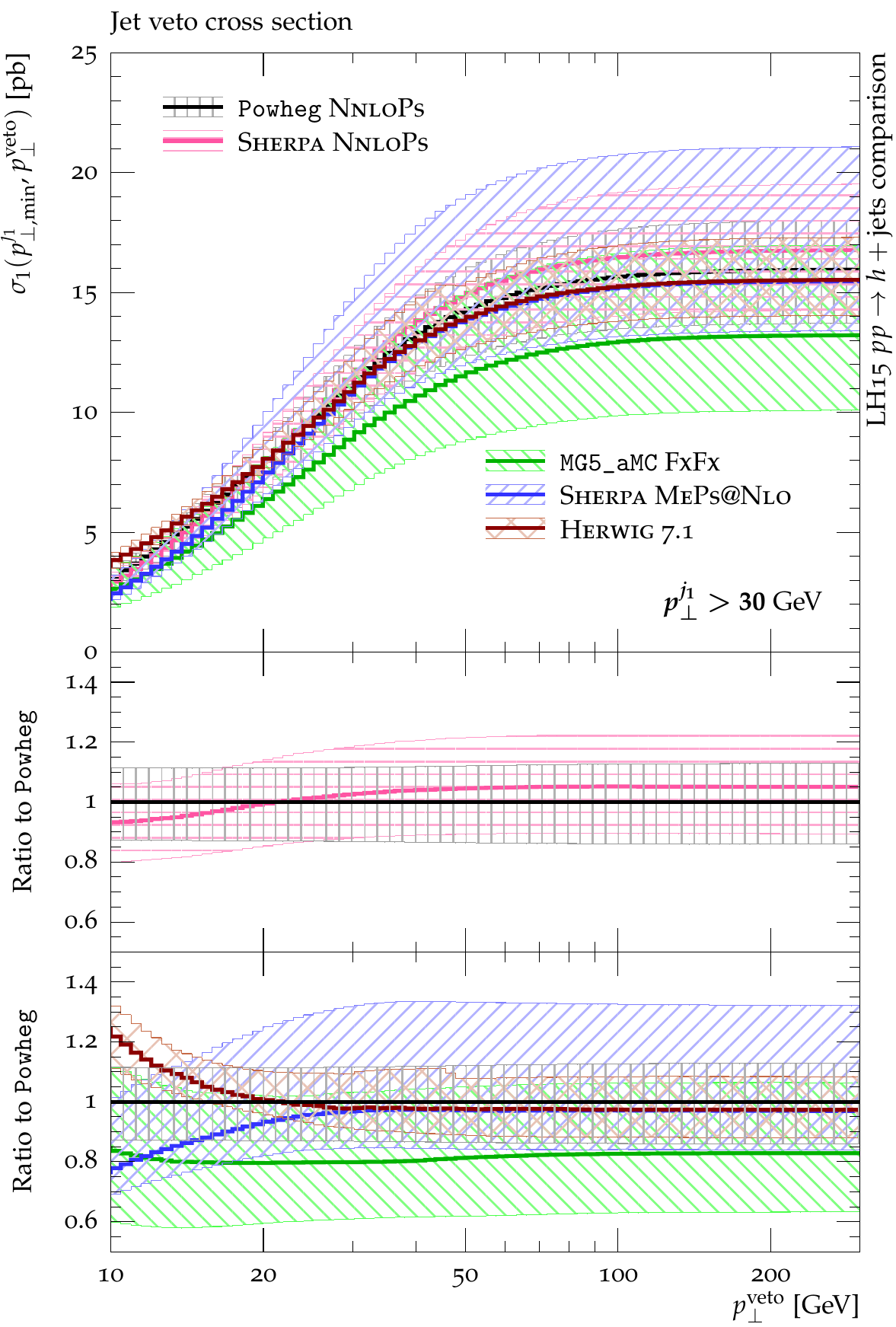}
  \caption{
    The cross section for events containing a Higgs boson 
    and one jet with $p_\perp>30\,\hjetscompgev$ as a function of
    the vetoed minimal second jet transverse momentum without
    (left) and with (right) uncertainties.
    \label{fig:hjetscomp:results:jvobs:jvxs1j30}
  }
\end{figure}

\begin{figure}[t!]
  \centering
  \includegraphics[width=0.47\textwidth]{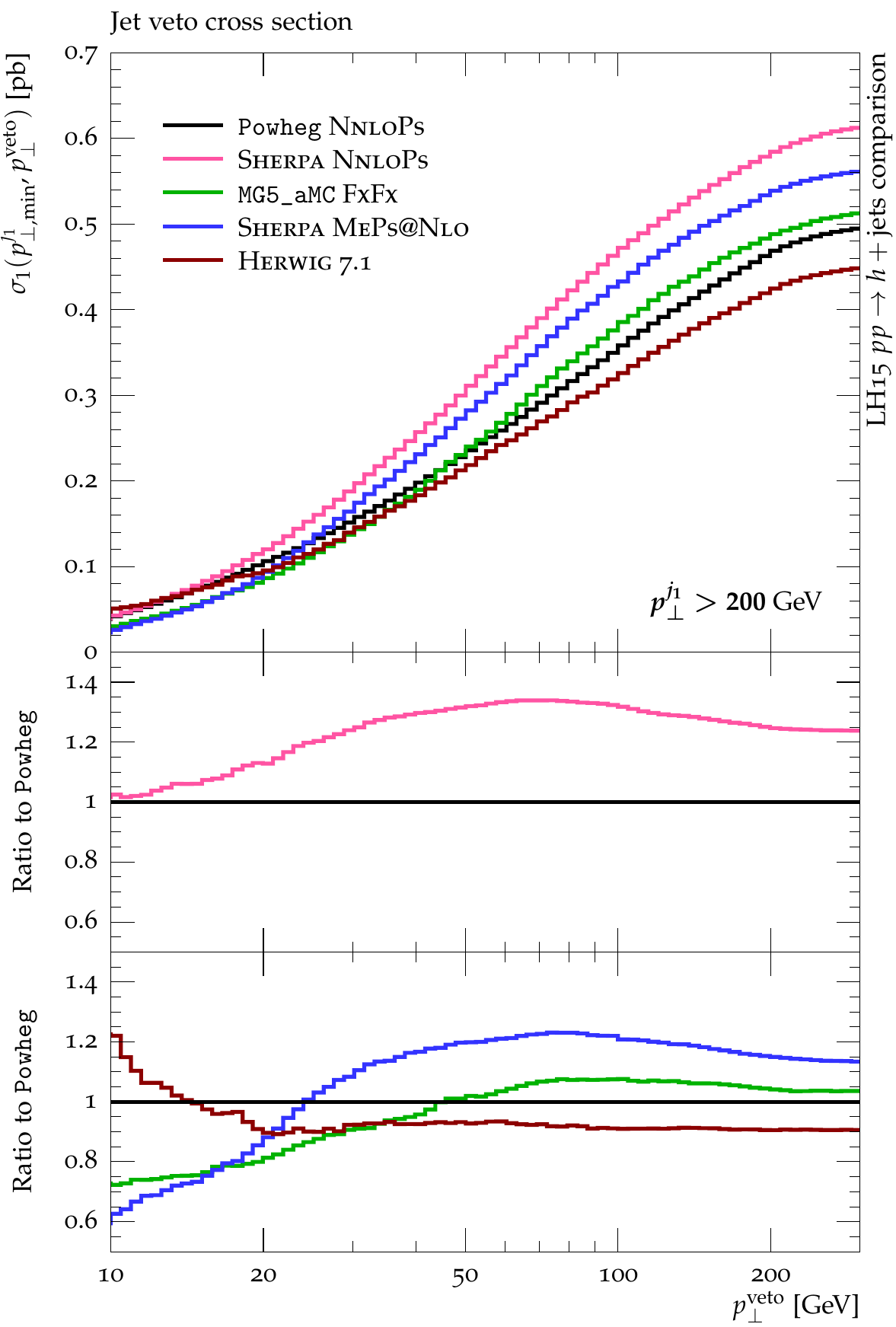}
  \hfill
  \includegraphics[width=0.47\textwidth]{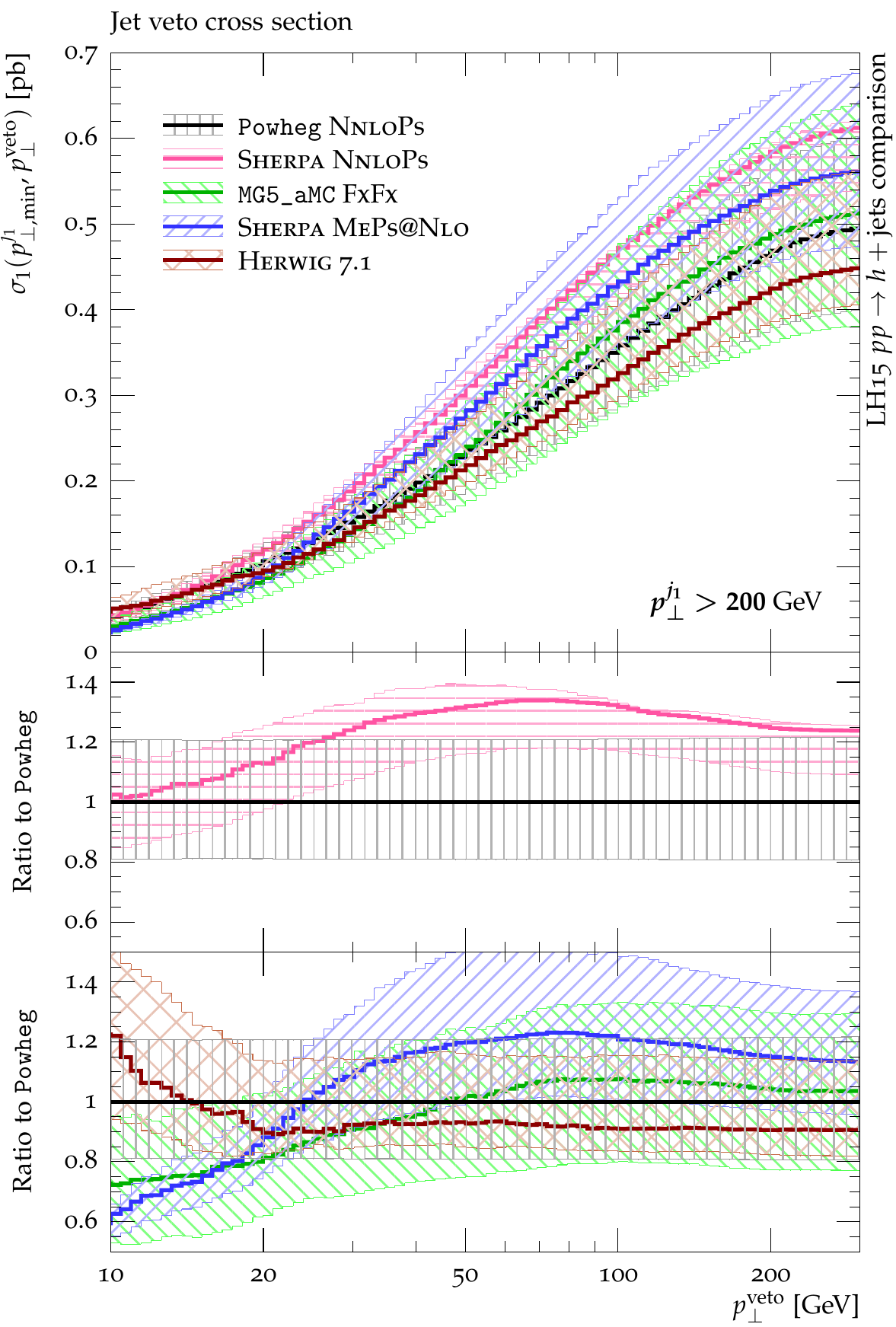}
  \caption{
    The cross section for events containing a Higgs boson 
    and one jet with $p_\perp>200\,\hjetscompgev$ as a function of
    the vetoed minimal second jet transverse momentum without
    (left) and with (right) uncertainties.
    \label{fig:hjetscomp:results:jvobs:jvxs1j200}
  }
\end{figure}

In this section, we investigate jet veto cross sections, where the phase space for
additional gluon radiation is suppressed by means of a jet veto. 
In  Figures 
\ref{fig:hjetscomp:results:jvobs:jvxs0}-\ref{fig:hjetscomp:results:jvobs:jvxs1j200}, 
additional radiation has been vetoed by the application of a maximal transverse 
momentum for the (sub)leading jet, $p_\perp^\text{veto}$. The observables plotted
in the figures recover the respective inclusive cross sections as 
$p_\perp^\text{veto}\to\infty$. In this region, the fixed order 
part of the respective calculations dominates the cross section 
and associated uncertainties. The opposite regime, where $p_\perp^\text{veto}\to 0$, 
is a classic example of a resummation-dominated observable. Here, the 
properties of the respective parton showers come fully into play and 
differences are largely due to their separate characteristics. 

We start by considering 
the cross section for the production of a Higgs boson and no 
additional jets, as a function of the minimum jet transverse momentum 
as shown in Figure~\ref{fig:hjetscomp:results:jvobs:jvxs0}. Remarkable, 
and possibly largely accidental, 
agreement between both \hjetscompNNLOPS simulations and the dedicated resummation 
calculation of STWZ is found, typically better than 5\% within the considered range. 
However, as the resummation accuracy for both \hjetscompNNLOPS' implementations is 
limited by their parton 
shower's accuracy and they do not (\hjetscompPowheg) or only partially (\hjetscompSherpa) 
assess their intrinsic resummation uncertainties and the interplay with 
the hard process' scale variations, their uncertainties are less 
well-determined than those of STWZ. Consequently, while both \hjetscompNNLOPS' uncertainties are largely 
independent of $p_\perp^\text{veto}$, those of the dedicated STWZ calculation 
show the expected comparably large uncertainty in the resummation dominated 
region gradually easing into the near contstant small fixed-order uncertainty 
at large $p_\perp^\text{veto}$ as the influence of the resummation decreases. 
For $p_\perp^\text{veto}\to\infty$ all vetoed cross sections revert to their 
inclusive NNLO accuracy, show-casing the improved convergence of the STWZ 
calculation thanks to its included $\pi^2$-resummation, cf.\ also Fig.\ 
\ref{fig:hjetscomp:results:inclobs:njets}. 
The multijet merged predictions have a wider variation, and have veto cross sections lower than
those provided by the STWZ and the \hjetscompNNLOPS predictions. In addition to 
suffering from their NLO normalisation in the $p_\perp^\text{veto}\to\infty$ 
limit, they also show different behavior as $p_\perp^\text{veto}\to 0$. For example, 
\hjetscompSherpa \hjetscompMEPSatNLO exhibits more QCD activity than the other computations. 

Next, we require the presence of at least one jet with 
a minimal transverse momentum of either $30$ or $200\,\hjetscompgev$. 
The cross sections as a function of 
the subleading jets' maximal transverse momentum are displayed in 
Figure~\ref{fig:hjetscomp:results:jvobs:jvxs1j30} and 
Figure~\ref{fig:hjetscomp:results:jvobs:jvxs1j200} for the two lead jet cuts. 
Note that, although all parton 
shower matched or merged calculations have the same accuracy both as 
$p_\perp^\text{veto}\to\infty$ and in the resummation dominated region, 
the multijet merged calculations possess a better description of the 
second jet emission, and thus should lead to more accurate results 
throughout the spectrum, provided the merging systematics are under control. 
Currently employed uncertainty estimates, however, will not reflect this, 
as resummation uncertainties are not assessed or only incompletely assessed.

If we put no special requirements on the leading jet, cf.\ Figure 
\ref{fig:hjetscomp:results:jvobs:jvxs1j30}, good agreement 
between all calculations is found. The \hjetscompNNLOPS predictions agree again within 
5\% of one another and have very similar uncertainties. This is noteworthy 
as, in comparison with the results of 
Figure~\ref{fig:hjetscomp:results:jvobs:jvxs0}, both calculations' 
accuracies have been degraded by one order. The multijet merged calculations 
show similar behavior as in the previous case: \hjetscompMGaMC exhibits a smaller cross section due to 
its scale choice while \hjetscompSherpa \hjetscompMEPSatNLO predicts more soft radiation. 
The relative lack of small $p_\perp$ radiation is more 
pronounced in \hjetscompHerwig for this observable. Again, the uncertainties of \hjetscompMGaMC and \hjetscompSherpa 
are of similar size while those of \hjetscompHerwig are somewhat smaller than those of the \hjetscompNNLOPS 
predictions, especially in the resummation-dominated region 
$p_\perp^\text{veto}\to 0$.

Raising the requirements on the leading jet to $200\,\hjetscompgev$, in Figure 
\ref{fig:hjetscomp:results:jvobs:jvxs1j200}, displays clear distinctions 
between the different calculations.  
\hjetscompSherpa \hjetscompNNLOPS and \hjetscompPowheg \hjetscompNNLOPS have noticeably different shapes, with \hjetscompSherpa having a 
much lower cross section for low values of the subleading jet veto requirement. 
The multijet merged calculations show \hjetscompHerwig largely agreeing with 
\hjetscompPowheg with a constant offset of $-10\%$, and the familiar lower 
probability of low-$p_\perp$ jet emissions. 
The asymptotic cross section for \hjetscompSherpa \hjetscompMEPSatNLO is similar to that from \hjetscompSherpa \hjetscompNNLOPS and, 
unsurprisingly, due to the use of the same parton shower, shows a 
similar radiation pattern. As before, it exhibits 
a relative overabundance of soft jet radiation. Lastly, the asymptotic cross section from \hjetscompMGaMC 
is the same level as \hjetscompPowheg's, despite 
its higher scale choice. The pattern of the uncertainties, however, 
remains the same as before.

\subsection{Conclusions}
\label{sec:hjetscomp:conclusions}

Precision Higgs boson measurements will soon be possible using the new
13 TeV data currently being collected during Run II of the LHC.
The largest Higgs boson production process is gluon fusion. A variety of theoretical 
tools exist for predictions for Higgs boson (+jets) production in this 
channel. As higher order corrections are especially sizable for $gg\to h$, 
it is important to understand the accuracies and regions of applicability 
of the various predictions. Too often, the comment has been that since 
the predictions agree within the theoretical uncertainties, all is well. 
However, a higher standard must be used, as the predictions have many of 
the theoretical uncertainties, such as scale uncertainties, in common. 

We have compared fixed order predictions at NLO and NNLO (including 
approximate NNLO), resummed predictions, NNLO predictions matched to parton 
showers, multjet merged predictions at NLO accuracy, and resummations 
in the BFKL limit. This allows a better understanding of two main issues: firstly how 
consistent are calculations which should be consistent, and secondly the impact 
of soft gluon radiation and higher order corrections. All predictions have 
been carried out without non-perturbative corrections to allow for a one-to-one 
comparisons. A few observations follow. 

NNLO effects can change not only the normalization of distributions, but 
also the shape. For example, NLO predictions of the Higgs rapidity 
distribution are all similar in normalization and shape; however, the 
NLO predictions fall off more rapidly at high rapidity than predictions 
at NNLO. The differences between NLO and NNLO, however, are only 
noticeable in regions beyond the kinematic cuts applied at the LHC. It is 
interesting that, for the scale choices used in this study, there is 
neither a shape nor normalization shift from the NNLO corrections for the 
inclusive lead jet $p_\perp$ distribution while they are present for 
the Higgs boson $p_\perp$ in the presence of at least one jet.. 

The highest precision for inclusive jet multiplicity distributions is 
present with fixed order predictions, either at NLO or NNLO; in general, 
predictions from resummed/parton shower programs agree well with the 
fixed order predictions within their expected accuracy. The most 
extensive comparisons in this study are with respect to the $p_\perp$ 
distribution for the lead jet in $h+\ge1$ jet events. With the recent 
NNLO calculation for this quantity, the uncertainties are less than 10\%. 
As mentioned above, the NNLO corrections are small. The \hjetscompNNLOPS 
predictions for this variable are in good agreement with each other 
as well as with the NLO/NNLO predictions for $p_\perp \le$ 100 GeV/c, with 
some separation between the \hjetscompNNLOPS results at higher transverse momentum. 
The multijet jet merged predictions agree well with NLO/NNLO at low $p_\perp$, 
but fall below by about 15-20\% at high $p_\perp$, an effect that can be attributed 
to the specific scale choices. The resummed prediction from STWZ and \hjetscompResbos 
agree well at low $p_\perp$, but rise above the fixed order results 
by about 20\% at high $p_\perp$, again dut to diffences in the scale choice. 
One of the key take-home points is that 
the introduction of a parton shower or a resummation should not greatly 
affect the fixed order results, for observables that are suitably 
inclusive. These conclusions are largely true for comparisons for the 
sub-leading and third-leading jet as well. The situation is more complex 
for exclusive final states, where a jet veto is applied to any additional 
jet. Here, there are jet veto logs that have to be resummed. We note, 
though, in general that there is still good agreement among the 
predictions used here. Although, this is not explicitly part of this 
study, we note that the impact of jet veto logarithms (after resummation) 
on the NNLO prediction for $h+\ge1$ jet are small, indicating again that 
the fixed order predictions for that quantity should be 
reliable~\cite{Banfi:2012jm,Banfi:2015pju}.

Resummed/parton shower predictions provide a better description for 
observables where Sudakov effects are important, such as the transverse 
momentum distribution for inclusive (exclusive) Higgs production. For 
the inclusive case, all predictions agree well with each other (and with 
the reference \hjetscompHqT), with some deviations observed at the lowest $p_\perp$ 
values. Differences are more evident for the exclusive case, where the 
high transverse momentum for the Higgs boson must be supplied by a combination of 
jets lower than the $30\,\hjetscompgev$ cutoff and soft gluon radiation. Although 
the fixed order predictions are unstable at low $p_\perp$, there is good 
agreement with the resummed/parton shower predictions at high $p_\perp$, 
where the bulk of the transverse momentum for the resummed/parton shower 
prediction is provided by the hard matrix element. 

In Sudakov regions involving multiple jets the situation gets more complex
and discrepancies are often more evident, as for example for the 
Higgs boson $p_\perp$ distribution in $h+\ge n$ jet production, or the 
system $p_\perp$ for $h+n$ jets. It is interesting to note that the fixed 
order predictions are more stable at low transverse momentum when more 
jets are involved (as expected). 

The rapidity interval between two jets for $h+\ge2$ jets might be thought 
of as a fairly robust variable. However, differences can be observed among 
the various multijet merged predictions at high $\Delta y$. This is especially true 
if the two most forward-backward (rather than the two leading) jets are 
chosen, indicating perhaps that the differences in evidence are a result 
of the parton shower. It is interesting to note that the \hjetscompPowheg \hjetscompNNLOPS 
and \hjetscompSherpa \hjetscompNNLOPS predictions agree well with each other and with the 
fixed order predictions (taking into account the NLO corrections present 
in the latter). The $\Delta \phi$ distribution between the two leading 
jets is an observable where basically all predictions agree. 

In order to measure the vector boson fusion process, additional kinematic 
cuts are necessary to reduce the gluon-gluon fusion background, requiring 
a dijet rapidity separation, and/or a dijet mass requirement, applied 
either to the leading jets (VBF) or to any pair of 
jets (VBF2). The cross sections for $\ge2$ and $\ge3$ jets from \hjetscompSherpa 
\hjetscompMEPSatNLO and from \hjetscompGoSam{}+\hjetscompSherpa (with either of the VBF cuts) are larger than 
those from \hjetscompPowheg \hjetscompNNLOPS, expected as the normalizations for the $\ge2$ 
and $\ge3$ jet cross section for the former are at NLO. In general, the 
predictions for the kinematic distributions are similar among the various 
programs for both VBF and VBF2 cuts, except that both \hjetscompSherpa \hjetscompMEPSatNLO and 
\hjetscompSherpa \hjetscompNNLOPS tend to predict a shallower dip that occurs in the $\Delta \phi$ 
distribution at values around 1.5. The effect is larger when the 
the second tagging definition is used, as might be expected if the effect was 
primarily due to the parton shower. 

A multi-jet quantity such as $H_{T, {\rm jets}}$ is sensitive to 
radiation/production of extra jets. Basically, all programs, with the 
exception of \hjetscompHerwig, predict larger cross section than \hjetscompPowheg \hjetscompNNLOPS. 
The largest deviations are from the NNLO  $h+\ge1$ jet prediction, at 
$H_{T, {\rm jets}}$ values roughly from $200\,\hjetscompgev$ to $400\,\hjetscompgev$. 
It is interesting that the \hjetscompLoopsim (nNLO) prediction for the same quantity 
is in good agreement with the exact NNLO prediction, as expected from the 
\hjetscompLoopsim procedure.

Finally, we conclude with a brief review of the results from the jet veto 
cross section comparisons. For $h$ + no jets, the \hjetscompNNLOPS predictions are 
in remarkable agreement with those obtained from STWZ. There is a wider 
variation from the multijet merged programs, with all predicting a smaller 
jet veto cross section than STWZ and the \hjetscompNNLOPS programs. For $h$ plus 
exactly one jet ($\ge 30\,\hjetscompgev$), the two \hjetscompNNLOPS predictions are still in 
agreement with each other, and with the multijet predictions, with some 
divergence as the jet veto threshold is reduced below $30 \,\hjetscompgev$. If the 
lead jet threshold is increased to $200\,\hjetscompgev$, there is a wide divergence 
of predictions, indicating the difficulties of dealing with such multi-scale 
situations.

\subsection*{Acknowledgements}

We thank the organisers.
The work of SH, YL and SP was supported by the U.S. Department of Energy 
under contract DE--AC02--76SF00515.
The work of FJT was supported by the German Science Foundation (DFG) 
through the Emmy-Noether Grant No.\ TA 867/1-1. 
MS acknowledges support by the Swiss National Science Foundation (SNF) 
under contract PP00P2--128552. 
KB would like to thank Keith Hamilton and Giulia Zanderighi for their 
advice and support.

\section{Photon isolation studies
  \texorpdfstring{\protect\footnote{
    L.~Cieri, G.~Heinrich
  }}{}}

We study the  effect of various photon isolation criteria in processes
involving single photon plus jet and diphoton production.

\subsection{Introduction}

 Final states involving isolated photons have played and still do play
 a very important role at the LHC. Due to the clean photon signal, 
the diphoton channel has been the discovery channel for the Higgs
boson, and it keeps intriguing us by exhibiting structures that may
hint to resonances existing around 750\,GeV.

Even within the Standard Model, photons in the final state can have
a number of  different origins.
They can either originate directly from the hard interaction
(in this case they are called {\it direct}), or they can come from a
fragmentation process of a QCD parton in the final state.
Further there is a (large but reducible) background of secondary photons 
coming from the decay of pions, $\eta$-mesons, etc.
In order to suppress the background, an 
 {\it isolation criterion} is usually applied, which suppresses both
 the secondary photons and the fragmentation component.

Beyond the leading order, the distinction between the direct and
the fragmentation contribution is not unique any more. 
For example, at NLO, the collinear splitting of a parent quark into a 
photon and quark leads to a singularity, which can be absorbed into
fragmentation functions, analogous to the absorption of initial
state QCD radiation into ``bare PDFs''. 
The fragmentation functions have a perturbative component, 
involving the splitting function $P_{q\to \gamma+q}(x)$, and a
non-perturbative component which has to be obtained from fits to data.
As these fits have been done based on LEP data, and are less refined than 
the PDF fits which are based on a wealth of data and which are constantly
improved, 
there is an inherent uncertainty in the fragmentation component which 
makes it desirable to suppress this contribution as much as possible. 
Another reason to suppress it, apart from the obvious experimental reasons, 
is given by the fact that higher
orders for this part are much more difficult to calculate than 
for the direct photon contribution, as it requires the calculation of 
jet cross sections, which are then convoluted with the fragmentation functions.

In the following we will study the behaviour of the direct and fragmentation
contributions with different isolation criteria and at different
orders in perturbation theory, not only for
diphotons, but also for photon plus jet final states. 
We will also study how the effects vary with the centre of mass energy 
by comparing results for $\sqrt{s}=8, 14$ and 100 TeV.
 
\subsection{Isolation criteria}

The  most commonly used isolation criterion at hadron collider 
experiments is a {\it cone-based isolation} prescription.
In this procedure, the photon candidate is called isolated if
in a cone of radius $R$ in rapidity $y$ and azimuthal angle $\phi$ around the
photon direction, 
the amount of  hadronic transverse energy $\sum E_{T}^{had}$
is smaller than some value 
$E_{T}^{max}$ chosen by the experiment:
\begin{eqnarray}\label{eq:coneisol}    
&\sum E_{T}^{had} \leq E_{T}^{max} \nonumber\\
&\mbox{inside a cone with}     
\left( y - y_{\gamma} \right)^{2} +    
\left(  \phi - \phi_{\gamma} \right)^{2}  \leq R^{2}  \;.   
\end{eqnarray} 
$E_{T}^{max}$ can be defined either in absolute terms, or as 
 fraction $\eps_c$ of $p_{T}^{\gamma}$ (typically $\eps_c$= 0.1 or below).  

\vspace*{4mm}

In a theoretical calculation, the need for a fragmentation contribution 
can be eliminated by using a {\it smooth  isolation}
criterion~\cite{Frixione:1998jh}, where
the  threshold on the hadronic energy inside the isolation cone
decreases with the
 radial distance from the photon. It is described by the cone size $R$, 
 a weight factor $n$ and 
 an isolation parameter $\epsilon_f$. With this criterion, the
 photon is isolated if
 \begin{eqnarray}\label{eq:frixisol}     
&\sum E_{T}^{had} \leq E_{T}^{max}~\chi(r)\nonumber\\
&\mbox{inside each cone with} \;\;      
r^{2}=\left( y - y_{\gamma} \right)^{2} +    
\left(  \phi - \phi_{\gamma} \right)^{2}  \leq R^{2}  \;,    
\end{eqnarray}  
where  the function $\chi(r)$ has to fulfill
\begin{equation}
\chi(r) \rightarrow 0 \;\mbox{if} \; r \rightarrow 0\; \mbox{and}\;
0<\chi(r)< 1 \;\mbox{if} \; 0<r<R\;.
\end{equation}
One possible choice is
\begin{equation}
\label{eq:chi}
\chi(r) = \left( \frac{1-\cos (r)}{1-\cos R} \right)^{n}\;,
\end{equation}
where usually $n=1$ and $E_{T}^{max}=\eps_f\,p_T^\gamma$ are chosen.

However, due to finite detector resolution, an 
 experimental realization of this criterion will only be possible to some minimal value of $r$, thereby 
 leaving potentially a residual collinear contribution. 

We would also like to point out that $\epsilon_f$ and $\epsilon_c$ are
not directly comparable, because $\epsilon_f$ is still multiplied by
$\chi(r)$ and therefore can be much larger than $\epsilon_c$ for
comparable values of $\sum E_{T}^{had}$.


\subsection{Results}

\subsubsection{Diphotons at NLO}
We compare the standard and the smooth cone~\cite{Frixione:1998jh} isolation prescriptions using an acceptance criterion based on Higgs boson searches and studies~\cite{aad:2012gk}. We present numerical results at NLO requiring $p_T^{\rm harder} \geq 40$~GeV and $p_T^{\rm softer}\geq 30$~GeV, and restricting the rapidity of both photons to $|y_\gamma|<2.37$.  We use the NNPDF23\_nlo\_as\_0119~\cite{Ball:2012cx} PDF set with densities and $\alpha_s$ evaluated at each corresponding order (i.e., we use $(n+1)$-loop $\alpha_s$ at N$^n$LO, with $n=0,1$) and we consider $N_f=5$ massless quarks/antiquarks as well as gluons in the initial state. The QED coupling constant $\alpha$ is fixed to $\alpha=1/137$. We show results for three different values for the centre of mass energy\,: $\sqrt{s} = 8$~TeV, $14$~TeV and $100$~TeV, where we consider the standard cone isolation criterion implemented in the numerical code {\tt DiPhox}~\cite{Binoth:1999qq}, and we use two different sets of fragmentation functions: the Bourhis-Fontannaz-Guillet (BFG)~\cite{Bourhis:1997yu} set\footnote{The BFG set is obtained from an NLO fit to LEP data.} and the LO fragmentation functions by Owens\footnote{The LO Owens fragmentation functions are used only in combination with the LO matrix elements.}~\cite{Owens:1986mp}. The smooth cone isolation prescription is implemented in the numerical code $2\gamma${\tt NNLO}~\cite{Catani:2011qz}. We compare both isolation criteria using the following isolation parameters: $\epsilon_c=\epsilon_f=0.05$,~$0.1$, and $0.5$. We impose a size of the cone $R=0.4$ and in the smooth cone isolation criterion we use $n=1$.

\begin{figure}[t!]
\includegraphics[width=0.47\textwidth]{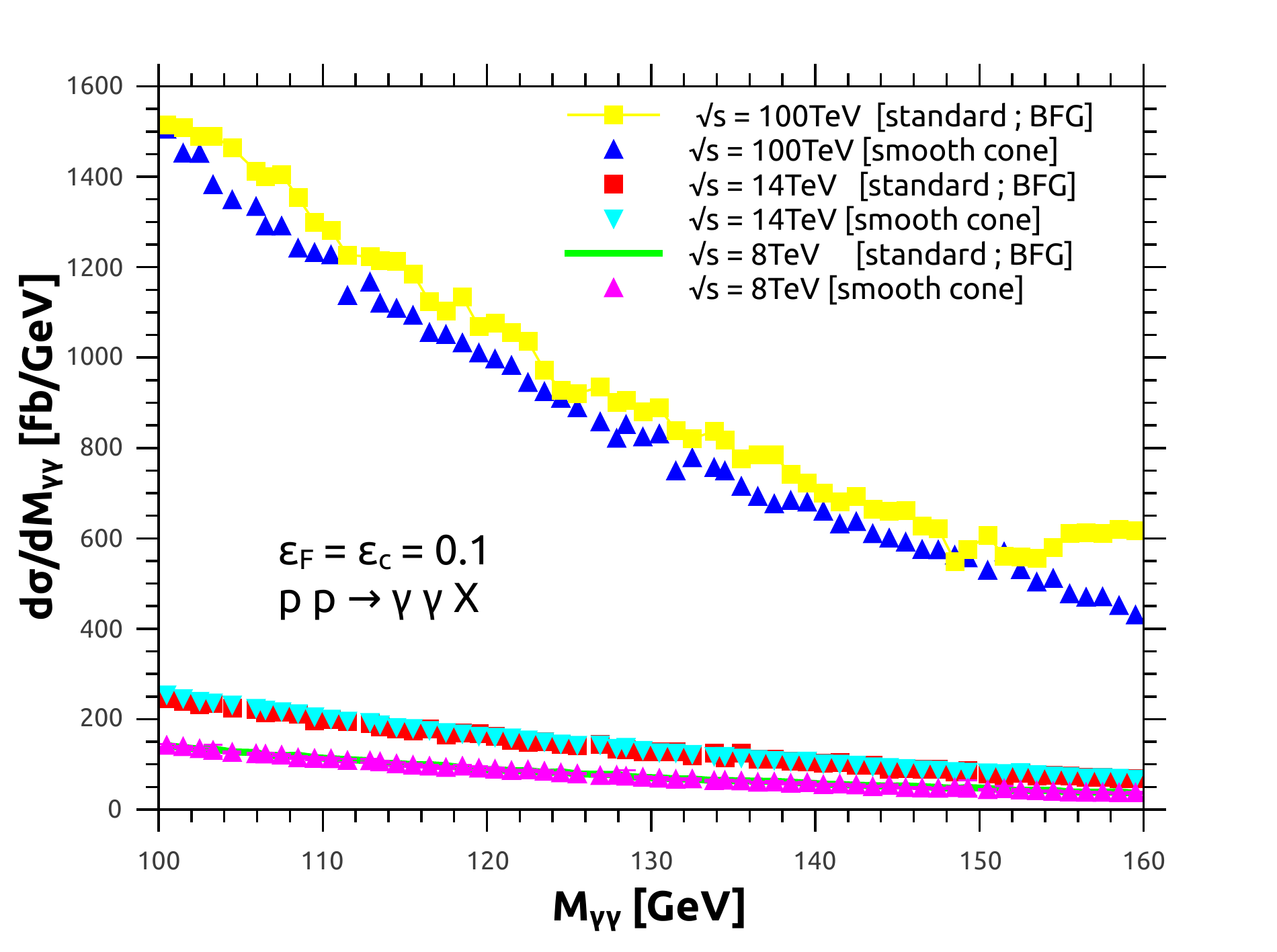}
\hfill
\includegraphics[width=0.47\textwidth]{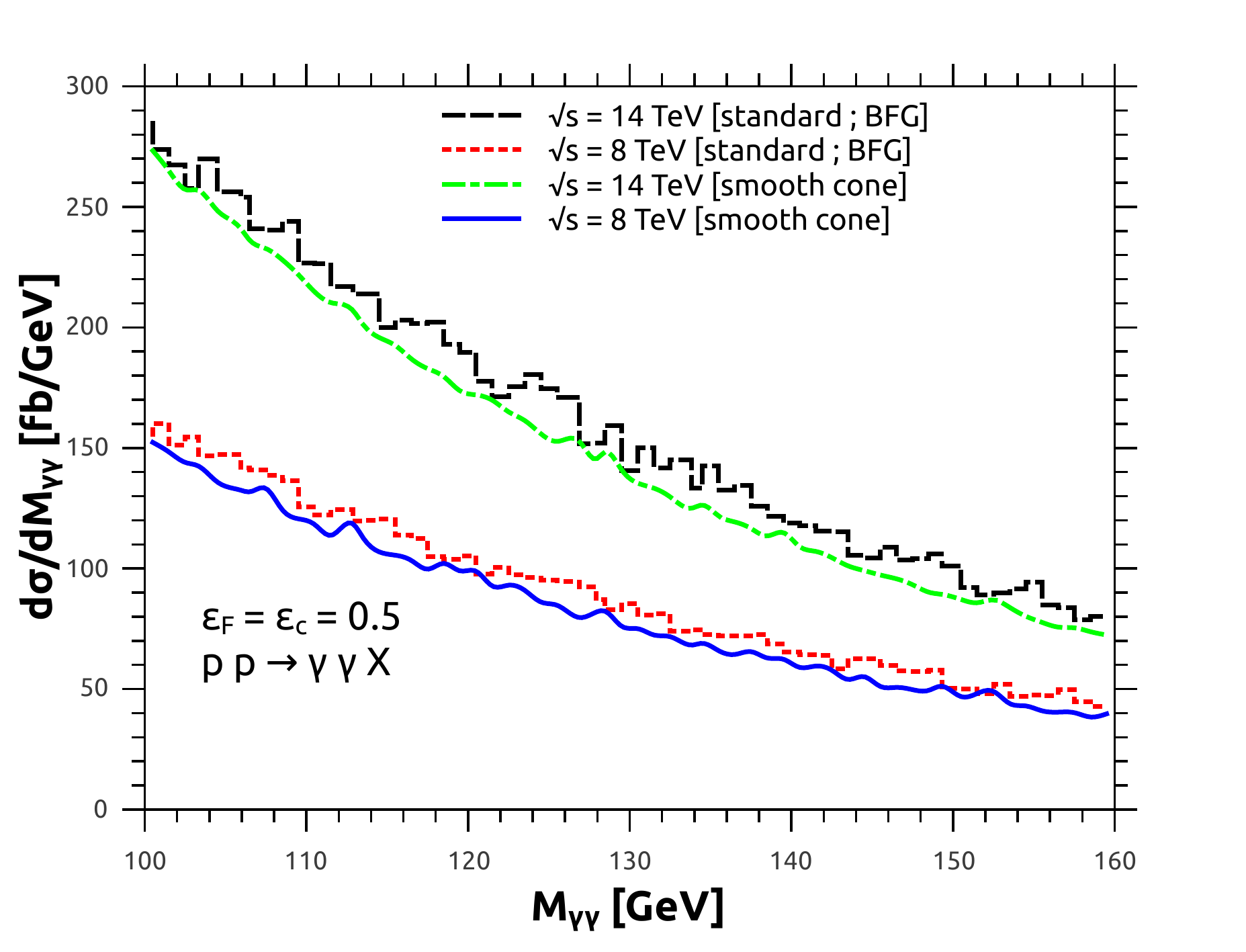}
\caption{Comparison of different isolation criteria at $\sqrt{s}=8$~TeV, 14~TeV and 100~TeV. In the left panel we show the invariant mass distribution obtained with $\epsilon = 0.1$ and in the right panel the corresponding one for $\epsilon = 0.5$ \label{fig:prelim}.}
\end{figure}

In Fig.~\ref{fig:prelim} we show the invariant mass distributions at the centre of mass energies $\sqrt{s}=8$~TeV, 14~TeV and 100~TeV for two values of the isolation parameter: $\epsilon = 0.1$ (left panel) and $\epsilon = 0.5$ (right panel). In Fig.~\ref{fig:prelim} we notice the different orders of magnitude of the cross-sections depending on the centre of mass energy. In the following we study in detail each centre of mass energy and whether it is possible to define a ``\textit{tight isolation prescription}''~\cite{Andersen:2014efa}.

\begin{figure}[t!]
\includegraphics[width=0.47\textwidth]{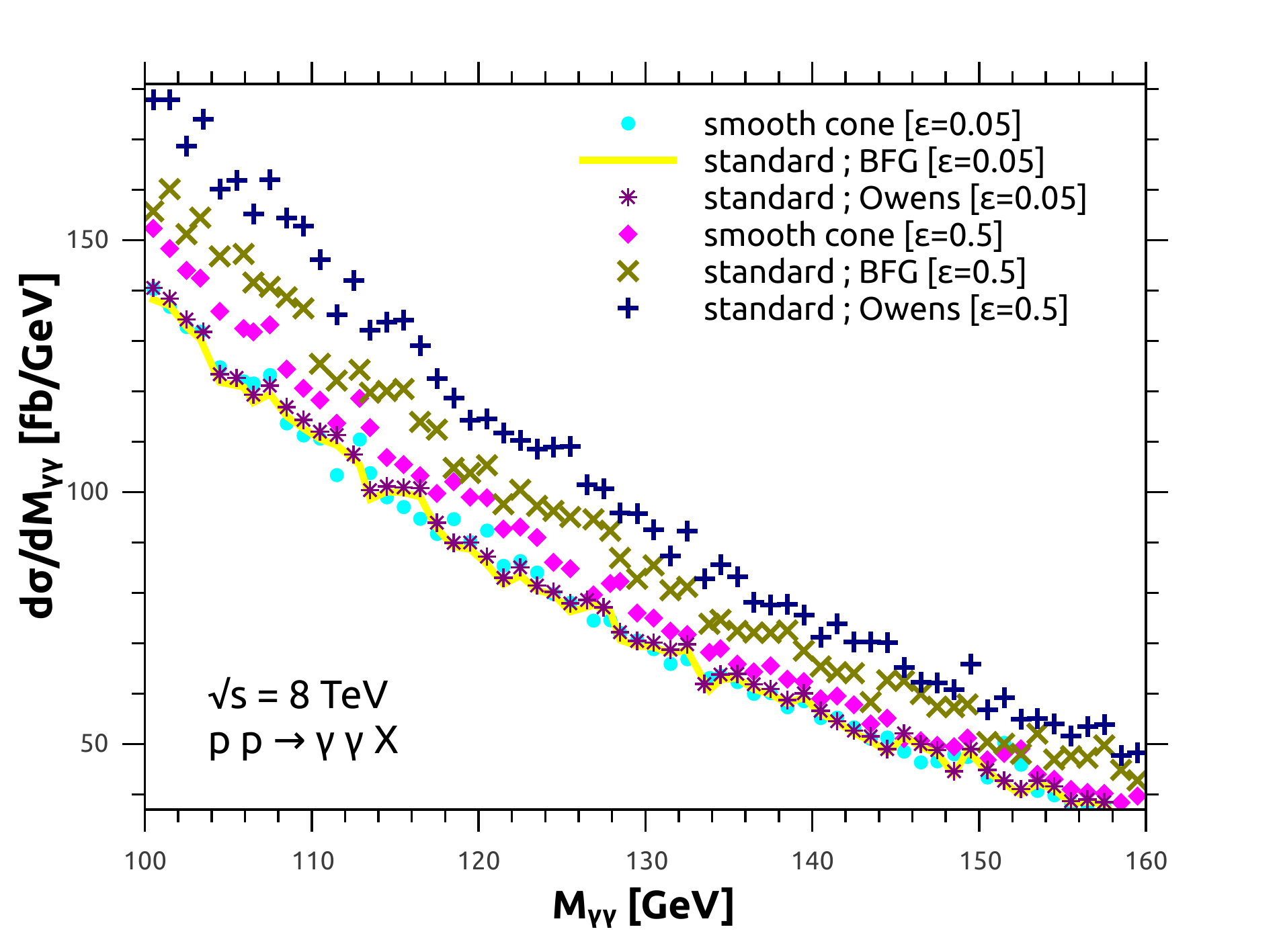}
\hfill
\includegraphics[width=0.47\textwidth]{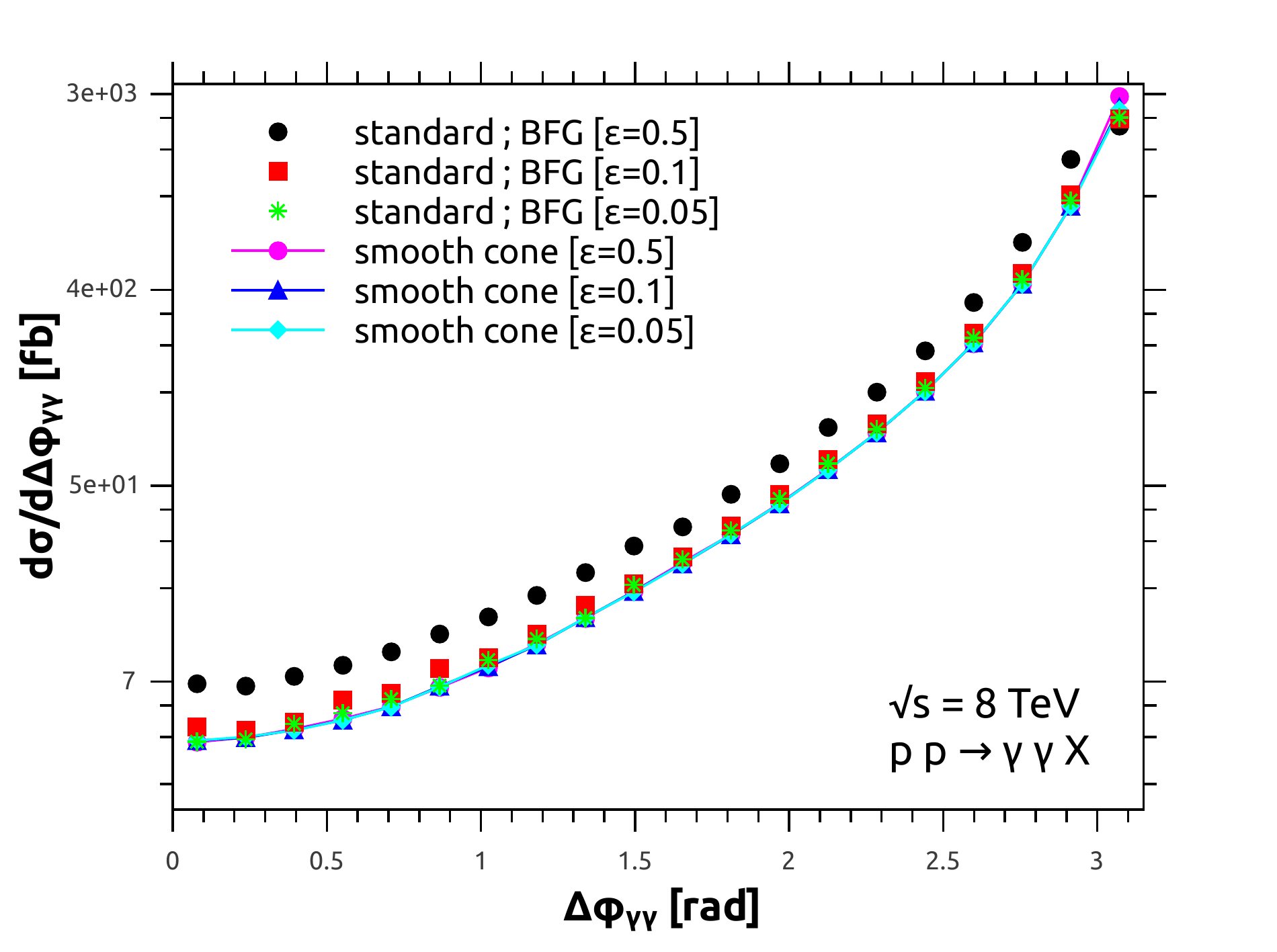}
\caption{Comparison of different isolation criteria at $\sqrt{s}=8$~TeV. In the left panel we show the invariant mass distribution and in the right panel the $\Delta \Phi_{\gamma\gamma}$ \label{fig:8TeVDphiMgg}.}
\end{figure}

In the left panel of Fig.~\ref{fig:8TeVDphiMgg} we show the invariant mass distribution of the diphoton pair at NLO for a centre of mass energy of $\sqrt{s}=8$~TeV. For tight isolation parameters ($\epsilon \leq 0.1$) we observe that the two criteria give essentially the same result (agreement at  percent level) considering the BFG fragmentation functions. While for \textit{loose} isolation parameters (in particular for $\epsilon = 0.5$) the result with the smooth cone is about $9\%$ smaller than the standard cone result (with BFG). The cross-section obtained with the Owens set of fragmentation functions coincides (within $1\%$) with the standard cone (BFG) and smooth cone differential cross-sections for the tight isolation parameter $\epsilon=0.05$. 
However, for $\epsilon=0.1$, the cross section obtained with the Owens  fragmentation functions starts to deviate (being about $2.3\%$ larger) from the result with the BFG fragmentation functions and the cross section obtained with the smooth cone isolation criterion. For $\epsilon=0.5$ the Owens result exhibits large deviations from the BFG ($10\%$) and smooth ($19\%$) cross-sections.

In the right panel of Fig.~\ref{fig:8TeVDphiMgg} we present the results for the $\Delta \Phi_{\gamma\gamma}$ distribution. In this differential distribution, $\Delta \Phi_{\gamma\gamma}=\pi$ is the Born-like kinematical region in which the two photons are back-to-back. All the other kinematical regions are away from the back-to-back configuration. Kinematical regions with $\Delta \Phi_{\gamma\gamma}\neq \pi$ receive contributions from NLO real radiation configurations to the direct photon matrix element, or from the fragmentation part. The standard cone cross-section obtained with the BFG set of fragmentation functions is about $4\%$ larger than the smooth cone cross-section ($\Delta \Phi_{\gamma\gamma}\neq \pi$) if tight isolation parameters are considered ($\epsilon \leq 0.1$).

For $\epsilon=0.5$, the distribution obtained with the smooth cone isolation is about $37\%$ smaller than the standard cone result with BFG fragmentation functions for $\Delta \Phi_{\gamma\gamma}\neq \pi$. 

The ratio between the smooth cone results with $\epsilon=0.05$ and $\epsilon=0.5$ is close to 1 (within $0.1\%$) for all the kinematical ranges with the exception of the last bin\,\footnote{Concerning the $\Delta \Phi_{\gamma\gamma}$ distribution in the last bin $\Delta \Phi_{\gamma\gamma}\simeq \pi$ the cross section with the smooth cone criterion using $\epsilon=0.5$ is about $12.8\%$ larger than the cross section obtained with $\epsilon=0.05$.} $\Delta \Phi_{\gamma\gamma}\simeq \pi$, which contains the Born-like contributions. This fact can be explained noticing that all  configurations where the extra QCD parton (which is present at NLO 
in the ral radiation part) lies outside the cone around each photon are independent of the value for $\epsilon$ or $E_{T}^{max}$, while the nearly collinear configurations 
are suppressed with smooth isolation, and the ones where the extra parton is soft or collinear to one of the photons contribute to the back-to-back situation 
$\Delta \Phi_{\gamma\gamma} \simeq \pi$.

\begin{figure}[t!]
\includegraphics[width=0.47\textwidth]{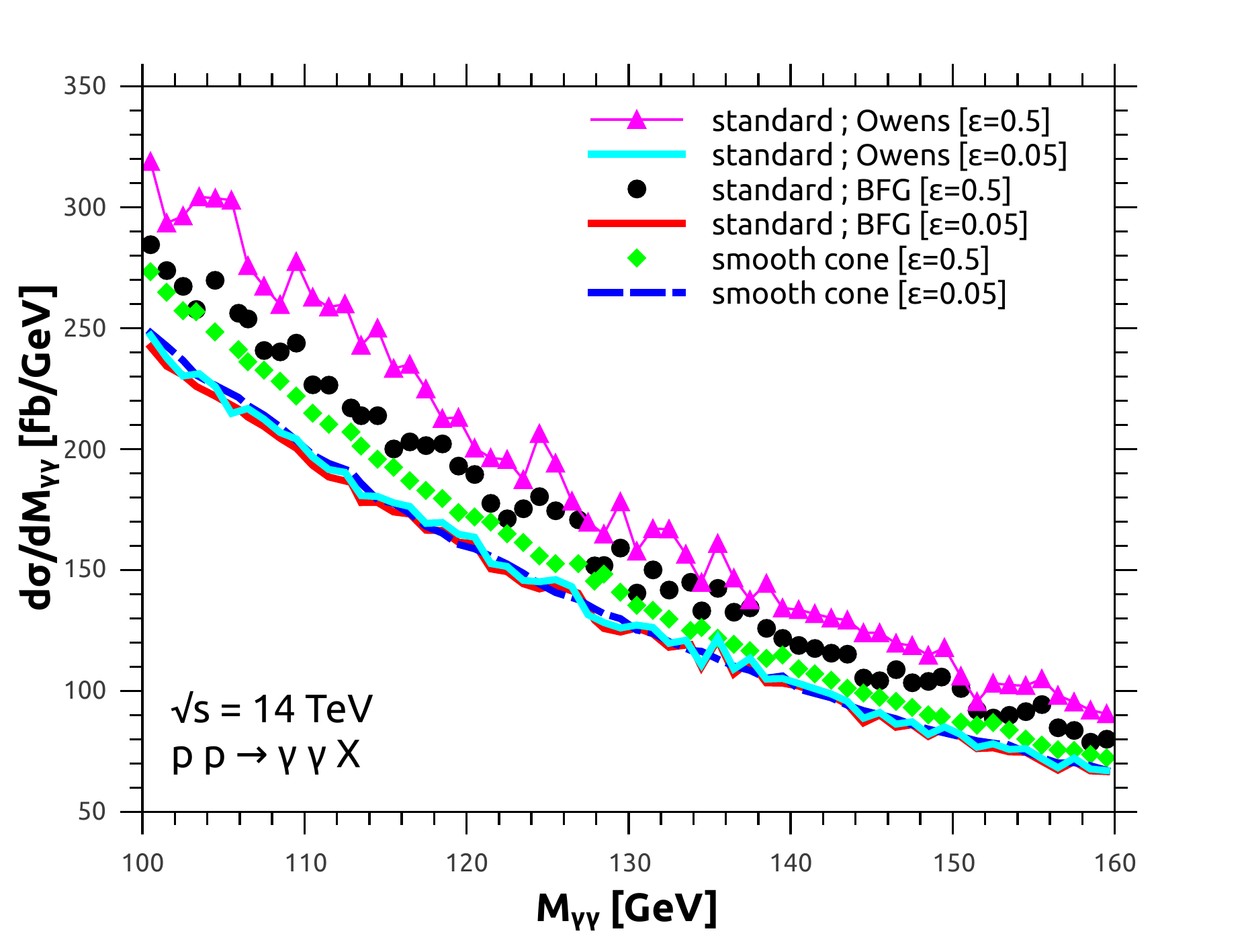}
\hfill
\includegraphics[width=0.47\textwidth]{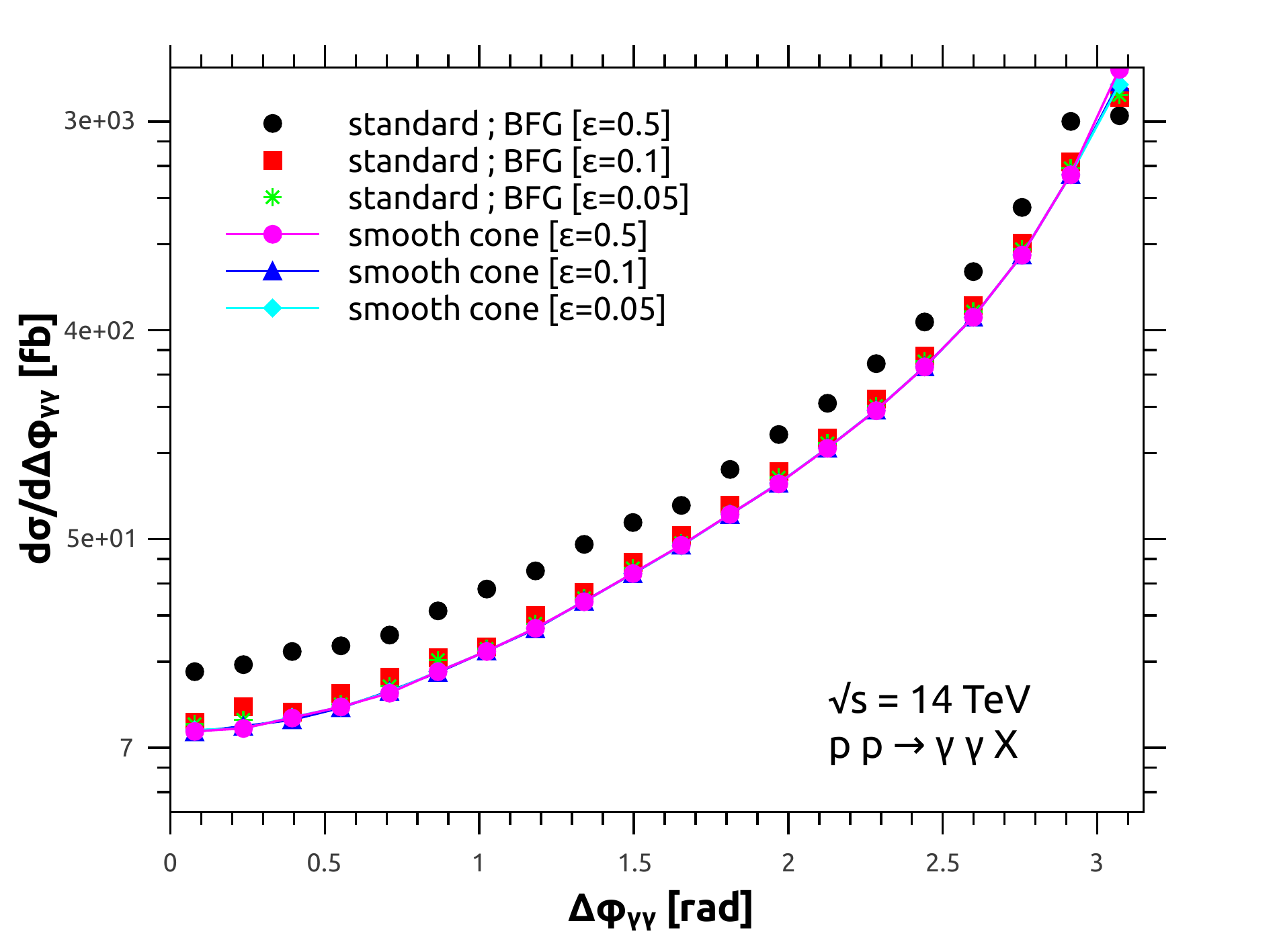}
\caption{Comparison of different isolation criteria at $\sqrt{s}=14$~TeV. In the left panel we show the invariant mass distribution and in the right panel the $\Delta \Phi_{\gamma\gamma}$ distribution.\label{fig:14TeVDphiMgg}}
\end{figure}

At $\sqrt{s}=14$~TeV (Fig.~\ref{fig:14TeVDphiMgg}), concerning the invariant mass and the $\Delta \Phi_{\gamma\gamma}$ distributions, we have essentially the same considerations as at $\sqrt{s}=8$~TeV. 

\begin{figure}[t!]
\includegraphics[width=0.47\textwidth]{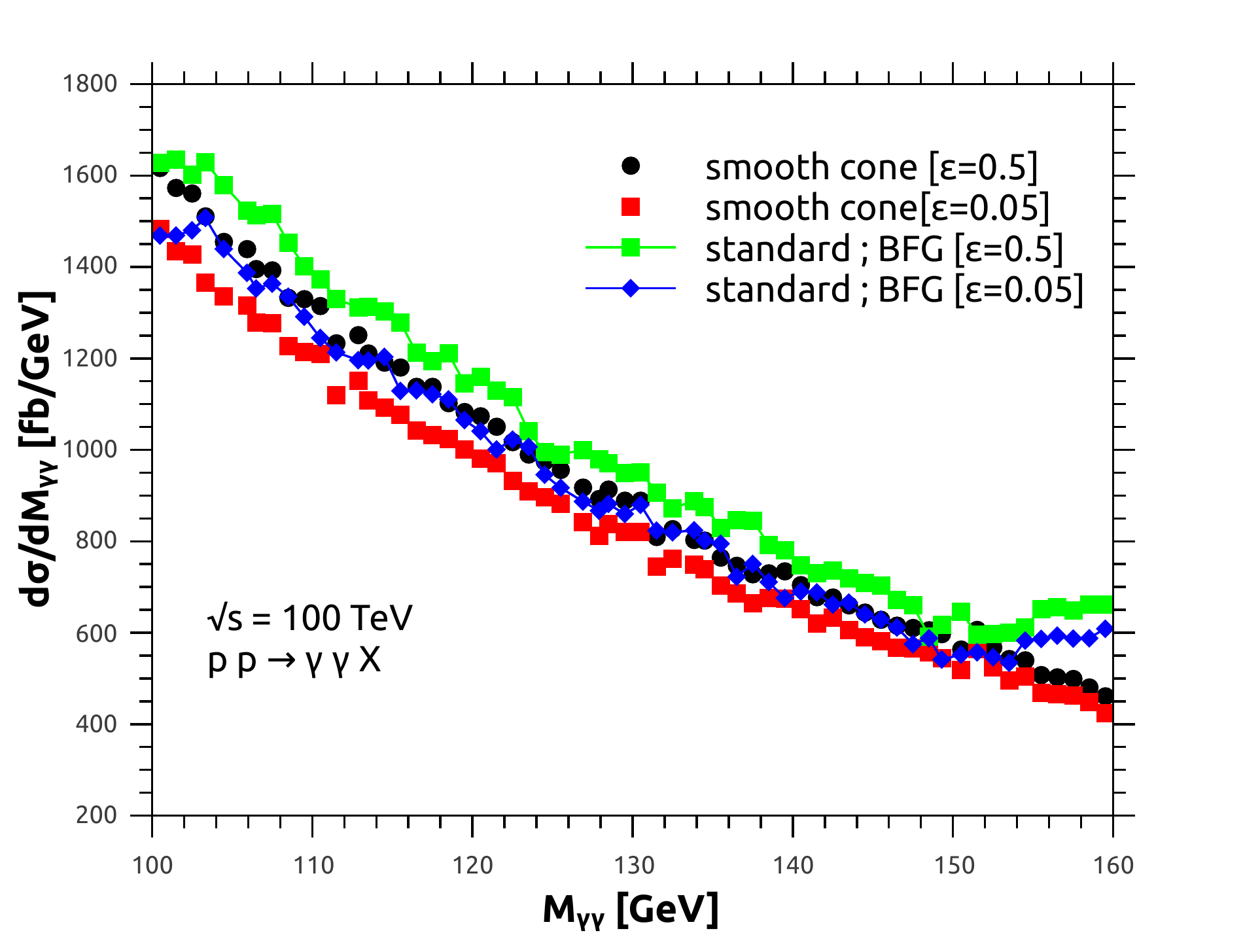}
\hfill
\includegraphics[width=0.47\textwidth]{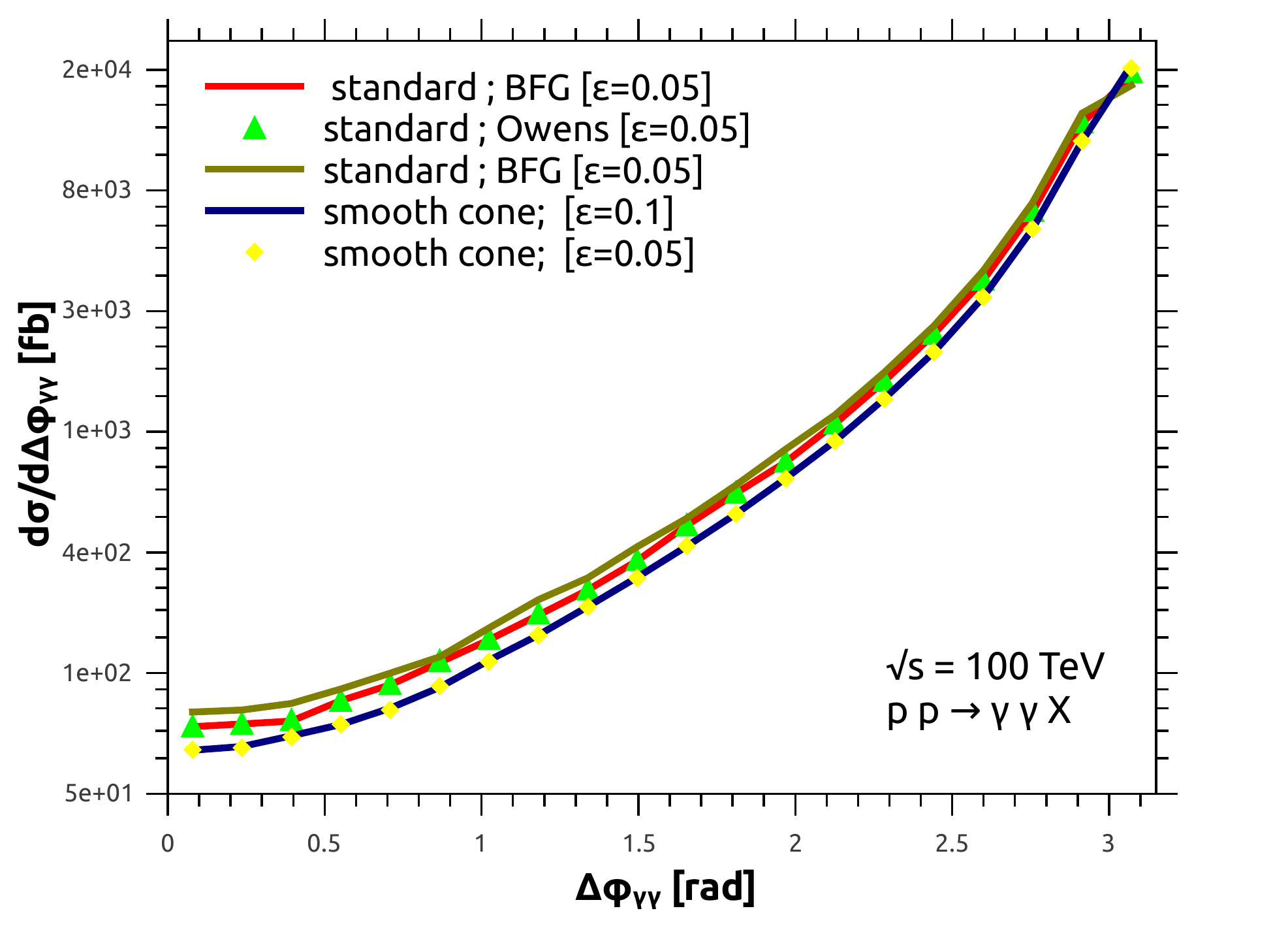}
\caption{Comparison of different isolation criteria at $\sqrt{s}=100$~TeV. In the left panel we show the invariant mass distribution and in the right panel the $\Delta \Phi_{\gamma\gamma}$ \label{fig:100TeVDphiMgg}.}
\end{figure}

In Fig~\ref{fig:100TeVDphiMgg} we present our results for $\sqrt{s}=100$~TeV. In the left panel we show the invariant mass distribution in which we compare the results using the smooth and standard cone isolation criteria. The cross section obtained using the smooth isolation criterion is about $7.5\%$ smaller than the standard cone result using BFG fragmentation functions for $\epsilon=0.05$. For $\epsilon=0.5$, the smooth cone result is about $8.4\%$ smaller than the standard cone result with BFG fragmentation. The cross section obtained using the Owens set of fragmentation functions is about $18\%$ larger than the BFG result with $\epsilon=0.5$, which is almost twice the same ratio at $8$~TeV. Regarding the smooth cone result with $\epsilon=0.5$, the cross section obtained with the Owens set is about $26.4\%$ larger.

In the right panel of Fig~\ref{fig:100TeVDphiMgg} we show the $\Delta \Phi_{\gamma\gamma}$ distribution for $\sqrt{s}=100$~TeV. The cross sections obtained with the standard cone isolation criterion with BFG and Owens fragmentation functions coincide (within $1\%$) if $\epsilon=0.05$. The standard result (with BFG) obtained using $\epsilon=0.1$ is about $23\%$ larger than the cross section with the smooth cone isolation (if $\Delta \Phi_{\gamma\gamma}\neq \pi$). 

Similar to the centre of mass energies already discussed, far away from $\Delta \Phi_{\gamma\gamma}\simeq \pi$ there is no difference between the smooth cone result with $\epsilon=0.05$ and $\epsilon=0.5$.

At $\sqrt{s}=100$~TeV  the fragmentation functions, extracted at lower energies (from LEP data), reach the limit of their validity range. Notice that the invariant mass distributions (left panel of Fig~\ref{fig:100TeVDphiMgg}), obtained with the standard cone isolation criterion, show a rising slope in the last bins, which most likely comes from the fact that the range of validity of the fragmentation functions (BFG quote a range of 10\,GeV $\leq Q\leq$ 100\,TeV) is surpassed.

At $\sqrt{s}=8$~TeV and 14~TeV it is possible to define a tight isolation prescription as done in the 2013 Les Houches proceedings~\cite{Andersen:2014efa}.

If we require $\epsilon\leq0.05$ the two isolation criteria give very similar cross sections, with agreement at the percent level, also at the level of  distributions. For $0.05 \leq \epsilon \leq 0.1$ we have agreement at the percent level, for all the kinematical regions in which a LO Born cross-section is present. For kinematical regions far way from the back-to-back configuration, in which the direct LO contribution vanishes (i.e $\Delta \Phi_{\gamma\gamma}\neq \pi$ in the $\Delta \Phi_{\gamma\gamma}$ distribution), the discrepancies between cross-sections obtained using the standard and the smooth cone isolation criteria are at the $4\%$ percent level. Given the size of the scale uncertainties at these energies and at this perturbative order ($\pm 6\%$ percent at the level of total cross-sections), a tight isolation prescription ($\epsilon < 0.1$) allows the comparison of cross sections obtained using these two different types of isolation criteria.

\subsubsection{Photon + Jet at NLO}

In this section we use the program {\tt
  JetPhox}~\cite{Aurenche:2006vj,Belghobsi:2009hx} to calculate cross
sections for the production of a photon and a jet. 
As {\tt JetPhox} allows to calculate the fragmentation component at
NLO, it can be used to assess the importance of this contribution 
for various isolation parameters and centre of mass energies.

Unless stated otherwise, we use the Bourhis-Fontannaz-Guillet (BFG)~\cite{Bourhis:1997yu}
fragmentation functions (set II).
For comparison, we also give results obtained with the (LO) fragmentation functions by Owens~\cite{Owens:1986mp},
which we use for consistency only in combination with the leading order matrix elements.

We use the following  settings and cuts: the NNPDF23\_nlo\_as\_0119~\cite{Ball:2012cx} 
PDF set with $\alpha_s$
taken from the PDFs.
For the renormalisation, factorisation and fragmentation scales we use
 $\mu=\mu_f=\mu_F=p_T^\gamma$.  The  photon is required to be in the rapidity range 
 $ |y^\gamma|\leq 2.37$ and to have a minimum transverse momentum of $p_{T,min}^\gamma=30$\,GeV.
For the jet we use the $k_T$-algorithm with $R=0.4$ and the rapidity and transverse momentum cuts 
$|y^{jet}|\leq 4.7$, $p_{T,min}^{jet}=25$\,GeV.

We will compare three different isolation prescriptions:
\begin{itemize}
\item[(a)] smooth isolation (``Frixione isolation''~\cite{Frixione:1998jh}) 
as defined in eq.~(\ref{eq:frixisol}), with $R=0.4$, $n=1$ and various values of $\eps_f$,
\item[(b)] standard cone isolation as defined in eq.~(\ref{eq:coneisol})
with $R=0.4$ and $E_{T}^{max}= \eps_c\, p_{T}^{\gamma}$,
\item[(c)] ``hybrid'' cone isolation with $E_{T}^{max}$ composed of both, 
a fixed amount of energy and a fraction of the photon transverse momentum, as used e.g. 
in Ref.~\cite{ATLAS-CONF-2015-081}: 
$E_{T}^{max}= \eps\, p_{T}^{\gamma}+ $6\,GeV with $\eps=0.05$.
\end{itemize}
In Table~\ref{tab:frix} we show results for the total NLO cross sections using criterion (a) 
at $\sqrt{s}=8, 14$ and 100 TeV.
Table \ref{tab:cone} shows results for the standard cone isolation criterion (b), 
Table \ref{tab:hybrid} for the hybrid cone isolation criterion (c). 
The results for the direct and fragmentation parts are shown separately. 
In particular, one can see that the NLO corrections to the fragmentation part are substantial, 
in particular for small values of $\epsilon_c$, with K-factors (for the fragmentation component) 
of about 2.2 (8 TeV) to 2.4 (100 TeV) 
for standard cone isolation, and about 1.7 for hybrid cone isolation. With the hybrid isolation criterion, 
this K-factor does not increase as the centre of mass energy increases. However, as the fragmentation 
component is about one order of magnoitude smaller than the direct component, 
the K-factor for the fragmentation part does not play a major role.

\begin{figure}[t!]
\includegraphics[width=0.47\textwidth]{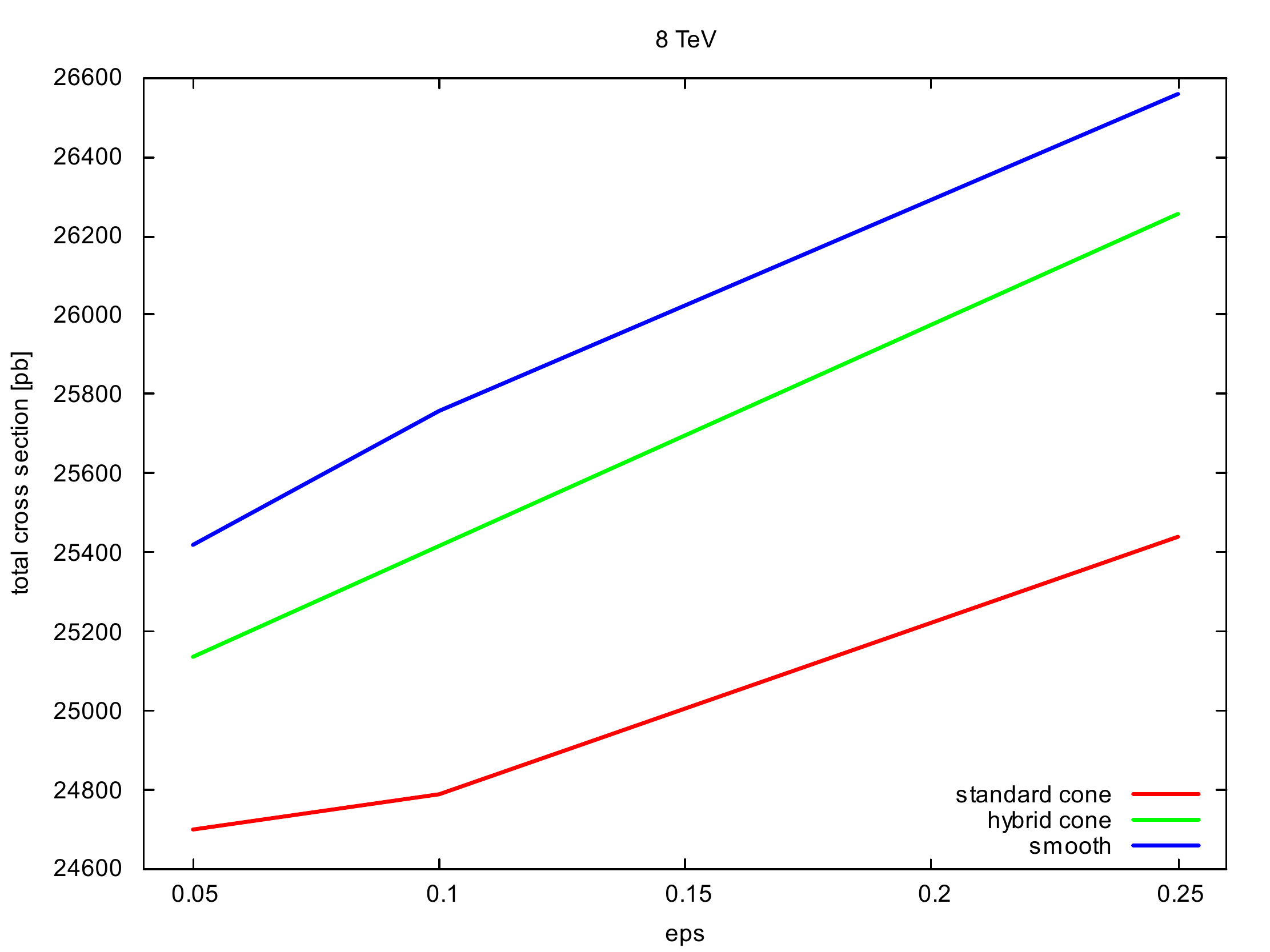}
\hfill
\includegraphics[width=0.47\textwidth]{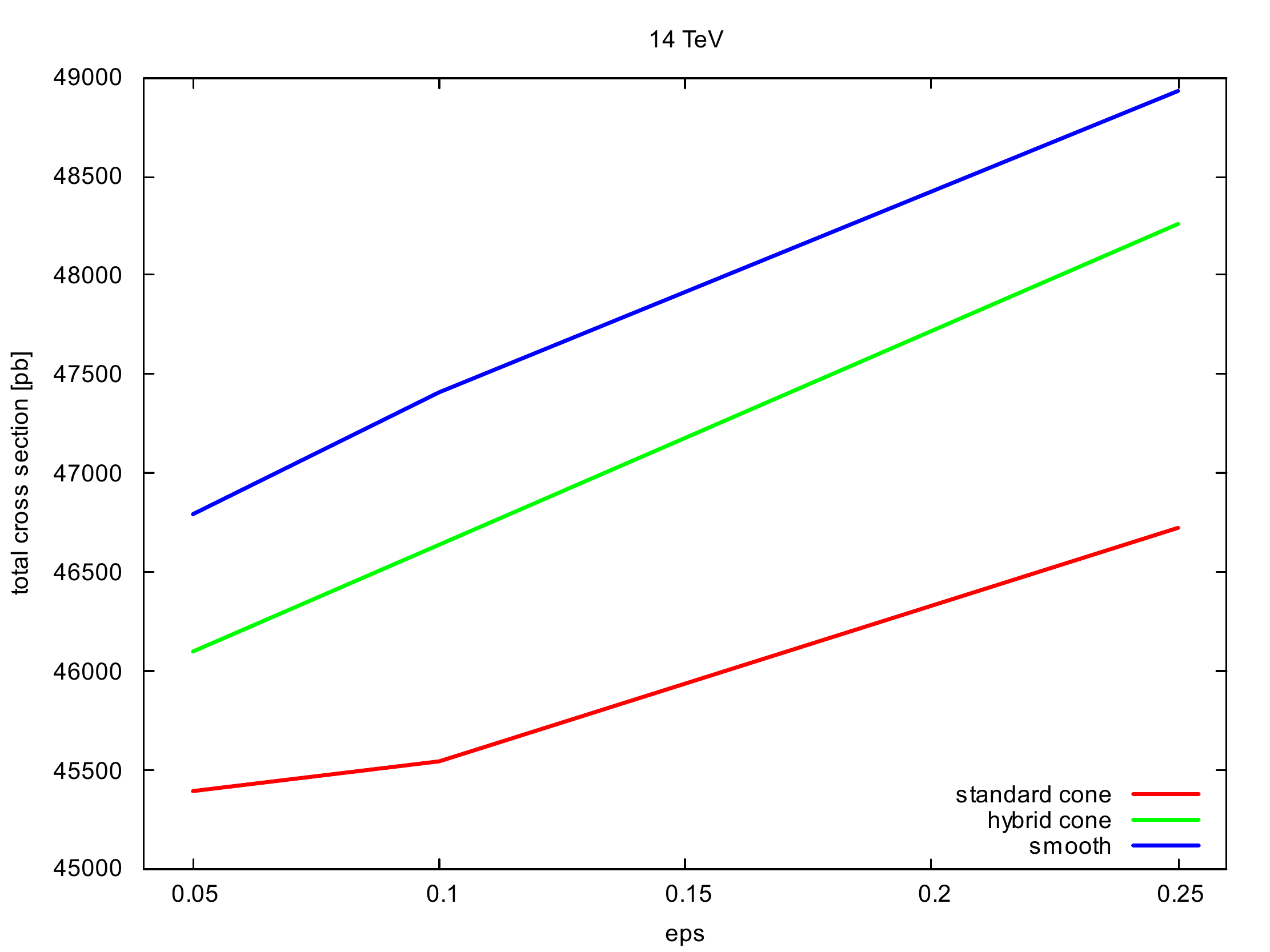}
\caption{Comparison of different isolation criteria at $\sqrt{s}=8$ and 14\,TeV.\label{fig:comp_conefrix}}
\end{figure}

The results for the total cross sections at $\sqrt{s}=8\,$TeV obtained with the 
three criteria are compared to each other in Fig.~\ref{fig:comp_conefrix}. 
As to be expected, the hybrid isolation leads to larger values of the cross section 
than the standard cone isolation for the same value of $\epsilon_c$, because it allows 
a larger fragmentation component.
It is interesting to observe that, in contrast to the diphoton case, the cross sections 
obtained with the smooth isolation criterion are always larger than the ones obtained with 
standard or hybrid cone isolation. This means that the subtraction terms for collinear configurations 
in the direct photon component  are more dominant in the photon plus jet case than in the diphoton case.

\begin{figure}[t!]
\includegraphics[width=0.47\textwidth]{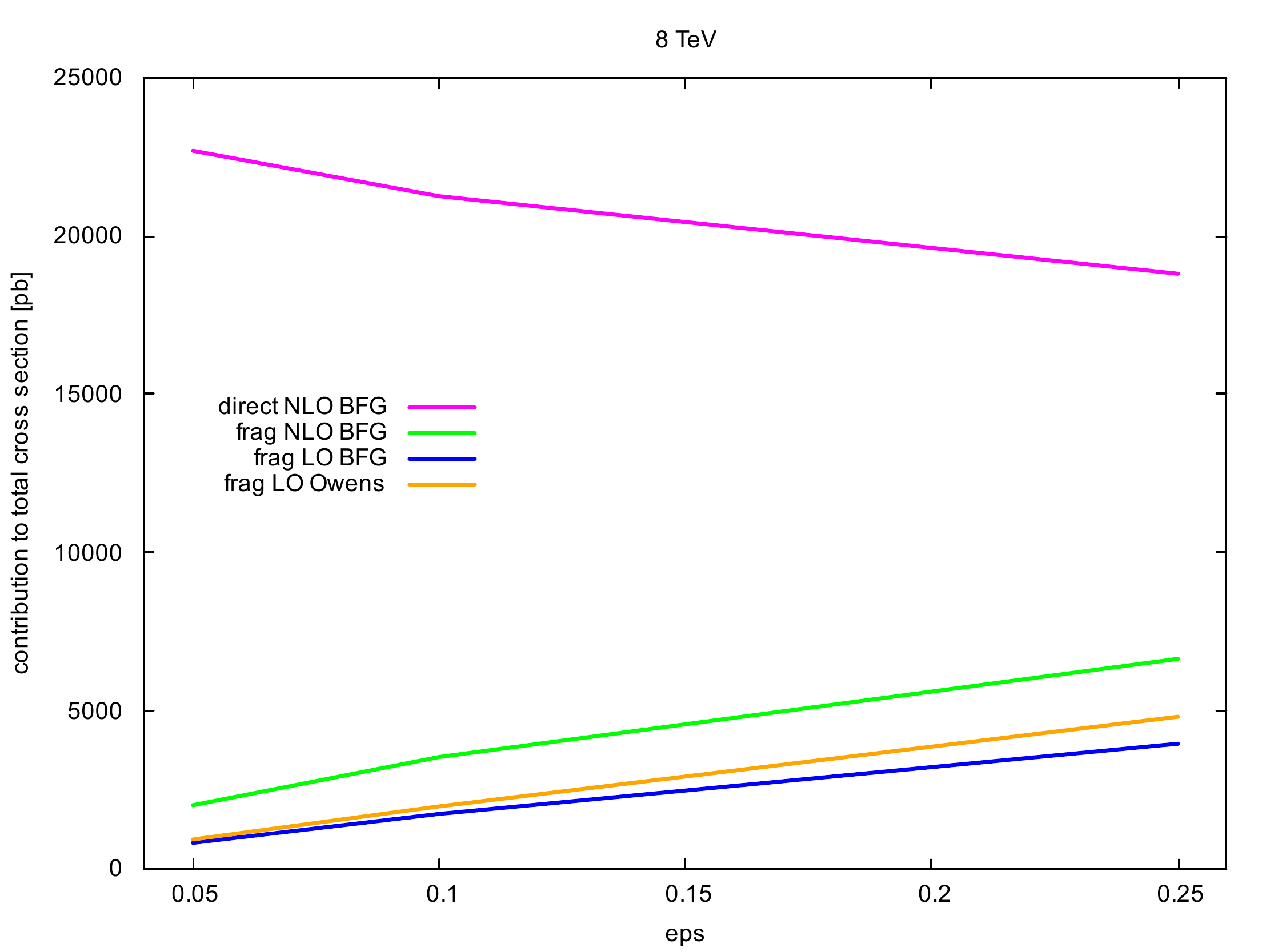}
\hfill
\includegraphics[width=0.47\textwidth]{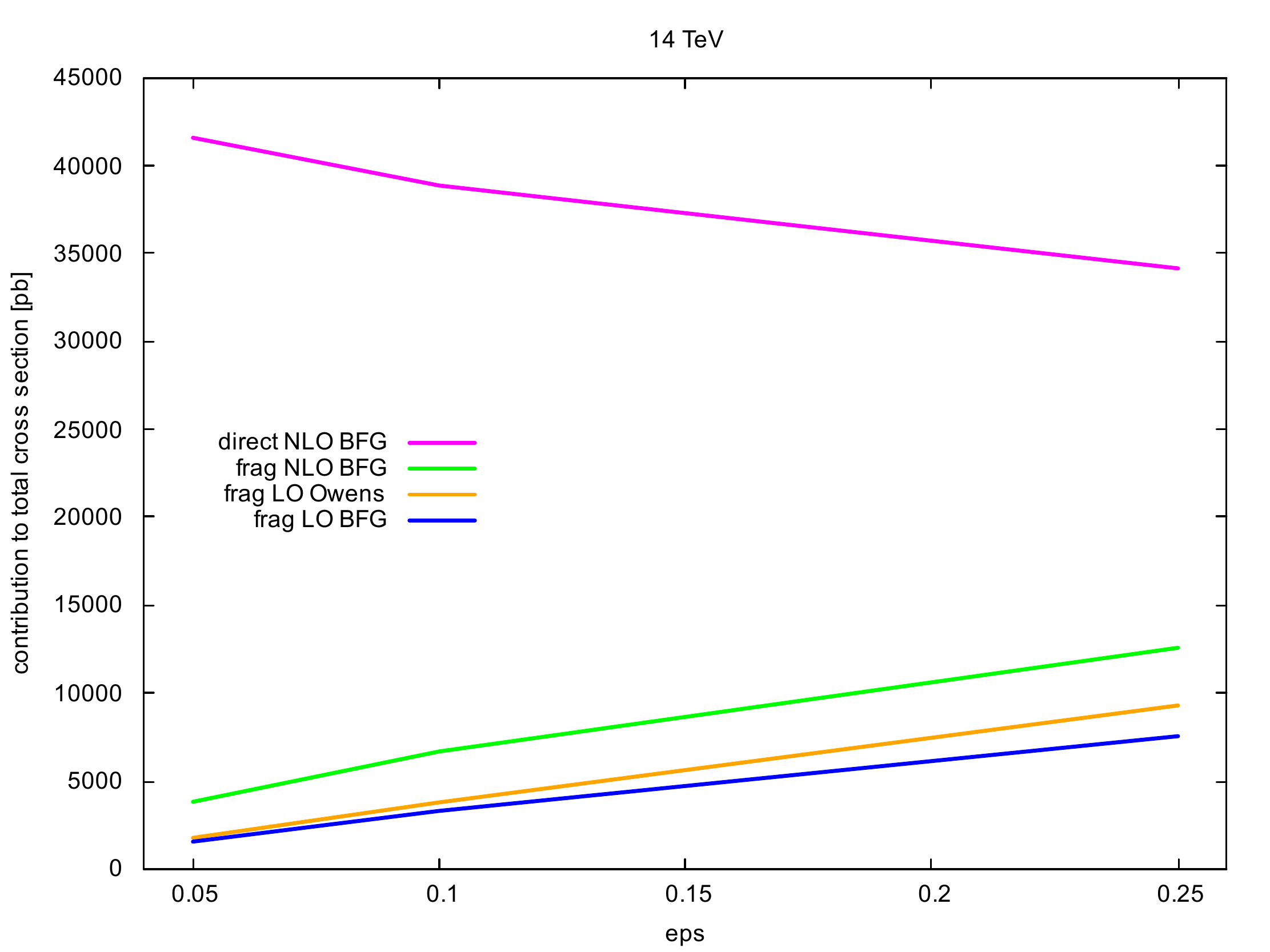}
\caption{Comparison of different fragmentation components at $\sqrt{s}=8$ and 14\,TeV.\label{fig:compfrag}}
\end{figure}

\begin{figure}[t!]
\centering
\includegraphics[width=0.95\textwidth]{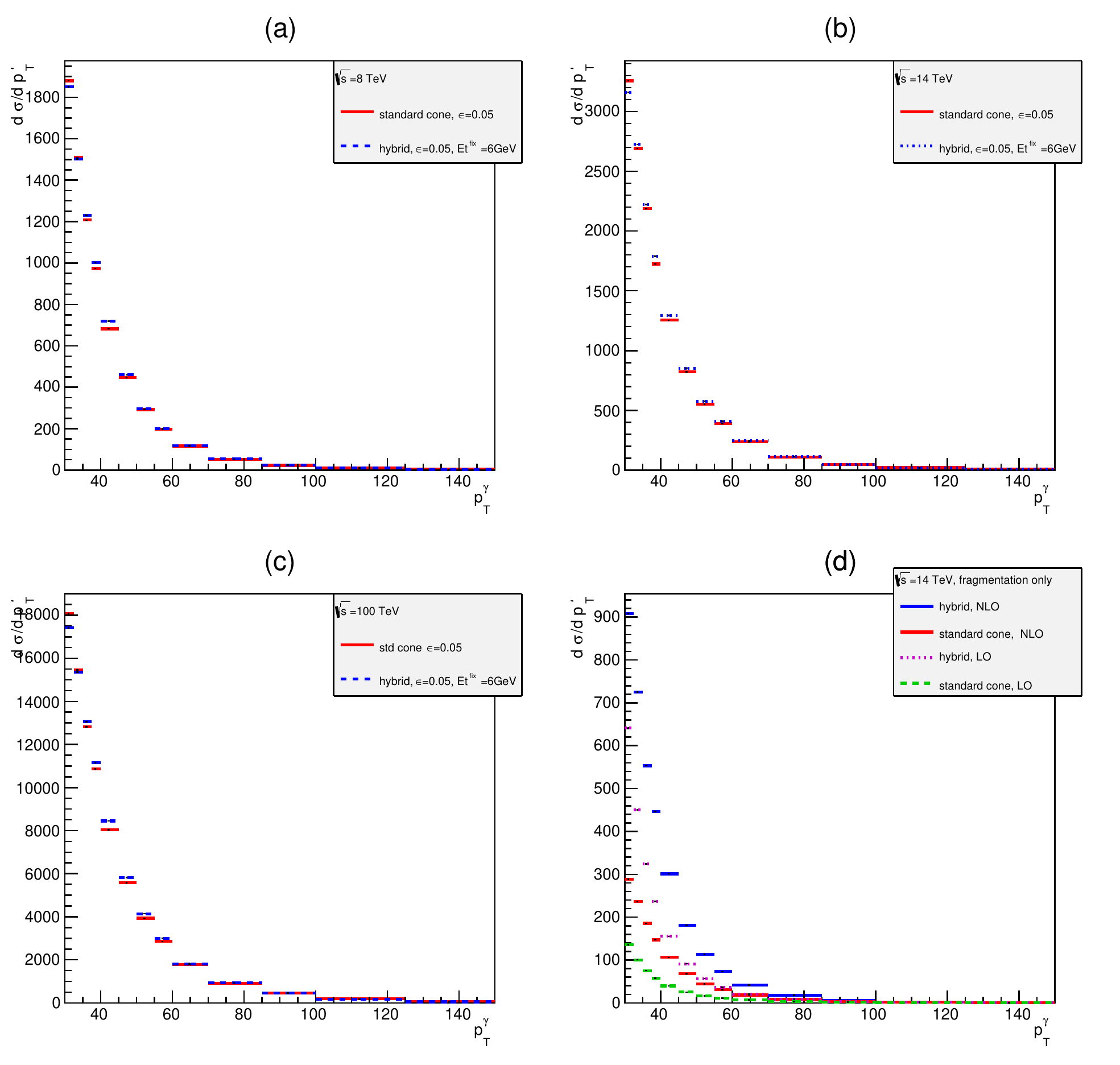}
\caption{Photon transverse momentum distribution for standard and hybrid cone isolation at (a) $\sqrt{s}=8$~TeV, 
(b) $\sqrt{s}=14$~TeV and (c) $\sqrt{s}=100$~TeV. 
Panel (d) compares different fragmentation contributions at $\sqrt{s}=14$~TeV, where the 
isolation parameters are the same as in the other panels.\label{fig:pTisohybrid}}
\end{figure}

\begin{figure}[t!]
\centering
\includegraphics[width=0.95\textwidth]{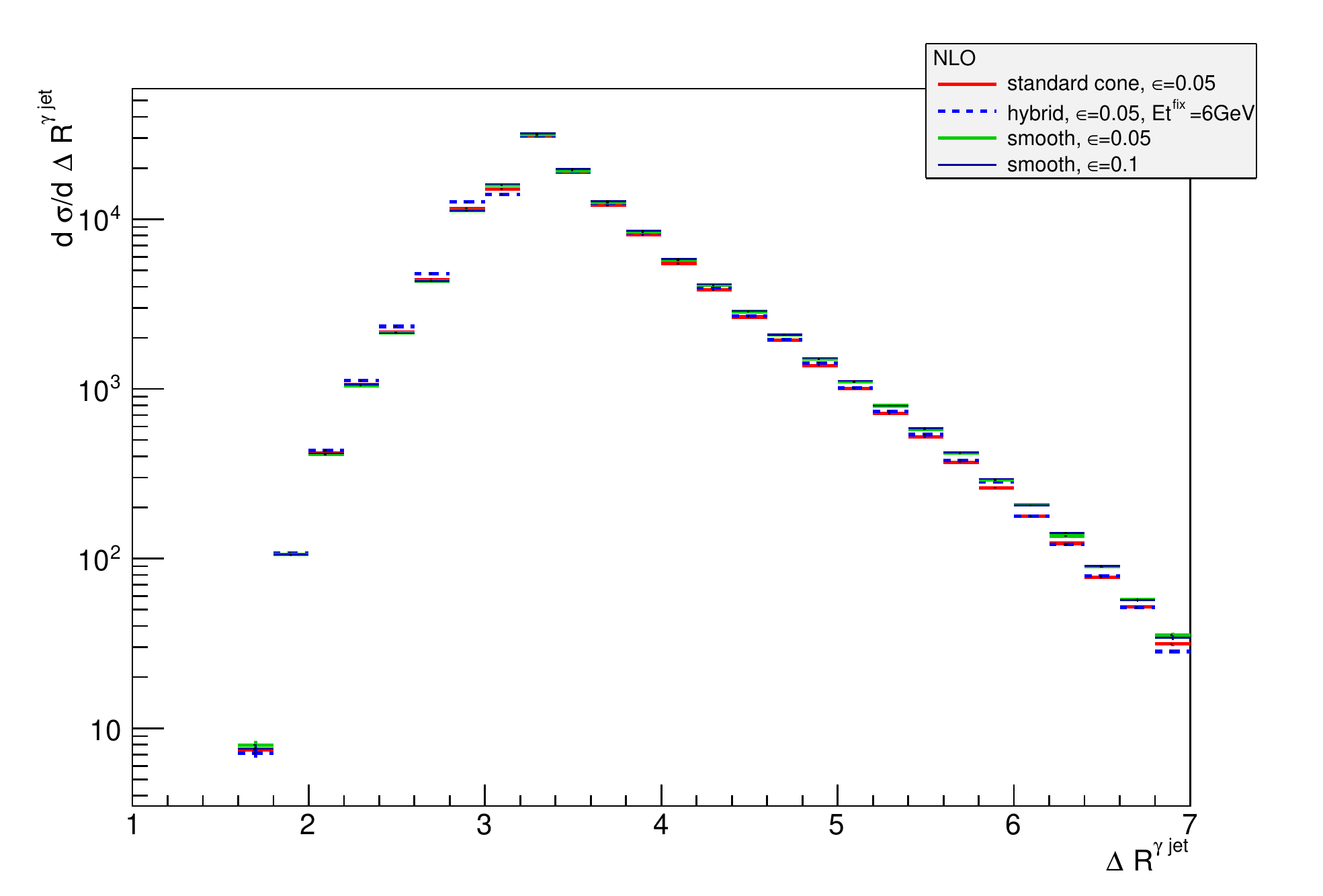}
\caption{Photon-jet $R$-separation  for standard and hybrid cone isolation at $\sqrt{s}=8$ TeV. \label{fig:deltaRisohybrid}}
\end{figure}

\begin{figure}[t!]
\centering
\includegraphics[width=0.95\textwidth]{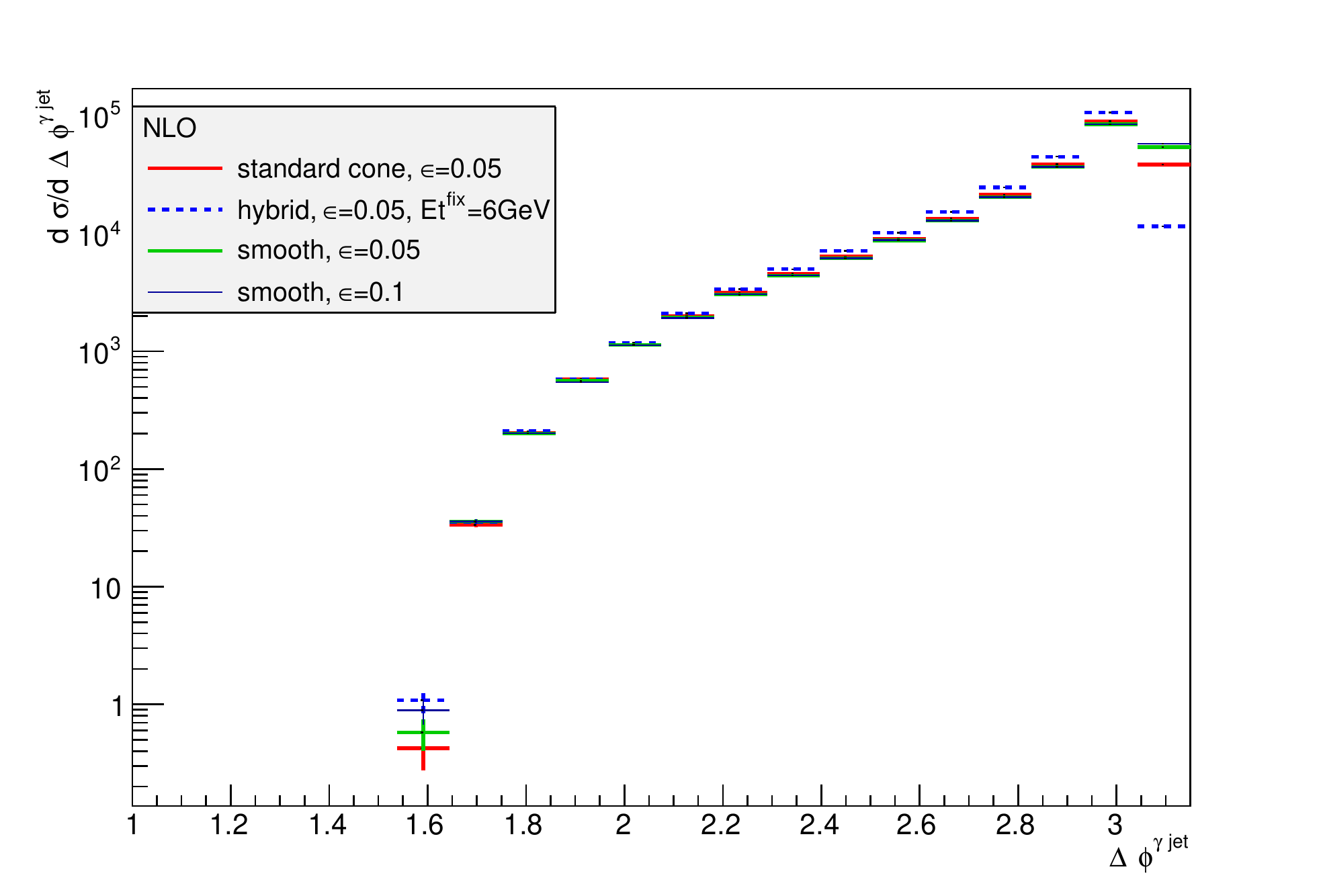}
\caption{Photon-jet azimuthal angle separation  for standard and hybrid cone isolation at $\sqrt{s}=8$ TeV. \label{fig:deltaPhiisohybrid}}
\end{figure}

Fig.~\ref{fig:compfrag} shows various options for the fragmentation contribution separately, 
for $\sqrt{s}=14\,$TeV.
In Figs.~\ref{fig:pTisohybrid}, \ref{fig:deltaRisohybrid}, \ref{fig:deltaPhiisohybrid} 
we compare the standard and the hybrid cone isolation for the distributions of 
$p_T^\gamma$, $\Delta R^{\gamma-jet}$ and $\Delta \phi^{\gamma-jet}$, respectively.

\begin{table}
\centering
\begin{tabular}{|c|l|r|r|r|}
\hline
$\sqrt{s}$ [TeV]&smooth isol.&$\eps_f=0.05$&$\eps_f=0.1$&$\eps_f=0.5$\\
\hline
8 &LO& 13.172&13.172&13.172\\
   &NLO& 25.418& 25.757& 27.501\\
\hline
14&LO&23.396&23.396&23.396\\
    &NLO& 46.793&  47.409&50.719 \\
\hline
100& LO& 13.607& 13.607& 13.607\\
&NLO& 300.70& 305.09&327.76\\
\hline
\end{tabular}
\caption{Total cross sections (in nanobarn) for the smooth isolation criterion~\cite{Frixione:1998jh} with
  different values of $\eps_f$.\label{tab:frix}}
\end{table}

\begin{table}
\centering
\begin{tabular}{|c|l|r|r|r|}
\hline
$\sqrt{s}$ [TeV]&standard cone isol.&$\eps_c=0.05$&$\eps_c=0.1$&$\eps_c=0.25$\\
\hline
8 &dir NLO& 22.682&21.249&18.803\\
   &frag LO (BFG)&0.824&1.738& 3.956\\
  & frag LO (Owens)& 0.930&1.980&4.808\\
  & frag NLO   &2.017&3.539&6.636\\
 & total NLO   &24.699&24.788& 25.439\\
\hline
14&dir NLO&41.577&38.861&34.145 \\
   &frag LO (BFG)&1.559&3.296&7.554 \\
   & frag LO (Owens)& 1.772& 3.788& 9.296\\
  & frag NLO  &3.815&6.682&12.577\\
 & total NLO&45.393&45.544&46.724\\
\hline
100&dir NLO&265.44&246.55&213.66 \\
   &frag LO  &10.593 &22.547&52.634\\
  & frag NLO&25.688&45.321&86.586\\
 & total NLO&291.13&291.87&300.25\\
\hline
\end{tabular}
\caption{Total cross sections (in nanobarn) for the cone isolation
  criterion with $R=0.4$ and
  different values of $\eps_c$.
  Note that ``frag LO'' refers to the matrix element being calculated at leading order, 
  while the BFG fragmentation functions are always NLO fits, the Owens fragmentation functions LO fits.\label{tab:cone}}
\end{table}

\begin{table}
\centering
\begin{tabular}{|c|l|r|r|r|}
\hline
$\sqrt{s}$ [TeV]&hybrid cone isol.&$\eps_c=0.05$&$\eps_c=0.1$&$\eps_c=0.25$\\
\hline
8 &dir NLO&19.392&18.769&17.504\\
   &frag LO&3.324&3.987&5.625\\
  & frag NLO&5.744&6.646&8.751\\
 & total NLO&25.136&25.416&26.255\\
\hline
14&dir NLO&35.302&34.111&31.631\\
   &frag LO&6.288&7.572&10.775\\
  & frag NLO&10.797&12.527&16.627\\
 & total NLO&46.099&46.638&48.259\\
\hline
100&dir NLO&223.201&214.327&195.901\\
   &frag LO&42.409&51.708&75.429\\
  & frag NLO&72.411&84.668&114.829\\
 & total NLO&295.61&298.99&310.73\\
\hline
\end{tabular}
\caption{Total cross sections (in nanobarn) for the cone isolation
  criterion which uses a ``hybrid'' isolation~\cite{ATLAS-CONF-2015-081} considering both fixed energy in the cone and a
  fraction of the photon transverse momentum, with $R=0.4$,
  $E_{T,fix}^{cone}=6$\,GeV  and
  different values of $\eps_c$.\label{tab:hybrid}}
\end{table}

\clearpage

\subsection{Summary}
We have studied various photon isolation criteria at $\sqrt{s}=8,14$ and 100 TeV for both diphoton production and photon plus jet production at NLO. 
In the diphoton case, we observe that, 
for tight isolation parameters ($\epsilon \leq 0.1$),  the smooth and standard cone isolation criteria give  results which agree  at the  percent level, 
where the standard cone isolation always gives a larger result. 
For looser isolation, the two criteria start to deviate, the differences being in the 10\% range for $\epsilon = 0.5$ for the $m_{\gamma\gamma}$ distribution, 
and even larger for the $\Delta \Phi_{\gamma\gamma}$ distribution.

At 100 TeV, the fragmentation functions may be called at  scales which are close to their limit of validity. 
Therefore a new fit of the photon fragmentation functions based on LHC data would be desirable.

The photon+jet case shows a different behaviour when comparing the smooth isolation with  cone isolation including the fragmentation part, 
in the sense that the smooth cone result is larger than the ``standard'' cone result, which is the opposite behaviour as in the diphoton case.
However, for tight isolation, the results with the two criteria also agree within a few percent.\\
In addition, a new isolation criterion, used recently by {\sc Atlas}~\cite{ATLAS-CONF-2015-081} has been implemented, 
where both a fixed energy in the cone and a fraction of the photon transverse momentum are used to determine the maximally allowed hadronic energy in the cone. 
This ``hybrid'' criterion leads to results which are very similar to the ``standard'' cone isolation where only a fraction of the photon momentum is used, 
except for low $p_T^\gamma$ values, where the spectrum with hybrid isolation is slightly shifted towards larger  $p_T^\gamma$ values.

\section{Systematics of quark/gluon tagging 
  \texorpdfstring{\protect\footnote{
   G.~Soyez, J.~Thaler,
   M.~Freytsis, P.~Gras, D.~Kar, 
   L.~L\"onnblad, S.~Pl\"atzer, A.~Si\'{o}dmok, P.~Skands, 
   D.~Soper
 }}{}}

By measuring the substructure of a jet, one can assign it a ``quark'' or ``gluon'' tag.  In the eikonal (double-logarithmic) limit, quark/gluon discrimination is determined solely by the color factor of the initiating parton ($C_F$ versus $C_A$).  In this section, we confront the challenges faced when going beyond this leading-order understanding, using parton shower generators to assess the impact of higher-order perturbative and nonperturbative physics.  Working in the idealized context of electron-positron collisions, where one can define a proxy for quark and gluon jets based on the Lorentz structure of the production vertex, we find a fascinating interplay between perturbative shower effects and nonperturbative hadronization effects.

\subsection{Overview}
\label{quarkgluon_sec:overview}

Jets are a robust tool for studying short-distance collisions involving quarks and gluons.  With a suitable jet definition, one can connect jet measurements made on clusters of hadrons to perturbative calculations made on clusters of partons.  More ambitiously, one can try to tag jets with a suitably-defined flavor label, thereby enhancing the fraction of, say, quark-tagged jets over gluon-tagged jets.  This is relevant for searches for physics beyond the standard model, where signals of interest are often dominated by quarks while the corresponding backgrounds are dominated by gluons.  A wide variety of quark/gluon discriminants have been proposed \cite{Gallicchio:2011xq,Gallicchio:2012ez,Krohn:2012fg,Pandolfi:1480598,Chatrchyan:2012sn,Larkoski:2013eya,Larkoski:2014pca,Bhattacherjee:2015psa}, and there is a growing catalog of quark/gluon studies at the Large Hadron Collider (LHC) \cite{Aad:2014gea,Aad:2014bia,Khachatryan:2014dea,Aad:2015owa,Khachatryan:2015bnx,Aad:2016oit}.\footnote{For an incomplete selection of earlier proposed discriminants, see Refs.~\cite{Nilles:1980ys,Jones:1988ay,Fodor:1989ir,Jones:1990rz,Lonnblad:1990qp,Pumplin:1991kc}.}

In order to achieve robust quark/gluon tagging, though, one needs theoretical and experimental control over quark/gluon radiation patterns.  At the level of eikonal partons, a quark radiates proportional to its $C_F = 4/3$ color factor while a gluon radiates proportional to $C_A = 3$.  In this section, we demonstrate that quark/gluon discrimination performance is highly sensitive to subleading perturbative effects beyond the eikonal limit, such as $g \to q \overline{q}$ splittings and color coherence, as well as to nonperturbative effects such as color reconnection and hadronization.   While these effects are modeled (to differing degrees) in parton shower generators, they are relatively unconstrained by existing collider measurements, especially in the gluon channel.  The goal of this section is to highlight these uncertainties, which then suggests a set of future LHC measurements that will improve the modeling of jets in general and quark/gluon tagging in particular.

A common misconception about quark/gluon tagging is that it is an intrinsically ill-defined problem.  Of course, quark and gluon partons carry color while jets are composed of color-singlet hadrons, so the labels ``quark'' and ``gluon'' are fundamentally ambiguous. 
But this is philosophically no different from the fact that a ``jet'' is fundamentally ambiguous and one must therefore always specify a concrete jet finding procedure.  As discussed in Sec.~\ref{quarkgluon_sec:def}, one can indeed create a well-defined quark/gluon tagging procedure based unambiguous hadron-level measurements.  In this way, even if what one means by ``quark'' or ``gluon'' is based on a naive or ambiguous concept (like Born-level cross sections or eikonal limits), quark/gluon discrimination is still a well-defined technique for enhancing desired signals over unwanted backgrounds.
  
There are a wide range of possible quark/gluon discriminants and a similarly large range of ways to quantify discrimination power.  As a concrete set of discriminants, we consider the generalized angularities $\lambda_\beta^\kappa$ \cite{Larkoski:2014pca} (see also \cite{Berger:2003iw,Almeida:2008yp,Ellis:2010rwa,Larkoski:2014uqa}),
\begin{equation}
\label{quarkgluon_eq:genang_intro}
\lambda^{\kappa}_{\beta} = \sum_{i \in \text{jet}} z_i^\kappa \theta_i^\beta,
\end{equation}
with the notation to be explained in Sec.~\ref{quarkgluon_sec:genang}.  We consider five different $(\kappa, \beta)$ working points, which roughly map onto five variables in common use in the literature:
\begin{equation}
\begin{array}{ccccc}
(0,0) & (2,0) & (1,0.5) & (1,1) & (1,2) \\
\text{multiplicity} &  p_T^D &  \text{LHA} & \text{width} & \text{mass}
\end{array}
\end{equation}
Here, multiplicity is the hadron multiplicity within the
jet, $p_T^D$ was defined in
Refs.~\cite{Pandolfi:1480598,Chatrchyan:2012sn}, LHA refers to the
``Les Houches Angularity'' (named after the venue of this workshop),
width is closely related to jet broadening
\cite{Catani:1992jc,Rakow:1981qn,Ellis:1986ig}, and mass is closely
related to jet thrust \cite{Farhi:1977sg}.  To quantify discrimination
performance, we focus on classifier separation (a default output of
TMVA \cite{2007physics...3039H}):
\begin{equation}
\label{quarkgluon_eq:deltadef_intro}
\Delta =  \frac{1}{2} \int \text{d} \lambda \, \frac{\bigl(p_q(\lambda) - p_g(\lambda)\bigr)^2}{p_q(\lambda) + p_g(\lambda)},
\end{equation}
where $p_q$ ($p_g$) is the probability distribution for $\lambda$ in a generated quark jet (gluon jet) sample.  This and other potential performance metrics are discussed in
Sec.~\ref{quarkgluon_sec:classsep}.

We begin our parton shower generator predictions for quark/gluon discrimination in Sec.~\ref{quarkgluon_sec:ee}, using an idealized setup with $e^+e^-$ collisions.  Here, we can use the following processes as proxies for quark and gluon jets:
\begin{align}
\text{``quark jets''}: \quad & e^+e^- \to (\gamma/Z)^* \to u \overline{u}, \\
\text{``gluon jets''}: \quad & e^+e^- \to h^* \to g g,
\end{align}
where $h$ is the Higgs boson.  These processes are physically
distinguishable by the quantum numbers of the associated color singlet
production operator, giving a way to define truth-level quarks and
gluons labels without reference to the final state.\footnote{Of course, the quantum numbers of the color singlet operator are not measurable event by event.  The idea here is to have a fundamental definition of ``quark'' and ``gluon'' that does not reference QCD partons directly.}  We
compare six different parton shower generators both before
hadronization (``parton level'') and after hadronization (``hadron
level''):
\begin{itemize}
\item \textsc{Pythia 8.205} \cite{Sjostrand:2006za,Sjostrand:2014zea},
\item \textsc{Herwig++ 2.7.1} \cite{Bahr:2008pv,Bellm:2013hwb},
\item \textsc{Sherpa 2.1.1} \cite{Gleisberg:2008ta},
\item \textsc{Vincia 1.201} \cite{Giele:2013ema},
\item \textsc{Deductor 1.0.2} \cite{Nagy:2014mqa} (with hadronization performed by \textsc{Pythia 8.212}),\footnote{Note that this
\textsc{Deductor} plus \textsc{Pythia} combination has not yet been tuned to data.}
\item \textsc{Ariadne 5.0.$\beta$} \cite{Flensburg:2011kk}.\footnote{This version of \textsc{Ariadne} is not yet public, but available from the author on request.  For $e^+ e^-$ collisions, the physics is the same as in \textsc{Ariadne 4} \cite{Lonnblad:1992tz}.}
\end{itemize}
In the future, we plan to make \textsc{Herwig 7} \cite{Bellm:2015jjp} and \textsc{Dire} \cite{Hoche:2015sya} predictions about quark/gluon discrimination as well as investigate predictions from analytic resummation \cite{Larkoski:2013eya,Larkoski:2014pca}.

As we will see, the differences between these generators arise from
physics at the interface between perturbative showering and
nonperturbative fragmentation.  One might think that the largest
differences between generators would appear for infrared-and-collinear
(IRC) unsafe observables like multiplicity and $p_T^D$, where
nonperturbative hadronization plays an important role.  Surprisingly, comparably-sized differences are also
seen for the IRC safe angularities, indicating that these generators
have different behavior even at the level of the perturbative final
state shower.  In Sec.~\ref{quarkgluon_sec:ee_scales}, we study these
differences as a function of collision energy $Q$, jet radius $R$,
and strong coupling constant $\alpha_s$, showing that the generators
have somewhat different discrimination trends.  In
Sec.~\ref{quarkgluon_sec:ee_settings}, we compare the default parton
shower configurations to physically-motivated changes, showing that
modest changes to the shower/hadronization parameters can give rather
large differences in quark/gluon separation power.

At the end of the day, most of the disagreement between generators is due to gluon radiation patterns.  This is not so surprising, since most of these generators have been tuned to reproduce distributions from $e^+ e^-$ colliders, and quark (but less so gluon) radiation patterns are highly constrained by event shape measurements at LEP \cite{Heister:2003aj,Abdallah:2003xz,Achard:2004sv,Abbiendi:2004qz}.  In Sec.~\ref{quarkgluon_sec:pp}, we suggest a possible analysis strategy at the LHC to specifically constrain gluon radiation patterns.  At a hadron collider, the distinction between quark jets and gluon jets is rather subtle, since radiation patterns depend on color connections between the measured final state jets and the unmeasured initial state partons.  That said, we suspect that much can be learned from hadron-level measurements, even without isolating ``pure'' quark or gluon samples.

We present our final recommendations and conclusions in
Sec.~\ref{quarkgluon_sec:conclude}.  The main take home message from
this study is that, contrary to the standard lore, existing
measurements at $e^+e^-$ colliders are insufficient to constrain
uncertainties in the final state shower.  Therefore, gluon-enriched
measurements at the LHC will be crucial to achieve robust quark/gluon
discrimination.

\subsection{What is a quark/gluon jet?}
\label{quarkgluon_sec:def}

\begin{figure}
\centering
\includegraphics[width=0.7 \columnwidth]{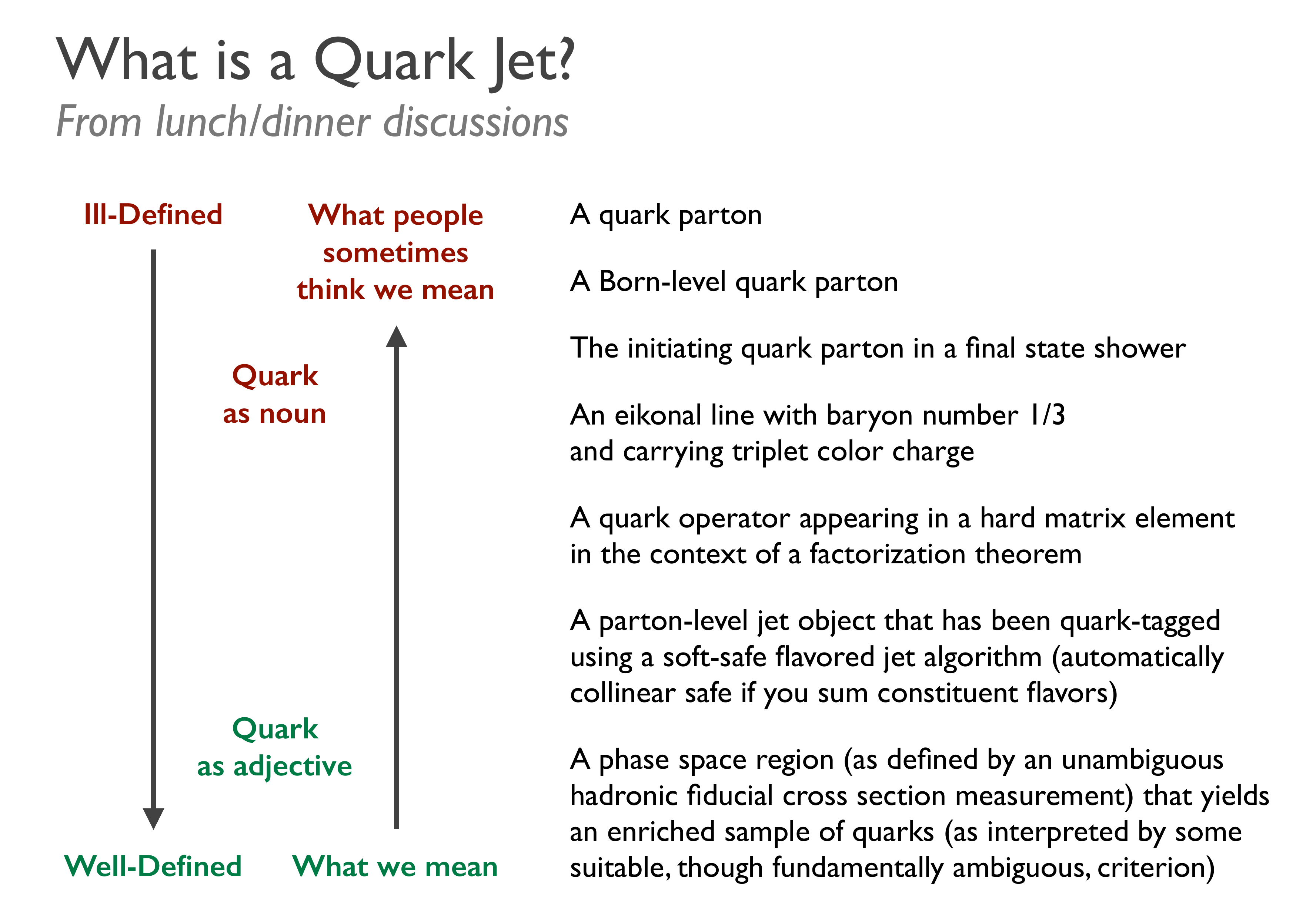}
\caption{Original slide from the June 10, 2015 summary report of the quark/gluon Les Houches subgroup.}
\label{quarkgluon_fig:summary_slide}
\end{figure}

As part of the 2015 Les Houches workshop, an attempt was made to
define exactly what is meant by a ``quark jet'' or ``gluon jet'' (see Fig.~\ref{quarkgluon_fig:summary_slide}).
Here are some suggested options for defining a quark jet, in
(approximate) order from most ill-defined to most well-defined.
Related statement can be made for gluon jets.

\noindent \textbf{A quark jet is...}
\begin{itemize}
\item \textbf{A quark parton.}  This definition (incorrectly) assumes that there is a one-to-one map between a jet and its initiating parton.  Because it neglects the important role of additional radiation in determining the structure of a jet, we immediately dismiss this definition.
\item \textbf{A Born-level quark parton.}  This definition at least acknowledges the importance of radiative corrections to jet production, but it leaves open the question of how exactly to define the underlying Born-level process from an observed final state.  (For one answer valid at the parton level, see flavored jet algorithms below.)
\item \textbf{An initiating quark parton in a final state parton
    shower.}  We suspect that this is the definition most LHC
  experimentalists have in mind.  This definition assumes that the parton shower history is meaningful, though,
  which may not be the case beyond the strongly-ordered or
  leading-logarithmic approximations.  Because the parton shower is
  semi-classical, this definition neglects the impact of genuinely
  quantum radiative corrections as well as nonperturbative
  hadronization.
\item \textbf{A maximum-$p_T$ quark parton within a jet in a final state parton shower.}  This ``max-$p_T$'' prescription is a variant on the initiating parton prescription above (see further discussion in \cite{Buckley:2015gua}).  It differs from the initiating parton by a calculable amount in a leading logarithmic shower \cite{Dasgupta:2014yra} and is based on the same (naive) assumption that the parton shower history is meaningful. 
\item \textbf{An eikonal line with baryon number 1/3 and carrying triplet color charge.}  This is another semi-classical definition that attempts to use a well-defined limit of QCD to define quarks in terms of light-like Wilson lines.  Philosophically, this is similar to the parton shower picture, with a similar concern about how to extrapolate this definition away from the strict eikonal limit.
\item \textbf{A parton-level jet object that has been quark-tagged using an IRC safe flavored jet algorithm.}  This is the strategy adopted in \cite{Banfi:2006hf}.  While this definition neglects the impact of hadronization, it does allow for the calculation of quark jet cross sections at all perturbative orders, including quantum corrections.
\end{itemize}
The unifying theme in the above definitions is that they try to identify a quark as an object unto itself, without reference to the specific final state of interest.  However, it is well-known that a ``quark'' in one process may not look like a ``quark'' in other process, due to color correlations with the rest of the event, especially the initial state in $pp$ collisions.  The next definition attempts to deal with the process dependence in defining quarks. 
\begin{itemize}
\item \textbf{A quark operator appearing in a hard matrix element in the context of a factorization theorem.}  This is similar to the attitude taken in \cite{Gallicchio:2011xc}.  In the context of a well-defined cross section measurement, one can (sometimes) go to a limit of phase space where the hard production of short-distance quarks and gluons factorizes from the subsequent long-distance fragmentation.  This yields a nice (gauge-covariant) operator definition of a quark jet.  That said, even if a factorization theorem does exist for the measurement of interest, this definition is potentially ambiguous beyond leading power.
\end{itemize}
The definition we adopt for this study is inspired by the idea that one should think about quark/gluon tagging in the context of a specific measurement, but it tries to avoid relying on the presence of a factorization theorem.
\begin{itemize}
\item \textbf{A phase space region (as defined by an unambiguous
    hadronic fiducial cross section measurement) that yields an
    enriched sample of quarks (as interpreted by some suitable, though
    fundamentally ambiguous, criterion).}  Here, the goal is to
  \emph{tag} a phase space region as being quark-like, rather than try
  to determine a truth definition of a quark.  This definition has the
  advantage of being explicitly tied to hadronic final states and to
  the discriminant variables of interest. \emph{The main
  challenge with this definition is how to determine the criterion
  that corresponds to successful quark enrichment.}  For that, we will
  have to rely to some degree on the other less well-defined notions
  of what a quark jet is.
\end{itemize}

To better understand this last definition, consider a quark/gluon discriminant $\lambda$.  Since $\lambda$ can be measured on any jet, one can unambiguously determine the cross section $\text{d} \sigma / \text{d} \lambda$ for any jet sample of interest.  But measuring $\lambda$ does not directly lead to the probability that the jet is a quark jet, nor to the probability distribution  $p_q(\lambda)$ for $\lambda$ within a quark jet sample.  Rather, the process of measuring $\lambda$ must be followed by a separate process of interpreting how the value of $\lambda$ should be used as part of an analysis.

For example, the user could choose that small $\lambda$ jets should be tagged as ``quark-like'' while large $\lambda$ jets should be tagged as ``gluon-like''. Alternatively, the user might combine $\lambda$ with other discriminant variables as part of a more sophisticated classification scheme.  The key point is that one first measures hadron-level discriminant variables on a final state of interest, and only later does one interpret exactly what those discriminants accomplish (which could be different depending on this physics goals of a specific analysis).  Typically, one might use a Born-level or eikonal analysis to define which regions of phase space should be associated with ``quarks'' or ``gluons'', but even if these phase space regions are based on naive or ambiguous logic, $\lambda$ itself is a well-defined discriminant variable.

In Sec.~\ref{quarkgluon_sec:ee}, we will consider the generalized
angularities $\lambda_{\beta}^\kappa$ as our discriminant variables
and we will assess the degree to which the measured values of
$\lambda_{\beta}^\kappa$ agree with a quark/gluon interpretation based
on Born-level production modes.  This is clearly an idealization,
though one that makes some sense in the context of $e^+e^-$
collisions, since the truth-level ``quark'' and ``gluon'' labels are
defined by the Lorentz structure of the production vertex.  In
Sec.~\ref{quarkgluon_sec:conclude}, we will recommend that the LHC
experiments perform measurements of $\lambda_\beta^\kappa$ in
well-defined hadron-level final states, without necessarily attempting
to determine separate $p_q(\lambda_\beta^\kappa)$ and
$p_g(\lambda_\beta^\kappa)$ distributions.  Eventually, one would want
to use these hadron-level measurements to infer something about
parton-level quark/gluon radiation patterns.  Even without that
interpretation step, though, direct measurements of $\text{d} \sigma /
\text{d} \lambda_\beta^\kappa$ would provide valuable information for
parton shower tuning.  This in turn would help $\lambda_\beta^\kappa$ become a more robust and powerful discriminant in searches for new physics beyond the standard model. 

\subsection{Generalized angularities}
\label{quarkgluon_sec:genang}

\begin{figure}
\centering
\includegraphics[scale = 0.7]{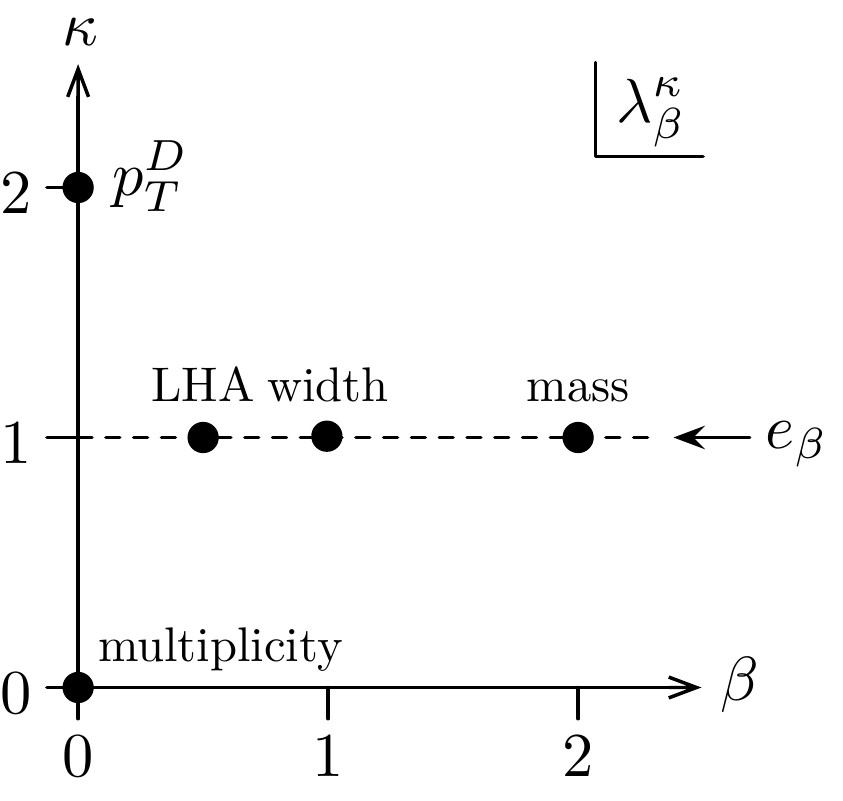}
\caption{Two-parameter family of generalized angularities, adapted from \cite{Larkoski:2014pca}.  The dots correspond to the five benchmark angularities used in this study.  The horizontal line at $\kappa = 1$ corresponds to the IRC safe angularities, $e_\beta = \lambda^{1}_{\beta}$.}
\label{quarkgluon_fig:lambda_space}
\end{figure}

A wide variety of quark/gluon discriminants have been proposed (see \cite{Gallicchio:2012ez} for an extensive catalog), but here we limit ourselves to a two-parameter family of generalized angularities \cite{Larkoski:2014pca}, shown in Fig.~\ref{quarkgluon_fig:lambda_space}.  These are defined as (repeating Eq.~\eqref{quarkgluon_eq:genang_intro} for convenience)
\begin{equation}
\label{quarkgluon_eq:genang}
\lambda^{\kappa}_{\beta} = \sum_{i \in \text{jet}} z_i^\kappa \theta_i^\beta,
\end{equation}
where $i$ runs over the jet constituents, $z_i \in [0,1]$ is a momentum fraction, and $\theta_i \in [0,1]$ is a (normalized) angle to the jet axis.  The parameters $\kappa \ge 0$ and $\beta \ge 0$ determine the momentum and angle weighting, respectively.  For $\kappa = 1$, the generalized angularities are IRC safe and hence calculable in perturbation theory \cite{Larkoski:2014uqa} (see also \cite{Ellis:2010rwa,Larkoski:2013paa,Larkoski:2014tva,Procura:2014cba,Hornig:2016ahz}).  For general $\kappa \not= 1$, there are quasi-perturbative techniques based on generalized fragmentation functions \cite{Larkoski:2014pca} (see also \cite{Krohn:2012fg,Waalewijn:2012sv,Chang:2013rca,Chang:2013iba}).  In our parton shower studies, we determine $\lambda^{\kappa}_{\beta}$ using all constituents of a jet, though one could also consider using charged-particle-only angularities to improve robustness to pileup (at the expense of losing some particle-level information).

For our $e^+ e^-$ study, we cluster jets with \textsc{FastJet 3.1.3} \cite{Cacciari:2011ma} using the $ee$-variant of the
anti-$k_t$ algorithm \cite{Cacciari:2008gp}, with $|\vec{p}|$-ordered
winner-take-all recombination
\cite{Larkoski:2014uqa,Bertolini:2013iqa,Salam:WTAUnpublished} to
determine the jet axis $\hat{n}$.  Unlike standard $E$-scheme
recombination \cite{Blazey:2000qt}, the winner-take-all scheme yields
a jet axis $\hat{n}$ that does not necessarily align with the jet
three-momentum $\vec{p}$; this turns out to be a desirable feature
for avoiding soft recoil effects
\cite{Larkoski:2013eya,Larkoski:2014uqa,Catani:1992jc,Dokshitzer:1998kz,Banfi:2004yd}.  We define
\begin{equation}
z_i \equiv \frac{E_i}{E_{\rm jet}}, \qquad \theta_i \equiv \frac{\Omega_{i \hat{n}}}{R},
\end{equation}
where $E_i$ is the particle energy, $\Omega_{i \hat{n}}$ is the opening angle to the jet axis, and $R$ is the jet radius (taken to be $R = 0.6$ by default).  To translate our $ee$ study to an eventual $pp$ study (left to future work), one would use the standard $pp$ version of anti-$k_t$ with $p_T$-ordered winner-take-all recombination, defining
\begin{equation}
z_i \equiv \frac{p_{Ti}}{\sum_{j \in \text{jet}} p_{Tj}}, \qquad \theta_i \equiv \frac{R_{i \hat{n}}}{R},
\end{equation}
where $p_{Ti}$ is the particle transverse momentum and $R_{i \hat{n}}$ is the rapidity-azimuth distance to the jet axis.

By adjusting $\kappa$ and $\beta$, one can probe different aspects of the jet fragmentation.  We consider five benchmark values for $(\kappa, \beta)$ indicated by the black dots in Fig.~\ref{quarkgluon_fig:lambda_space}:
\begin{equation}
\label{quarkgluon_eq:benchmarkang}
\begin{aligned}
(0,0) &= \text{hadron multiplicity},\\
(2,0) &\Rightarrow p_T^D \text{  \cite{Pandolfi:1480598,Chatrchyan:2012sn} (specifically $\lambda^{2}_{0} = (p_T^D)^2$)},\\
(1,0.5) & = \text{Les Houches Angularity (LHA)},\\
(1,1) &= \text{width or broadening \cite{Catani:1992jc,Rakow:1981qn,Ellis:1986ig}},\\
(1,2) & \Rightarrow \text{mass or thrust \cite{Farhi:1977sg}
  (specifically $\lambda^{1}_{2} \simeq m_{\rm jet}^2 / E_{\rm
    jet}^2$)}.
\end{aligned}
\end{equation}
Except for the LHA, these angularities (or their close cousins) have already been used in quark/gluon discrimination studies.  The LHA has been included to have an IRC safe angularity that weights energies more heavily than angles, similar in spirit to the $\beta = 0.2$ value advocated in Ref.~\cite{Larkoski:2013eya}.

For the IRC safe case of $\kappa = 1$, there is an alternative version
of the angularities based on energy correlation functions \cite{Larkoski:2013eya} (see also \cite{Banfi:2004yd,Jankowiak:2011qa}),
\begin{equation}
\text{ecf}_\beta = \sum_{i < j \in \text{jet}} z_i z_j \theta_{ij}^\beta \simeq \lambda^{1}_{\beta},
\end{equation}
where equality holds in the extreme eikonal limit.\footnote{This equality also relies on using a recoil-free axis choice $\hat{n}$ to define $\theta_i$.  Amusingly, $\lim_{\beta \to 0} \text{ecf}_\beta = (1 - \lambda^{2}_{0})/2$ (i.e.~$\kappa = 2$, $\beta = 0$), such that the $\beta \to 0$ limit of the IRC safe energy correlation functions corresponds to the IRC unsafe $p_T^D$.}  For the $e^+ e^-$ case, the pairwise angle $\theta_{ij}$ is typically normalized to the jet radius as $\theta_{ij} \equiv \Omega_{ij}/R$.   To avoid a proliferation of curves, we will not show any results for $\text{ecf}_\beta$.  We will also neglect quark/gluon discriminants that take into account azimuthal asymmetries within the jet, though observables like the covariance tensor \cite{Gallicchio:2012ez} and 2-subjettiness \cite{Thaler:2010tr,Thaler:2011gf} can improve quark/gluon discrimination.

\subsection{Classifier separation}
\label{quarkgluon_sec:classsep}

Since we will be testing many parton shower variants, we need a way to quantify quark/gluon separation power in a robust way that can easily be summarized by a single number.  For that purpose we use classifier separation  (repeating Eq.~\eqref{quarkgluon_eq:deltadef_intro} for convenience),
\begin{equation}
\label{quarkgluon_eq:deltadef}
\Delta =  \frac{1}{2} \int \text{d} \lambda \, \frac{\bigl(p_q(\lambda) - p_g(\lambda)\bigr)^2}{p_q(\lambda) + p_g(\lambda)},
\end{equation}
where $p_q$ ($p_g$) is the probability distribution for the quark jet (gluon jet) sample as a function of the classifier $\lambda$.  Here, $\Delta = 0$ corresponds to no discrimination power and $\Delta = 1$ corresponds to perfect discrimination power.

A more common way to talk about discrimination power is in terms of receiver operating characteristic (ROC) curves.  At a point ($q$,$g$) on the ROC curve, where $q,g \in [0,1]$, one can define a selection that yields $q$ efficiency for quarks and $g$ mistag rate for gluons, or equivalently, a $(1-g)$ efficiency for gluons for a $(1-q)$ mistag rate for quarks.  To turn the ROC curve into a single number, it is common to report the gluon rejection rate at, say, 50\% quark efficiency.  Since we are more interested in understanding the relative performance between parton showers rather than the absolute performance, we will not show ROC curves here, though they can be easily derived from the $p_q$ and $p_g$ distributions.  If one observable has an everywhere better ROC curve than another (i.e.~it is Pareto optimal), then it will also have a larger $\Delta$ value.  The converse is not true, however, since depending on the desired working point, a ``bad'' discriminant as measured by $\Delta$ might still be ``good'' by another metric.  In that sense, $\Delta$ contains less information than the full ROC curve.

An alternative way to quantify discrimination power is through mutual information, which counts the number of ``bits'' of information gained from measuring a discriminant variable (see \cite{Larkoski:2014pca}).  Given a sample with quark fraction $f \in [0,1]$ and gluon fraction $(1-f)$, the mutual information with the truth (a.k.a. the truth overlap) is
\begin{equation}
I(T; \Lambda) = \int \text{d} \lambda \left(f \, p_q(\lambda) \log_2 \frac{p_q(\lambda)}{p_{\rm tot}(\lambda)} + (1-f) \, p_g(\lambda) \log_2 \frac{p_g(\lambda)}{p_{\rm tot}(\lambda)}   \right),
\end{equation}
where $T = \{q,g\}$ is the set of truth labels, $\Lambda = \{\lambda\}$ is the (continuous) set of discriminant values, and 
\begin{equation}
p_{\rm tot}(\lambda) = f \, p_q(\lambda) + (1-f) \, p_g(\lambda).
\end{equation}
The choice $f = 1/2$ was used in Ref.~\cite{Larkoski:2014pca}, though other $f$ choices are plausible.  Though we will not use mutual information in this study, it is amusing to note that the second derivative of $I(T;\Lambda)$ with respect to $f$ is related to classifier separation as
\begin{equation}
\label{quarkgluon_eq:altdeltadef}
- \frac{\log 2}{4} \frac{\partial^2 I(T;\Lambda)}{\partial f^2} \Big|_{f = \frac{1}{2}} = \Delta.
\end{equation}

One advantage of $\Delta$ over $I(T;\Lambda)$ is that the integrand in Eq.~\eqref{quarkgluon_eq:deltadef} is easier to interpret, since it tracks the fractional difference between the signal and background at a given value of $\lambda$.\footnote{Another advantage of $\Delta$ over $I(T; \Lambda)$ arises when trying to assign statistical uncertainties to finite Monte Carlo samples.  Since $\Delta$ is defined as a simple integral, one can use standard error propagation to assign uncertainties to $\Delta$.  By contrast, because of the logarithms in the $I(T; \Lambda)$ integrand, one has to be careful about a potential binning bias \cite{Larkoski:2014pca}.}  Specifically, by plotting 
\begin{equation}
\label{quarkgluon_eq:deltaintegrand}
\frac{\text{d} \Delta}{\text{d} \lambda} = \frac{1}{2} \frac{\bigl(p_q(\lambda) - p_g(\lambda) \bigr)^2}{p_q(\lambda) + p_g(\lambda)},
\end{equation}
one can easily identify which regions of phase space contribute the most to quark/gluon discrimination.  One can then ask whether or not the regions exhibiting the most separation power are under sufficient theoretical control, including both the size of perturbative uncertainties and the impact of nonperturbative corrections.  

\subsection{Idealized quark/gluon discrimination}
\label{quarkgluon_sec:ee}

Our parton shower studies are based on the idealized case of discriminating quark and gluon jets in $e^+ e^-$ collisions.  As we will see, this example demonstrates the importance of final state evolution for quark/gluon discrimination, separate from initial state complications arising in $pp$ collisions.  The analysis code used for this study is available as a \textsc{Rivet} routine \cite{Buckley:2010ar}, which can be downloaded from \verb|https://github.com/gsoyez/lh2015-qg| under \verb|MC_LHQG_EE.cc|.

To define the truth-level jet flavor, we use a simple definition:  a quark jet is a jet produced by a parton shower event generator in $e^+ e^- \to (\gamma/Z)^* \to u \bar{u}$ hard scattering, while a gluon jet is a jet produced in $e^+ e^- \to h^* \to gg$.  Of course, an $e^+e^- \to u \bar u$ event can become a $e^+e^- \to u \bar u g$ event after one step of shower evolution, just as $e^+e^- \to g g$ can become $e^+e^- \to g u \bar u$.  This illustrates the inescapable ambiguity in defining jet flavor.\footnote{In an $e^+e^-$ context, our definition at least respects the Lorentz structure of the production vertex, so in that sense it is a fundamental definition that does not reference (ambiguous) quark or gluon partons directly.}  To partially mitigate the effect of wide-angle emissions, we restrict our analysis to jets that satisfy
\begin{equation}
\label{quarkgluon_eq:Erestrict}
\frac{E_{\rm jet}}{Q/2} > 0.8,
\end{equation}
with up to two jets studied per event.  There is also the ambiguity of which parton shower to use, and we investigate this ambiguity by looking at results from several event generators:  \textsc{Pythia 8.205} \cite{Sjostrand:2006za,Sjostrand:2014zea}, \textsc{Herwig++ 2.7.1} \cite{Bahr:2008pv,Bellm:2013hwb}, \textsc{Sherpa 2.1.1} \cite{Gleisberg:2008ta}, \textsc{Vincia 1.201} \cite{Giele:2013ema}, \textsc{Deductor 1.0.2} \cite{Nagy:2014mqa} (with hadronization performed by \textsc{Pythia}), and \textsc{Ariadne 5.0.$\beta$} \cite{Flensburg:2011kk}.  

\begin{figure}
\centering
\subcaptionbox{\label{quarkgluon_fig:LHA_hadron_quark}}{
\includegraphics[width = 0.45\columnwidth]{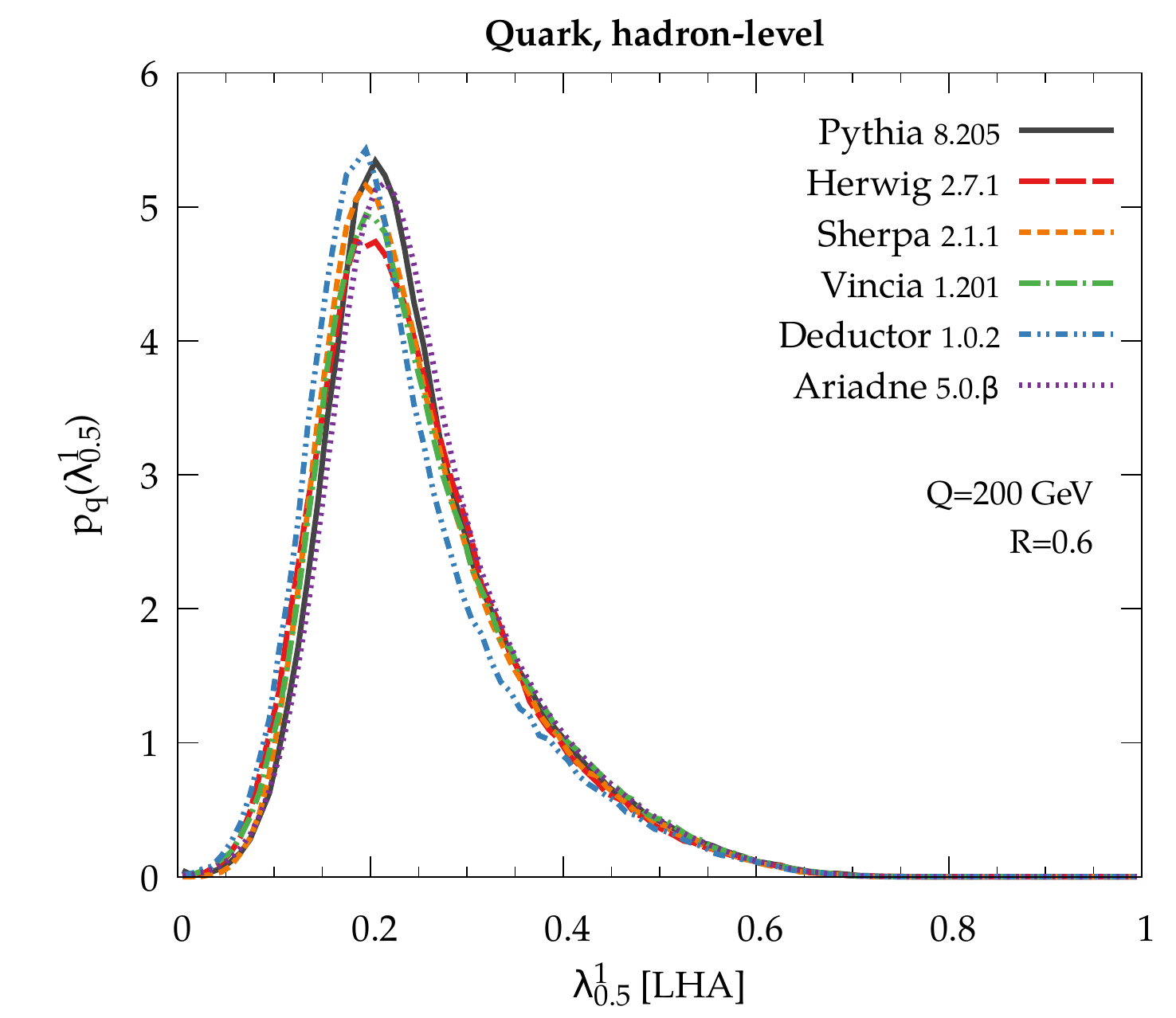}
}
$\qquad$
\subcaptionbox{\label{quarkgluon_fig:LHA_hadron_gluon}}{
\includegraphics[width = 0.45\columnwidth]{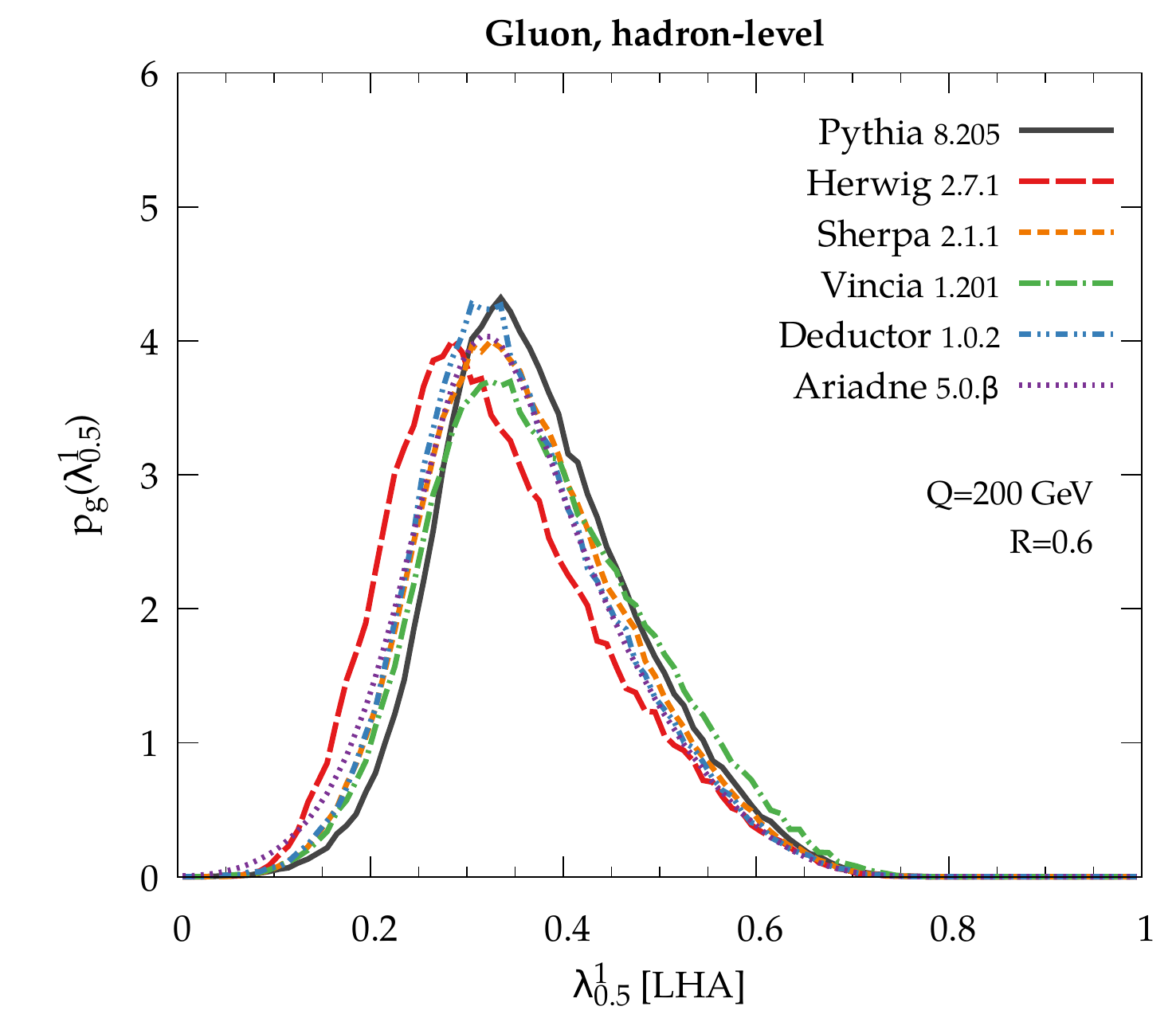}
 }

\subcaptionbox{\label{quarkgluon_fig:LHA_hadron_separation}}{
 \includegraphics[width = 0.65\columnwidth]{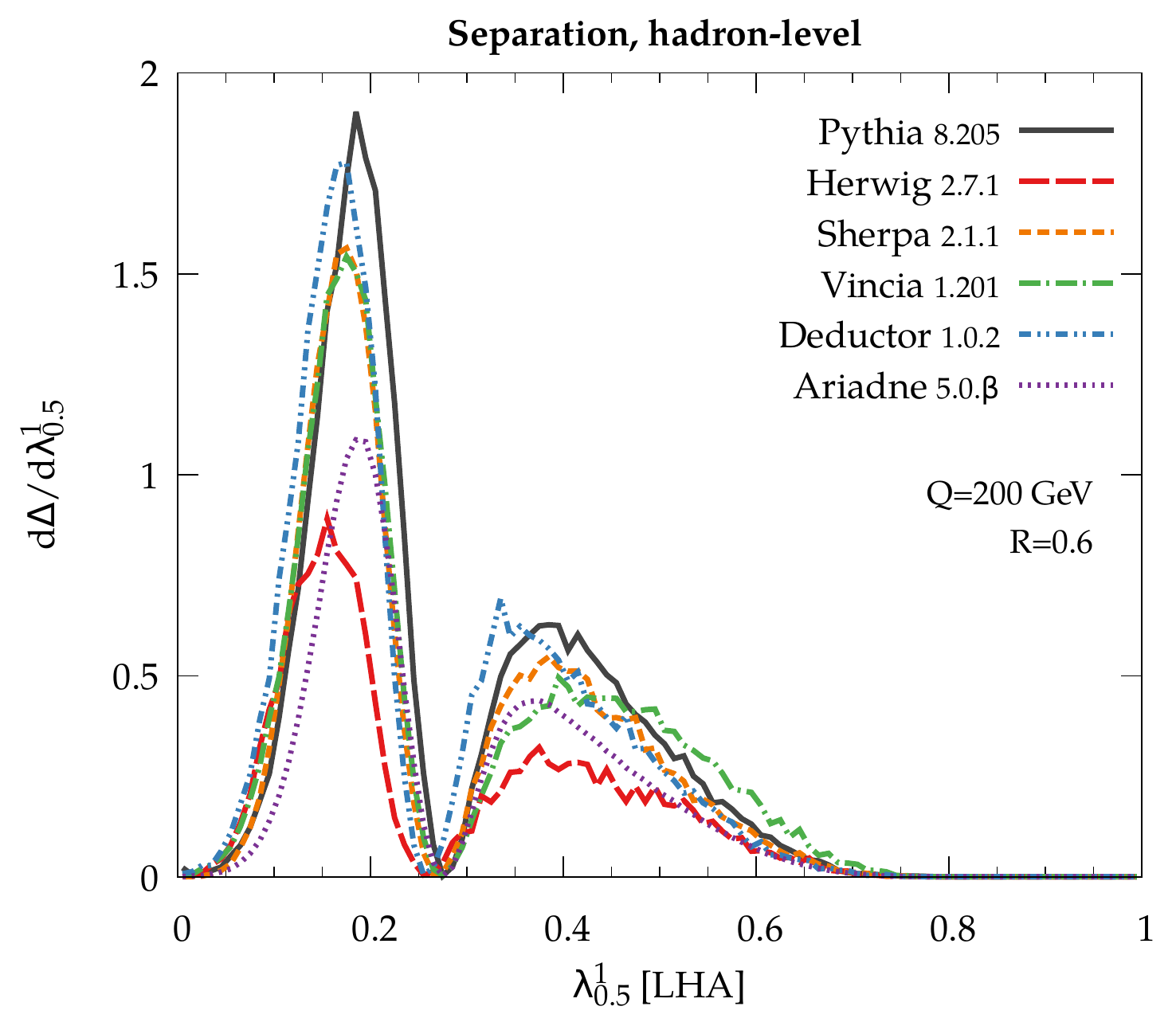}
 }
\caption{Hadron-level distributions of the LHA for (a) the $e^+ e^- \to u \bar{u}$ (``quark jet'') sample, (b) the $e^+ e^- \to gg$ (``gluon jet'') sample, and (c) the classifier separation integrand in Eq.~\eqref{quarkgluon_eq:deltaintegrand}.  Six parton shower generators---\textsc{Pythia 8.205}, \textsc{Herwig++ 2.7.1}, \textsc{Sherpa 2.1.1}, \textsc{Vincia 1.201}, \textsc{Deductor 1.0.2}, and \textsc{Ariadne 5.0.$\beta$}---are run at their baseline settings with center-of-mass energy $Q = 200~\GeV$ and jet radius $R= 0.6$.}
\label{quarkgluon_fig:LHA_hadron}
\end{figure}

In Fig.~\ref{quarkgluon_fig:LHA_hadron}, we show hadron-level distributions of the LHA (i.e.~$\lambda_{0.5}^1$) in the quark sample ($p_q$) and gluon sample ($p_g$), comparing the baseline settings of six different parton shower generators with a center-of-mass collision energy of $Q = 200~\GeV$ and jet radius $R = 0.6$. In the quark sample in Fig.~\ref{quarkgluon_fig:LHA_hadron_quark}, there is relatively little variation between the generators, which is not surprising since most of these programs have been tuned to match LEP data (though LEP never measured the LHA itself).  Turning to the gluon sample in Fig.~\ref{quarkgluon_fig:LHA_hadron_gluon}, we see somewhat larger variations between the generators; this is expected since there is no data to directly constrain $e^+ e^- \to gg$.  In Fig.~\ref{quarkgluon_fig:LHA_hadron_separation}, we plot the integrand of classifier separation, $\text{d} \Delta / \text{d} \lambda$ from Eq.~\eqref{quarkgluon_eq:deltaintegrand}. This shows where in the LHA phase space the actual discrimination power lies, with large values of the integrand
corresponding to places where the quark and gluon distributions are
most dissimilar.  Now we see considerable differences between the
generators, reproducing the well-known fact that \textsc{Pythia} is
more optimistic about quark/gluon separation compared to
\textsc{Herwig} \cite{Aad:2014gea}.  The predicted discrimination power from the other four generators are intermediate between these
extremes.

\begin{figure}
\centering
\subcaptionbox{\label{quarkgluon_fig:LHA_parton_quark}}{
 \includegraphics[width = 0.45\columnwidth]{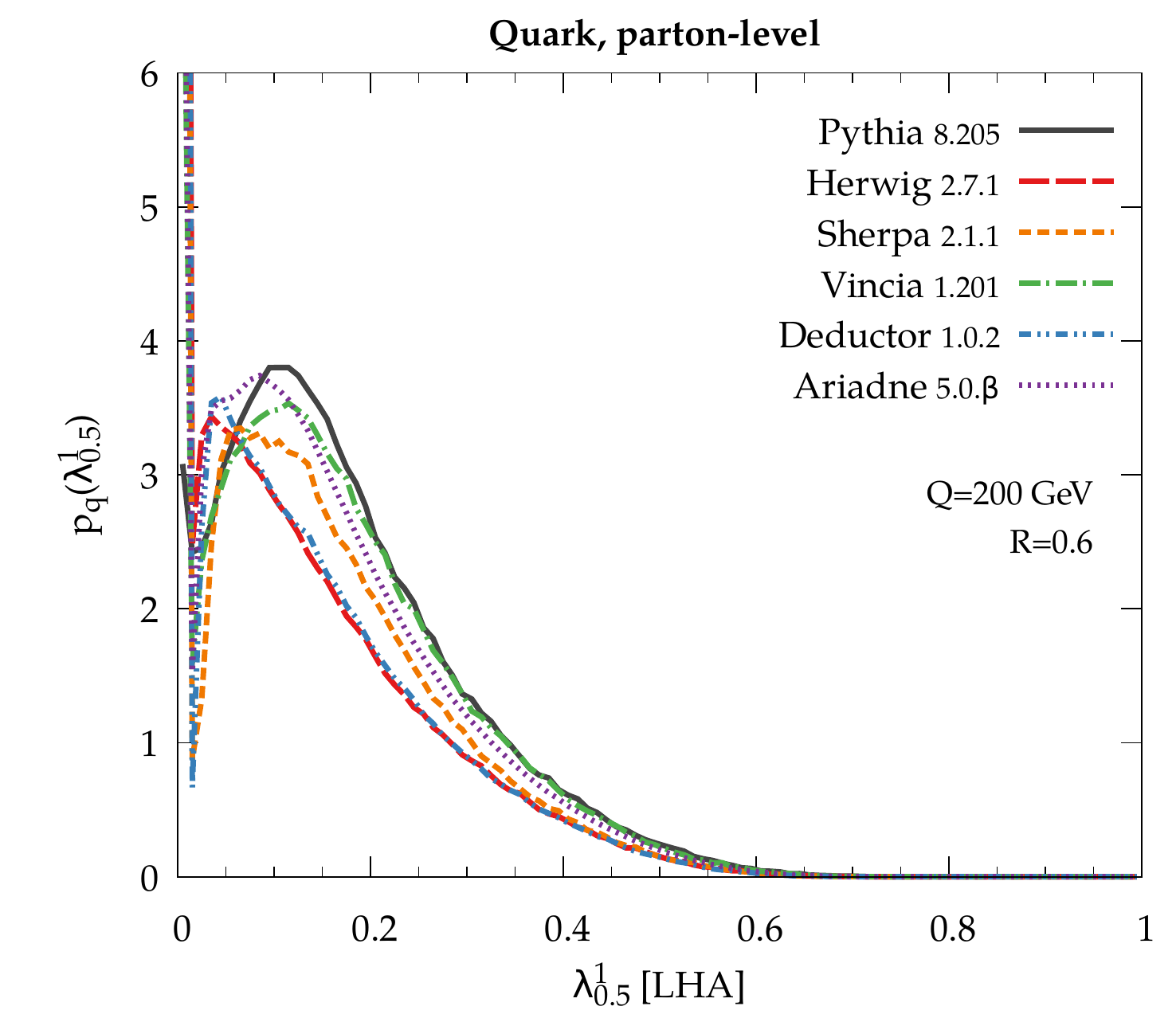}
 }
$\qquad$
\subcaptionbox{\label{quarkgluon_fig:LHA_parton_gluon}}{
 \includegraphics[width = 0.45\columnwidth]{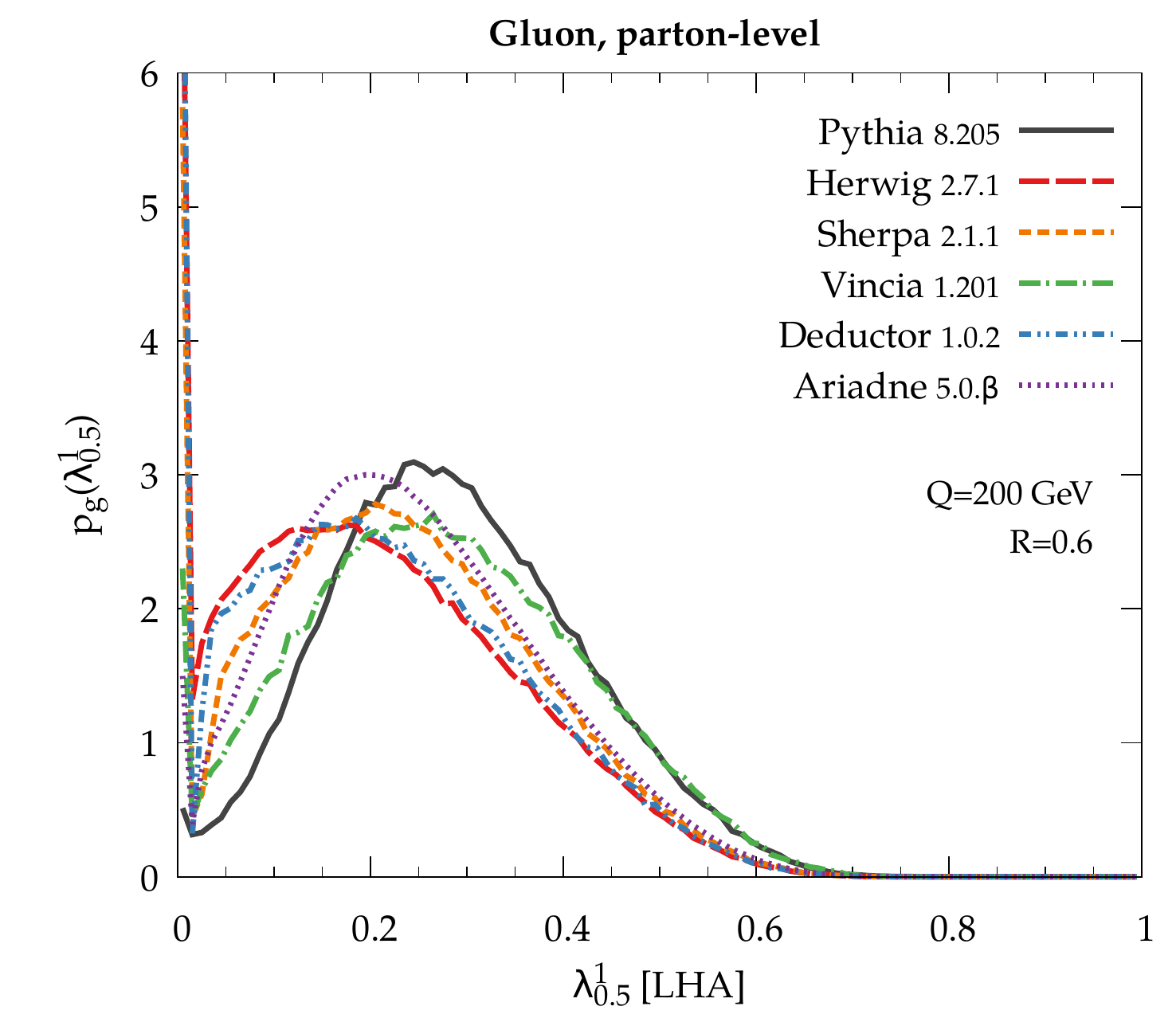}
 }

\subcaptionbox{\label{quarkgluon_fig:LHA_parton_separation}}{
 \includegraphics[width = 0.65\columnwidth]{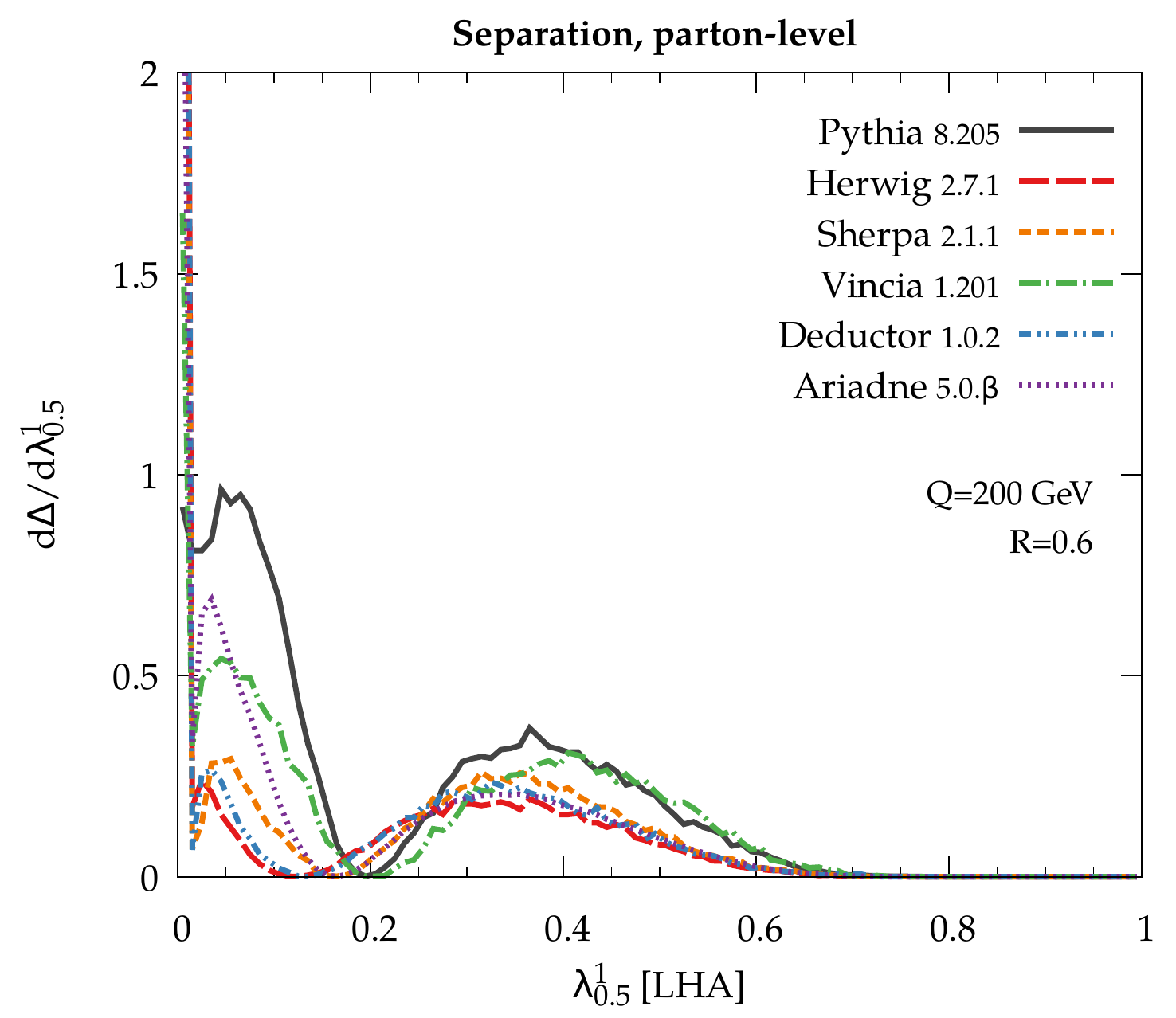}
 }
\caption{Same as Fig.~\ref{quarkgluon_fig:LHA_hadron}, but at the parton level.  Note that \textsc{Herwig}, \textsc{Sherpa}, and \textsc{Deductor} all have cross section spikes at $\lambda_{0.5}^1 = 0$ that extend above the plotted range.}
\label{quarkgluon_fig:LHA_parton}
\end{figure}

One might expect that the differences between generators are due
simply to their having different hadronization models.  It seems,
however, that the differences already appear at the parton level prior
to hadronization. We should say at the outset that it is nearly impossible to do a true apples-to-apples comparison of parton-level results, since these generators are interfaced to different hadronization models, and only the hadron-level comparison is physically meaningful.  In particular, the crossover between the perturbative and nonperturbative regions is ambiguous and each of these showers has a different effective shower cutoff scale, resulting in different amounts of radiation being generated in the showering versus hadronization steps.\footnote{In general, generators based on string hadronization tend to use a lower shower cutoff scale ($\sim 0.5$ MeV) compared to those based on cluster hadronization ($\sim 1$ GeV).}

With that caveat in mind, we show parton-level results in Fig.~\ref{quarkgluon_fig:LHA_parton}.  One immediately notices that three of the
generators---\textsc{Herwig}, \textsc{Sherpa}, and
\textsc{Deductor}---yield a large population of events where the
perturbative shower generates no emissions.  This gives
$\lambda_{0.5}^1 = 0$ such that non-zero values of the LHA are
generated only by the hadronization model.  By contrast,
\textsc{Pythia} and \textsc{Vincia} give overall larger values of the
LHA from the perturbative shower alone.  As mentioned above, some of this difference can be explained simply by the different shower cutoff scales used in each generator, but it probably also reflects a difference in how semi-perturbative gluon splittings are treated.  Since Fig.~\ref{quarkgluon_fig:LHA_hadron_quark} shows that all generators give similar distributions for quark jets after hadronization, we
conclude that understanding quark/gluon discrimination is a challenge
at the interface between perturbative showering and nonperturbative
hadronization.

\begin{figure}
\centering
\subcaptionbox{\label{quarkgluon_fig:summary_hadron_all}}{
 \includegraphics[width = 0.45\columnwidth]{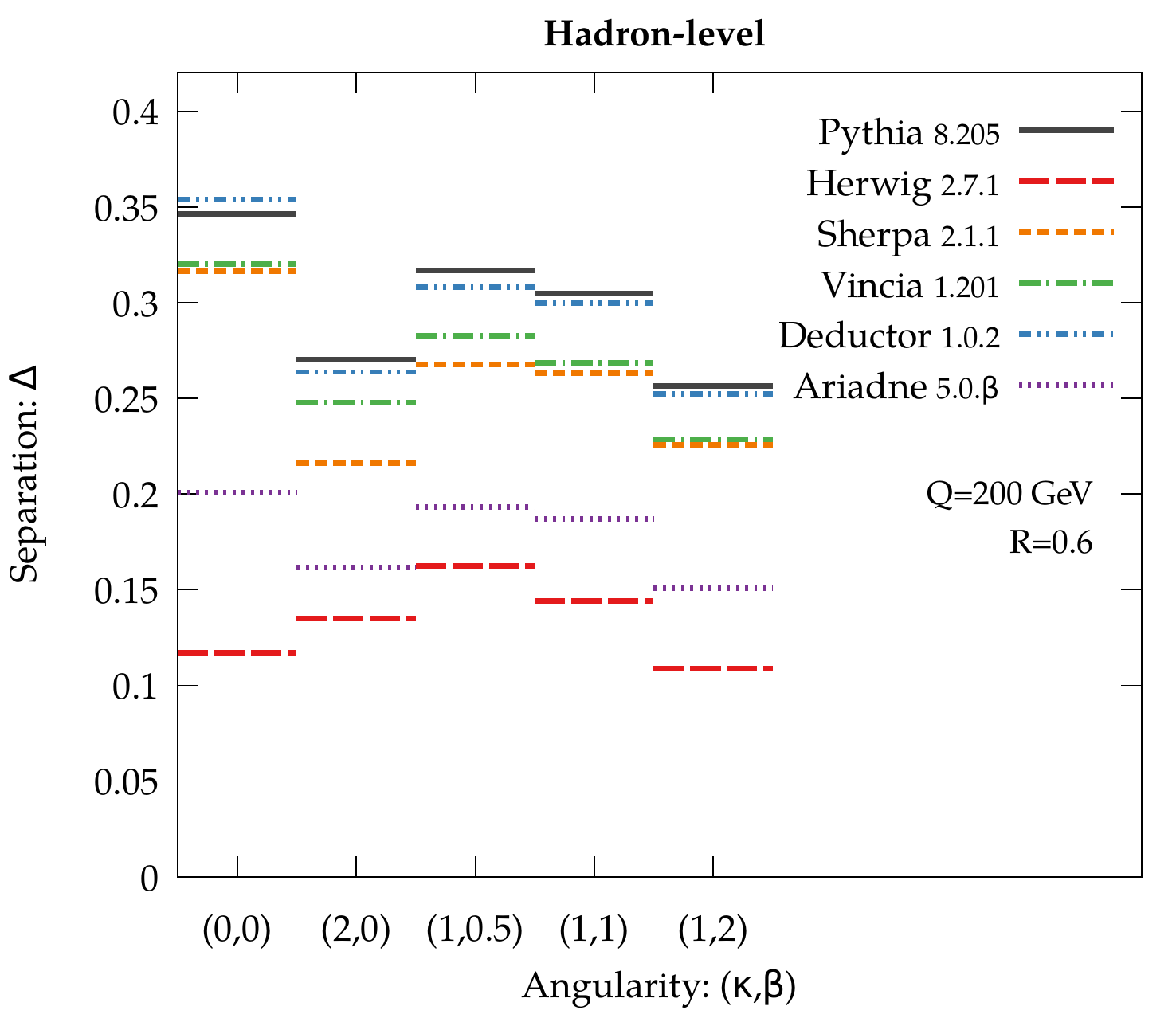}
 }
$\qquad$
\subcaptionbox{\label{quarkgluon_fig:summary_parton_all}}{
 \includegraphics[width = 0.45\columnwidth]{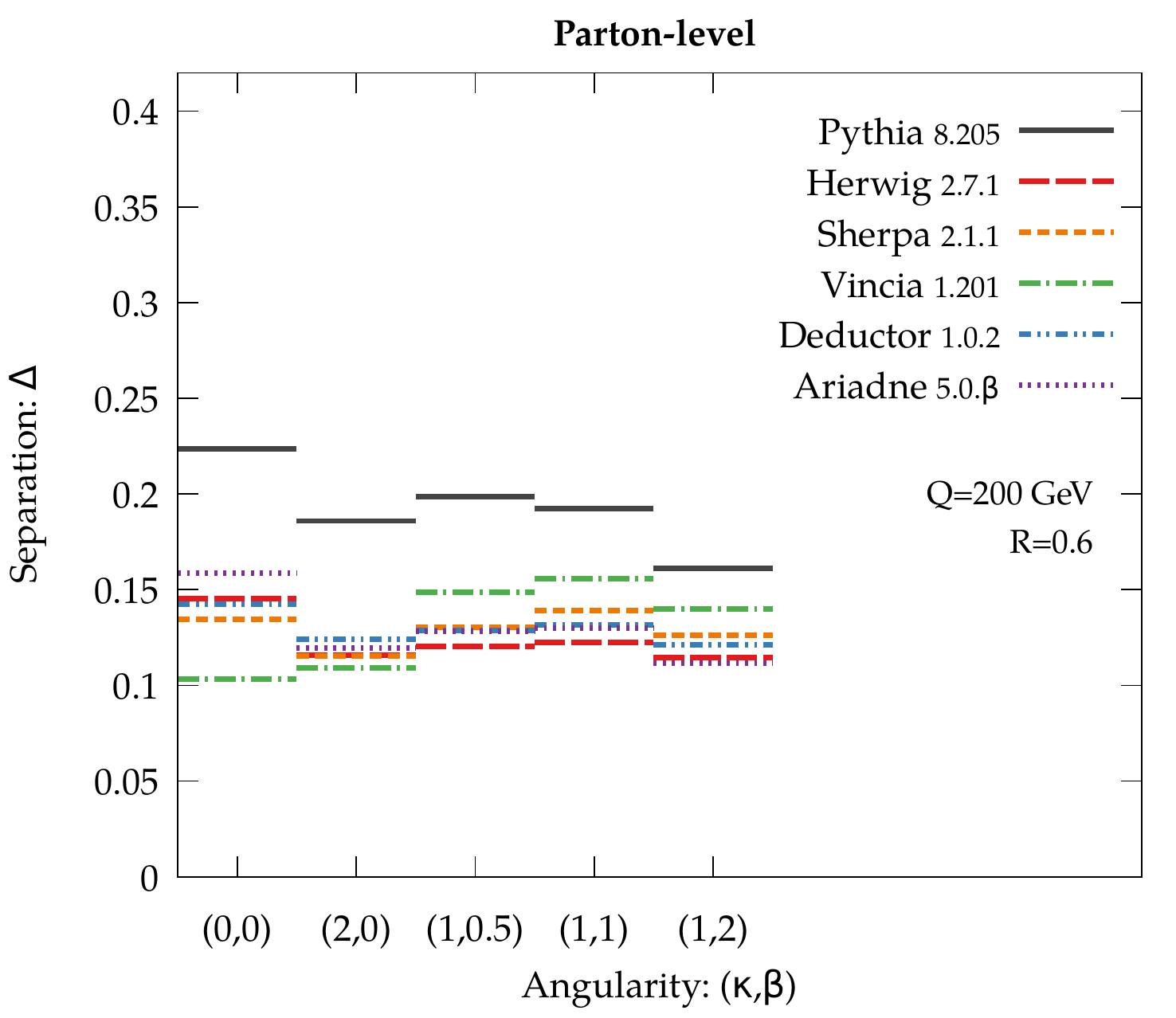}
 }
\caption{Classifier separation $\Delta$ for the five benchmark angularities in Eq.~\eqref{quarkgluon_eq:benchmarkang}, determined from the various generators at (a) hadron level and (b) parton level.  The first two columns correspond to IRC unsafe distributions (multiplicity and $p_T^D$), while the last three columns are the IRC safe angularities.  The LHA (i.e.~$\kappa = 1$, $\beta = 1/2$) is shown in the middle column.}
\label{quarkgluon_fig:summary_all}
\end{figure}

To summarize the overall discrimination power, we integrate
Eq.~\eqref{quarkgluon_eq:deltaintegrand} to obtain the value of
classifier separation $\Delta$ for the LHA.  This is shown in
Fig.~\ref{quarkgluon_fig:summary_all}, which also includes the four
other benchmark angularities from
Eq.~\eqref{quarkgluon_eq:benchmarkang}.  There is a rather large
spread in predicted discrimination power between the generators,
especially at hadron level in
Fig.~\ref{quarkgluon_fig:summary_hadron_all}.  While such differences
might be expected for IRC unsafe angularities (multiplicity and
$p_T^D$) which depend on nonperturbative modeling, these differences
persist even for the IRC safe angularities at parton level (see
Fig.~\ref{quarkgluon_fig:summary_parton_all}).\footnote{It is interesting that four of the generators---\textsc{Herwig}, \textsc{Sherpa}, \textsc{Deductor}, and \textsc{Ariadne}---have a comparatively narrow spread in predicted discrimination power at parton level, though this spread increases dramatically at hadron level.}  This suggests a more
fundamental difference between the generators that is already present
in the perturbative shower

For the IRC safe angularities with $\kappa = 1$, there is a generic trend seen by all of the hadron-level generators that discrimination power decreases as $\beta$ increases.  This trend agrees with the study performed in Ref.~\cite{Larkoski:2013eya}, but disagrees with the ATLAS study in Ref.~\cite{Aad:2014gea}, which found flat (or even increasing) discrimination power with increasing $\beta$.  Understanding this $\beta$ trend will therefore be crucial for understanding quark/gluon radiation patterns.

\subsection{Parameter dependence}
\label{quarkgluon_sec:ee_scales}

\begin{figure}
\centering
\subcaptionbox{\label{quarkgluon_fig:sweep_Q_hadron}}{
 \includegraphics[width = 0.45\columnwidth]{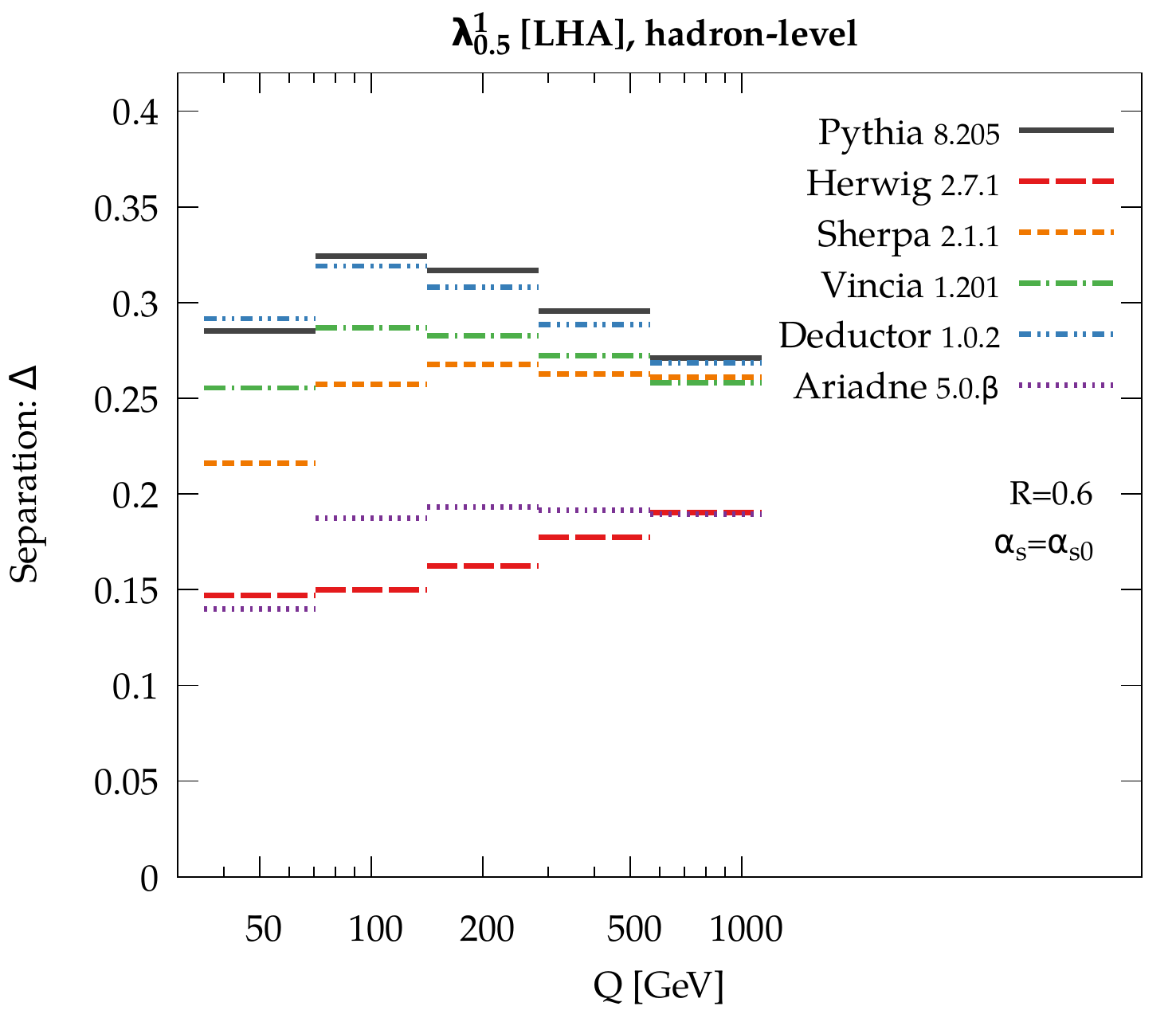}
 }
$\qquad$
\subcaptionbox{\label{quarkgluon_fig:sweep_Q_parton}}{
 \includegraphics[width = 0.45\columnwidth]{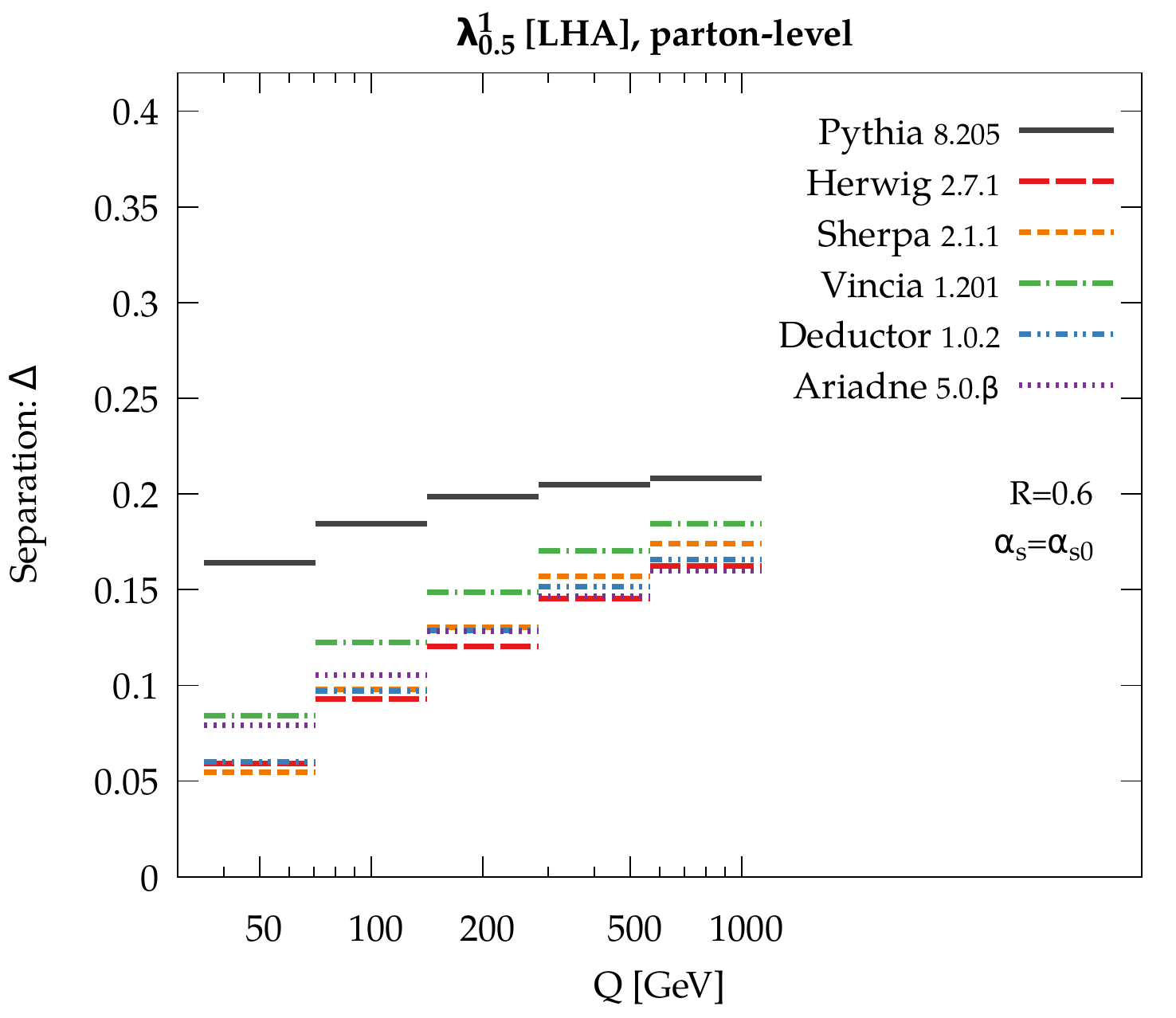}
 }

\subcaptionbox{\label{quarkgluon_fig:sweep_R_hadron}}{
 \includegraphics[width = 0.45\columnwidth]{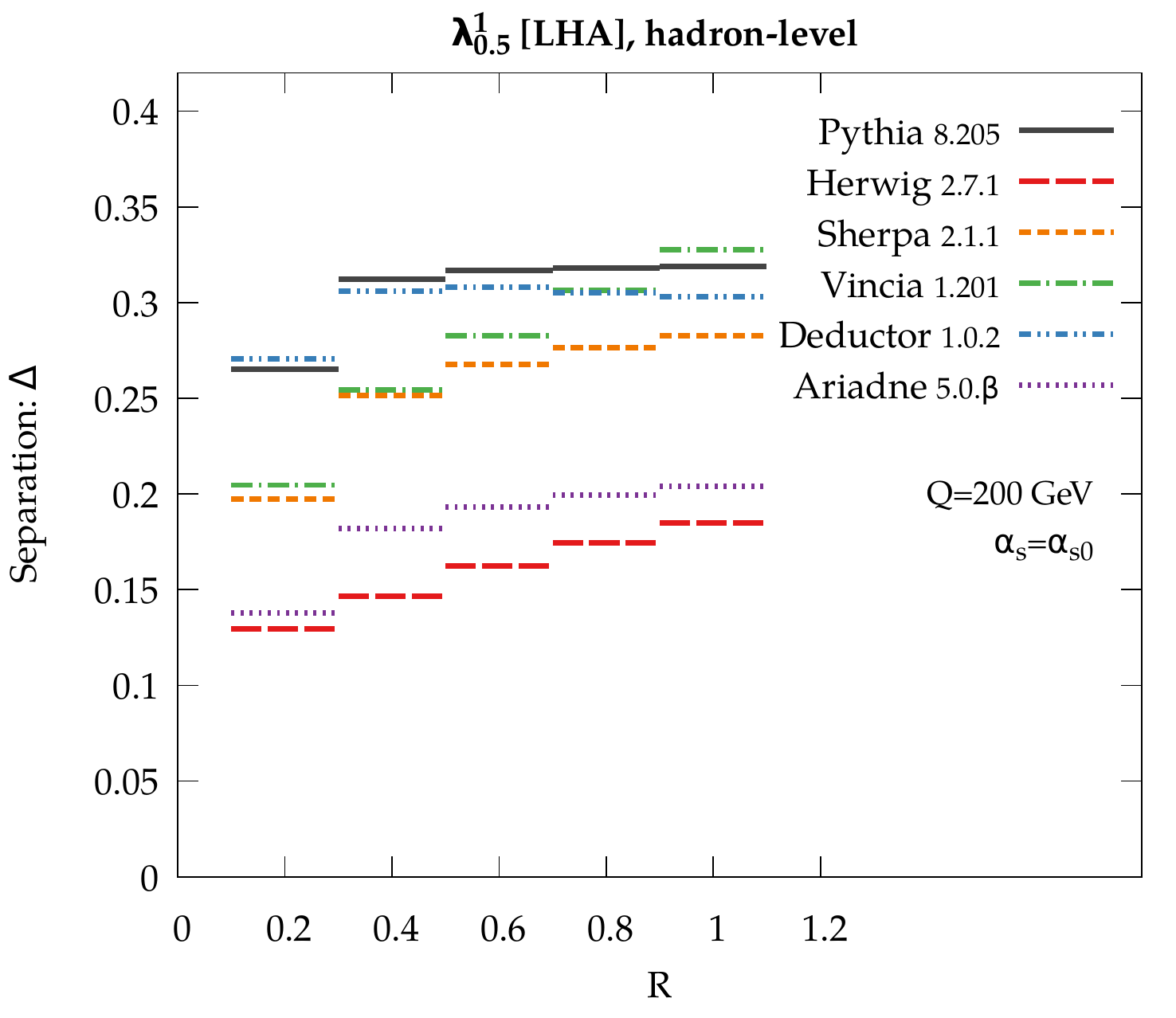}
 }
$\qquad$
\subcaptionbox{\label{quarkgluon_fig:sweep_R_parton}}{
 \includegraphics[width = 0.45\columnwidth]{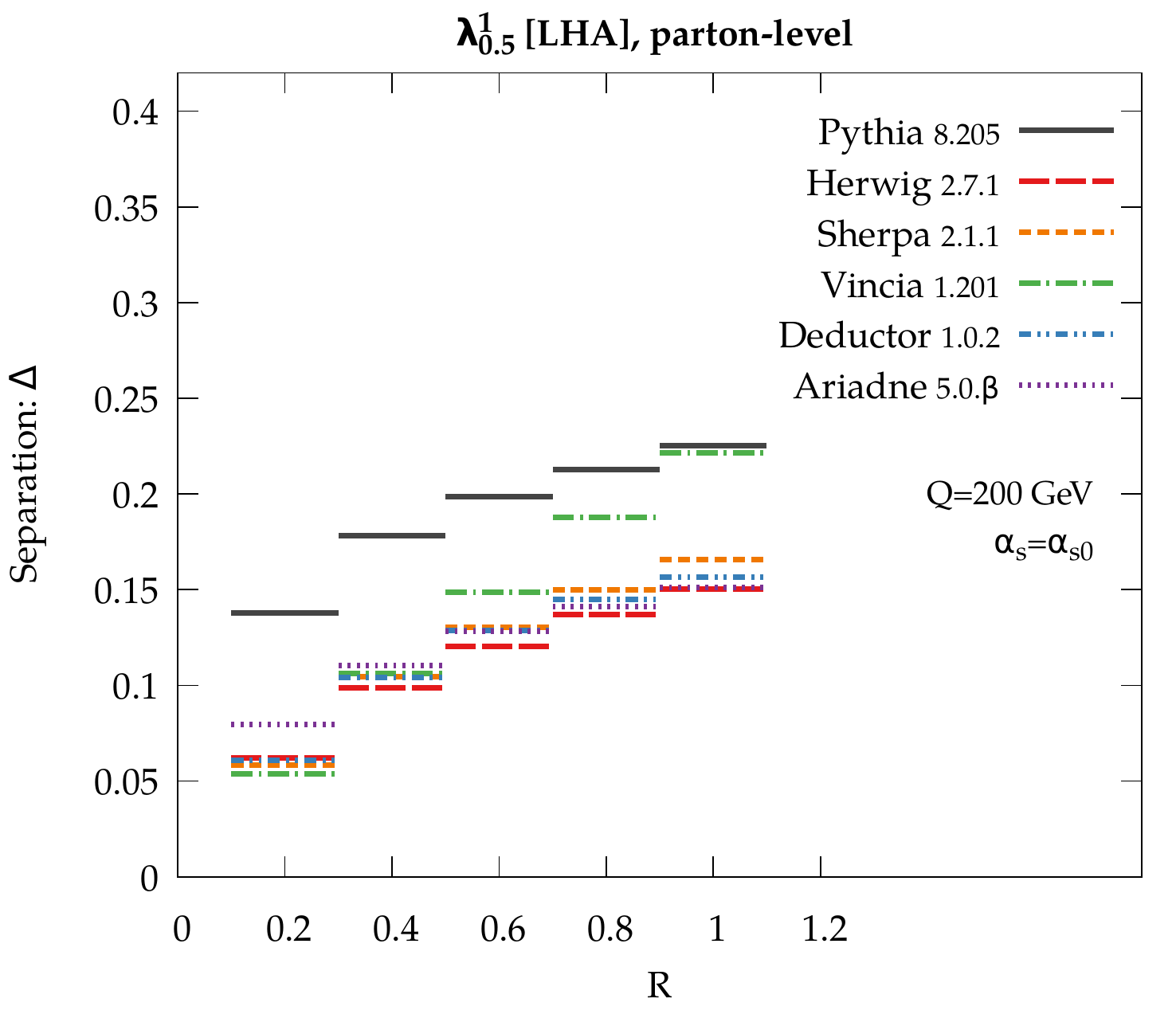}
 }

\subcaptionbox{\label{quarkgluon_fig:sweep_as_hadron}}{
 \includegraphics[width = 0.45\columnwidth]{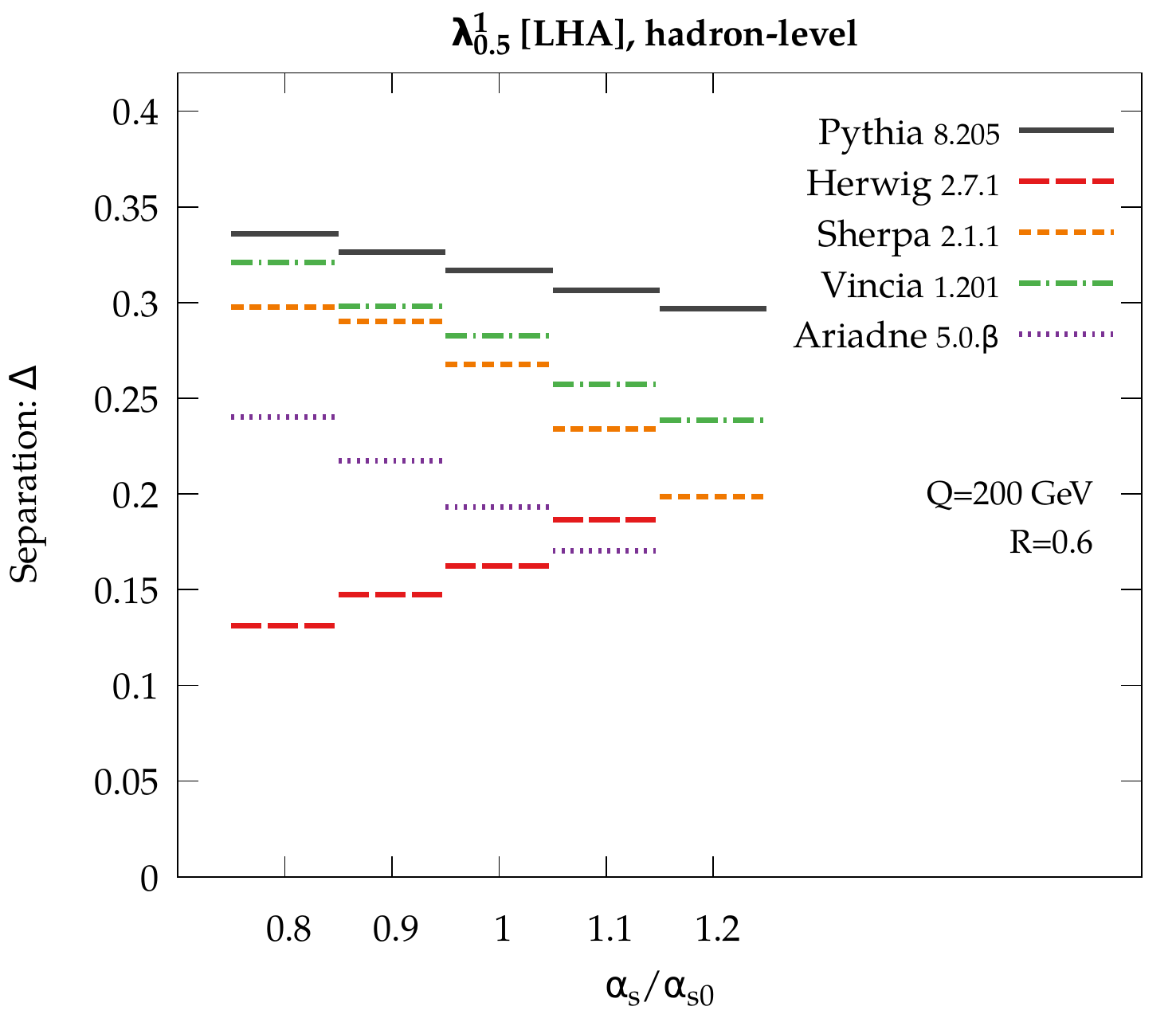}
 }
$\qquad$
\subcaptionbox{\label{quarkgluon_fig:sweep_as_parton}}{
 \includegraphics[width = 0.45\columnwidth]{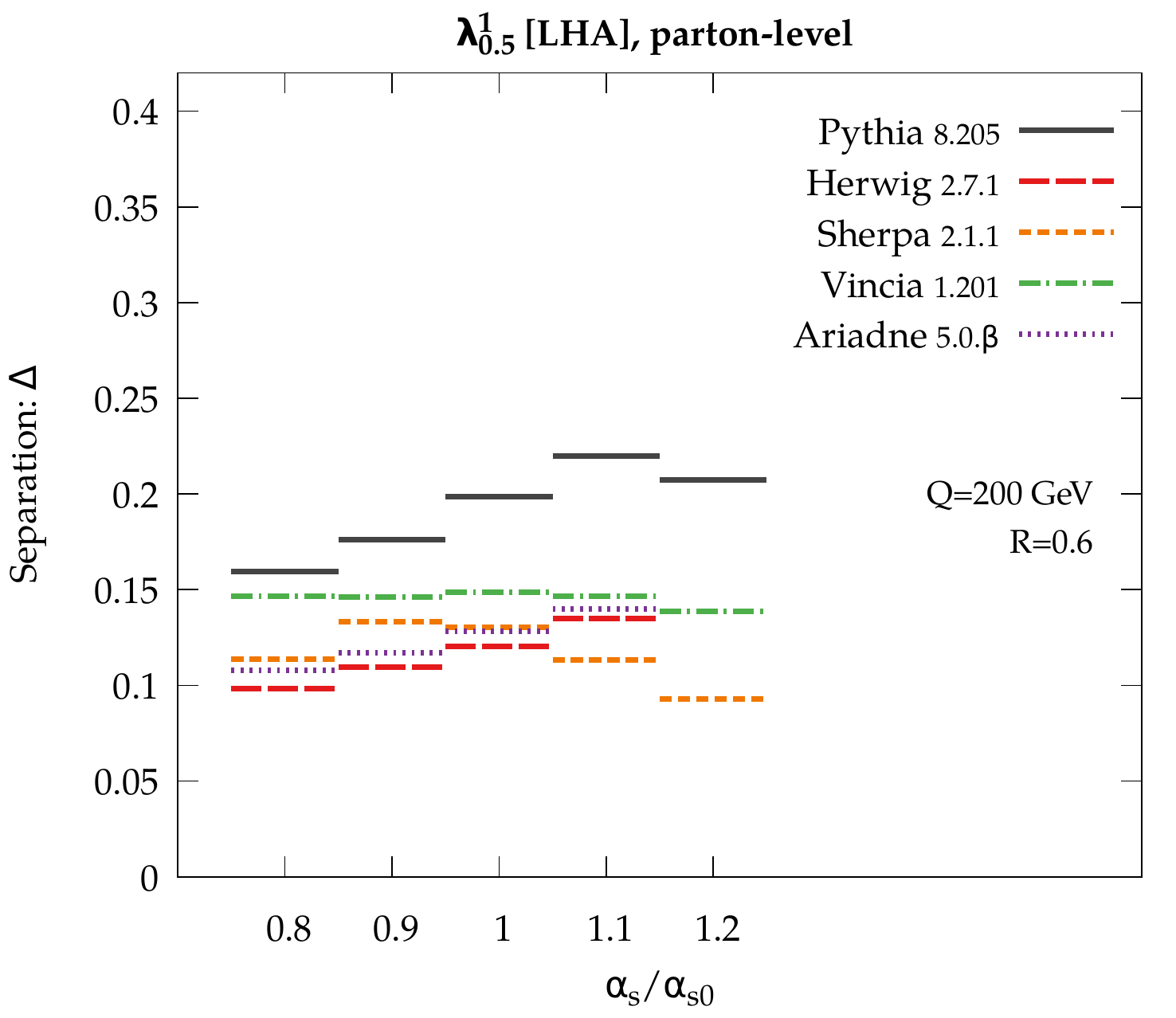}
 }
\caption{Classifier separation $\Delta$ for the LHA, sweeping the collision energy $Q$ (top row), jet radius $R$ (middle row), and coupling constant $\alpha_s/\alpha_{s0}$ (bottom row).  Results are shown at hadron level (left column) and parton level (right column).}
\label{quarkgluon_fig:ee_sweep}
\end{figure}

Given the large absolute differences in discrimination power seen above, we next want to check if the parton shower generators exhibit similar or dissimilar trends as parameters are varied.  We perform three parameter sweeps, using the boldface values below as defaults:
\begin{equation}
\begin{aligned}
\text{Collision Energy}: Q &= \{ 50, 100, \mathbf{200}, 400, 800\}~\GeV, \\
\text{Jet Radius}: R &= \{ 0.2, 0.4, \mathbf{0.6}, 0.8, 1.0\}, \\
\text{Strong Coupling}: \alpha_s / \alpha_{s0} &= \{0.8,0.9,\mathbf{1.0},1.1,1.2\}, \\
\end{aligned}
\end{equation}
where $\alpha_{s0}$ is the default value of the strong coupling, which is different between the generators (and sometimes different between different aspects of the same generator).

The resulting values of $\Delta$ for the LHA are shown in Fig.~\ref{quarkgluon_fig:ee_sweep}, at both the hadron level and parton level.   There are number of surprising features in these plots.  Perhaps the most obvious (and seen already in Fig.~\ref{quarkgluon_fig:summary_all}) is that even for the IRC safe angularities, the effect of hadronization is rather large, both on the absolute scale of discrimination and the trends.  The main exception to this is \textsc{Herwig}, which does not exhibit as much of an effect from hadronization, though an effect is still present.

The next surprising feature is that the parton-level trends for sweeping $\alpha_s$ do not necessarily correspond to those for sweeping $Q$ and $R$.  According to the perturbative next-to-leading-logarithmic (NLL) logic in Ref.~\cite{Larkoski:2013eya}, quark/gluon discrimination should depend on $\alpha_s$ evaluated at the scale $Q R / 2$, with larger values of $\alpha_s(Q R / 2)$ leading to improved discrimination power.  Indeed, \textsc{Pythia}, \textsc{Herwig}, and \textsc{Ariadne} do show improved performance with larger $\alpha_s$.  However, larger values of $Q$ and $R$ correspond to smaller values of $\alpha_s$, so the NLL logic would predict that increasing $Q$ or $R$ should lead to worse discrimination power.  Instead, all of the generators show the opposite trend.

One reason to expect quark/gluon discrimination to improve as higher energies is that that phase space available for shower evolution increases as $Q$ increases.  The scale $\mu$ of the shower splitting is $\mu_0^2 < \mu^2 < Q^2$, where $\mu_0 = \mathcal{O}(\GeV)$ is the shower cutoff scale.  With more range for shower evolution at higher $Q$, there is a greater possibility to see that a quark jet is different from a gluon jet.  Similarly, larger values of $R$ allow for more emissions within a jet, and from scaling symmetry, one expects that parton-level discrimination power should depend on the combination $Q R$.\footnote{At small values of $R$, one has to worry about the flavor purity of a jet sample, since scale evolution can change the leading parton flavor \cite{Dasgupta:2014yra,Dasgupta:2016bnd}.  Similarly, the restriction in Eq.~\eqref{quarkgluon_eq:Erestrict} can impose a non-trivial bias on the jet flavor at small $R$.}  By contrast, the NLL logic says that quark/gluon discrimination should be dominated by the leading emission(s) in a jet, and since $\alpha_s$ is smaller at higher values of $Q R$, those leading emissions are more similar between quarks and gluons.  Given these two different but equally plausible logics, both of which probably play some role in the complete story, this motivates experimental tests of quark/gluon separation as a function of $Q$ and $R$.

For many of the generators, going from parton-level to hadron-level reverses or flattens the $Q$ and $\alpha_s$ trends, though the $R$ trends are more stable.  The study in Ref.~\cite{Larkoski:2013eya} did not include nonperturbative hadronization corrections, so we do not yet have a theoretical expectation for the impact of hadronization.  In future work, we plan to follow Ref.~\cite{Larkoski:2013paa} and include nonperturbative hadronization corrections via shape functions~\cite{Manohar:1994kq, Dokshitzer:1995zt, Korchemsky:1999kt, Korchemsky:2000kp, Salam:2001bd, Lee:2006nr, Mateu:2012nk} as well as test the impact of imposing a hard IR cutoff scale.  With confusions already at parton level, though, further perturbative calculations beyond NLL accuracy are also needed.  For example, at NLL accuracy, one does account for the fact that a jet can contain multiple perturbative emissions, but those emissions are treated as if they themselves do not radiate.  By contrast, parton showers (while formally only LL accurate) allow every emission to reradiate, which might be driving the $Q$ and $R$ discrimination trends.

\subsection{Impact of generator settings}
\label{quarkgluon_sec:ee_settings}

Formally, parton shower generators are only accurate to modified leading-logarithmic (MLL) accuracy, though they include physically important effects like energy/momentum conservation and matrix element corrections that go beyond MLL.  We can assess the impact of these higher-order effects by changing the baseline parameter settings in each parton shower generator.  

Because each generator is different, we cannot always make the same changes for each generator.  Similarly, the spread in discrimination power shown below should \emph{not} be seen as representing the intrinsic uncertainties in the shower, since many of these changes we explore are not physically motivated.  The goal of these plots is to demonstrate possible areas where small parameter changes could have a large impact on quark/gluon discrimination.  Ultimately, collider data and higher-order calculations will be essential for understanding the origin of quark/gluon differences.  In all cases, we show both hadron-level and parton-level results, even if a setting is only expected to have an impact at the hadron level.  

\begin{figure}
\centering
\subcaptionbox{\label{quarkgluon_fig:summary_hadron_pythia}}{
 \includegraphics[width = 0.45\columnwidth]{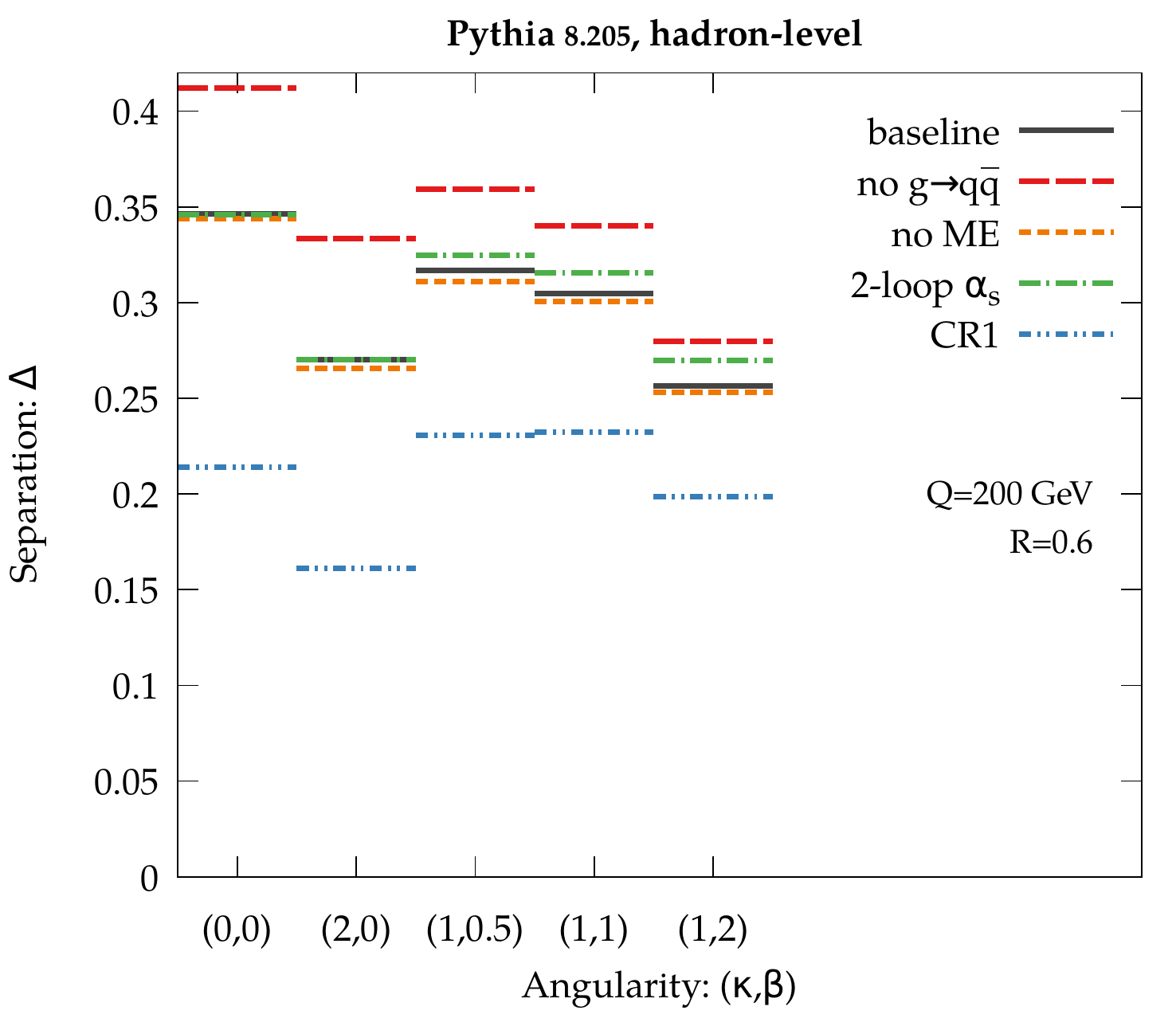}
 }
$\qquad$
\subcaptionbox{\label{quarkgluon_fig:summary_parton_pythia}}{
 \includegraphics[width = 0.45\columnwidth]{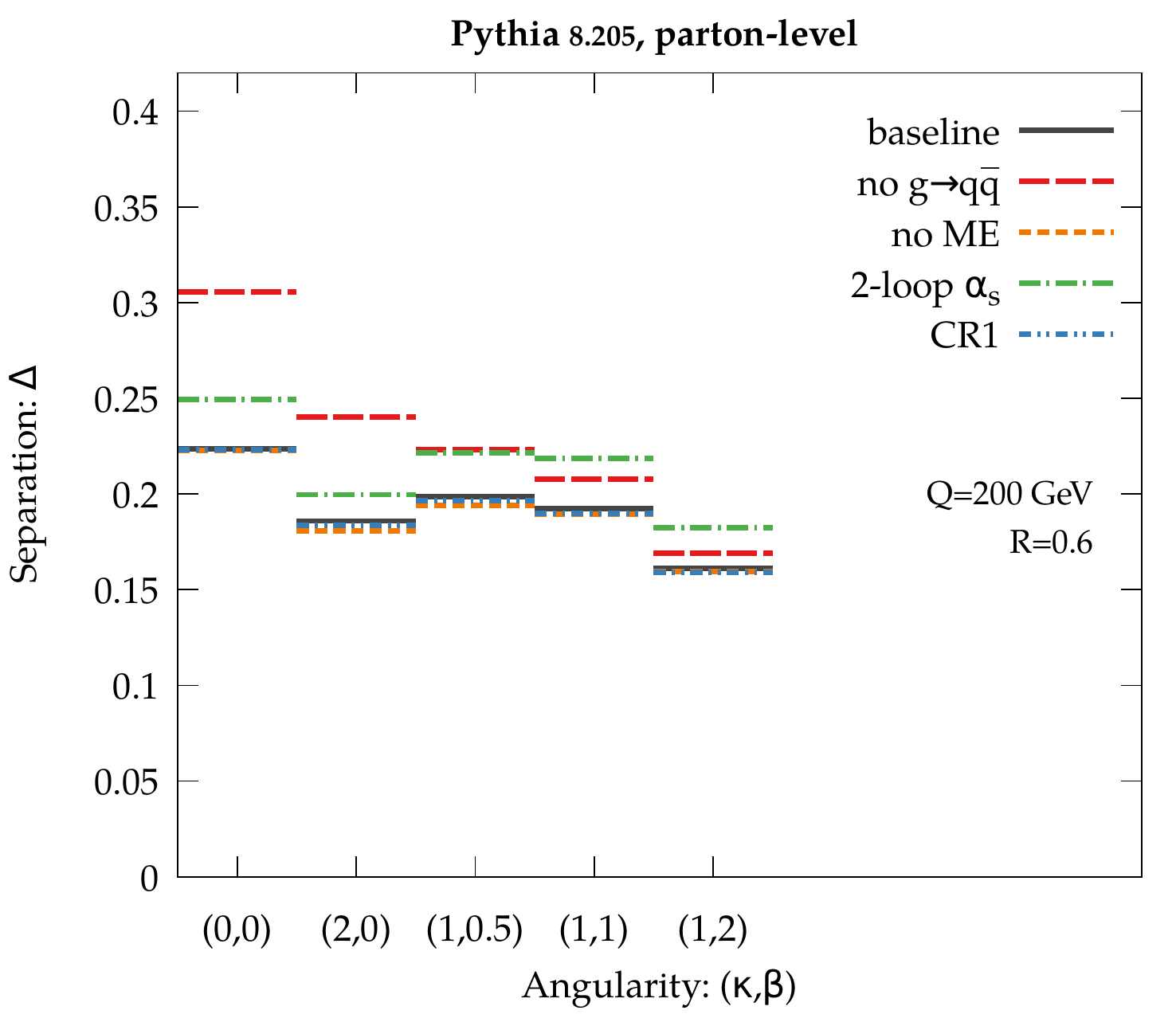}
 }
\caption{Settings variations for \textsc{Pythia 8.205}.  Shown are (a) hadron-level and (b) parton-level results for the classifier separation $\Delta$ derived from the five benchmark angularities.}
\label{quarkgluon_fig:settings_variation_pythia}
\end{figure}

Our $\textsc{Pythia}$ baseline is based on the Monash 2013 tune, with parameters described in Ref.~\cite{Skands:2014pea}.  In Fig.~\ref{quarkgluon_fig:settings_variation_pythia}, we consider the following \textsc{Pythia} variations:
\begin{itemize}
\item \textsc{Pythia: no $g \to q\bar{q}$}.  While the dominant gluon splitting in the parton shower is $g \to gg$, \textsc{Pythia}---and every other shower in this study---also generates the subleading $g \to q \bar{q}$ splittings by default.  This variation turns off $g \to q \bar{q}$, which makes gluon jets look more gluon-like, thereby increasing the separation power.
\item \textsc{Pythia: no ME}.  The first emission in \textsc{Pythia} is improved by applying a matrix element correction \cite{Miu:1998ju}, but this variation turns those corrections off, showing the impact of non-singular terms.  No matrix element correction is available for $h^* \to g g$, though, so the true impact of these corrections might be larger than the relatively small effect seen for this variation.
\item \textsc{Pythia: 2-loop $\alpha_s$}.  The default \textsc{Pythia} setting is to use 1-loop running for $\alpha_s$.  This variation turns on 2-loop running for $\alpha_s$, which has a small (beneficial) effect at parton level which is washed out at hadron level.
\item \textsc{Pythia: CR1}.  Often, one thinks of color reconnection as being primarily important for hadron colliders, but even at a lepton collider, color reconnection will change the Lund strings used for hadronization.  Compared to the baseline, this variation uses an alternative ``\text{SU}(3)''-based color reconnection model~\cite{Christiansen:2015yqa} (i.e.~\texttt{ColourReconnection:mode = 1}).  No attempts were made to retune any of the other hadronization parameters (as would normally be mandated in a tuning context), so this change simply illustrates the effect of switching on this reconnection model with default parameters, leaving all other parameters unchanged.  At parton level, this variation has no effect as expected.  At hadron level, this variation considerably decreases quark/gluon separation compared to the baseline.
\end{itemize}
The most surprising \textsc{Pythia} effect is the large potential impact of the color reconnection model, which is also important for the \textsc{Herwig} generator described next.

\begin{figure}
\centering
\subcaptionbox{\label{quarkgluon_fig:summary_hadron_herwig}}{
 \includegraphics[width = 0.45\columnwidth]{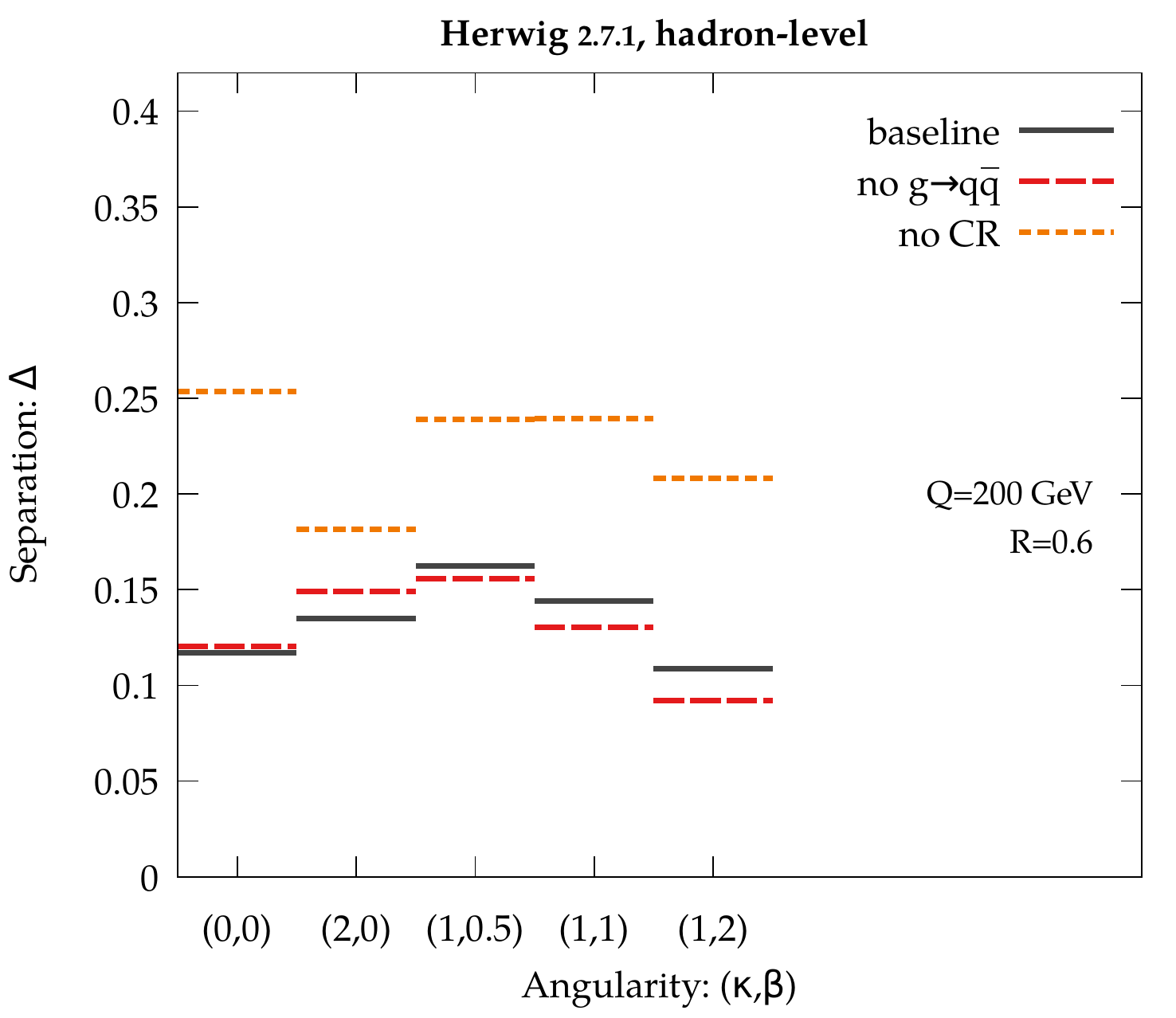}
 }
$\qquad$
\subcaptionbox{\label{quarkgluon_fig:summary_parton_herwig}}{
 \includegraphics[width = 0.45\columnwidth]{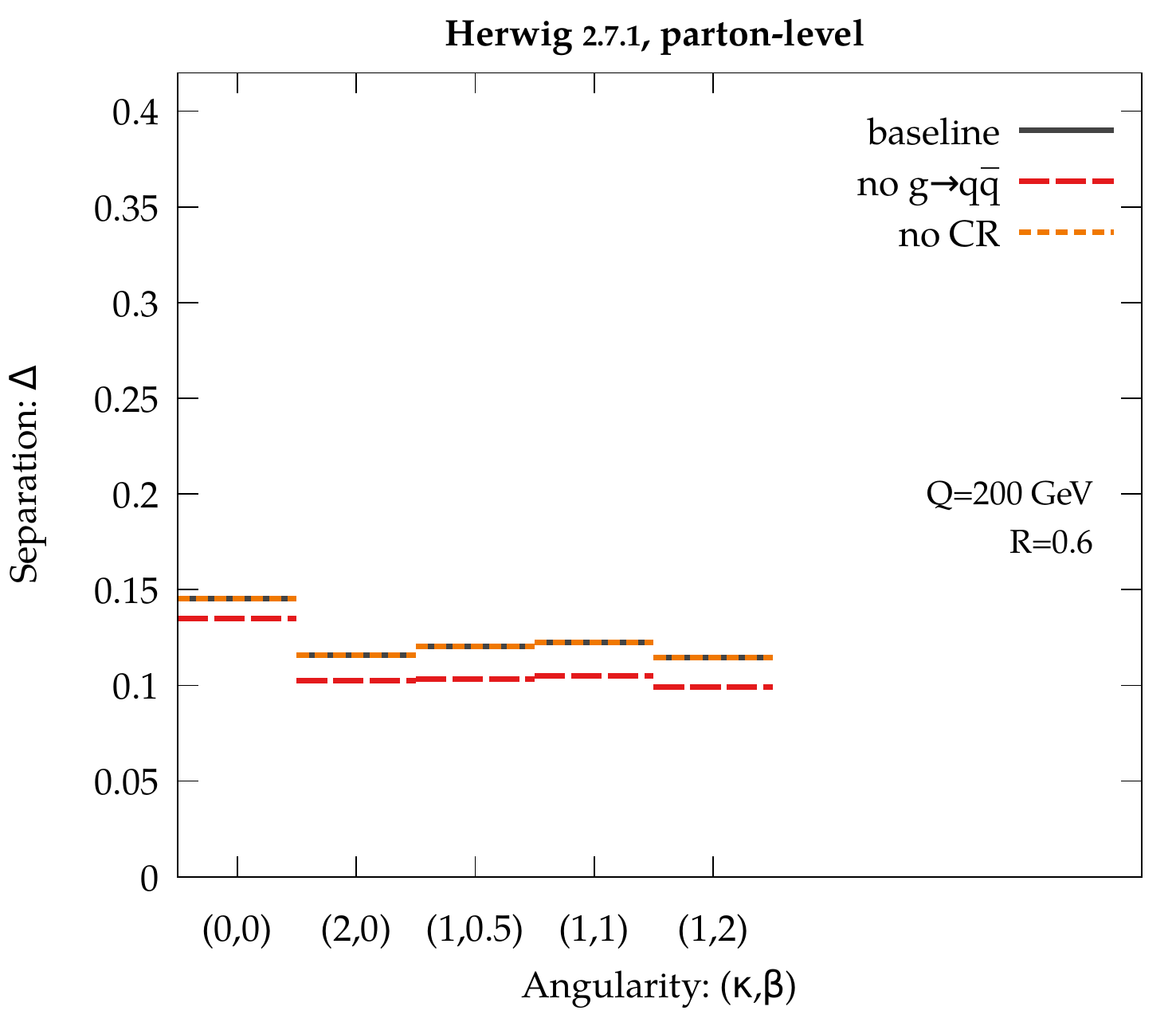}
 }
\caption{Same as Fig.~\ref{quarkgluon_fig:settings_variation_pythia}, but for \textsc{Herwig 2.7.1}.}
\label{quarkgluon_fig:settings_variation_herwig}
\end{figure}

Our \textsc{Herwig++} baseline uses version 2.7.1, with improved modeling of underlying event \cite{Gieseke:2012ft} and the most recent UE-EE-5-MRST tune~\cite{Seymour:2013qka}, which is able to describe the double-parton scattering cross section~\cite{Bahr:2013gkj} and underlying event data from $\sqrt{s} = 300$~GeV to $\sqrt{s} = 7$~TeV.  In Fig.~\ref{quarkgluon_fig:settings_variation_herwig}, we consider the following  \textsc{Herwig} variations:
\begin{itemize}
\item \textsc{Herwig: no $g \to q\bar{q}$}.  Turning off $g \to q \bar{q}$ splittings in \textsc{Herwig} has the reverse behavior as seen in \textsc{Pythia}, leading to slightly worse discrimination power, though the effect is modest.
\item \textsc{Herwig: no CR}.  The variation turns off color reconnections in \textsc{Herwig}.  This has no effect at parton level, as expected.  At hadron level, this variation for \textsc{Herwig} gives a rather dramatic improvement in quark/gluon discrimination power.  We think this arises since color reconnection in \textsc{Herwig} allows any color-anticolor pair to reconnect, even if they arose from an initially color octet configuration.  By turning off color reconnection, the gluons look more octet-like, explaining the improvement seen.
\end{itemize}
The importance of color reconnections in \textsc{Herwig} is a big surprise from this study, motivating future detailed studies into which color reconnection models are most realistic when compared to data.  In the future, we also plan to test the default angular-ordered \textsc{Herwig} shower against an alternative dipole shower \cite{Platzer:2011bc}.

\begin{figure}
\centering
\subcaptionbox{\label{quarkgluon_fig:summary_hadron_sherpa}}{
 \includegraphics[width = 0.45\columnwidth]{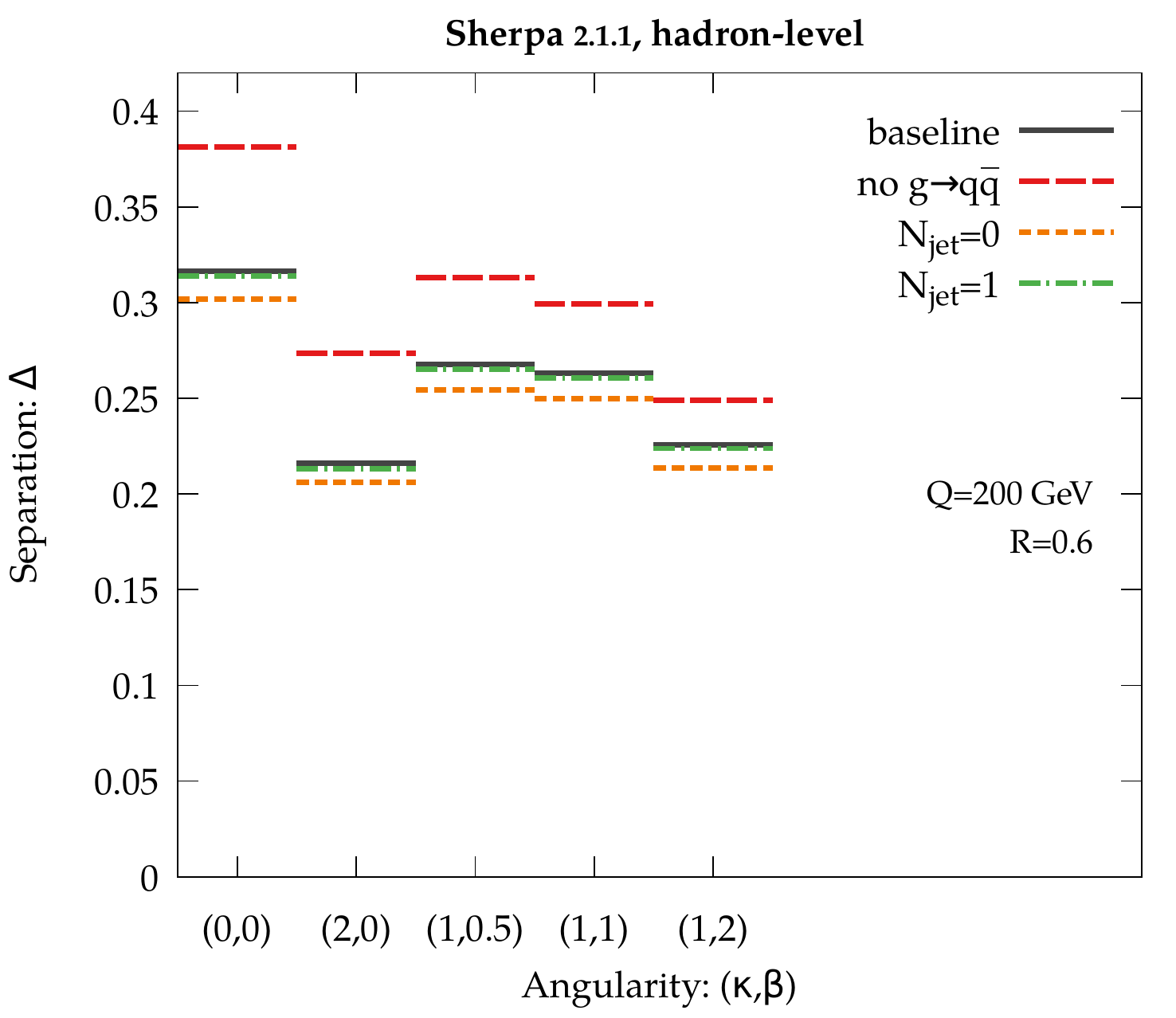}
 }
$\qquad$
\subcaptionbox{\label{quarkgluon_fig:summary_parton_sherpa}}{
 \includegraphics[width = 0.45\columnwidth]{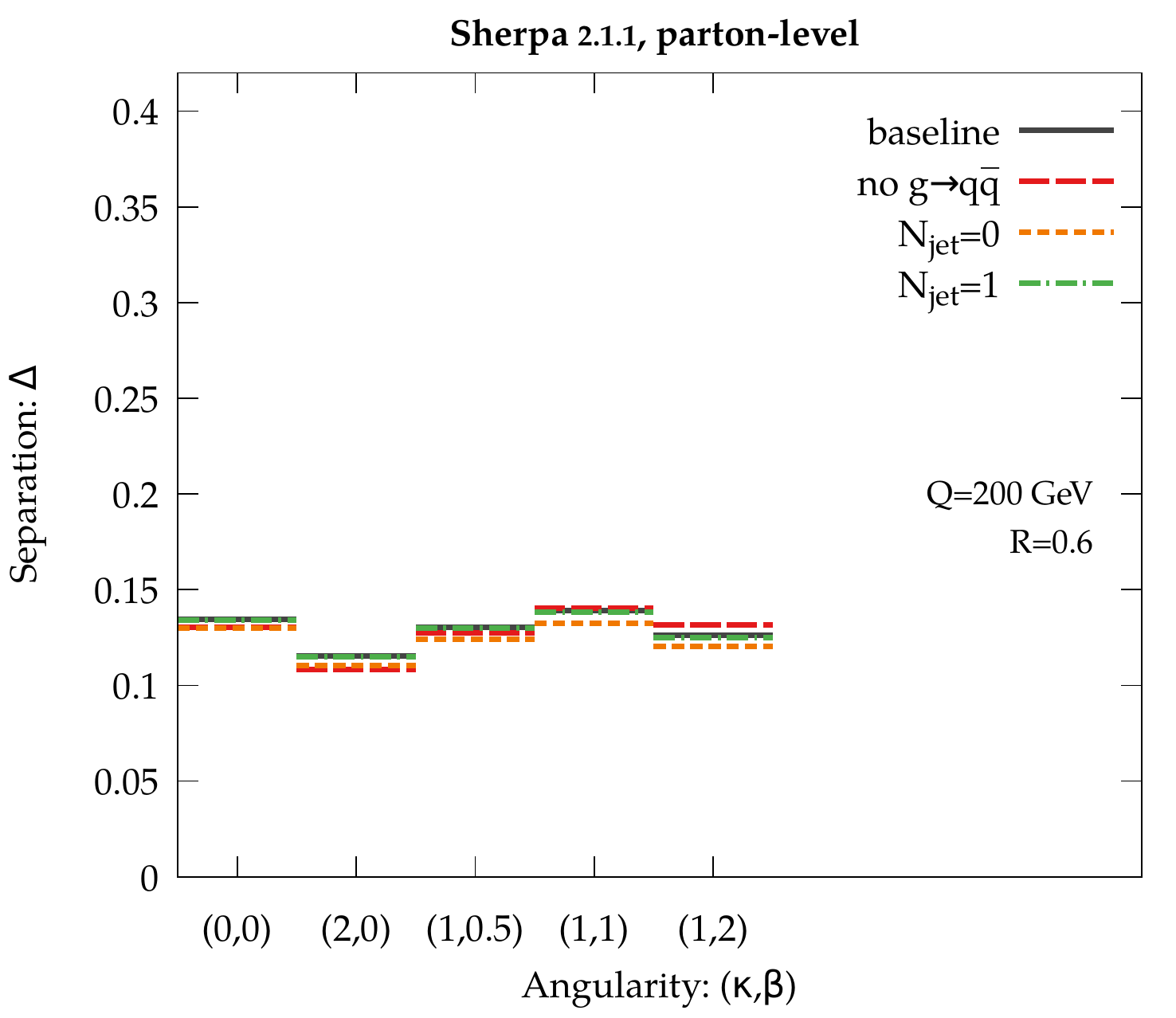}
 }
\caption{Same as Fig.~\ref{quarkgluon_fig:settings_variation_pythia}, but for \textsc{Sherpa 2.1.1}.}
\label{quarkgluon_fig:settings_variation_sherpa}
\end{figure}

Our  \textsc{Sherpa} baseline uses matrix element corrections for the first two emissions ($N_\text{jet} = 2$) with CKKW-style matching \cite{Catani:2001cc}.  In Fig.~\ref{quarkgluon_fig:settings_variation_sherpa}, we consider the following \textsc{Sherpa} variations:
\begin{itemize}
\item \textsc{Sherpa: No $g \to q\bar{q}$}.  Turning off $g \to q \bar{q}$ splittings in \textsc{Sherpa} has a negligible effect at parton level, but it leads to a large jump in discrimination power at hadron level, again due to an interplay between the perturbative shower and nonperturbative hadronization.
\item \textsc{Sherpa: $N_\mathrm{jet} = 1$}.  This variation only performs CKKW matching for the first emission, leading to negligible changes in the discrimination performance.
\item \textsc{Sherpa: $N_\mathrm{jet} = 0$}.  Turning off all matrix element corrections in \textsc{Sherpa} slightly decreases the predicted quark/gluon discrimination power, in agreement with the behavior of \textsc{Pythia}.
\end{itemize}
Within \textsc{Sherpa}, matrix element corrections appear to have a very small effect at parton level.  The large changes seen at hadron level from turning off $g \to q \bar{q}$ splittings motivates further investigations into the shower/hadronization interface.

\begin{figure}
\centering
\subcaptionbox{\label{quarkgluon_fig:summary_hadron_vincia}}{
 \includegraphics[width = 0.45\columnwidth]{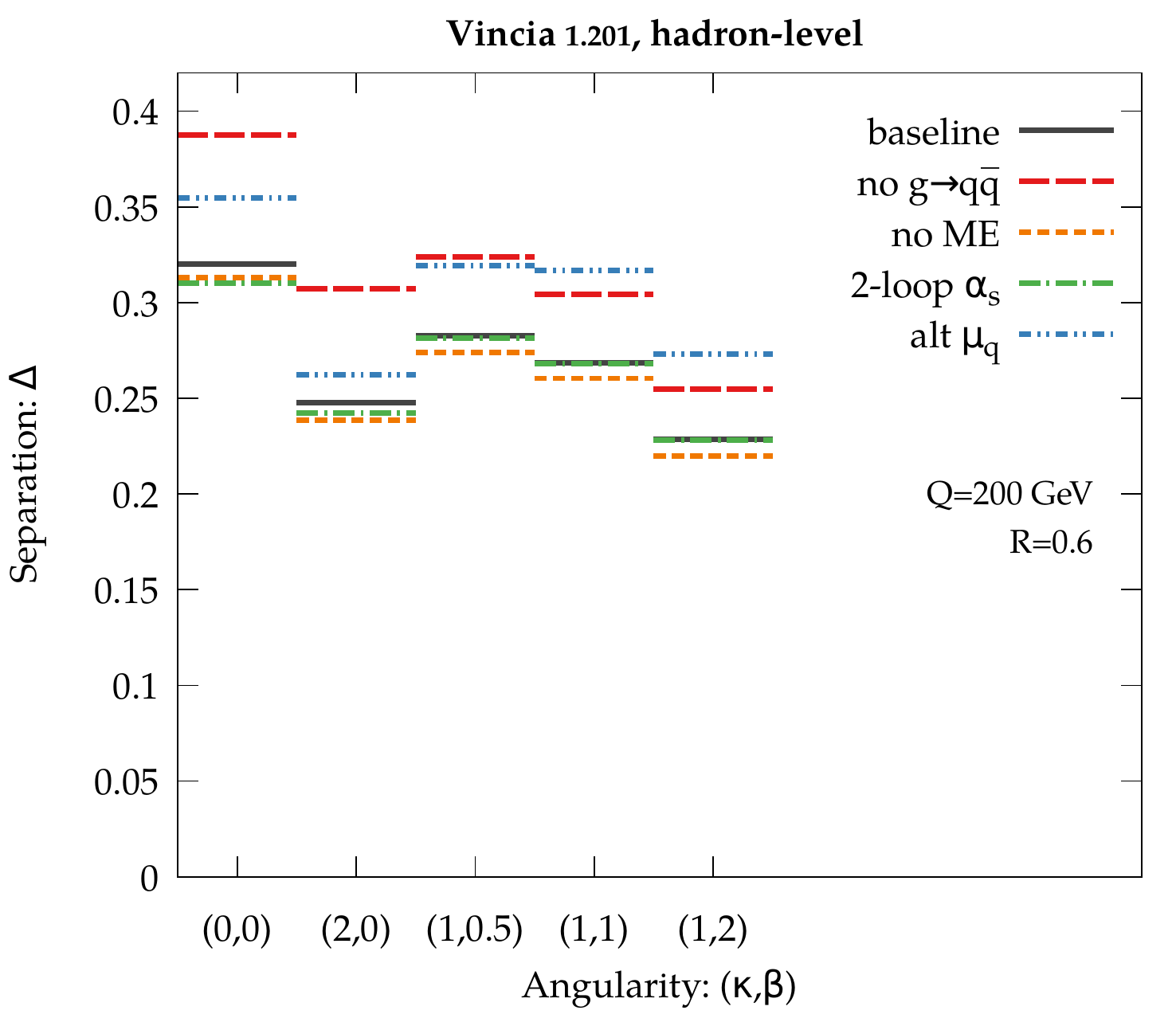} 
 }
$\qquad$
\subcaptionbox{\label{quarkgluon_fig:summary_parton_vincia}}{
 \includegraphics[width = 0.45\columnwidth]{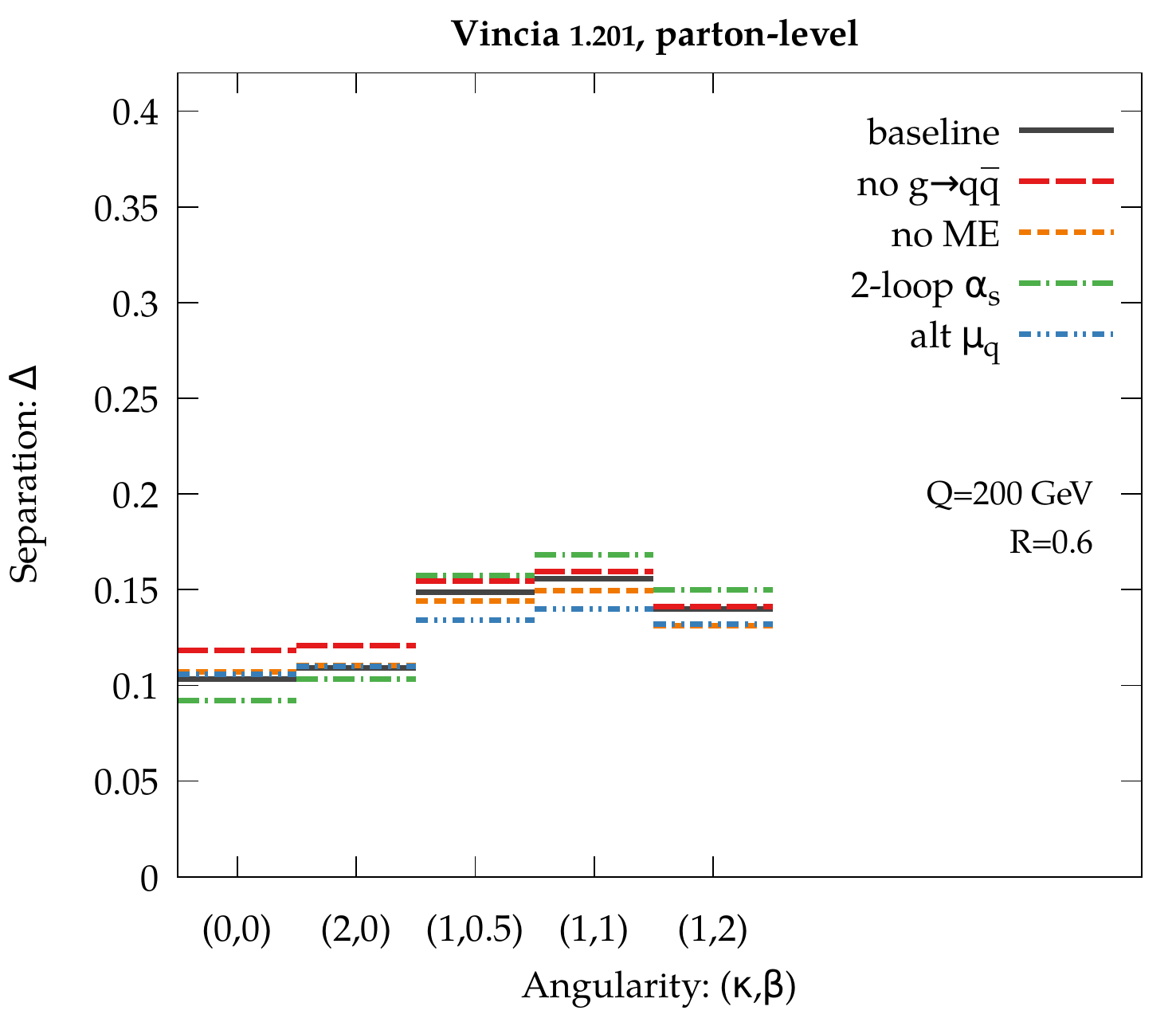}
 }
\caption{Same as Fig.~\ref{quarkgluon_fig:settings_variation_pythia}, but for \textsc{Vincia 1.201}.}
\label{quarkgluon_fig:settings_variation_vincia}
\end{figure}

Our \textsc{Vincia} baseline is based on the ``jeppsson5'' tune.  While \textsc{Vincia} has NLO matrix elements for $e^+ e^- \to q \bar{q}$, it does not have them for $e^+ e^- \to gg$, so we will use LO \textsc{Vincia} throughout.  In Fig.~\ref{quarkgluon_fig:settings_variation_vincia}, we consider the following \textsc{Vincia} variations:
\begin{itemize}
\item \textsc{Vincia:  no $g \to q\bar{q}$}.  This variation turns off $g \to q \bar{q}$, leading to the expected increase in separation power as seen in \textsc{Pythia}.
\item \textsc{Vincia: no ME}.  By default, each $2 \to 3$ antenna in \textsc{Vincia} has an associated matrix element correction factor.  Since the antenna are already rather close to the true matrix elements, turning off these matrix elements has a modest effect on quark/gluon discrimination power.
\item \textsc{Vincia: 2-loop $\alpha_s$}.  Like for \textsc{Pythia}, this variation switches from 1-loop to 2-loop $\alpha_s$ running, yielding a modest parton-level difference and almost no hadron-level difference. 
\item \textsc{Vincia: alt $\mu_q$}.  By default, the $\alpha_s$ scale used in \textsc{Vincia} is determined by the transverse momentum of the corresponding antenna.  In this variation, the $\alpha_s$ scale is set to half the invariant mass of the mother antenna.  This slightly decreases the discrimination power at hadron level, but increases the discrimination power at parton level, again showing the complicated interplay of perturbative and nonperturbative effects.
\end{itemize}
Since \textsc{Vincia} and \textsc{Pythia} share the same underlying engine, it is not surprising that they exhibit similar behaviors as parameters are changed.  The biggest surprise is the way that changing the $\alpha_s$ scale for the antenna can lead to different trends at parton and hadron level.

Our \textsc{Deductor} baseline uses leading color plus (LC+) showering, which includes some subleading color structures. We find that switching from LC+ to LC showering at parton level has a negligible impact on quark/gluon discrimination power.  When \textsc{Deductor} interfaces with the default tune of \textsc{Pythia 8.212} for hadronization, only leading color is used in the showering, such that partons with their LC color information can be directly passed to the Lund string model.  No \textsc{Deductor} variations are shown here, but we anticipate studying the effect of $g \to q \bar{q}$ splitting in future work.

\begin{figure}
\centering
\subcaptionbox{\label{quarkgluon_fig:summary_hadron_ariadne}}{
 \includegraphics[width = 0.45\columnwidth]{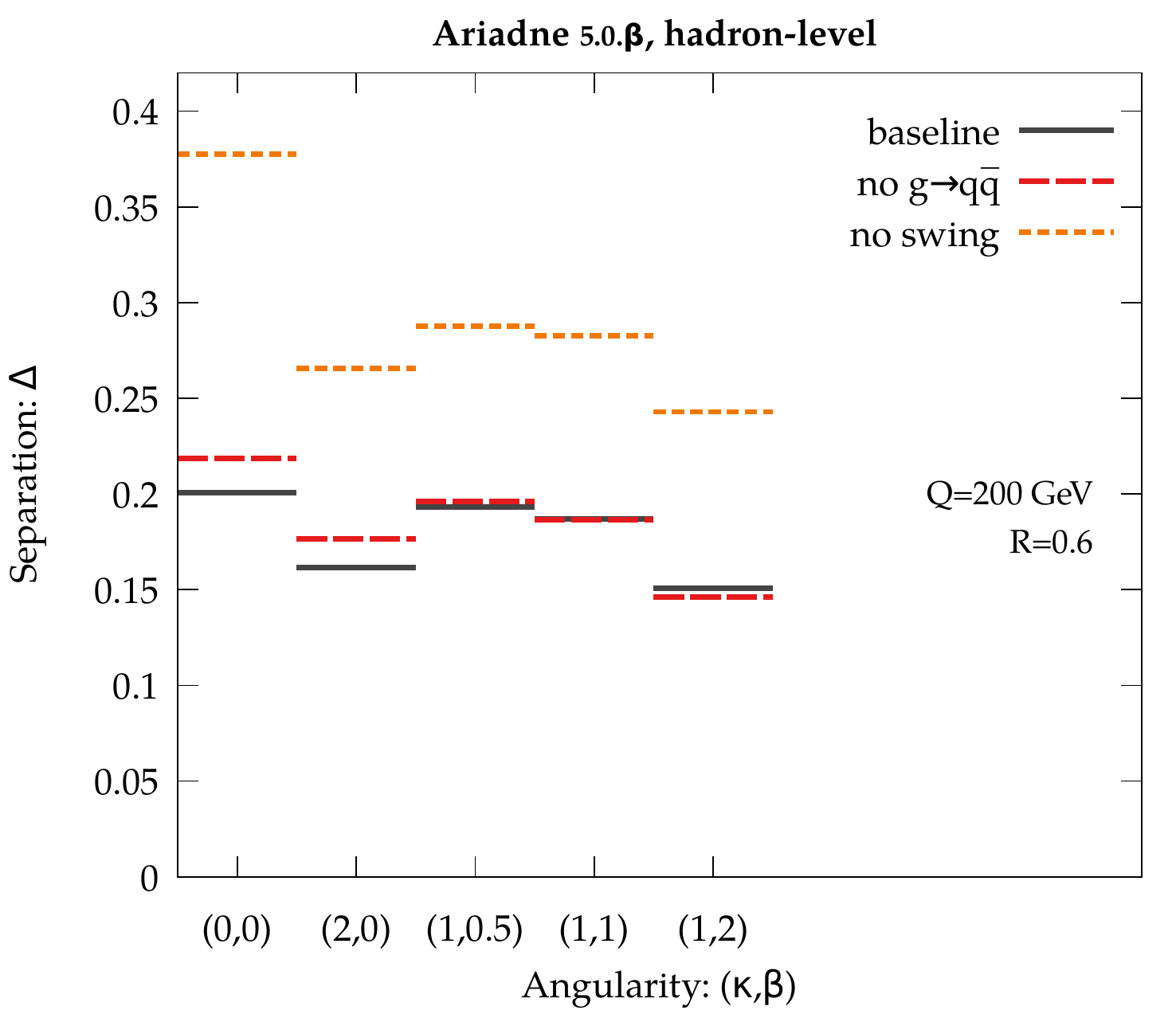}
 }
$\qquad$
\subcaptionbox{\label{quarkgluon_fig:summary_parton_adriadne}}{
 \includegraphics[width = 0.45\columnwidth]{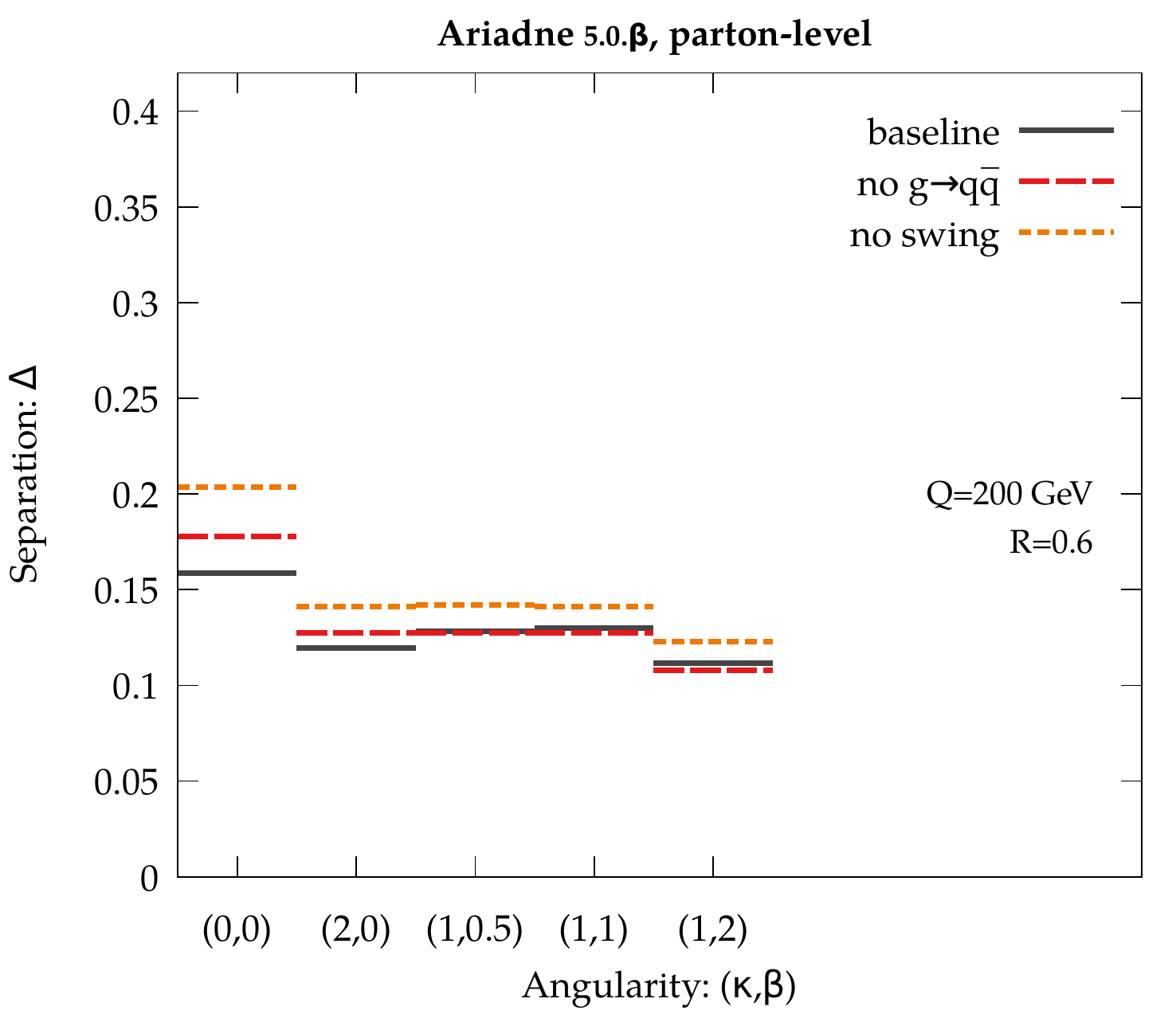}
 }
\caption{Same as Fig.~\ref{quarkgluon_fig:settings_variation_pythia}, but for \textsc{Ariadne 5.0.$\beta$}.}
\label{quarkgluon_fig:settings_variation_ariadne}
\end{figure}

Finally, our \textsc{Ariadne} baseline is based on a beta release of version 5.  In Fig.~\ref{quarkgluon_fig:settings_variation_ariadne}, we consider the following \textsc{Ariadne} variation:
\begin{itemize}
\item \textsc{Ariadne:  no $g \to q\bar{q}$}.  This variation turns off $g \to q \bar{q}$, leading to modest change in separation power, similar in magnitude to \textsc{Herwig}.
\item \textsc{Ariadne: no swing}.  Swing refers to color reconnections performed during the perturbative cascade, where dipoles in the same color state are allowed to reconnect in a way which prefers low-mass dipoles  \cite{Flensburg:2011kk,Bierlich:2014xba}.  Turning off swing has an effect already at parton level, which is amplified at hadron level, leading to improved quark/gluon separation.
\end{itemize}
Like for \textsc{Pythia} and \textsc{Herwig}, color reconnections play a surprisingly important role in \textsc{Ariadne}.

\subsection{Looking towards the LHC}
\label{quarkgluon_sec:pp}

It is clear from our $e^+e^-$ study that quark/gluon radiation patterns face considerable theoretical uncertainties, as seen from the differing behaviors of parton shower generators.  This is true even accounting only for final state physics effects, so additional initial state complications can only increase the uncertainties faced in $pp$ collisions.  Beyond just the application to quark/gluon tagging, this is an important challenge for any analysis that uses jets.  For example, a proper experimental determination of jet energy scale corrections requires robust parton shower tools that correctly model effects like out-of-cone radiation.  Eventually, one would like to perform improved analytic calculations to address these radiation pattern uncertainties.  In the near term, though, measurements from the LHC will be essential for improving the parton shower modeling of jets.

\begin{figure}
\centering
\includegraphics[width = 0.45\columnwidth]{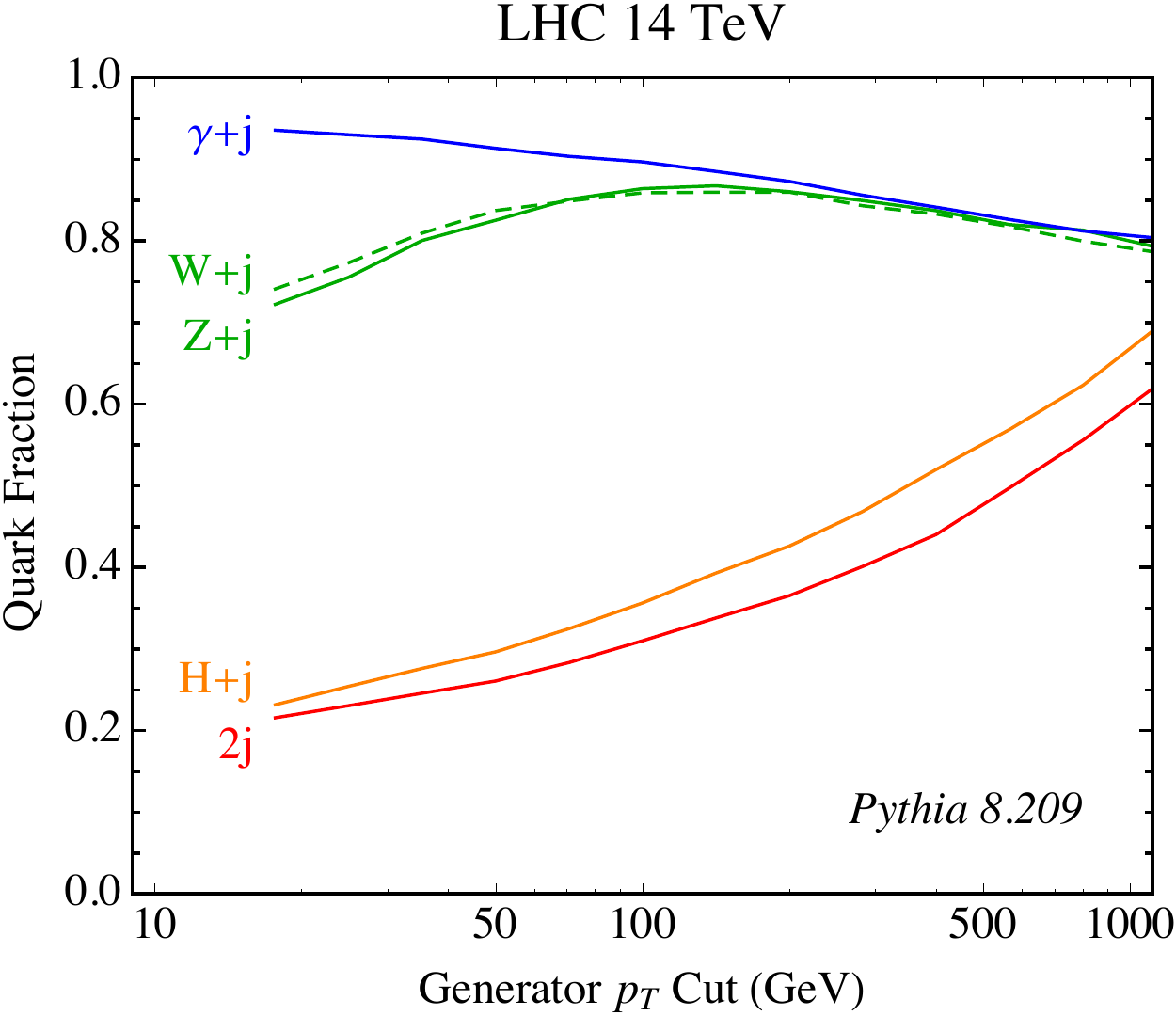}
\caption{Quark fraction of jets at parton level, as defined by the Born-level parton flavor.}
\label{quarkgluon_fig:parton_level_qg_composition}
\end{figure}

At the LHC, there is no way to isolate pure samples of quark or gluon jets, but one can isolate quark/gluon-enriched samples, as defined by the flavor label of the jet in the corresponding Born-level partonic process.  As shown in Fig.~\ref{quarkgluon_fig:parton_level_qg_composition}, the Born-level jet in $W/Z/\gamma + \text{jet}$ is more than 70\% quark enriched over the entire jet $p_T$ range of interest.  For jets softer than around 200 GeV, the Born-level jet in dijets or $H+\text{jet}$ is more than 60\% gluon enriched, with that fraction decreasing as the jet $p_T$ increases.  More sophisticated enrichment procedures are described in Ref.~\cite{Gallicchio:2011xc}.

In principle, one could try to ``diagonalize" some combination of vector boson plus jet and dijet samples in order to define separate quark or gluon samples (see e.g.~\cite{Aad:2014gea}).  In the spirit of Sec.~\ref{quarkgluon_sec:def}, though, we think it is more beneficial for the LHC experiments to perform process-specific measurements without trying to directly determine their quark and gluon composition.  For example, instead of quark/gluon separation, one can ask the more well-defined question about whether one can tell ``the jet in $Z$ plus jets" (quark-enriched) apart from ``the jet in dijets" (gluon-enriched).  Similarly, one could test for differences within a jet sample, such as comparing central jets versus forward jets in dijet production.  This process-based strategy is also helpful for sidestepping the known process dependance of defining quarks and gluons at the LHC, where color correlations to the rest of the event impede a universal definition of quark and gluon jets.

Already, one would learn a lot from unfolded measurements at the LHC  of the generalized angularities in a variety of quark/gluon enriched samples.  Even detector-level measurement comparing a wide range of generators (and generator modifications) would help in understanding how to improve quark/gluon modeling.   The five benchmark angularities in Eq.~\eqref{quarkgluon_eq:benchmarkang} probe both the perturbative and nonperturbative structure of jets and are therefore a good starting point for a more comprehensive quark/gluon jet shape analysis.  In this spirit, we are encouraged by the track multiplicity study of Ref.~\cite{Aad:2016oit}, though for parton shower tuning is it is important to have measurements not only of jet shape averages but also of the full jet shape probability distributions.

Finally, we think it would be useful to perform LHC jet shape measurements after jet grooming (see e.g.~\cite{Butterworth:2008iy,Ellis:2009su,Ellis:2009me,Krohn:2009th}).  Often jet grooming is described as a pileup mitigation strategy, but even independent of pileup, grooming modifies the observed jet radiation patterns in ways that are interesting from the quark/gluon perspective.  Using techniques from Refs.~\cite{Dasgupta:2013ihk,Larkoski:2014wba}, one can calculate the quark/gluon discrimination power for angularities after jet grooming.  Those calculations generically predict that groomed jet shapes should have reduced quark/gluon discrimination power compared to ungroomed jet shapes.  That said, because techniques like soft drop \cite{Larkoski:2014wba} remove soft radiation from jets, they tend to reduce the process dependence in quark/gluon radiation patterns, and may therefore yield a more robust theoretical definition for quark and gluon jets.

\subsection{Summary and recommendations}
\label{quarkgluon_sec:conclude}

By measuring the substructure of jets, one can gain valuable information about the quark/gluon composition of a jet sample.  The challenge we have identified in this study is that the precise radiation patterns of quark and gluon jets is poorly understood, in the sense that parton shower generators give rather different predictions for absolute quark/gluon discrimination power as well as relative trends as a function of the jet kinematics.  At the moment, analytic calculations are not at a level of accuracy where they can offer a useful guide.  Therefore, LHC measurements are the best near-term strategy to constrain quark/gluon radiation patterns and enable quark/gluon discrimination to become a robust experimental tool.

In terms of specific measurements that should be highest priority for ATLAS and CMS, our study has not revealed a silver bullet.  Rather, all of the generalized angularities studied here showed similar levels of disagreement between generators, so a systematic LHC study of even one observable is likely to offer crucial new information.  What is essential is to make measurements at multiple jet $p_T$ scales with multiple jet radii $R$ in multiple different jet samples.  Unfolded distributions would be most useful for parton shower tuning, but even detector-level measurements compared to detector-simulated parton showers would help spot troubling trends.  For the IRC safe angularities in particular, studying the $\beta$ dependence would help separate information about collinear and soft radiation patterns, especially given the fact that the $\beta$ trends seen in the parton shower generators here disagree with those seen in Ref.~\cite{Aad:2014gea}.

If possible, it would be interesting to study the LHA ($\beta = 1/2$) on archival LEP data, since this angularity probes the core of jets in a new way, distinct from broadening-like ($\beta = 1$) or thrust-like ($\beta = 2$) observables.  Among the IRC safe angularities studied here, LHA has the best predicted discrimination power, making it (and other $0<\beta < 1$ angularities) a well-motivated target for future lepton collider measurements.  Similarly, it would be worthwhile to improve our analytic understanding of the LHA.  From Fig.~\ref{quarkgluon_fig:LHA_hadron_separation}, we see that the LHA has discrimination power both at small values of $\lambda_{0.5}^1$ (where non-perturbative corrections play an important role) as well as at larger values of $\lambda_{0.5}^1$ (where fixed-order corrections are important), so one must go beyond an NLL understanding to understand the quark/gluon performance of the LHA.

The key lesson to parton shower authors is that, contrary to some standard lore, existing LEP measurements \emph{do not} constrain all of the relevant aspects of the final state parton shower.  While we have extensive information about quark jet radiation patterns from LEP event shapes, gluon jet radiation patterns are largely unconstrained.  This has important implications for parton shower tuning strategies, since LHC data can and should be used to adjust final state shower parameters.  For example, the ATLAS A14 tune of \textsc{Pythia} has a 10\% lower value of $\alpha_s$ in the final state shower compared to the Monash tune, which yields better agreement with charge particle multiplicity distributions \cite{Aad:2016oit}.  However, A14 has not been tested on LEP event shapes, suggesting that a global tuning strategy is needed.   In addition, it is worth mentioning that similar quark/gluon studies have been carried out in deep inelastic electron-proton scattering \cite{Chekanov:2004kz}, which offer an intermediate step between $pp$ and $e^+ e^-$ collisions, and this $ep$ data could also be valuable for parton shower tuning.

Based on this study, we have identified three aspects of the final state parton shower that deserve closer scrutiny.
\begin{itemize}
\item \textit{Gluon splitting to quarks}.  Some of the largest differences between generators came from turning on and off the $g \to q \overline{q}$ splitting process.  While \textsc{Pythia}, \textsc{Sherpa}, \textsc{Vincia}, and \textsc{Ariadne} suggest that (unphysically) turning off $g \to q \overline{q}$ would improve quark/gluon separation, \textsc{Herwig} (and the analytic calculation in Ref.~\cite{Larkoski:2013eya}) suggests the opposite conclusion.  Beyond quark/gluon discrimination, it would be helpful to identify other contexts where $g \to q \overline{q}$ might play an important role.
\item \textit{Color reconnection in the final state}.  Color reconnection is often thought of as an issue mainly at hadron colliders, but we have seen that it can have an impact in $e^+ e^-$ collisions as well.  This is particularly the case with the default color reconnection model in \textsc{Herwig}, since it allows the reconnection of color/anticolor lines even if they originally came from an octet configuration.  We also saw large changes from \textsc{Pythia: CR1} and  \textsc{Ariadne: no swing}, suggesting that one should revisit color reconnection physics when tuning to LEP data.
\item \textit{Reconsidering $\alpha_s$ defaults}:  In the context of parton shower tuning, the value of $\alpha_s$ used internally within a code need not match the world average value, since higher-order effects not captured by the shower can often be mimicked by adjusting $\alpha_s$.  That said, one has to be careful whether a value of $\alpha_s$ tuned for one process is really appropriate for another.  For example, \textsc{Pythia} uses a relatively large value of $\alpha_s$ in its final state shower, which allows it to match LEP event shape data.  The same value of $\alpha_s$, though, probably also leads to too much radiation within gluon jets.
\end{itemize}
Finally, we want to emphasize that despite the uncertainties currently present in parton shower generators, parton showers in particular (and QCD resummation techniques more generally) will be essential for understanding quark/gluon discrimination.  Fixed-order QCD calculations cannot reliably probe the very soft and very collinear structure of jets, which is precisely where valuable information about quark/gluon radiation patterns reside.  Given the ubiquity and value of parton shower generators, improving the understanding of quark/gluon discrimination will assist every jet study at the LHC.

\subsection*{Acknowledgments}

We thank Samuel Bein, Andy Buckley, Jon Butterworth, Mario Campanelli, Andrew Larkoski, Peter Loch, Ben Nachman, Zoltan Nagy, Chris Pollard, Salvatore Rappoccio, Gavin Salam, Alexander Schmidt, Frank Tackmann, and Wouter Waalewijn for helpful discussions and feedback.
DK acknowledges support from the Science Faculty Research Council, University of Witwatersrand.
The work of LL, SP, and AS is supported in part by the MCnetITN FP7 Marie Curie Initial Training Network, contract PITN-GA-2012-315877.
LL is also support by the Swedish Research Council (contracts 621-2012-2283 and 621-2013-4287).
SP acknowledges support from a FP7 Marie Curie Intra European Fellowship under Grant Agreement PIEF-GA-2013-628739.
The work of DS is supported by the U.S. Department of Energy (DOE) under Grant DE-SC-0011640.
The work of GS is supported by the Paris-Saclay IDEX under the IDEOPTIMALJE grant. 
The work of JT is supported by the DOE under cooperative research agreement DE-SC-00012567, by the DOE Early Career research program DE-SC-0006389, and by a Sloan Research Fellowship from the Alfred P. Sloan Foundation.

\section{Study of associated production of vector bosons and b-jets 
  at the LHC 
  \texorpdfstring{\protect\footnote{
    V.~Ciulli,
    M.~Bell, J.~Butterworth, G.~Hesketh, D.~Grellscheid, F.~Krauss, 
    G.~Luisoni, G.~Nail, D.~Napoletano, C.~Oleari, L.~Perrozzi, S.~Platzer, 
    C.~Reuschle, B.~Waugh
  }}{}}
\label{thisLH_Vbb}

\subsection{Introduction}

The vector boson production in association with one and two b jets at the
CERN Large Hadron Collider is important for many different experimental and
theoretical reasons. Bottom quarks have a peculiar signature which allows one
to easily identify them thanks to a displaced decay vertex. The associated
production with vector bosons is an important backgrounds to VH production
with the Higgs boson decaying to b quarks, and many new physics
searches. Theoretically, it offers an interesting testing ground for
predictions involving heavy quarks.

There are two possible options for the calculation of processes with b-quarks
in the final state at hadron colliders. In the four-flavour scheme (4F)
b-quarks are not present in the parton density of the incident protons. They
can only be generated in the final state and they are usually massive. In the
five-flavour scheme (5F) the b-quark mass is considered small with respect to
the scale of the process $Q$ and powers of logarithms of the type
$\log(Q^2/m_b^2)$ are resummed into a b parton density function. The
b-quark is therefore massless in this approach, though higher order mass
effects can be included in the calculation. A critical review of the
different flavour number schemes and of the status of theoretical
calculations is available in Ref.~\cite{Maltoni:2012pa}. To all orders
in perturbation theory the two approaches give identical results up to
power suppressed mass terms. At finite order, however, they may give
different results. In the 4F scheme the computation is more
complicate, but the full kinematics of the heavy quarks is taken into
account. Furthermore it can be easily interfaced to parton showers, even at
NLO using the \MCatNLO~\cite{Frixione:2002ik} or the
\Powheg~\cite{Nason:2004rx} formalisms.  On the other hand logarithms in the
initial state are not resummed and could lead to large discrepancies in the
inclusive quantities like the total cross-section. In the 5F approach, on the
opposite, calculations for the inclusive quantities are highly simplified and
generally more accurate, but differential distributions and exclusive
observables are technically more involved.

The goal of this study is to compare the most recent measurements with the
predictions of the state of the art generators using 4F and 5F scheme. The
report is organised as follows. In Section \ref{sec:vbb:rivet} we provide a short
description of the ATLAS and CMS measurements, available in the Rivet
framework, for $V+b+X$ and $V+b\bar{b}+X$, where $V$ is a $Z$ or a $W$ boson.
In Section \ref{sec:vbb:generators} we describe the generator setups used to obtain
the predictions, which are compared to the measurements in Section \ref{Zbb}
for the $Z$ and \ref{Wbb} for the $W$, before conclusions are drawn in
Section \ref{sec:vbb:concl}.

\subsection{Rivet Routines}
\label{sec:vbb:rivet}

Results in this study were produced using three Rivet routines to compare to
published ATLAS and CMS data.

\begin{description}
\item[ATLAS Z+b(b)] Measurement of differential production cross-sections for a $Z$\ boson
  in association with b-jets in proton-proton collisions at $\sqrt{s} =
  7$~TeV with the ATLAS detector~\cite{Aad:2014dvb} (Rivet routine
  ATLAS\_2014\_I1306294). A pair of opposite-sign charge dressed
  leptons\footnote{Leptons are dressed by adding the four-vectors of all
    photons within $\Delta R<0.1$\ to the lepton 4-vector} (i.e. electrons or
  muons) with $\pt>20$~GeV and $|\eta|<2.5$ are required, with a dilepton
  mass between 76 and 106~GeV. Anti-k$_{t}$\ 0.4 jets are reconstructed from
  all final state particles, and required to have $\pt>$20~GeV, $|y|<2.4$ and
  not overlap with the leptons used to make the $Z$~candidate ($\Delta R(jet,
  l)> 0.5$). Jets are labelled as $b$-jets based on matching with $\Delta
  R<0.3$ to a weakly decaying $b$-hadron with $\pt>5$~GeV.

  Distributions include the \pt and rapidity of $b$-jets and of the
  $Z$-boson, and for each $b$-jet, the $y_{boost}$ of the $b$-jet and
  $Z$. For events with $Z\, \pt>20$~GeV, the $\Delta R, \Delta\phi$, and
  $\Delta y$\ between the $Z$ and all $b$-jets are plotted. For events with
  at least two $b$-jets, the $\Delta R$\ and di-$b$-jets mass for the two
  leading $b$-jets, along with the $Z$ \pt and rapidity are plotted.


\item[CMS Z+BB] Cross-section and angular correlations in $Z$ boson with b-hadrons
  events at $\sqrt{s} = 7$ TeV~\cite{Chatrchyan:2013zja} (Rivet routine
  CMS\_2013\_I1256943). A pair of opposite-sign charge dressed lepton with
  $\pt>20$~GeV and $|\eta|<2.4$ are required, with dilepton mass between 81
  and 101~GeV. Exactly two weakly decaying b-hadrons with $\pt>15$~GeV and
  $|\eta|<2$\ are then required.

  Distributions include the $Z$\ \pt, the $\Delta R$\ and
  $\Delta\phi$\ between $b$-hadrons, $\Delta R$\ between the $Z$ and closest
  $b$-hadron, and the asymmetry of the $\Delta R$\ between the $Z$ and
  closest $b$-hadron, and the $Z$\ and the furthest $b$-hadron. The angular
  distributions are repeated with a requirement of $Z$\ $\pt>50$~GeV.

\item[ATLAS W+b] Measurement of the cross-section for W boson production in association
  with b-jets in pp collisions at $\sqrt{s} = 7$~TeV with the ATLAS
  detector~\cite{Aad:2013vka} (Rivet routine ATLAS\_2013\_I1219109). A
  dressed lepton with $\pt>25$~GeV and $|\eta|<2.5$\ and a same-flavour
  neutrino with $\pt>25$~GeV are used to form a $W$~candidate, which is
  required to have a transverse mass greater than $60$~GeV. Anti-k$_{t}$\ 0.4
  jets are reconstructed from all final state particles, and required to have
  $\pt>$25~GeV, $|y|<2.1$ and not overlap with the charged lepton used to
  make the $W$~candidate ($\Delta R(jet, l)> 0.5$). Events with more than two
  selected jets are vetoed, and the at least one of the selected jets is
  required to be labelled as $b$-jet, based on matching with $\Delta R<0.3$
  to a weakly decaying $b$-hadron with $\pt>5$~GeV.

  Distributions include the number of jets, and the $b$-jet \pt in events
  containing exactly one or two selected jets.

\end{description}

\subsection{Event generators}
\label{sec:vbb:generators}

\subsubsection*{\protect\Sherpa \label{sec:sherpa}}
In this section we present the setups that are used in this study for the \Sherpa event
generator~\cite{Gleisberg:2008ta}. In particular for Z+b(b) we consider three different
classes of samples: 4F~\MCatNLO, 5F~\MEPS and a 5F~\MEPSatNLO one.
\begin{description}
\item[4F \MCatNLO : ]{ This first set of results is obtained in the
  four-flavour scheme, and based on the \MCatNLO
  technique~\cite{Frixione:2002ik}, as implemented in
  \Sherpa~\cite{Hoeche:2011fd}. In a four-flavour scheme calculation,
  $b$--quarks can only be produced as final state massive particles. They
  are, therefore, completely decoupled from the evolution of the strong
  coupling, $\alpha_S$ and that of the PDFs. In this scheme the associated
  production at tree-level starts from processes such as $jj \to b\bar{b}Z$
  where $j$ can be either a light quark or a gluon. No specific cuts are
  applied on the $b$--quarks, their finite mass regulates collinear
  divergences that would appear in the massless case. In most cases,
  therefore, a $b$-jet actually originates from the parton shower evolution
  and hadronization of a $b$--quark produced by the matrix element.}
\item[5F~\MEPS :]{ In a 5F scheme $b$--quarks are treated as massless
  partons. Collinear logs are resummed into a $b$--PDF and they can appear as
  initial state particles as well as final state ones. In order to account
  for 0 and 1 $b$--jets bins as well as to cure the collinear singularity
  that would arise with a massless final state parton, we use multi-jet
  merging. In \Sherpa, the well-established mechanism for combining into one
  inclusive sample towers of matrix elements with increasing jet multiplicity
  at tree--\-level is the CKKW~\cite{Catani:2001cc}.  For this sample we
  merge together LO samples of $jj \to Z$, $jj \to Z+j$, $jj \to Z+jj$, $jj
  \to Z+jjj$ where now $j$ can be a light quark, a $b$--quark or a gluon, and
  all these samples are further matched to the \Sherpa parton shower
  \CSS~\cite{Schumann:2007mg}.  Merging rests on a jet-criterion, applied to
  the matrix elements.  As a result, jets are being produced by the
  fixed--order matrix elements and further evolved by the parton shower.  As
  a consequence, the jet criterion separating the two regimes is typically
  chosen such that the jets produced by the shower are softer than the jets
  entering the analysis.  This is realised here by a cut-off of $\mu_{\rm
    jet}\,=\, 10 $ GeV.}

\item[5F~\MEPSatNLO : ]{ In this scheme we use the extension to next--to
  leading order matrix elements, in a technique dubbed
  \MEPSatNLO~\cite{Hoeche:2012yf}.  In particular, we merge $jj \to Z$, $jj
  \to Z+j$, $jj \to Z+jj$ calculated with NLO accuracy and we further merge
  this sample with $jj \to Z+jjj$ at the LO.  As in the previous case
  matching criterion has to be chosen, and this is realised by a cut-off of
  $\mu_{\rm jet}\,=\, 10 $ GeV.}
\end{description}
In \Sherpa, tree--\-level cross sections are provided by two matrix element
generators, \Amegic~\cite{Krauss:2001iv} and \Comix~\cite{Gleisberg:2008fv},
which also implement the automated infrared
subtraction~\cite{Gleisberg:2007md} through the Catani--\-Seymour
scheme~\cite{Catani:1996vz,Catani:2002hc}.  For parton showering, the
implementation of~\cite{Schumann:2007mg} is employed with the difference that
for $g\to b\bar{b}$ splittings the invariant mass of the $b\bar{b}$ pair,
instead of their transverse momentum, is being used as scale.  NLO matrix
elements are instead obtained from \OpenLoops~\cite{Cascioli:2011va,
  Cascioli:2014wya}.

The \Sherpa W+b sample is the same one used for the study presented in 
Section~\ref{thisLH_bkgsub}. It is generated with AMEGIC at LO in the
5F scheme, using NNPDF~\cite{Ball:2014uwa} library.
The $b$-quark is massive, with the mass value set to 4.75~GeV, and the
$W$ boson is treated through the narrow width approximation. The order
of the electroweak couplings is fixed to 2, therefore removing single-top
contribution. Multi-parton interactions (MPI) are switched on/off to
estimate this contribution.

\subsubsection*{\Herwig \label{subsubsec:herwigsetup}}

In this section we present the setup for those results obtained with the
\Herwig event generator~\cite{Bellm:2015jjp,Bahr:2008pv}.

Based on extensions of the previously developed \Matchbox
module~\cite{Platzer:2011bc}, \Herwig facilitates the automated setup of all
ingredients necessary for a full NLO QCD calculation in the subtraction
formalism: an implementation of the Catani--Seymour dipole subtraction
method~\cite{Catani:1996vz,Catani:2002hc}, as well as interfaces to a list of
external matrix--element providers -- either at the level of squared matrix
elements, based on extensions of the BLHA
standard~\cite{Binoth:2010xt,Alioli:2013nda,Andersen:2014efa}, or at the
level of color--ordered subamplitudes, where the color bases are provided by
an interface to the \ColorFull~\cite{Sjodahl:2014opa} and
\CVolver~\cite{Platzer:2013fha} libraries.

For this study the relevant tree--level matrix elements are taken from \MGaMC
\cite{Alwall:2014hca,Alwall:2011uj} (via a matrix--element interface at
the level of color--ordered subamplitudes), whereas the relevant
tree--level/one--loop interference terms are provided by
\OpenLoops~\cite{Cascioli:2011va,Cascioli:2014wya} (at the level of squared
matrix elements).

Fully automated NLO matching algorithms are available, henceforth referred to
as subtractive (\textit{NLO$\oplus$}) and multiplicative (\textit{NLO$\otimes$})
matching -- based on the \MCatNLO~\cite{Frixione:2002ik} and
\Powheg~\cite{Nason:2004rx} formalism respectively -- for the systematic and
consistent combination of NLO QCD calculations with both shower variants (the
angular--ordered \textit{QTilde} shower~\cite{Gieseke:2003rz} and the
\textit{Dipole} shower~\cite{Platzer:2009jq}) in \Herwig.

We consider four different classes of samples, for varying combinations of
matching and shower algorithms (a selection of plots can be found in sections
\ref{sec:ZHerwig} and \ref{Wbb}):
\begin{description}
\item[4F, Zbb] For this set we consider the subtractive and multiplicative
  matching together with the \textit{QTilde} shower. The tree--level process of
  the underlying hard sub--process in this case is
  $pp \to e^+ e^- b\bar{b}$. For this sample the $b$ quark is
  considered massive and $p$ only consists of light quarks or a gluon, not a $b$
  quark.
\item[5F, Zbb] For this set we consider the subtractive and multiplicative
  matching together with the \textit{QTilde} and \textit{Dipole} shower. The
  tree--level process of the underlying hard sub--process in this case is
  $pp \to e^+ e^- b\bar{b}$. For this sample the $b$ quark is
  treated as massless, and $p$ may also include a $b$ quark. Generator--level
  cuts on the $b$ quarks have thus been applied. Only in the shower evolution of
  the \textit{QTilde} shower is the $b$ quark assumed massive.
\item[5F, Zb] For this set we consider the subtractive and multiplicative
  matching together with the \textit{QTilde} and \textit{Dipole} shower. The
  tree--level process of the underlying hard sub--process in this case is
  $pp \to e^+ e^- j_b$, where $j_b\ni\{b,\bar{b}\}$. For this sample
  the $b$ quark is treated as massless, and $p$ may also include a $b$ quark.
  Generator--level cuts on the $b$ quark have thus been applied. Only in the
  shower evolution of the \textit{QTilde} shower is the $b$ quark assumed
  massive. For single $b$--quark production only one $p$ must contribute a $b$
  quark at a time, at the level of the hard sub--process at hand.
\item[4F, Wbb] For this set we consider the subtractive and multiplicative
  matching together with the \textit{QTilde} shower. The tree--level process of
  the underlying hard sub--process in this case is
  $pp' \to W b\bar{b} \to l \nu_l b\bar{b}$, where
  $l \in \{e^+, e^-, \mu^+, \mu^-\}$ and $\nu_l$ the associated (anti-)neutrino.
  For this sample the $b$ quark is considered massive and $p,p'$ only consist of
  light quarks or a gluon, not a $b$ quark; $p'$ simply denotes the contribution
  of pair--wise different quark flavours in the intial state, as a result from
  the $Wdu$ or $Wsc$ vertex.
\end{description}

\medskip

In all samples the uncertainty bands are purely from scale variations by
simultaneously varying all scales in the hard sub--process and in the shower by
factors of two up and down, i.e. factorization and renormalization scale in the
hard sub--process, as well as scales related to $\alpha_s$ and the PDFs
in the shower, as well as the hard shower scale. The central scale
choice is always a fixed scale (the $Z$ mass in associated $Z$ production; the
$W$ mass in associated $W$ production).

The PDF sets being used are MMHT2014lo68cl and
MMHT2014nlo68cl~\cite{Harland-Lang:2014zoa}, i.e. the default PDF sets
to which the showers are currently tuned. An internal study showed that using
different PDF sets (a different $n_f\!\!=\!\!5$ PDF set for the 5F runs or
$n_f\!\!=\!\!4$ PDF sets for the 4F runs) results in only minor differences,
within the scale variation uncertainties.

In case the $b$ quark is assumed massive, its mass is set to the default value
in \Herwig. All other relevant parameters, like $W$ and $Z$ mass and width,
etc., are set to their respective default values in \Herwig as well.

For the 5F, Zbb sample we cut on the final state $b$ quarks by including them
into the jet definition and requiring at least two jets,
with a min.~$p^\bot$ of 18~GeV and 15~GeV for the first and second jet
respectively. The statistics for this sample are 100k unweighted events.

For the 5F, Zb sample we cut on the final state $b$ quark (similarly to above)
by requiring at least one jet, with a min.~$p^\bot$ of 18~GeV for the first jet.
The statistics for this sample are 100k unweighted events.

For both 5F samples we apply generator level cuts on the invariant mass of the
charged--lepton pair, with a min.~invariant mass of 60~GeV and a max.~invariant
mass of 120~GeV.

For the 4F, Zbb sample we apply no cuts on the $b$ quarks. However, we require
the same generator level cuts as for the 5F, Zbb and Zb samples on the invariant
mass of the charged--lepton pair again. In addition we cut slightly on the
charged leptons, with a min.~$p^\bot$ of 5~GeV and a rapidity range between -4
and 4. The statistics for this sample are 100k unweighted events.

For the 4F, Wbb sample we also apply no cuts on the $b$ quarks. However, we
require a slight generator level cut on the transverse mass of the $W$, with a
min.~transverse mass of 20~GeV. We also cut slightly on the charged lepton, with
a min.~$p^\bot$ of 5~GeV and a rapidity range between -4 and 4. The statistics
for this sample are 100k unweighted events.

\subsubsection*{\POWHEGBOX \label{sec:powheg}}

The results obtained with the \POWHEGBOX{} framework are based on the
generators presented in ref.~\cite{Luisoni:2015mpa}. The tree-level
amplitudes, which include Born, real, spin- and colour-correlated Born
amplitudes, were automatically generated using an
interface~\cite{Campbell:2012am} to
\MadGraphfour{}~\cite{Stelzer:1994ta,Alwall:2007st}, whereas the
one-loop amplitudes were generated with
\GOSAM{}~\cite{Cullen:2011ac,Cullen:2014yla} via the
Binoth-Les-Houches (BLHA)
interface~\cite{Binoth:2010xt,Alioli:2013nda}, presented for the
\POWHEGBOX{} and \GOSAM{} in~\cite{Luisoni:2013cuh}.  The version
of \GOSAM{} is the 2.0: it uses
\QGRAF{}~\cite{Nogueira:1991ex}, \FORM~\cite{Kuipers:2012rf} and
\SPINNEY{}~\cite{Cullen:2010jv} for the generation of the Feynman
diagrams. These diagrams are then computed at running time with
\NINJA{}~\cite{vanDeurzen:2013saa,Peraro:2014cba}, which is a
reduction program based on the Laurent expansion of the
integrand~\cite{Mastrolia:2012bu}, and using
\ONELOOP~\cite{vanHameren:2010cp} for the evaluation of the scalar
one-loop integrals. For unstable phase-space points, the reduction
automatically switches to \GOLEM{}~\cite{Cullen:2011kv}, that allows
to compute the same one-loop amplitude evaluating tensor integrals.

Further details can be found in ref.~\cite{Luisoni:2015mpa}. Here we
briefly summarize the most important features.
\begin{enumerate}
\item We have used a mixed renormalization scheme~\cite{Collins:1978wz},
  generally known as decoupling scheme, in which the $n_{\rm lf}$ light flavours
  are subtracted in the usual \MSB{} scheme, while the heavy-flavour loop is
  subtracted at zero momentum. In this scheme, the heavy flavour decouples at
  low energies. To make contact with other results expressed in terms of the
  $\MSB$ strong coupling constant, running with 5 light flavours, and with
  pdfs with 5 flavours, we have switched our scheme using the procedure
  discussed in ref.~\cite{Cacciari:1998it}.

\item We have generated $Wb\bar{b}j$ events using the
  \MINLO~\cite{Hamilton:2012np} prescription, that attaches a suitable
  Sudakov form factor to the $Wb\bar{b}j$ cross section at NLO, and subtracts
  its expansion (not to have double counting of the Sudakov logarithms), in
  order to get a finite cross section down to small transverse momentum of
  the hardest jet.  The scales of the primary process (i.e.~the process
  obtained by the attempt to cluster a $Wb\bar{b}j$ event with a procedure
  similar to CKKW~\cite{Catani:2001cc}) have been chosen as follows:

  \vspace{2mm}
  \begin{enumerate}
    \item if there has been a clusterization, then the scales are set to
      \begin{equation}
        \label{eq:Wbb_scales}
        \mur=\muf= \mu\equiv\frac{\sqrt{\hat{s}}}{4}\,, \qquad\hat{s} = (p_{\sss\rm W}
        + p_{\sss\rm b}+p_{\sss\rm \bar b})^2,
      \end{equation}
      where $p_{\sss \rm W}$, $p_{\sss \rm b}$ and $p_{\sss\rm \bar b}$
      are the momenta of the $W$, $b$ and $\bar{b}$ in the primary process
    \item If the event has not been clustered by the \MINLO{} procedure,
      i.e.~if the underlying Born $Wb\bar{b}j$ process is not clustered by
      \MINLO{}, we take as scale the partonic center-of-mass energy of the
      event.
  \end{enumerate}
  \vspace{2mm}

The bands in the plots of Figures~\ref{fig:wbb-njet}
and~\ref{fig:wbb-pt} of this section are the envelope of the
distributions obtained by varying the renormalization and factorization
scales by a factor of 2 around the reference scale $\mu$ of
eq.~(\ref{eq:Wbb_scales}), i.e.~by multiplying the factorization and the
renormalization scale by the scale factors $\KFA$ and $\KRA$, respectively,
where
\begin{equation}
\label{eq:KRA_KFA}
(\KRA,\KFA)=(0.5,0.5),  (0.5,1), (1,0.5), (1,1),(2,1),(1,2),(2,2).
\end{equation}
These variations have been computed using the \POWHEGBOX{} reweighting
procedure, that recomputes the weight associated with an event in a fast way.
\end{enumerate}

\clearpage

\subsection{Z+b(b) production \label{Zbb}}

\subsubsection*{Z+b(b) with \Sherpa}

\begin{figure}[htbp]
\begin{center}
   \includegraphics[scale=0.65]{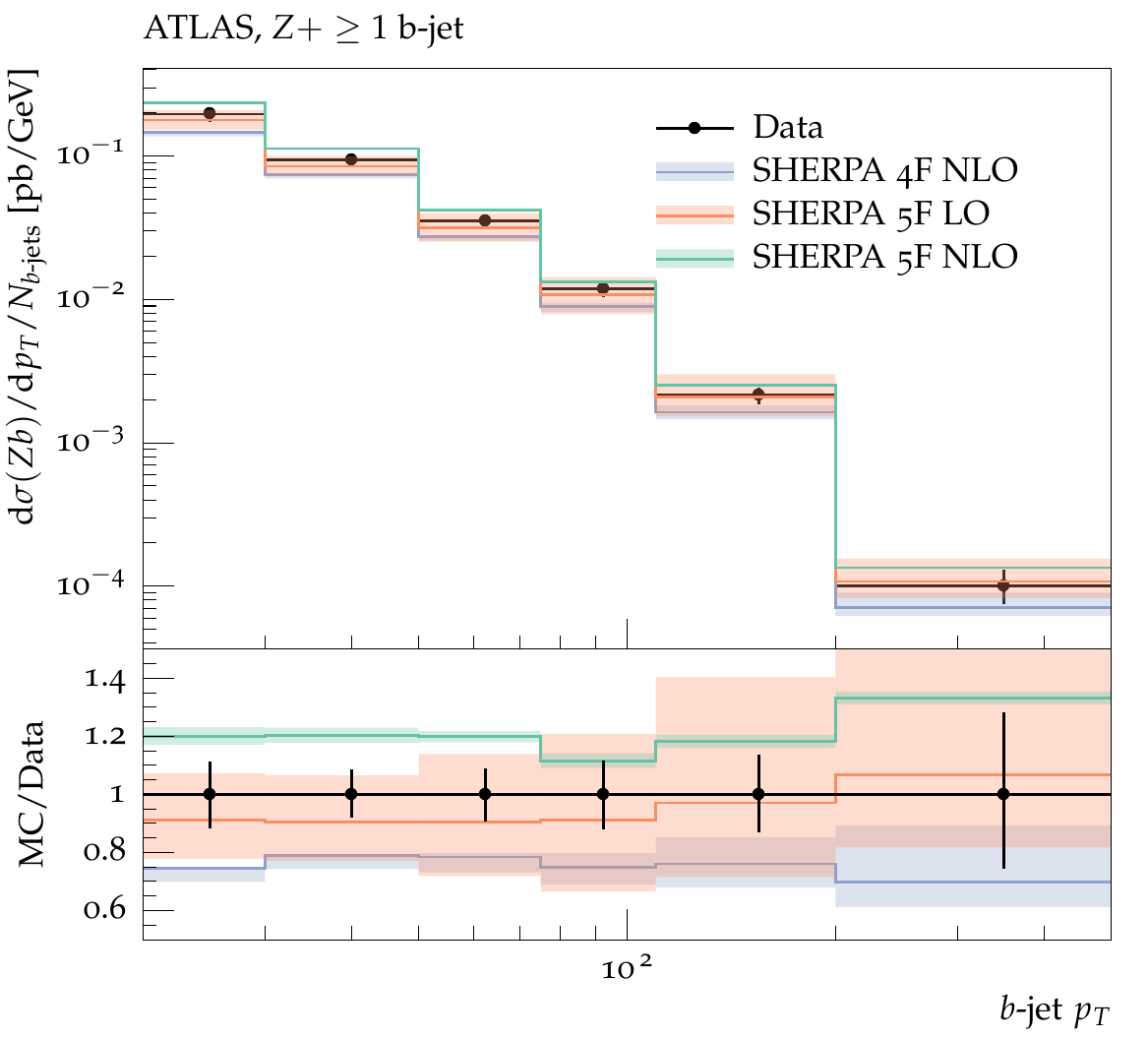}
   \includegraphics[scale=0.65]{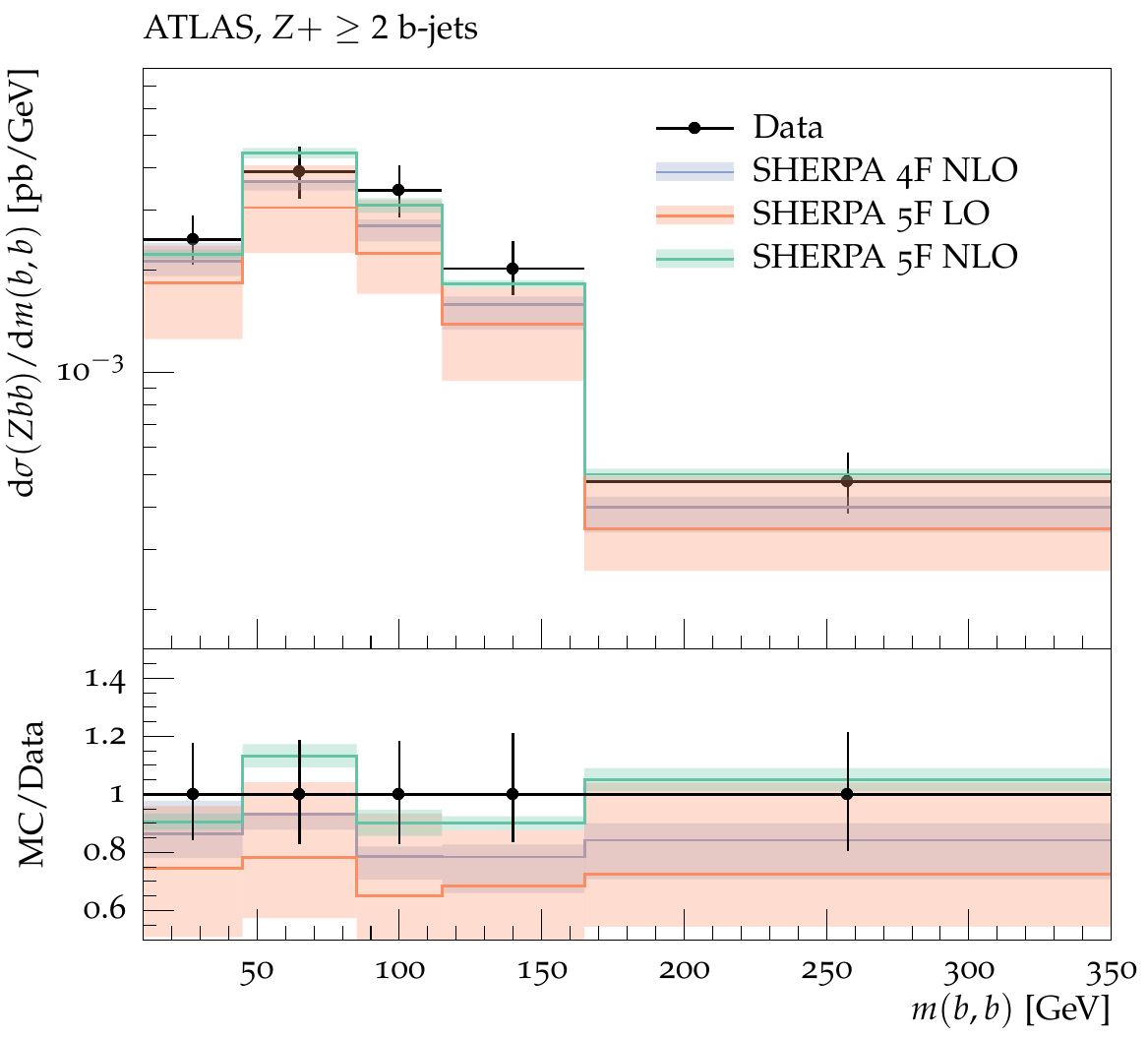} \\
   \includegraphics[scale=0.65]{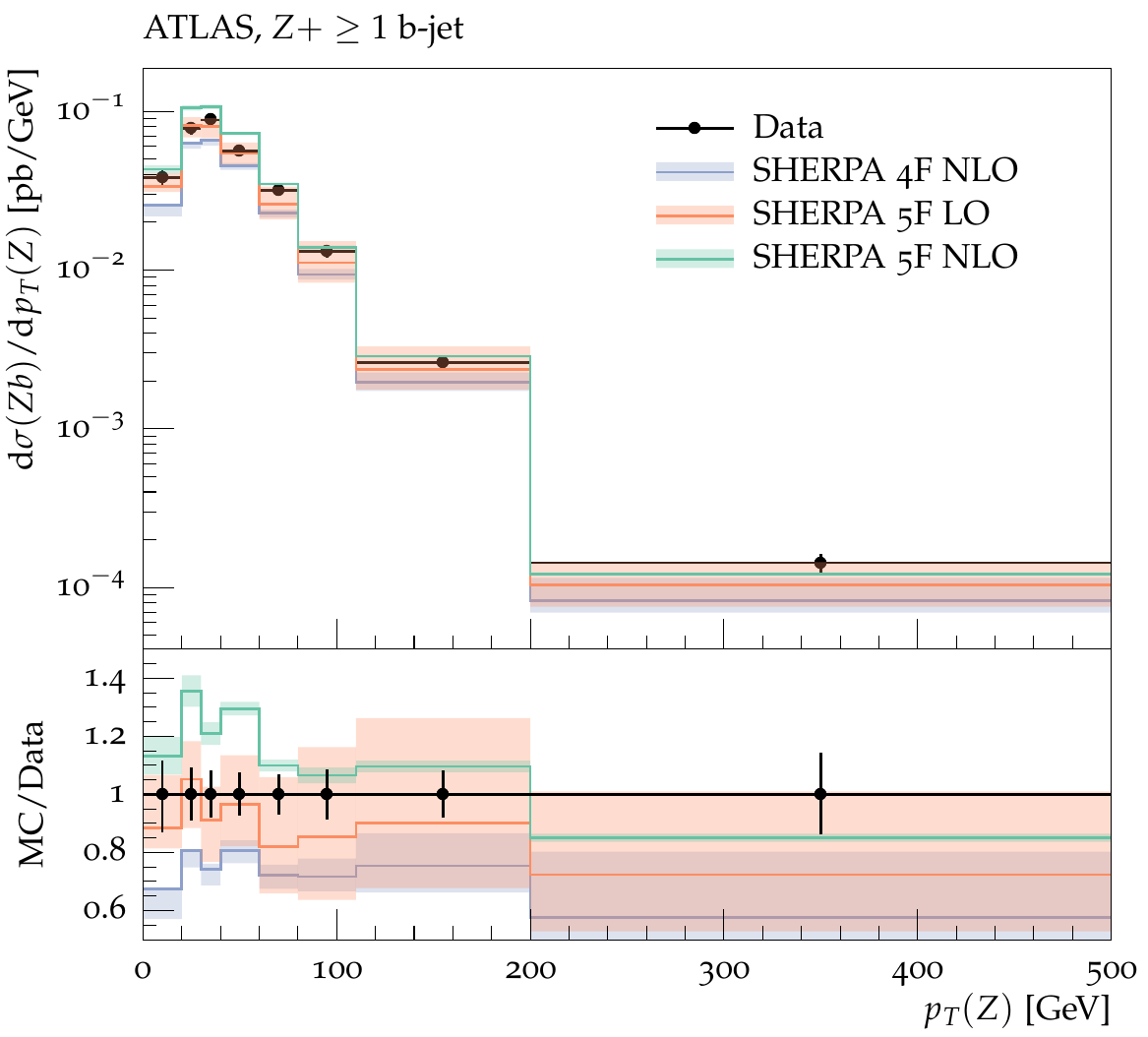}
   \includegraphics[scale=0.65]{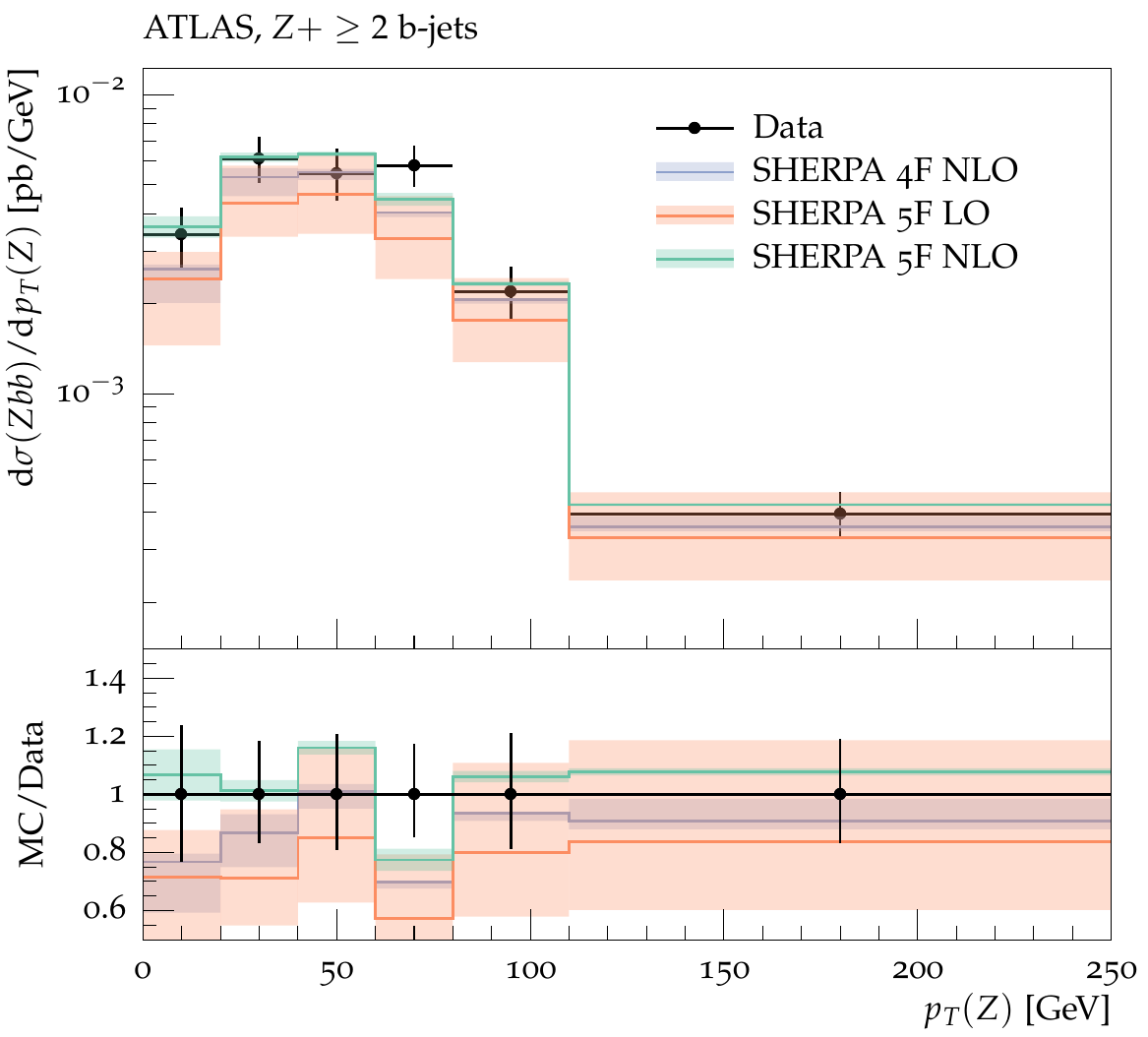}
\caption{A selection of the plots comparing Sherpa predictions to ATLAS results.}
\label{zbb-sherpa-atlas}
\end{center}
\end{figure}

\begin{figure}[htbp]
   \includegraphics[scale=0.65]{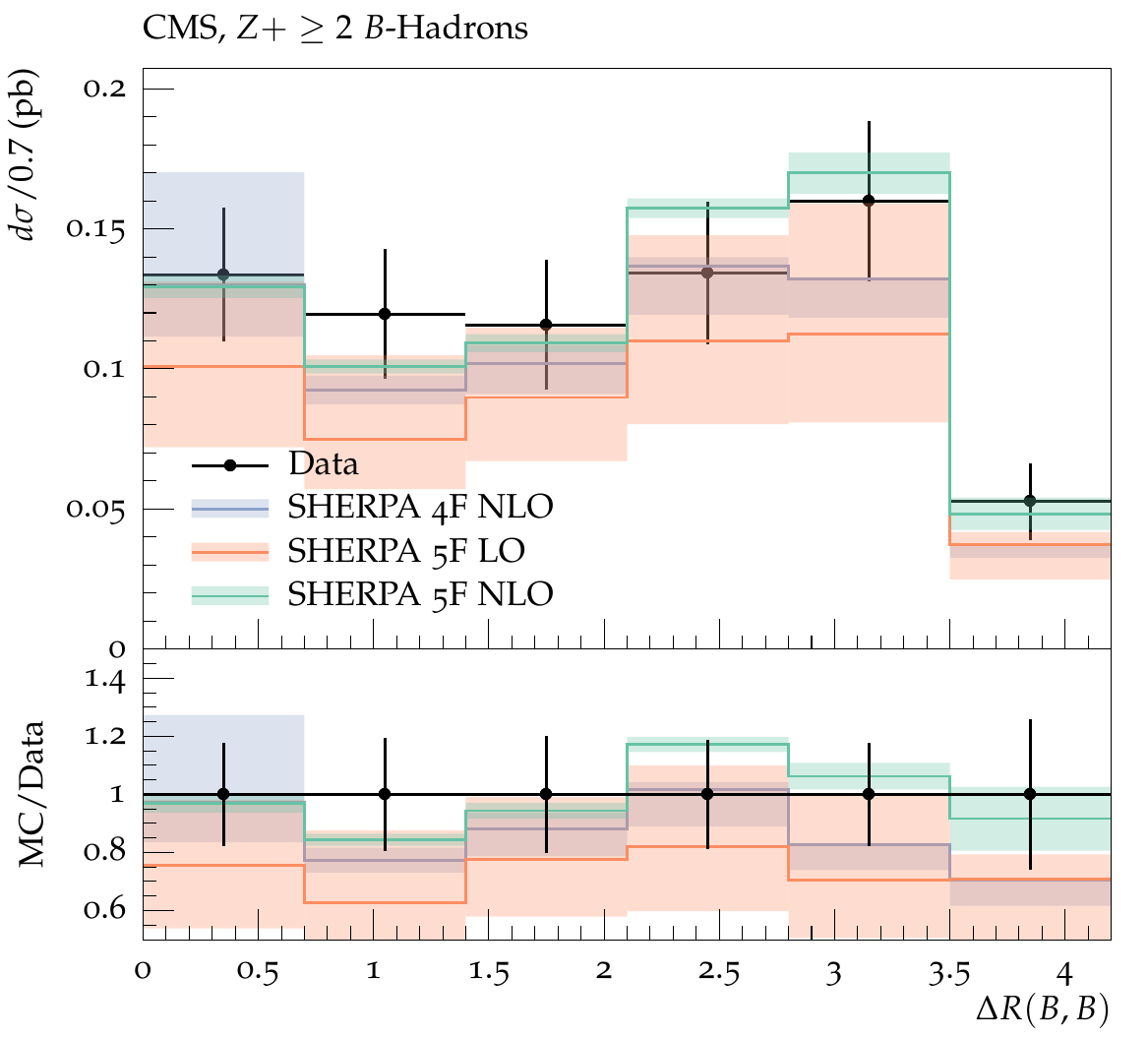}
   \includegraphics[scale=0.65]{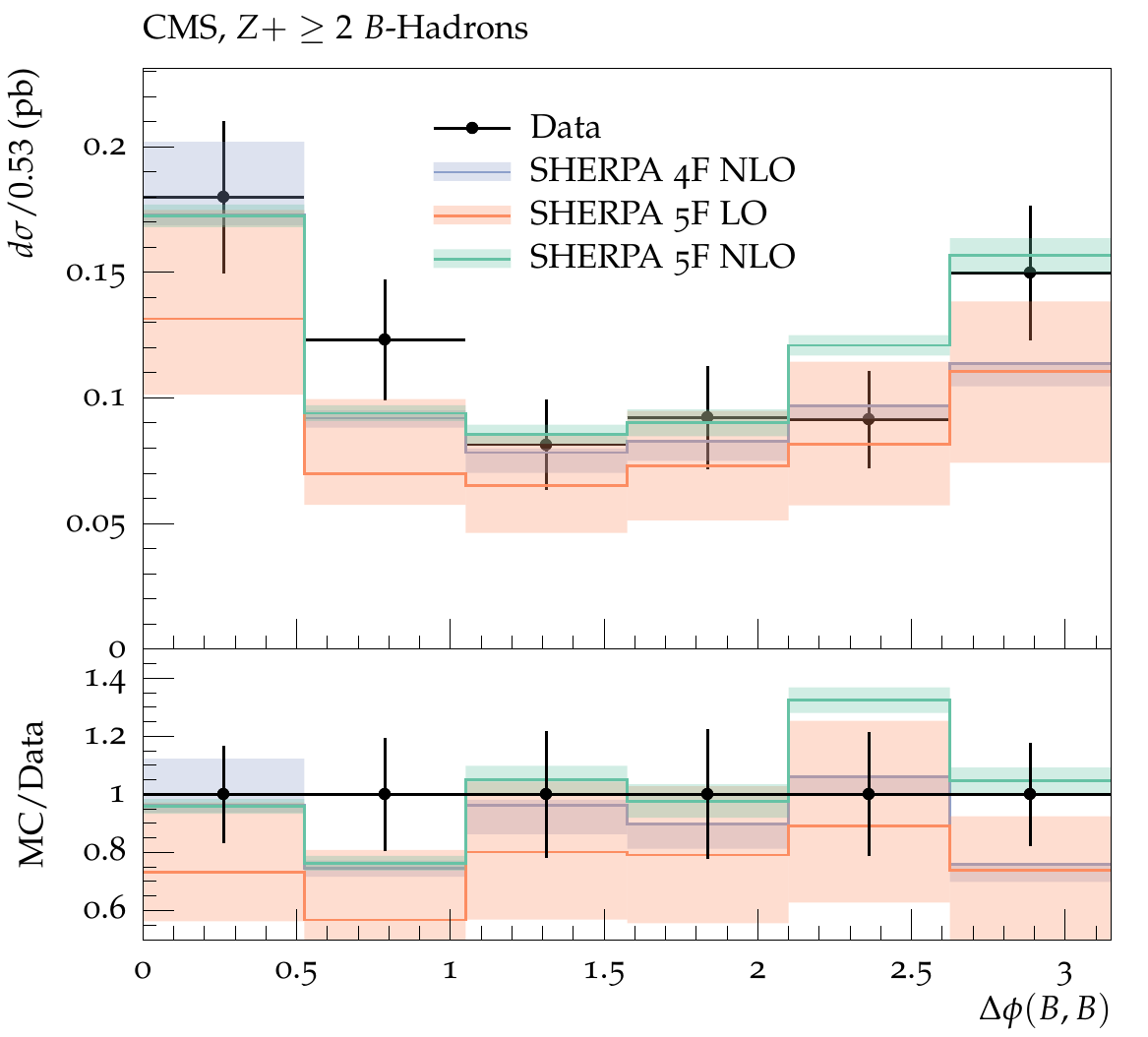}
\caption{A selection of the plots comparing Sherpa predictions to CMS results.}
\label{zbb-sherpa-cms}
\end{figure}

Figures~\ref{zbb-sherpa-atlas} and ~\ref{zbb-sherpa-cms} show a selection of
the plots comparing Sherpa predictions to data.  There is overall a good
agreement, but for the normalization. The 5F LO order predictions are
generally below the data, though compatible within the large scale
uncertainty. For NLO predictions this uncertainty is smaller and some
patterns can be observed. Both the 5F and the 4F NLO are in good agreement
with distributions for events with two b-tagged jets.  But when a single
b-jet is tagged, the 5F and 4F results have an opposite behaviour: the 5F is
~20\% above the data (except for high $Z$ \pt), while 4F is ~20\% below.

\begin{figure}[htbp]
\begin{center}
   \includegraphics[scale=0.65]{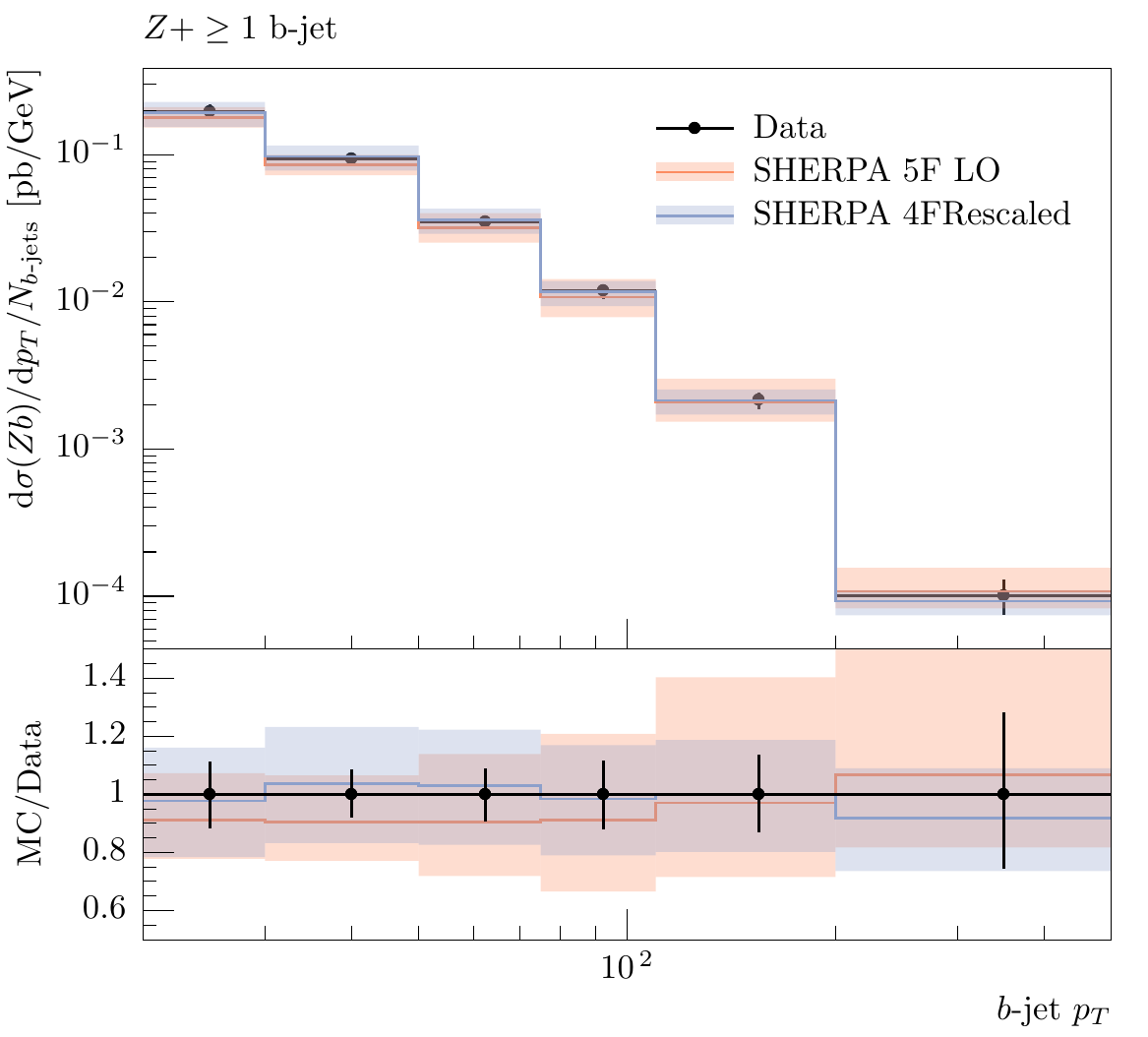}
   \includegraphics[scale=0.65]{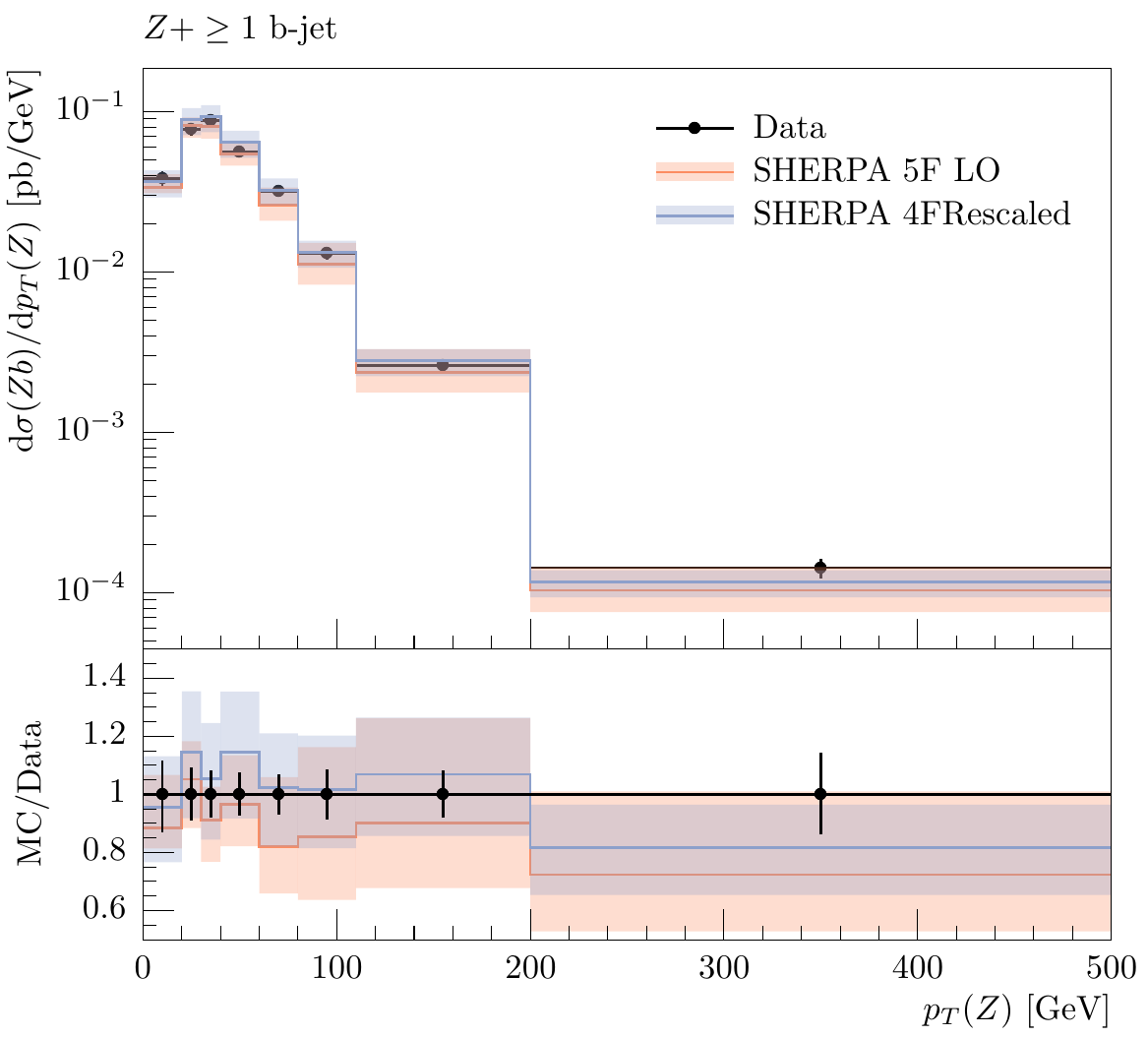} \\
   \includegraphics[scale=0.65]{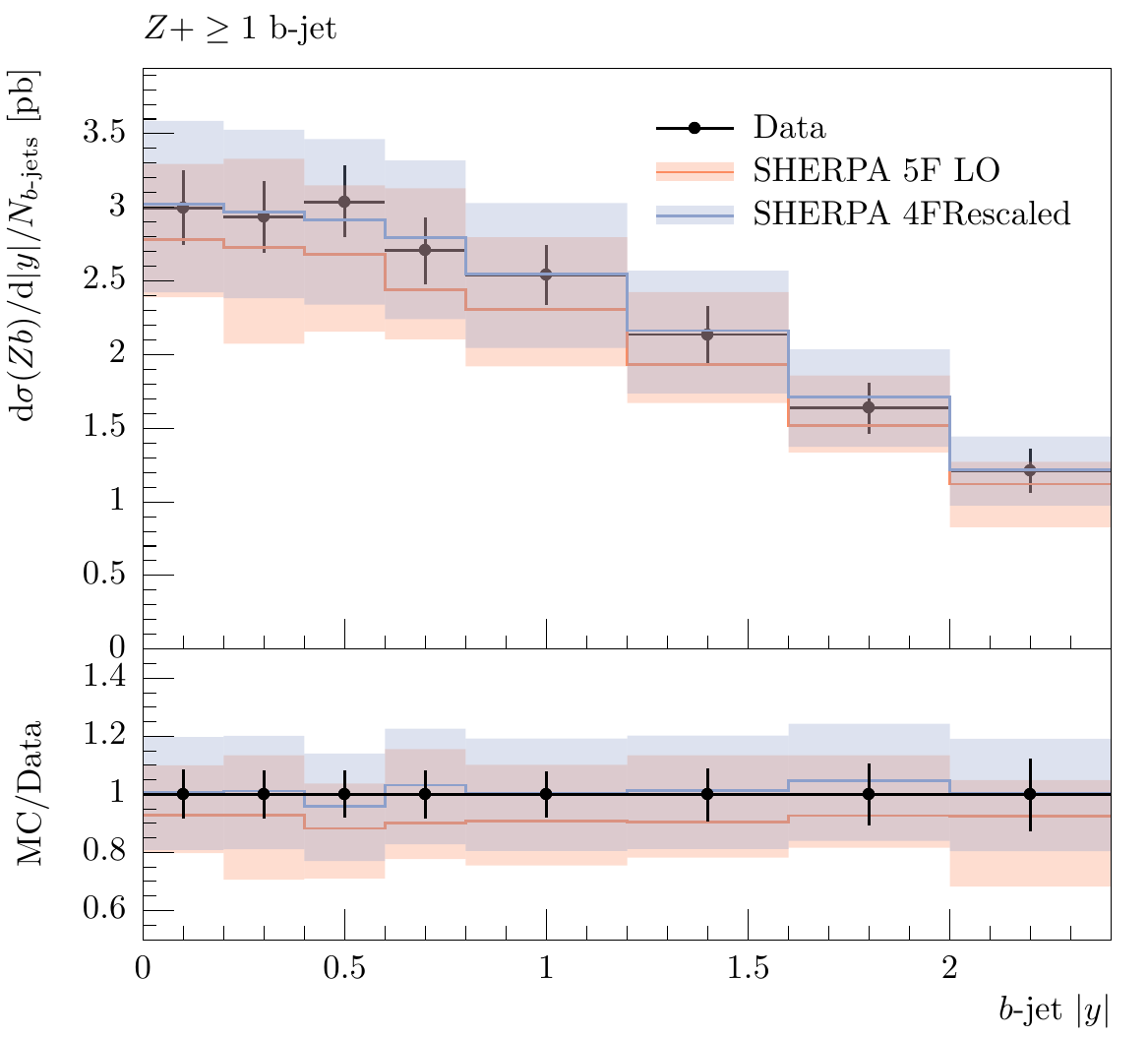} 
   \includegraphics[scale=0.65]{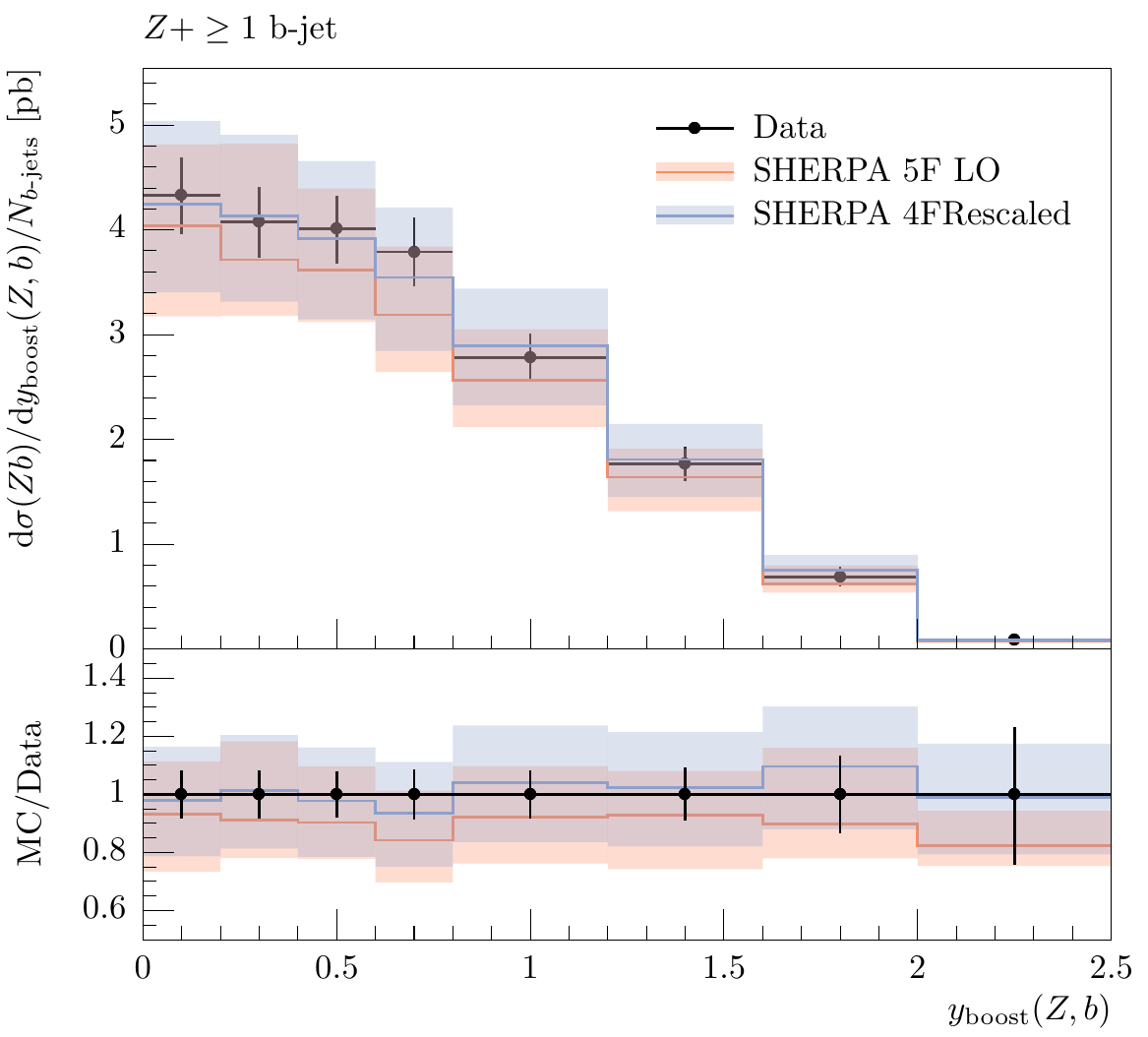} \\
   \includegraphics[scale=0.65]{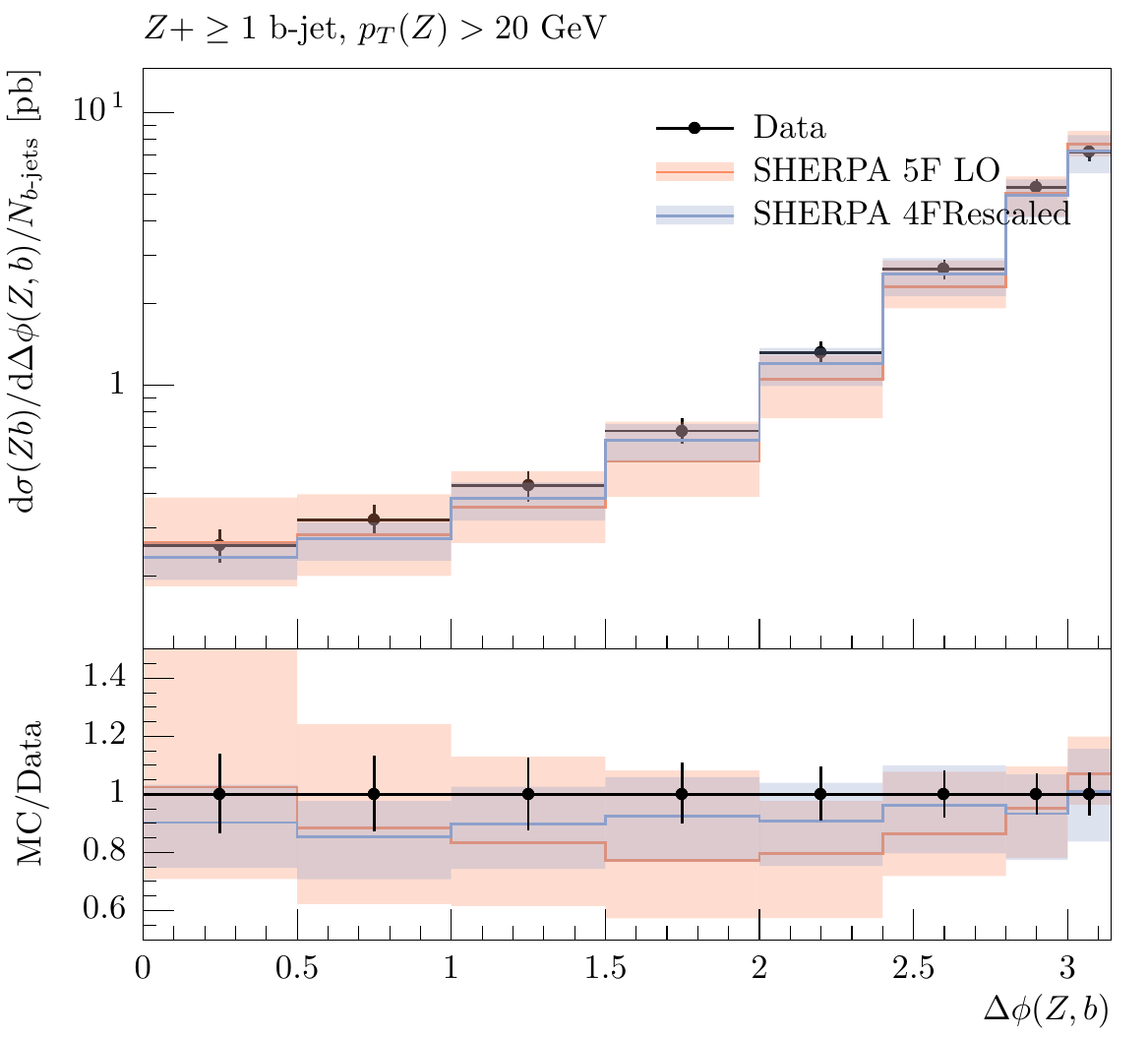} 
   \includegraphics[scale=0.65]{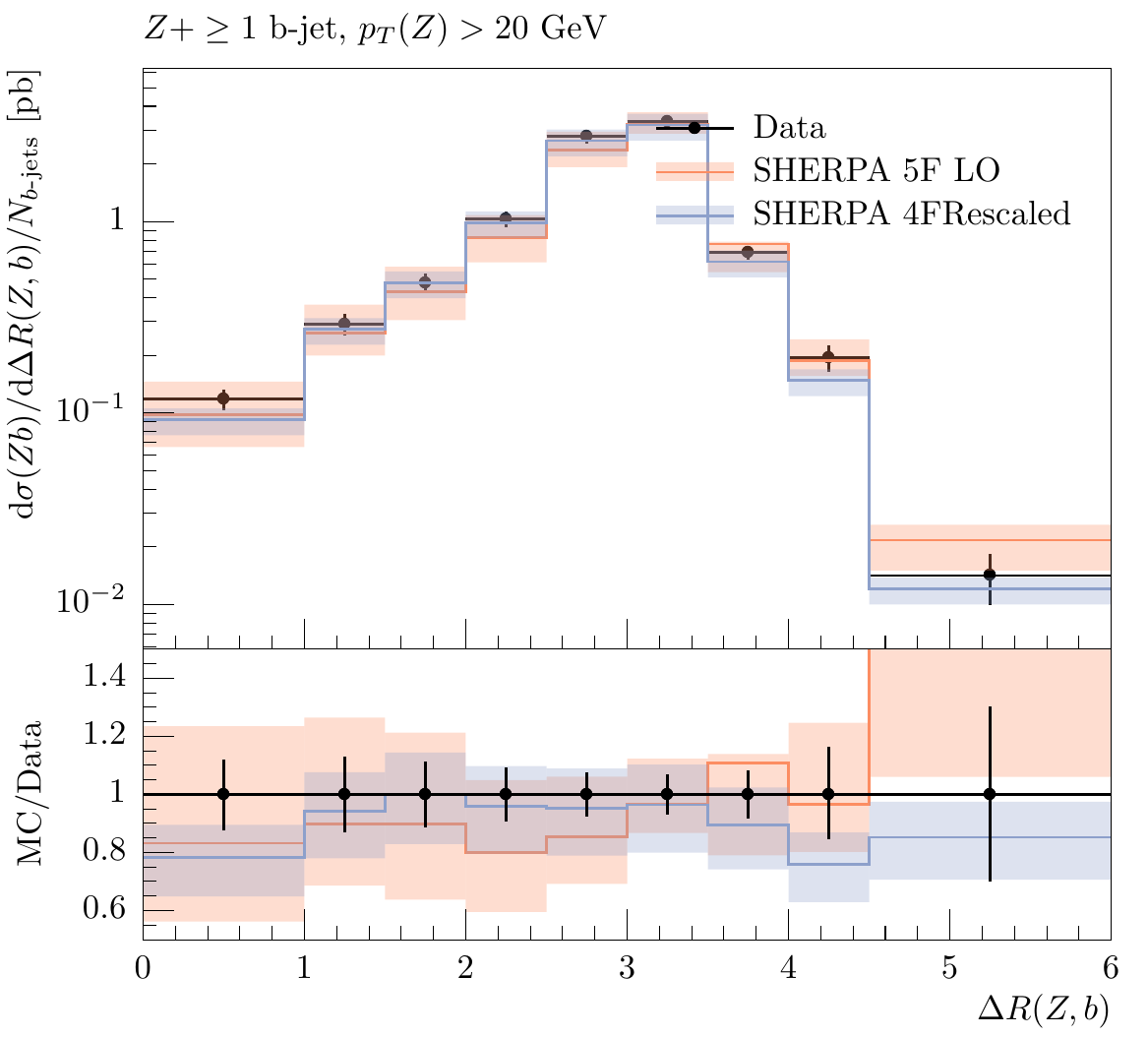}
\caption{A selection of the plots
  comparing rescaled Sherpa 4F NLO predictions to ATLAS results.}
\label{zbb-sherpa-scaled}
\end{center}
\end{figure}

It is nevertheless remarkable that the ratio of 4F NLO predictions to data is
flat for all the observables. This is particularly interesting, since it is
more efficient to generate a sample of $Z+b\bar{b}$ events with the 4F scheme
than with the 5F. The reason why an overall normalization factor is needed
could lie in the large logarithms, that in the 5F scheme are resummed in the
$b$ parton distribution function. However they might not affect the shape of
the distributions. To check this hypothesis the 4F NLO predictions have been
rescaled to the integrated cross-sections calculated with
\MCFM~\cite{Campbell:2010ff}. Depending on the observable and the applied
selection, four different cross-sections are defined, as explained
in~\cite{Aad:2014dvb}. The value is corrected for QED final-state
radiation, hadronisation, underlying event and multi-parton
interactions. The uncertainty is given by the envelope of
the results obtained with several PDFs, taking for each the sum in quadrature of all theory uncertainties.
A selection of the plots is shown in
Figure~\ref{zbb-sherpa-scaled}. The results are very encouraging but further
studies are needed to understand if this approach fails for other
observables, e.g. those related to the presence of additional light-quark
jets.


\subsubsection*{Z+b(b) with \Herwig}
\label{sec:ZHerwig}

\begin{figure}[htbp]
\begin{center}
   \includegraphics[scale=0.65]{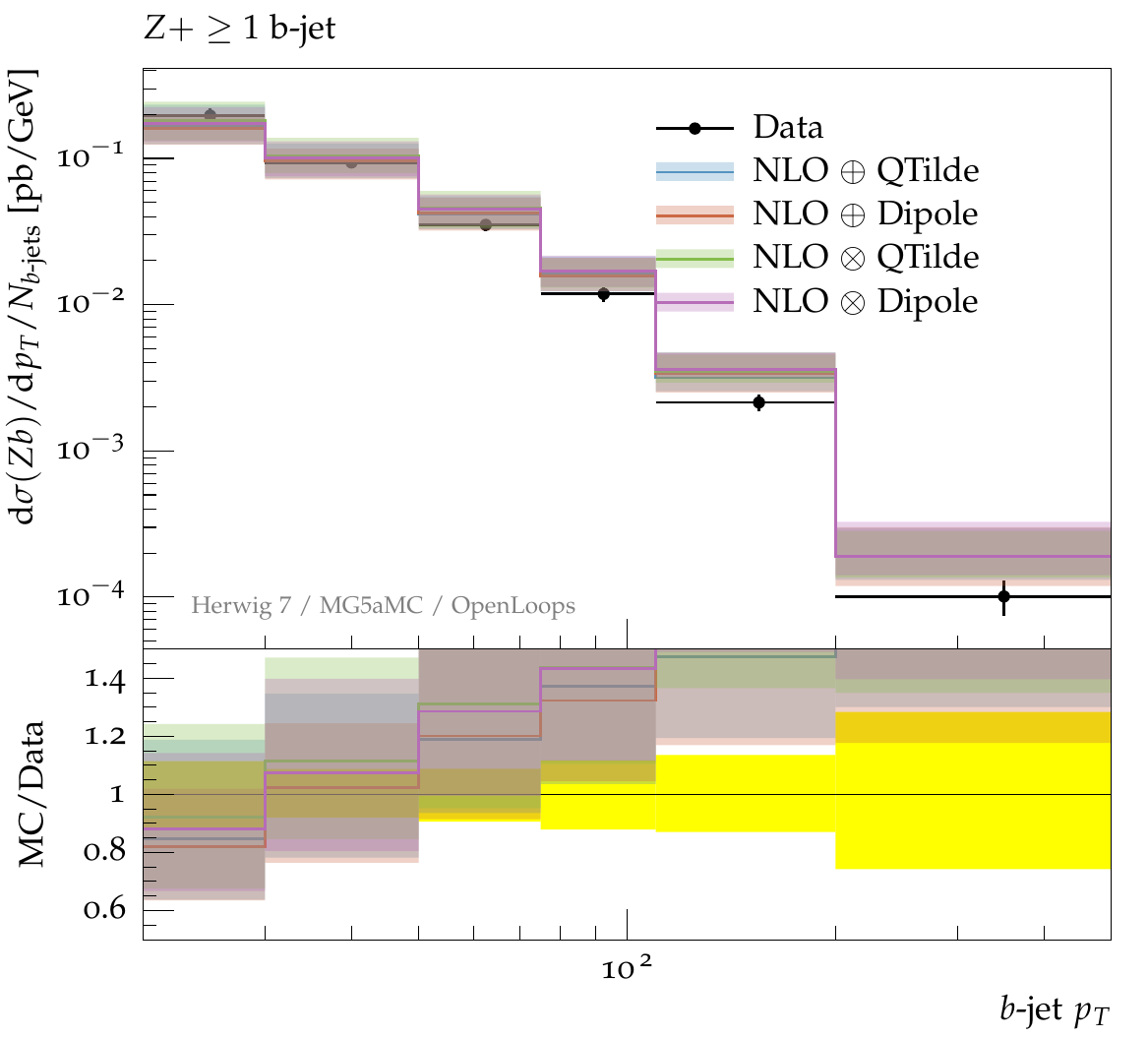}
   \includegraphics[scale=0.65]{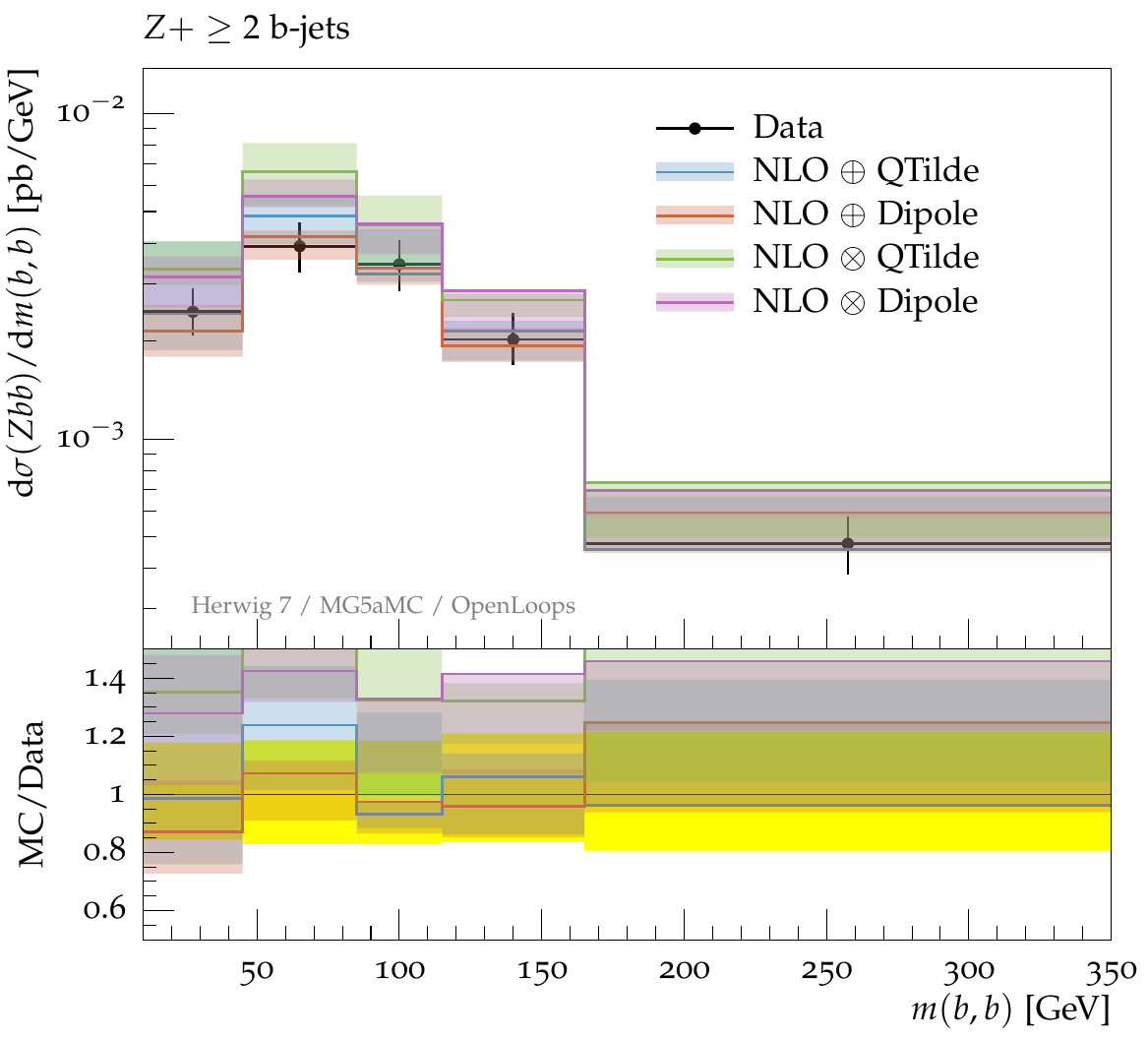} \\
   \includegraphics[scale=0.65]{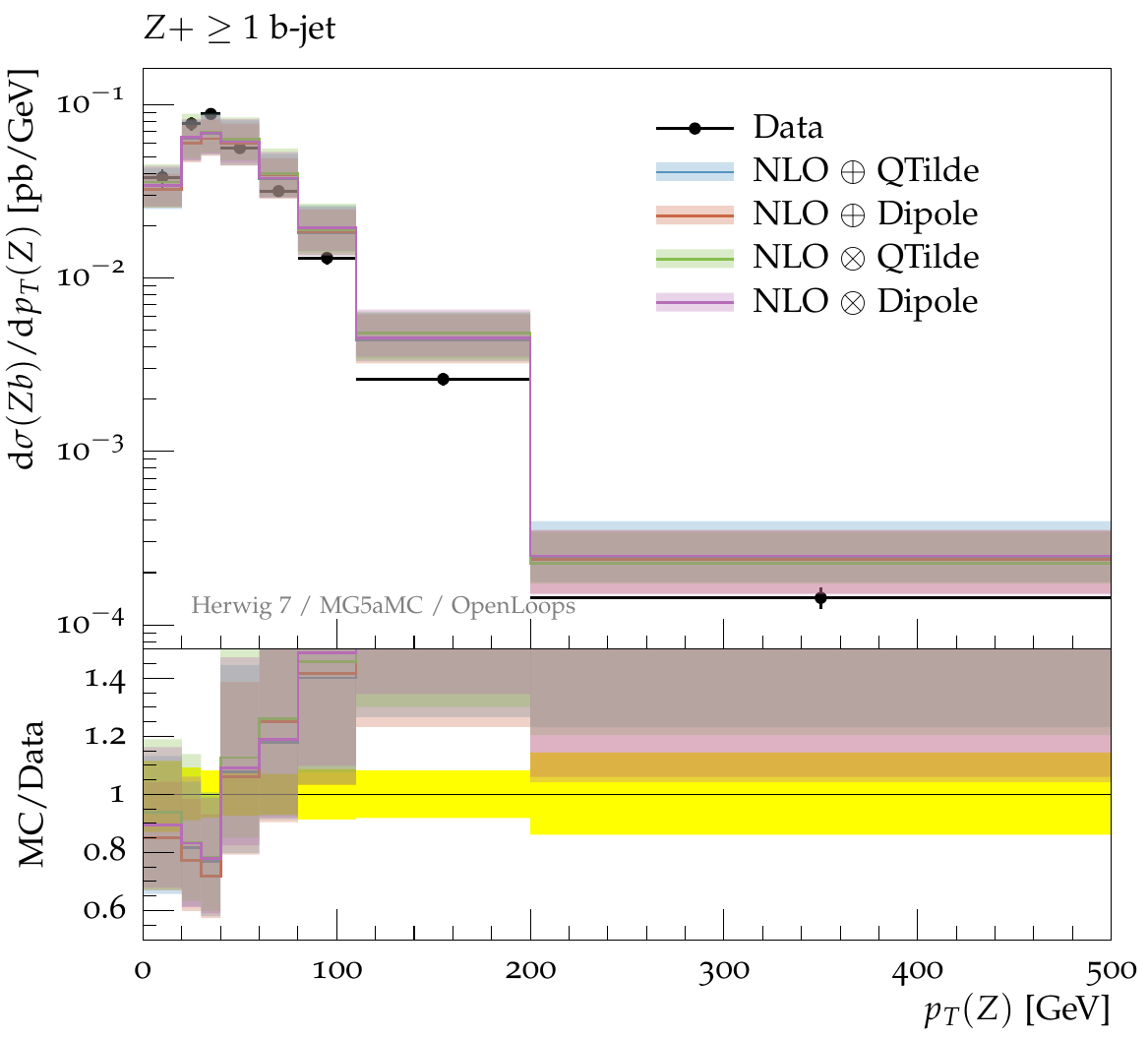}
   \includegraphics[scale=0.65]{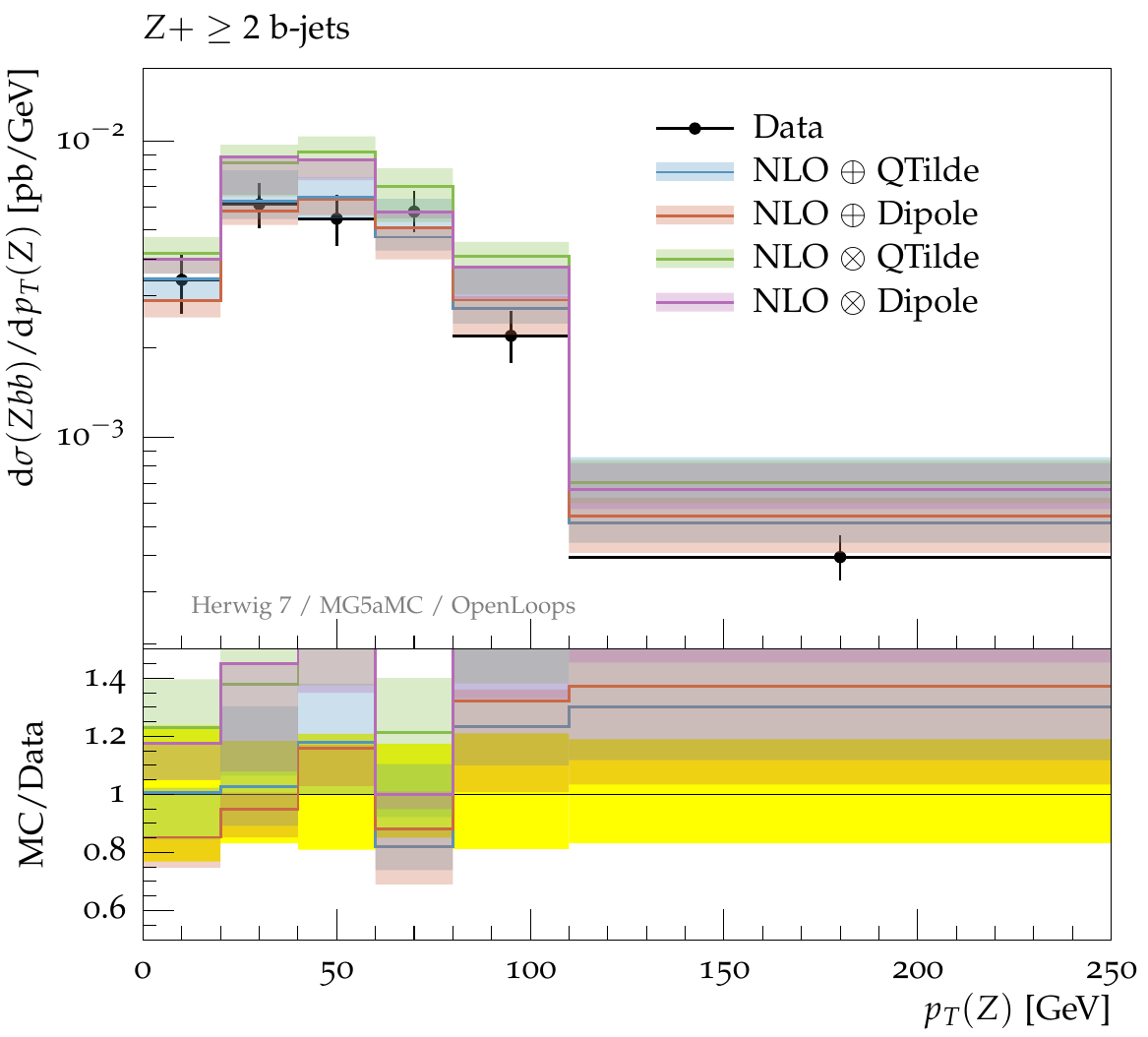}
\caption{A selection of the plots comparing \Herwig 5F, Zbb predictions to ATLAS results.}
\label{zbb-herwigzbb-atlas}
\end{center}
\end{figure}

\begin{figure}[htbp]
\begin{center}
   \includegraphics[scale=0.65]{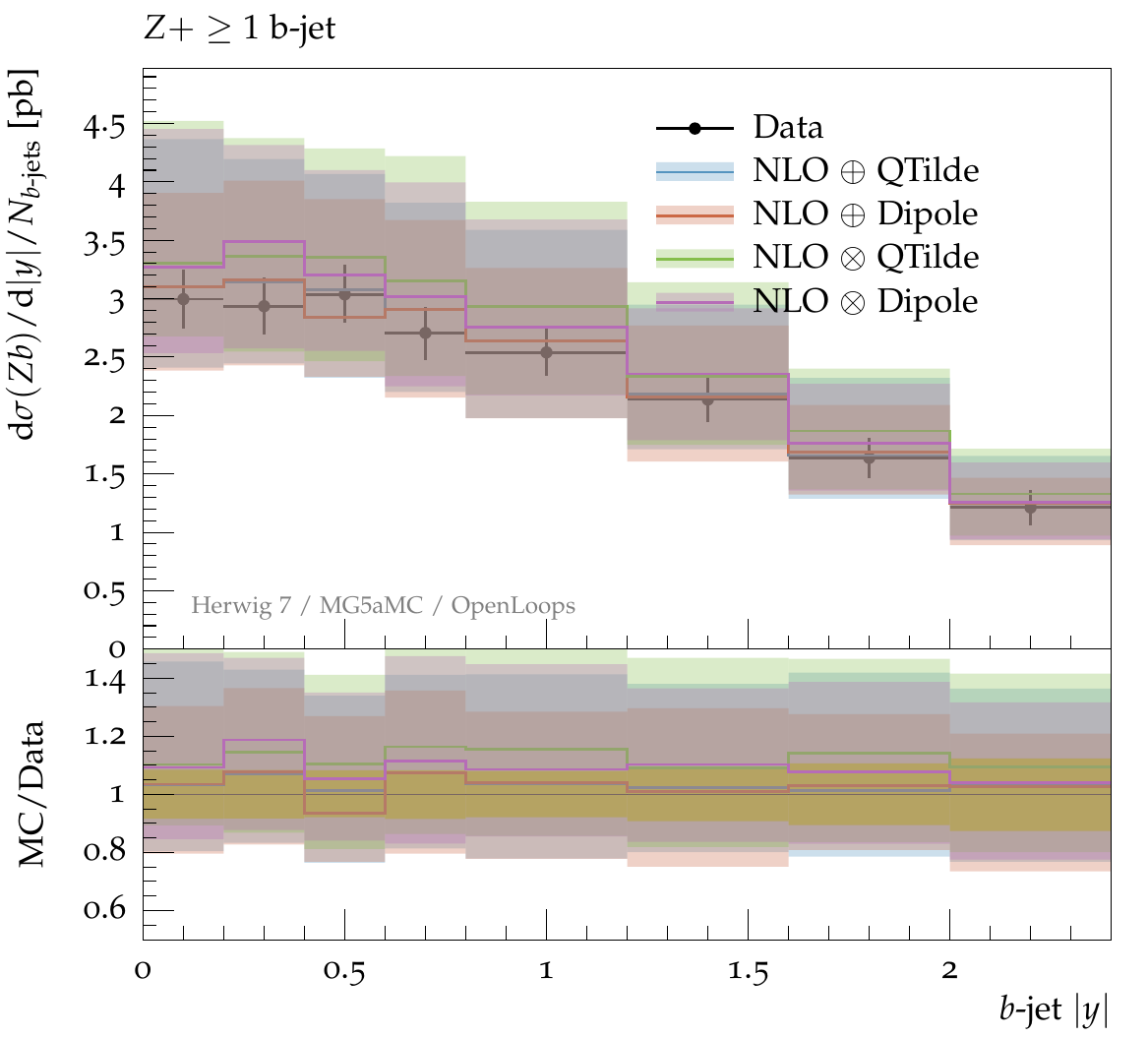}
   \includegraphics[scale=0.65]{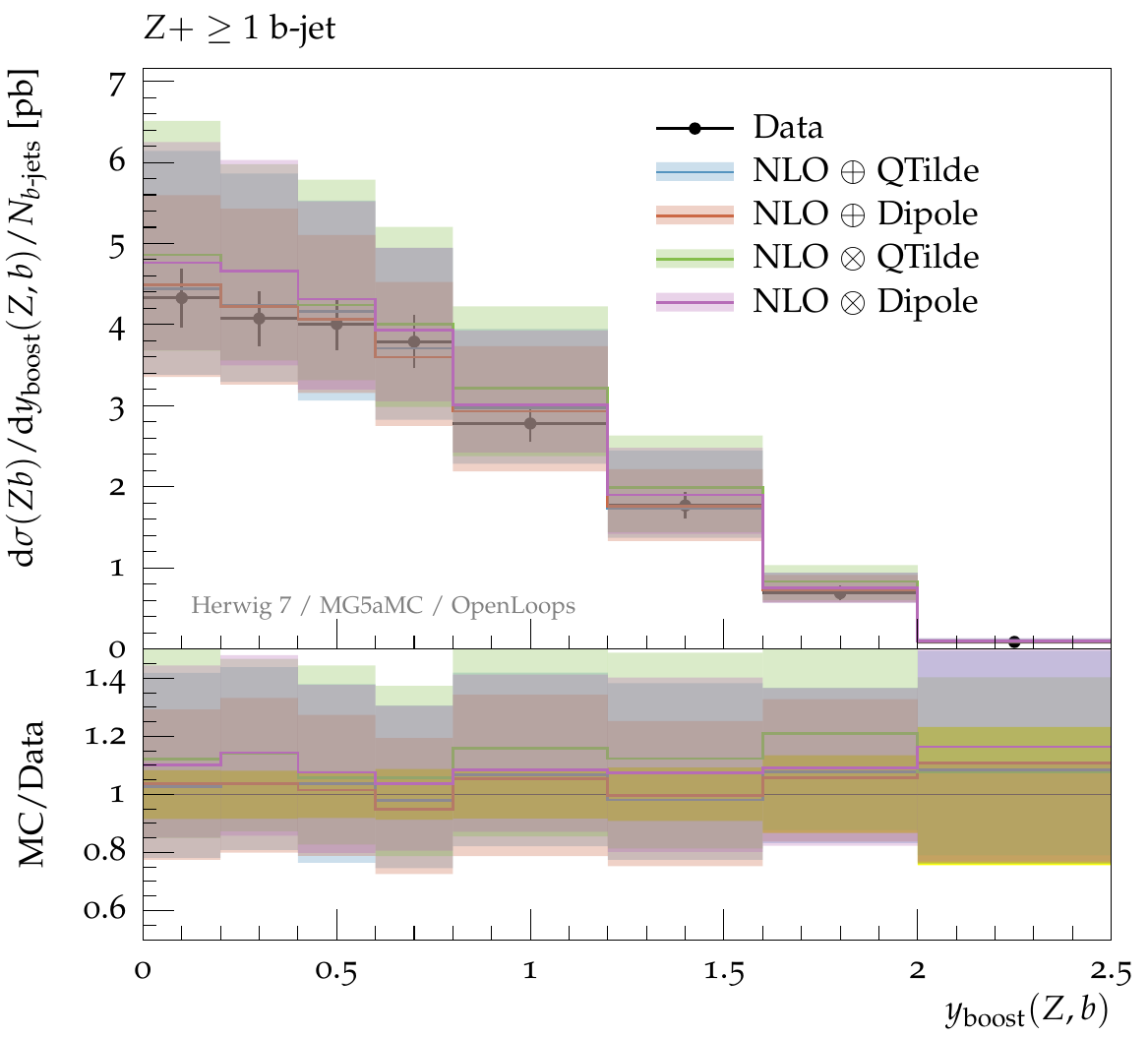} \\
   \includegraphics[scale=0.65]{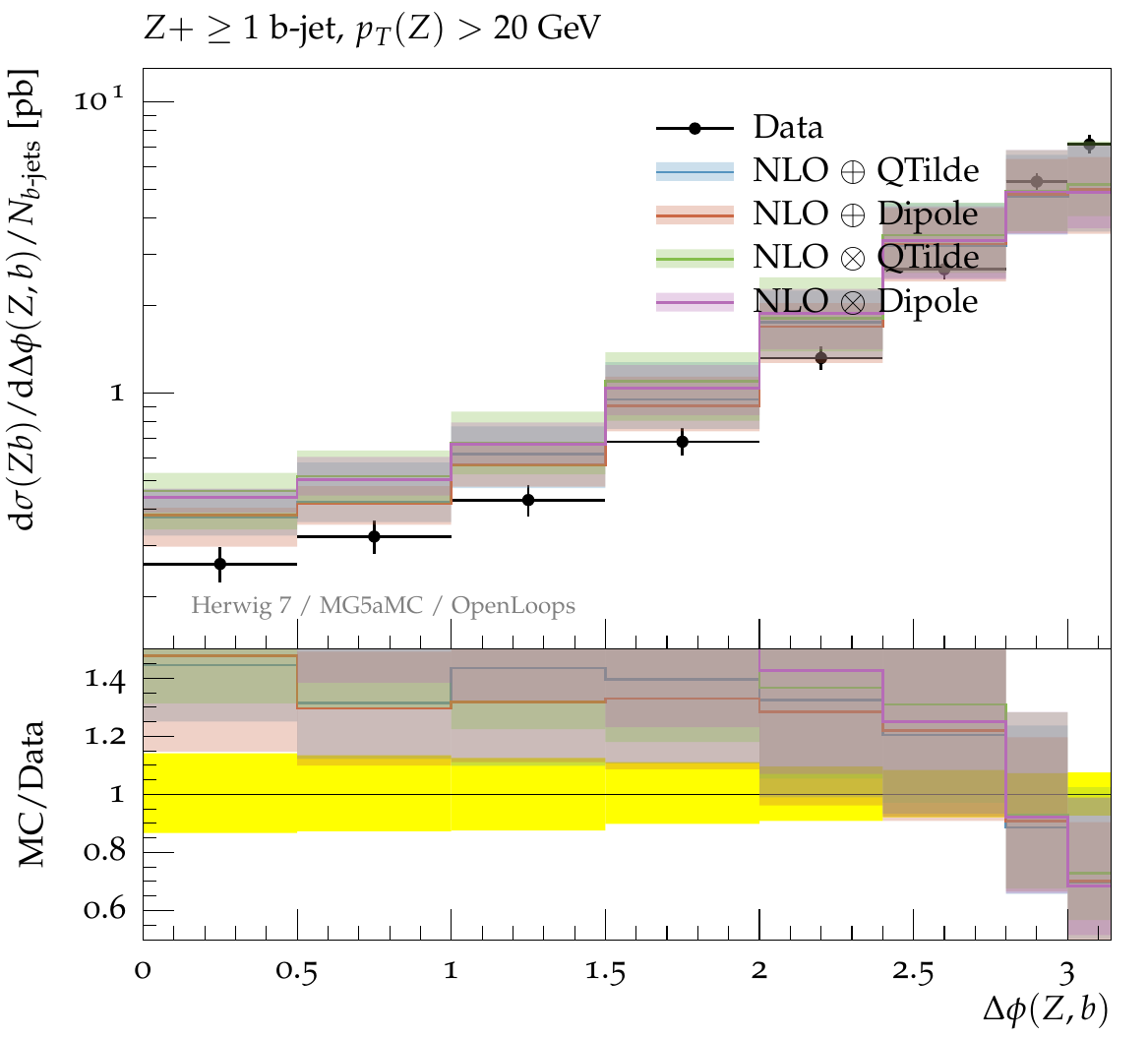}
   \includegraphics[scale=0.65]{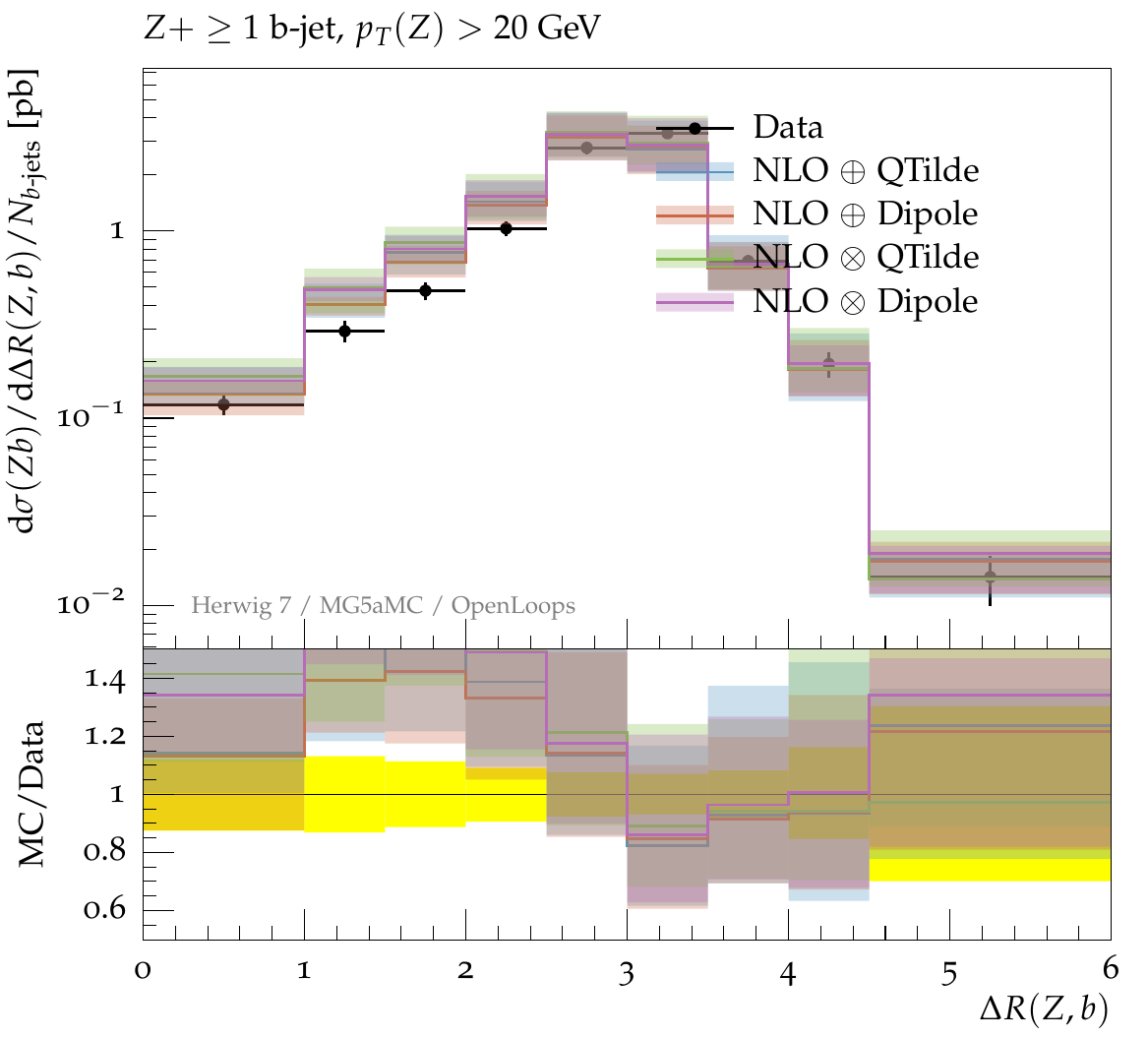}
\caption{A selection of the plots comparing \Herwig 5F, Zbb predictions to ATLAS
  results; together with the plots in figure~\ref{zbb-herwigzbb-atlas}
  a comparison to the rescaled \Sherpa 4F NLO predictions in
  Figure~\ref{zbb-sherpa-scaled} can be made.}
\label{zbb-herwigzbb-atlas-2}
\end{center}
\end{figure}

\begin{figure}[htbp]
   \includegraphics[scale=0.65]{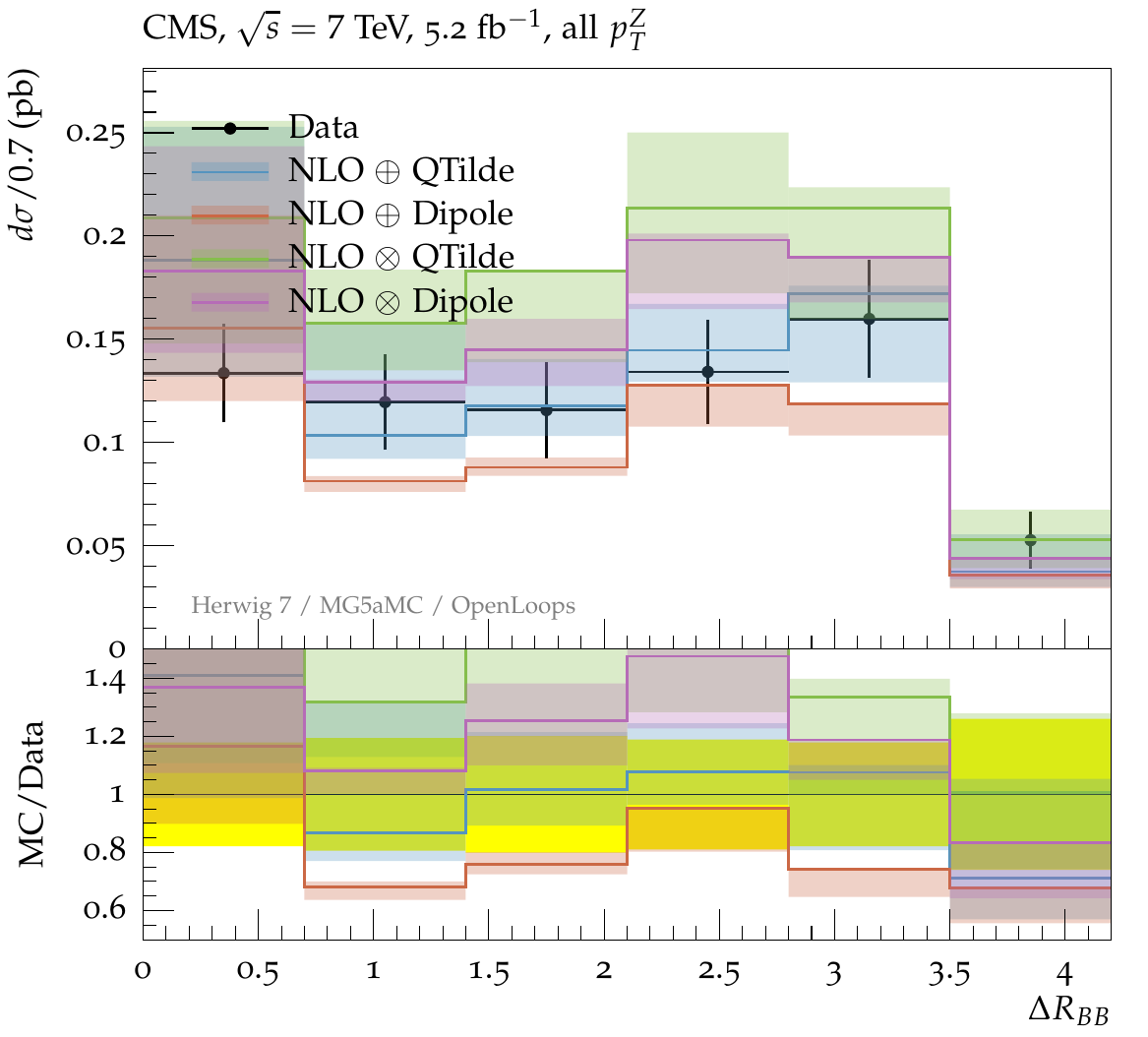}
   \includegraphics[scale=0.65]{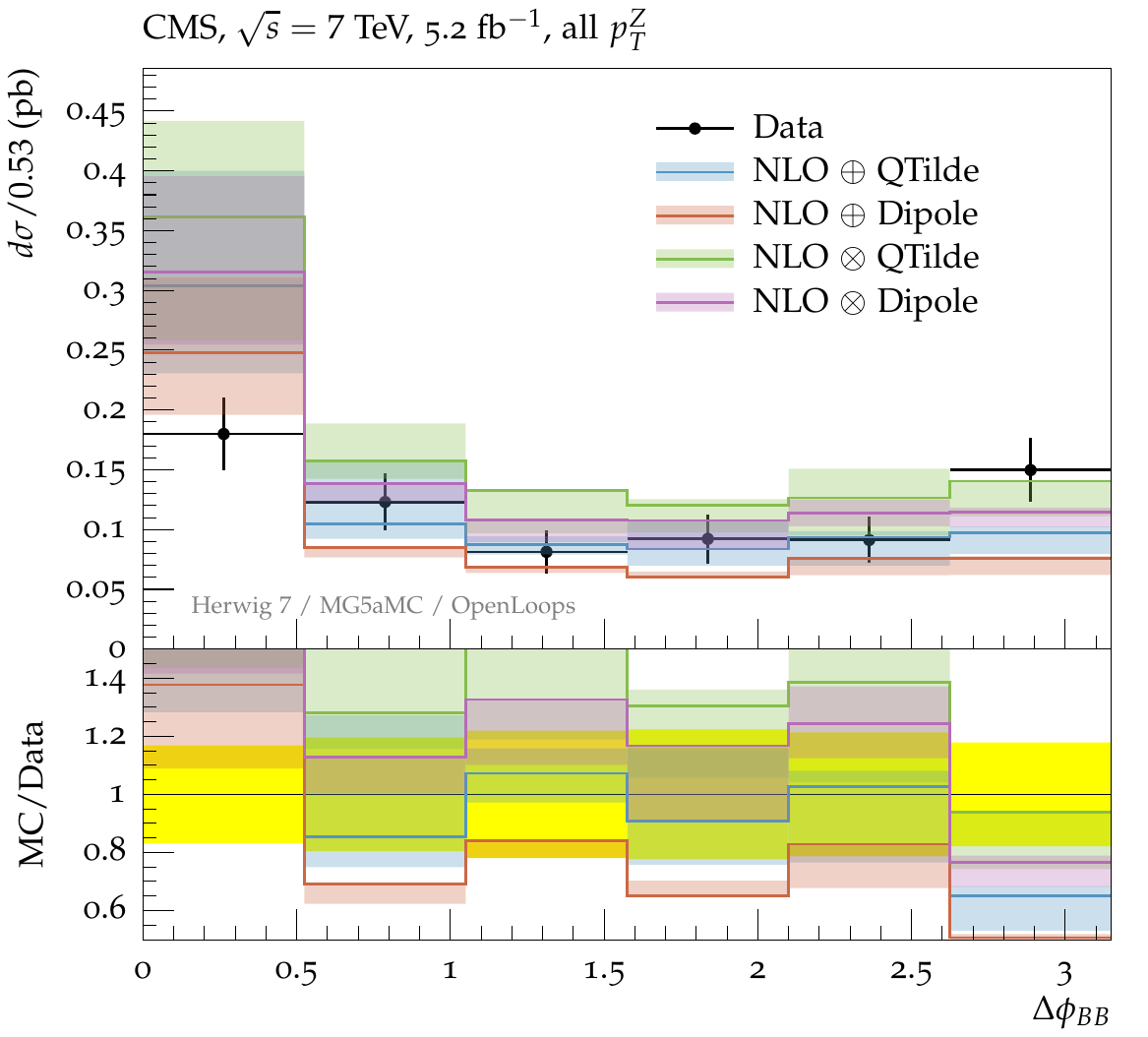}
\caption{A selection of the plots comparing \Herwig 5F, Zbb predictions to CMS results.}
\label{zbb-herwigzbb-cms}
\end{figure}

\begin{figure}[htbp]
\begin{center}
   \includegraphics[scale=0.65]{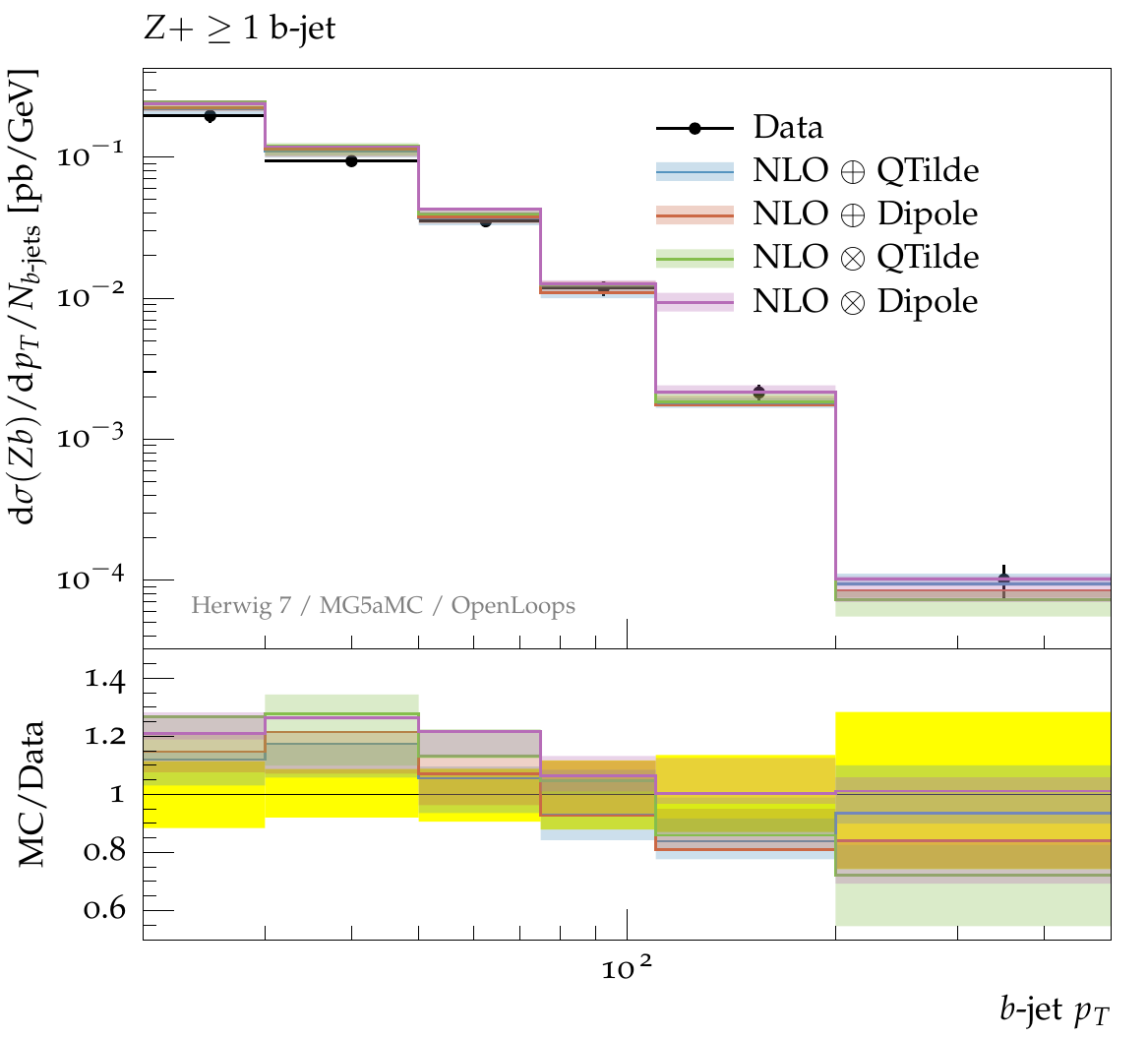}
   \includegraphics[scale=0.65]{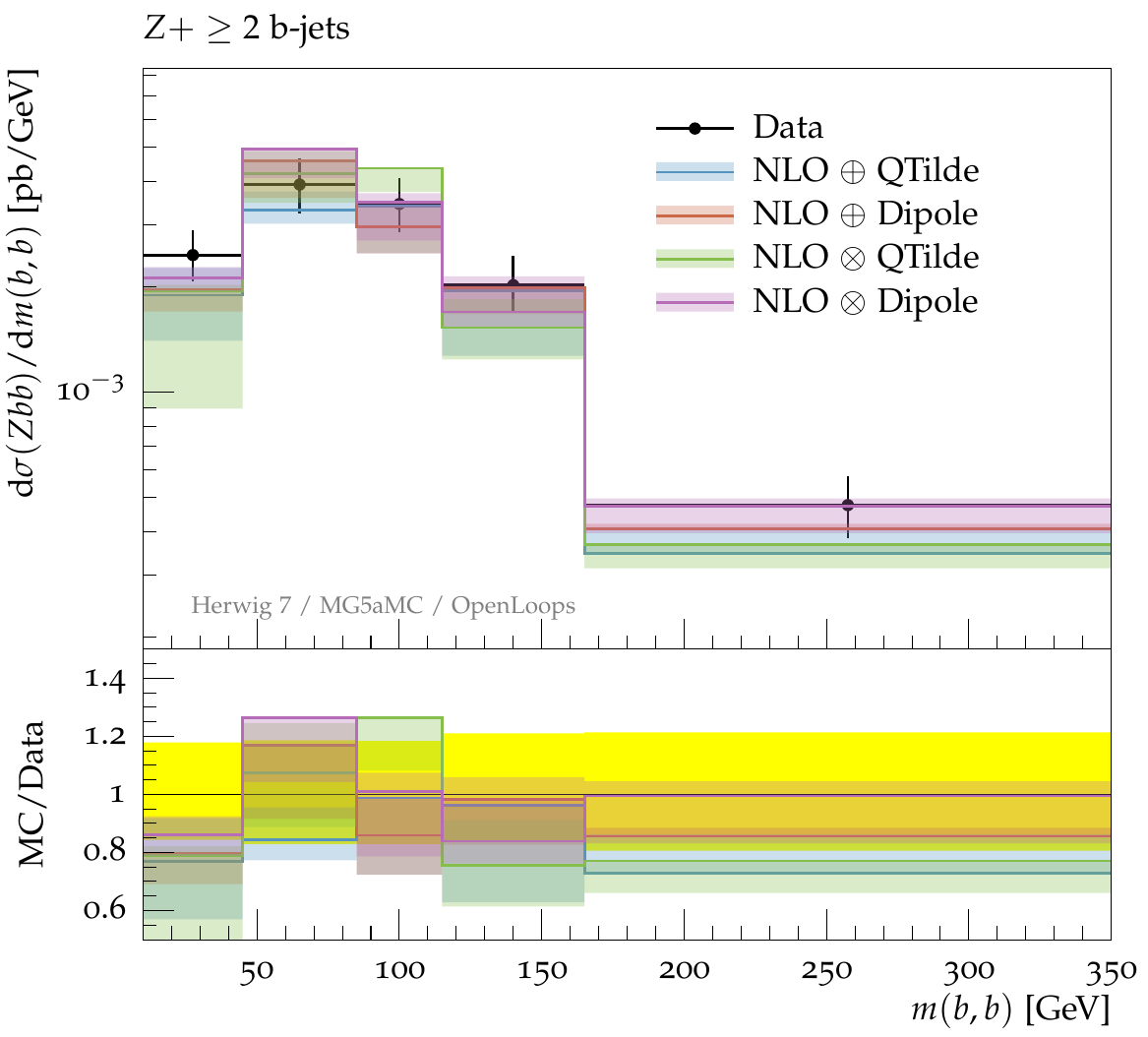} \\
   \includegraphics[scale=0.65]{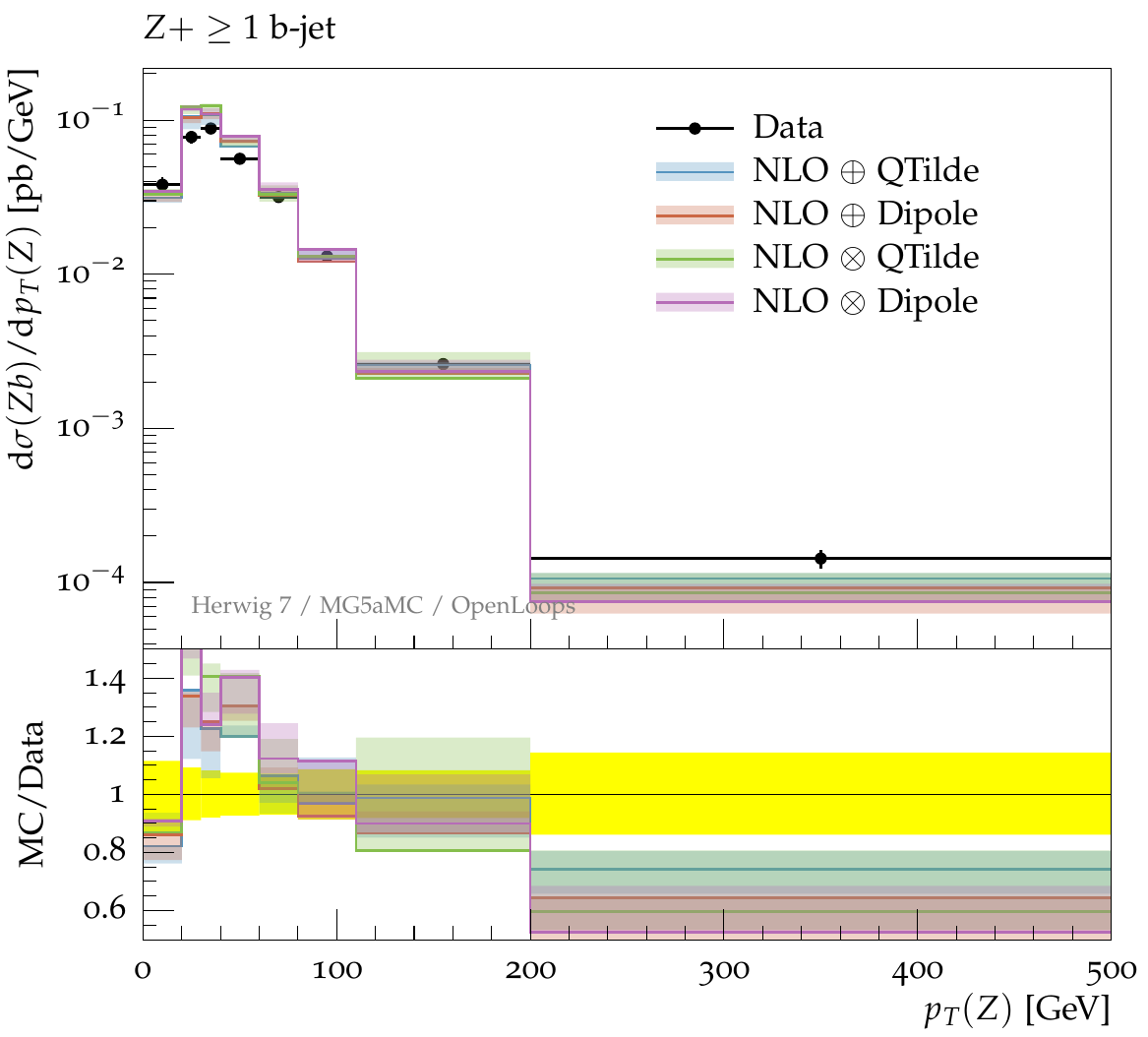}
   \includegraphics[scale=0.65]{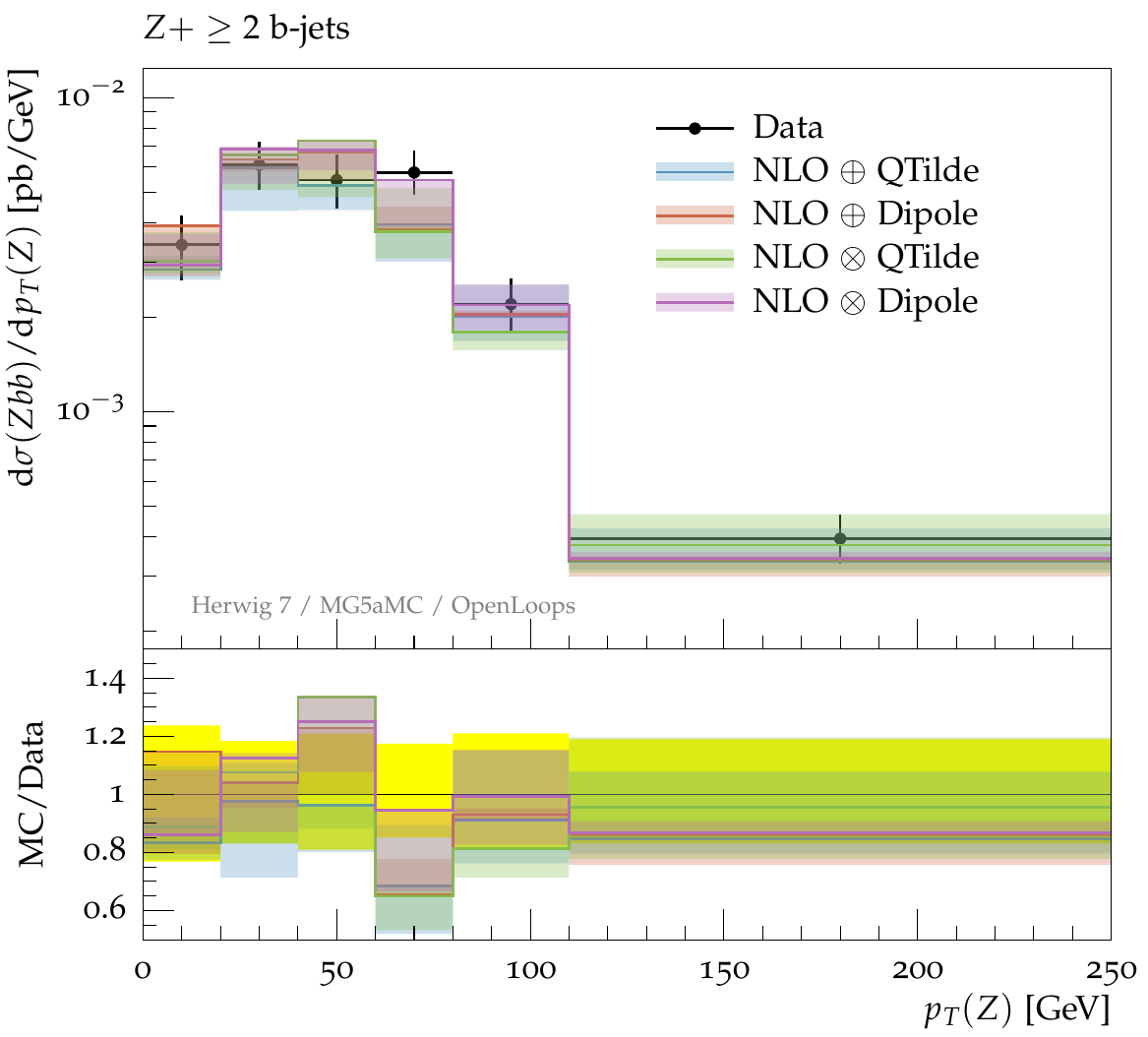}
\caption{A selection of the plots comparing \Herwig 5F, Zb predictions to ATLAS results.}
\label{zbb-herwigzb-atlas}
\end{center}
\end{figure}

\begin{figure}[htbp]
\begin{center}
   \includegraphics[scale=0.65]{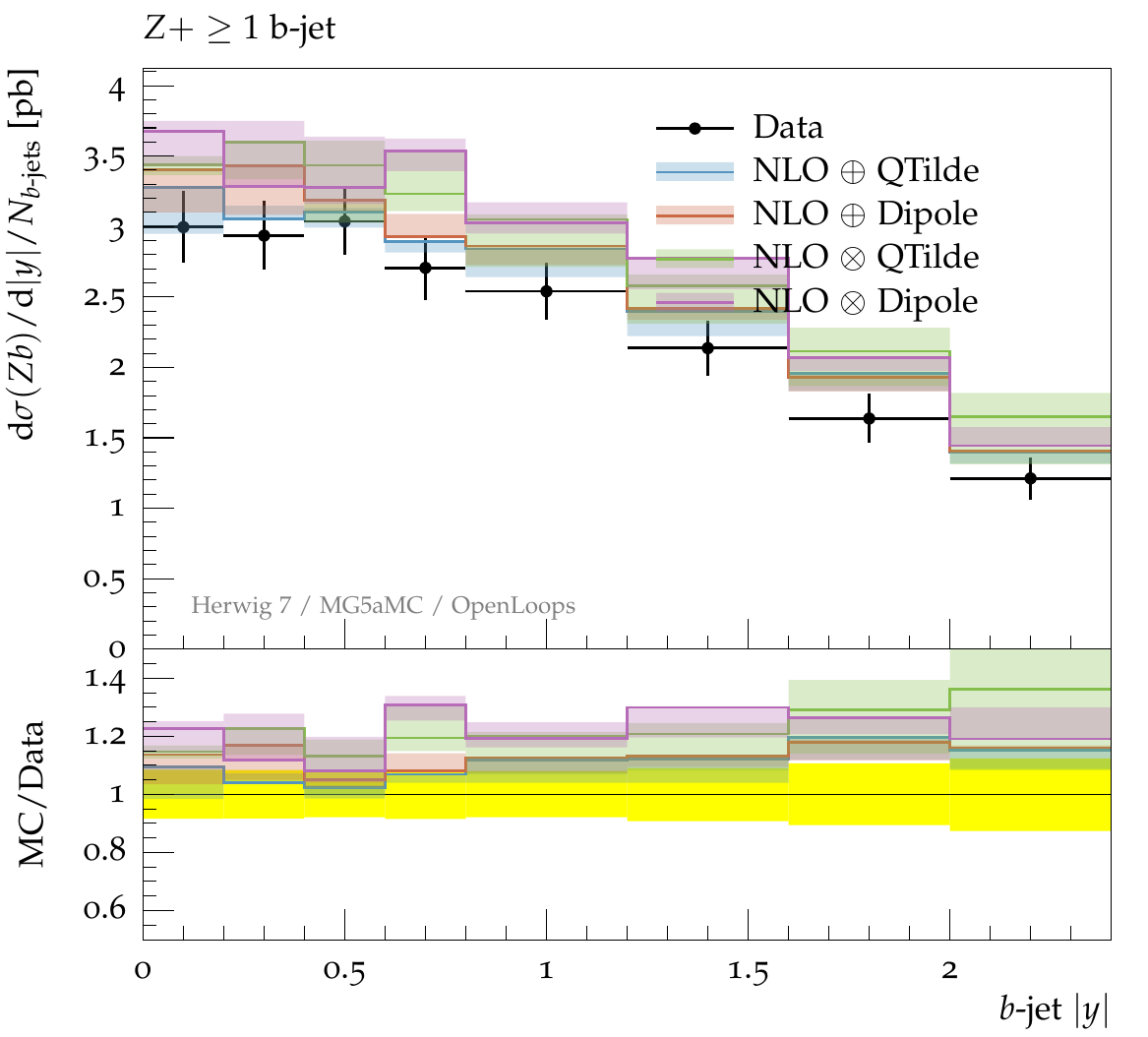}
   \includegraphics[scale=0.65]{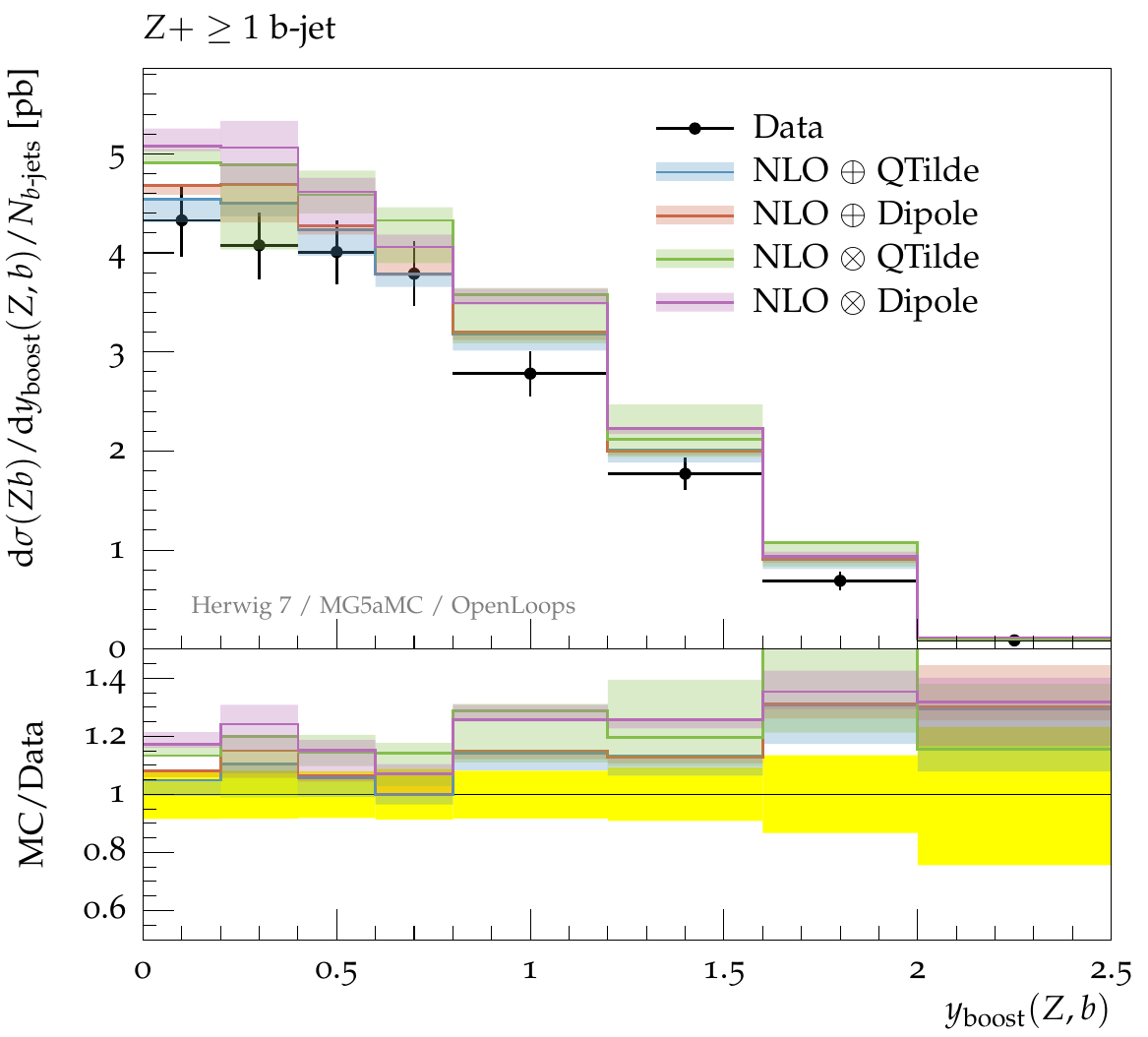} \\
   \includegraphics[scale=0.65]{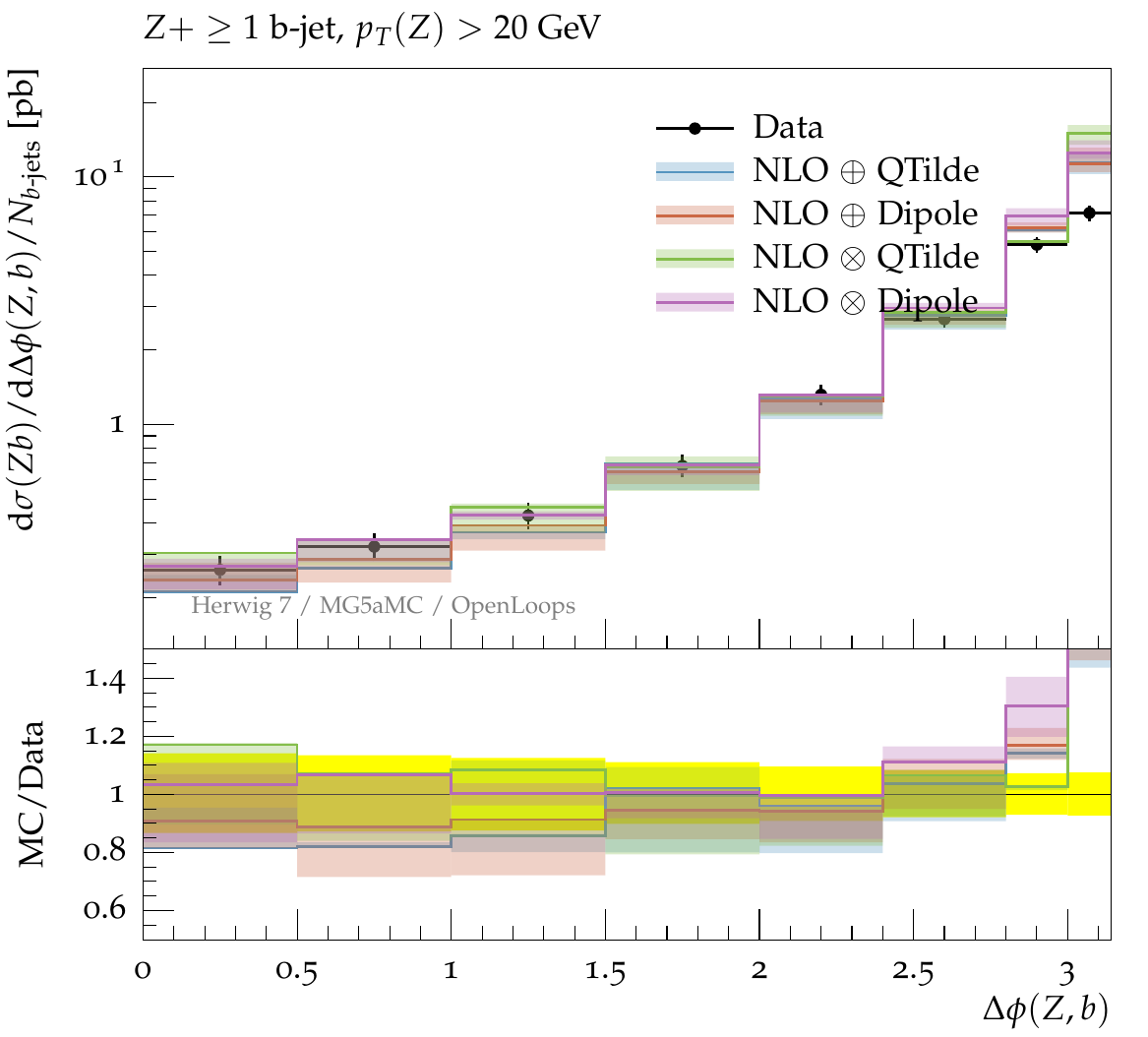}
   \includegraphics[scale=0.65]{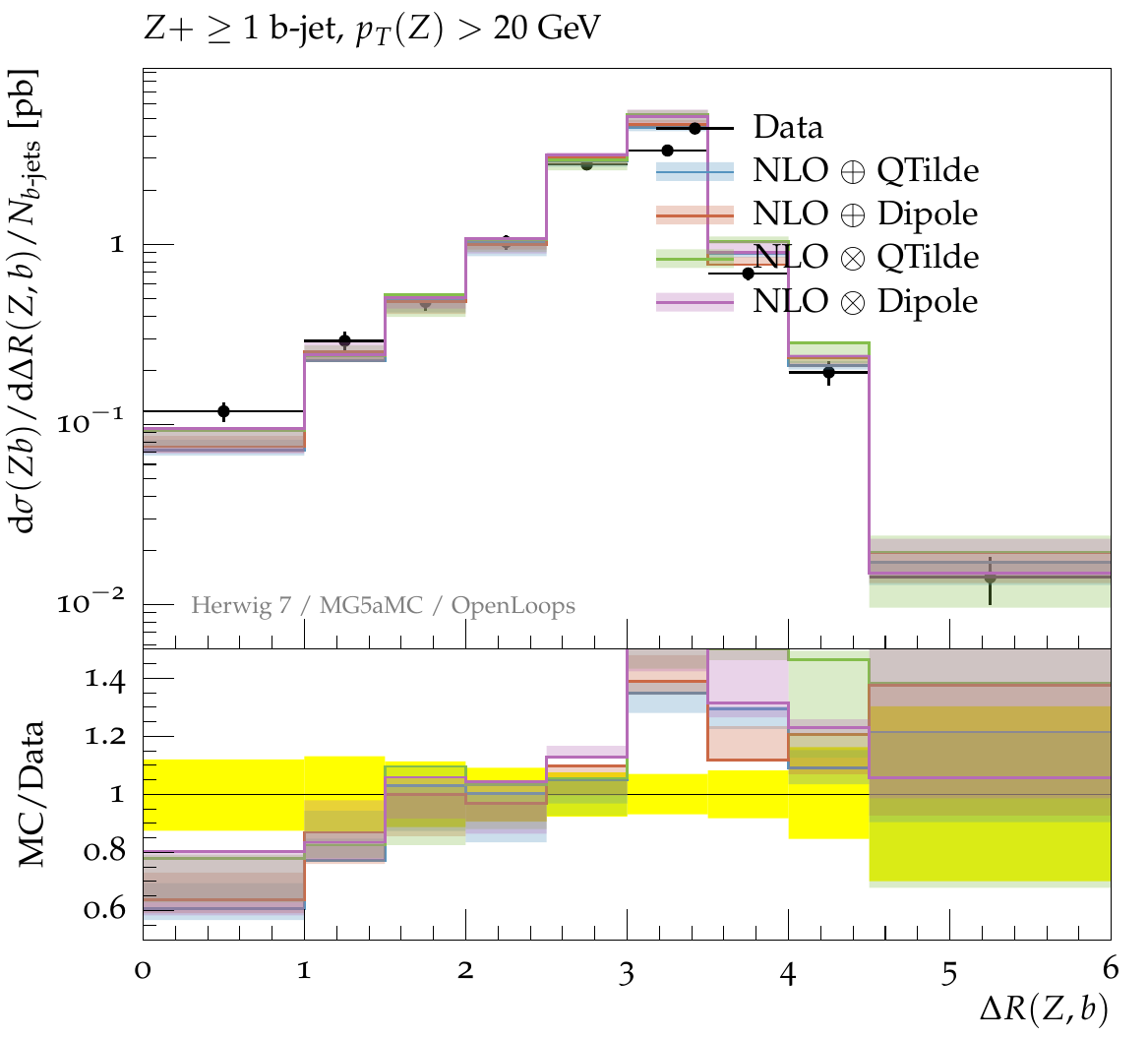}
\caption{A selection of the plots comparing \Herwig 5F, Zb predictions to ATLAS
  results; together with the plots in Figure~\ref{zbb-herwigzb-atlas}
  a comparison to the rescaled \Sherpa 4F NLO predictions in
  Figure~\ref{zbb-sherpa-scaled} can be made.}
\label{zbb-herwigzb-atlas-2}
\end{center}
\end{figure}

\begin{figure}[htbp]
   \includegraphics[scale=0.65]{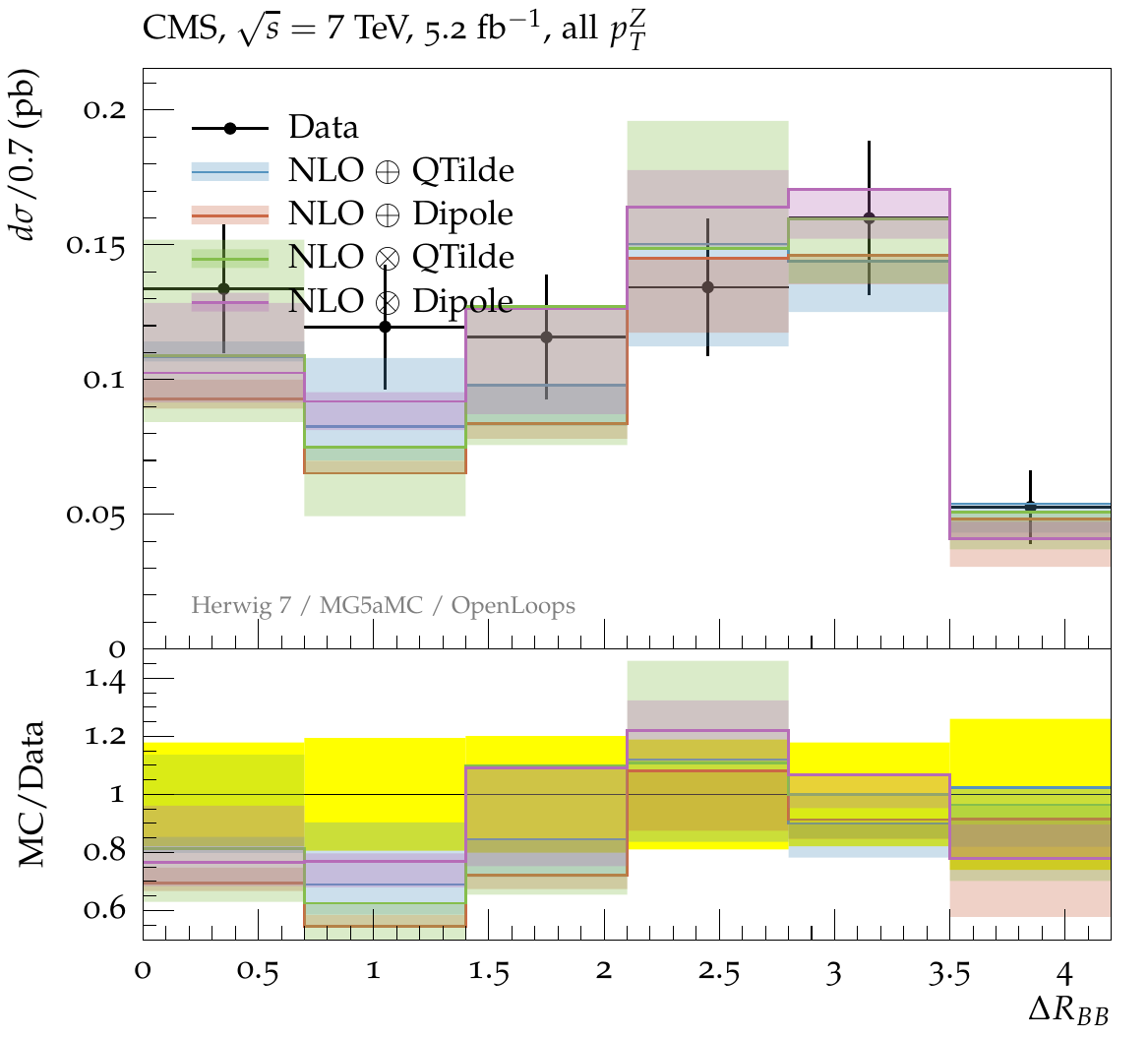}
   \includegraphics[scale=0.65]{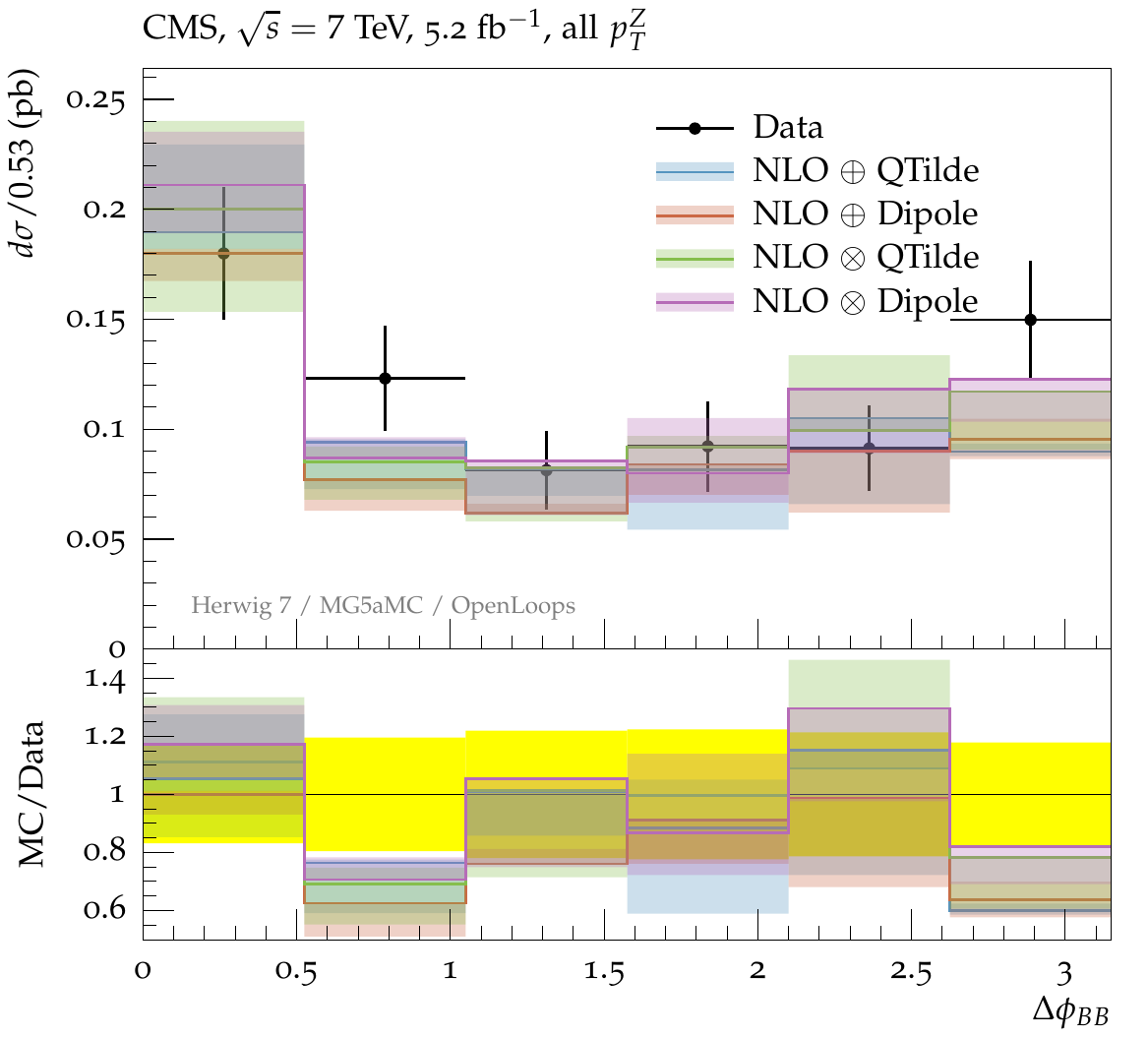}
\caption{A selection of the plots comparing \Herwig 5F, Zb predictions to CMS results.}
\label{zbb-herwigzb-cms}
\end{figure}

\begin{figure}[htbp]
\begin{center}
   \includegraphics[scale=0.65]{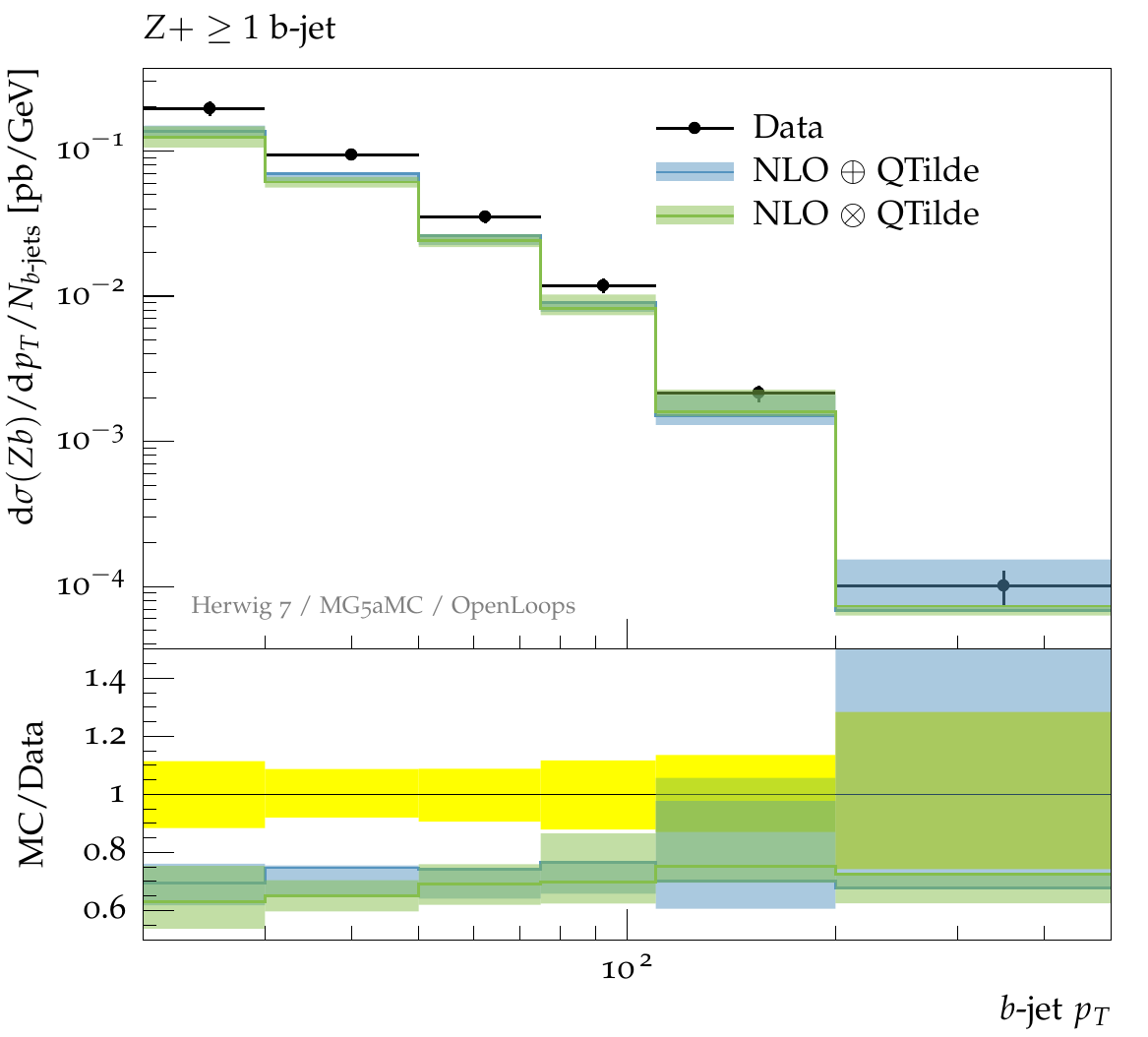}
   \includegraphics[scale=0.65]{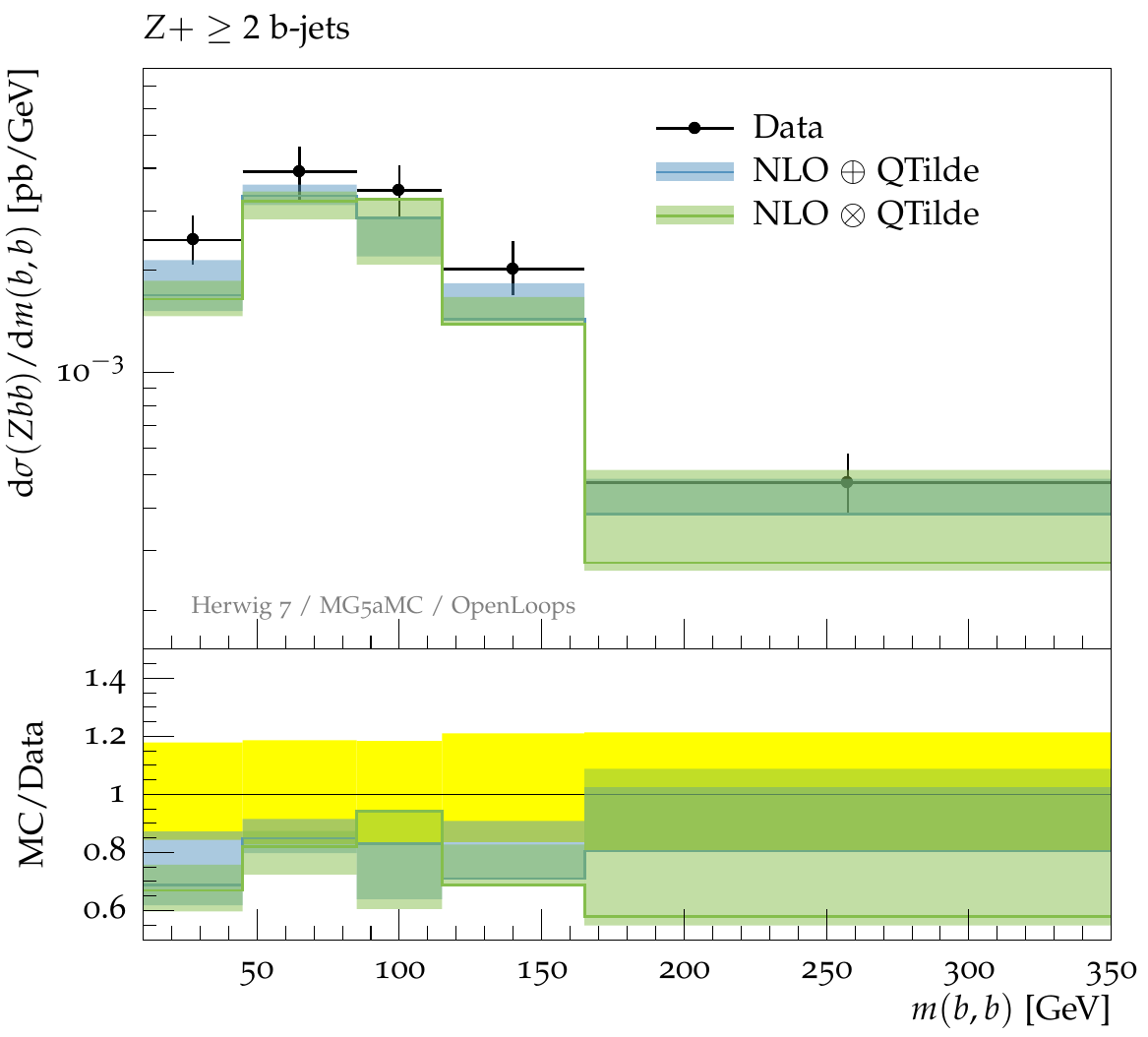} \\
   \includegraphics[scale=0.65]{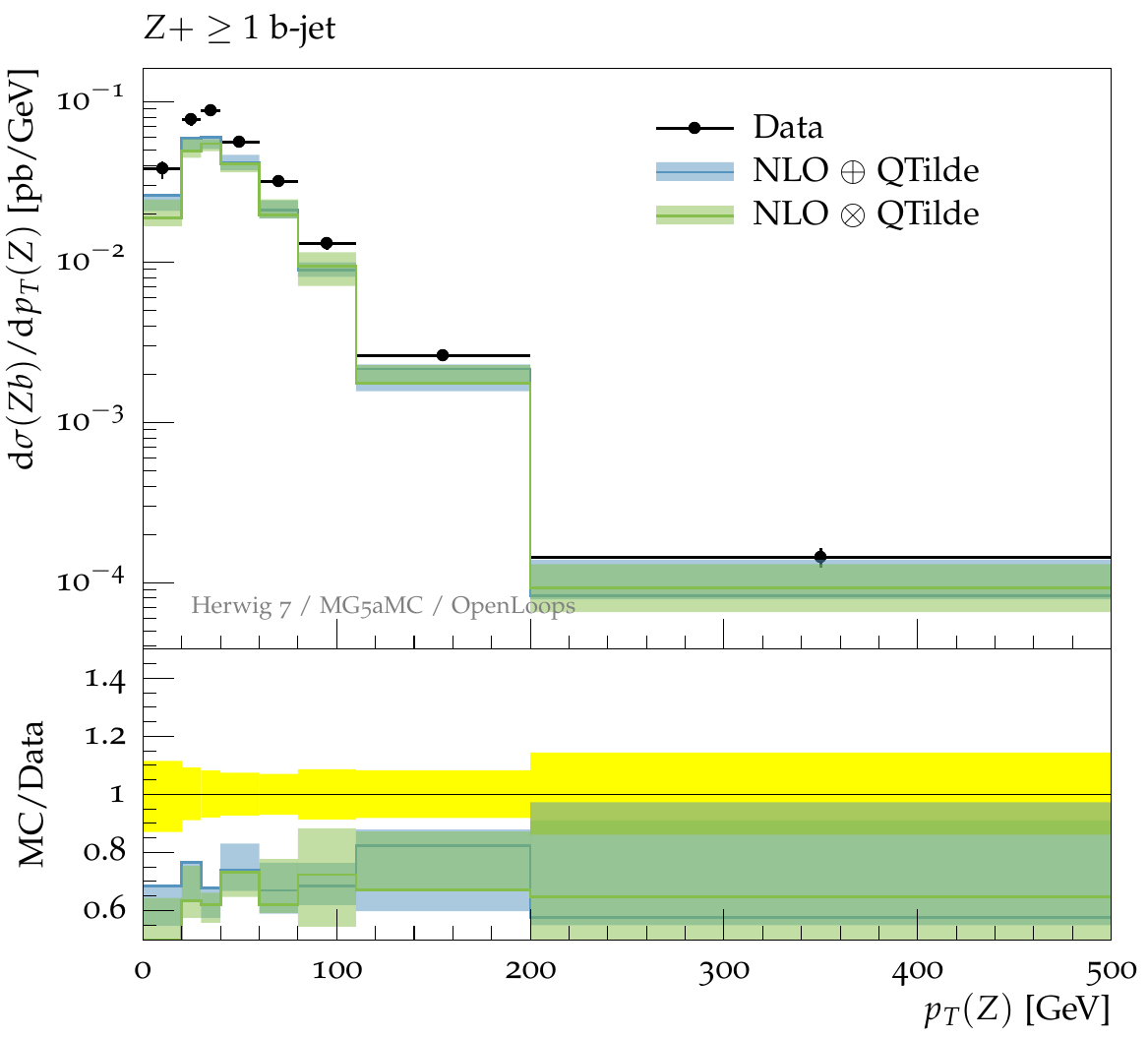}
   \includegraphics[scale=0.65]{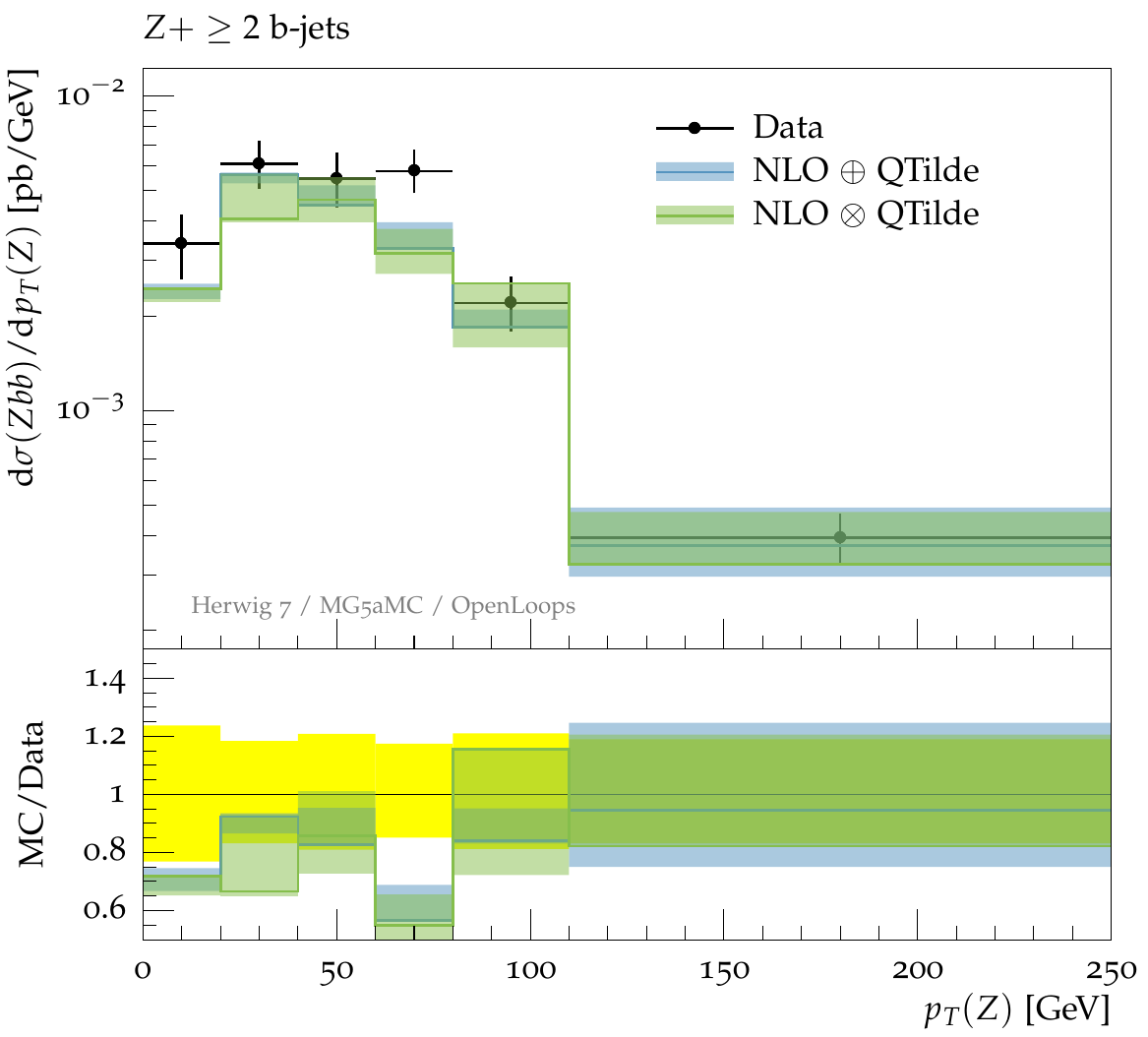}
\caption{A selection of the plots comparing \Herwig 4F, Zbb predictions to ATLAS results.}
\label{zbb-herwig4F-atlas}
\end{center}
\end{figure}

\begin{figure}[htbp]
\begin{center}
   \includegraphics[scale=0.65]{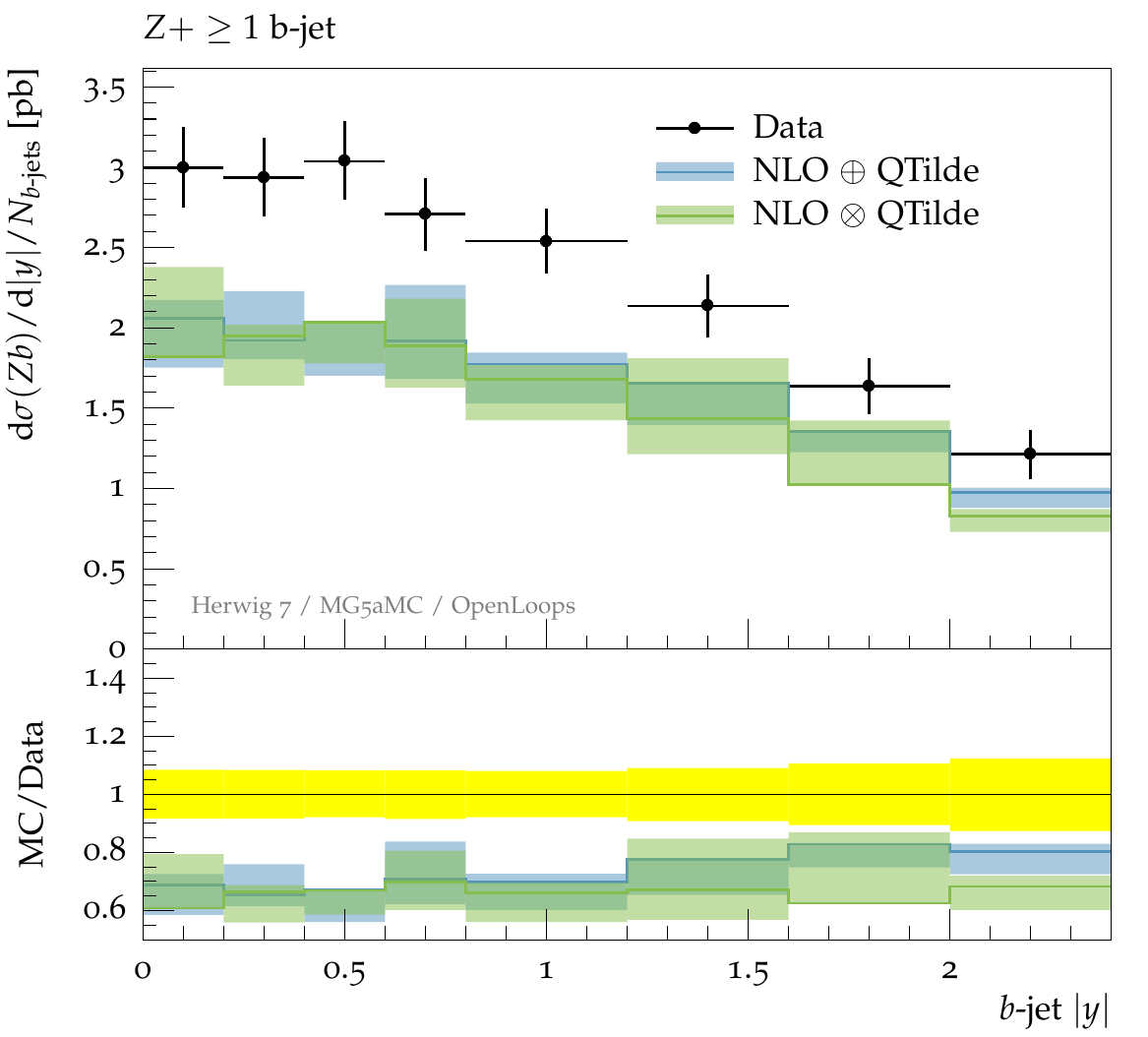}
   \includegraphics[scale=0.65]{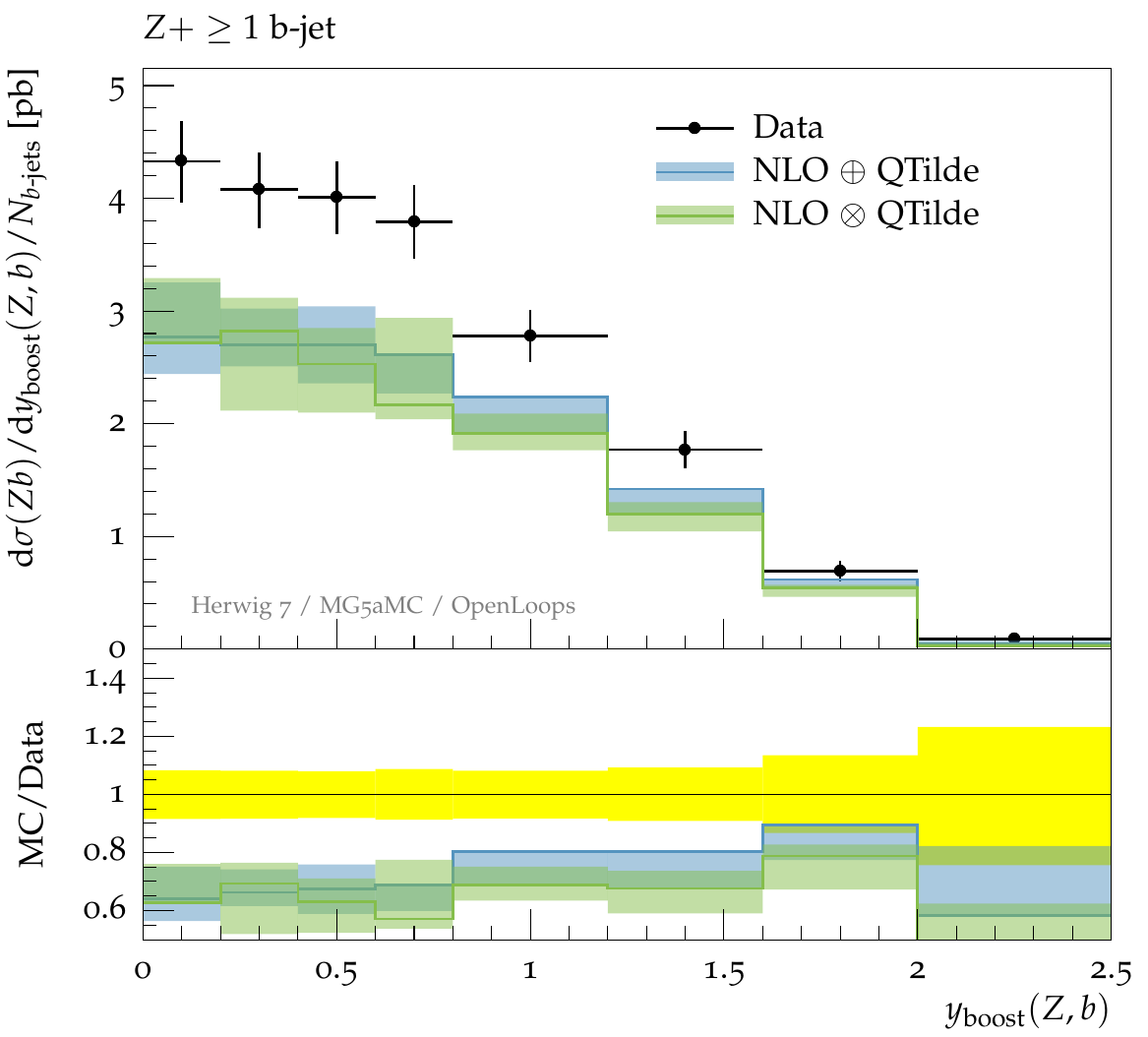} \\
   \includegraphics[scale=0.65]{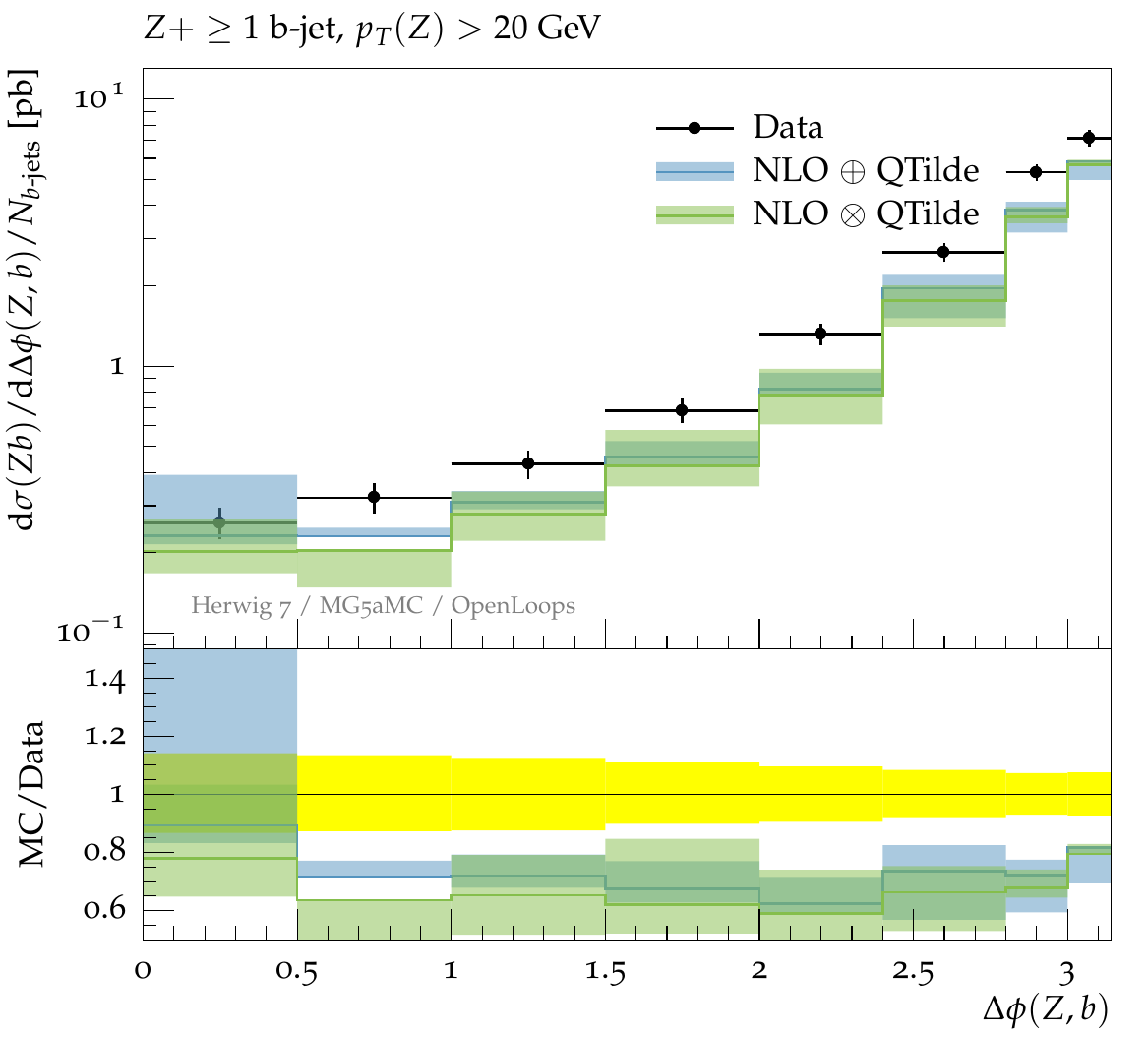}
   \includegraphics[scale=0.65]{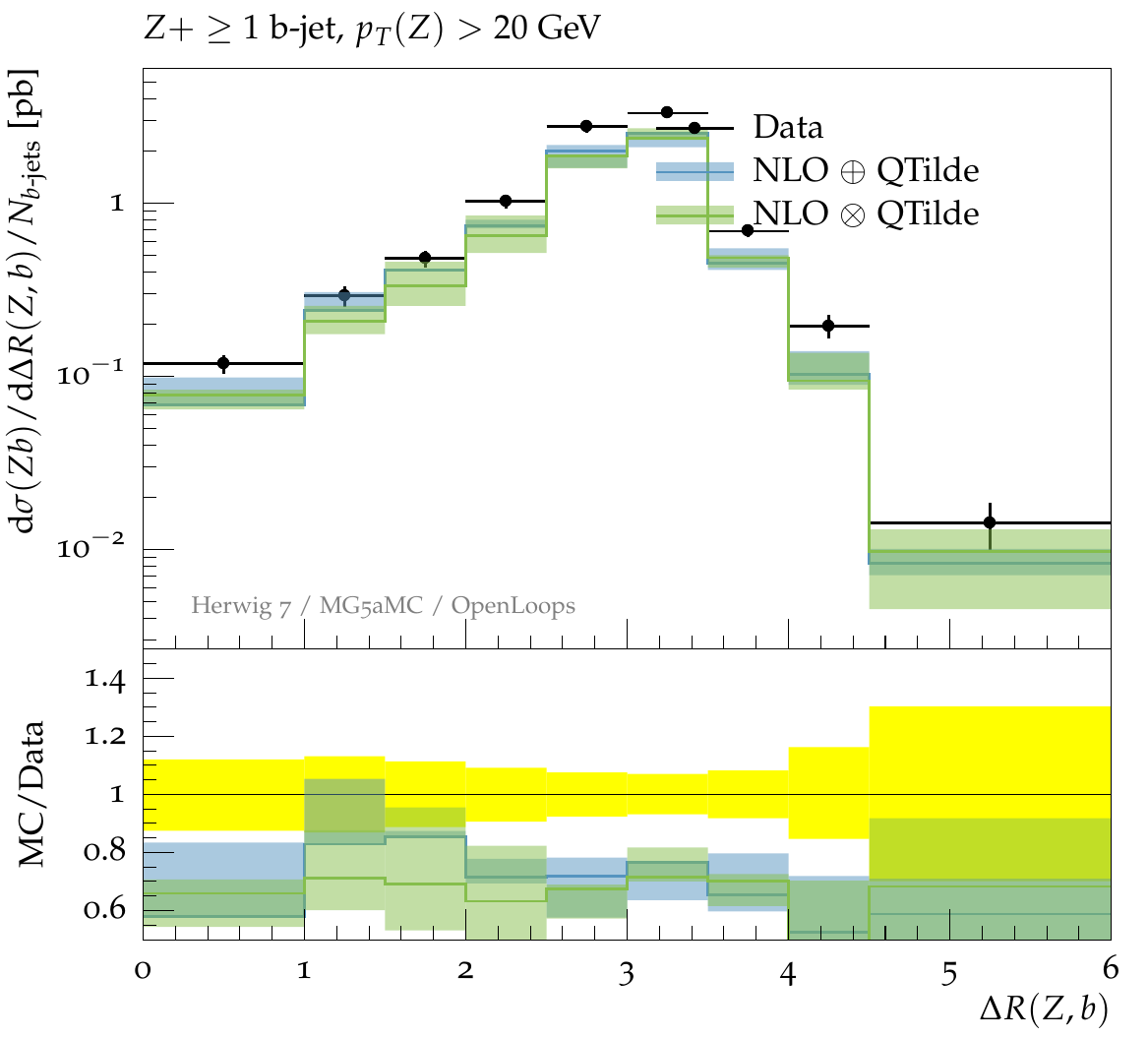}
\caption{A selection of the plots comparing \Herwig 4F, Zbb predictions to ATLAS
  results; together with the plotss in Figure~\ref{zbb-herwig4F-atlas}
  a comparison to the rescaled \Sherpa 4F NLO predictions in
  Figure~\ref{zbb-sherpa-scaled} can be made.}
\label{zbb-herwig4F-atlas-2}
\end{center}
\end{figure}

\begin{figure}[htbp]
   \includegraphics[scale=0.65]{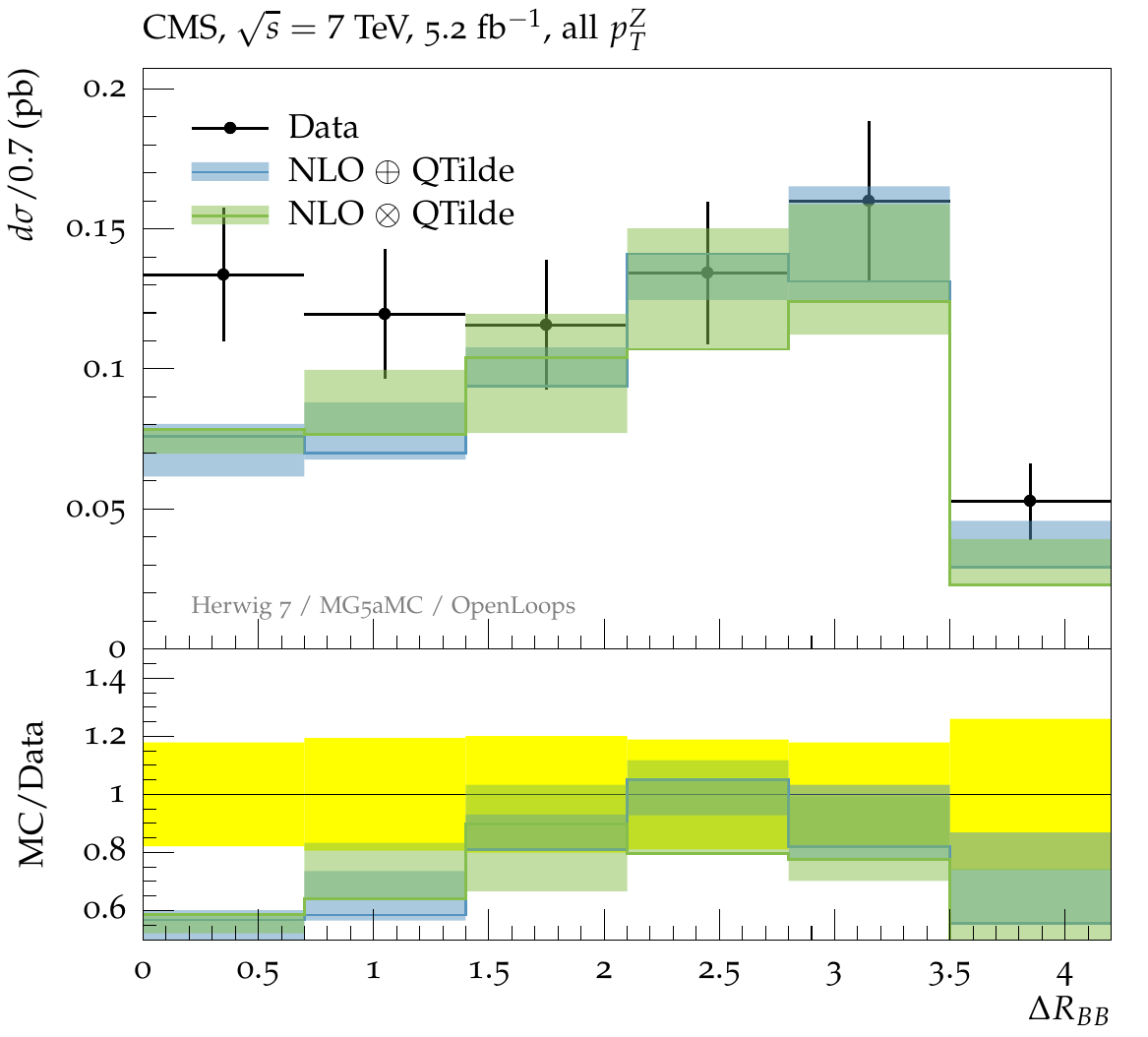}
   \includegraphics[scale=0.65]{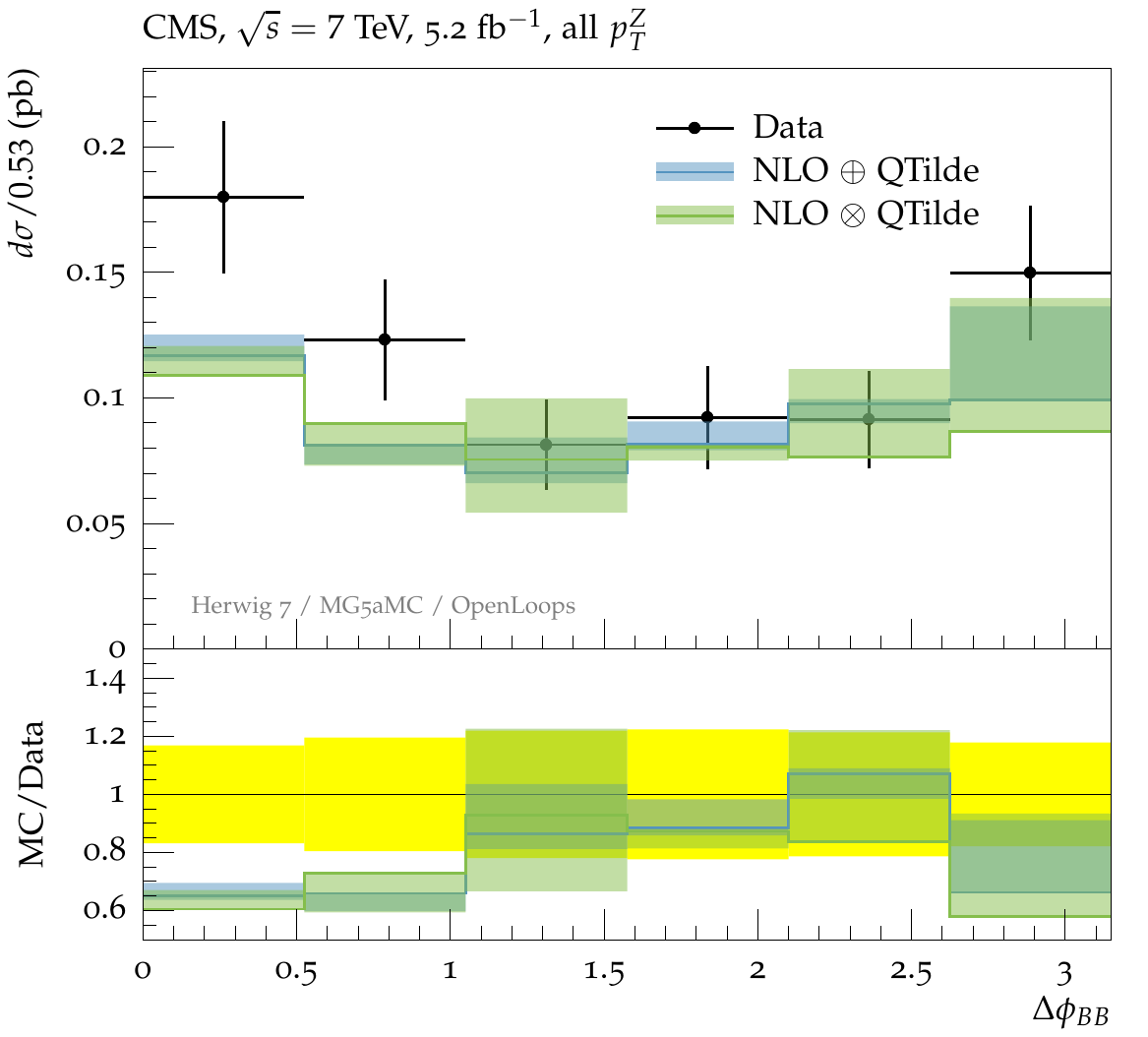}
\caption{A selection of the plots comparing \Herwig 4F, Zbb predictions to CMS results.}
\label{zbb-herwig4F-cms}
\end{figure}

A selection of results obtained with \Herwig is shown in
Figures~\ref{zbb-herwigzbb-atlas}, \ref{zbb-herwigzbb-atlas-2}
and~\ref{zbb-herwigzbb-cms} for the 5F, Zbb setup,
in Figures~\ref{zbb-herwigzb-atlas}, 
\ref{zbb-herwigzb-atlas-2} and~\ref{zbb-herwigzb-cms} for the 5F, Zb
setup,
and in Figures~\ref{zbb-herwig4F-atlas}, 
\ref{zbb-herwig4F-atlas-2} and~\ref{zbb-herwig4F-cms} for the 4F,
Zbb setup.
We refer to section~\ref{subsubsec:herwigsetup} for the process setups of
the 5F, Zb and Zbb samples, as well as for the 4F, Zbb sample.

Regarding the 5F, Zbb setup we note that we should expect somewhat large
uncertainties in observables which are also sensitive to events with 1 $b$ jet,
since at the level of the hard sub--process events with only 1 $b$ jet are only
described by the real emission in this sample. Within those uncertainties, the
prediction describes the data, with the exception of the $\Delta \phi(Z,b)$ and
$\Delta R(Z,b)$ observables in Figure~\ref{zbb-herwigzbb-atlas-2}, where the
prediction shows a slight tendency to be systematically above the data towards
the lower values of $\Delta \phi(Z,b)$ and $\Delta R(Z,b)$. Looking at the CMS
data comparisons the \textit{Dipole} shower together with the subtractive matching
seems to undershoot the data in the
$\Delta R_{BB}$ and $\Delta \phi_{BB}$ observables, which, however, seems not to
be the case in the corresponding observables (for $Z+\geq2b$-jets) in the ATLAS
data comparisons (not shown here); however, here the combinations with the
multiplicative matching seem to overshoot the data.

A brief internal study with the \textit{NLO}$\oplus$\textit{QTilde} combination
showed that a 5F, Zbj setup (where at the level of the hard sub--process events
with only 1 $b$ jet are already produced at the Born level) yields the expected
reduction in the uncertainty bands.

The predictions from the 5F, Zb setup describes the data overall well. There
seems to be a slight tendency, though, to systematically overshoot the data in
the $b\mathrm{-jet}\,|y|$ and $y_{\mathrm{boost}}(Z,b)$ observables for
$Z+\geq 1\,b$-jet in Figure~\ref{zbb-herwigzb-atlas-2} (the same holds for the
$\Delta y(Z,B)$ observable, not shown here).

With the 4F, Zbb setup we see that the predictions are generally below the
data. However, also here we notice a similar behaviour as already pointed out
for the \Sherpa results, in that the ratio of this 4F NLO prediction to data is
flat -- except for the $\Delta R_{BB}$ and $\Delta \phi_{BB}$ observables in
Figure~\ref{zbb-herwig4F-cms}, where we notice that the predictions fall below
the data at the lowest end of the distributions.

\subsection{W+b production \label{Wbb}}

\begin{figure}[htbp]
\begin{center}
   \includegraphics[scale=0.65]{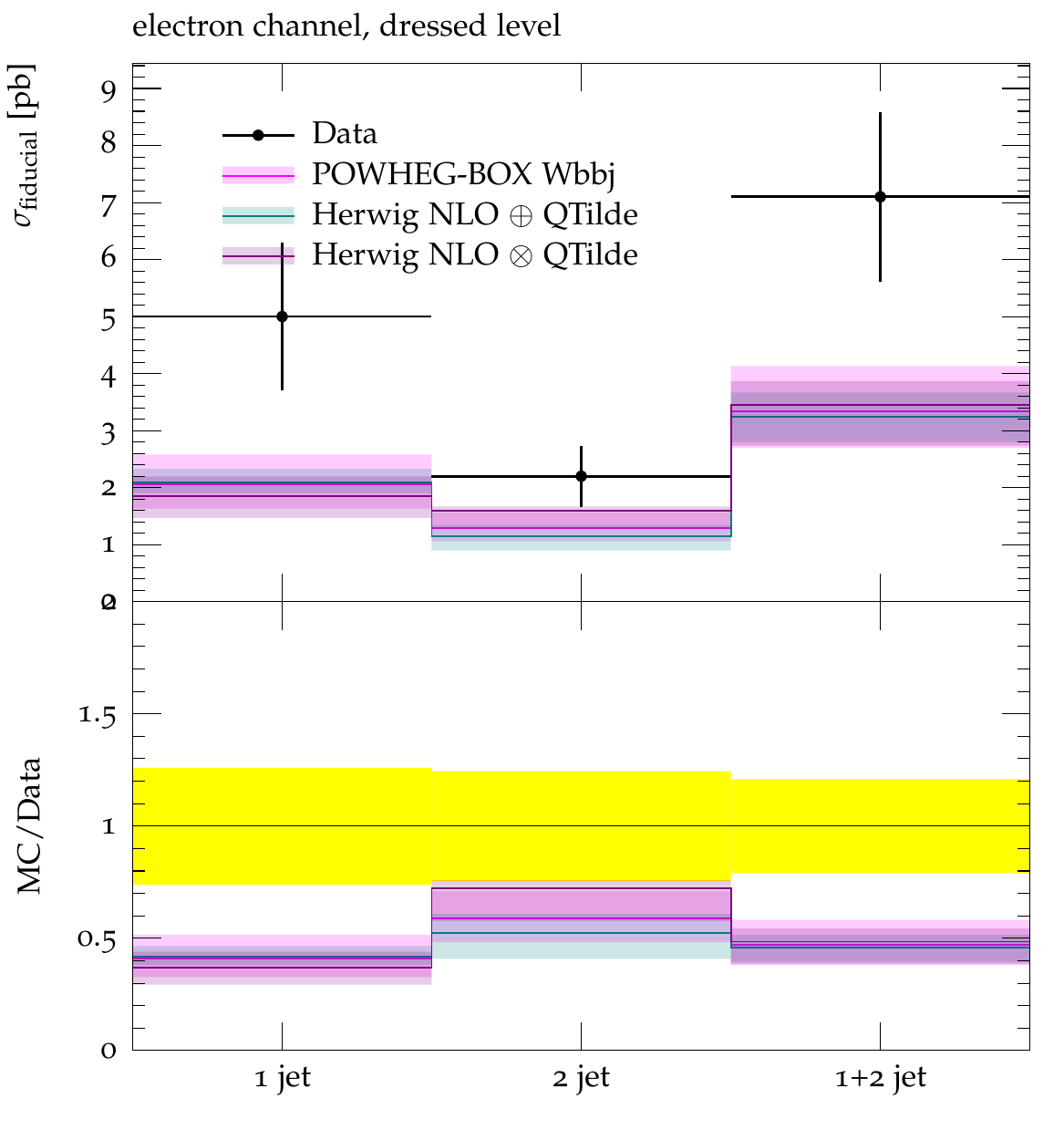}
\end{center}
\caption{ATLAS measured cross sections for $Wb$ production with only a
  b-tagged jet (``1 jet''), one b-tagged and at least an additional
  jet (``2 jet''), or both (``1+2 jet'') .  The theoretical results
  are at the full shower+hadron level. No DPI corrections are
  included. }
\label{fig:wbb-njet}
\end{figure}

\begin{figure}[htbp]
\begin{center}
   \includegraphics[scale=0.65]{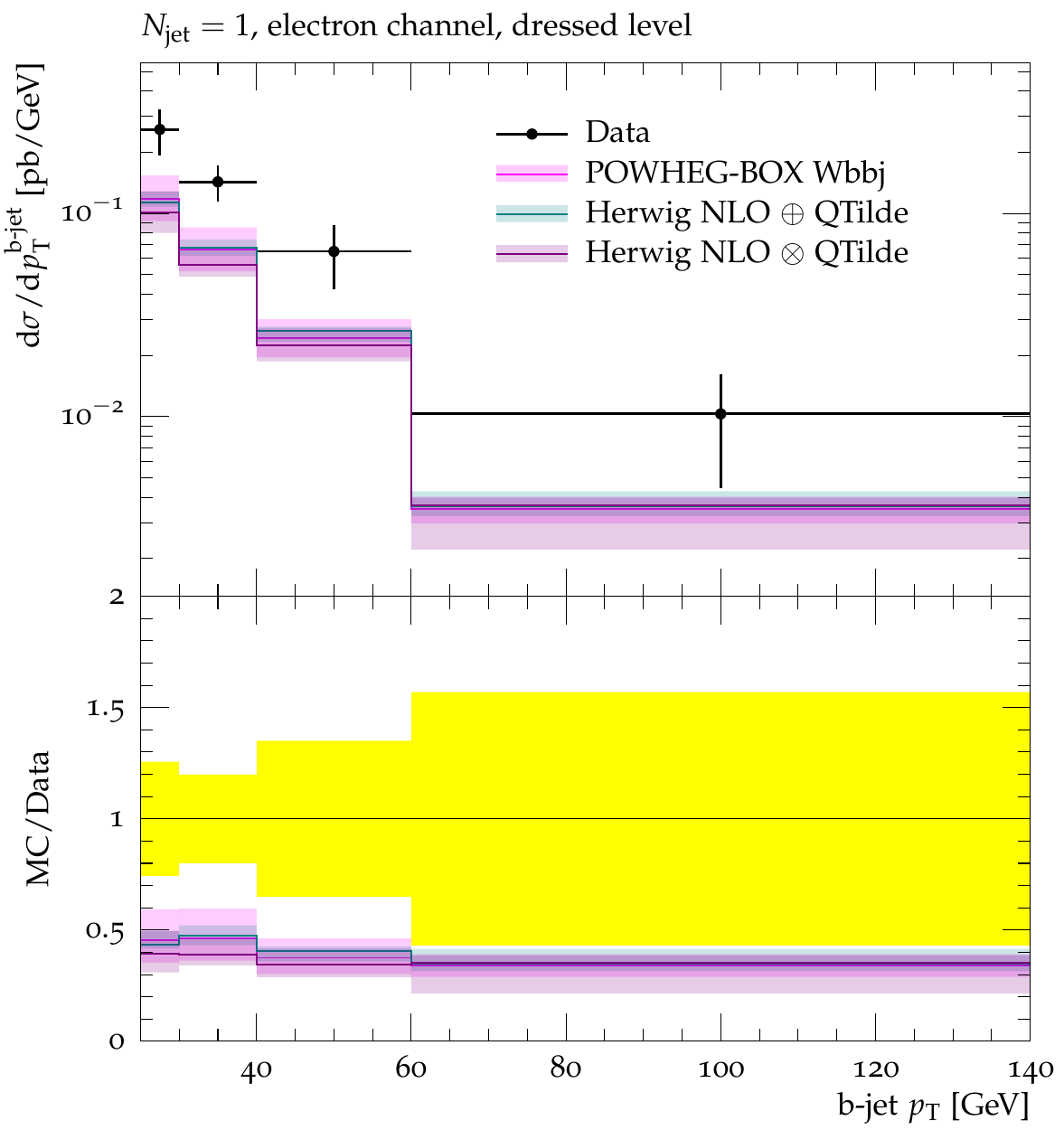}
   \includegraphics[scale=0.65]{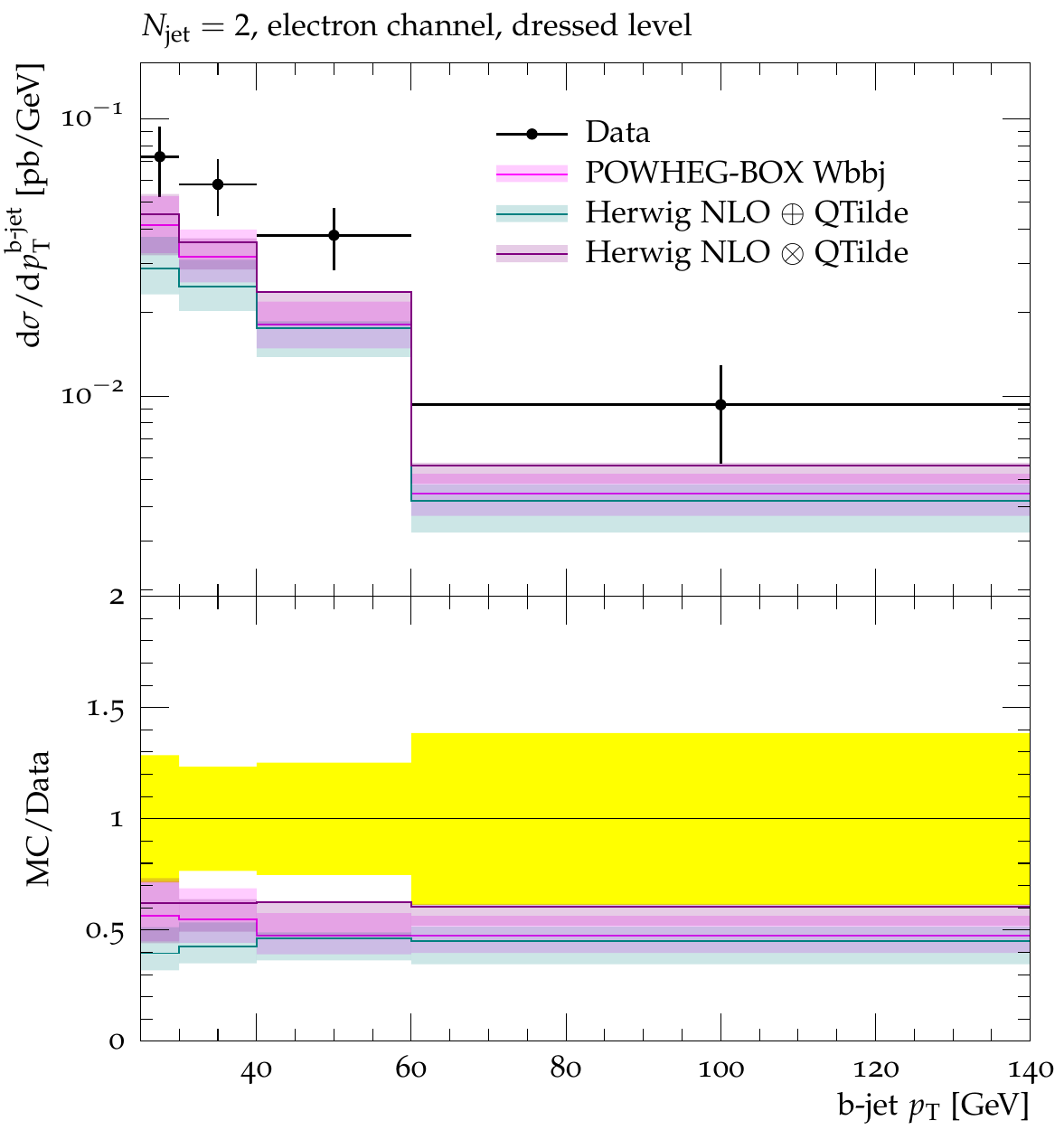}
\end{center}
\caption{ATLAS measured differential \pt distribution of the
  $b$-tagged jet in $W+b$ events with a single jet (left) or with at
  least one additional jet (right). The theoretical results are at the
  shower+hadron level. No DPI corrections are included.}
\label{fig:wbb-pt}
\end{figure}

\begin{figure}[htbp]
\begin{center}
   \includegraphics[scale=0.65]{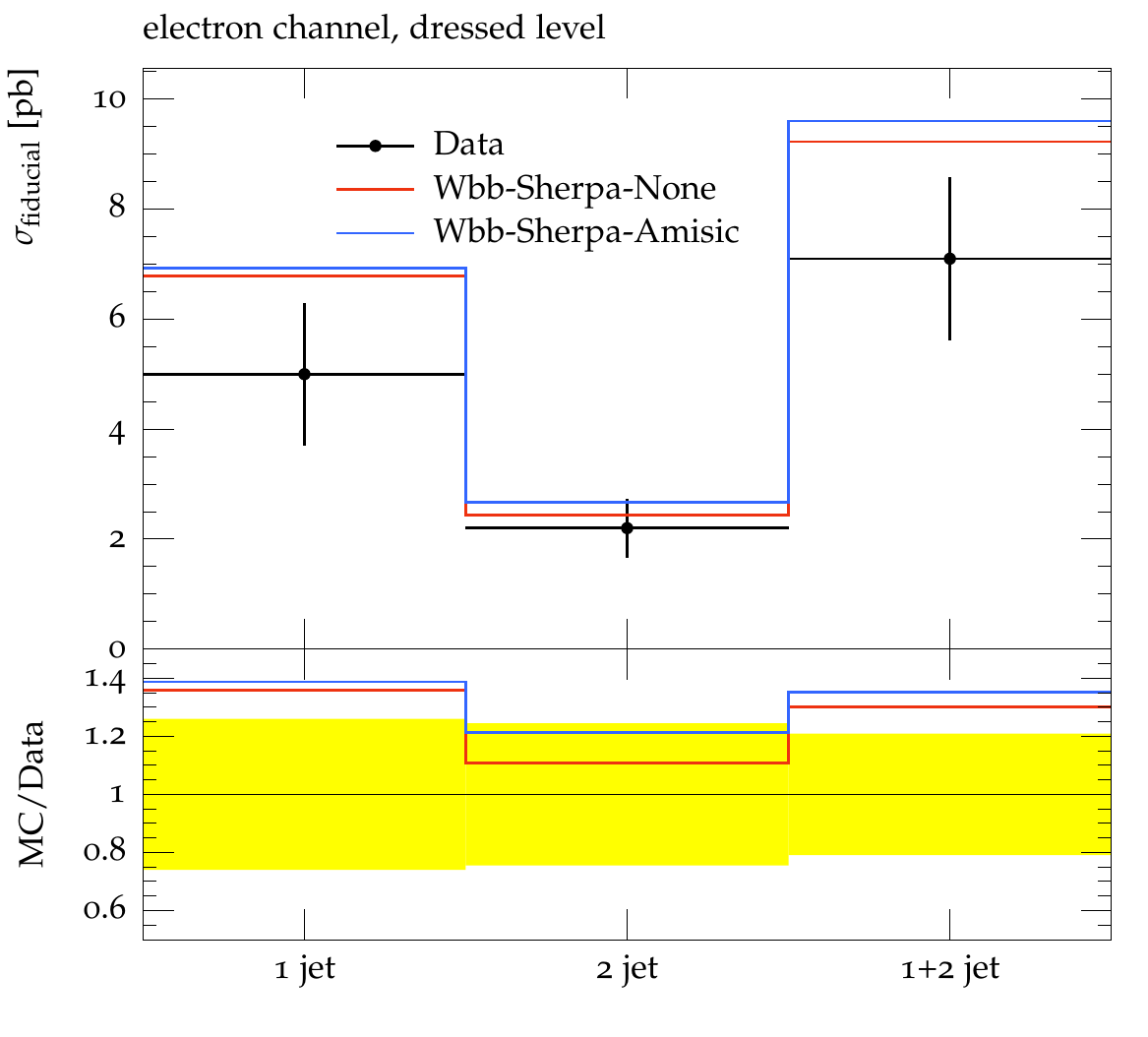}
\end{center}
\caption{ATLAS measured cross sections for $Wb$ production with only a b-tagged jet (``1
  jet''), one b-tagged and at least an additional jet (``2 jet''), or both (``1+2 jet'').
  Superimposed are shown the theoretical results obtained with \Sherpa
  at LO in the 5F scheme, both with and without MPI.} 
\label{fig:wbb-njet-sherpa}
\end{figure}

\begin{figure}[htbp]
\begin{center}
   \includegraphics[scale=0.65]{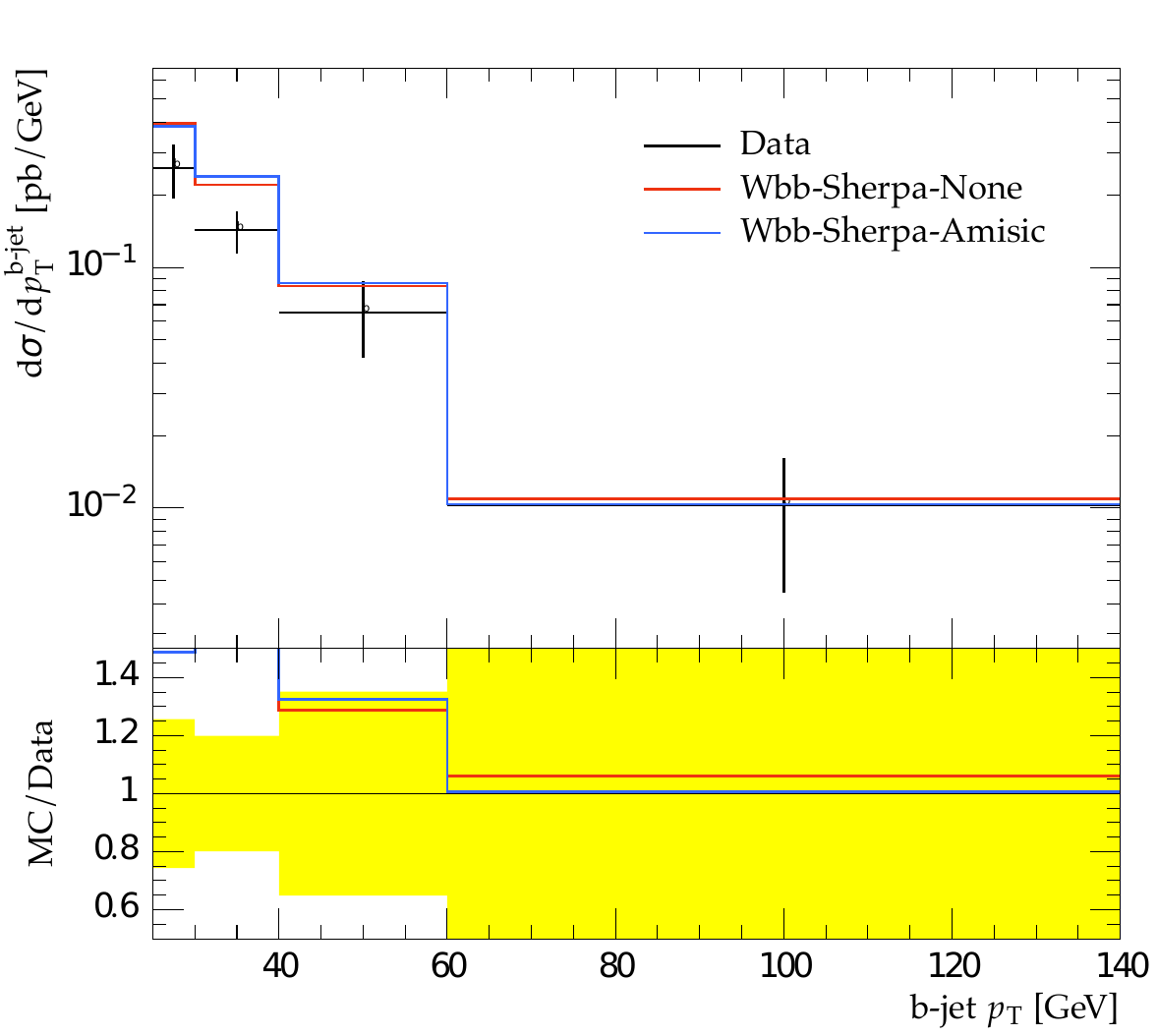}
   \includegraphics[scale=0.65]{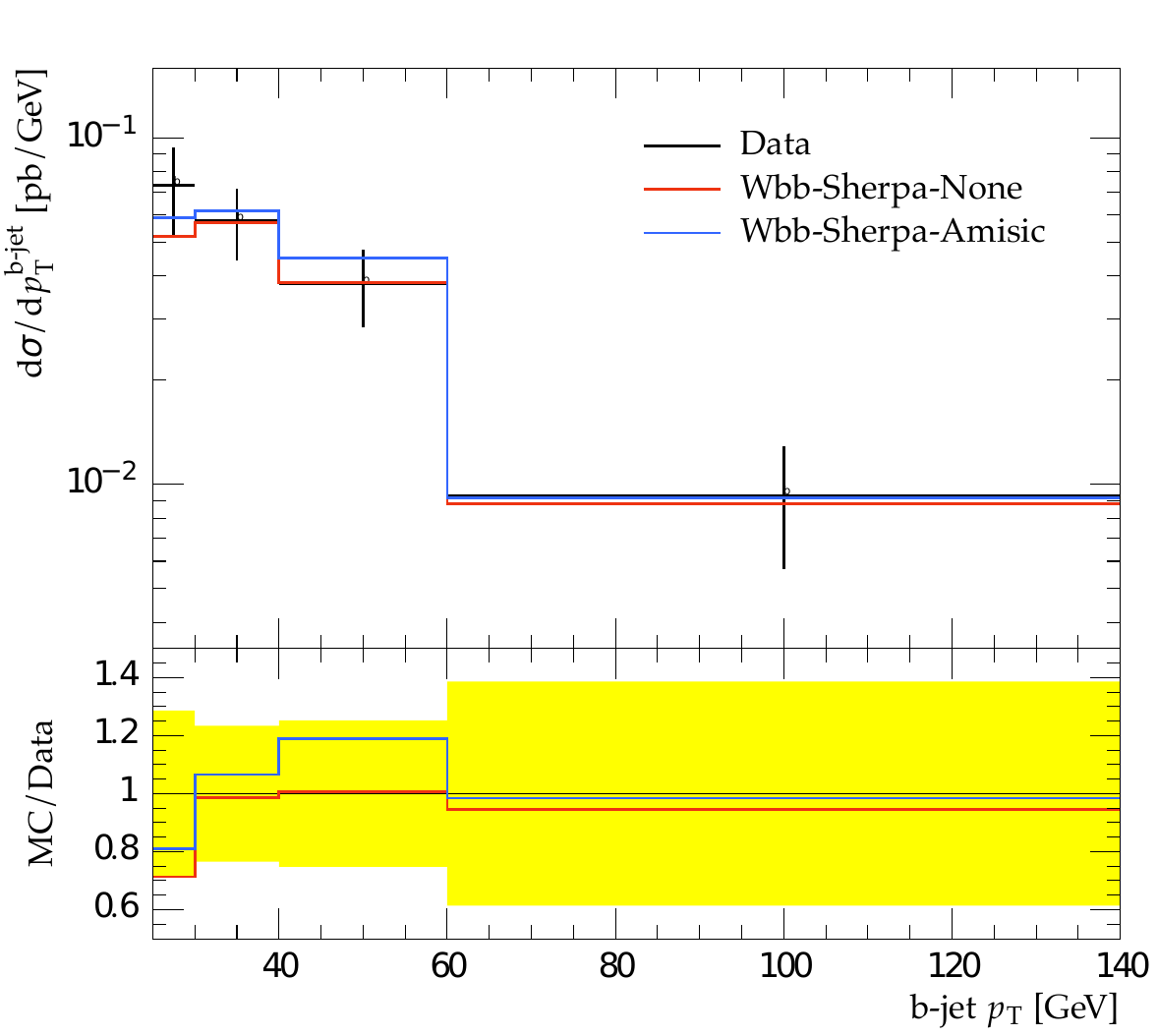}
\end{center}
\caption{ATLAS measured differential \pt distribution of the $b$-tagged jet in $W+b$ events with a
  single jet (left) or with at least one additional jet
  (right).Superimposed are shown the theoretical results obtained with
  \Sherpa at LO in the 5F scheme, both with and without MPI.}
\label{fig:wbb-pt-sherpa}
\end{figure}

Figures~\ref{fig:wbb-njet} and~\ref{fig:wbb-pt} present a comparison
between the data of the ATLAS W+b measurement and 4F NLO predictions obtained
with the \POWHEGBOX{} generator for $Wb\bar{b}j$ presented in
sec.~\ref{sec:powheg} and the \Herwig generator for $Wb\bar{b}$
discussed in sec.~\ref{subsubsec:herwigsetup}. Figures~\ref{fig:wbb-njet-sherpa} and~\ref{fig:wbb-pt-sherpa} present the same comparison but
with 5F LO predictions obtained with the \Sherpa generator as explained in sec.~\ref{sec:sherpa}.


The first Figure present results for the total cross section in the
case of only one b-tagged jet (``1 jet''), one b-tagged and at least
an additional jet (``2 jet''), or both (``1+2 jet''). Despite the
underlying generators consider two different underlying partonic
processes, $Wb\bar{b}j$ in the \POWHEGBOX{} and $Wb\bar{b}$ in the
case of \Herwig, the agreement between the two 4F NLO theoretical
predictions is very good and the uncertainties due to scale variations
are very similar.  The general good agreement between the $Wb\bar{b}$
and the $Wb\bar{b}j$ predictions is not trivial a priori due to the
different choices of scales in the two codes, and the fact that in the
$Wb\bar{b}j$ code the transverse momentum of the jet goes to zero and
the divergent behavior is regulated by the \MINLO{}
prescription~\cite{Luisoni:2015mpa}


The agreement of these 4F NLO predictions with the experimental data is however not so
good. As shown in~\cite{Luisoni:2015mpa} this can only partially be
compensated by the missing double-parton interaction (DPI)
corrections, which are estimated and applied as an additive correction
factor in the original ATLAS publication~\cite{Aad:2013vka}.
This is confirmed by the results obtained with \Sherpa with and without MPI.  
Compared to the estimate of DPI corrections given in the paper, the MPI contribution determined by \Sherpa is smaller for the ``1 jet''
case, but not incompatible given the large uncertainty: 0.2 pb vs $1.0^{+0.4}_{-0.3}$ pb; while a similar result is obtained
for the ``2 jet'' case: 0.4 vs $0.3\pm0.1$. 

The 5F LO prediction for the cross section agrees very well with data for the ``2-jet'' final state, while overshoot the data for ``1 jet'' by
more than one standard deviation (taking into account the experimental uncertainty only). Previous studies~\cite{Caola:2011pz}
have shown that the contribution from $qb \to Wbq'$ can be sizeable in the 5F scheme, while in the 4F scheme it appears at tree
level only in the NLO corrections to $q\bar{q'}\to Wb\bar{b}$. Nevertheless it is quite surprising that the 5F predictions at LO are in better
agreement with the data with respect to 4F at NLO. As pointed out in the introduction, this might arise from the presence of
large logarithms in the initial state, which can be properly resummed in the b-quark PDF. 

Turning to less inclusive distributions, the ATLAS analysis presented
above also contains measurements for the transverse momentum spectrum
of the $b$-tagged jet in the ``1 jet'' and ``2 jet'' samples. A
comparison of the data with the \POWHEGBOX{} and \Herwig predictions
for these observables is shown in the plots of
Figure~\ref{fig:wbb-pt}. Similarly to the inclusive case, also for
these distribution the different theoretical predictions are in very
good agreement among each other, but roughly a factor of two smaller
that the measurements. It is likely that also in this case, only part of
the difference can be explained with missing DPI corrections (not given in the paper for these differential distributions). 
It is interesting to note that, as for the Zb(b), the results are off by a constant factor, i.e. the shape is correct and only
the normalization seems to be wrong. 

The b-jet $p_T$ spectrum predicted by the 5F LO calculation is in very good agreement with data in the ``2 jet'' case, but it
is steeply falling in the ``1 jet'' sample, substantially overshooting the data for low $p_T$ values.

\subsection{Conclusions}
\label{sec:vbb:concl}

We presented a comparison of generators predictions using 4F and 5F scheme to most recent
measurements of vector boson production in association with b-jets at the LHC. In the 4F scheme a good agreement is found among the
different generators at NLO accuracy, and among different matrix-element to parton-shower matching algorithms. The
agreement with data however is good only when two b-jets are tagged in the final state or, when one b-jet only is
required, if a rescaling to the 5F integrated cross-section is applied. For Wbb, even taking into account the contribution from MPI,
predictions seem to significantly undershoot the data. The Zb(b) production has been
compared with predictions obtained in the 5F scheme with different setups, i.e. explicitely requiring one or two b-jets
in the final state or a b-quark in the incoming proton when calculating the matrix element; or with no requirements on
b-jets (treating them as light quarks) and combining final states with additional jets with a merging technique at LO
and, where possible, also at NLO. Pro and cons of the different approaches are more difficult to pin down. In some case
the scale uncertainty is quite large and not all distributions shows a nice agreement with data, especially if two
b-tagged jets are present in the final state. For the Wb final state results have been also compared to a 5F calculation at LO:
even if the precision is limited, because higher order corrections are missing, the cross section
agrees better with data than for the 4F NLO calculation, as observed for the Zb(b) process and as previous NLO studies at parton
level suggested. 

Overall these results show that the associated production of vector bosons and b-jets is still an important benchmark for
perturbative QCD at hadron colliders. More measurements and additional theoretical studies are definitely needed.

\subsection*{Acknowledgements}

S.~Platzer acknowledges support by a FP7 Marie Curie Intra European 
Fellowship under Grant Agreement PIEF-GA-2013-628739. The work of 
C.~Reuschle is supported in part by the U.S. Department of Energy 
under grant DE-FG02-13ER41942.

\section{Irreducible backgrounds and measurement uncertainties
\texorpdfstring{\protect\footnote{
  J.~Butterworth, F.~Krauss,
  V.~Ciulli, P.~Francavilla, V.~Konstantinides, 
  P.~Lenzi, C.~Pandini, L.~Perrozzi, L.~Russo, 
  M.~Sch\"onherr, U.~Utku, L.~Viliani, B.~Waugh
  }}{}}
\label{thisLH_bkgsub}



\subsection{Introduction}
\label{sec:intro}

The general principle of minimising the model-dependence of results from particle colliders by making measurements of 
well-defined final states in fiducial regions is by now widely accepted, and implemented by the LHC collaborations. 
The fiducial regions are designed to reflect the acceptance of the detectors and data-selection. 
The final states are defined in terms of stable, or quasi-stable,
particles. Increasingly impressive theoretical calculations are able to implement the appropriate kinematic cuts, and
modulo some uncertainty associated with soft physics (for example hadronisation), can predict precisely what 
is actually being measured, without the need for additional assumptions or extrapolations into unmeasured regions of 
phase space.

This represents great progress. One area, however, where the principle of defining a measurement in terms of the final state
is not so widely implemented, is in the consideration of background processes and their subtraction. 
Often backgrounds are subtracted using a mixture of theoretical and data-driven techniques, 
even though in some cases the backgrounds are strictly speaking ``irreducible'', in that they produce final states 
identical to the ``signal'' final state (even in a perfect detector) and thus should be added to the signal 
at the amplitude, rather than cross-section, level. These subtractions are also often carried out before, or intermingled with, 
the unfolding and correction for detector effects such as efficiency and resolution, and thus are 
impossible to revert or reproduce once applied.

In practice, the uncertainty introduced by such subtractions is often insignificant compared to other uncertainties in the measurements, 
for example because the kinematic overlap
is in fact small and interference terms are negligible. 
Nevertheless, as precision of both experiment and theory increase, such considerations can become important in some processes.
In this contribution we highlight some such cases in an attempt 
to raise awareness of the issues for future studies. 

\subsection{Single top and $W+b$-jet production}

\begin{figure}
\begin{center}
\includegraphics[width=0.48\textwidth]{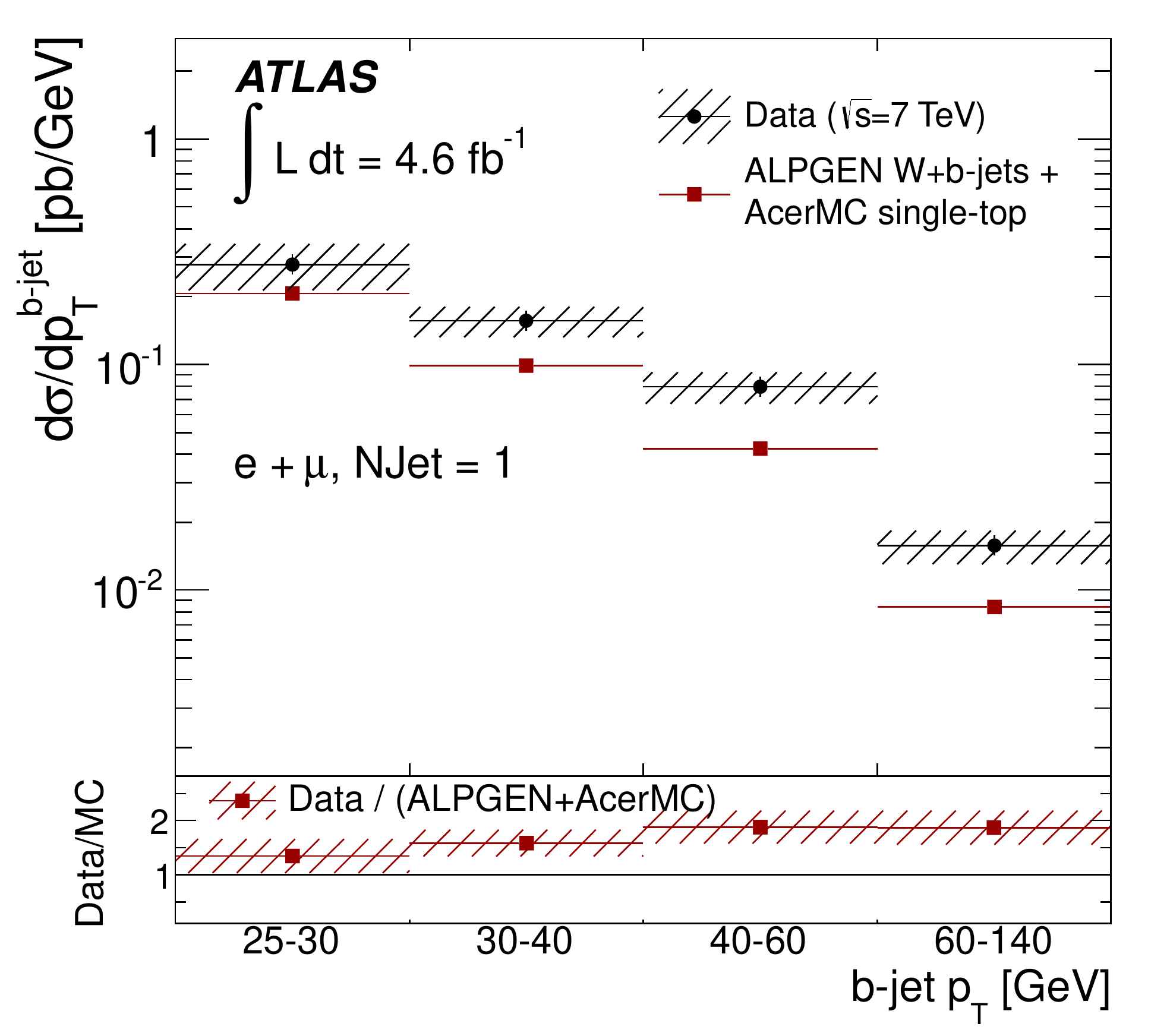}
\includegraphics[width=0.48\textwidth]{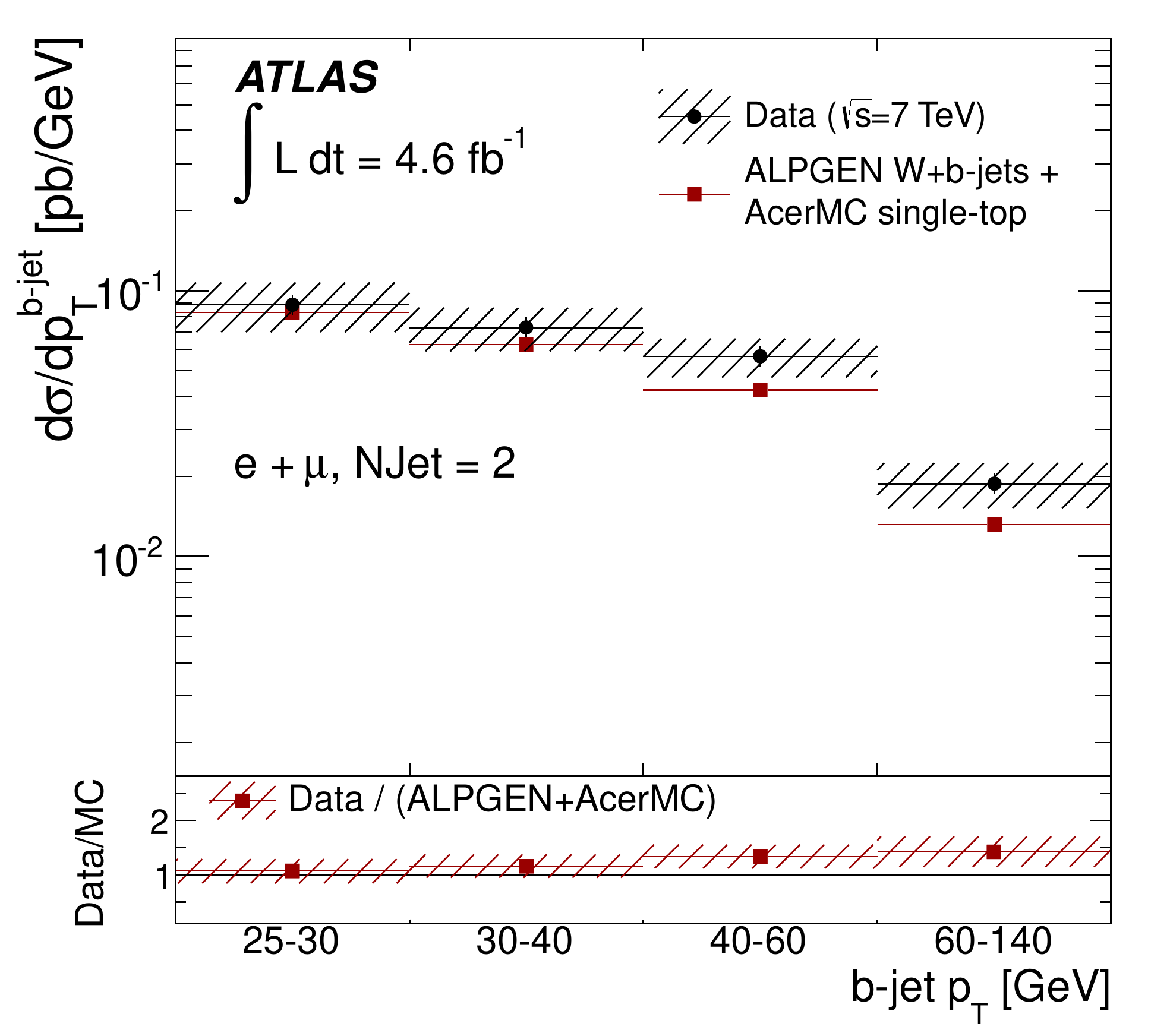}
\caption{\label{fig:wb}
Measured differential $W+b$-jet cross-section before single-top subtraction as a function of the transverse momentum of the $b$-jet, in the 
1-jet bin (left) and 2-jet bin (right). 
The measurements are compared to the sum of separate $W+b$-jet and single-top predictions 
obtained using ALPGEN interfaced to HERWIG and JIMMY and scaled by a NNLO inclusive $W$ normalization factor, and ACERMC interfaced to PYTHIA and scaled to a 
NLO single-top cross-section. 
The ratios between measured and predicted cross-sections are also shown. From~\protect\cite{Aad:2013vka}.}
\end{center}
\end{figure}

\begin{figure}
\centering
	\includegraphics[width=0.48\textwidth]{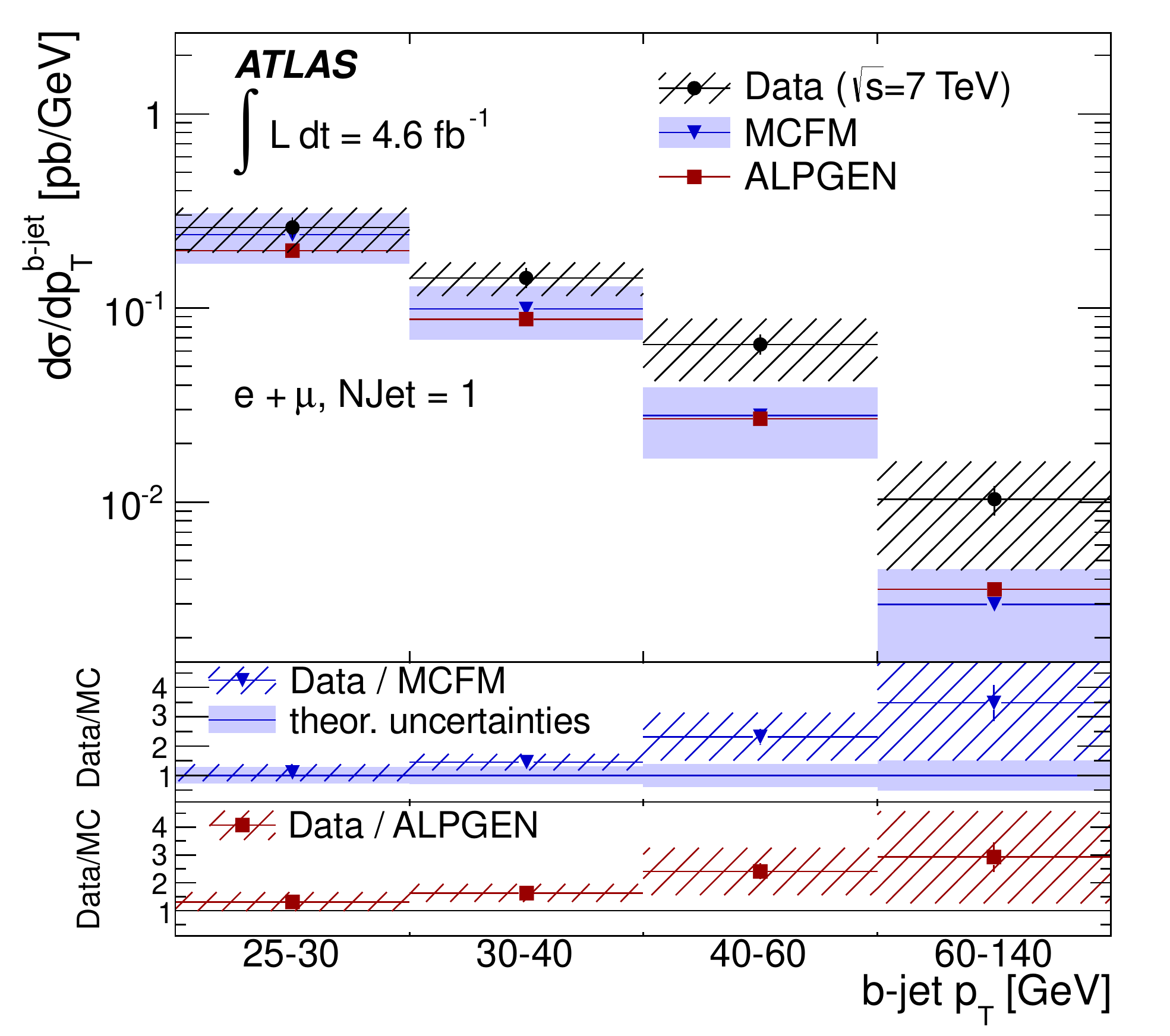}
	\includegraphics[width=0.48\textwidth]{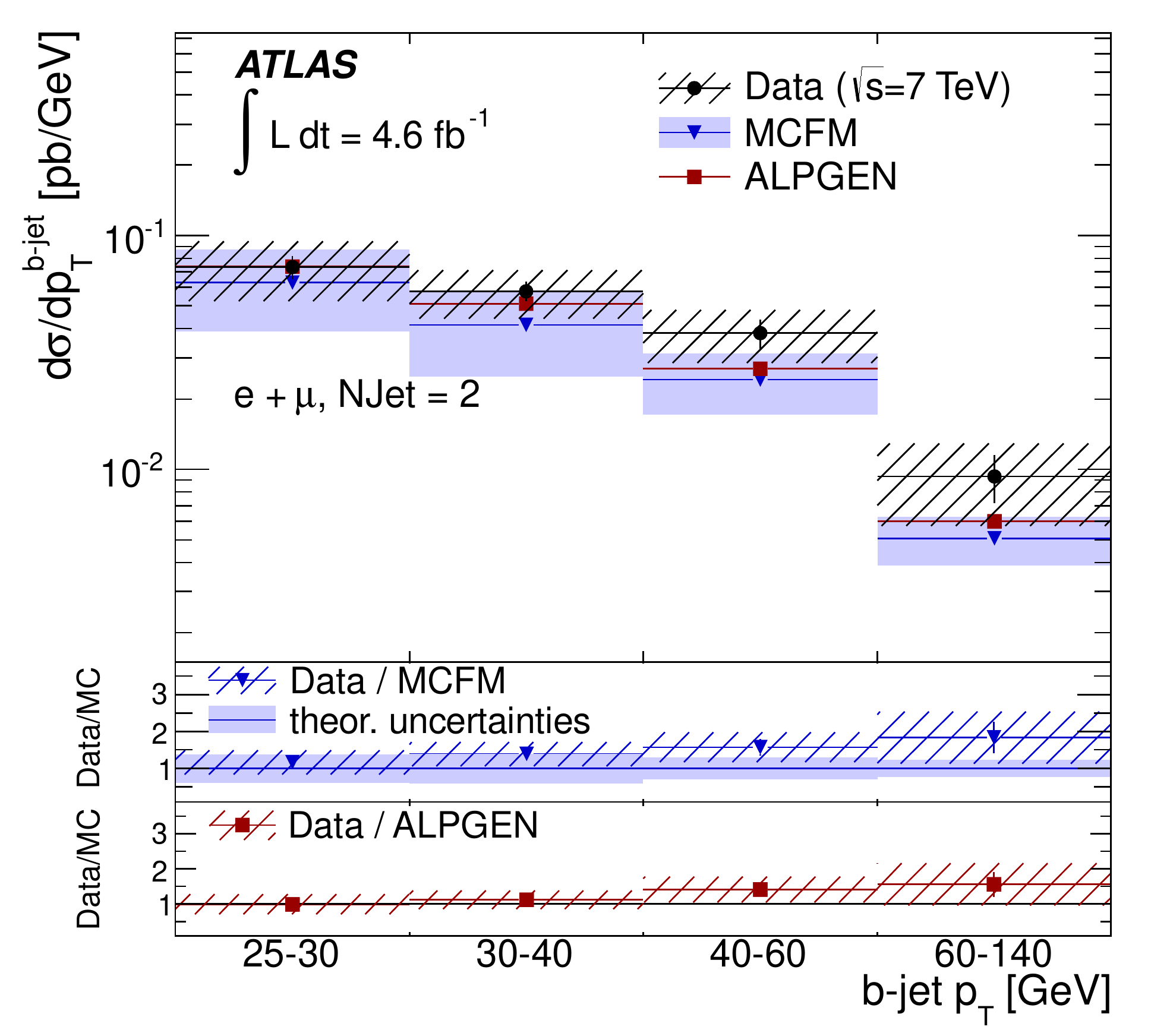}
\caption{\label{fig:notop}
Measured differential $W+b$-jet cross-section after single-top subtraction as a function of the transverse momentum of the $b$-jet, in the 
1-jet bin (left) and 2-jet bin (right). 
The measurements are compared to the a calculation of $W+b$-jet production in the absence of top quark propagators obtained using ALPGEN interfaced to HERWIG and JIMMY and scaled by a NNLO inclusive $W$ normalization factor, and ACERMC interfaced to PYTHIA and scaled to a NLO single-top cross-section. 
The ratios between measured and predicted cross-sections are also shown. From~\protect\cite{Aad:2013vka}.}
\end{figure}

An example of a final state in which two contributions are often treated as distinct processes is the measurement of a 
leptonically-decaying $W$ boson (that is, charged lepton plus missing transverse energy) in association with a $b$-tagged hadronic jet. 
The publication of the ATLAS analysis of 7~TeV LHC collision data~\cite{Aad:2013vka} contains a measurement of the fiducial $W+b$-jet
cross section, 
presented as a function of jet multiplicity and of the transverse momentum of the leading $b$-jet. 
The $W+b$-jet cross-section, corrected for all known detector effects, 
is quoted in a limited kinematic range, using jets reconstructed with the anti-kt
clustering algorithm with transverse momentum above 25 GeV and absolute rapidity within 2.1.
The measurement is presented before and after the subtraction of the single-top contribution to the identical final state. Both
versions are available in HEPDATA~\cite{hepdata} and Rivet~\cite{Buckley:2010ar}\footnote{The Rivet analysis was modified to add the histograms 
for the unsubtracted data.}. 
The unsubtracted version is shown in Fig.~\ref{fig:wb}, and the subtracted version in Fig.~\ref{fig:notop},
in the case where the b-jet is the only jet in the ﬁducial region (1-jet bin) and when there is an additional jet (2-jet bin).
 
Several things may be noted:
\begin{itemize}
\item In neither case does the theory describe the data especially well. This is a challenging
final state to predict and the theory is likely to be superseded by more sophisticated and 
accurate predictions in future (indeed, NLO implementations of this
process in MC are already available, as discussed in
Section~\ref{cha:pheno}.\ref{thisLH_Vbb} of these proceedings). This strongly mitigates against embedding in a dependency 
on the theory in the experimental analysis - as is the case if the background is subtracted at detector-level - 
and is a strong motivation for the unsubtracted version of the
measurement. 
\item The contributions from diagrams 
with and without top are comparable (as can be noted from the cross section in the highest $p_T$ bin).
\item The data uncertainties on the unsubtracted version are smaller.
\end{itemize}

Integrated over $p_T$, the unsubstracted fiducial cross section is
$9.6 \pm 0.2 (stat) \pm 1.7 (syst)$ pb, a relative systematic uncertainty of 18\%.
The corresponding subtracted measurement is 
$7.1 \pm 0.5 (stat) \pm 1.4 (syst)$ pb, a relative systematic uncertainty of 20\%
- a small but noticeable decrease in precision. Looking in more detail, the main contributions 
to the systematic errors are:
\begin{itemize}
\item Jet energy scale: 10-50\%
\item Modelling of initial and final state QCD radiation ($Wb$, single top, $t\bar{t}$): 2-30\%
\item $b$-tagging: 1-8\%
\item MC modelling of the $Wb$ process: 2-8\%
\end{itemize}

The fact that jet energy scale dominates masks, to a large extent, the effect of the modelling uncertainties introduced by
the background subtraction.
The uncertainty due to the modelling of QCD radiation varies strongly with jet $p_T$. 
This is exactly the kind of model dependence which one would expect to increase if a theory-based background
subtraction is made, and indeed, in the highest $p_T$ bin the systematic uncertainty goes from 16\% before subtraction 
to 54\% after it (compare Table~4 with Table~9 of Ref.~\cite{Aad:2013vka}).

The comparisons were repeated using Sherpa 2.2~\cite{Gleisberg:2008ta} and Herwig7~\cite{Bellm:2015jjp}. 
\\
For Sherpa, all intermediate particles in the matrix element are kept on-shell and the AMEGIC 
ME generator is used for LO calculations~\cite{Krauss:2001iv}. Only decays of the $W$ boson to the 
electron channel are allowed. Multi-parton interactions are switched off. 
The Sherpa default 5-flavour pdf library (NNPDF~\cite{Ball:2014uwa}) is used. In $Wb$ production
without tops, the $b$-quark is treated as massive with a mass of
4.75~GeV and the $W$ boson is treated through the narrow width approximation. The order of the electroweak couplings is fixed to 2. 
For single top production, the $b$-quark is treated as massless in the matrix element calculation
but retains its mass settings in the rest of the simulation. QCD and EW order couplings are not
fixed. Production modes include all channels: $s$-channel, $t$-channel and $tW$ single-top channels.
\\
For Herwig, the built-in matrix elements for $W$+jet and single top were used. All leptonic decays
were generated, but the electron only channel was selected in Rivet, with a normalisation factor
of three applied post-hoc. Production includes $s$-channel,
$t$-channel and $tW$ single-top channels. The pdf MMHT2014~LO~\cite{Harland-Lang:2014zoa} is used. 

\begin{figure}
\centering
	\includegraphics[width=0.48\textwidth]{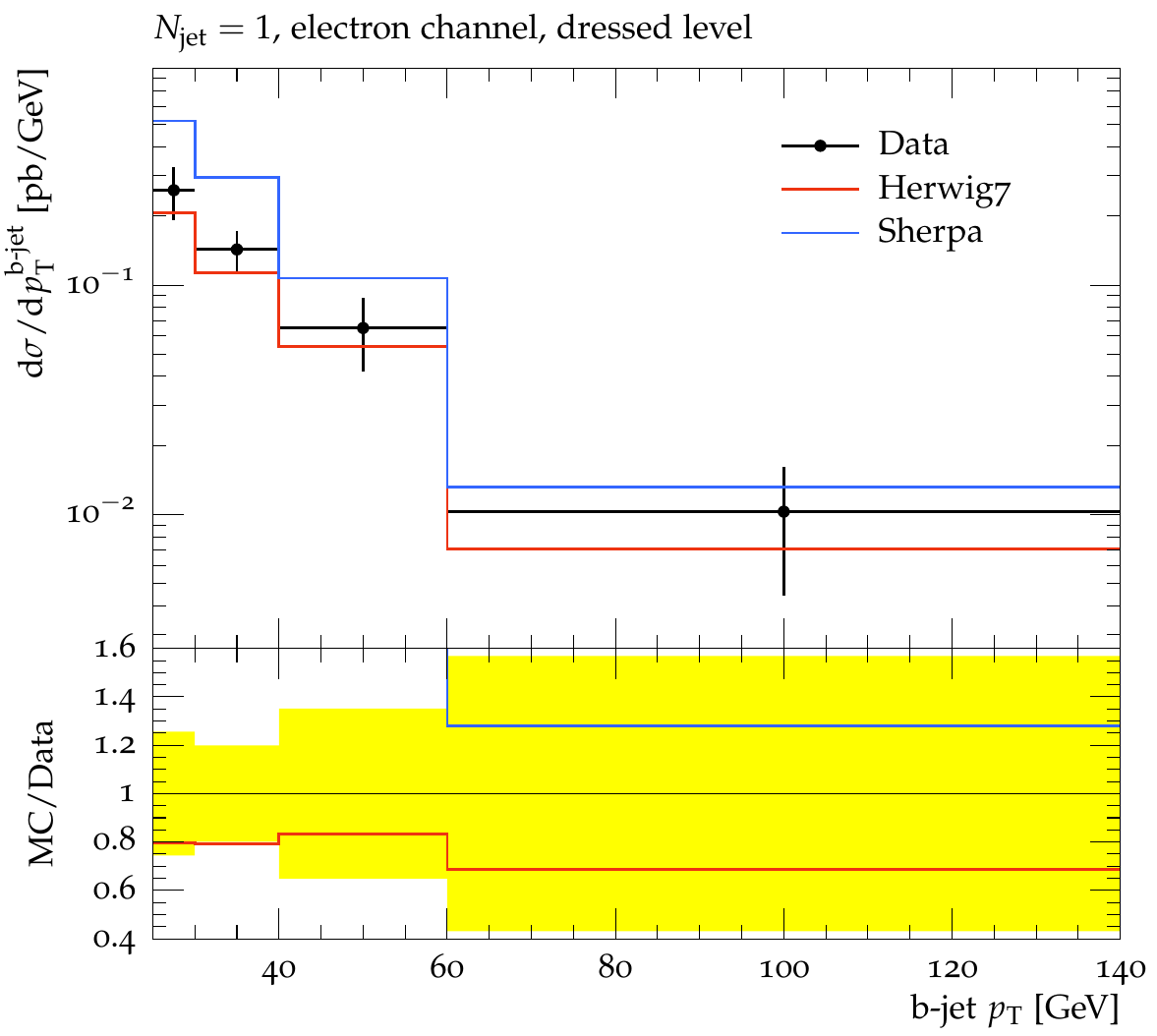}
	\includegraphics[width=0.48\textwidth]{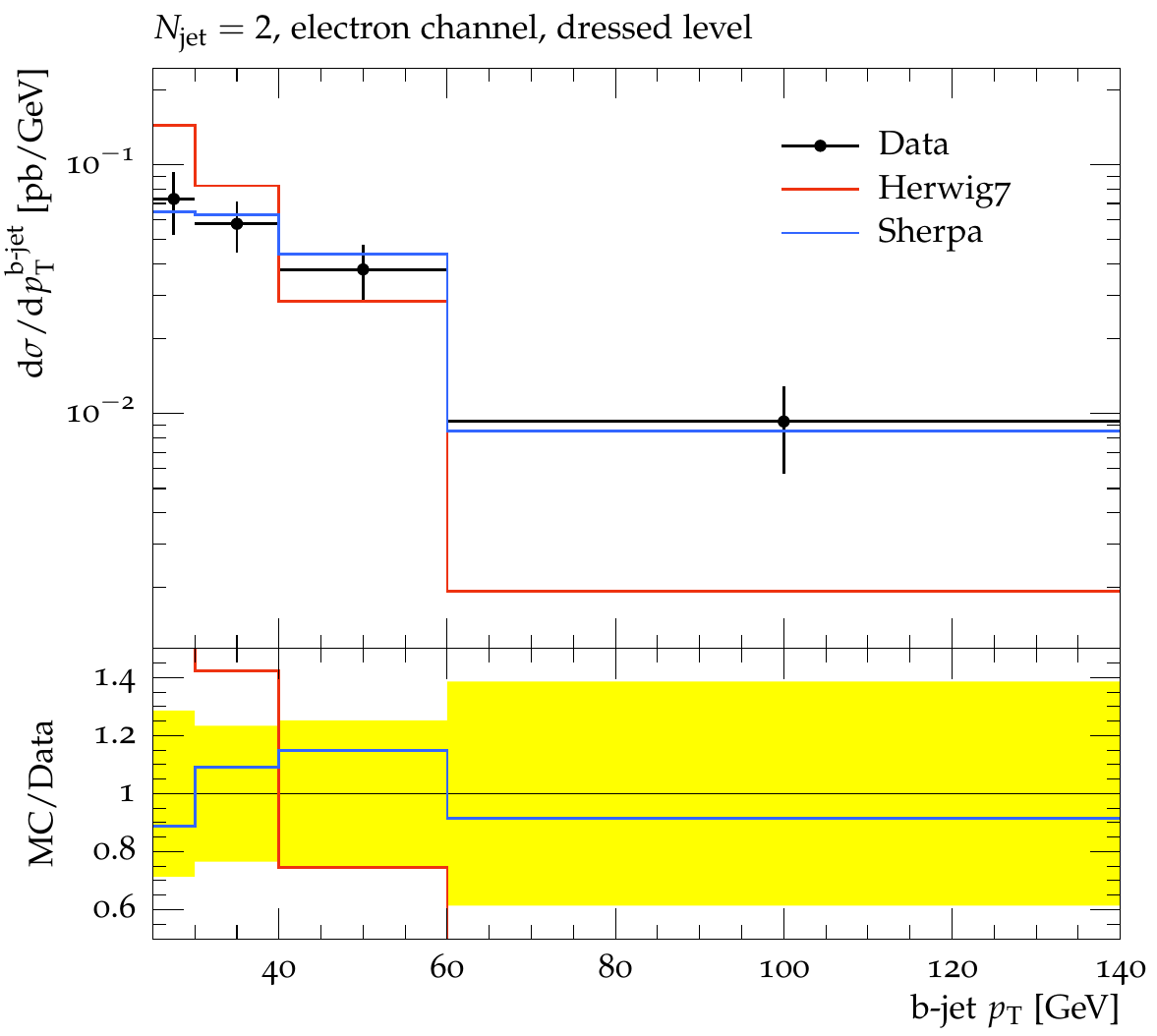}
\caption{\label{fig:subtracted}
  Measured differential $W+b$-jet cross-section after single-top subtraction as a function of the
  transverse momentum of the $b$-jet, in the 1-jet bin (left) and 2-jet bin (right). 
  The measurements are compared to the expectations of
  Sherpa and Herwig, for $Wb$ production processes excluding diagrams containing top quarks.}
\end{figure}

The comparison of the non-top diagrams only to the subtracted data is shown in Fig~\ref{fig:subtracted}. 
With these settings, Herwig agrees with the data normalisation both in the 1-jet and 2-jet bins.
It correctly models the shape of the $b$-jet $p_T$ in the 1-jet bin, but fails to describe it in the 2-jet bin.
Sherpa models well the $p_T$ dependence in both jet bins, but overestimates the normalisation of the 1-jet bin 
by nearly a factor of 2.

\begin{figure}
\centering
	\includegraphics[width=0.48\textwidth]{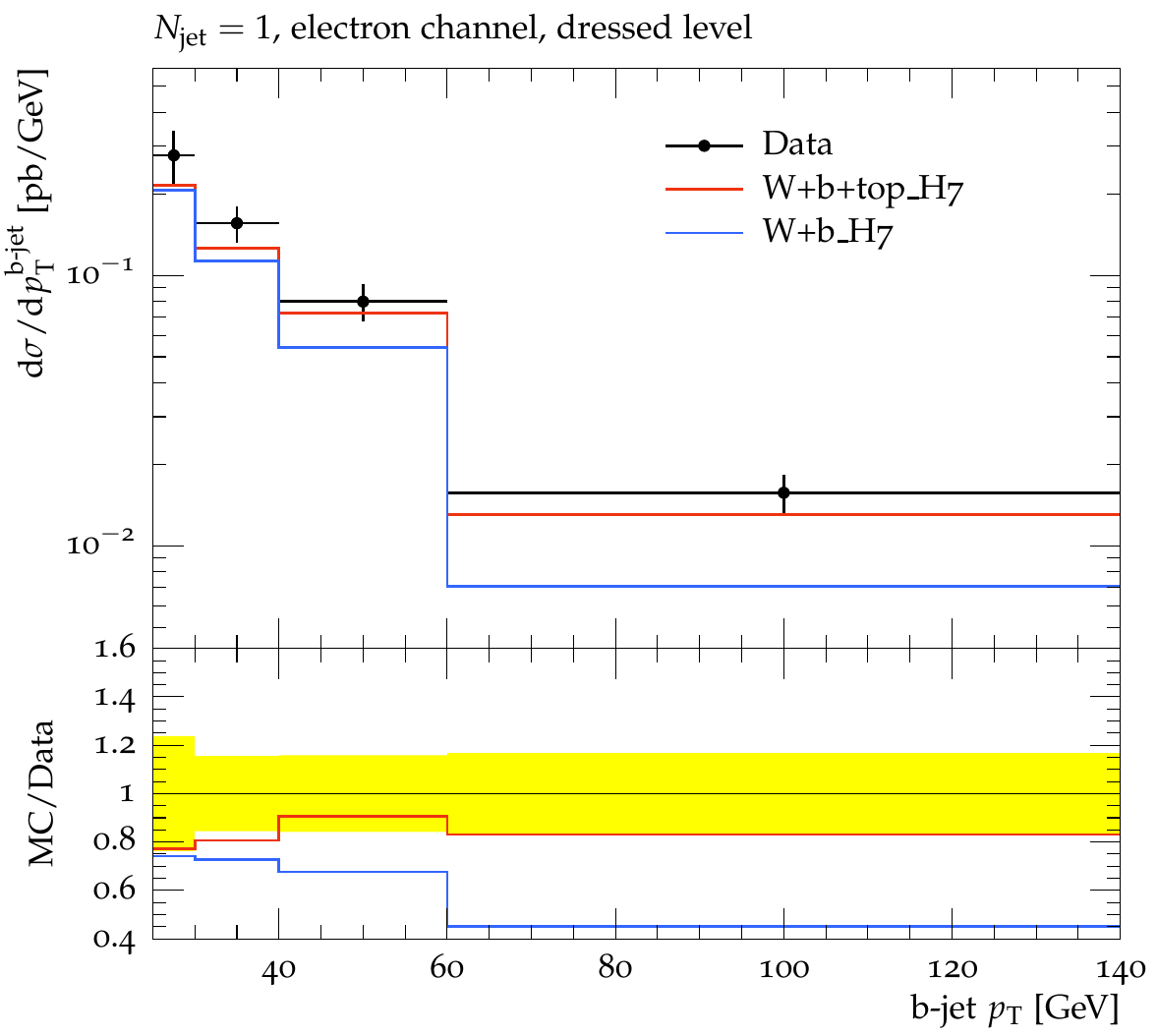}
	\includegraphics[width=0.48\textwidth]{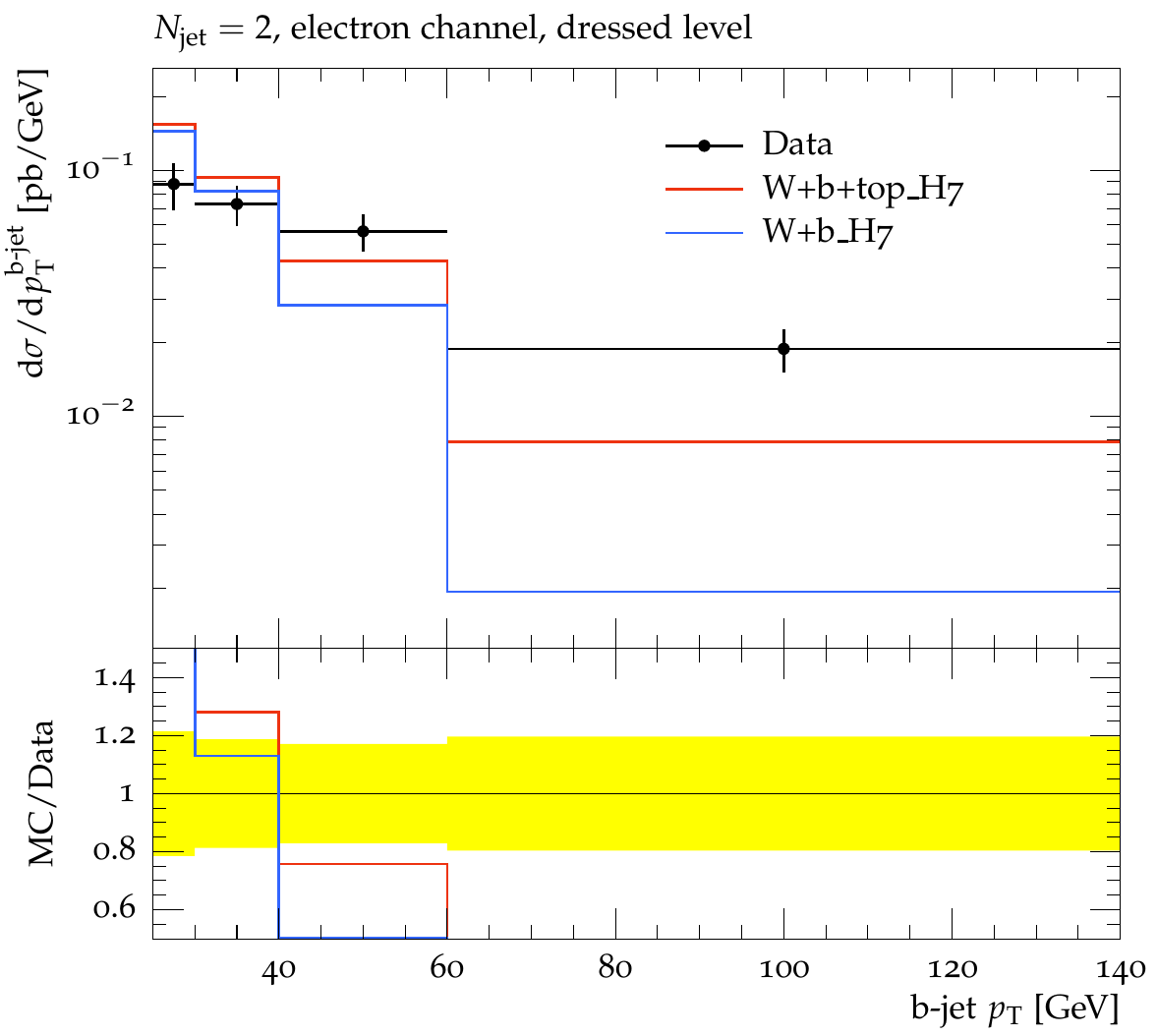}
	\includegraphics[width=0.48\textwidth]{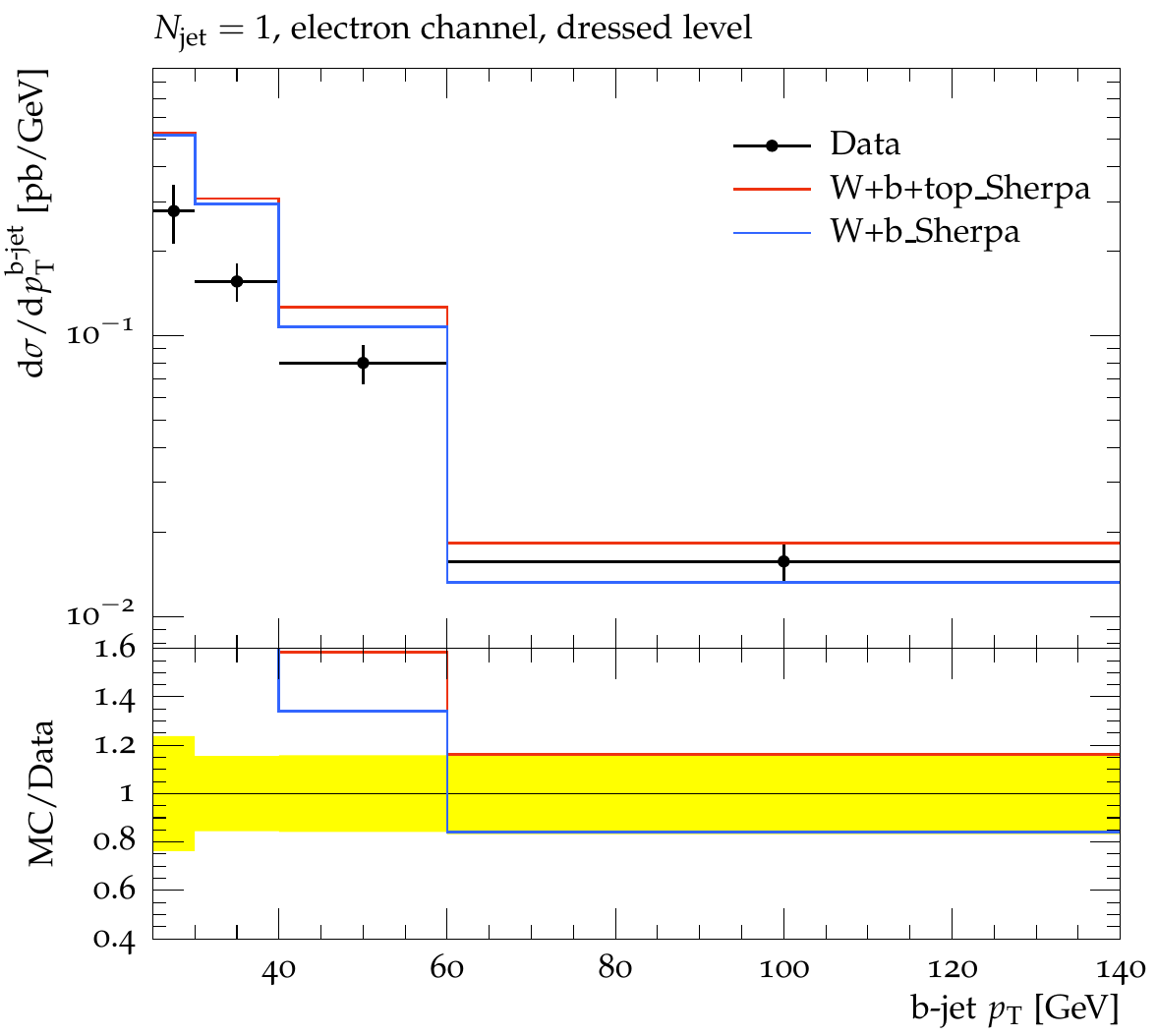}
	\includegraphics[width=0.48\textwidth]{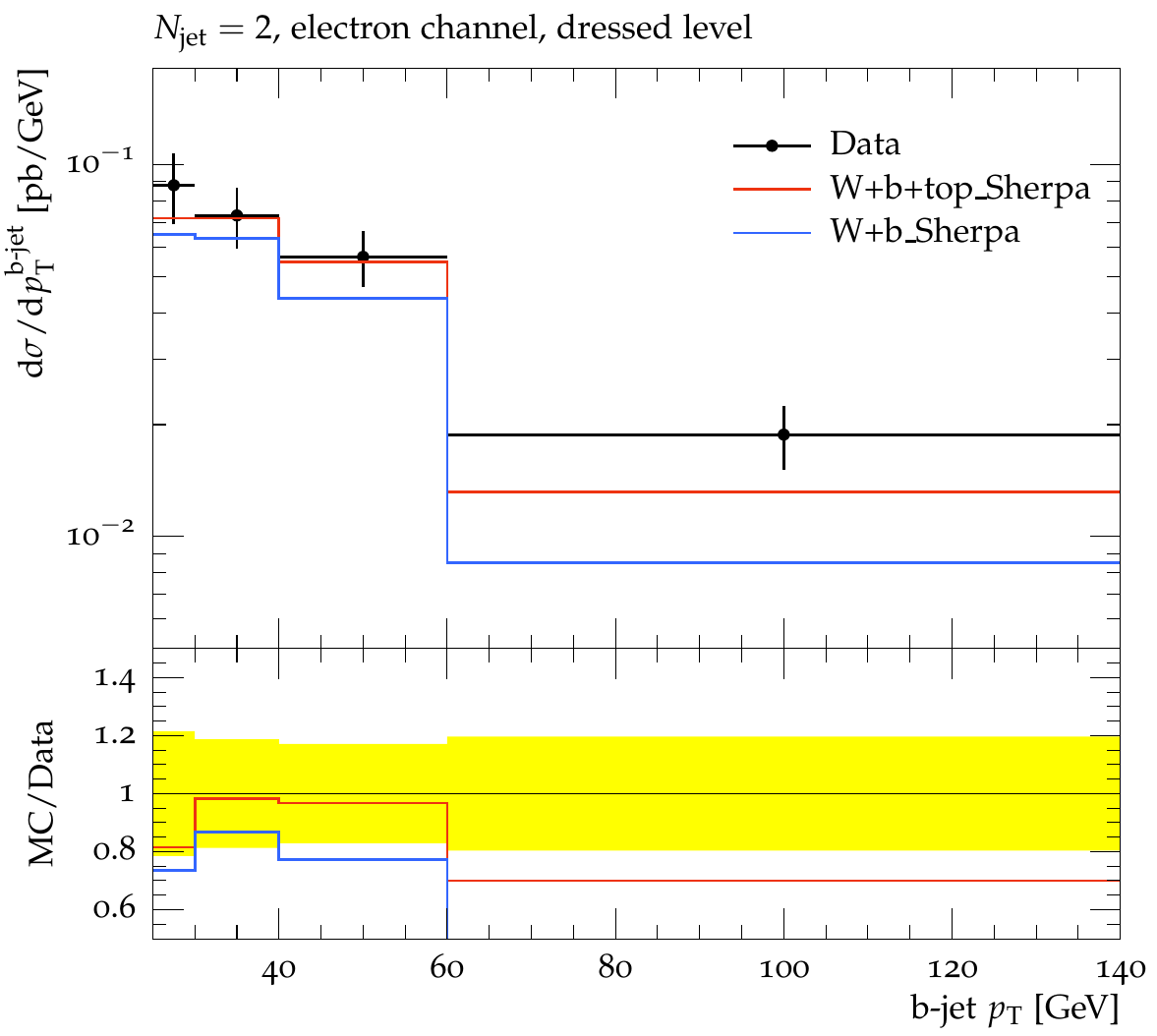}
\caption{\label{fig:unsubtracted}
  Measured differential $W+b$-jet cross-section, including single-top contributions, as a function of the
  transverse momentum of the $b$-jet, in the 1-jet bin (left) and 2-jet bin (right).
  The measurements are compared to the Sherpa (top) and Herwig (bottom)
  calculations of $W+b$-jet production including resonant top contributions, but excluding finite width effects
  and interference terms between top and non-top diagrams. The contribution from non-top diagrams alone is
  also shown.}
\end{figure}

In Fig.~\ref{fig:unsubtracted} the unsubtracted measurement is shown, compared to Herwig and Sherpa.
The predicted distributions from $W+b$-jet diagrams are shown without and with top contribution (in the latter case the interference terms are neglected).
Once more, Herwig agrees quite well with the data normalisation, and models well the $b$-jet $p_T$ distribution in the 1-jet bin, 
while struggling with this dependences in the 2-jet bin. Also in this case Sherpa overestimates the normalisation of
the 1-jet bin (this time by about $65\%$), but correctly models the normalisation of the 2-jet bin
and the $p_T$ dependence in both jet bins, within the data uncertainties.

\begin{figure}
\centering
	\includegraphics[width=0.48\textwidth]{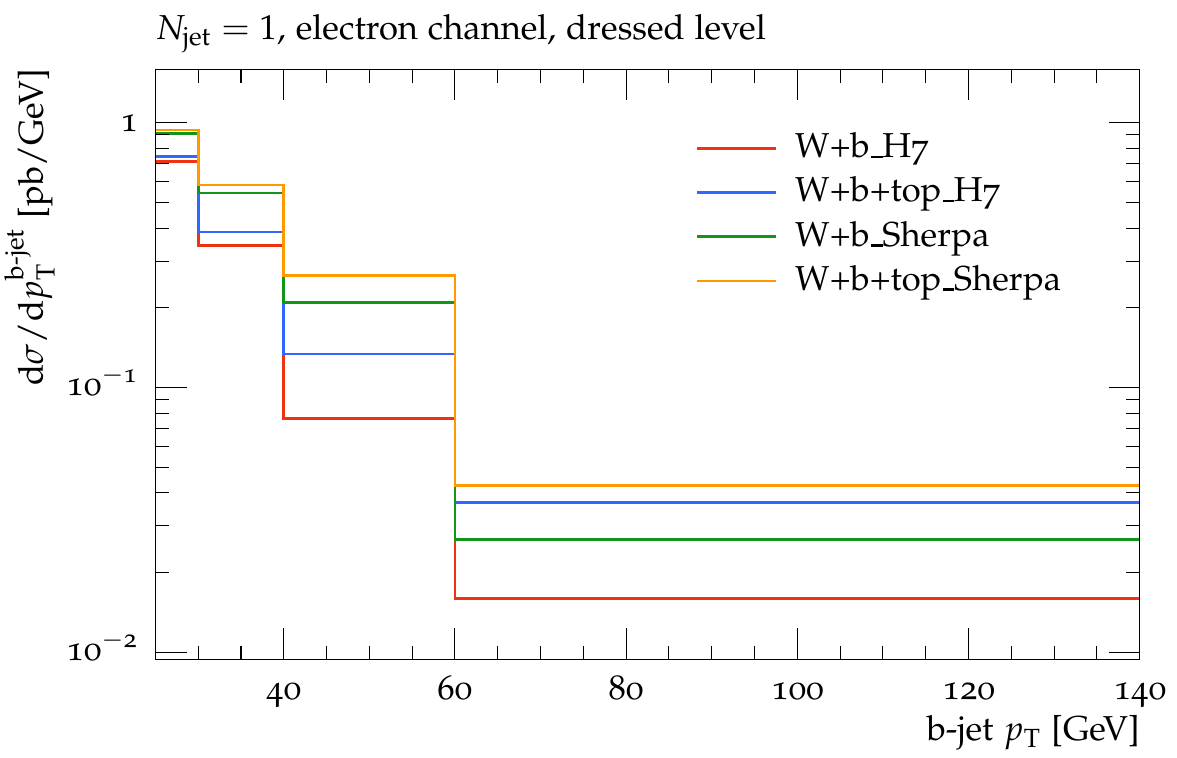}
	\includegraphics[width=0.48\textwidth]{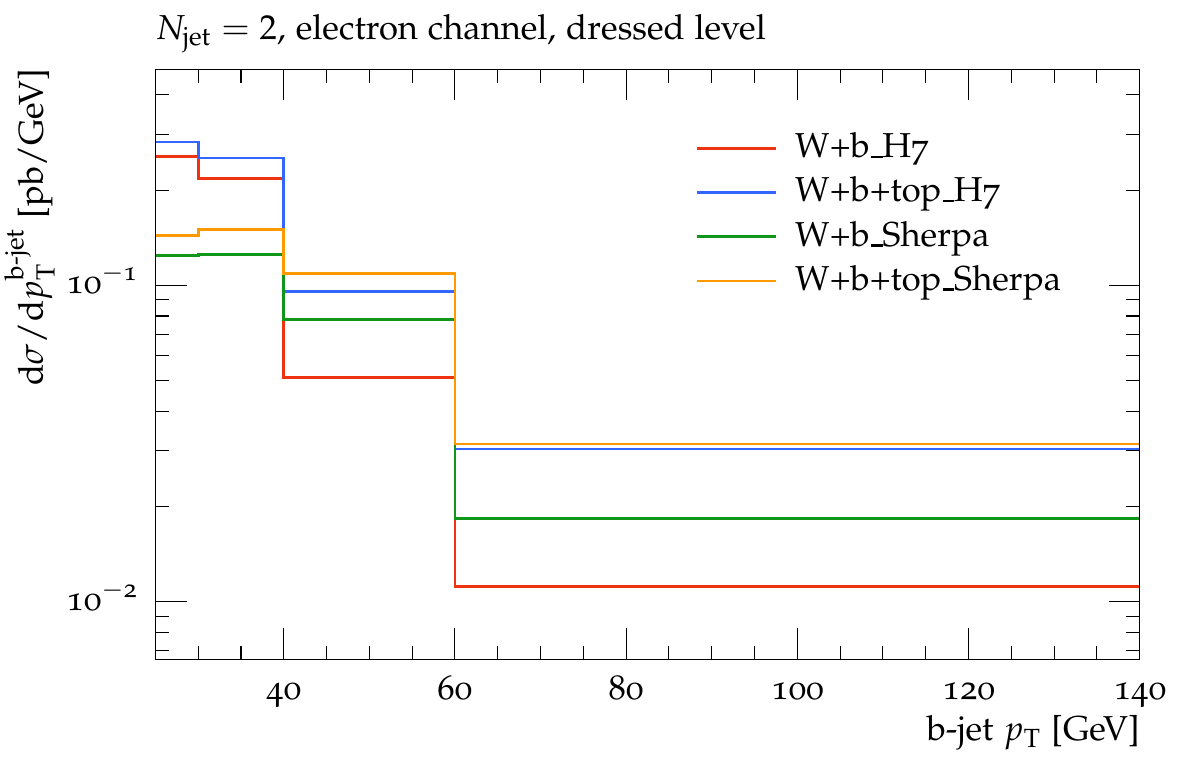}
\caption{\label{fig:13tev}
  Differential $W+b$-jet cross-section at 13 TeV as a function of the
  transverse momentum of the $b$-jet, in the 1-jet bin (left) and 2-jet bin (right).
  In both Sherpa and Herwig calculations of $W+b$-jet production including resonant top contributions, finite width effects 
  and interference terms between top and non-top diagrams are not
  taken into account. The contributions from non-top diagrams alone are also shown.}
\end{figure}

In Fig.~\ref{fig:13tev} we show the MC predictions for 13~TeV collisions. These measurements have yet to be
performed, but the main point to be made here is that the total top contribution rises from 15\% (32\%)
to 23\% (42\%) according to Sherpa (Herwig), with greater effects in some regions. This shows that any
problems and uncertainties associated with the subtraction of irreducible backgrounds are likely to become
more severe at higher energies. The differences between the generators themselves is another indication of the 
challenges associated with predicting these cross sections, and thus the need to minimise the theory
dependence of the measurement.

\subsection{Diboson plus jet production}

Processes in which two $W$-bosons and two jets (including possibly $b$-jets) are produced are of great interest
at the LHC. Contributing amplitudes include:
\begin{itemize}
\item $t\bar{t}$ (with on- or off-shell top quarks),
\item genuine QCD processes with $b$ quarks already entering from the initial state or being
  pair-produced in the final state through gluon splitting amplitudes, 
\end{itemize}
and, to a lesser extent,
\begin{itemize}
\item electroweak processes such as vector-boson fusion diagrams including the Higgs boson
  as a propagator and $b$-associated Higgs boson production.
\end{itemize}
In Sherpa, the leading order processes for $t\bar{t}$ and $tWb$ (both with on-shell tops), and $WWb\bar{b}$
(excluding all top contributions) were generated separately, and the full leading order $WWb\bar{b}$ process,
including all top contributions was also generated for comparison. All processes were generated for centre-of-mass
energies of 13~TeV. 

\begin{figure}
\centering
	\includegraphics[width=0.48\textwidth]{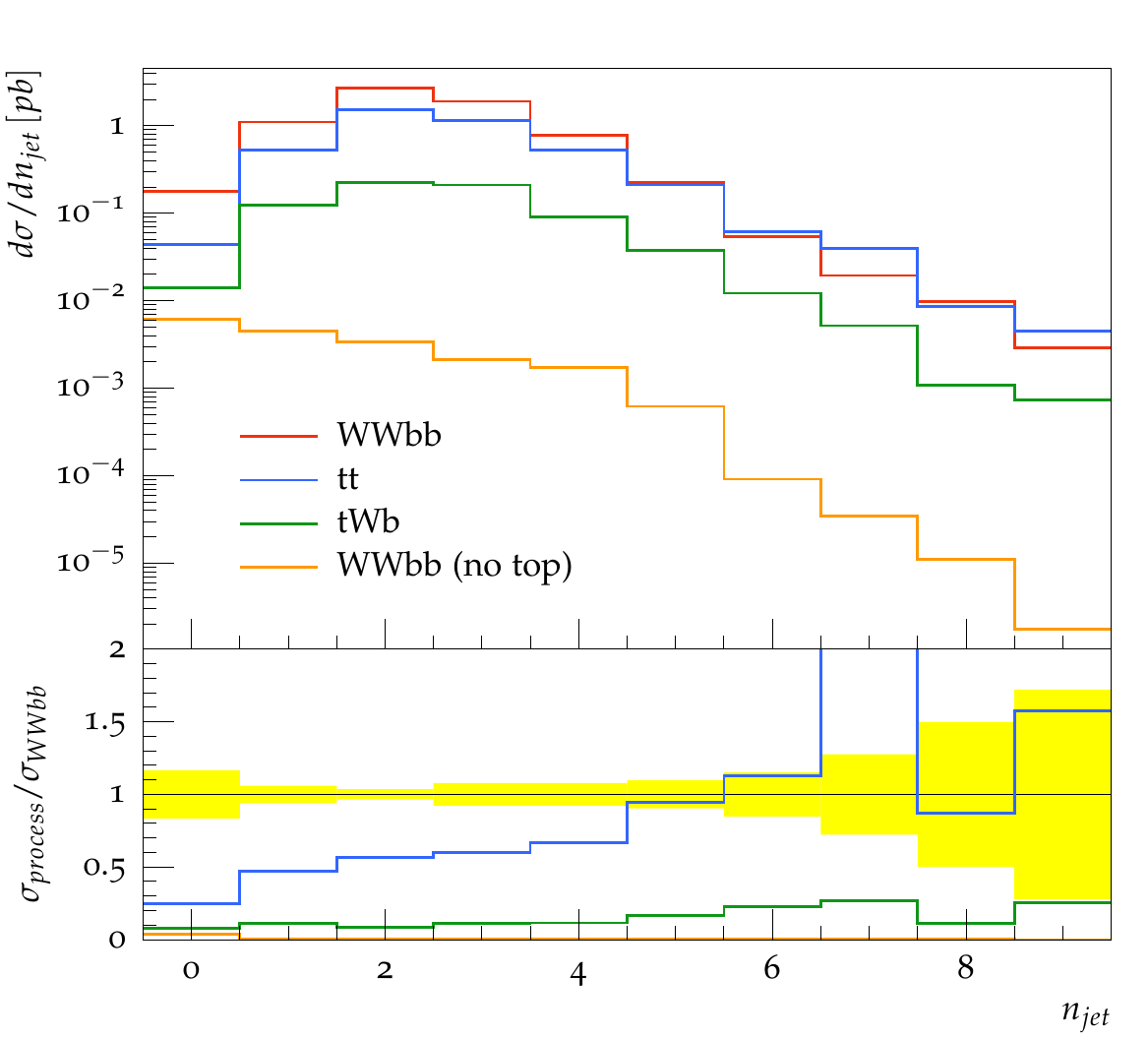}
	\includegraphics[width=0.48\textwidth]{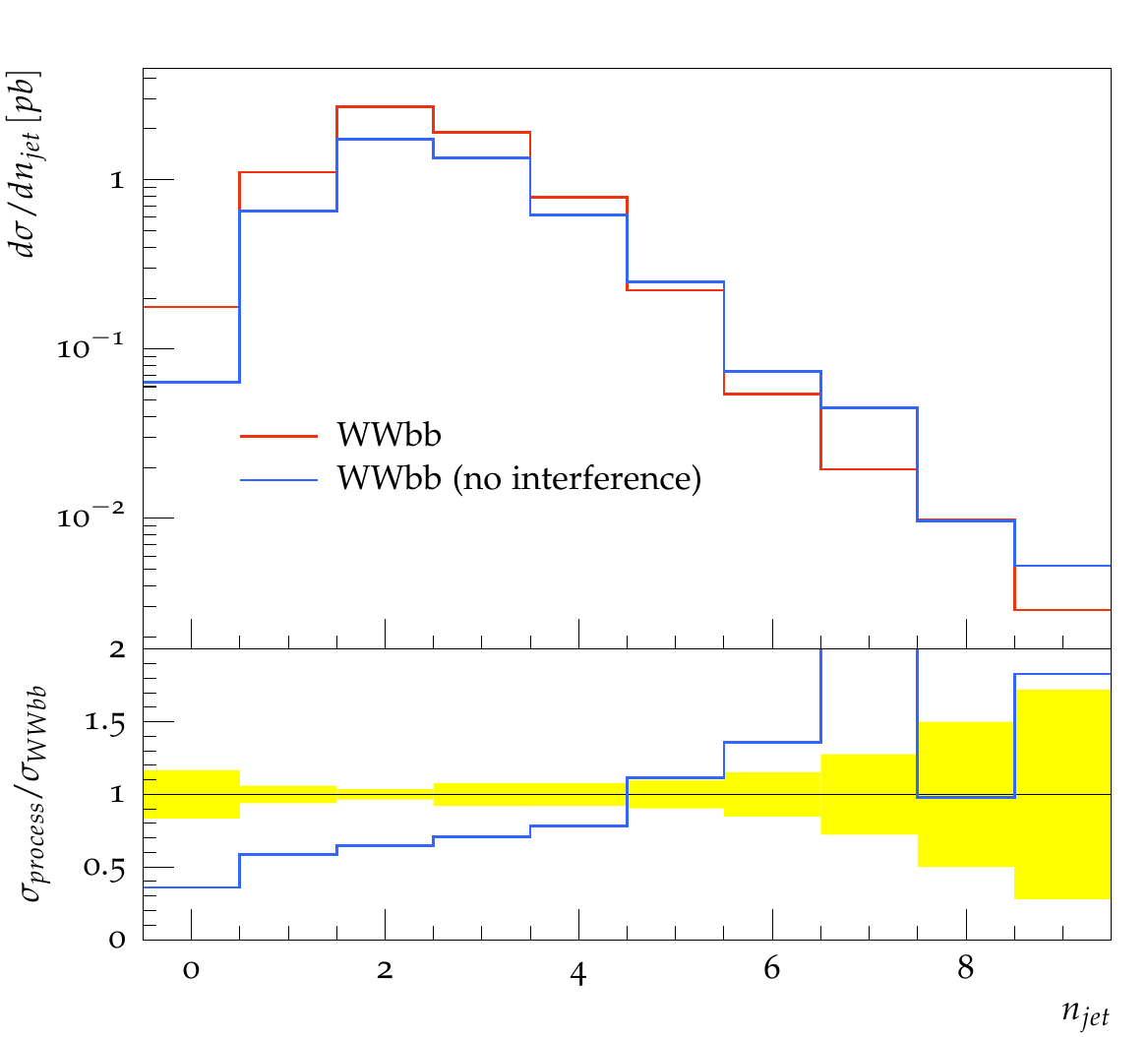}
\caption{\label{fig:wbf_before}
  Simulated $WWb\bar{b}$ events at 13 TeV using Sherpa.
  Individual contributions (left) and the comparison between incoherent and coherent sums (right) are shown.}
\end{figure}

An initial set of basic selection cut was applied, requiring two isolated leptons with $|\eta| < 2.5, p_T > 25$~GeV and
missing $E_T > 25$~GeV, typical of an experimental analysis. The multiplicity of jets (identified with the anti-$k_T$
algorithm, $R=0.4$, $p_T > 25$~GeV, $|\eta| < 4.5$) in events passing these cuts is shown in Fig.~\ref{fig:wbf_before}.
It can be seen that the diagrams involving at least one top quark dominate, though the contribution of non-top diagrams
is significant at low jet multiplicities. Fig.~\ref{fig:wbf_before} also compares the incoherent sum of the different
contributions with the coherently generated $WWb\bar{b}$ process. It can be seen that the interference terms are largely
positive.

\begin{figure}
\centering
	\includegraphics[width=0.48\textwidth]{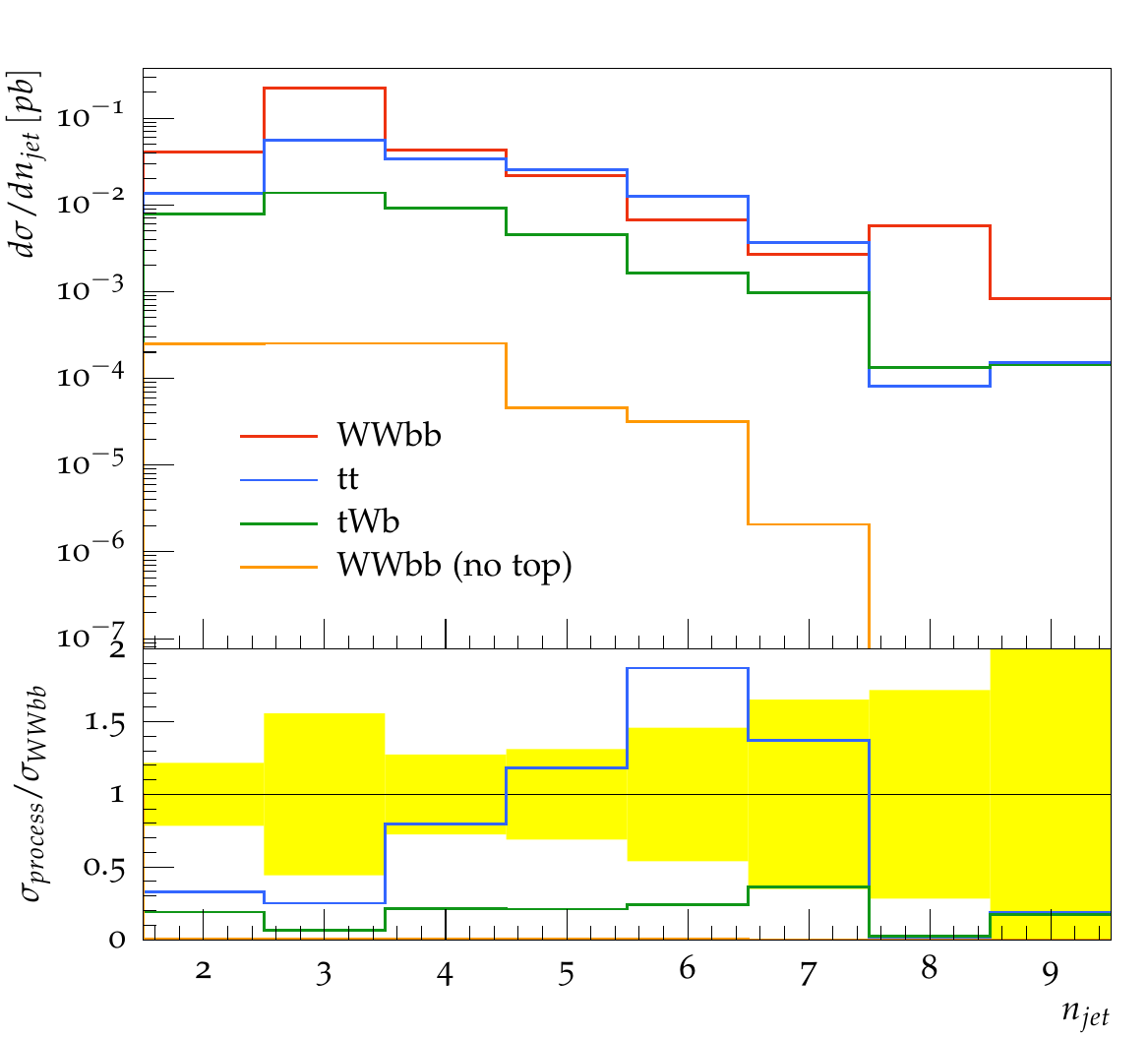}
	\includegraphics[width=0.48\textwidth]{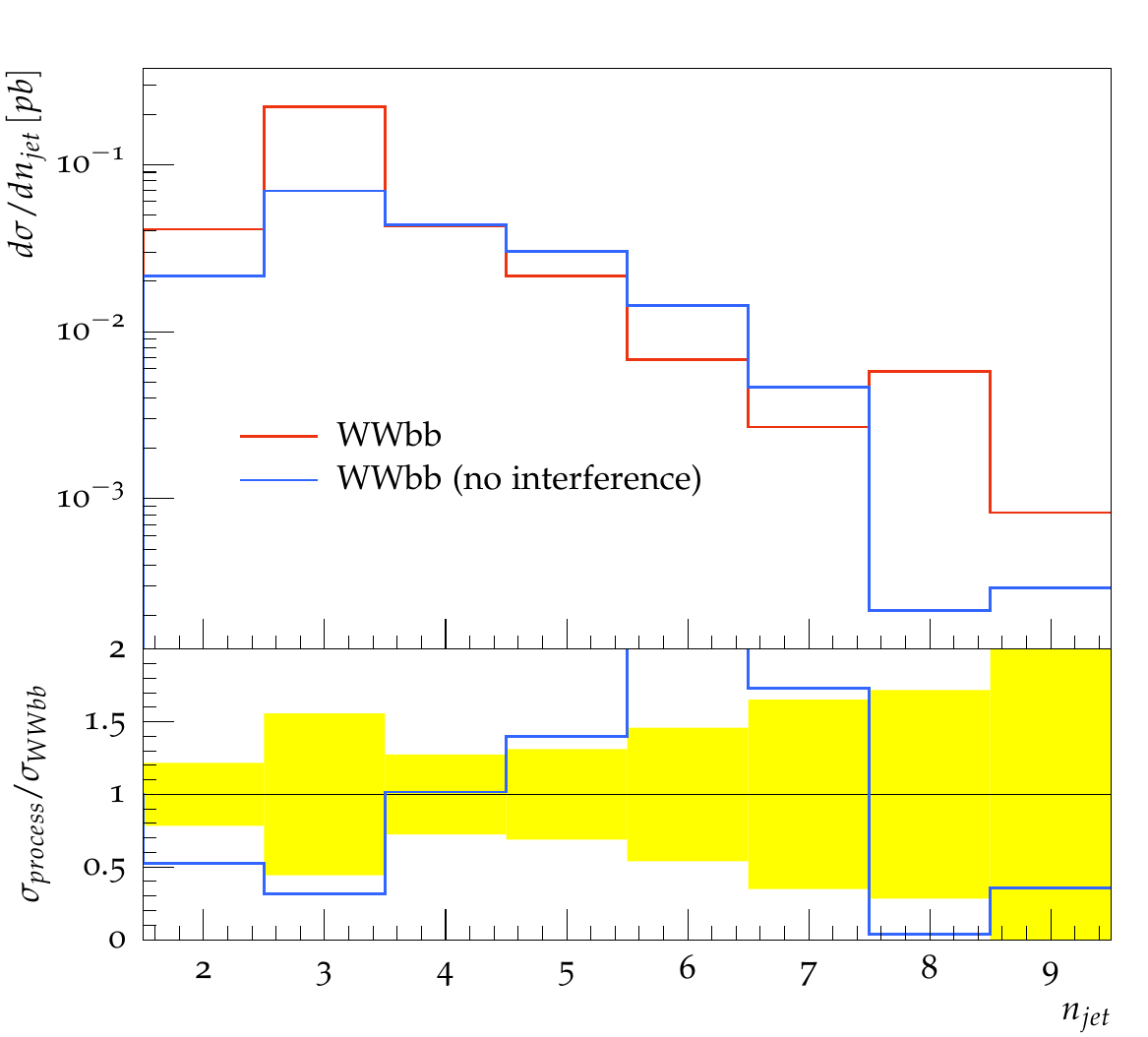}
\caption{\label{fig:wbf_after}
  Simulated $WWb\bar{b}$ events at 13 TeV using Sherpa, after vector-boson-fusion selection cuts. 
  Individual contributions (left) and the comparison between incoherent and coherent sums (right) are shown.}
\end{figure}

Further cuts were applied to mimic a vector-boson-fusion like analysis, requiring that there are at least two jets in
opposite hemispheres of rapidity, with a rapidity difference between them
$\Delta y_{jj} > 2.4$ and a dijet invariant
mass $m_{jj}>500$~GeV, and after additional cuts on the transverse mass
of the dilepton $m_{\ell\ell} > 20$~GeV and missing $E_T > 40$~GeV.
Fig.~\ref{fig:wbf_after} shows that the same general features persist, with the interference terms being large and
positive for jet multiplicities below five.

\subsection{Conclusions}\label{sec:conclusions}

This brief study exploits the $Wb$ measurement made by ATLAS at 7~TeV, 
and the multi-process capabilities of Sherpa, to illustrate 
how distinct processes and their interference contribute to the same measurable $W+b$-jet final state.
The discussion draws the attention on the subtraction of irreducible backgrounds which, although commonly used,
can increase the model-dependence of systematic uncertainties and lead to unphysical results,
since the interference terms are not treated correctly. 
The ATLAS measurement shows increased systematic
uncertainties in the region of high $b$-jet $p_T$ when single top background contributions are subtracted.
The simulation shows that the background will become more significant at 13~TeV, therefore expecting further increase of uncertainties.
Furthermore, the studies performed with Sherpa indicate that, even after realistic selection of vector-boson-fusion-like topologies, 
interference terms are significant in $WWbb$ production. 
In conclusion, a careful treatment of the irreducible background in future measurements at the LHC 
will become more and more relevant when presenting the results.

\subsection*{Acknowledgements}

V.~Ciulli and L.~Russo acknowledge support by MIUR Italy under 
Grant Agreement 2012Z23ERZ of PRIN 2012.

\clearpage

\chapter{MC uncertainties and output formats}
\label{cha:mc}

\section{Towards parton shower variations \texorpdfstring{\protect\footnote{S.\ H\"oche, A.\ Jueid, G.\ Nail, S.\ Pl\"atzer,  M.\ Sch\"onherr, A.\ Si\'odmok, P.\ Skands, D.\ Soper, K.\ Zapp}}{}}
\subsection{Introduction}
\label{sec:psunc:intro}

Parton showers are the back-bone of Monte-Carlo event generation and 
particle level theory predictions for collider experiments for decades
\cite{Fox:1979ag,Mazzanti:1980jj,Field:1982dg,Gottschalk:1982yt,
  Marchesini:1983bm,Field:1982vz,Odorico:1983yf,Sjostrand:1985xi,
  Ellis:1986bv,Marchesini:1987cf}, their exclusive and fully-differential 
resummation being indespensible for the experimental collaborations. 
With the recent advances in high precision perturbative calculations 
finding their way into such particle level predictions through 
parton shower matchings \cite{Frixione:2002ik,Nason:2004rx,Frixione:2007vw,
  Lavesson:2008ah,Hoeche:2011fd,Hamilton:2013fea,Hoeche:2014aia,Alioli:2013hqa}
the question regarding not only the well understood central value of 
the parton showers' prediction but its uncertainties arises. However, 
such uncertainty estimates are not yet generally available and no 
systematic treatment on resummation properties is thus commonly evaluated 
for LHC physics predictions.\footnote{With the notable exception of the 
\textsc{Vincia} shower \cite{Giele:2007di,Giele:2011cb}.}

In this contribution we will summarise the status and methods available 
in different parton shower programs. No attempt is made to harmonise the 
respective settings of each parton shower to produce comparable results. 
Due to the different evolution parameters and, thus, resummation variables, 
such a common parameter set would not produce an identical baseline. In the 
following we shortly describe the systematic variations available in each 
parton shower algorithm considered before discussing selected results on 
idealised observables in different kinematic regimes.

\subsection{Tools}
\label{sec:psunc:tools}

\subsubsection{\texorpdfstring{\textsf{Deductor}}{Deductor}}
\label{sec:psunc:tools:decuctor}

\textsf{Deductor} \cite{Nagy:2014mqa,Nagy:2014oqa,Nagy:2014nqa,Nagy:2015hwa} is a
dipole shower with virtuality based ordering. It uses a color aproximation
that goes beyond leading color (LC), namely the LC+ approximation. This
includes some -- but certainly not all -- corrections to order $1/N_c^4$. For
$e^+e^-$ collissions at $500\ {\rm GeV}$ no substantial difference for the
observables considered between LC and LC+ has been found. \textsf{Deductor} uses a
two-loop running of $\alpha_s$ with $\alpha_s(M_Z)=0.118$ and a parton shower
cutoff of $1\ {\rm GeV}$, and implements the CMW prescription
\cite{Catani:1990rr} through a scaling of the argument of $\alpha_s$.

\subsubsection{\texorpdfstring{\textsf{Herwig}}{Herwig}}
\label{sec:psunc:tools:herwig}

The \textsf{Herwig}~7 event generator \cite{Bellm:2015jjp} is currently offering two
shower modules: the traditional, angular ordered parton shower as set out in
\cite{Gieseke:2003rz}, as well as a dipole-type shower based on the work
presented in \cite{Platzer:2009jq,Platzer:2011bc}.
Though very different in their nature, both showers guarantee coherent
evolution and reach a similar level of description of data when interfaced to
the cluster hadronization model. The CMW prescription \cite{Catani:1990rr} is
not implemented directly, but the relevant change in $\Lambda$ is considered
to be absorbed into using a tuned value of $\alpha_S(M_Z)$ throughout the
simulation.

We do not include terms which modify the splitting kernels in the presence of
variations of the argument of the strong coupling. For the parton level
comparison at hand, which will have to be related to the cross-feed with
non-perturbative models, we chose to use shower settings which put the
\textsf{Herwig}~7 showers onto a similar-as-possible level using the same two-loop
running of $\alpha_s$, with $\alpha_s(M_Z)=0.118$, a similar cutoff
prescription and cutoff value ($p_{\perp,\text{min}}=1\ {\rm GeV}$) and not
intrinisic $p_\perp$ generated. The difference to the tuned settings may serve
as a first indicator of the impact of non-perturbative corrections.

\subsubsection{\texorpdfstring{\textsf{Pythia}}{Pythia}}
\label{sec:psunc:tools:pythia}
The parton showers in \textsf{Pythia}-8 are based on the dipole type $p_\perp$-ordered
evolution which has been available since \textsf{Pythia}-6.3 \cite{Sjostrand:2004ef}. 
This is used for both ISR and FSR algorithms. Furthermore, \textsf{Pythia}-8 \cite{Sjostrand:2007gs,Sjostrand:2014zea}
contains an implementation of $\gamma \to f \bar{f}, f = l,q$ splittings 
as a part of the parton-shower machinery, as well as weak gauge boson emissions \cite{Christiansen:2014kba}
(The latter are however not included in this study.) \\
The evolution of parton showers is based on the DGLAP splitting kernels. For the case of one flavour, they are
given by :
\begin{eqnarray}
 P_{q \to q g} (z) = C_F \frac{1 + z^2}{1 - z} \\
 P_{g \to g g} (z) = C_A \frac{(1 - (1 - z) z)^2}{z (1-z)} \\
 P_{g \to q \bar{q}} = T_R (z^2 + (1 - z)^2)
\end{eqnarray}
where $z$ is the energy fraction and $C_F = 4/3$, $C_A = 3$
and $T_R=1/2$ are the QCD factors. For QED radiation, there are only 
two splitting functions, $P_{f\to f \gamma} \text{ and } P_{\gamma \to f\bar{f}}$ which are given by :
\begin{eqnarray}
 P_{f \to f \gamma} (z) = Q_f^2 \frac{1 + z^2}{1 - z} \\
 P_{\gamma \to f \bar{f}} = Q_f^2 N_C (z^2 + (1 - z)^2)
\end{eqnarray}
Where $Q_f$ is the electric charge of the fermion involved in the shower 
and $N_C = 3$ for quarks and $N_C=1$ for leptons is the number of color degrees 
of freedom. \\
The ISR and FSR algorithms are cast as integro-differential equations whose solution
is the probability of showering as one goes from one (shower-evolution) scale to a lower scale. 
The differential equations driving the shower evolution 
in \textsf{Pythia} are given by :
\begin{eqnarray}
 \frac{\text{d} \mathcal{P}_{\text{FSR}}}{\text{d} p_\perp^2} = \frac{1}{p_\perp^2} \int dz \frac{\alpha_s}{2 \pi} P(z) \\
 \frac{\text{d} \mathcal{P}_{\text{ISR}}}{\text{d} p_\perp^2} = \frac{1}{p_\perp^2} \int dz \frac{\alpha_s}{2 \pi} P(z) 
 \frac{f(x/z,p_\perp^2)}{z f(x,p_\perp^2)}
\end{eqnarray}
For ISR, $z=x/x'$ and $f(x/z,p_\perp^2)$ is the PDF of the new parent parton carrying a fraction $x/z$
of the parent hadron at factorisation scale $p_\perp^2$. The evolution variable $p_\perp$ is defined in 
\textsf{Pythia} as:
\begin{eqnarray}
 p_{\perp,\text{evol}}^2 = p_\perp^2 = \bigg\{ \begin{array}{c}
                            (1-z) Q^2 \quad \text{ for ISR} \\
                            z (1-z) Q^2 \quad \text{ for FSR} \\
                           \end{array}
\end{eqnarray}
where $Q^2 > 0$ is the virtuality of the branching parton. \\
The strength of the radiation is controlled by the effective value of the 
strong coupling constant $\alpha_s(M_Z)$. In \textsf{Pythia}, $\alpha_s$ can be 
set separetely for ISR and FSR. The shower evolution scale ($p_{\perp;\mathrm{evol}}$) is used as the default renormalization scale
for the evaluation of $\alpha_s$. Furthermore,
in order to make uncertainty variations of the parton showers in \textsf{Pythia}, a 
multiplicative prefactor can be applied $\mu_R^2 = k_{\mu_R} p_{\perp, \text{evol}}^2$. The default 
value is $k_{\mu_R} = 1$. \\
We should note that in \textsf{Pythia}, the value of $\alpha_s(M_Z)$ is not comparable
to the $\alpha_s^{\overline{\text{MS}}}=0.118$ value. This is due to two reasons:
\begin{itemize}
 \item In the limit of soft-gluon emission, the dominant splitting function
   term, can be absorbed into the LO splitting kernel by a translation to the
   Catani-Marchesini-Webber (CMW) scheme \cite{Catani:1990rr}:
 \begin{eqnarray}
  \alpha_s^{\text{CMW}} = \alpha_s^{\overline{\text{MS}}} \bigg(1 + K \frac{\alpha_s}{2 \pi} \bigg)
 \end{eqnarray}
where $K = C_A \bigg(\frac{67}{18} - \frac{1}{6} \pi^2 \bigg) - \frac{5}{9} n_f$
\item The effective value of $\alpha_s(M_Z)$ tends to be a $10\%$ larger when tuned to the experimental
data, i.e $\alpha_s(M_Z)^{\text{\textsf{Pythia}}} \sim 0.139$ which is chosen as the default value.
\end{itemize}
The translation from the $\overline{\text{MS}}$ to the CMW scheme is equivalent
to a specific shift of the renormalization scale, $\mu_R \to \mu_R \exp(-K/4\pi\beta_0)$. \\
Finally, we note that corrections for parton masses are also avalaible for ISR \cite{Sjostrand:2004ef} 
and FSR \cite{Norrbin:2000uu}.

\subsubsection{\texorpdfstring{\textsc{Sherpa}}{Sherpa}}
\label{sec:psunc:tools:sherpa}

The \textsc{Sherpa} Monte Carlo event generator \cite{Gleisberg:2008ta} in its latest
release, \textsc{Sherpa}-2.2.0, comprises two parton shower algorithms: \textsc{Csshower}
\cite{Schumann:2007mg} and \textsc{Dire} \cite{Hoche:2015sya}. While the \textsc{Csshower} is based on
Catani-Seymour \cite{Catani:1996vz,Catani:2002hc} dipole splitting functions,
\textsc{Dire} combines the standard treatment of collinear configurations in parton
showers with the resummation of soft logarithms in color dipole cascades. A
third parton shower, \textsc{Ants} \cite{ants}, is under development. \textsc{Ants} is based on
dipole splitting functions in the spirit of
\cite{Winter:2007ye,Lonnblad:1992tz}.

In general the splitting functions of all three algorithms take the 
form 
\begin{equation}
  \begin{split}\label{eq:psunc:tools:sherpa:sf}
    \mathrm{D}_{ijk}(t,z,\phi)
    \,=\;& \frac{\alpha_s(k_\text{tune}b\,t)}{2\pi t}\;\mathrm{P}_{ijk}(z)
  \end{split}
\end{equation}
where $b=1$ for \textsc{Dire}. For \textsc{Csshower} and \textsc{Ants}, $b=k_\text{CMW}$, where
\begin{equation}
  \begin{split}\label{eq:psunc:tools:sherpa:kcmw}
    k_\text{CMW}
    \,=\;&\exp\left[-\frac{67-3\pi^2-\tfrac{1}{3}\,n_f(t)}{33-2n_f(t)}\right]
  \end{split}
\end{equation}
is the Catani-Marchesini-Webber scale factor \cite{Catani:1990rr}
to incorporate dominating higher-logarithmic contributions and $n_f(t)$ 
is the number of active flavours at scale $t$. The correction originating from
$k_\text{CMW}$ is included in \textsc{Dire} by multiplying the soft enhanced term of the
splitting functions with $1+\alpha_s/(2\pi)\beta_0k_\text{CMW}$. The $k_\text{tune}$ 
are manually inserted scale factors to accommodate one more degree-of-freedom 
in a tuning context. However, in all current versions of the programs 
$k_\text{tune}$ is fixed to 1 for final state splittings and $\tfrac{1}{2}$ 
for initial state splittings in the \textsc{Csshower}, while \textsc{Ants} and \textsc{Dire} employ 
$k_\text{tune}=1$ throughout. The $\mathrm{P}_{ijk}(z)$ are the shower 
dependent splitting functions. 
Since PDF uncertainties are not addressed here, cf.\ Sec.\ \ref{cha:mc}.\ref{sec:psrew}, ratios of 
PDFs are absorbed into the $\mathrm{P}_{ijk}(z)$.

For the present study the showers run in their default setting.
When now varying the argument of the strong coupling constant, i.e.\ 
replacing $t\to ct$ the higher-logarithmic structure induced by the running 
of the coupling constant in the presence of the CMW scale factor needs to be 
preserved in order not to upset the resummation quality of the parton 
shower. Thus, the following counter term is introduced
\begin{equation}
  \begin{split}\label{eq:psunc:tools:sherpa:asct}
    \alpha_s(k_\text{tune}k_\text{CMW}\,t)
    \,\to\;& \alpha_s(k_\text{tune}k_\text{CMW}\, c\cdot t)\cdot f(c,t)
  \end{split}
\end{equation}
where two forms for the counter term $f(c,t)$ are implemented,
\begin{equation}
  \begin{split}\label{eq:psunc:tools:sherpa:asctfac}
    f(c,t)
    \,=\;&\left\{\begin{array}{ll}
                  1-\sum_{i=0}^{n_\text{th}+1}\frac{\alpha_s}{2\pi}\,\beta_0(n_f(t))\,\log\frac{t_i}{t_{i-1}} & \text{additive threshold treatment}\\
                  \prod_{i=0}^{n_\text{th}+1}\left(1-\frac{\alpha_s}{2\pi}\,\beta_0(n_f(t))\,\log\frac{t_i}{t_{i-1}}\right) & \text{multiplicative threshold treatment.}
                 \end{array}\right.
  \end{split}
\end{equation}
Therein, the sum and product run over the number $n_\text{th}$ of parton mass
thresholds in the interval $[t,c\cdot t]$ with $t_0=t$,
$t_{n_\text{th}+1}=c\cdot t$ and $t_i$ are the encompassed parton mass
thresholds. If $c<1$, then the ordering is reversed recovering the correct
sign. $\beta(n_f)$ is the QCD beta function. Obviously, both forms coincide if
the interval $[t,c\cdot t]$ contains zero or one parton mass thresholds. For
this study the additive threshold treatment was used. We vary the
renormalization scale by a factor two, i.e.\ $c=4$ and $c=1/4$
in Eq.~\eqref{eq:psunc:tools:sherpa:asct}.

\subsection{Results}
\label{sec:psunc:results}
The events generated by the tools of Sec.~\ref{sec:psunc:tools} were analysed
with \textsc{Rivet} \cite{Buckley:2010ar} using three custom analyses. The
analysis for $e^+e^-$, is closely derived from \textsc{ALEPH\_2004\_S5765862}
\cite{Heister:2003aj} which contains descriptions of the observables
studied. The $pp$ results were produced using analyses that closely follow
content of \textsc{MC\_HINC}/\textsc{MC\_HJETS}. The jets definitions used in
all of these analyses are implemented in \textsc{FastJet}
\cite{Cacciari:2011ma}.

We investigate three separate processes at leading order with two different
scenarios. In sections \ref{sec:psunc:results:ee}-\ref{sec:psunc:results:z} we
display a selection of plots for discussion. The content of the uncertainty
bands for each generator are described in Sec.~\ref{sec:psunc:tools}

\subsubsection{$e^+e^-\to\text{jets}$}
\label{sec:psunc:results:ee}
This process is a natural starting point to explore shower uncertainties as we
only encounter final state radiation. For this process, the two scenarios
correspond to CoM energies of $\sqrt{s}=91\,\textrm{GeV}$ and $\sqrt{s}=500\,\textrm{GeV}$. The
jets are defined by the Durham jet algorithm with $y_\mathrm{cut}=0.7$. From
the analysis we present a subset of the event shape observables and jet
fractions. The results of both scenarios yielded similar conclusions and
we will omit the $500\,\textrm{GeV}$ results with one exception.
\begin{figure}[t!]
  \centering
  \begin{subfigure}[t]{0.49\textwidth}
    \includegraphics[width=\textwidth]{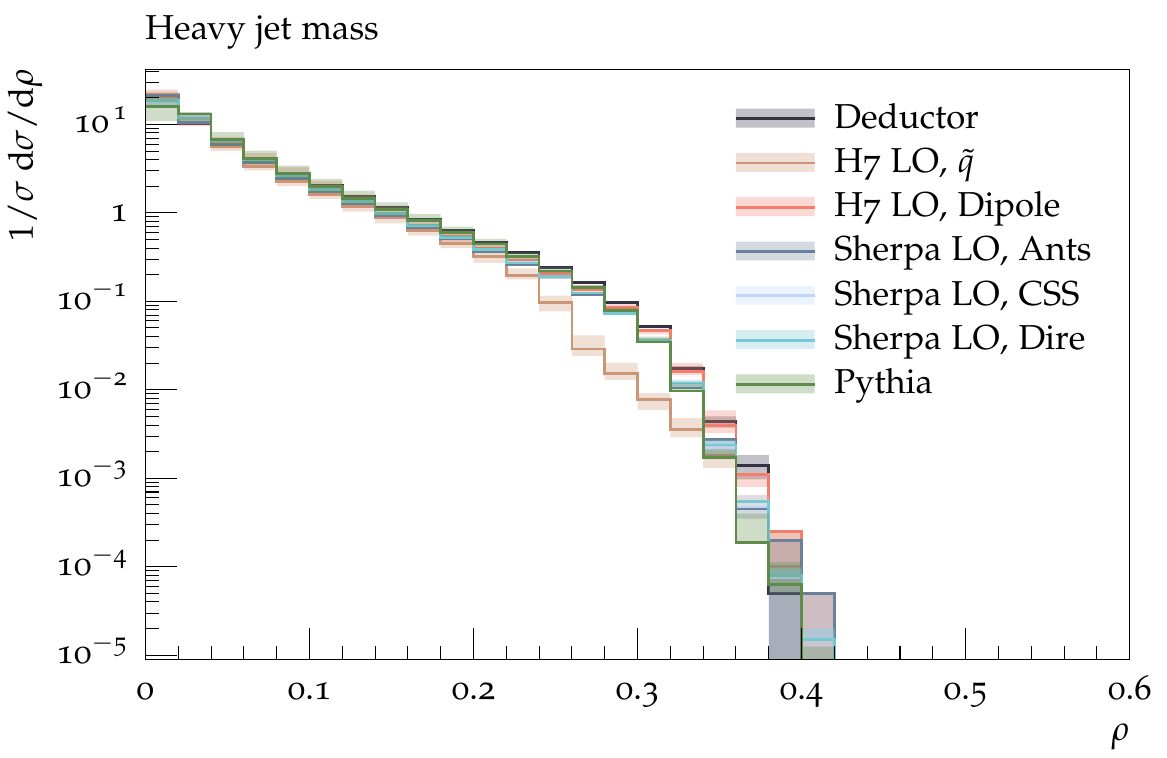}
    \caption{$\sqrt{s} = 91\,\textrm{GeV}$}
    \label{fig:ee:heavyjetmass:91}
  \end{subfigure}
  \begin{subfigure}[t]{0.49\textwidth}
    \includegraphics[width=\textwidth]{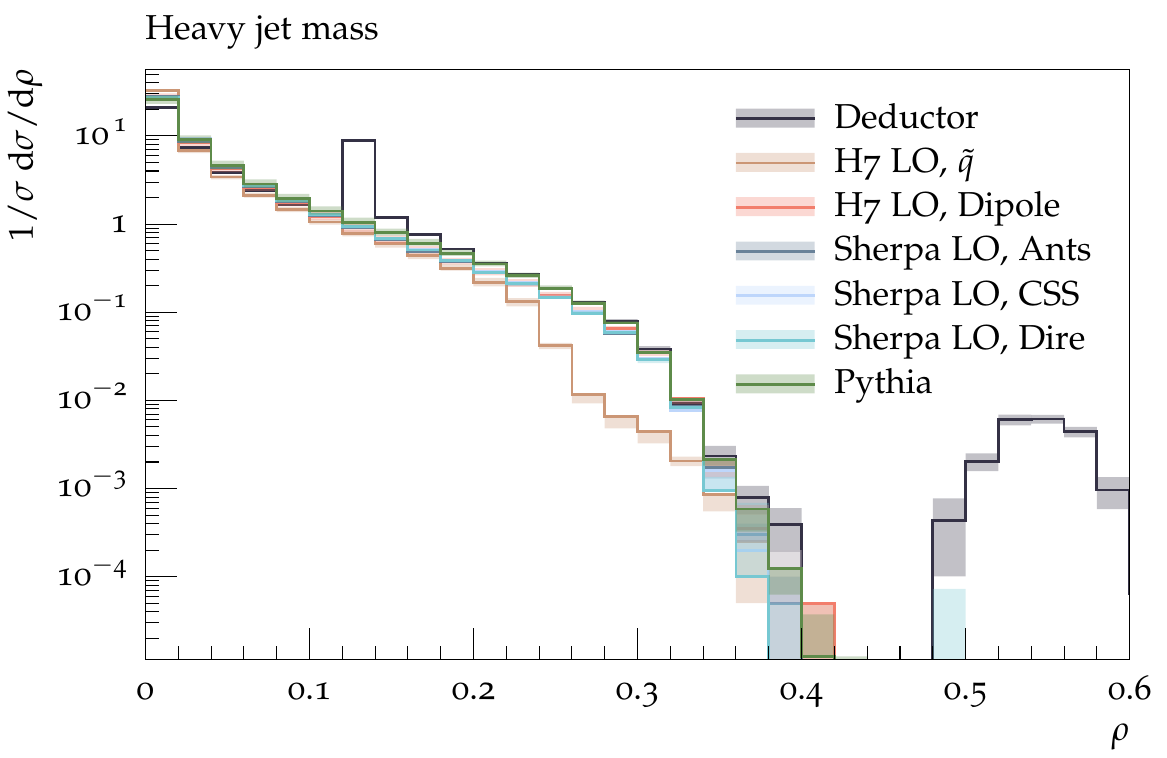}
    \caption{$\sqrt{s} = 500\,\textrm{GeV}$}
    \label{fig:ee:heavyjetmass:500}
  \end{subfigure}
  \caption{Comparison of Heavy Jet mass at two different energies.}
  \label{fig:ee:heavyjetmass}
\end{figure}

The predictions for the heavy jet mass, Fig.~\ref{fig:ee:heavyjetmass}, show
consistency over the majority of the distribution as well as consistent
predictions for the size of the uncertainty bands.  The \textsf{Deductor} curve for
$\rho$ at $500\,\textrm{GeV}$ has two features of interest; these stem from the
$t\bar{t}$ contribution without the top decay. The new features emerge from
the heavy jet containing 1 or 2 of the un-decayed top quarks.

\begin{figure}[t!]
  \centering
  \begin{minipage}[t]{0.49\textwidth}
    \includegraphics[width=1\textwidth]{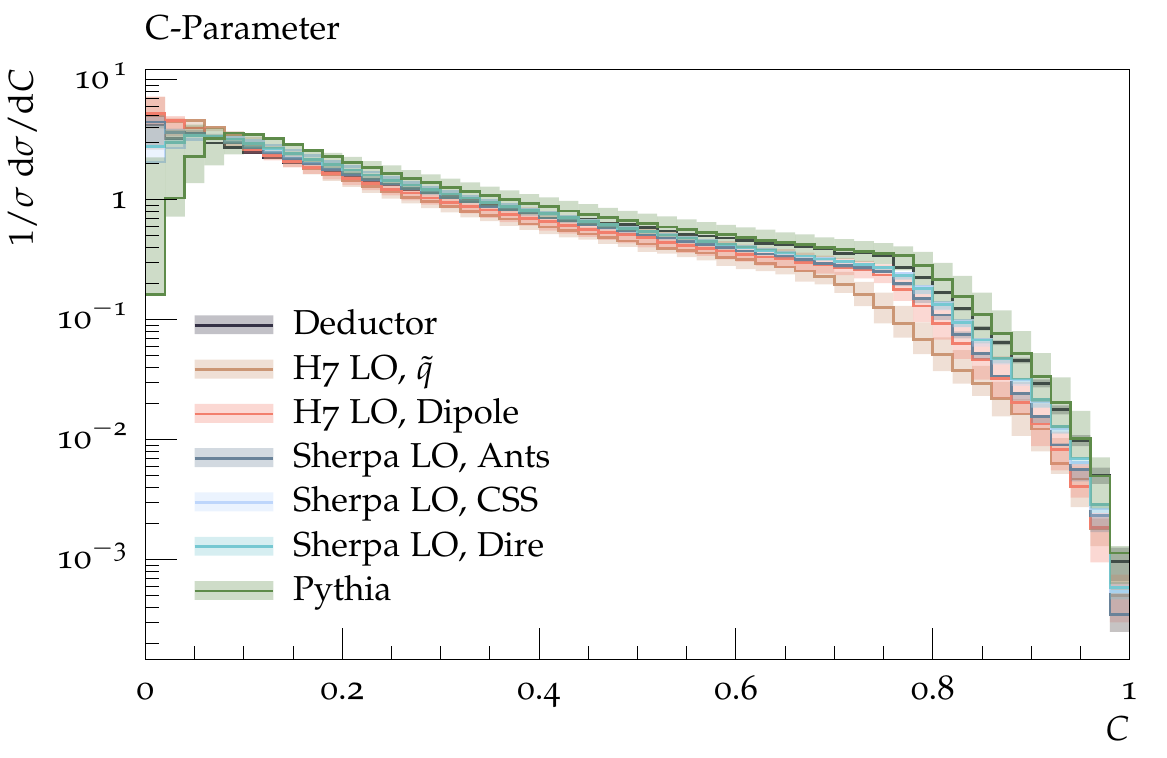}
    \caption{C-parameter at $\sqrt{s}=91\,\textrm{GeV}$.}
    \label{fig:ee:cparam:91}
  \end{minipage}
  \begin{minipage}[t]{0.49\textwidth}
    \includegraphics[width=1\textwidth]{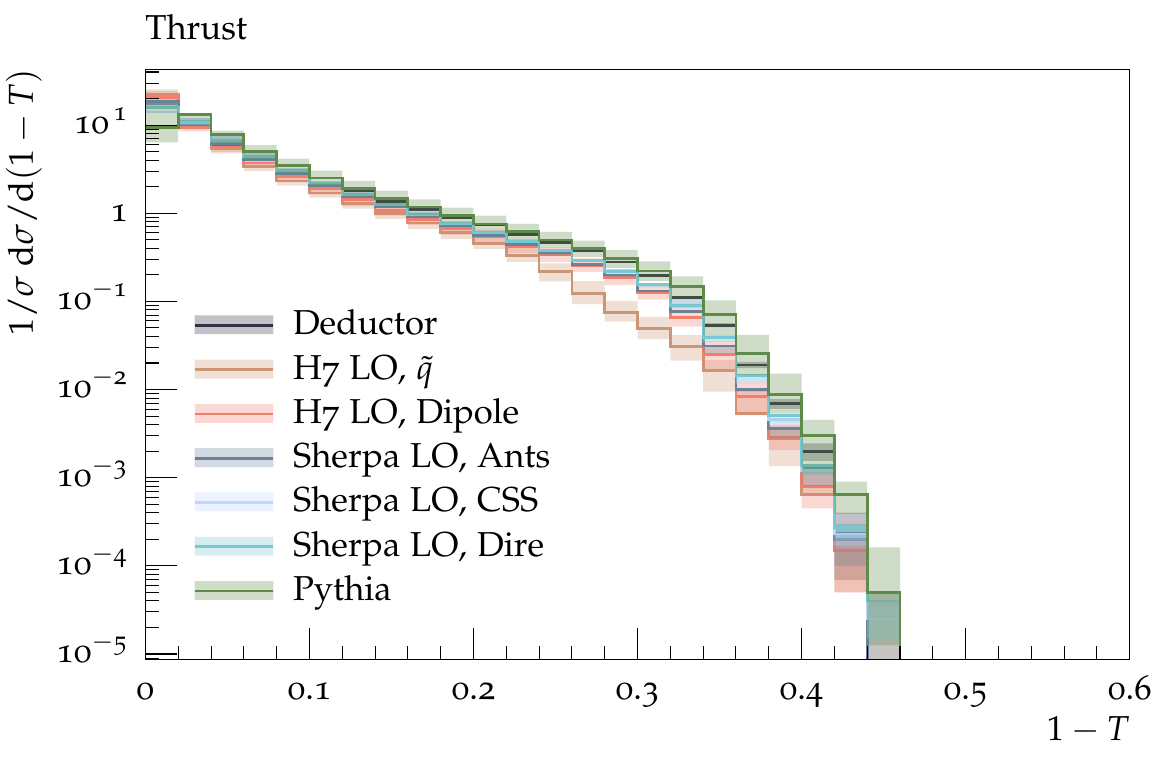}
    \caption{Thrust at $\sqrt{s}=91\,\textrm{GeV}$.}
	\label{fig:ee:thrust:91}
  \end{minipage}
\end{figure}

\begin{figure}[t!]
  \centering
  \begin{subfigure}[t]{0.49\textwidth}
    \includegraphics[width=1\textwidth]{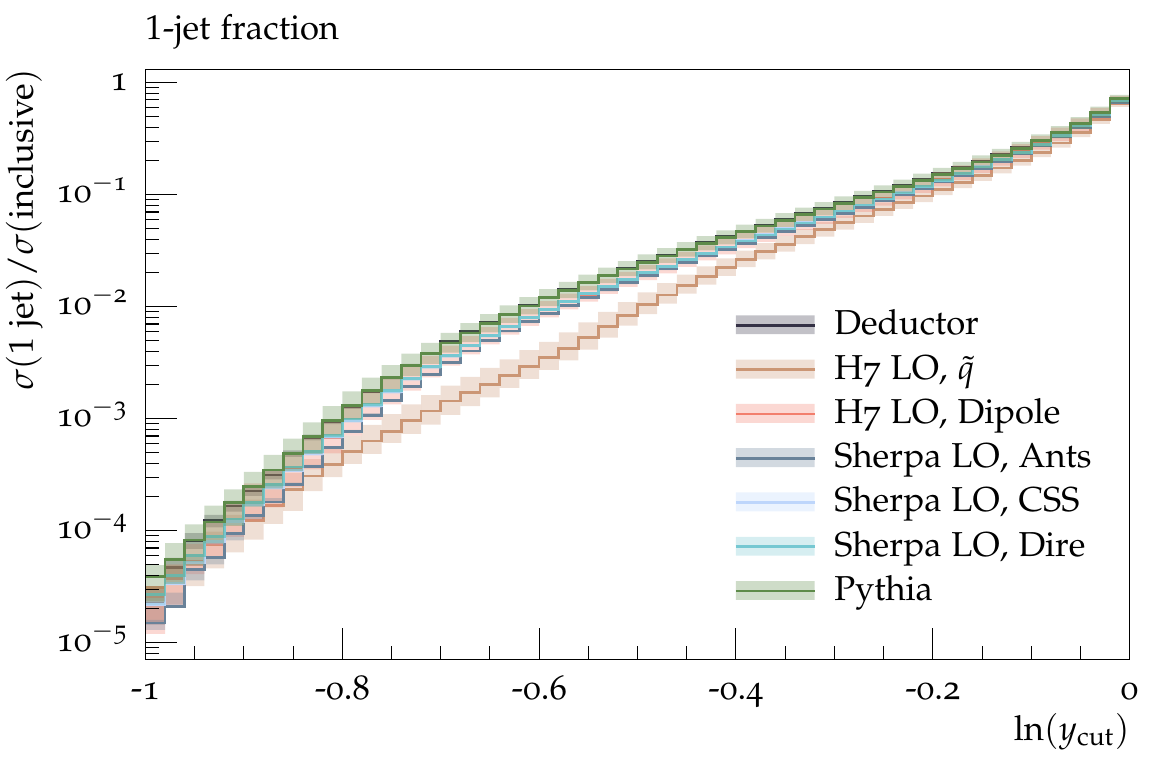}
    \caption{1-jet fraction}
    \label{fig:ee:r1:91}
  \end{subfigure}
  \begin{subfigure}[t]{0.49\textwidth}
    \includegraphics[width=1\textwidth]{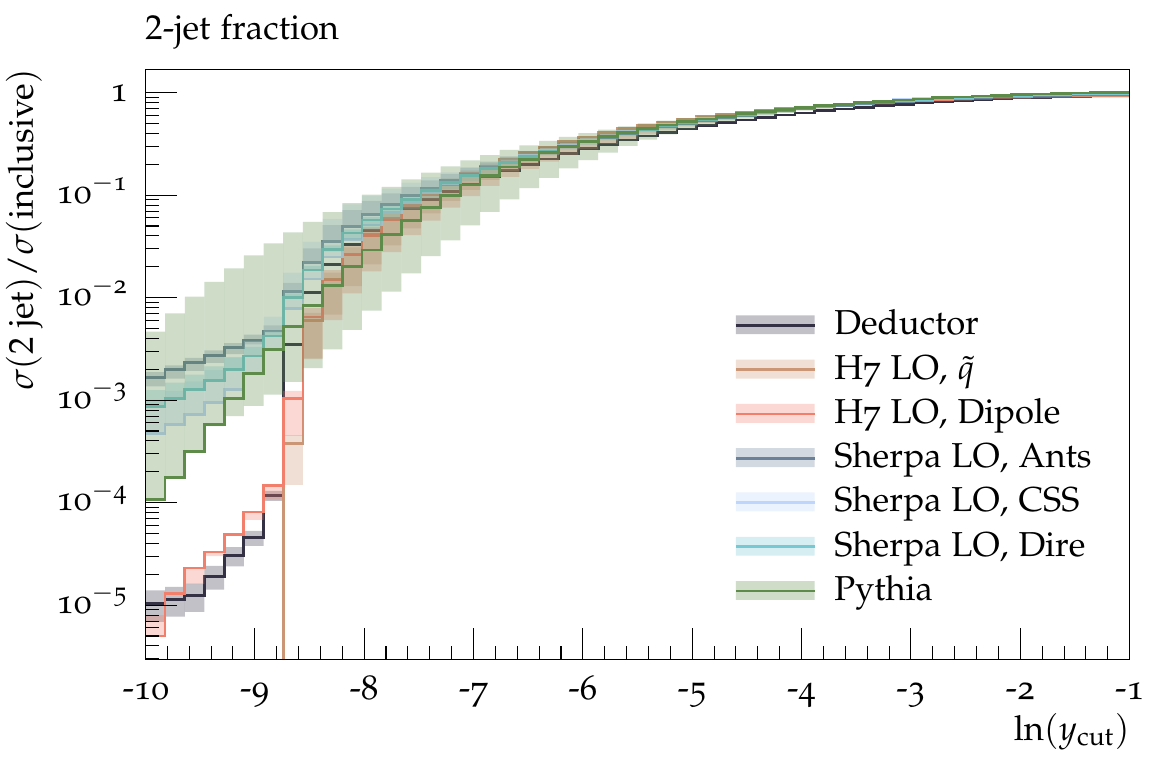}
    \caption{2-jet fraction}
    \label{fig:ee:r2:91}
  \end{subfigure}
  \caption{Jet fractions at $\sqrt{s}=91\,\textrm{GeV}$.}
  \label{fig:ee:r:91}
\end{figure}

The \textsf{Herwig} \textsf{QTilde Shower} ``dead-zone'' can be seen in 
Fig.~\ref{fig:ee:heavyjetmass}-\ref{fig:ee:thrust:91} and
Fig.~\ref{fig:ee:r1:91}.  The C-parameter, Fig.~\ref{fig:ee:cparam:91}, has
reasonable agreement over the majority of the distribution but differences are
evident in its behaviour near $C=0$, and in the region $C>3/4$. As the region
$C>3/4$ is driven by non-planar events, which contain multiple emissions, it
is sensitive to the shower implementation and differences should be expected.
This same is true for Thrust, Fig.~\ref{fig:ee:thrust:91}, which again shows
good agreement over the majority of the distribution, and differences that do
emerge are in regions that are sensitive to multiple emissions, or as we
approach the dijet configuration.
Aside from \textsf{Herwig} \textsf{QTilde Shower}, the results for the 1-jet fraction are in good
agreement within uncertainty bands. For the 2-jet fraction we see the
differences that come from the implementation of the shower cutoff. There is
also a large disagreement on the size of the uncertainty bands reported; here
the choice of evolution variable can cause the shower to probe much smaller
values of $p_\perp$. In this region we should expect some interplay with
hadronization and, though beyond the aim of this study, this merits further
investigation.

\subsubsection{$pp\to h$}
\label{sec:psunc:results:h}
Higgs production (via HEFT) presents two new aspects: initial-state radiation,
and the gluon PDF. The simulations take place at $\sqrt{s}=13\,\textrm{TeV}$,
and the analysis considers anti-$k_T$ jets with $p_\perp\geq 40\,\textrm{GeV}$. We
explore two scenarios $m_H=125\,\textrm{GeV}$ and $m_H=500\,\textrm{GeV}$ to quantify the impact of
the hard scalex. However, we neglect to show the heavy Higgs results, as they
offer no new insight.

\begin{figure}[t!]
  \centering
  \begin{subfigure}[t]{0.49\textwidth}
    \includegraphics[width=\textwidth]{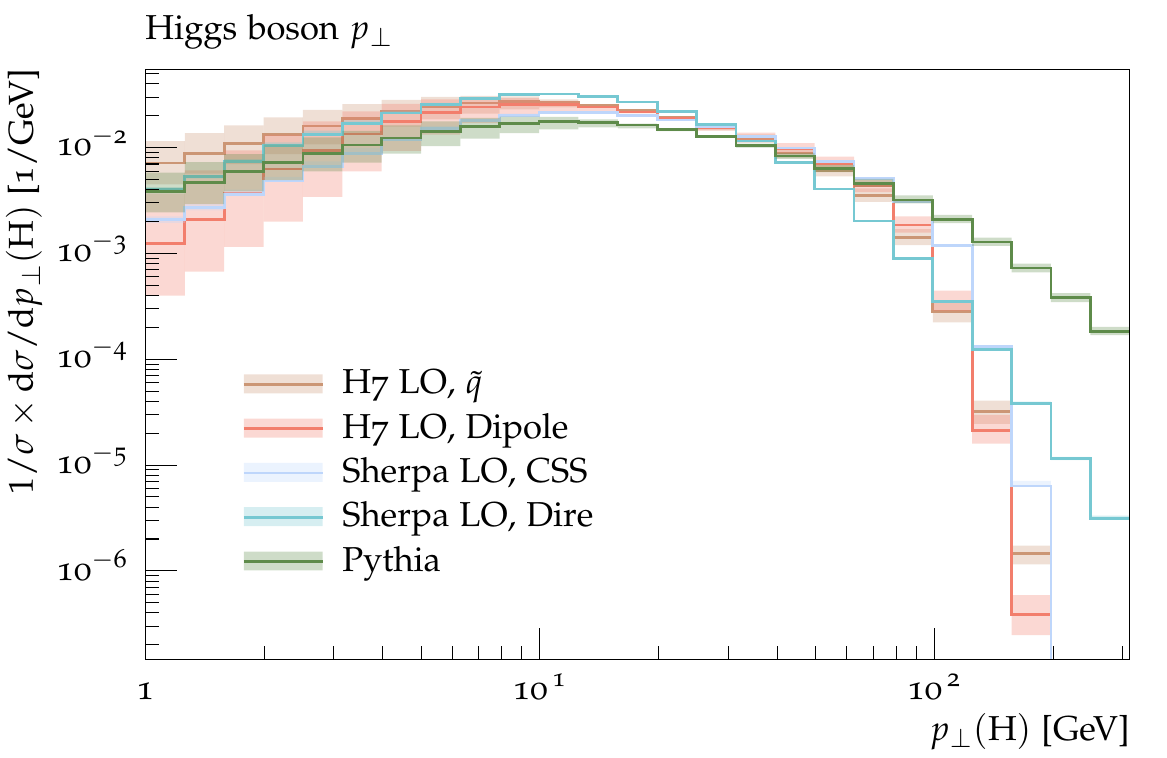}
    \caption{Entire Region}
    \label{fig:h:pt_full}
  \end{subfigure}
  \begin{subfigure}[t]{0.49\textwidth}
    \includegraphics[width=\textwidth]{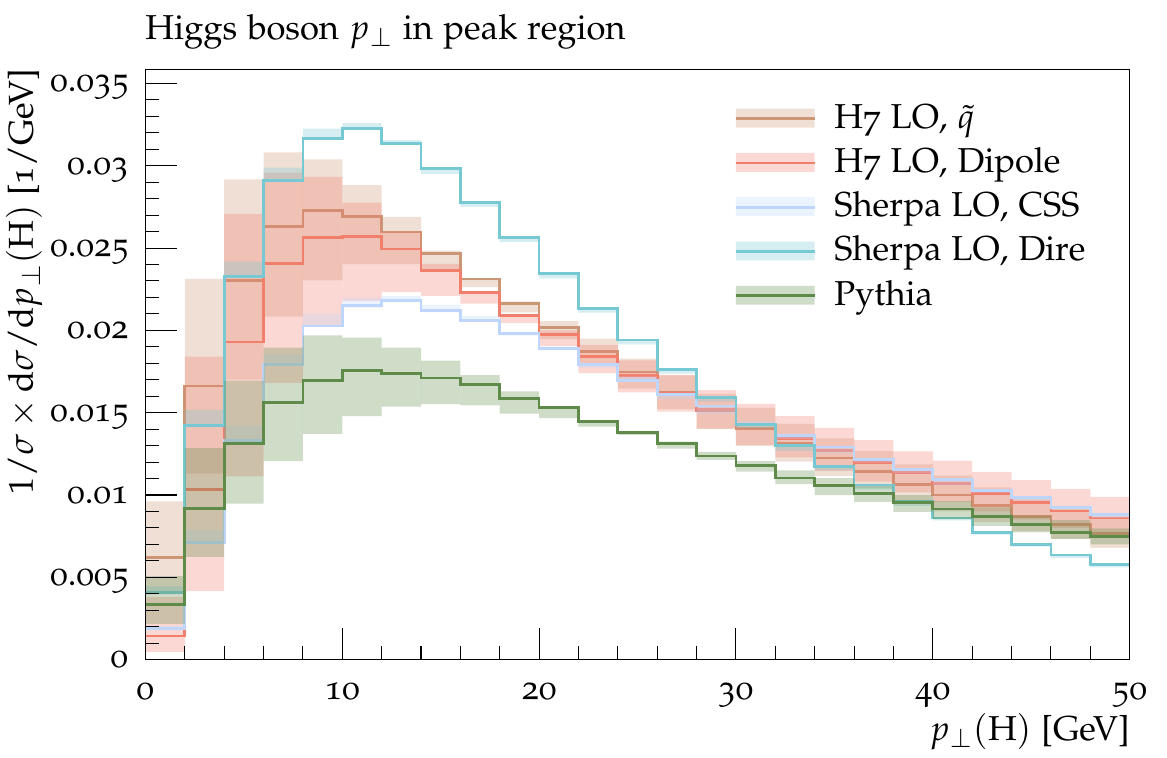}
    \caption{Peak Region}
    \label{fig:h:pt_peak}
  \end{subfigure}
  \caption{Generator comparison of the $H$ $p_\perp$ showing the overall
    behaviour \ref{fig:h:pt_full} as well as the behaviour in the peak region
    \ref{fig:h:pt_peak}}
  \label{fig:h:pt}
\end{figure}

\begin{figure}[t!]
  \centering
  \begin{minipage}[t]{0.49\textwidth}
    \includegraphics[width=1\textwidth]{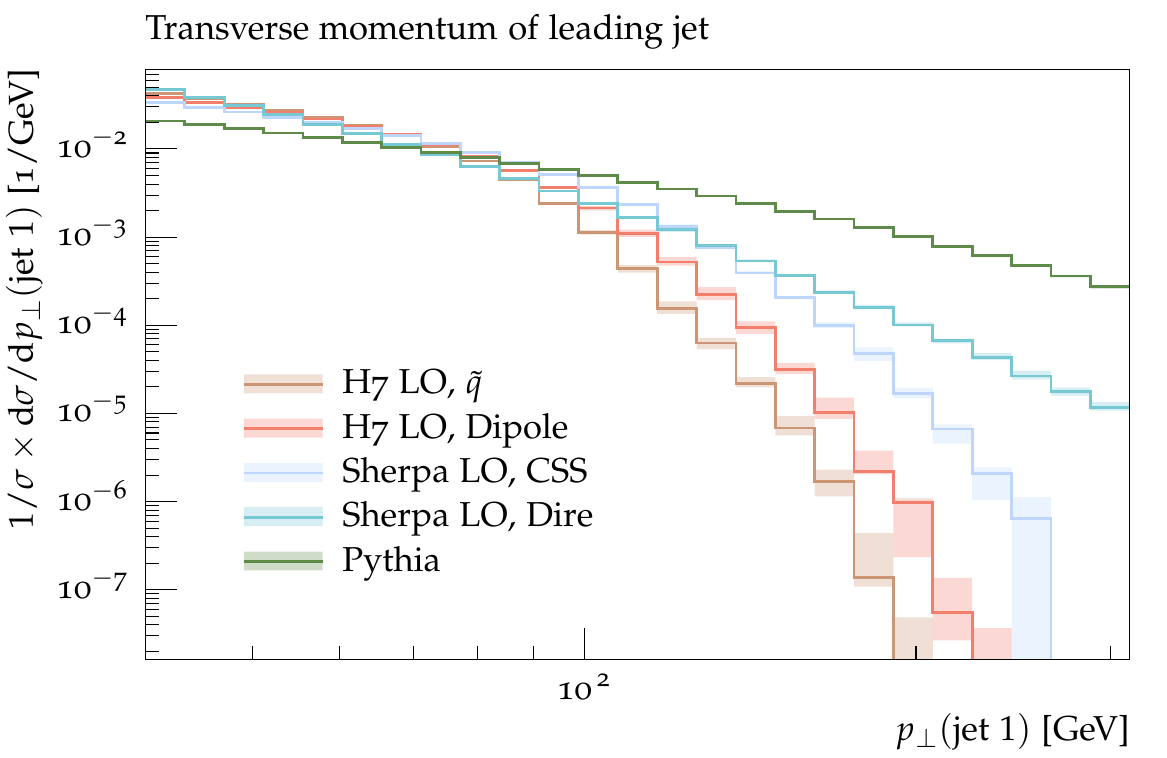}
    \caption{$p_\perp$ of the leading jet}
    \label{fig:h:jet1_pt}
  \end{minipage}
  \begin{minipage}[t]{0.49\textwidth}
    \includegraphics[width=1\textwidth]{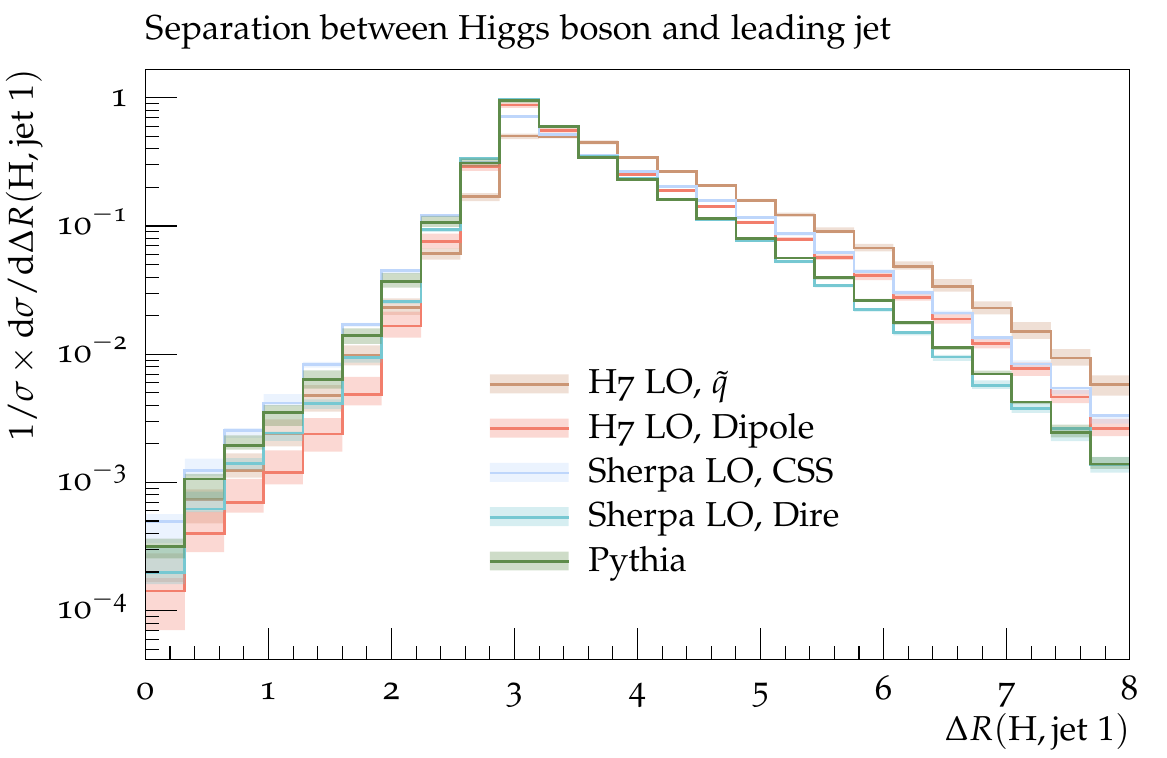}
    \caption{Comparison for the lego-plot distance between the Higgs and the leading jet}
    \label{fig:h:deltaR}
  \end{minipage}
\end{figure}

\begin{figure}[t!]
  \centering
  \begin{minipage}[t]{0.49\textwidth}
    \includegraphics[width=1\textwidth]{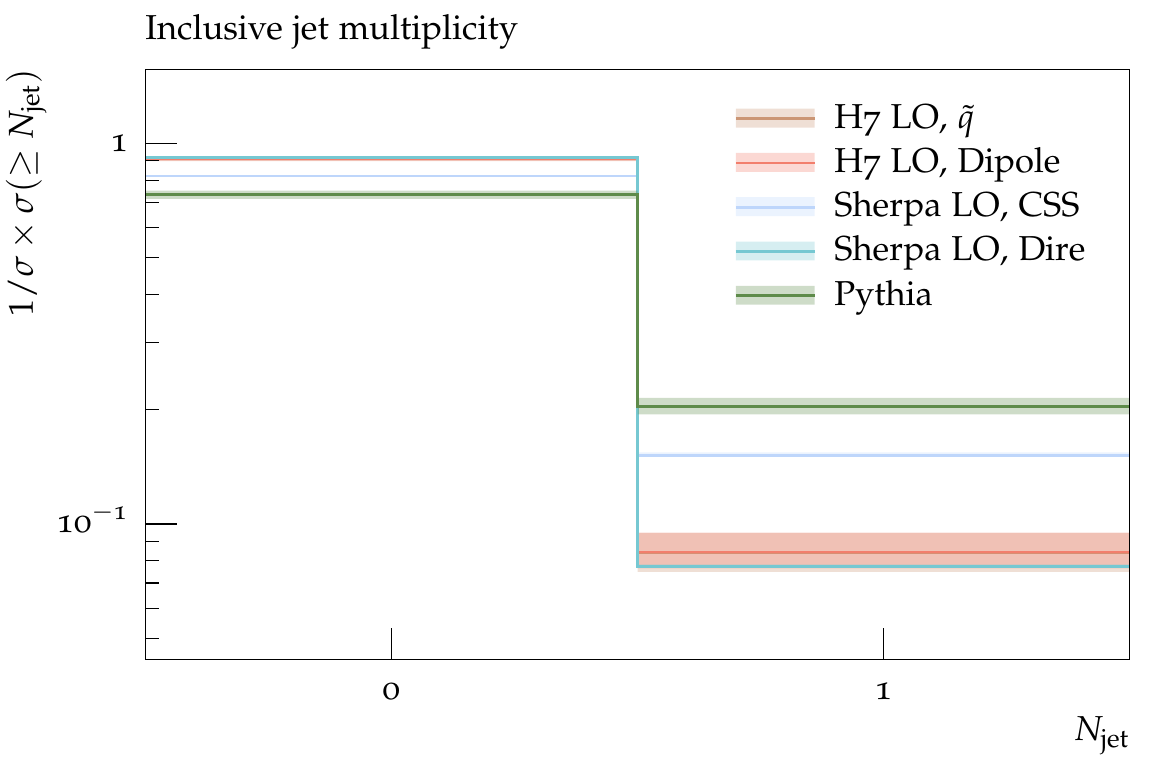}
    \caption{Comparison for inclusive jet multiplicity}
    \label{fig:h:jet_multi_inc}
  \end{minipage}
  \begin{minipage}[t]{0.49\textwidth}
    \includegraphics[width=1\textwidth]{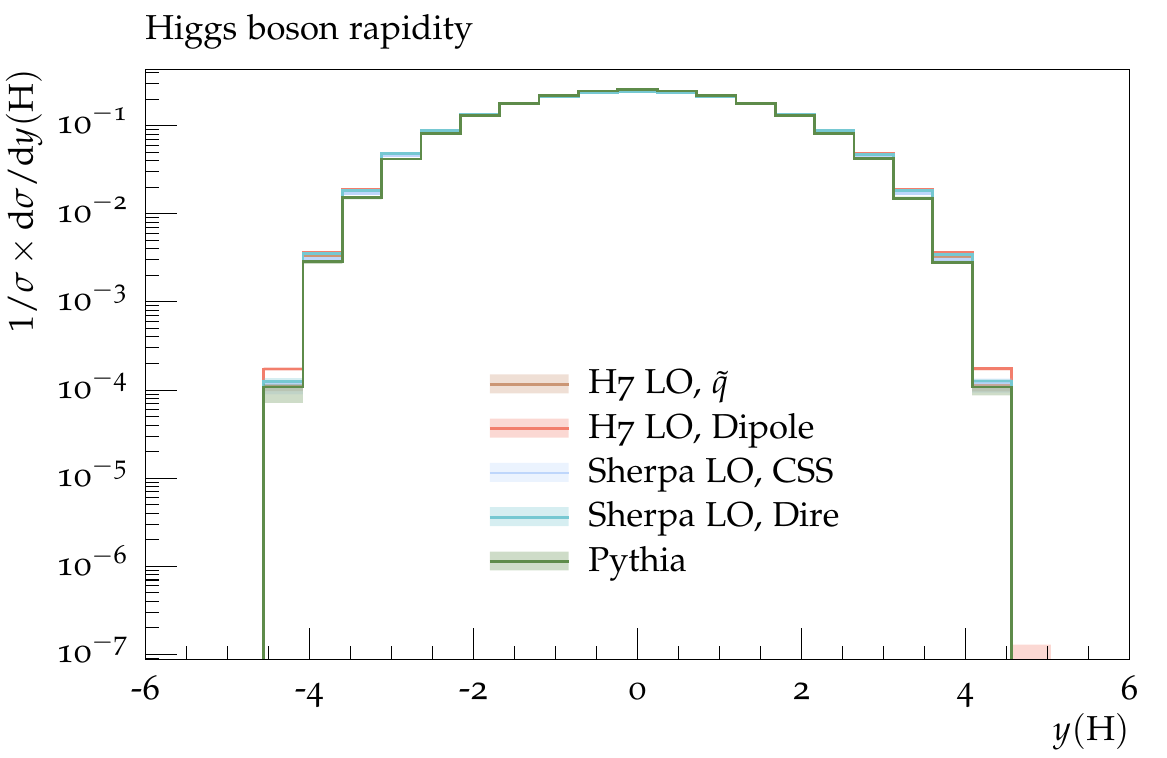}
    \caption{Higgs rapidity}
    \label{fig:h:y}
  \end{minipage}
\end{figure}

The $p_\perp$ distributions for the Higgs, Fig.~\ref{fig:h:pt}, have good
agreement at low $p_\perp$ and expectedly disagree at high $p_\perp$, where
the differences are not reflected in the uncertainty bands. In the peak region
further differences are also apparent, and can be attributed to different
shower implementations: evolution variable, $\alpha_s$ and intrinsic $k_T$,
and recoil. The high $p_\perp$ behaviour of \textsf{Herwig} and \textsc{Sherpa} \textsc{Csshower} are
closer to the scale $m_H$, while \textsf{Pythia} and \textsc{Sherpa} \textsc{Dire} extend further and
probe the phase space of a power-shower setup. The agreement displayed between
the different \textsf{Herwig} showers is expected from the setup, nevertheless it is a
satisfying result given their different implementations.  The leading jet
$p_\perp$, Fig.~\ref{fig:h:jet1_pt}, is consistent at low $p_\perp$ with the
exception of \textsf{Pythia}, which has a notably harder distribution. Though the high
$p_\perp$ predictions are markedly different, \textsf{Herwig} and \textsc{Sherpa} \textsc{Csshower} do
again show the same suppression near the scale $m_H$; the harder jets produced
by a power-shower can clearly be seen here.  The lego-distance,
Fig.~\ref{fig:h:deltaR}, has two notable regions $\Delta R \geq\pi$ and
$\Delta R < \pi$. For the first region, we see that the generators give
differing predictions with small uncertainty bands, this region is typically
driven by few hard-emissions. The second region, driven by multiple emissions,
displays better agreement and, expectedly, has larger uncertainty bands.  The
inclusive jet multiplicity shows some agreement between generators, but
notably \textsc{Sherpa} \textsc{Csshower} and \textsf{Pythia} produce a smaller fraction of $0$-jet
events. Finally, we show the resilience of an inclusive observable to the
shower, and its uncertainties. This can be seen in the Higgs rapidity,
Fig.~\ref{fig:h:y}.

\subsubsection{$pp\to Z$}
\label{sec:psunc:results:z}
To complement the previous process a quark-initiated setup is presented, again
using two different hard scales of $91\ {\rm GeV}$ and $500\ {\rm GeV}$. The
analysis and findings of this section are similar to that of
Sec.~\ref{sec:psunc:results:h}. The predictions for the inclusive jet
multiplicity, Fig.~\ref{fig:z:jet_multi_inc}, are much less consistent
Fig.~\ref{fig:h:jet_multi_inc}.
\begin{figure}[t!]
  \centering
    \begin{subfigure}[t]{0.49\textwidth}
    \includegraphics[width=\textwidth]{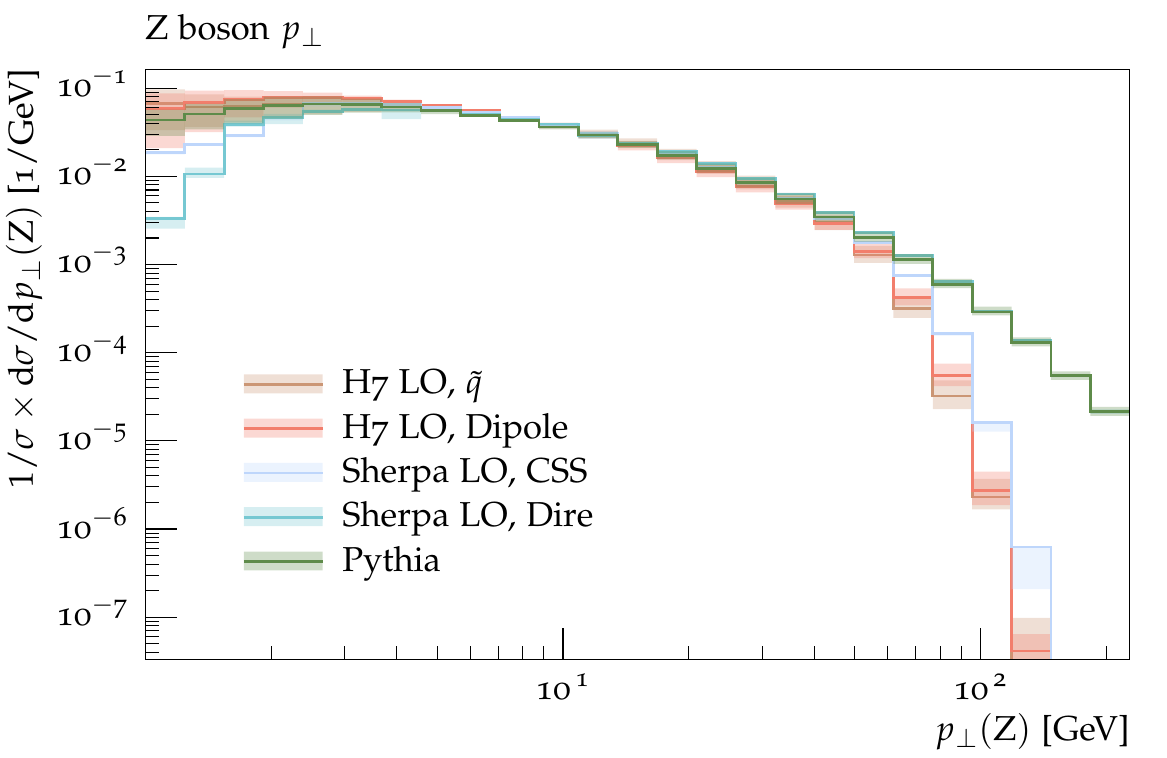}
    \caption{Entire region}
    \label{fig:z:pt_full}
  \end{subfigure}
  \begin{subfigure}[t]{0.49\textwidth}
    \includegraphics[width=\textwidth]{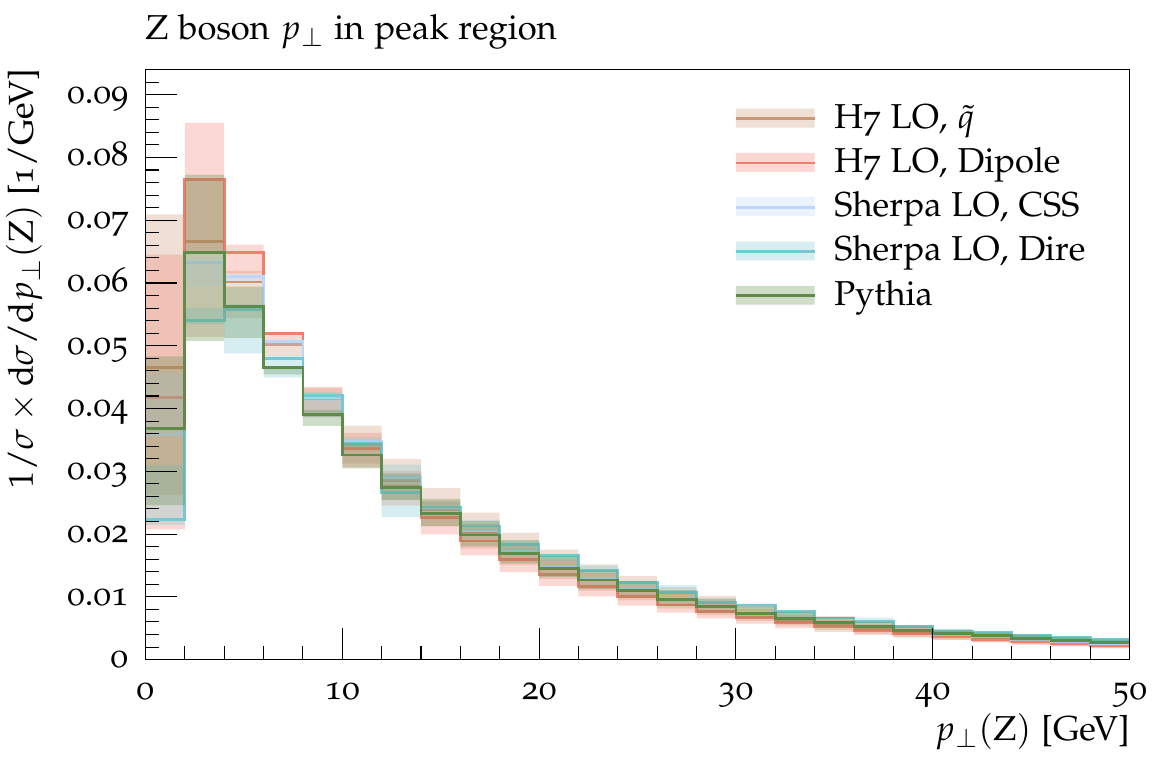}
    \caption{Peak region}
    \label{fig:z:pt_peak}
  \end{subfigure}
  \caption{Generator comparison of the $Z$ $p_\perp$ showing the overall behaviour \ref{fig:z:pt_full} as well as the behaviour in the peak region \ref{fig:z:pt_peak}.}
  \label{fig:z:pt}
\end{figure}

\begin{figure}[t!]
  \centering
  \begin{minipage}[t]{0.49\textwidth}
    \includegraphics[width=1\textwidth]{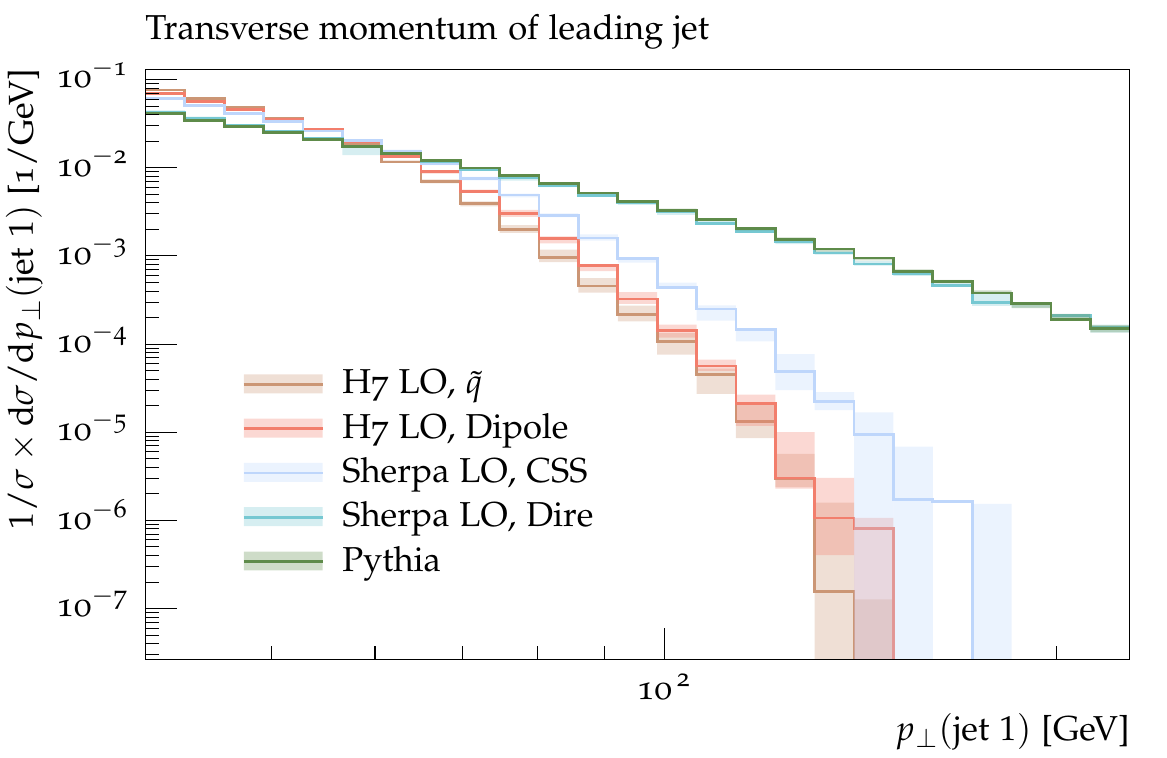}
    \caption{$p_\perp$ of the leading jet}
    \label{fig:z:jet1_pt}
  \end{minipage}
  \begin{minipage}[t]{0.49\textwidth}
    \includegraphics[width=1\textwidth]{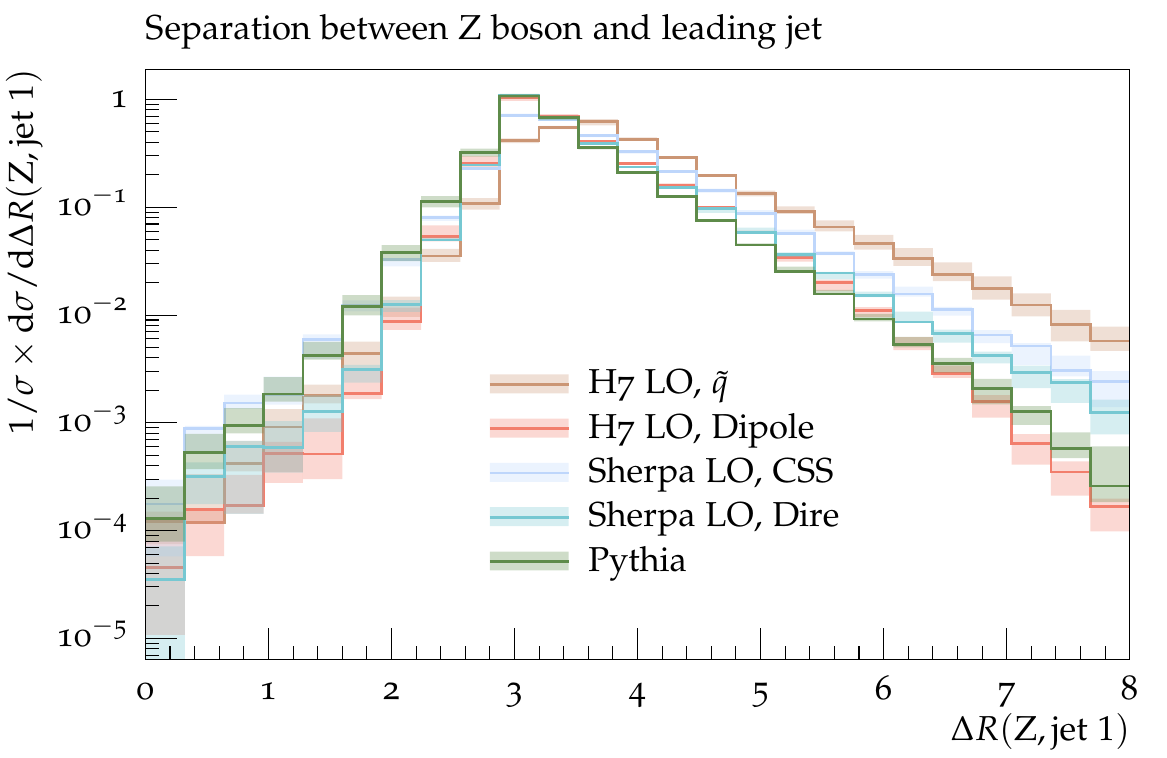}
    \caption{Lego-plot distance between the $Z$ and the leading jet}
    \label{fig:z:deltaR}
  \end{minipage}
\end{figure}

\begin{figure}[t!]
  \centering
  \begin{minipage}[t]{0.49\textwidth}
    \includegraphics[width=1\textwidth]{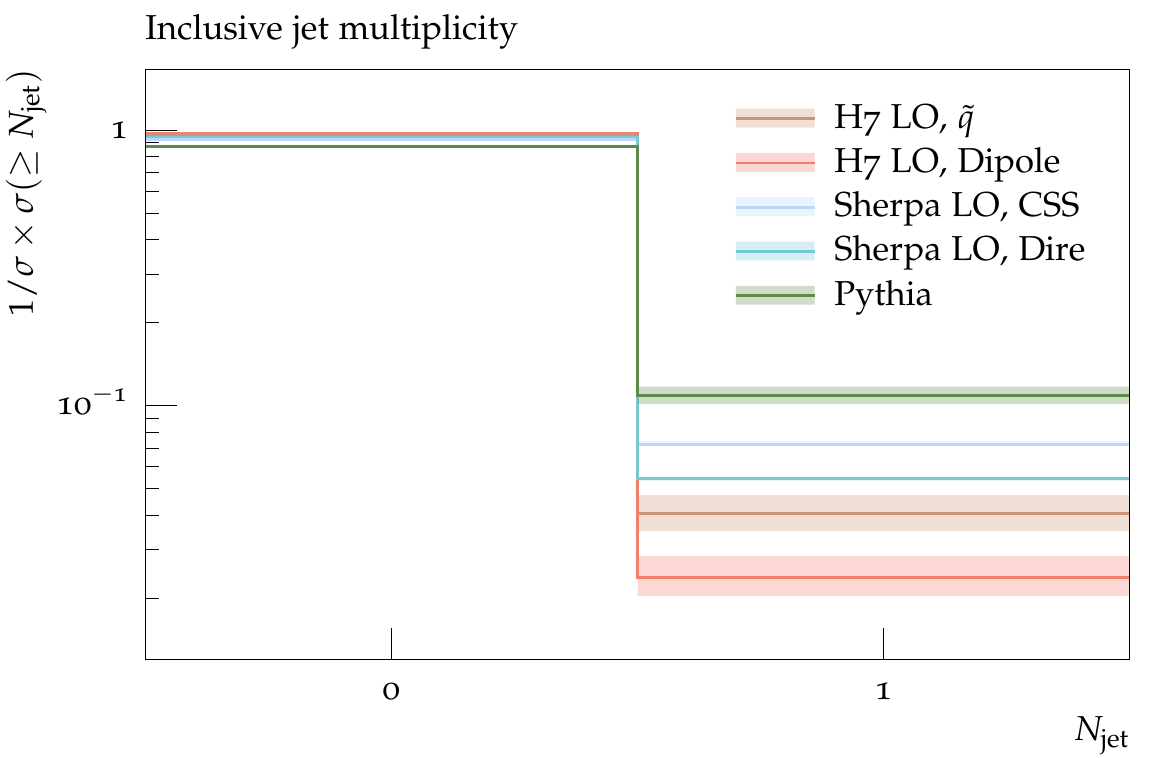}
    \caption{Inclusive jet multiplicity}
    \label{fig:z:jet_multi_inc}
  \end{minipage}
  \begin{minipage}[t]{0.49\textwidth}
    \includegraphics[width=1\textwidth]{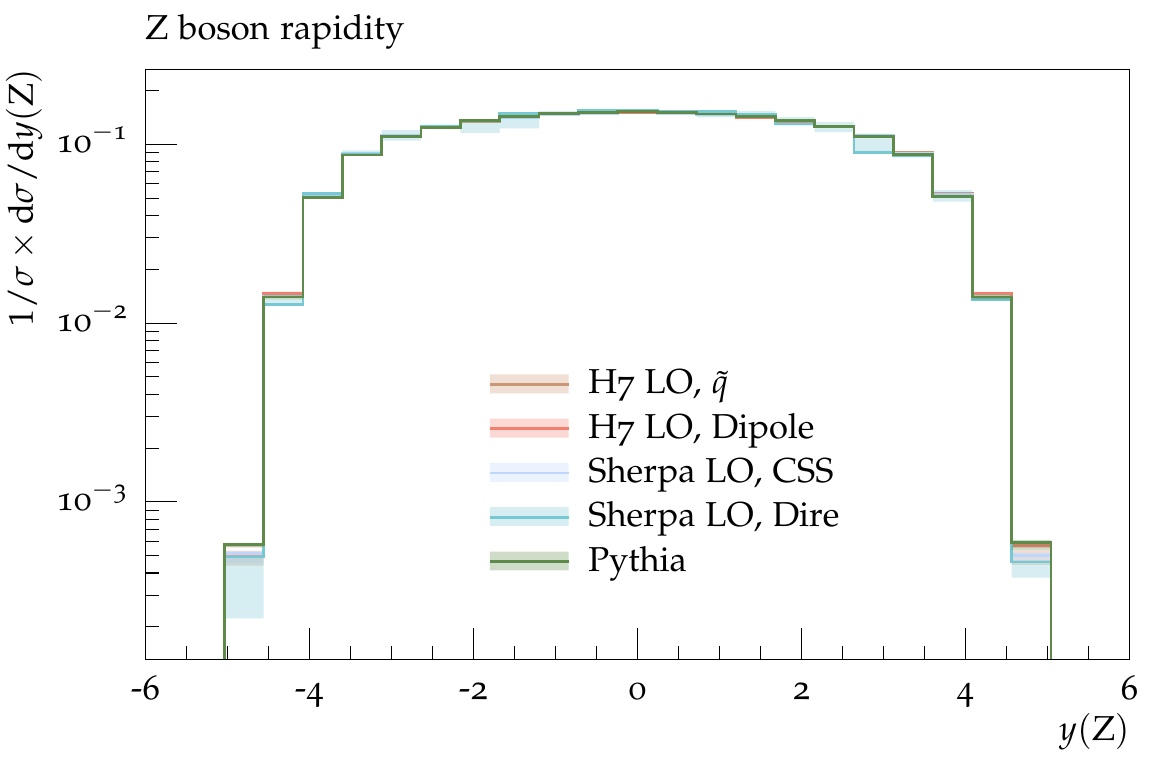}
    \caption{$Z$ rapidity}
    \label{fig:z:y}
  \end{minipage}
\end{figure}

\subsection{Conclusions}
\label{sec:psunc:conclusions}
Clearly there is a need to find a robust prescription to produce variations
that we can definitively call `shower-uncertainties'. Such a prescription
should not only be theoretically grounded, but also practical in its use. This
will require further studies with more detailed appraisals of the different
shower components and their interplay with rest of the event generation
machinery. All of the multi-purpose event generators are able to assign
uncertainties to their prescription and so will enable more detailed studies
in the future.

\subsection*{Acknowledgments}

We thank the organisers.  The work of DS is been supported by the
U.S. Department of Energy under Grant DE-SC0011640.  SP acknowledges support
by a FP7 Marie Curie Intra European Fellowship under Grant Agreement
PIEF-GA-2013-628739.  MS acknowledges support by the Swiss National Science
Foundation (SNF) under contract PP00P2-128552.  The work of SH was supported
by the U.S. Department of Energy under contract DE--AC02--76SF00515.

\section{Reweighting the \texorpdfstring{S\protect\scalebox{0.8}{HERPA}}{SHERPA} parton shower \texorpdfstring{\protect\footnote{E.\ Bothmann, M.\ Sch\"onherr, S.\ Schumann}}{}}
\label{sec:psrew}

\subsection{Introduction}

The last ten years of improving Monte-Carlo event generators (MCEG) enabled
quantitative predictions at an unprecedented accuracy. These developments 
led to a considerable growth in computational cost per event. This can become
limiting, as calculations must often be performed repeatedly, 
e.g.\ to assess theory uncertainties. For QCD calculations such 
re-calculations are necessary for each combination of QCD input parameters. 
Those include parton density functions (PDF), the strong coupling 
($\alpha_S$) or the renormalisation/factorisation scale. This adds up 
to $\mathcal{O}(10^3)$ re-calculations quickly. For PDF fits, the 
demand for re-calculations is even higher.

This issue is addressed by \emph{not doing the whole calculation again}.
Instead, parameter-independent bits of the calculation are stored
as weights and can then be recombined with different parameter values.
This can either be done directly for each event,
or after projecting weights to interpolation grids first,
with one grid per observable bin and dependence
structure~\cite{Carli:2010rw,Kluge:2006xs}.
PDF fitters use such interpolation grids for cross section predictions,
as the time needed for a recombination is of the order of milliseconds only.

For leading-order (LO) events, getting the \emph{parameter-indendent weight} is easy:
divide the event weight
by the PDFs $f_a f_b$ of the incoming partons
and by $\alpha_S^p$, with $p$ being the perturbative order of the process.
Then multiply with $f'_a f'_b \alpha_S'^p$ to get the new event weights
for a different set of input PDFs $f'$ and a different value of the strong 
coupling $\alpha_S'$. For the proper evaluation of these {\em variational
weights} the partonic momentum fractions $x_{1}$ and $x_{2}$ as well as
the factorisation and renormalisation scale need to be book-kept.
At next-to-leading order (NLO)
the procedure is more complicated, because of the additional scale-dependence 
of the virtual corrections and the different kinematics and initial states of the 
subtraction parts~\cite{Bern:2013zja}.

The newly added internal reweighting of \textsc{Sherpa}~\cite{Gleisberg:2008ta}
as of version 2.2.0 performs \emph{on-the-fly}
during event generation~\cite{Bothmann:2015woa}. The result for each combination is inserted into the
{\small \tt HepMC::WeightContainer} object and thus can be either saved to disk,
programmatically accessed or directly passed via the internal interface to the
{\small \tt Rivet} analysis framework~\cite{Buckley:2010ar}.
The supported event generation modes are (N)LO, (N)LO+PS (i.e. matched to a parton shower)
and MEPS@LO (i.e. merged LO matrix elements with a shower).
The recently introduced interfaces {\small\tt MCgrid}~\cite{DelDebbio:2013kxa} and
{\small\tt aMCfast}~\cite{Bertone:2014zva} on the other hand
allow for the automated creation of interpolation grids using
general MCEGs. (N)LO and (N)LO+PS are supported.

For both reweighting approaches, event generations
involving a parton shower come with a caveat:
The shower takes no part
in the reweighting procedure, and is kept as-is. Its dependences
on PDF, scale and $\alpha_S$ are thus not taken into account.
Therefore the reweighting leads to an \emph{inconsistent result},
where dependences in the hard process have been updated,
but not in the shower.
There are indications that this does not affect major parts of
phase space~\cite{Buckley:2016caq,Gieseke:2004tc}.
A more thorough study is however still missing.
We address this by presenting ongoing work to extend the internal 
\textsc{Sherpa} reweighting to include all dependences of the 
parton shower emissions, using a property of the
veto algorithm that allows for a simple multiplicative reweighting.

\subsection{Reweighting the veto algorithm}

The default shower of \textsc{Sherpa}, \textsc{Csshower} \cite{Schumann:2007mg}, 
uses the veto algorithm to
numerically integrate the Sudakov form factors
\begin{equation}
  \Delta(t_0, t_c) = e^{- \int_{t_c}^{t_0} \Gamma(t) \mathrm{d} t}\,,
\end{equation}
which give the no-branching probabilities between the starting scale
$t_0$ and the cut-off scale $t_c$.
In the veto algorithm the splitting kernels $\Gamma$ are replaced with
integrable overestimates $\hat{\Gamma}$. This is balanced by only accepting a
proposed emission with the probability $P_\mathrm{acc} = \Gamma / \hat{\Gamma}$.
A multiplicative factor in $\Gamma$ is therefore equivalent to a
multiplicative factor in $P_\mathrm{acc}$~\cite{Hoeche:2009xc}.
This observation is for example used to apply matrix element 
corrections~\cite{Hoeche:2011fd}, where the splitting kernels are replaced 
with a real-emission-like kernel $R/B$. This is done a-posteriori, i.e. the event weight is multiplied
by $(R/B)/\Gamma$. The emission itself is unchanged.
The same method is also used in the {\small\tt Vincia} parton shower to calculate
uncertainty variations for different scales, finite terms of the antenna functions,
ordering parameters and sub-leading colour corrections~\cite{Giele:2011cb}.

The emission kernels $\Gamma$ depend linearly on $\alpha_S$ and
on a ratio of PDF values $f_a(x/z)/f_b(x)$\footnote{The exact definition of
the flavours $a$, $b$ and the momentum fractions $x$, $x/z$ depend on the
dipole configuration of the emission. There is no PDF dependence in the case
of dipoles that consist only of final-state partons.}.
A change of PDFs $f \rightarrow f'$ and the strong coupling
$\alpha_S \rightarrow \alpha_S'$
is equivalent to modifying the emission probability
accordingly\footnote{Both $\alpha_S$ and $f$ depend on the
emission scale. Although the emission scales can not be
reweighted themselves using the presented method, the functional
form of these dependences can be changed.}:
\begin{equation}
  P_\mathrm{acc} \rightarrow q\,P_\mathrm{acc}\,,\qquad
  q \equiv \frac{\alpha_S'}{\alpha_S} \cdot
  \frac{f'_a(x/z)/f'_b(x)}{f_a(x/z)/f_b(x)}\,,
\end{equation}
i.e. we need to multiply the event weight for each accepted emission by the 
corresponding $q$ in order to reweight the event with a new choice of PDFs.
For the rejected emissions we obtain
\begin{equation}
  P_\mathrm{rej} = 1 - P_\mathrm{acc} \rightarrow 1 - q P_\mathrm{acc}
  = \left[ 1 + \left( 1 - q \right) \frac{P_\mathrm{acc}}{1 - P_\mathrm{acc}}
    \right]\,P_\mathrm{rej}\,,
\end{equation}
so for each of those we need to multiply the event weight with the expression
in square brackets.

Our current implementation supports PDF and $\alpha_S$ reweighting of parton shower emissions for
both LO+PS and NLO+PS events.
One can choose a maximum number
of reweighted emissions per event.
This is useful because some observables are
sensitive only to the first few emissions.
The reduced amount of reweighting per event then
allows for faster event generations.

\subsection{Validation}

We present a selection of plots to validate our implementation.
All contain comparisons between ``dedicated'' and ``reweighted'' results.
The ``dedicated'' results are produced through normal runs for the varied
input parameters, whereas the ``reweighted'' ones are obtained by reweightings
from a run with a central parameter choice.
The reweighting is done for the matrix element exclusively (``ME''),
including all parton shower emissions (``ME+PS'') or just
the first emission (``ME+PS(1st em.)'').

\begin{figure}[htbp]
\centering
\includegraphics[width=0.49\textwidth]{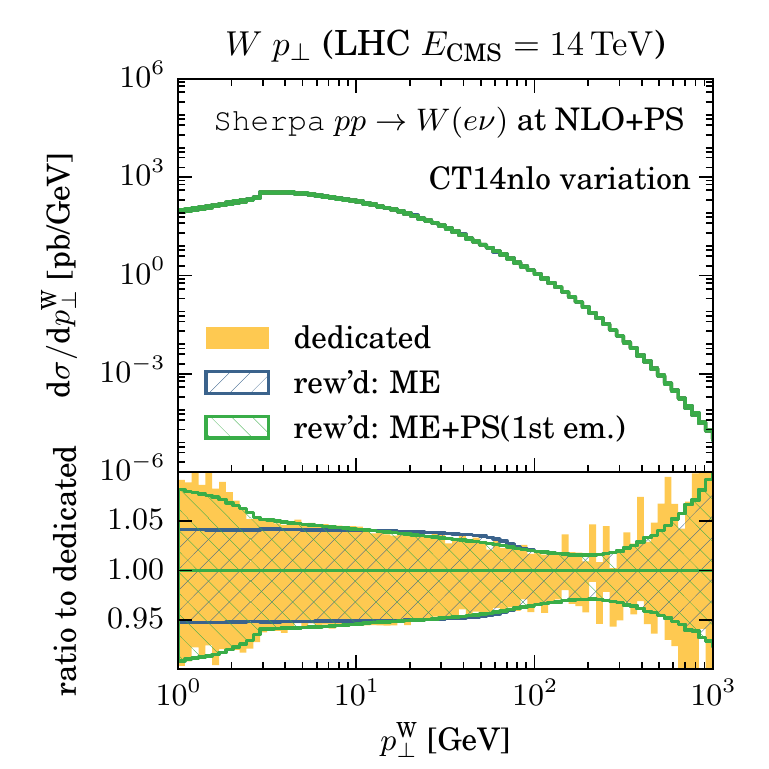}
\includegraphics[width=0.49\textwidth]{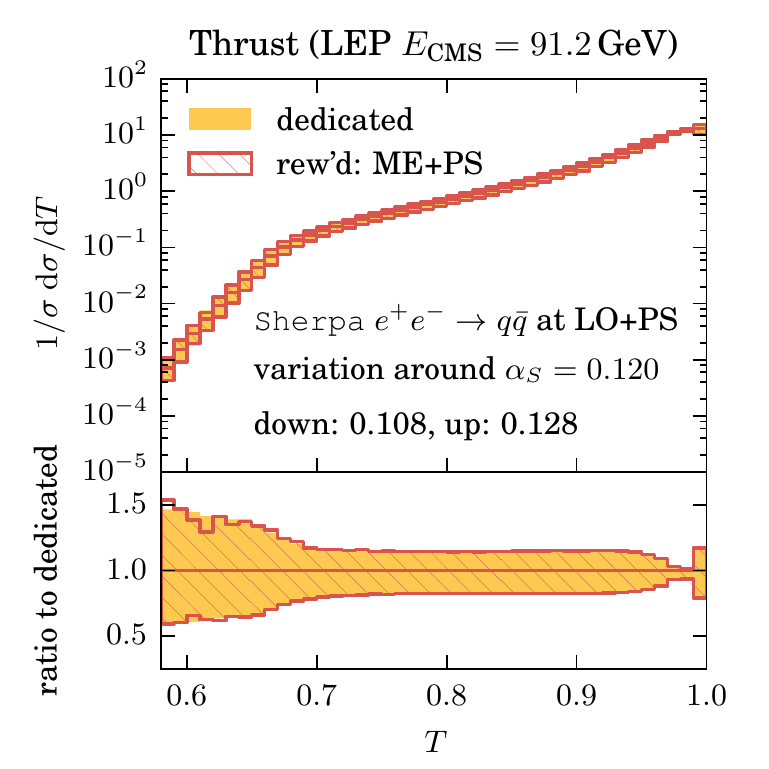}
\caption{A PDF variation for the $W$-boson $p_\perp$ distribution at the LHC 
  at NLO+PS (left) and an $\alpha_S$ variation for the Thrust for $q\bar{q}$ 
  production at LEP at LO+PS (right). Predictions generated by single 
  reweighting runs (``rew'd'') are compared to predictions from dedicated runs.}
\label{psrew_fig:appl}
\end{figure}

In Fig.~\ref{psrew_fig:appl}, we present uncertainty bands for PDF and 
$\alpha_S$ variations. The left-hand plot features a 
{\small\tt CT14nlo}~\cite{Dulat:2015mca} PDF error band for the $W$-boson
transverse momentum distribution in $W$ production
at the LHC at NLO+PS. The band is a combination of the central member and 56
Hessian eigenvector members. So for the ``dedicated'' band,
57 independent event generation runs had to be performed,
whereas each of the reweighting bands is generated by
a single run.\footnote{We are yet to do a detailed timing
study (and to optimise the code), but for the case of doing 57 variations
and reweighting the ME and up to one parton shower emission,
the run took about a factor of 6 longer than a run
without any reweighting.
Comparing the time needed to generate the entire error band,
the reweighting run thus needed one order of magnitude
less than the combination of dedicated runs.}

A few observations can be made for the $W$ $p_\perp$ plot: (i) For
$p_\perp < 10$ GeV, solely reweighting the ME underestimates 
the error found with the dedicated calculation by about 4\,\% 
in both directions. The ME+PS reweighting is able to reproduce the error.
(ii) For $p_\perp$ values between 20\,GeV and the $W$~mass,
the ME reweighting \emph{overestimates} the positive error slightly,
by about 1\,\%. Again, the ME+PS reweighting correctly
reproduces the error.
(iii) The reweighted bands are much smoother because the prediction
for each PDF member shares the statistics of the central run.
Unfortunately there is no way to guarantee this for the dedicated
runs when varying input parameters of the parton shower,
which makes closure tests between reweighting and dedicated runs a
statistical exercise.

The right-hand plot in Fig.~\ref{psrew_fig:appl}
contains a band for an $\alpha_S$ variation for the thrust
observable for $e^+e^- \to q\bar{q}$ at LEP at LO+PS.
The variations considered around the central values of 
$\alpha_S(m_Z)=0.120$ include $\alpha_S(m_Z)=0.108$ and 
$\alpha_S(m_Z)=0.128$. The running of $\alpha_S$ is evaluated 
at the one-loop order. The plot shows the envelope for the three values.
The reweighted band is obtained by reweighting
all emissions in the run with the central $\alpha_S$ value.
It reproduces the dedicated band within statistical errors.
For this we note that the variance of the reweighted prediction
increases. In some bins, the sum of squared weights is a factor
of 100 higher compared to the dedicated result.

\begin{figure}[htbp]
\centering
\includegraphics[width=0.49\textwidth]{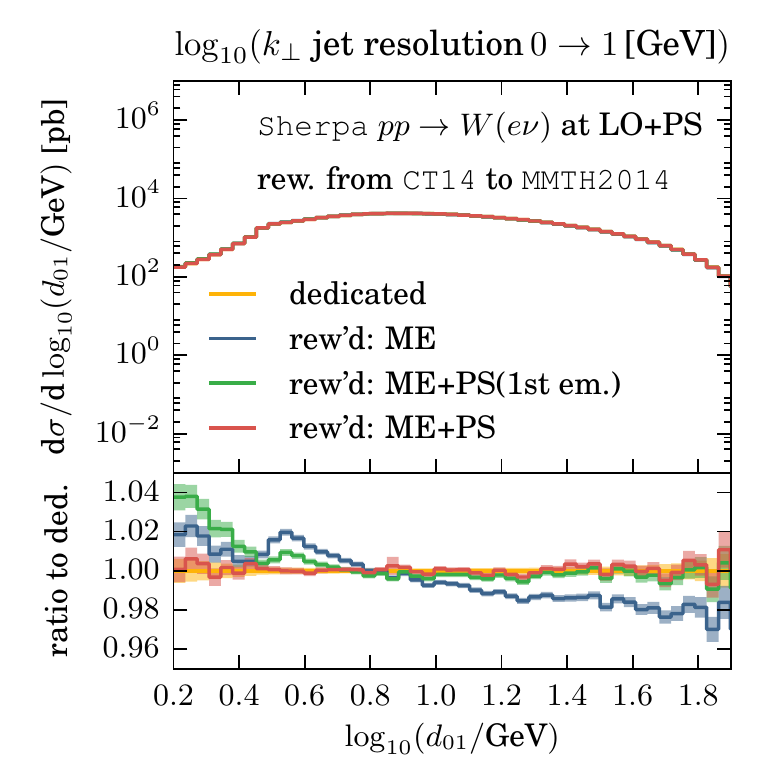}
\includegraphics[width=0.49\textwidth]{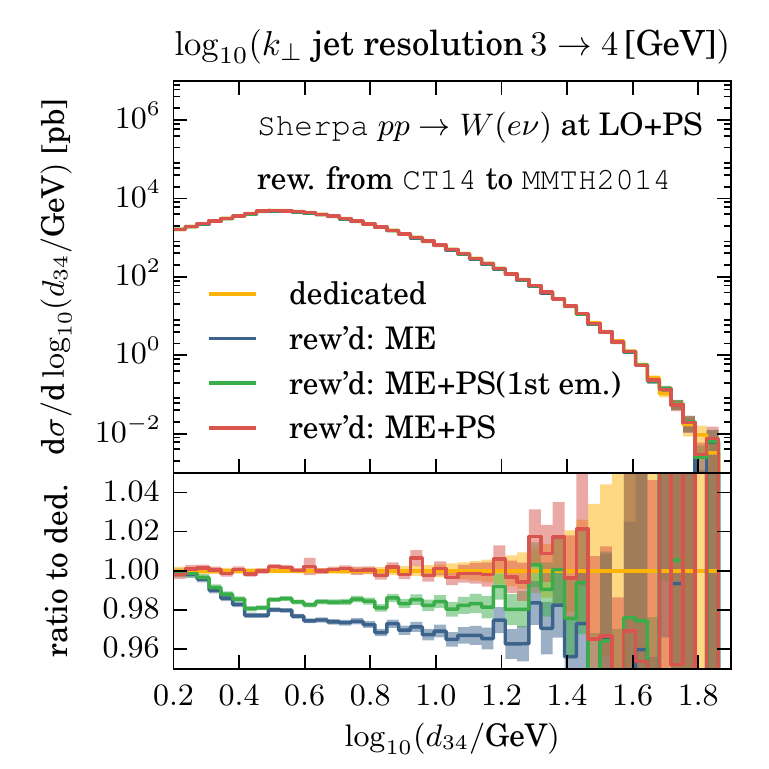}
\caption{Predictions for jet resolutions for $W$-boson productions at the LHC 
  at LO+PS. Results from reweighting runs {\small\tt CT14} $\to$ {\small\tt MMHT2014} 
  PDF are compared to the dedicated result for direct use of the {\small\tt MMHT2014} 
  PDF.}
\label{psrew_fig:nrew}  
\end{figure}

Jet resolutions are sensitive to the specifics of parton shower emissions,
which is why we have in Fig.~\ref{psrew_fig:nrew} a series of closure tests
for this observable for $W$-boson production at the LHC at LO+PS.
The reweightings are done from the central {\small\tt CT14}
to the central {\small\tt MMHT2014} set~\cite{Harland-Lang:2014zoa}.
The maximum number of shower emissions
being reweighted is varied between 0, 1 and $\infty$.
The corresponding reweighted
predictions are compared to the dedicated {\small\tt MMHT2014} result.
This is done for the $0\to1$ and for the $3\to4$ jet resolution.
We observe deviations for the ME-only reweighting prediction of about $\pm2\,\%$.
The 1-emission-reweighting fixes the deviation for $d_{01} > 10$ GeV, but can not
fix the deviations elsewhere. The all-emissions-reweighting is perfect within statistics
for all bins of both jet resolutions.

\subsection{Conclusions}

We have presented the extension of the internal reweighting of \textsc{Sherpa}
to include the $\alpha_S$ and PDF dependences of parton shower emissions.
It has been shown that the ME-only reweighting has shortcomings compared to the newly
implemented full reweighting.
We did this by comparing theory uncertainty bands generated by reweighting runs
with dedicated calculations for $\alpha_S$ and PDF variations at LO+PS and (N)LO+PS.
Now with the support for these input parameters in place, the next step is to
allow for varying the scales at which $\alpha_S$ and the PDFs are evaluated
for a given emission. This is different from varying the actual emission scale
(e.g.\ indirectly by choosing a new starting scale), which is not covered
by the presented reweighting method.

The full direct reweighting for parton showers
will leave us with a complete picture of
how to reweight when all information is still available.
This will help us when bringing parton shower reweighting
also to grids, where individual event information is \emph{not} available.
For this we are considering an approximate approach.
Such shower-aware grids would allow for including more data in PDF fits,
namely data that is currently omitted because its theoretical prediction
is sensitive to higher-order corrections.

\subsection*{Acknowledgements}
We acknowledge financial support from BMBF under contract 05H15MGCAA, 
the Swiss National Foundation (SNF) under contract PP00P2--128552 and 
from the EU MCnetITN research network funded under Framework Programme 
7 contract PITN--GA--2012--315877.

\section{Ntuples for NNLO events produced by \texorpdfstring{\texttt{EERAD3}}{EERAD3} \texorpdfstring{\protect\footnote{G.~Heinrich, D.~Ma{\^i}tre}}{}}
\label{sec:NNLOntuples}

We study the production of Ntuples based on the program {\tt EERAD3}
which produces parton-level QCD events to calculate event shapes and
jet rates in electron-positron annihilation through to order $\alpha_s^3$.
The aim of this study is to assess the viability of Ntuples as a general
way to have NNLO results stored and made available to the experimental community.

\subsection{Introduction}

High precision calculations will be vital in the next phase(s) of the LHC
in order to be able profit from the high quality data being collected.
In order to further explore the Higgs sector and distinguish
BSM effects from higher order effects within the Standard model,
next-to-next-to leading order (NNLO) predicitions
are necessary for a number of proccesses.
However, such predictions are the results of complex calculations,
which may take a considerable amount of time and computing resources.
Running such programs for various scale choices, parton distribution functions
and sets of cuts  is a tedious, time consuming task.

For processes with multi-particle final states at NLO, one is faced with similar
problems.
A possible solution, described in detail in Ref.~\cite{Bern:2013zja},
is to store the phase space points and the corresponding matrix elememt weights,
together with other relevant information,
in {\tt Root Ntuple} files.
This has the following advantages:
\begin{enumerate}
\item the results are flexible for (tighter) cuts to be applied at a later stage,
\item scale and PDF variations can be performed without lengthy runs of  the original program,
\item it allows to make the results readily available to experimantalists.
\end{enumerate}
The Ntuple format for NLO results already has been used successfully for various
high-multiplicity processes, e.g. W+4\,jets~\cite{Bern:2011ep},
W+5\,jets~\cite{Bern:2013gka}, or H+3\,jets~\cite{Cullen:2013saa,Greiner:2015jha}.

In this note we investigate whether the Ntuple format is also a viable option to store NNLO
results. The limitation here can come only from the size of the files containing the Ntuples.
Total sizes of the order of a TeraByte have proven practical for NLO calculations.

To get a first idea of the file sizes needed to store events which allow to produce reasonably smooth
histograms for observables calculated at NNLO, we take the program {\tt EERAD3}~\cite{Ridder:2014wza} as an example.
{\tt EERAD3} is a parton level event generator which computes the QCD corrections
to event shape observables  and
jet rates in electron-positron annihilation to order $\alpha_s^3$~\cite{GehrmannDeRidder:2007hr,GehrmannDeRidder:2008ug}.
As this example does not require any parton distribution functions, and therefore
does not contain any factorization scale dependence,
it lends itself for a first study,
which can be extended to the case of hadronic collisions later.

\subsection{Ntuples at NNLO}

\subsubsection{Brief description of the {\tt EERAD3} program}

The perturbative expansion for the distribution of an
event shape observable $\Obs$ up to NNLO at the centre-of-mass energy $\sqrt{s}$
and renormalisation scale $\mu^2 = s$, with
$\alpha_s\equiv \alpha_s(\sqrt s)$,  is given by
\begin{eqnarray}
\frac{1}{\sigma_{{\rm had}}}\, \frac{\dbox\sigma}{\dbox \Obs} &=&
\left(\frac{\alpha_s}{2\pi}\right) \frac{\dbox \bar A}{\dbox \Obs} +
\left(\frac{\alpha_s}{2\pi}\right)^2 \frac{\dbox \bar B}{\dbox \Obs}
+ \left(\frac{\alpha_s}{2\pi}\right)^3
\frac{\dbox \bar C}{\dbox \Obs} + {\cal O}(\alpha_s^4)\;.
\label{eq:NNLO}
\end{eqnarray}
In Eq.~(\ref{eq:NNLO}) the event shape distribution
is normalised to the total hadronic cross section $\sigma_{\rm{had}}$.
The latter can be expanded as
\begin{equation}
  \sigma_{\rm{had}}=\sigma_0\,
\left(1+\frac{3}{2}C_F\,\left(\frac{\alpha_s}{2\pi}\right)
+K_2\,\left(\frac{\alpha_s}{2\pi}\right)^2+{\cal O}(\alpha_s^3)\,
\right) \;,
\end{equation}
where the Born cross section for $e^+e^- \to q \bar q$ is
$\sigma_0 = \frac{4 \pi \alpha}{3 s} N  e_q^2$,
assuming massless quarks.
The constant $K_2$ is given by~\cite{Chetyrkin:1996ia}
\begin{equation}
  K_2=\frac{1}{4}\left[- \frac{3}{2}C_F^2
+C_FC_A\,\left(\frac{123}{2}-44\zeta_3\right)+C_FT_RN_F\,(-22+16\zeta_3)
 \right] \;,
\end{equation}
with
$C_A = N$, $C_F = (N^2-1)/(2N)$,
$T_R = {1}/{2}$,  and $N_F$ light quark flavours.
The program {\tt EERAD3} computes the perturbative coefficients $A$, $B$ and $C$, which are
normalised to $\sigma_0$:
\begin{eqnarray}
\frac{1}{\sigma_0}\, \frac{\dbox\sigma}{d \Obs} &=&
\left(\frac{\alpha_s}{2\pi}\right) \frac{\dbox  A}{\dbox \Obs} +
\left(\frac{\alpha_s}{2\pi}\right)^2 \frac{\dbox  B}{\dbox \Obs}
+ \left(\frac{\alpha_s}{2\pi}\right)^3
\frac{\dbox  C}{\dbox \Obs} + {\cal O}(\alpha_s^4)\,.
\label{eq:NNLOsigma0}
\end{eqnarray}
$A$, $B$ and $C$ are straightforwardly related to $\bar{A}$, $\bar{B}$
and $\bar{C}$:
\begin{eqnarray}
&&\bar{A} = A\;,\;
\bar{B} = B - \frac{3}{2}C_F\,A\;,\;
\bar{C} = C -  \frac{3}{2}C_F\,B+ \left(\frac{9}{4}C_F^2\,-K_2\right)\,A
\;.\label{eq:ceff}
\end{eqnarray}
As these coefficients are computed at a renormalisation scale fixed to
the centre-of-mass energy, they
depend only on the value of the observable $\Obs$.

The QCD coupling constant evolves according to the renormalisation group
equation, which reads to NNLO:
\begin{equation}
\label{eq:running}
\mu^2 \frac{\dbox \alpha_s(\mu)}{\dbox \mu^2} = -\alpha_s(\mu)
\left[\beta_0 \left(\frac{\alpha_s(\mu)}{2\pi}\right)
+ \beta_1 \left(\frac{\alpha_s(\mu)}{2\pi}\right)^2
+ \beta_2 \left(\frac{\alpha_s(\mu)}{2\pi}\right)^3
+ {\cal O}(\alpha_s^4) \right]\,
\end{equation}
The coefficients $\beta_i$ can be found e.g. in \cite{Ridder:2014wza}.
Eq.~(\ref{eq:running})
is solved by introducing $\Lambda$ as integration constant
with $L= \log(\mu^2/\Lambda^2)$, yielding the running coupling constant:
\begin{equation}
\alpha_s(\mu) = \frac{2\pi}{\beta_0 L}\left( 1-
\frac{\beta_1}{\beta_0^2}\, \frac{\log L}{L} + \frac{1}{\beta_0^2 L^2}\,
\left( \frac{\beta_1^2}{\beta_0^2}\left( \log^2 L - \log L - 1
\right) + \frac{\beta_2}{\beta_0}  \right) \right)\;.
\end{equation}

In terms of the running coupling $\alpha_s(\mu)$, the
NNLO (non-singlet) expression for event shape distributions therefore becomes
\begin{eqnarray}
\frac{1}{\sigma_{{\rm had}}}\, \frac{\dbox\sigma}{\dbox \Obs} (s,\mu^2,\Obs) &=&
\left(\frac{\alpha_s(\mu)}{2\pi}\right) \frac{\dbox A}{\dbox \Obs} +
\left(\frac{\alpha_s(\mu)}{2\pi}\right)^2 \left(
\frac{\dbox  B}{\dbox \Obs} + \frac{\dbox  A}{\dbox \Obs} \left[\beta_0
\log\frac{\mu^2}{s} -\frac{3}{2}\,C_F\right]\right)
\nonumber \\ &&
+ \left(\frac{\alpha_s(\mu)}{2\pi}\right)^3
\Bigg(\frac{\dbox  C}{\dbox \Obs} +  \frac{\dbox  B}{\dbox \Obs}
\left[ 2\beta_0\log\frac{\mu^2}{s}-\frac{3}{2}\,C_F\right]\nonumber\\
&&+ \frac{\dbox  A}{\dbox \Obs} \left[\beta_0^2\,\log^2\frac{\mu^2}{s}
+ \beta_1\, \log\frac{\mu^2}{s}  -3\beta_0 C_F \, \log\frac{\mu^2}{s}-K_2+\frac{9}{4}C_F^2\right]\Bigg)
\nonumber \\ &&
 + {\cal O}(\alpha_s^4)\;.
\label{eq:NNLOmu}
\end{eqnarray}
The program {\tt EERAD3} computes the perturbative coefficients
$A$, $B$ and $C$ defined in eq.~(\ref{eq:NNLOsigma0}), where the renormalisation scale
has been fixed to the centre-of-mass energy.
In order to be able to perform variations of the renormalization scale,
it is therefore sufficient to store the coefficients $A$, $B$ and $C$
in the Ntuples and apply Eq.~(\ref{eq:NNLOmu}) at the level of the Ntuple analysis.

\subsubsection{Details about the {\tt EERAD3} Ntuples}

The input file {\tt eerad3.input} contains the field {\tt nshot3\,  nshot4\,  nshot5}.
They denote the numbers of sampling points per iteration to be used by
the Monte Carlo integrator Vegas for the 3-parton, 4-parton, 5-parton channels, respectively.
The field {\tt itmax1\, itmax2} contains the number of iterations
for the ``warm up run'' (grid construction, {\tt itmax1} iterations)
and the ``production run'' (event generation and integration, {\tt itmax2} iterations).
The total number of events in each channel is then roughly given by
{\tt nshot} $\times$ {\tt itmax2}, modulo events which do not pass the
cuts.

\begin{figure}
  \begin{center}
  \includegraphics[scale=0.5]{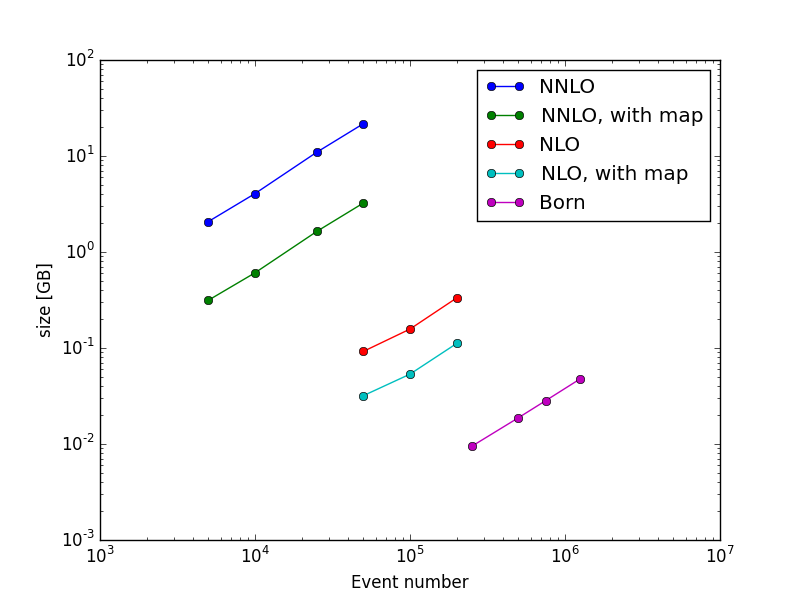}
  \end{center}
\caption{File size as a function of the number of events.}\label{fig:filesizes}
\end{figure}

For the implementation of this experimentation we used the same format as for the BlackHat+Sherpa NLO Ntuples~\cite{Bern:2013zja},
removing however the unnecessary information about the partonic initial states.
Figure~\ref{fig:filesizes} shows the size of the Ntuple file as a function of the number of events.
In Table~\ref{tab:size}, we list the size of the Ntuple files
for a million events. The Ntuples contain the NNLO
results for the sum of all colour factors, i.e. {\tt icol=0}.
Further, we used {\tt y0=1d-7, iaver=0, cutvar=1d-5}.
The event numbers given in Table~\ref{tab:size} are summed over all partonic channels.

\begin{table}
\begin{center}
\begin{tabular}{|l|c|c|}
\hline
run & size/Mevents [GB] & size/Mevents [GB] with mapping \\
\hline
LO  & 0.038 & 0.038  \\
NLO & 1.66& 0.56   \\
NNLO & 433.9 & 64.7\\
\hline
\end{tabular}
\end{center}
\caption{Sizes of the Ntuple files for various choices of the total number of events.\label{tab:size}}
\end{table}
The first column of Table~\ref{tab:size} shows the size of the Ntuple file if all phase-space information is recorded in the Ntuple.
An alternative strategy is to use the fact that there is a clear relationship between the kinematical information of a real phase-space point
(``real'' refers to the phase-space with the largest final state multiplicity)  and its subtractions.
Starting from one phase-space configuration for the real radiation one can get to all the kinematical configurations for the subtractions
using a fixed number of phase-space mappings. If we can enumerate these mappings and record the mapping label instead of the actual mapped momenta
we can save a significant amount of storage, at the cost of having to perform the mappings when the Ntuple file is analysed.
The second column of Table~\ref{tab:size} shows the size of an Ntuple file if the mapping label is recorded instead of the momenta for the subtractions.
However, apart from the increased CPU time to read an Ntuple file produced in this way there is another disadvantage:
the Ntuple is more tightly bound to the program that produced it, as the implementation of the mappings has to be provided along with the Ntuple file.

We should mention that the reconstruction of observables from the information contained in the
Ntuples is
only applicable to observables which are rotation-invariant. This is due to the fact that
the phase space in {\tt EERAD3} is constructed assuming rotational invariance.

\subsection{Summary and outlook}

We have produced Root Ntuple files based on the program {\tt EERAD3}, a partonic Monte Carl program  to calculate  event shapes and
jet rates in electron-positron annihilation up to NNLO.
We have found that the size of the Ntuple files needed to produce
reasonably smooth histograms is of the order of 2-3 TeraBytes.
Therefore it seems that Ntuples can be a convenient way to store NNLO results in hadronic collisions as well.

Apart from the obvious advantages of Ntuples to be able to perform
scale and PDF variations without having to run the full program,
there are a number of other aspects which make the use of Ntuples for
NNLO results additionally appealing:
\begin{itemize}
\item the results can be made available to experimentalists,
where they can apply different sets of cuts, PDFs, etc.
 without having to deal with the NNLO code
\item it is easy to change value of $\alpha_s(M_Z)$,
such that fits of  $\alpha_s$ could be readily performed.
\end{itemize}
The file sizes presented here are obtained using the EERAD3 phase-space generator that was designed to optimize the CPU time.
One can easily imagine that, if the focus was shifted to optimise the storage space, improvements for the file size could be achieved.

We are looking forward to the first NNLO results in hadronic collisions being available in the form of Ntuples.


\clearpage



\bibliographystyle{atlasnote}
\bibliography{LH15}

\begin{thebibliography}{100}
\bibitem{munich}
{{\sc Munich} is the abbreviation of ``MUlti-chaNnel Integrator at Swiss (CH)
  precision''---an automated parton level NLO generator by S.~Kallweit, in
  preparation}.
\end{thebibliography}



\end{document}